# Interpretable and Frugal Learning Systems Employing Multiresolution Pyramids and Volterra Kernels

Submitted in partial fulfillment of the requirements of the degree of

Ph.D.

by

**Kishore Kumar Tarafdar**
**(Roll No. 204070017, PMRF ID. 1301592)**

Advisor:
**Prof. Vikram M. Gadre**

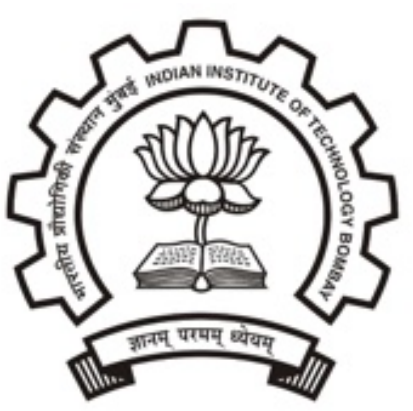

DEPARTMENT OF ELECTRICAL ENGINEERING
INDIAN INSTITUTE OF TECHNOLOGY BOMBAY
2026

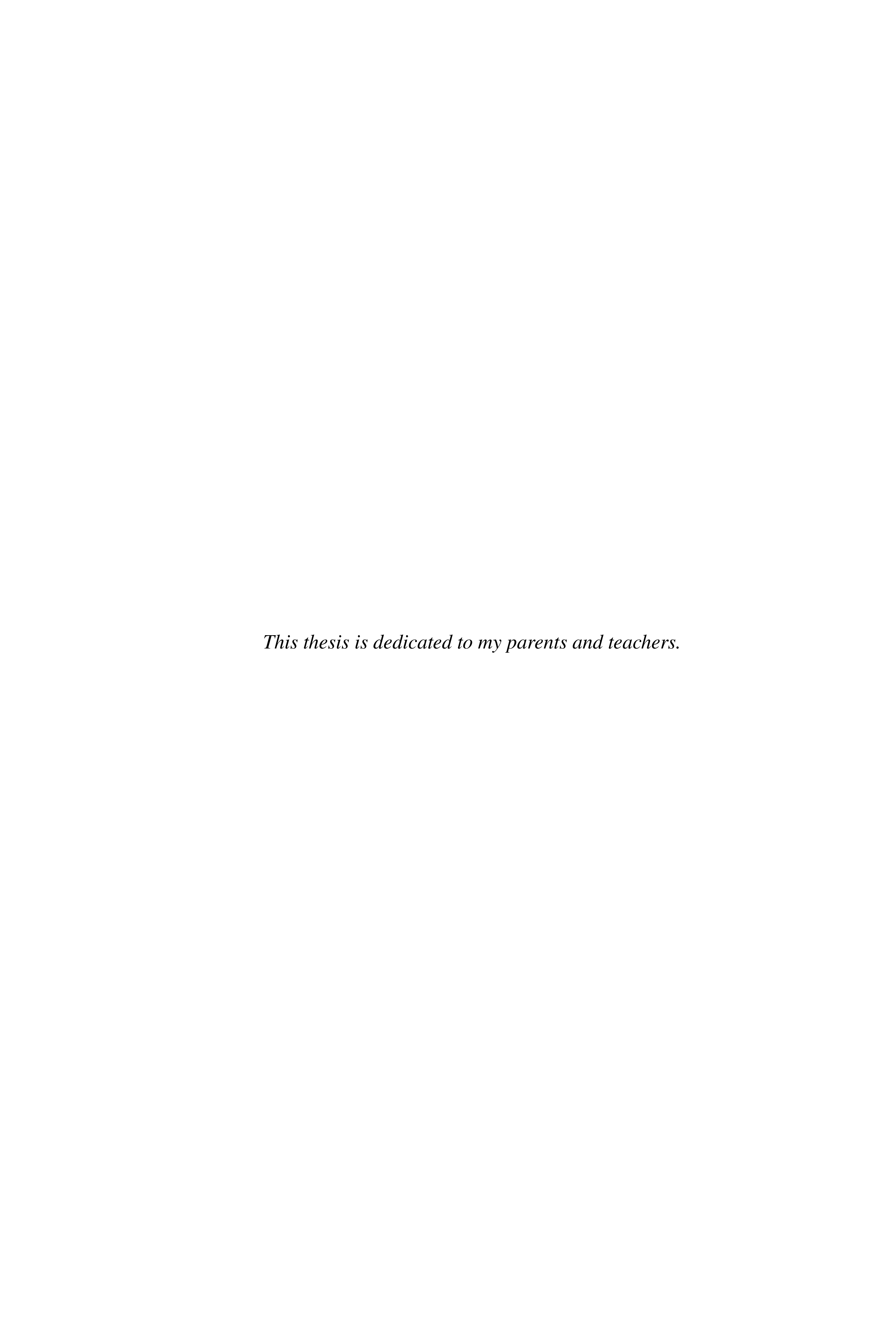

*This thesis is dedicated to my parents and teachers.*

## Approval Sheet

This thesis entitled **Interpretable and Frugal Learning Systems Employing Multiresolution Pyramids and Volterra Kernels** by **Mr. Kishore Kumar Tarafdar** is approved for the degree of **Doctor of Philosophy (PhD)**.

---

**External Examiner**
Dr. M.K. Bhuyan
Professor, Department of
Electronics and Electrical
Engineering, IIT Guwahati

---

**Internal Examiner**
Prof. Rajbabu Velmurugan,
Electrical Engineering

---

**Supervisor**
Prof. Vikram Manohar Gadre,
Electrical Engineering

---

**Chairperson**
Prof. Alankar Alankar,
Mechanical Engineering

Date : 11-06-2026

Place : Mumbai

# Declaration

I, Kishore Kumar Tarafdar, declare that this written submission represents my ideas in my own words and where others' ideas or words have been included, I have adequately cited and referenced the original sources. I also declare that I have adhered to all principles of academic honesty and integrity and have not misrepresented or fabricated or falsified any idea/data/fact/source in my submission. I understand that any violation of the above will be cause for disciplinary action by the Institute and can also evoke penal action from the sources which have thus not been properly cited or from whom proper permission has not been taken when needed.

(Signature of the candidate)
Kishore Kumar Tarafdar,
Roll No. 204070017, PMRF ID. 1301592

# Abstract

Deep learning models are widely used to process multidimensional signals such as time series, images, and volumetric medical images, but their learned representations often lack explicit signal structure and are difficult to inspect. This thesis develops model-based, signal-theoretic learning systems guided by data and task objectives, combining multiresolution analysis, wavelets and filterbanks, multirate and frequency-domain representations, nonlinear Volterra systems, and neural computation graphs. Scale, directional geometry, memory, and nonlinear input-output interactions are represented as differentiable operator modules trainable by backpropagation. This design keeps key intermediate variables tied to kernels, subbands, recursions, and transform-domain coefficients rather than only to opaque feature channels. The thesis formulates fast GPU-compatible realizations of $D$-dimensional convolution layers (`convd`), especially for $D \geq 4$, multirate sampling layers, higher-order Volterra-kernel layers in natural and biorthogonal wavelet coefficient domains (`VolterraSys`), and rational polynomial linear-nonlinear cascade heads. It further develops stability-constrained multidimensional infinite impulse response filters (`IIRD`), perfect-reconstruction Fast Discrete Wavelet Transform (`FDWT`), and Fast Digital Shearlet Transform (`FDST`) filterbank layers with learnable gains for end-to-end architectures. These operator modules are then composed into task-specific neural architectures for inverse modeling, classification, and segmentation across atmospheric, audio, texture, and medical-imaging problems. For atmospheric microwave radiometric inversion, learnable Volterra kernels represented in wavelet bases are used in the neural architecture `InVeRt M/N` to retrieve vertical atmospheric temperature and humidity profiles from microwave brightness temperature observations. Efficient classifiers are constructed using multiresolution filter-bank encoders and Volterra kernel heads. The thesis also introduces two multiresolution subband vision transformers, `WaveletViT` and `ShearViT`, built with wavelet and learnable shearlet filter banks. These transformers serve as building blocks of `WaveNETR` and `ShearNETR`, direction-sensitive encoder-decoder segmenters for image and volumetric MRI segmentation. Trained 3D T1-weighted brain MRI segmenter checkpoints are deployed in `MRILong`, an inference pipeline for automatic segmentation and longitudinal analysis of brain MRI volumes of ischemic stroke patients undergoing intervention.

In the microwave radiometric inverse problem, the `InVeRt 3/3` model with the bior1.3 wavelet achieves $R^2 \approx 0.99$ on 28,052 unseen all-weather test profiles, using 384 parameters for water vapor density with root mean square error (RMSE) 0.61, g, m$^{-3}$, and 528 parameters for temperature with RMSE 1.15, K. In classification, the spectrogram filter-bank encoder im-proves the ESC-20 audio baseline from 22% to 82% mean accuracy, and Volterra heads provide comparable audio and texture classification with fewer parameters than dense MLP heads. In volumetric MRI segmentation, `ShearNETR3D` reports intersection over union (IoU) scores of 0.98 for skull stripping, 0.86 for grey matter, 0.90 for white matter, 0.71 for cerebrospinal fluid, and 0.60 for lesion segmentation using 4.52 million parameters. `MRILong` is validated on unlabeled MRI volumes of ischemic stroke patients undergoing Ayurvedic and Yogic interventions, using the trained segmenter checkpoints for automatic tissue annotation. The combined results show that multiresolution filter banks, Volterra kernels, rational nonlinearities, IIR recursions, and subband attention modules embedded in trainable neural graphs support microwave radiometric inversion, classification, and segmentation while retaining compact and inspectable coefficient-level representations. Across the evaluated datasets, the resulting intermediate multiresolution feature spaces enable economical model designs in diverse application settings. Several computational packages developed alongside this thesis are organized as reusable Python deep-learning libraries for TensorFlow and PyTorch. Selected packages are already publicly available, and the remaining packages are planned for release through PyPI and public source GitHub repositories.

# Contents

# List of Tables

# List of Figures

# Notations, Symbols and Abbreviations

The following notation is used throughout this thesis and local notation paras are included in the individual chapters where additional clarity is required with chapter–specific symbols. For example, the symbol $z$ is the z-transform variable throughout but with an exception in Chapter 3, where it represents altitude.

|  |  |
|---|---|
| $\mathbb{R}, \mathbb{C}, \mathbb{N}, \mathbb{Z}$ | Real numbers, complex numbers, positive integers, and integers. |
| $\mathrm{L}^p(\mathbb{R}^D)$ | Lebesgue space of $p$–integrable functions on $\mathbb{R}^D$. |
| $\ell^2(\mathbb{Z}^D)$ | Square–summable sequences on the $D$–dimensional integer lattice. |
| $\lVert\cdot\rVert_2, \lVert\cdot\rVert_1, \lVert\cdot\rVert_\infty$ | Euclidean, $\ell_1$, and maximum norms. |
| $\lVert\cdot\rVert_\mathrm{F}$ | Frobenius norm for matrices or tensors. |
| $\mathcal{H}$ | Generic Hilbert space (for example $\mathrm{L}^2(\mathbb{R}^D)$). |
| $\Omega$ | Discrete spatial or frequency grid, with the interpretation specified locally. |
| $\widetilde{\Omega}$ | Padded spatial grid used by WaveletViT before cropping back to $\Omega$. |
| $\mathcal{X}, \mathcal{Y}$ | Input space and label space in classification or segmentation settings. |
| $\Delta_{K-1}$ | Probability simplex of dimension $K-1$ for $K$ classes. |
| $f, f(x), f[\boldsymbol{n}]$ | Signal or image viewed as a function or as discrete samples indexed by $\boldsymbol{n}$. |
| $\hat{f}(\boldsymbol{\omega}), \widehat{x}[\boldsymbol{\omega}]$ | Fourier transform of $f$ or $x$ at frequency vector $\boldsymbol{\omega}$. |
| $x, y$ | Input and output signals or fields (classification or segmentation context explained locally). |
| $\boldsymbol{n}, \boldsymbol{k}, \boldsymbol{m} \in \mathbb{Z}^D$ | Multi–indices for samples on a $D$–dimensional grid. |
| $i, j, k, \ell, m$ | Integer indices for scale, band, orientation, rotation, spatial shift, or other local roles. |
| $D$ | Spatial or spatio–temporal dimension. |
| $B$ | Batch size in learning-model tensor notation. |
| $N_i, \widetilde{N}_i, N_{\text{core}}$ | Spatial side length, padded side length, and square or cubic core side length. |
| $H, W$ | Image height and width (number of pixels). |
| $C_{\text{in}}, C_{\text{out}}$ | Number of input and output channels. |
| $*$ | Convolution in one or more dimensions (circular or linear as specified). |
| $\star$ | Cross–correlation. |
| $\langle f, g\rangle$ | $L^2$ inner product or discrete sum of products. |
| $\mathcal{F}, \mathcal{F}^{-1}$ | Fourier transform and inverse Fourier transform. |
| $\text{Range}(\cdot)$ | Range/column space of a linear operator or matrix. |
| $\delta[\boldsymbol{n}]$ | Kronecker impulse at the origin of the discrete grid. |
| $\psi, \psi_{j,k}, \psi_{j,d,k}$ | Wavelet or bandpass atom, possibly indexed by scale, orientation, and spatial shift. |
| $\phi, \phi_J$ | Scaling function and coarse–scale lowpass component. |
| $h, g$ | One dimensional analysis and synthesis filters in wavelet filter banks. |
| LL, LH, HL, HH | Separable two dimensional wavelet subbands (low–low and three detail bands). |
| $\text{LP}_j, \text{HP}_j^{(g)}$ | Lowpass and highpass DWT coefficient bands at level $j$ and HP group $g$. |
| $W_j, W_j^{-1}$ | Separable DWT analysis operator and matched IDWT synthesis operator at level $j$. |
| $G = 2^D - 1$ | Number of highpass orientation groups per separable $D$-dimensional DWT level. |
| $u_j^{(g)}, q_L$ | Compact HP token grid and deepest LP token grid in WaveletViT. |
| $\mathcal{F}_{j,r}^{\text{tok}}, P_{\text{LP}}$ | HP tokenization operator and pointwise LP projection in WaveletViT. |
| $\mathcal{A}_{j,\text{HP}}, \mathcal{A}_{\text{axial}}^{\text{LP}}, \mathcal{A}_{\text{global}}$ | HP axial, LP axial, and deepest-LP global attention operators in WaveletViT. |
| $\alpha_{\text{HP}}, \alpha_{\text{LP}}, \beta_{\text{HP}}, \beta_{\text{LP}}$ | Scalar gates for WaveletViT attention and feed-forward token updates. |
| $\rho, \eta_j, s$ | Level-decay factor, level weight, and residual strength in wavelet attention or WaveletViT. |
| $\Theta_j^{\text{HP}}, \Theta_L^{\text{LP}}, \gamma_j, \gamma_L^{\text{LP}}$ | WaveletViT gain heads and bounded HP or LP gain fields. |
| $R_j(\lvert\boldsymbol{\omega}\rvert)$ | Radial window at scale $j$ in steerable or related frequency tilings. |
| $A_d(\theta)$ | Angular window centered at orientation index $d$. |
| $W_j^{\text{curv}}, V_{j,d}^{\text{curv}}$ | Curvelet radial and angular window functions when curvelet notation is used locally. |
| $M_{j,k}(\omega), M_{\text{low}}(\omega)$ | Wave atom band masks and lowpass mask in the frequency plane. |
| $H_k[\omega], G_k[\omega]$ | Shearlet analysis and synthesis windows in the frequency domain. |
| $W_k[\omega], \mathcal{I}_k$ | FDST band window and sparse support of band $k$ on the discrete frequency grid. |
| $D_0[\omega], D[\omega]$ | Pre-gain partition-of-unity profile and dual synthesis normalizer in FDST. |
| $A_k[\omega], \phi_k[\omega], g_k[\omega], \beta_k[\omega]$ | Amplitude, phase, complex gain, and effective rotation-weighted gain in learnable FDST. |
| $p_k[\omega], q_k[\omega], z_m^{\text{fdst}}, \alpha_m$ | Real gain parameters, rotation logit, and rotation softmax weight in learnable FDST. |
| $P_{\text{low}}, P_{\text{high}}$ | Nonsubsampled pyramid lowpass and highpass filters. |
| $D_d$ | Directional fan filter at orientation index $d$ (NSDFB and NSCT). |
| $G_j, B_j, D_{j,d}$ | Coarse, bandpass, and oriented subbands at scale $j$ in the contourlet transform. |
| $A_{2^j}, S_s$ | Shearlet anisotropic scaling matrix and shear matrix for slope $s$. |
| $R_i, R_{i,j}$ | Riesz transform component $i$ at scale $j$ in the monogenic pyramid. |
| $A_j, \phi_j, \theta_j$ | Monogenic amplitude, phase related quantity, and local orientation at scale $j$. |
| $\psi_a$ | Approximate analytic wavelet in the dual tree complex wavelet transform. |
| $W_{j,o}$ | Complex DTCWT coefficient at scale $j$ and orientation $o$. |
| $\xi_{j,k}$ | Central frequency associated with a bandpass atom (for example a wave atom). |
| $L_j$ | Number of directional subbands or wedges at scale $j$. |
| $s, S$ | Slope or shear index and the set of all slopes used in directional schemes. |
| $\theta, \theta_B$ | Orientation angle and dominant orientation in block or patch $B$. |
| $m_i$ | Binary mask that selects patches or pixels assigned to orientation index $i$. |
| $T_k, T_k^{\text{out}}$ | ShearViT FDST tile before and after band scaling. |
| $t_k, \mu_k, u_k, \widehat{u}_k, U$ | ShearViT band descriptor, token, attended token, and stacked token matrix. |
| $j(k), \nu(k), \ell(k), r(k), \boldsymbol{\xi}_k$ | Scale, cone-slot, in-cone orientation, rotation id, and central frequency metadata for band $k$. |
| $\phi_{\text{meta}}, \psi_k$ | Metadata embedding map and metadata embedding vector in ShearViT. |
| $\eta_k, \lambda_g, \delta_k$ | ShearViT per-band gain, gain-range hyperparameter, and static band gate. |
| $\gamma_\nu, \tilde{\alpha}_m$ | ShearViT cone-slot gate and rotation gate. |
| $s^{(0)}, s^{(1)}$ | Even and odd components in a polyphase or lifting splitting of a one dimensional signal. |
| $P_s, U_s$ | Prediction and update operators for a lifting scheme along slope $s$. |
| $H(z)$ | Two channel polyphase analysis matrix as a function of complex frequency $z$. |
| $P_k(z), U_k(z)$ | Lifting prediction and update polynomials in the factorization of $H(z)$. |
| $K$ | Scaling factor in lifting or a generic count, including atoms, classes, or transform bands. |
| $\downarrow_2, \uparrow_2$ | Dyadic decimation and dyadic zero insertion along a discrete axis. |
| $d_k, z_k$ | Convolutional dictionary filter and associated sparse feature map in CSC. |
| $d_k^{\Lsh}$ | Filter $d_k$ rotated by 180° (correlation equivalent). |
| $r^{(t)}$ | Residual at ISTA iteration $t$ in convolutional sparse coding. |
| $\mathcal{S}_\tau(x)$ | Soft–threshold function $\operatorname{sign}(x)\max(\lvert x\rvert - \tau, 0)$. |
| $U_1 f, U_2 f$ | Intermediate scattering feature maps of order one and two. |
| $S_0 f, S_1 f, S_2 f$ | Scattering coefficients of order zero, one, and two. |
| $j_1 < j_2$ | Scale ordering constraint for second order scattering paths. |
| $x \in \mathcal{X}$ | Input sample (signal, image, or volume) in the learning or inference setting. |
| $y \in \mathcal{Y}$ | Label or structured output associated with $x$. |
| $p_\theta(y \mid x)$ | Parametric conditional model with parameters $\theta$ (for example class posterior). |
| $f_\theta(x)$ | Decision rule or predictor obtained from $p_\theta$ or from a discriminative model. |
| $\theta$ | Collection of trainable parameters in a model. |
| $\mathcal{L}$ | Overall loss or objective function used for training. |
| $\mathcal{E}(f)$ | Population risk (expected loss) of a decision rule $f$. |
| $\mathcal{R}(\theta)$ | Regularization functional applied to parameters $\theta$. |
| $\lambda$ | Regularization or penalty weight (for example in sparse coding or regularized training). |
| $\alpha$ | Step size or Lipschitz related parameter in gradient based iterations. |
| $h_m$ | $m$th order Volterra kernel in Volterra system models. |
| $m$ | Volterra order index or generic model order, interpreted from context. |
| $O(\cdot)$ | Big–O asymptotic order notation. |
| $\mathbb{E}[\cdot]$ | Expectation with respect to the relevant probability distribution. |

# Summary of abbreviations

| | | |
|---|---|---|
| 1D-Var | : | One-dimensional variational retrieval |
| ADNI | : | Alzheimer’s Disease Neuroimaging Initiative |
| AI | : | Artificial Intelligence |
| AProcS | : | Analysis–Processing–Synthesis |
| ATLAS | : | Anatomical Tracing of Lesions After Stroke |
| BraTS | : | Brain Tumor Segmentation |
| CNN | : | Convolutional Neural Network |
| CPU | : | Central Processing Unit |
| CRF | : | Conditional Random Field |
| CSF | : | Cerebrospinal Fluid |
| CT | : | Computed Tomography |
| CVAE | : | Convolutional Variational Autoencoder |
| CWT | : | Continuous Wavelet Transform |
| DL | : | Deep Learning |
| DFT | : | Discrete Fourier Transform |
| DTCWT | : | Dual-Tree Complex Wavelet Transform |
| DWT | : | Discrete Wavelet Transform |
| ECA | : | Efficient Channel Attention |
| ELBO | : | Evidence Lower Bound |
| FDST | : | Fast Digital Shearlet Transform |
| FFN | : | Feed-Forward Network |
| FFT | : | Fast Fourier Transform |
| FIR | : | Finite Impulse Response |
| FISTA | : | Fast Iterative Shrinkage–Thresholding Algorithm |
| GAP | : | Global Average Pooling |
| GAN | : | Generative Adversarial Network |
| GM | : | Grey Matter |
| GMAO | : | Global Modeling and Assimilation Office |
| GPU | : | Graphics Processing Unit |
| HCP | : | Human Connectome Project |
| IBSR | : | Internet Brain Segmentation Repository |
| IDWT | : | Inverse Discrete Wavelet Transform |
| IFFT | : | Inverse Fast Fourier Transform |
| IKS | : | Indian Knowledge System |
| IIR | : | Infinite Impulse Response |
| IoU | : | Intersection-over-Union |
| I/O | : | Input–Output |
| ISTA | : | Iterative Shrinkage–Thresholding Algorithm |
| InVeRt | : | **In**verse model with **V**olterra k**e**rnel and **R**ational func**t**ions |
| KL | : | Kullback–Leibler divergence |
| LP/HP | : | Low-Pass and High-Pass |
| LSI | : | Linear Shift Invariant |
| MAE | : | Mean Absolute Error |
| MAP | : | Maximum a Posteriori |
| MBD | : | Mean Bias Deviation |
| MEDNet | : | Multiresolution Encoder–Decoder Convolutional Neural Network |
| MEV | : | Multiresolution Encoder–Volterra classifier |
| MERRA-2 | : | Modern-Era Retrospective Analysis for Research and Applications, Version 2 |
| MHSA | : | Multi-Head Self-Attention |
| ML | : | Machine Learning |
| MLP | : | Multilayer Perceptron |
| MRA | : | Multiresolution Analysis |
| MRI | : | Magnetic Resonance Imaging |
| MRILong | : | Longitudinal MRI analysis pipeline developed in this thesis |
| MSE | : | Mean Squared Error |
| MWR | : | Microwave Radiometer |
| NLSI | : | Nonlinear Shift Invariant |
| NSCT | : | Nonsubsampled Contourlet Transform |
| NSDFB | : | Nonsubsampled Directional Filter Bank |
| NSDL | : | Nonsubsampled Directional Lifting |
| NSMAPE | : | Normalized Symmetric Mean Absolute Percentage Error |
| NWP | : | Numerical Weather Prediction |
| OEM | : | Optimal Estimation Method |
| PR | : | Perfect Reconstruction |
| PoU | : | Partition of Unity |
| QMF | : | Quadrature Mirror Filter |
| QKV | : | Query-Key-Value |
| QSI | : | Quadratic Shift Invariant |
| $R^2$ | : | Coefficient of determination |
| Re/Im | : | Real and Imaginary |
| RF | : | Random Forest |
| RH | : | Relative Humidity |
| RMSE | : | Root Mean Squared Error |
| RoI | : | Region of Interest |
| RTTOV | : | Radiative Transfer for TIROS Operational Vertical sounder |
| SAM | : | Segment Anything Model |
| SDS | : | Sum of Dilated Spectra |
| ShearNETR | : | ShearViT encoder-decoder segmenter |
| ShearViT | : | Shearlet subband Vision Transformer |
| SwiGLU | : | Swish-Gated Linear Unit |
| STFT | : | Short-Time Fourier Transform |
| U-Net | : | U-shaped Encoder–Decoder CNN |
| VAE | : | Variational Autoencoder |
| ViT | : | Vision Transformer |
| WaveNETR | : | WaveletViT encoder-decoder segmenter |
| WaveletViT | : | Wavelet subband Vision Transformer |
| WM | : | White Matter |

# Chapter 1

# Introduction and Background

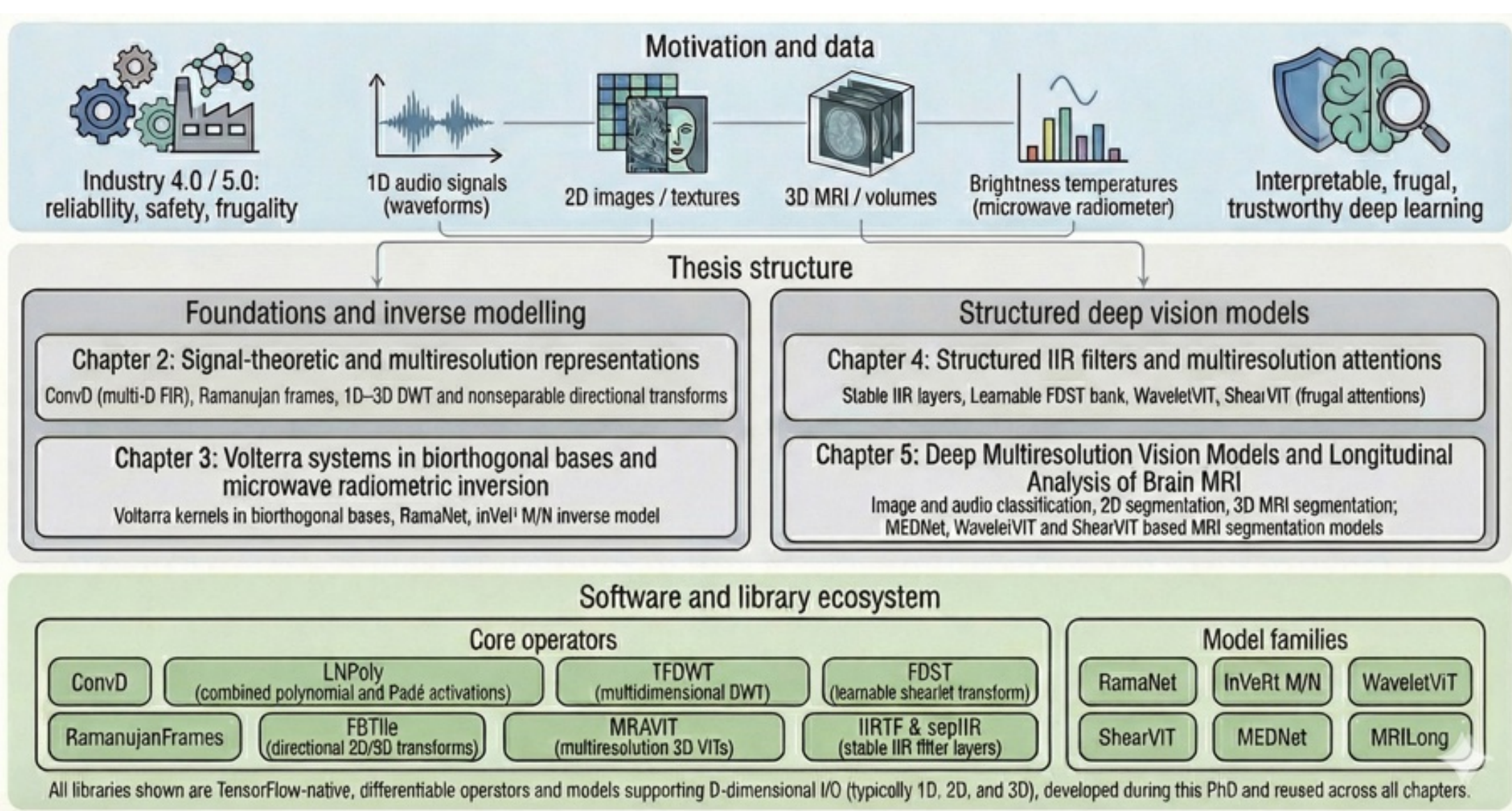

The Fourth Industrial Revolution (Industry 4.0) envisioned a digital transformation for a smart, connected, and automated industrial ecosystem fueled by data and machine learning. Here, the key enabling technologies are robots, Internet of Things, cyber-physical systems, big data, and Artificial Intelligence (AI). The technologically advanced nations in the world are largely operating in the 4.0 industrial phase at present, and some very advanced industries are transitioning to the Fifth Industrial Revolution (Industry 5.0). Interestingly, this transition to Industry 5.0 complements and extends 4.0, leveraging the same technologies for human-machine synergy, green manufacturing, and resilient industrial automation. This vision promotes sustainable industrial value creation by placing human and societal needs at the center of industrial innovation. India is largely transitioning from Industry 3.0 to 4.0, with a few leading sectors operating on 4.0 and automation with a human in the loop (i.e., Industry 5.0) is emerging as a vision of the near future. The present development of powerful parallel processing electronic hardware continuously boosts the research and development of intelligent machines of varying capacities all around us. These capacities can range from a simple automated on-off transistor gate that prevents a rooftop water tank from overflowing to solving highly nonlinear problems with algorithmic deep inference models that understand human emotions, forecast weather patterns, or unfold protein structures for drug discovery. The theme of this thesis is to unify machine learning and deep learning with multiresolution signal processing and nonlinear systems theory primitives to design interpretable [1, 2], frugal [3, 4] data-driven intelligent inference models. Now, let us establish the definition and scope of intelligence here.

**Definition 1.1** (Intelligence in AI)**.** *Intelligence is the capacity of an agent (model), natural or artificial, to achieve goals across a range of environments by acquiring, representing, and using knowledge and skills. An intelligent agent learns from data, performs inference and planning, and selects actions that improve expected performance under uncertainty, while operating with finite data, time, and computational resources.*

The above definition is adopted for AI and is limited from the perspective of natural intelligence that arises in living beings from the mind, memory, knowledge, and experience. Intelligence in AI is viewed as a model's ability to achieve goals in many possible situations. An agent or model that turns an input signal into informed actions using its internal linear and/or nonlinear systems after learning by minimizing a loss is considered intelligent to the extent that it can ensure success across a wide variety of environments, with a principled preference for simpler environments (an Occam-style weighting) [5–7]. This view is substrate-neutral and highlights adaptability, prediction, planning, and control, while acknowledging practical limits and the dependence on how objectives are defined and constraints from the real world on data, time, and computation. Thus, it provides a precise yardstick for comparison and measurement, focusing on the outcome without claiming to resolve the larger philosophical question of what intelligence "is."

Natural intelligence in living beings is much more than just achieving goals, and its formal connection to AI used in science and engineering has not yet been discovered. Before we dive into the main technical content starting from section 1.3.2.3 onward, let us have a quick look at how intelligence is viewed in the non-engineering fields dealing with biological beings (out of the scope of this thesis). Psychology defines human or animal intelligence as the general mental ability that involves learning, reasoning, and adaptation, which is the basis of psychometrics. This mental ability arises from the mind, which is a mystery in modern science. Interestingly, the schools of thought in Indian Knowledge System describe perception, mind, memory, knowledge, intelligence, and their sources with a rich Sanskrit vocabulary.

## 1.1 Mind and intelligence in the Indian Knowledge System (out of scope):

Various schools of thought in the *Bhāratiya* civilization formally describe mind, knowledge, and intelligence and their sources. In this thesis, these ideas are not technically used, but are worth acknowledging as part of a long history of reflection on cognition and agency in the Indian Knowledge System (IKS). In *Advaita Vedānta*, a central starting point is the distinction between pure consciousness and its appearance as a world of bodies and minds. *Ātman* or *Brahman* is described as pure, self-luminous consciousness and as the witness (*sākṣin*) of changing mental states. Consciousness is not an object, but it is the light in whose presence objects, mental states, and acts of knowing manifest. The apparent manifold arises through *Māyā*, a power of appearance characterised by three *guṇa*: *sattva* (clarity and harmony), *rajas* (activity and propulsion), and *tamas* (inertia and opacity). When Brahman is considered together with this ordered potency, it is referred to as *Īśvara*, the principle that sustains regularity so that experience is law-like rather than chaotic. In cosmological accounts shared across *Sāṅkhya* and *Vedānta*, the basic contents of possible experience are introduced as five *tanmātra* or subtle qualities: *śabda* (sound), *sparśa* (touch), *rūpa* (form or light), *rasa* (taste or fluidity), and *gandha* (smell). Each subtle quality is associated with a corresponding subtle element. The quality *śabda* is linked with *ākāśa* (space), *sparśa* with *vāyu* (air), *rūpa* with *tejas* (fire), *rasa* with *āpas* (water), and *gandha* with *pṛthvī* (earth). At this stage, these are not gross, measurable substances. They are proto-qualitative seeds that will later be taken up by the cognitive apparatus. Through *pañcīkaraṇa*, or five-fold combination, the subtle elements are said to combine into *mahābhūta* or gross elements. A standard Advaita description states that each gross element contains one-half of its own subtle element and one-eighth of each of the other four. This structure is used to explain why tangible objects exhibit multiple sensible qualities in graded measure. At this level, one obtains a shared external world made of bodies, instruments, and environments in which cognition and action take place. On the cognitive side, Vedānta introductions often describe an *antaḥkaraṇa* or inner instrument that arises primarily from the *sāttvika* aspect of the subtle elements. The inner instrument is typically analysed into four functional components. *Manas* is the mind in the sense of an attentive coordinator that samples inputs, switches focus, and composes preliminary cognitions. It is the locus where doubt and options arise. *Buddhi* is the determinative intellect that discriminates, evaluates, and issues *niścaya* or settled judgment. Figure 1.1 provides a compact schematic of this description.

In this tradition, **intelligence** corresponds most directly to this faculty of *buddhi*. *Ahaṃkāra*, sometimes translated as the I-maker or self-appropriation, attributes ownership and agency through thoughts such as “I know” and “I act”. *Citta* is the memory store that retains *saṃskāra* or impressions and supports *smṛti* or recollection. Through these stored traces, it gives continuity to learning and behaviour. The inner instrument is interfaced with the world through ten *indriya*. The *jñānendriya* are five sense capacities, namely hearing, touch, sight, taste, and smell, while the *karmendriya* are five action capacities, namely speech, grasp, locomotion, excretion, and reproduction. Their operation is energised and coordinated by five *prāṇa* or vital functions, named *prāṇa*, *apāna*, *vyāna*, *udāna*, and *samāna*. Together, the body, the *indriya*, the *antaḥkaraṇa*, and the *prāṇa* form a psychophysical apparatus through which experience, decision, and action occur in these classical accounts.

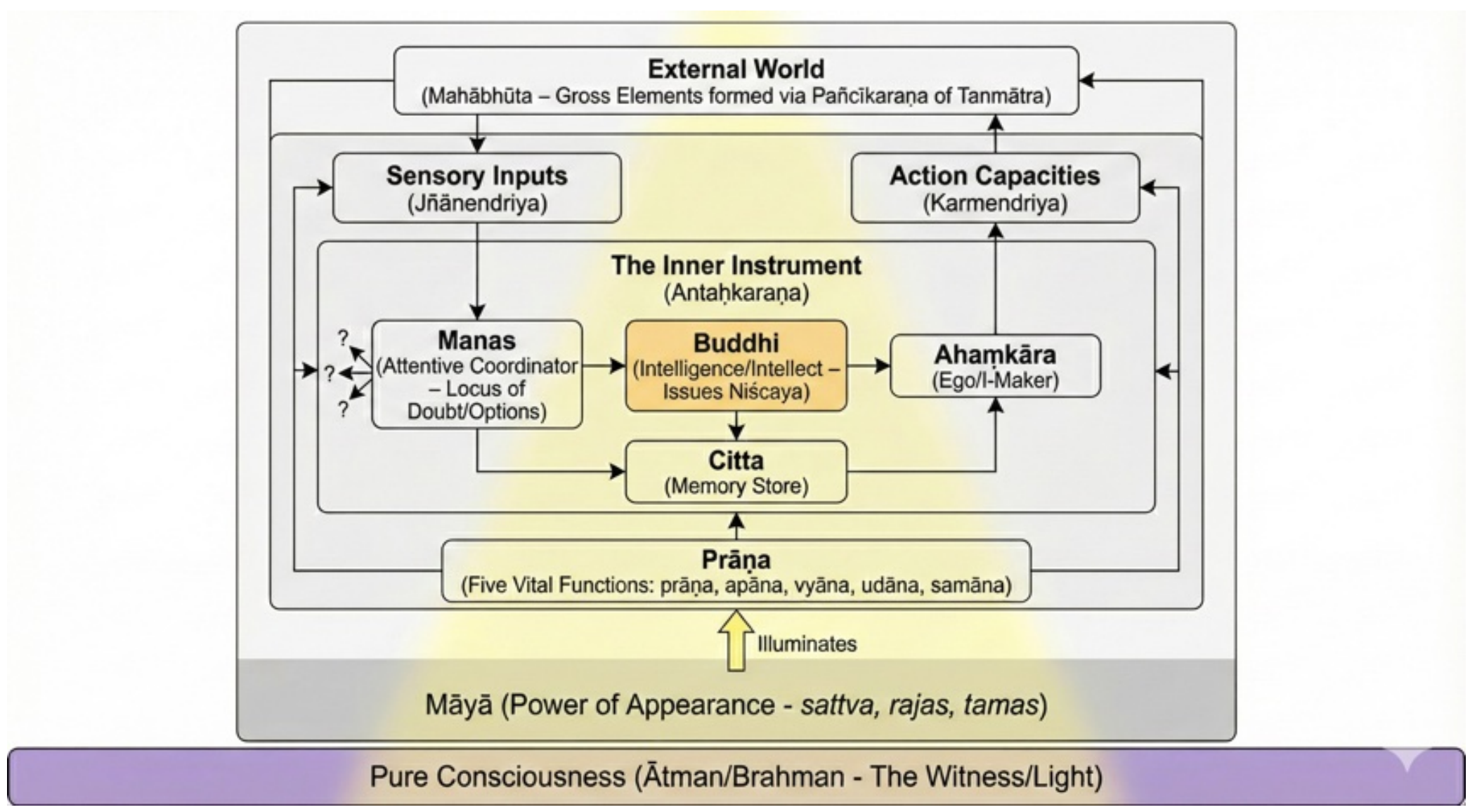

Figure 1.1: Architecture of cognition described in IKS

**Knowledge** is distinguished from mere appearance by *pramāṇa*, usually translated as a reliable means of cognition. Most classical systems recognise at least three *pramāṇa*: *pratyakṣa* or perception, *anumāna* or inference, and *śabda* or reliable testimony. Several systems also admit further *pramāṇa* such as *upamāna* or comparison, *arthāpatti* or postulation, to best explain a datum, and *anupalabdhi* or knowledge of absence and non-cognition. In Advaita Vedānta, a particular piece of knowledge, or *jñāna*, is described as a *buddhi-vṛtti*, that is, a specific modification of the determinative intellect. This modification is said to be illuminated by consciousness and supported by some appropriate *pramāṇa*. A cognition that is thus supported is treated as valid until *bādha*, or sublation, occurs, when a better grounded cognition defeats the earlier deliverance. Classical Indian epistemology also analyses **error** and **update** with considerable care. From the psychological angle, one may describe error as arising from mis-selection in *manas*, from bias introduced by the stored traces of *citta*, from over-appropriation by *ahaṃkāra*, or from misapplication of a *pramāṇa* in *buddhi* or **intelligence**. Figure 1.2 sketches a high-level view of knowledge update in this setting.

More formally, different schools propose different *khyāti-vāda* or theories of erroneous appearance. *Nyāya* develops *anyathā-khyāti*, where error occurs when a real object or property remembered from elsewhere is mislocated into the present perceptual situation. *Prābhākara Mīmāṃsā* speaks of *akhyāti*, a failure to distinguish what is immediately perceived from what memory supplies. *Bhāṭṭa Mīmāṃsā* is often grouped under *anyathā-khyāti* or *viparīta-khyāti* and emphasises that the erroneous cognition is a positive but misdirected appearance that arises under defective conditions. Advaita Vedānta defends *anirvacanīya-khyāti*, according to which the illusory item is neither absolutely real nor absolutely unreal and is therefore called indefinable until a superior cognition cancels it. Buddhist authors propose further analyses. *Yogācāra* speaks of *ātma-khyāti*, which explains illusion as a projection or misrecognition of consciousness itself, while *Mādhyamika* develops *asat-khyāti*, which treats the illusory object as sheer non-existence. Across these accounts, **correction** is again understood as *bādha*, where a cognition backed by a stronger *pramāṇa* cancels or sublates the earlier appearance.

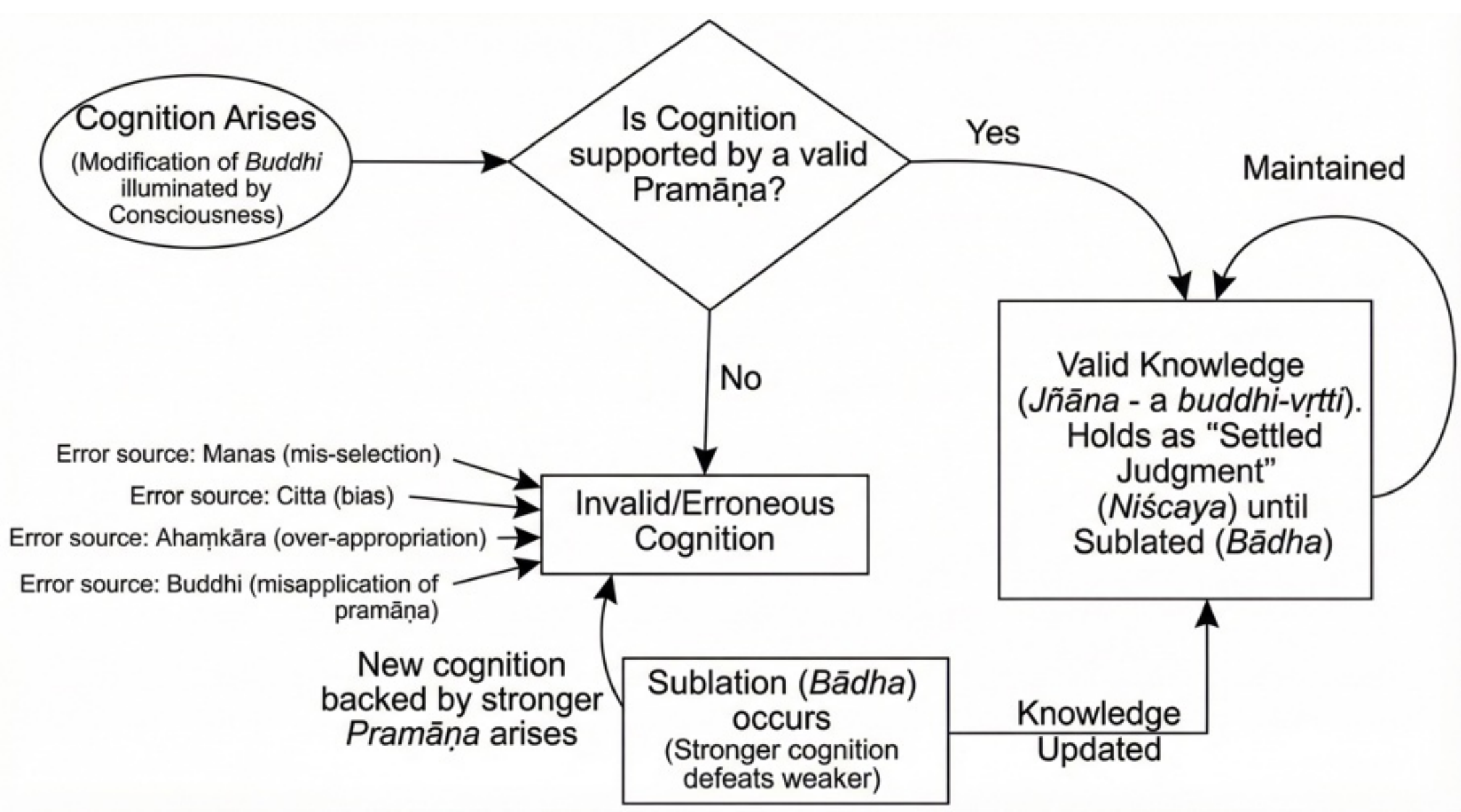

Figure 1.2: Flowchart or updating knowledge according to IKS

The *Bhagavad-Gītā* [8] presents a compact summary of the embodied situation in a verse that states that the indwelling Lord causes beings to wander "as if mounted on a *yantra*", often translated as a machine constituted of material nature (Bhagavad-Gītā 18.61). In this reading, the *yantra* can be understood as the coordinated apparatus of *indriya* or sense and action capacities, *antaḥkaraṇa* or inner instrument, and *prāṇa* or vital functions. The point is not that life is literally mechanical. Rather, the verse suggests that seeing, deciding, and acting occur within an ordered system whose parts have assignable roles and constraints. The *Bhāgavata Purāṇa* [9] offers more detailed expositions of this system. In the teachings attributed to the sage *Kapila* in Canto 3, the cognitive apparatus is analysed in a *Sāṅkhya*-like manner. The text enumerates *buddhi* or determinative intellect, *ahaṃkāra* or ego-sense, *manas* or mind, the five *tanmātra*, the ten *indriya*, the vital energies *prāṇa*, and the gross elements, and explains how their coordinated operation yields composite experience. In another passage, the allegory of *Purañjana* in Canto 4 describes the human body as a "city of nine gates". The city stands for the body, its gates for the sensory openings, the king for the embodied knower, and the queen is identified with *buddhi* or intelligence, while attendants represent *manas* and *prāṇa*. The story is used to clarify inputs, control, regulation, memory, agency, and the way attachment and error arise within that system.

The notion of **non-human intelligence** also appears in several strands of this literature as part of systematic accounts of mind and knowing. *Nyāya* develops a detailed theory of *pramāṇa* and presents an explicit five-membered form of inference, or *anumāna*. The five members are *pratijñā* or proposition, *hetu* or reason, *udāharaṇa* or *dṛṣṭānta*, which states a universal rule together with an example, *upanaya* or application of the rule to the case at hand, and *nigamana* or conclusion. Classic examples such as "The hill has fire because there is smoke. Wherever there is smoke, there is fire, as in a kitchen. This hill has smoke. Therefore, this hill has fire." are structurally similar to rule-based inference schemes in early symbolic Artificial Intelligence. *Sāṅkhya* distinguishes *puruṣa* or witness consciousness from *prakṛti* or the evolving cognitive and material apparatus, and thereby raises early questions about agency and selfhood. *Advaita Vedānta* analyses awareness and subjective experience in depth, which resonates with some contemporary discussions of consciousness in neuroscience and AI. Descriptions of constructed agents also appear in medieval Sanskrit and related technical texts. The architectural treatise *Samarāṅgaṇa-Sūtradhāra*, attributed to King Bhoja and dated to around the eleventh century, includes a chapter on *yantra* or mechanical devices. This chapter describes moving figures such as mechanical birds and human-shaped automata that pour oil into lamps or play instruments. A narrative preserved in the Pāli compilation *Lokapaññatti* recounts an army of automaton guardians, called *bhūta-vāhana-yantra* or "spirit movement machines", that protect a hidden stūpa containing the Buddha's relics. Later retellings describe them as intricate and sometimes water-driven mechanisms whose construction plans are stolen from a distant land. These provide a long-standing philosophical record of descriptions about the mind, knowledge, and both natural and man-made intelligent systems from a different era.

## 1.2 Basic definitions

**Notations** Scalars, discrete signal values, and scalar-valued functions are written in lightface italic. Examples include $a_0$, $x[\boldsymbol{n}]$, and $f(t)$. Bold italic symbols such as $\boldsymbol{n}$, $\boldsymbol{k}$, and $\boldsymbol{m}$ denote multi-indices or coordinate tuples in $D$ dimensions. Bold upright symbols such as $\mathbf{x}$, $\mathbf{y}$, and $\mathbf{W}$ denote finite-dimensional vectors, matrices, or linear operators that act on discrete signals. Calligraphic letters such as $\mathcal{X}$, $\mathcal{Y}$, and $\mathcal{H}$ denote sets or function spaces. Blackboard-bold symbols such as $\mathbb{R}$, $\mathbb{C}$, and $\mathbb{Z}$ denote standard number systems. We write norms as $\|\cdot\|$ and inner products as $\langle\cdot,\cdot\rangle$ and add subscripts when a specific space is important, for example $\|\cdot\|_{\ell^2}$ or $\langle\cdot,\cdot\rangle_{\mathrm{L}^2}$. Sums over multi-indices are written as $\sum_{\boldsymbol{n}\in\Omega_d}$ and are sometimes shortened to $\sum_{\boldsymbol{n}}$ when the index set is clear from context; when we need to make the nested structure explicit we write $\boldsymbol{\sum}_{\boldsymbol{n}}$ to denote $\sum_{n_1}\cdots\sum_{n_D}$. Transposes of matrices are written with the symbol $\top$ and conjugate transposes are written with the symbol $^*$.

**Signal or Data** A $D$-dimensional discrete signal (or data) is defined below.

**Definition 1.2** (Discrete $D$-dimensional signal)**.** *A discrete signal is a function*

$$x[\boldsymbol{n}] : \Omega_d \subseteq \mathbb{Z}^D \to \mathbb{R}^C \text{ or } \mathbb{C}^C, \tag{1.1}$$

*where $\boldsymbol{n} = (n_1, \ldots, n_D)$ is a multi-index of integer coordinates, $\Omega_d$ is a finite or countable index set (support) in the discrete lattice $\mathbb{Z}^D$, $D$ denotes the signal dimension, and $C$ denotes the number of channels.*

Typical cases in this thesis include

- $D = 1$: sequences or time series $x[n_1]$,
- $D = 2$: images $x[n_1, n_2]$,
- $D = 3$: volumetric data (e.g. 3D MRI) $x[n_1, n_2, n_3]$.

Digital data arrays correspond to signals whose domain is restricted to a finite, usually rectangular, grid such as

$$\Omega_d = \{0, \ldots, N_1 - 1\} \times \cdots \times \{0, \ldots, N_D - 1\}. \tag{1.2}$$

Intuitively, we view $x[\boldsymbol{n}]$ as a collection of samples (or pixels/voxels) indexed by integer coordinates on a regular grid.

**Linear and nonlinear systems** A general $D$-dimensional discrete system is defined below.

**Definition 1.3** (General discrete $D$-dimensional system)**.** *A* system *or* operator *is a mapping*

$$\mathcal{S} : \mathcal{X} \to \mathcal{Y}, \qquad y[\boldsymbol{n}] = (\mathcal{S}x)[\boldsymbol{n}], \tag{1.3}$$

*that assigns to each input signal $x[\boldsymbol{n}] \in \mathcal{X}$ an output signal $y[\boldsymbol{n}] \in \mathcal{Y}$.*

Here $\mathcal{X}$ and $\mathcal{Y}$ are sets of discrete $D$-dimensional signals (scalar or vector valued) defined on a common index set $\Omega_d \subseteq \mathbb{Z}^D$. Informally, the system specifies a rule that uses the input samples $\{x[\boldsymbol{m}]\}$ to produce output samples $y[\boldsymbol{n}]$. In many cases, the output at position $\boldsymbol{n}$ depends only on the input in a neighborhood $\mathcal{N}(\boldsymbol{n}) \subseteq \Omega_d$, which can be written as

$$y[\boldsymbol{n}] = F\big(\boldsymbol{n}, \{x[\boldsymbol{m}] : \boldsymbol{m} \in \mathcal{N}(\boldsymbol{n})\}\big), \tag{1.4}$$

for some (possibly nonlinear) function $F$.

**Definition 1.4** (Linear systems)**.** *A system $\mathcal{S}$ is called* linear *if it satisfies the superposition principle. For all input signals $x_1, x_2 \in \mathcal{X}$ and all scalars $\alpha, \beta$,*

$$\mathcal{S}(\alpha x_1 + \beta x_2) = \alpha\, \mathcal{S}x_1 + \beta\, \mathcal{S}x_2. \tag{1.5}$$

Equivalently, linearity can be stated as the combination of

$$\mathcal{S}(x_1 + x_2) = \mathcal{S}x_1 + \mathcal{S}x_2 \quad \text{and} \quad \mathcal{S}(\alpha x) = \alpha\, \mathcal{S}x. \tag{1.6}$$

**Discrete shifts and shift invariance:** For a discrete shift vector $\boldsymbol{k} \in \mathbb{Z}^D$, the shift operator $T_{\boldsymbol{k}}$ acts on a signal $x[\boldsymbol{n}]$ as

$$(T_{\boldsymbol{k}}x)[\boldsymbol{n}] \triangleq x[\boldsymbol{n} - \boldsymbol{k}]. \tag{1.7}$$

A system $\mathcal{S}$ is called *shift invariant* (or translation invariant) if shifting the input and then applying the system is equivalent to applying the system and then shifting the output, that is,

$$\mathcal{S}(T_{\boldsymbol{k}}x) = T_{\boldsymbol{k}}(\mathcal{S}x), \qquad \forall\, \boldsymbol{k} \in \mathbb{Z}^D,\ \forall\, x \in \mathcal{X}. \tag{1.8}$$

**Definition 1.5** (Linear shift-invariant system)**.** *A system that is both linear and shift invariant is called a linear shift-invariant (LSI) system. For discrete $D$-dimensional signals, the output of a LSI system is a discrete convolution:*

$$y[\boldsymbol{n}] = (\mathcal{S}x)[\boldsymbol{n}] = \sum_{\boldsymbol{k} \in \mathbb{Z}^D} h[\boldsymbol{k}]\, x[\boldsymbol{n} - \boldsymbol{k}], \tag{1.9}$$

*where $h[\boldsymbol{k}]$ is the impulse response of the system.*

Special cases include $D = 1$ (classical discrete-time LSI systems), $D = 2$ (2D convolutions for images), and $D = 3$ (3D convolutions for volumetric data).

**Nonlinear systems:** A system that does not satisfy the linearity condition above is called *nonlinear*. Formally, $\mathcal{S}$ is nonlinear if there exist signals $x_1, x_2$ and scalars $\alpha, \beta$ such that

$$\mathcal{S}(\alpha x_1 + \beta x_2) \neq \alpha\, \mathcal{S}x_1 + \beta\, \mathcal{S}x_2. \tag{1.10}$$

In practice, many systems of interest in this thesis are constructed by combining linear convolutional operators with pointwise or local nonlinearities, such as activation functions in deep neural networks or multiplicative interactions between channels or features.

**Model** A *model* is a parameterised system that maps input signals to outputs.

**Definition 1.6** (Model as parameterised system)**.** *A model is a family of systems*

$$\{\mathcal{S}_{\boldsymbol{\theta}} : \mathcal{X} \to \mathcal{Y} \mid \boldsymbol{\theta} \in \Theta\}, \tag{1.11}$$

*where $\mathcal{X}$ and $\mathcal{Y}$ are sets of discrete $D$-dimensional signals, $\boldsymbol{\theta}$ denotes a vector (or collection) of parameters, $\Theta$ is the corresponding parameter space, and each $\mathcal{S}_{\boldsymbol{\theta}}$ is a system in the sense defined above.*

Given a training dataset $\{(x^{(i)}, y^{(i)})\}$, a learning algorithm selects a parameter value $\hat{\boldsymbol{\theta}} \in \Theta$ (for example, by minimizing a loss function). The resulting *trained model*

$$\hat{\mathcal{S}}(\,\cdot\,) \triangleq \mathcal{S}_{\hat{\boldsymbol{\theta}}}(\,\cdot\,) \tag{1.12}$$

is then treated as a fixed system for inference: for a new input signal $x \in \mathcal{X}$, the model produces an output $y = \hat{\mathcal{S}}x \in \mathcal{Y}$. In informal discussion, we use the words "system" and "model" almost interchangeably, and reserve "model" when we want to emphasise the presence of learned parameters.

**Knowledge in AI models** Knowledge in practical AI is defined below.

**Definition 1.7** (Knowledge)**.** *Given a family of models $\{\mathcal{S}_{\boldsymbol{\theta}} : \mathcal{X} \to \mathcal{Y} \mid \boldsymbol{\theta} \in \Theta\}$ and a trained instance $\hat{\mathcal{S}} = \mathcal{S}_{\hat{\boldsymbol{\theta}}}$, the* knowledge *of the system (with respect to a task and data domain) is the organized internal representation of the relevant structure of the task that $\hat{\mathcal{S}}$ uses to map inputs to outputs and generalize beyond training samples. Here, $\mathcal{X}$ and $\mathcal{Y}$ denote the input and output signal spaces, $\boldsymbol{\theta} \in \Theta$ is the parameter vector of the model family, and $\hat{\boldsymbol{\theta}}$ denotes the parameter value obtained after training.*

Concretely, this knowledge is embodied in:

- the learned parameters $\hat{\boldsymbol{\theta}}$, which encode regularities, constraints, and priors about the input–output relationship; and
- the intermediate latent variables or feature maps $\{z_\ell(x)\}_\ell$ produced when processing an input $x \in \mathcal{X}$, which re-express the raw signal $x$ in structured forms that are useful for inference.

Together, these internal structures transform discrete signals (data) into meaningful internal representations on which the model performs reasoning, prediction, and decision-making.

The descriptions of some other terminology frequently used here are given below.

- **Learning**: processes that update knowledge from data (supervised, unsupervised, self-supervised, reinforcement).
- **Inference or reasoning**: drawing conclusions from knowledge (deductive, inductive, and probabilistic).

- **Generalization**: competent performance on novel cases beyond the training or past experience.
- **Efficiency**: Learning, inference and generalization with bounded data, time, and computation (realistic agents or models).
- **Layer**: A layer is a learnable system block inside an AI model.

## 1.3 Evolution of AI from symbolic reasoning to deep learning

Early visions of mechanized reasoning can be traced to the work of Leibniz, who imagined a universal symbolic language and designed early calculating machines in the 17th century, and to Charles Babbage and Ada Lovelace, whose 19th-century Analytical Engine anticipated the idea of programmable computation. While these early accounts were mythological or philosophical, they grappled with enduring questions of cognition, reasoning, and autonomy. The modern scientific foundations of artificial intelligence began to take shape in the 19th and early 20th centuries through advances in mathematical logic, formal reasoning,and computational theory. George Boole introduced binary logic (Boolean algebra, 1854), forming the basis of digital computation, while Gottlob Frege developed symbolic logic (Begriffsschrift, 1879), essential for knowledge representation. In the 1930s, Alan Turing proposed the universal Turing machine (1936) [10], offering a formal model of computation and raising the possibility of machine-based reasoning. Kurt Gödel's incompleteness theorems (1931) [11] highlighted fundamental limits of formal systems, anticipating key questions about the capabilities and boundaries of intelligent machines. In the 1940s, Warren McCulloch and Walter Pitts proposed a logical model of artificial neurons (1943) [12], anticipating the structure of modern neural networks. The 1940s marked a turning point as theoretical ideas in logic and computation began to materialize into physical systems. John von Neumann introduced the stored-program architecture (1945), which unified memory and processing within a single system, a blueprint that became the foundation of modern digital computers and enabled programmable, general-purpose computation. Norbert Wiener's theory of cybernetics (1948) [13] introduced the principles of feedback, control, and communication in machines and living systems. It described how systems use feedback and control to adapt to their environments, and his theory played a key role in shaping early AI, robotics, and modern theories of learning systems with goal-directed behavior. These foundational ideas also influenced early control systems and robotics, where feedback loops and adaptive behavior remain essential, laying conceptual groundwork for modern autonomous agents. Claude Shannon's information theory(1948) [14] provided a mathematical basis for data transmission and encoding, crucial for AI systems. These interdisciplinary contributions culminated in the formal inception of artificial intelligence as a research field, marked by the 1956 Dartmouth Conference.

The period from the 1950s to the late 2000s is often referred to as the Central Processing Unit (CPU) era of artificial intelligence, characterized by the dominance of general-purpose central processing units as the primary computational platform for AI research and applications. During this time, developments in AI largely relied on symbolic reasoning, statistical learning,and shallow neural models, all of which were constrained by the computational and memory limitations of CPUs. In the early decades, symbolic AI or good old-fashioned AI leveraged formal logic, rule-based systems, and handcrafted knowledge representations. Expert systems such as MYCIN and DENDRAL, and natural language models like ELIZA, operated entirely on serial CPU architectures. These systems demonstrated early success in narrow domains, but were ultimately limited by their brittleness, inability to generalize, and dependence on manually engineered rules. Parallel to symbolic AI, machine learning and connectionist models gradually gained traction. The rediscovery of backpropagation in the 1980s, and the development of statistical models such as support vector machines, decision trees, Bayesian networks,and ensemble methods, marked a shift toward data-driven approaches. However, the training and inference of these models were still confined to CPU architectures, which offered limited parallelism and scalability. As a result, deep neural networks, while theoretically promising, remained computationally prohibitive in practice. Even notable advances like LeNet-5 by Yann LeCun (1998) [15], which applied Convolutional Neural Networks (CNNs) to handwritten digit recognition, were constrained by available CPU performance and limited dataset sizes. During this era, AI progress was predominantly algorithmic, with hardware playing a passive and constraining role. The field's reliance on CPUs,coupled with restricted access to large datasets and high-performance computing, contributed to modest empirical performance and delayed the adoption of more compute-intensive architectures. By the early 2000s, with modest improvements in CPU capabilities and increasing data availability, statistical learning approaches became widely deployed across domains such as document classification, medical diagnosis,and financial modeling. However, the limitations of CPU-bound computation became increasingly evident as models grew in complexity and datasets scaled.

The CPU era was long and encompassed a wide range of foundational developments. Notably, it saw increasing cross-pollination between signal processing and machine learning, particularly in applied domains such as speech recognition, image compression, biomedical signal analysis,and communication systems. Signal processing problems often involved high-dimensional, structured, and noisy data, making them ideal testbeds for early learning algorithms. Conversely, signal processing concepts such as filtering, feature extraction, dimensionality reduction,and spectral analysis informed the design of interpretable, efficient ML models. Classical tools like Principal Component Analysis (PCA), Least Mean Square (LMS), Kalman filters, and wavelet transforms were frequently integrated into learning pipelines, while statistical models like Hidden Markov Models (HMMs) and Support Vector Machines (SVMs) found widespread use in signal and time-series applications. Despite rich and extensive theoretical developments in the CPU era, the scalability and practical implementation of many methods were limited by CPU-bound time and memory constraints. Techniques such as high dimensional transform-domain machine learning (e.g., multidimensional Fast Fourier Transform based convolution), high-dimensional adaptive filtering, and structured representations were often computationally prohibitive, restricting their deployment in large-scale or real-time settings. These limitations delayed the realization of theoretically promising models until more powerful computational platforms, particularly Graphics Processing Units (GPUs) in the late 2000s and early 2010s became available.

With the advent of GPU-accelerated computing around 2009–2012, AI entered a new era. Initially driven by advances in graphics hardware for parallel processing, GPUs proved exceptionally well-suited for the matrix and tensor operations central to training deep neural networks. One of the early conceptual breakthroughs came with the development of Deep Belief Networks by Hinton et al. (2006) [16], which used unsupervised pretraining of stacked Restricted Boltzmann Machines to make deep architectures trainable on CPUs. While these models rekindled interest in deep learning, they remained limited by computational constraints. The true turning point came with AlexNet (Krizhevsky et al., 2012 [17]), a deep CNN trained on GPUs that dramatically improved performance on the ImageNet benchmark. This achievement demonstrated the feasibility of end-to-end deep learning at scale and catalyzed the GPU-driven deep learning revolution. Many theories developed during the CPU era though rich and well-established remained practically stagnant due to computational constraints. The advent of GPU-accelerated computing broke these bottlenecks, making large-scale deep learning feasible and sparking a revolution that transformed fields such as computer vision, natural language processing, speech recognition, and beyond. In this present AI era, this convergence of scalable computation, abundant data, and open-source tools created a powerful ecosystem accelerating AI research and deployment. To put this in perspective, the volume and diversity of problems solved in just the past decade of the GPU era far exceed the cumulative progress made throughout the much longer CPU era, reflecting a paradigm shift in both capability and speed of innovation.

The GPU era, beginning around the early 2010s, catalyzed a rapid succession of breakthroughs that had long remained bottlenecked by computational limits. Foundational ideas from the CPU era, such as backpropagation (Rumelhart et al., 1986 [18]), convolutional neural networks (LeCun et al., 1989–1995 [19,20]), and recurrent memory-based models such as Long Short-Term Memory (LSTM) (Hochreiter & Schmidhuber, 1997 [21]), became practically scalable. This triggered rapid growth in deep learning research and applications. Subsequent years saw the rise of generative models such as GANs (Goodfellow et al., 2014 [22]), sequence-to-sequence learning with attention (Bahdanau et al., 2014 [23]), and the transformer architecture (Vaswani et al., 2017 [24]), which replaced recurrent processing in many sequence modeling settings with scalable self-attention. Meanwhile, architectural developments such as residual connections and normalization layers improved the trainability of deep networks, while U-Net improved their applicability to biomedical image segmentation. Pretrained large language models such as BERT (Devlin et al., 2019 [25]) and GPT models (Radford et al., 2019 [26]; Brown et al., 2020 [27]) further marked a shift toward large-scale transfer learning. In parallel, more structured and mathematically grounded approaches emerged. Fourier features, scattering transforms, spectral graph CNNs, and Volterra neural networks introduced ways to incorporate spectral structure, multiscale stability, geometric priors, and nonlinear system-theoretic modeling into learning architectures. More recent models, such as Neural ODEs, Fourier Neural Operators, Implicit Neural Representations (INRs), and Equivariant Neural Networks, extended this trend toward continuous-depth modeling, operator learning, coordinate-based representations, and symmetry-aware learning. Collectively, these developments show a growing effort to combine the empirical success of deep learning with explicit mathematical structure.

Table 1.1: Historical evolution of model architectures in machine or deep learning

| Year | Breakthrough and seminal works | Learning System Description | Learning Dynamics Innovation |
|---|---|---|---|
| 1958–1969 | Perceptron [28], Early NN [29] | Linear inputs + threshold; single-layer; hard-thresholded | Thresholded linear classification |
| 1986 | Backprop for deep NN [18] | Feedforward MLP; gradient descent; chain rule | Chain rule-based weight updates |
| 1989 | Early CNNs [19] | Convolution + pooling + FC; gradient descent + backprop | Backprop through spatial filters |
| 1992 | Wavelet Neural Nets [30] | Localized wavelet basis $\psi(ax + b)$; gradient over wavelet params | Gradient descent on wavelet coefficients |
| 1997 | LSTM [21] | Recurrent memory cells; gated updates; BPTT | Gated feedback; BPTT on memory |
| 1998 | LeNet-5 [15] | Conv + Subsampling + MLP; sigmoid/tanh; backprop | Supervised spatial hierarchy learning |
| 2006 | Compressed Sensing (CS) [31, 32] | Sparse recovery from undersampled linear measurements; $\ell_1$ regularization; wavelet/sparsifying bases | $\ell_1$-regularized optimization; sparse thresholding |
| 2006 | Deep Belief Nets [16] | RBM stack; layer-wise unsupervised pretraining + fine-tuning | Contrastive divergence pretraining |
| 2008 | t-SNE [33] | KL-divergence loss; stochastic gradient; low-dim embedding | Loss-based nonlinear projection |
| 2012 | AlexNet [17] | Deep conv + ReLU + pooling + dropout; SGD w/ momentum | SGD over deep pipelines |
| 2013 | Variational Autoencoders [34] | Latent-variable generative models with amortized variational inference | Stochastic gradient variational Bayes |
| 2013 | Scattering Networks [35] | Non-learned multiscale wavelet modulus cascades with stable invariants | Fixed wavelet transform + learned classifier |
| 2014-2015 | GANs, Seq2Seq, Attention [22, 23, 36] | GAN: min–max loss; Seq2Seq: encoder–decoder RNN; Attention: soft weights | Adversarial, encoder–decoder, soft alignment |
| 2015 | Adam Optimizer [37] | Per-parameter adaptive learning; first-order optimizer | Momentum + adaptive moving averages |
| 2015-2016 | ResNet, U-Net, BN [38–40] | Skip connections; encoder–decoder; batch-wise normalization | Residual learning; BN training stability |

*Table 1.1 (continued)*

| Year | Breakthrough and seminal works | Learning System Description | Learning Dynamics Innovation |
|---|---|---|---|
| 2016-2021 | Equivariant Neural Nets [41, 42] | Group convolutions; symmetry-preserving filters | Group-constrained weight learning |
| 2017 | Transformer [24] | Self-attention; position encoding; FFN blocks; no recurrence | Scaled dot-product attention |
| 2017 | Graph Convolutional Networks (GCNs) [43] | Message passing on graphs via neighborhood aggregation layers | Gradient-based end-to-end learning on graphs |
| 2017 | NN approximation theory [44] | Depth/width trade-offs; memory/complexity-aware approximation bounds | Theory of expressive classes under constraints |
| 2018 | Neural Tangent Kernel [45] | Infinite-width limit; training as kernel evolution | NTK-based linearized functional updates |
| 2018 | Scattering networks (hybrid learning) [46] | Scattering features paired with trainable heads for representation learning | Hybrid fixed transform + backprop-trained head |
| 2019–2020 | Shearlet–DL hybrids [47, 48] | Shearlet features or model-based priors combined with deep networks | Hybrid model-based + data-driven training |
| 2019–2021 | BERT, GPT [25–27, 49] | Transformer-based LM; masked prediction; autoregression | Masked/auto-regressive pretraining |
| 2020 | Fourier Features / SIREN [50, 51] | Sinusoidal encodings; MLP learns spectral mapping | Frequency-encoded MLP training |
| 2020 | Volterra Neural Nets [52, 53] | Cascaded Volterra polynomials; BPTT or backprop learning | Polynomial series via gradients |
| 2020 | Fourier Neural Operators [54] | Spectral convolution; operator learning via Fourier updates | Gradient-based spectral operator learning |
| 2020–2021 | Implicit Neural Reps (INRs) [51, 55] | MLPs map coordinates to signal; image/SDF losses | Coord-encoded MLP supervision |
| 2020–2022 | Diffusion, ViT, RLHF [56–58] | U-Net-based denoising; patchified vision transformer | Score-based learning; human reward tuning |
| 2021-2022 | PDE Learning via NN [59, 60] | Feedforward net minimizing PDE residuals; supervised loss | Loss-driven PDE map training |
| 2023 | Fractional Graph Laplace ODEs [61] | Neural ODEs with fractional Laplacian; continuous-time graph learning | Fractional gradient flows on graphs |

As deep learning models grew in complexity, their opaque nature raised critical concerns about transparency, explainability, and trust. This challenge is especially pressing in high-stakes domains such as healthcare, finance, and autonomous industrial systems, where understanding and validating AI decisions is essential. In response, explainable AI (XAI) has developed methods for interpreting the behavior of complex trained models, while a complementary line of work has focused on architectures that are interpretable by construction, including model-based and algorithm-unrolled networks. Researchers are increasingly revisiting foundational principles from signal processing and system theory, which have long offered transparent and modular frameworks, to guide the development of interpretable learning architectures. Constructs such as filters, transforms, transfer functions, and feedback loops, originally central to analog and digital systems, are now being reimagined within neural frameworks. These structured approaches offer mathematical tractability and physically meaningful interpretations, helping to bridge the gap between powerful learning and human understanding.

### 1.3.1 Structured knowledge representations and their interpretations

At the core of most AI systems, whether supervised, unsupervised, or semi-supervised, one can distinguish two tightly coupled components. The first is **knowledge representation**, which encodes information in an explicit and structured form. The second is **search or inference mechanisms**, which operate on that representation to produce decisions or predictions. Together, they enable intelligent behavior by organizing domain knowledge and exploring candidate solutions in a principled way. Despite this clean decomposition, representing knowledge is inherently difficult because real-world knowledge is often voluminous, heterogeneous, incompletely specified, and continuously evolving. It is therefore useful to distinguish several complementary forms of knowledge.

- *Procedural knowledge* captures **how** tasks are carried out through recipes, strategies, and ordered procedures.
- *Declarative knowledge* captures **what** is true in a domain, including facts, concepts, objects, and their relationships.
- *Meta-knowledge* is knowledge about other knowledge sources or representations, and helps a system choose or combine them for a given task.
- *Heuristic knowledge* captures practical **rules of thumb** that guide decisions when information is incomplete, uncertain, or costly to process exactly.
- *Structural knowledge* organizes relationships among concepts and entities, including taxonomic ("kind-of"), compositional ("part-of"), and set-based ("group-of") associations.

Turning these heterogeneous forms of knowledge into something a machine can use requires **knowledge acquisition**, which involves eliciting, extracting, organizing, and formalizing information from domain experts, documents, or data. When coupled with an inference engine, such formalized knowledge gives rise to knowledge-based systems (KBS). These systems provided early examples of transparent and modular AI architectures.

#### 1.3.1.1 Symbolic knowledge acquisition and early representations

Early AI research treated knowledge as something that could be explicitly modeled, stored, and manipulated in symbolic form. This view led to representational paradigms designed to capture entities, relationships, and event structures in human-readable formats. In early KBS, knowledge engineers translated informal expertise from domain experts, documents, or observations into structured representations that were computationally usable. These systems typically separated a *knowledge base*, which stored rules, facts, and structured descriptions of a domain, from an *inference engine*, which applied logical or heuristic procedures to derive conclusions. This separation provided a transparent organization of representation and inference in early AI systems.

#### 1.3.1.2 Symbolic representations

Symbolic paradigms offered several ways of encoding acquired knowledge in a structured and interpretable form.

- *Semantic networks* represent knowledge as labeled graphs whose nodes denote concepts or objects and whose edges denote relations such as "is-a" or "part-of". They support inheritance and path-based reasoning.
- *Frame-based representations*, in the classical AI sense, describe typical entities or situations through attributes, values, default assumptions, and constraints.
- *Scripts* represent familiar event sequences through roles, preconditions, and ordered sub-events, allowing systems to infer implicit steps when interpreting narratives.

These early paradigms clarify what is meant by *symbolic* knowledge in AI. Such knowledge consists of internal representations built from discrete symbols, such as names for objects, concepts, and relations, together with explicit structures, such as graphs, attribute-value records, or event templates, on which rule-based inference can operate. The symbols themselves have linguistic or conceptual meaning that a human can read directly from the representation.

By contrast, modern representation learning models, such as multilayer perceptrons and convolutional neural networks, directly learn numerical representations of real-valued high-dimensional tensors rather than explicit symbols and rules. However, the underlying objective remains related. In both cases, the aim is to capture regularities, structures, and dependencies in data in a form that supports reliable prediction and decision-making. Early symbolic formalisms are therefore relevant as a **reference point** for what it means for a representation to be structured and interpretable, even when the concrete mechanisms in later chapters are numerical and signal-theoretic rather than symbolic.

#### 1.3.1.3 The present paradigm of representation learning

Despite the interpretability and modularity of symbolic representations such as semantic networks, frames, and scripts [62–64], these classical techniques struggled with noisy, high-dimensional, and evolving data. Building and maintaining large knowledge bases was labor-intensive, required domain experts, and often became brittle when uncertainty or missing information was present. These limitations encouraged the development of data-driven representation learning, where statistical and connectionist methods infer useful structure directly from examples [65]. With the availability of large datasets, powerful GPUs, and scalable optimization methods, it has become increasingly feasible to build AI solutions for highly nonlinear tasks such as inverse problems, classification, and segmentation directly from data. Deep neural networks, including multilayer perceptrons and convolutional neural networks, learn task-relevant representations through gradient-based learning [18, 66], where successive layers compute increasingly abstract features for a given task [67]. In these architectures, knowledge is no longer stored as explicit rules but as patterns of weights, which improves accuracy and scalability. Automatic representation learning directly from data, for example, in U-Net [39], Transformer [24], and Vision Transformer [57] (ViT), marks a major milestone in the present AI ecosystem. However, this also makes the internal states of a model difficult to interpret because of nonlinear multilayer processing with millions or billions of parameters, raising concerns about the robustness and trustworthiness of black-box algorithms. To address such concerns, some recent hybrid systems unify classical ideas with modern learning. For example, neuro-symbolic concept learners combine neural perception with symbolic reasoning [68, 69], and generative flow networks model structured discrete spaces using learned stochastic policies [70, 71]. These approaches use explicit structure to improve transparency and help connect learned models with human reasoning.

Our signal and system-theoretic view sees data as a general multidimensional signal, and these representation learning architectures as multidimensional systems (linear or nonlinear). This perspective broadens the design view, where prior knowledge can be embedded through linear operators, multiresolution transforms, and stability constraints on the underlying filters and recursions [35, 72]. Designing models in this way keeps each block in the model analyzable in the language of systems theory while still allowing its parameters to be learned from data. Some examples of such fundamental blocks are as follows:

- *Fourier and Wavelet bases* decompose the signals into frequency or time-frequency atoms. These bases introduce sparsity and shift invariance into learning pipelines.
- *Scattering Transforms* [35] and Wavelet CNNs introduce fixed or trainable filters modeled after wavelet bases, capturing multi-resolution structures in images and audio. Their mathematical grounding ensures robustness and partial explainability.
- *Shearlet filter banks* [73] capture the directional and anisotropic structures in images and multidimensional inputs, making them suitable for vision and biomedical image processing.
- *Ramanujan periodicity transforms* [74], inspired by number-theoretic constructs, capture periodicities and harmonic structures in time-series and audio data, offering highly compact and interpretable feature sets.

- *Polyphase sampling* [75], introduces true shift invariance through principled downsampling and aliasing-aware filters in decimated convolutional architectures by preserving structural fidelity across spatial scales.
- *Learnable infinite impulse response (IIR) filters* introduce recursive filtering operations into deep models. Their feedback coefficients provide a compact parametric mechanism for modeling long-range temporal or spatial dependencies, while their poles, stability, and frequency responses remain analyzable using classical system-theoretic methods.
- *Volterra neural networks* [76] extend classical nonlinear system theory into deep learning. They introduce polynomial kernel expansions to model interpretable interactions among input dimensions. These architectures generalize convolutions with Volterra series expansions and offer structured nonlinear representations in a controlled Hilbert space.

The above system blocks emphasize that interpretability need not be sacrificed for performance; rather, by enforcing structure, sparsity, and domain priors, it is possible to build models that are not only powerful and data-efficient but also explainable and trustworthy. Collectively, they also represent a convergence between symbolic reasoning, signal processing, system theory, and modern deep learning. This evolution reflects a broader shift in AI models for Industry 4.0 and 5.0 from purely data-driven pipelines to architectures that integrate prior knowledge, system constraints, and symbolic logic, laying the foundation for the next generation of robust and transparent intelligent systems.

### 1.3.2 A quick review of essential elementary ideas in machine learning

#### 1.3.2.1 Perceptron and Wavelon

**Perceptron** A Perceptron is a single unit of a feedforward neural network. A forward propagation in $i^{th}$ Perceptron defined by equation (1.13)

$$\hat{y}_i = g\left(w_{0,i} + \mathbf{x}^\top \mathbf{w}_i\right) = g\left(q_i\right) \tag{1.13}$$

Here, $q_i = w_{0,i} + \mathbf{x}^\top \mathbf{w}_i$ the decision boundary line that defines this Perceptron.

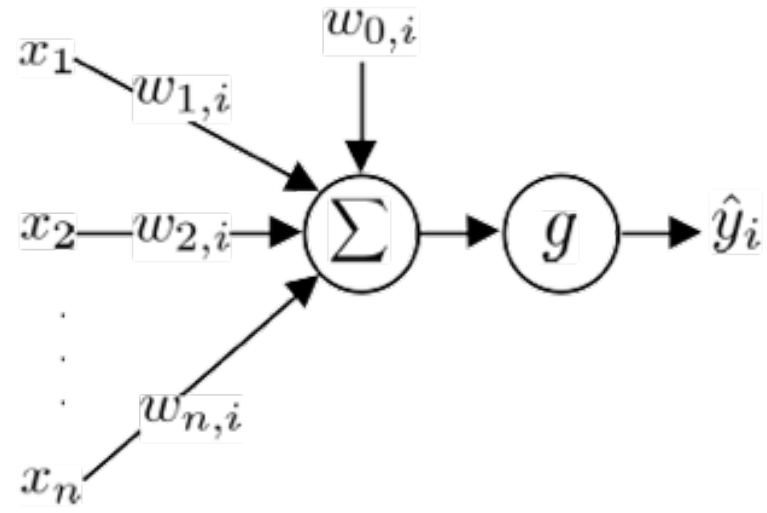

Figure 1.3: Perceptron

$w_{0,i} \in \mathbb{R}$ is bias, $\mathbf{w}_i \in \mathbb{R}^n$ is weight vector, $g(\cdot)$ is a nonlinear activation e.g. sigmoid, ReLU [77], Mish [78] etc. $\mathbf{x} \in \mathbb{R}^n$ is an input vector and $\hat{y}_i \in \mathbb{R}$ is the inferred output. Figure 1.3 shows a Perceptron.

**Wavelon** A wavelon is a special Perceptron with translation, dilation parameters and wavelet activation function. They are the fundamental units of wavelet networks [30]. The number of dilated and translated wavelet basis functions is realised by the number of wavelons in a wavelet network. Figure 1.4 shows the $i^{th}$ wavelon of a wavelet network.

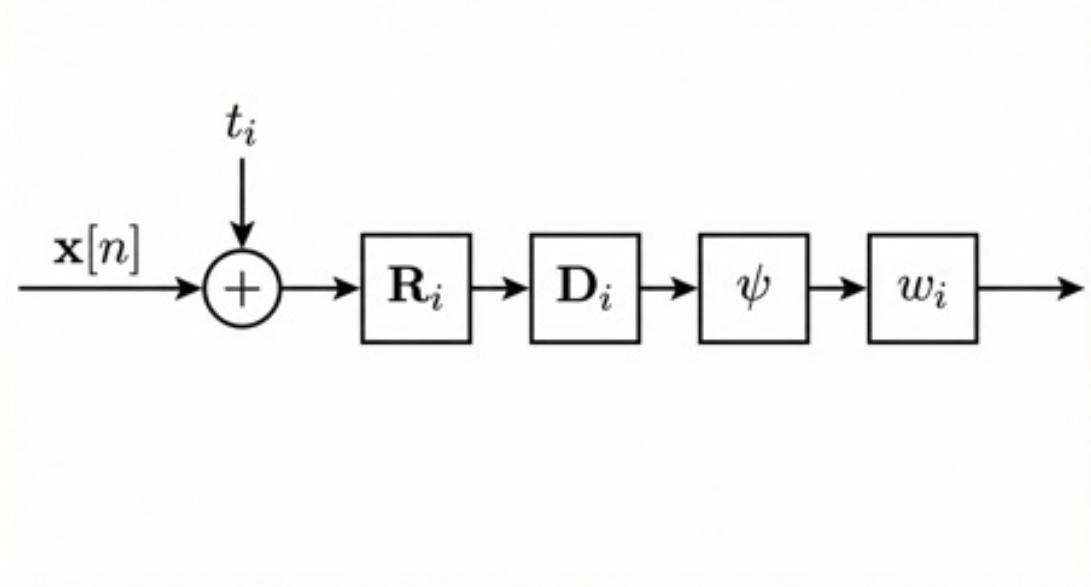

Figure 1.4: Wavelon

#### 1.3.2.2 Function approximation

A feedforward neural network with an input layer, a single hidden layer, and an output layer is usually classified as a shallow feedforward neural network. For an input $\mathbf{x} \in \mathbb{R}^n$, the output at the $k^{\text{th}}$ output node is given by

$$\hat{y}_k = g\left(w_{0,k}^{(2)} + \sum_i g(q_i) w_{i,k}^{(2)}\right), \tag{1.14}$$

where $q_i = w_{0,i}^{(1)} + \mathbf{x}^\top \mathbf{w}_i^{(1)}$ with $\mathbf{w}_i^{(1)} \in \mathbb{R}^n$ is a weight vector and $i, k \in \mathbb{N}$ are indices of the hidden layer and output neurons respectively. Here, $w_{0,k}^{(2)}$, $w_{i,k}^{(2)}$, $q_i$, $\hat{y}_k \in \mathbb{R}$.

A single hidden layer feedforward neural network is a powerful architecture for approximating nonlinear functions by mapping two algebraic spaces. The Stone–Weierstrass theorem [79], states that any continuous function defined on a closed interval can be precisely approximated using a polynomial function. In the context of neural networks this is further stated in the *universal approximation theorem* [80].

**$1 + \frac{1}{2}$ layer wavelet network** Zhang and Benveniste [30] introduced wavelet networks for function approximation by replacing standard hidden neurons with wavelet-based units called wavelons. Each wavelon applies an affine transformation to the input and then evaluates a multidimensional wavelet. In Eq. 1.15, the functions $\psi[\mathbf{R}_i\mathbf{D}_i(\mathbf{x} - \mathbf{t}_i)]$ represent the nonlinear wavelet units, while the weighted summation $\sum_{i=1}^{N} w_i(\cdot)$ with the offset $\bar{g}$ represents the additional half layer. It is called a half layer because this output stage only performs a linear combination of the wavelon responses and does not introduce another nonlinear activation. For an input $\mathbf{x} \in \mathbb{R}^n$, the input-output relation is written as

$$g(\mathbf{x}) = \sum_{i=1}^{N} w_i \psi \left[ \mathbf{R}_i \mathbf{D}_i (\mathbf{x} - \mathbf{t}_i) \right] + \bar{g}. \tag{1.15}$$

Here, $w_i \in \mathbb{R}$ is a scalar output weight, $\mathbf{t}_i \in \mathbb{R}^n$ is a translation vector, $\mathbf{D}_i \in \mathbb{R}^{n \times n}$ is a dilation matrix, and $\mathbf{R}_i \in \mathbb{R}^{n \times n}$ is a rotation matrix. The integer $N \in \mathbb{N}$ denotes the number of wavelons. The function $\psi : \mathbb{R}^n \to \mathbb{R}$ is the mother wavelet, and the offset $\bar{g} \in \mathbb{R}$ accounts for the constant component of the approximated function when the wavelet has zero mean. Example 1.1 demonstrates the Hysteresis function learning using a $1 + \frac{1}{2}$ layer wavelet network.

**Example 1.1.** Learning hysteresis loop using $1 + \frac{1}{2}$ layer wavelet network.

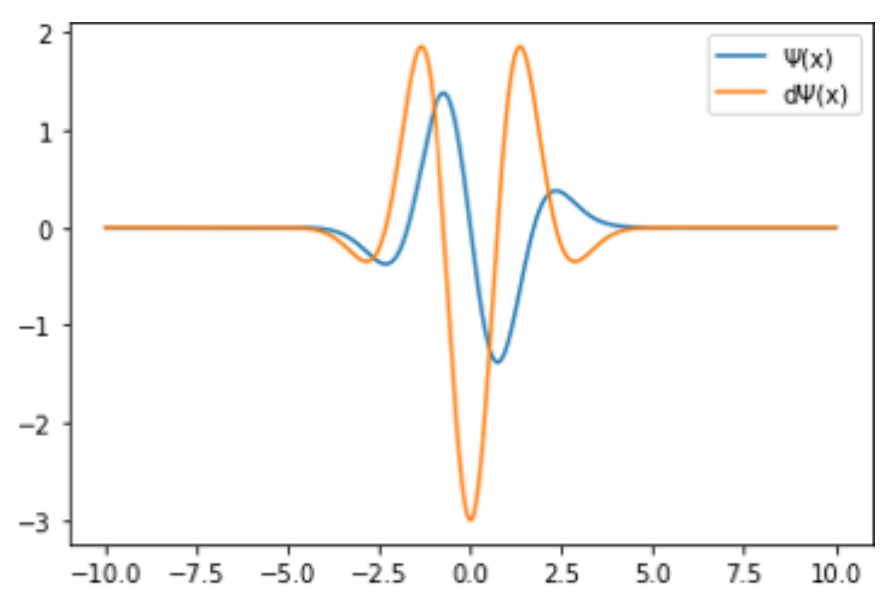

(a) First derivative of Mexican Hat wavelet and its derivative

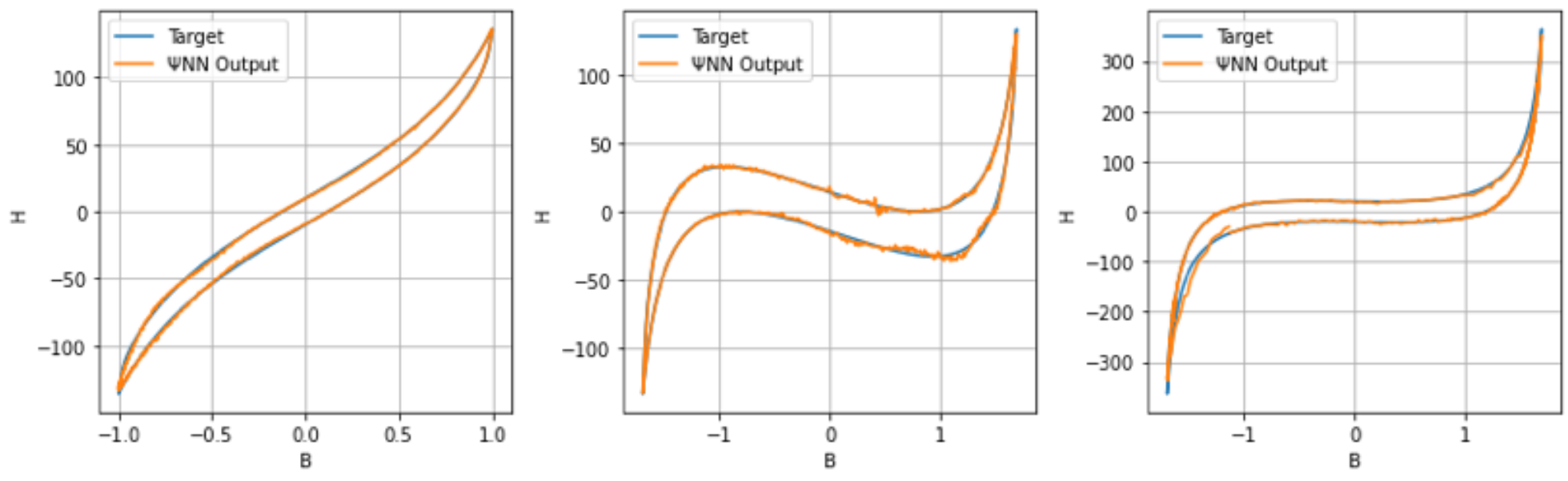

(b) Learning three different hysteresis curves

Figure 1.5: Learning hysteresis model $H = f(B, n)$ using $1 + \frac{1}{2}$ layer wavelet network

Hysteresis is the nonlinear phenomenon observed in electromagnetic devices, where the present state of a system depends on its previous states. It is caused due to different loading and unloading paths. A $1 + \frac{1}{2}$ layer wavelet network successfully learns to map the two algebraic spaces, magnetic flux density $\mathbf{B}$ (Tesla) and magnetic field intensity, $\mathbf{H}$ (Ampere/meter) with substantial computational efficiency.A parametric variable, time $n$ is used to keep track of the loading and unloading phases and the model is defined as $\mathbf{H} = f(\mathbf{B}, \boldsymbol{n})$. This one-to-many function $f(.)$ is successfully learnt using a $1 + \frac{1}{2}$ layer wavelet network. The first derivative of the Mexican hat mother wavelet as shown in Figure 1.5a very accurately approximated three different types of hysteresis curves as shown in Figure 1.5b.

#### 1.3.2.3 Deep neural network

Although shallow neural networks have important function approximation capabilities, they may require a large number of units to model highly nonlinear, high-dimensional, and compositional functions efficiently. Practical tasks such as pattern recognition, image segmentation, object recognition, speech recognition, and large-scale classification often involve hierarchical structure across different levels of abstraction. This motivated the development of deeper architectures, where multiple hidden layers can learn progressively more abstract feature representations from the input data. A deep neural network therefore, consists of several hidden layers that transform the input signal through a sequence of learned nonlinear mappings. The pre-activation state of the $i^{\text{th}}$ unit at depth $d$ can be written as

$$z_{d,i} = w_{0,i}^{(d)} + \sum_{m} g\left(z_{d-1,m}\right) w_{m,i}^{(d)}. \tag{1.16}$$

Here, $z_{d-1,m}$ denotes the state of the $m^{\text{th}}$ unit in the previous layer, $g(\cdot)$ is the activation function, $w_{m,i}^{(d)}$ is the weight connecting unit $m$ at depth $d-1$ to unit $i$ at depth $d$, and $w_{0,i}^{(d)}$ is the corresponding bias term.
During training, the objective is to estimate the network parameters that minimize the empirical loss over the training dataset. Let $\{(\mathbf{x}^{(i)}, y^{(i)})\}_{i=1}^{N}$ denote a training set of $N$ input-output pairs. The optimization problem is written as

$$\mathbf{W}^* = \arg\min_{\mathbf{W}} \frac{1}{N} \sum_{i=1}^{N} \mathcal{L}\left(f\left(\mathbf{x}^{(i)}; \mathbf{W}\right), y^{(i)}\right) = \arg\min_{\mathbf{W}} J(\mathbf{W}). \tag{1.17}$$

Here, $\mathbf{W} = [\mathbf{w}^{(0)}, \mathbf{w}^{(1)}, \ldots]$ denotes the collection of trainable parameters across all layers of the network, including weights and biases. The function $f(\mathbf{x}^{(i)}; \mathbf{W})$ denotes the network prediction for the $i^{\text{th}}$ training input, and $\mathcal{L}(f(\mathbf{x}^{(i)}; \mathbf{W}), y^{(i)})$ denotes the loss associated with the $i^{\text{th}}$ training example. The quantity $J(\mathbf{W})$ is the empirical risk or average training loss.

**Backpropagation and computing gradients** The network weights are optimized by computing gradients using backpropagation and chain rule as in equation (1.18) to minimize a loss function. A backpropagation for the simplest possible neural network having three layers with one neuron in each layer is as follows:

$$\mathbf{x} \xrightarrow{\mathbf{w}_1} \mathbf{z}_1 \xrightarrow{\mathbf{w}_2} \hat{\mathbf{y}} \rightarrow J(\mathbf{W})$$

$$\frac{\partial J(\mathbf{W})}{\partial \mathbf{w}_1} = \frac{\partial J(\mathbf{W})}{\partial \hat{\mathbf{y}}} \frac{\partial \hat{\mathbf{y}}}{\partial \mathbf{z}_1} \frac{\partial \mathbf{z}_1}{\partial \mathbf{w}_1} \tag{1.18}$$

The learning rate, $\eta$ determines the gradient step size during backpropagation. A common practice to set $\eta$ is by (i) hit and trial and (ii) explicit adaptive learning rate design that adapts to the weight landscape. The gradient descent and stochastic gradient descent algorithms used in backpropagation are explained in Algorithm 1.1 and Algorithm 1.2 respectively.

**Algorithm 1.1:** Gradient descent

1. Initialize weights randomly $\sim N(0, \sigma^2)$
2. Loop until convergence
   (a) Compute gradient, $\frac{\partial J(\mathbf{W})}{\partial \mathbf{W}}$
   (b) Update weights, $\mathbf{W} \leftarrow \mathbf{W} - \eta \frac{\partial J(\mathbf{W})}{\partial \mathbf{W}}$
3. Return weights

**Algorithm 1.2:** Stochastic Gradient Descent

1. Initialize weights randomly $\sim N(0, \sigma^2)$
2. Loop until convergence
   (a) Pick batch of $B$ data points
   (b) Compute gradient, $\frac{\partial J(\mathbf{W})}{\partial \mathbf{W}} = \frac{1}{B} \sum_{k=1}^{B} \frac{\partial J_k(\mathbf{W})}{\partial \mathbf{W}}$
   (c) Update weights, $\mathbf{W} \leftarrow \mathbf{W} - \eta \frac{\partial J(\mathbf{W})}{\partial \mathbf{W}}$
3. Return weights

**Regularization** Neural network training seeks to reduce empirical loss while maintaining good generalization on unseen data. Two common failure modes are underfitting and overfitting. Underfitting occurs when the model or the training procedure fails to capture the relevant structure in the data. Overfitting occurs when the model fits sample-specific variations in the training set and therefore generalizes poorly on unseen data. Regularization methods address overfitting by controlling the effective complexity of the learned function. Common examples include weight decay, dropout, and early stopping. Weight decay penalizes large parameter values, dropout randomly suppresses a subset of activations during training, and early stopping terminates training when validation performance no longer improves. In practice, overfitting is often indicated when training loss continues to decrease while validation loss stagnates or increases.

## 1.4 Signal and system primitives for interpretable AI models

**Interpretability** [1, 2] in machine learning (ML) and deep learning (DL) refers to the ability to understand and reason about a model's internal mechanisms and the rationale behind its predictions. As deep neural networks grow in depth and complexity, they consist of multiple layers of nonlinear transformations, often involving millions of parameters. These layers typically apply operations such as convolution, pooling, normalization, and activation functions (e.g., ReLU, sigmoid, tanh), creating highly nonlinear mappings from input to output. While these mappings possess strong approximation capabilities, they obscure how individual features propagate, interact, and influence final decisions. Moreover, the distribution of learned weights across layers makes it non-trivial to directly attribute a decision to a specific input feature or processing stage. As a result, these models form a **black box** architecture that is capable of producing accurate outputs, but lacking transparency in their internal computations. For example, a regression is interpretable computation. Now when we use a hidden layer neural network to realize this regression with support of Universal approximation theorem [80], a black box model is introduced. This black box nature is further intensified by the combined effects of architectural complexity and data-driven optimization. During training, model parameters are updated using gradients computed by backpropagation and optimized through gradient-based methods, often with stochastic or mini-batch estimates of the loss gradient, to reduce a high-dimensional nonconvex objective. Although this process effectively captures statistical patterns, it can produce internal representations that are difficult to align with human-interpretable or physically meaningful structures. These latent features are typically high-dimensional, abstract, and difficult to relate directly to the input domain without additional decoding or visualization. Furthermore, they are frequently sensitive to small input perturbations, including adversarial noise, due to the absence of explicit structural constraints or domain-informed priors in the model's design. Without such constraints, the learned representations may rely on fragile, high-frequency patterns or spurious correlations, features that are statistically discriminative but semantically or physically ungrounded. Since the optimization objective is typically limited to minimizing prediction error on the training set, there is no inherent mechanism to prefer robust or interpretable solutions. As a result, even imperceptible input changes can propagate unpredictably through deep layers, producing unstable or incorrect outputs. This instability reflects the lack of architectural mechanisms enforcing invariance, continuity, or interpretability in the learned function. To whiten these black box models, here we adopt a more primitive design approach using signal and system-theoretic methods powered by modern GPU parallelism.

### 1.4.1 A system-theoretic path to trustworthy and frugal architectures

The principles from signal processing, nonlinear systems theory, and multiresolution analysis offer a structured foundation for introducing interpretability and regularity into deep learning models to meet the goals of Industry 4.0. In particular, wavelet transforms provide localized, multiscale representations of input data, enabling networks to capture both low- and high-frequency components in a compact and structured form. This decomposition promotes sparsity, enhances noise suppression, and enables scale-selective feature learning, which is critical in classification and segmentation tasks where spatial and spectral coherence is essential. Wavelet-domain operations also constrain learned filters to exhibit specific frequency responses, making internal representations more analyzable and robust to perturbations. From a system-theoretic standpoint, most deep learning architectures, including MLP neural networks, CNNs, recurrent models, and attention-based systems, are constructed from simple computational units such as perceptrons and convolutional filters. These apply deterministic mappings composed of linear transformations followed by nonlinear activation functions and can thus be viewed as nonlinear, time-invariant input–output systems. A useful abstraction for analyzing such architectures is the Linear–Nonlinear (LN) cascade, also known as the Wiener system, which is a linear operator such as a summation or convolution in cascade with a static nonlinearity. This configuration underlies the computational mechanism of many neural layers, including convolutional layers with ReLU activations and fully connected layers with sigmoids or tanh functions. Table 1.2 lists representative learnable LN cascade systems that recur throughout modern deep models.

Table 1.2: Examples of nonlinear models with various learnable Linear–Nonlinear cascade systems

| # | System / Model and paper(s) | Linear Component | Nonlinearity | Key Contribution & Learning Method |
|---|---|---|---|---|
| 1 | Perceptron (1958) [28] | Weighted sum of inputs | Threshold (Heaviside step) | First neural model: learnable linear weights with a fixed hard threshold; perceptron learning rule for linearly separable data |
| 2 | ADALINE/MADALINE (1960–1962) [81] | Weighted sum (adaptive linear filter) | Threshold / sign | Early trainable LN units; LMS-style adaptive updates for weights; foundational to adaptive filtering and pattern recognition |
| 3 | GMDH polynomial networks (1970) [82] | Polynomial expansions | Polynomial (e.g., quadratic) | Layer-wise polynomial cascades with model selection/pruning (self-organization) |
| 4 | MLPs with sigmoid/tanh (1986–1989) [18, 83, 84] | Affine transforms | Sigmoid or tanh | Differentiable activations enable backpropagation training; universal approximation. |
| 5 | Wavelet Neural Network (1992) [30] | Weighted sum of inputs | Morlet / Haar wavelet | First use of wavelet functions as activation; trained via backpropagation |
| 6 | ReLU in deep nets (2010–2012) [17, 85] | Convolutions / linear layers | ReLU ($\max(0, x)$) | Non-saturating activation that stabilizes deep training; enabled practical deep CNN scaling |
| 7 | PReLU (2015) [86] | Convolutions / linear layers | Parametric ReLU (learned slope) | Learns a parameterized nonlinearity (negative slope) per channel to improve accuracy |
| 8 | Approximation theory for deep ReLU nets (2017) [87] | Deep linear compositions | ReLU | Theoretical error bounds showing depth-efficiency for piecewise-linear LN cascades |
| 9 | Polynomial activations (2019) [88] | Standard linear layers | Learned polynomial (order $\geq 2$) | Learns expressive nonlinearities directly from data within standard layers |
| 10 | Padé activations (2019) [89] | Standard layers | Rational functions (learned) | More expressive and trainable than polynomials; enables fine-grained function fitting |

Volterra theory provides a classical framework for modelling nonlinear input-output sys-tems, especially systems with memory or multidimensional interactions. It may be viewed as a functional extension of the Taylor series from static nonlinear functions to dynamic systems. In a Taylor expansion, scalar coefficients multiply powers of an instantaneous variable. In a Volterra expansion, these coefficients are replaced by kernels that weight products of input samples taken at different time instants, spatial locations, or feature coordinates. The first-order Volterra kernel represents the linear part of the input-output map. In the shift-invariant case, this kernel corresponds to the usual linear impulse response. Higher-order kernels represent nonlinear interactions among delayed, shifted, or spatially indexed input samples. In practice, the infinite Volterra expansion is usually truncated, so a model up to order $m$ gives an approximate description of nonlinear behavior over the operating range of interest. For an LN cascade consisting of a linear dynamic or spatial operator followed by a static polynomial nonlinearity of degree $m$, the corresponding Volterra representation contains kernels only up to order $m$. If the linear operator is shift invariant, the resulting kernels inherit structure from the underlying LSI operator. For a single LSI block followed by a scalar polynomial nonlinearity, the higher-order kernels have constrained product forms, up to symmetrization, rather than the fully general form of an unconstrained Volterra kernel. This formalism also helps interpret common neural units such as perceptrons and convolutional units, since both apply a linear or affine operation followed by a pointwise nonlinearity. When the activation is polynomial, the corresponding mapping admits an exact Volterra representation of the same order. For smooth non-polynomial activations such as sigmoid, tanh, and GELU, the Volterra interpretation is usually local or truncated. For piecewise activations such as ReLU, it should be understood as an approximate system-theoretic interpretation rather than a global finite-order identity. Thus, the affine transformation or convolution supplies the linear operator, while the activation introduces nonlinear interaction terms without explicitly constructing full Volterra kernels. This gives a compact and interpretable framework for studying nonlinear transformations, memory effects, and interactions across features, layers, and scales.

In view of an AI model as a composition of signals and systems, the lack of interpretability in deep networks reflects the absence of structured constraints in both signal transformations and internal state evolution. Unlike traditional signal processing pipelines or dynamical systems defined by explicit transfer functions, deep learning models learn arbitrary mappings without control over frequency content, energy preservation, or time–space localization. Moreover, they often lack formal guarantees on stability or robustness, such as those provided by bounded-input–bounded-output (BIBO) analysis. Without embedded priors, such as sparsity, multiscale decomposition, or system-theoretic invariants, the internal operations of these models become analytically opaque and formally unverifiable. To improve interpretability, it is essential to **reintroduce** structured signal representations, system dynamics, and multiresolution principles into ML architectures. Signal processing tools such as wavelet transforms, sparse bases, and time-frequency decompositions enable hierarchical, localized data encoding, making learned features more transparent and physically meaningful. Concurrently, systems theory provides tools to analyze information flow, ensure stability, and enforce structural constraints. Together, these methods promote interpretability, reliability, and computational efficiency, enabling models that are more aligned with the structure and constraints of real-world signals.

#### 1.4.1.1 Exploring conventional design paradigms and beyond

In two-dimensional settings, the **3×3 convolution filter** has long been an architectural workhorse in CNNs because it is small-support, (optionally) symmetric when explicitly constrained, and computationally efficient. Such kernels underpin seminal architectures such as VGGNet [90], ResNet [38], U-Net [39], and EfficientNet [91], and they remain a default choice in both research and deployment pipelines. From a signal-processing standpoint, this prevalence is not arbitrary: a 3×3 convolution is a linear local operator, consistent with the language of **linear systems**, **spatial filtering**, and **localized frequency decomposition**. With stride one (and away from boundary effects), convolution is **translation/shift equivariant** in the sense that spatial shifts in the input induce corresponding shifts in the feature maps. If the kernel is additionally constrained to be even-symmetric, the *linear filtering step* is approximately **zero/$\pi$-phase** (i.e., it does not introduce spatial shift), which helps preserve geometric alignment of local structures at that layer. A 3×3 convolution can be viewed as a 9-tap finite impulse response (FIR) filter applied over a discrete grid. Finite-support FIR filters are **non-recursive** and are BIBO-stable for bounded coefficients, making them structurally simple and analytically transparent. CNNs built from 3×3 kernels inherit the same non-recursive, local filtering structure at each linear layer; when stacked with nonlinearities, they remain tractable at the layer level while still being expressive enough to approximate highly nonlinear functions. In the frequency domain, the 2D Fast Fourier Transform (FFT) of a 3×3 kernel yields a coarse but meaningful frequency response. A single 3×3 filter cannot realize sharp spectral selectivity, but it can implement localized low-pass, band-pass, high-pass, or directional responses, which are well suited for extracting **edges**, **textures**, and other **structural primitives**. When layered, 3×3 filters act as **cascaded FIR stages**, expanding the network's **effective receptive field** and enabling progressively richer selectivity—conceptually analogous to multistage filtering in classical DSP. For example, three stacked 3×3 convolutions (with unit stride) yield an effective receptive field comparable to a 7×7 kernel, while retaining parameter efficiency and inserting additional nonlinearities that improve expressivity. This architectural modularity introduces a powerful inductive bias: complex structure in natural data can be decomposed hierarchically using local operations. This bias has translated into strong generalization across domains such as **computer vision**, **biomedical imaging**, and **remote sensing**. In vision tasks, 3×3 filters serve as robust local operators for edge and texture extraction, supporting state-of-the-art performance on datasets like ImageNet [92] in models such as VGG and ResNet. In biomedical imaging, U-Net and related encoder–decoder variants rely heavily on repeated 3×3 convolutions at each stage to extract fine-grained features essential for **cell segmentation**, **tumor boundary detection**, and **organ delineation** [39, 93]. In remote sensing, CNNs employing 3×3filters have achieved success in land-cover classification, satellite image segmentation, and aerial object detection, effectively capturing **localized spectral–spatial patterns** while remaining computationally tractable [94]. In remote sensing, CNNs employing 3×3 filters have achieved strong results in satellite image semantic segmentation (e.g., cloud and vegetation/forest mapping), while remaining computationally tractable [94] The consistency of these results underscores the **domain-agnostic utility** of the 3×3 kernel as a broadly adaptable, theoretically interpretable building block.

When $3 \times 3$ kernels are constrained to be **symmetric about the kernel center**, they can approximate **linear-phase FIR** behavior, or **zero/$\pi$-phase FIR** behavior under a centered filtering convention, at the level of the linear convolution. This helps maintain spatial alignment of features through that filtering step and is useful in tasks such as **denoising**, **segmentation**, and **restoration**. Beyond engineering convenience, the $3 \times 3$ filter also admits a degree of **biological plausibility**. Studies have reported that early CNN layers often learn **Gabor-like**, **Sobel**-like, or **Laplacian**-like patterns reminiscent of oriented and center-surround operators associated with early vision. For example, Vemuru (2020) [95] demonstrated spiking networks that learn **orientation-selective Gabor-like filters**, akin to those found in early cortical processing. Analyses of trained CNNs report that first-layer filters often resemble oriented edge and center-surround operators [17,96]. More recently, Doutsi et al. (2018) [97] developed a **retina-inspired filter**, further reinforcing links between learned operators and classical structures in visual neuroscience. The practical success of the $3 \times 3$ kernel is also supported by **optimized backend implementations**. Libraries and inference engines such as cuDNN and TensorRT provide efficient convolution kernels for many small-filter cases, using implementation strategies such as Winograd variants when applicable, **vectorized execution**, and **memory-aware tiling**. As a result, $3 \times 3$ convolutions remain well suited to settings where compactness and computational efficiency are important, including **real-time vision**, **mobile inference**, and **low-power embedded deployment**. In summary, the $3 \times 3$ convolutional filter is not merely a historical design choice. It reflects a convergence of signal-theoretic structure, biological motivation, and computational efficiency. As the field progresses toward more **interpretable**, **trustworthy**, and **efficient** AI models, the $3 \times 3$ kernel remains a foundational primitive that is compact, analyzable, and suitable as a structural core for efficient convolutional architectures.

Despite their foundational role, 3×3 filters exhibit inherent **spectral limitations**. A single 3×3 convolution filter, while effective at extracting local features, has a **limited frequency response** due to its small spatial support. It lacks the sharp spectral selectivity necessary for capturing **very narrowband** or **high angular resolution** textures, and it provides limited context for modeling **long-range dependencies**. Classical digital signal processing (DSP) addresses such limitations using **longer FIR filters**, **recursive IIR systems**, or **wavelet-based multiresolution transforms**. In deep learning, similar needs have prompted the exploration of architectural alternatives, such as **ViTs**, which replace localized convolutional operations with **global self-attention**, allowing **non-local**, **token-wise interactions** across the entire spatial domain. However, ViTs often sacrifice inductive bias, require substantial pretraining, and introduce interpretability challenges. Notably, modernized convolutional designs such as ConvNeXt [98] reaffirm the strength of pure ConvNet backbones, while hybrid architectures such as MobileViT [99] combine 3×3 convolutions with transformer blocks to inject non-local interactions, thereby retaining the **efficiency and structure** of small convolutions in real-world deployments. From a signal processing and system-theoretic perspective, the standard convolutional layers used in CNNs can be interpreted as **FIR** filters. They process a fixed window of input values through linear convolution, thereby encoding **finite memory** without internal state recurrence. For bounded coefficients, this design guarantees **BIBO stability**, structural simplicity, and analytical tractability, but it makes long-context modeling less parameter-efficient and can limit **narrowband frequency selectivity** without increased depth, dilation, or multiscale structure. In contrast, **IIR filters**, which introduce recursion through feedback, offer a more compact and expressive representation of systems with temporal or spatial continuity (subject to stability constraints). IIR-based models can achieve longer effective memory with fewer parameters, aligning more naturally with continuous-time or autoregressive system dynamics. Similarly, **Volterra series** offer a powerful mathematical framework for representing **nonlinear dynamical systems**. Volterra models describe system outputs as a **functional expansion over higher-order input terms**, incorporating **linear**, **quadratic**, and higher-degree interactions across input delays. A second-order Volterra kernel, for instance, captures both **linear** and **bilinear** interactions across input delays, offering a principled way to model nonlinear yet causal input-output dynamics. Recent research, including Volterra neural networks [76] and VolterraNet [53], demonstrates the feasibility of incorporating such models into **differentiable architectures** that blend nonlinearity with interpretability.

Despite the richness of these classical models with high iterpretability, their integration into deep learning frameworks was historically constrained by several **technical and practical barriers**:

1. *Recursive gradient instability*: IIR filters require internal state propagation, which introduces challenges in computing stable gradients through feedback loops, often resulting in training instability.
2. *Combinatorial complexity*: Full-rank higher-order Volterra kernels require **multidimensional parameter tensors** whose size scales exponentially with input dimension and kernel order, making naïve implementations of high-dimensional computationally expensive without enforced sparsity or structural constraints.
3. *Wavelet incorporation challenge*: Classical wavelet transforms are differentiable linear filter-bank operations with respect to the input, but discrete wavelet analysis and synthesis were not originally included as common built-in neural-network operators in mainstream deep learning libraries. Their use in neural networks, therefore, requires careful implementation of filtering, decimation, upsampling, boundary handling, and, for learnable wavelets, constraints that preserve valid filter-bank behavior and well-defined gradient flow.
4. *Tools and hardware constraints*: Until recently, mainstream libraries like TensorFlow and PyTorch have limited native support for recursive filtering, structured memory mecha-nisms, or multiscale system kernels with automatic differentiation. This led most deep learning architectures to default to widely supported operations, such as convolutions, ReLUs, and self-attention.

With recent advancements in GPU-accelerated computation, **differentiable signal processing modules**, and new architectural innovations such as **structured state-space models** (SSMs), including Mamba [100] and related Mamba-based architectures [101], many of these historical barriers are now being addressed. Such models support scalable training of **state-space and recurrence-based dynamics**, thereby enabling the practical incorporation of memory-bearing operators into deep networks. In parallel, differentiable implementations of **wavelet filter banks**, structured Volterra approximations, and nonlinear kernel operators have renewed interest in combining **larger or even-sized FIR filters**, **recursive IIR systems**, and **multiscale representations**. This convergence provides a pathway for addressing the **spectral**, **spatial**, and **temporal** limitations of conventional deep architectures. Moreover, these developments are aligned with current efforts in XAI and Reliable AI (RelAI). Recent studies note that the interpretability, robustness, and semantic alignment of modern neural networks can be limited by **unstructured and opaque internal representations** [102, 103]. XAI addresses this issue through **post hoc explanations** as well as through **intrinsically interpretable model design**. In parallel, RelAI emphasizes **predictive reliability**, **robustness to perturbations**, **distributional generalization**, and **confidence calibration** [104, 105]. Incorporating **system-theoretic constraints** into modern ML architectures offers one principled route toward these goals. **Wavelet-based decompositions** yield sparse, energy-localized, and scale-adaptive features, thereby supporting interpretability. **IIR-inspired recurrence** can enable efficient modeling of dependencies over extended horizons. Meanwhile, **Volterra-based representations** provide structured descriptions of nonlinear transformations with analytical tractability. Collectively, these advances indicate a shift toward architectures that are more expressive and data-efficient, while also being more interpretable, reliable, and scientifically grounded.

#### 1.4.1.2 Multiresolution pyramids and system-theoretic learnable graphs

In this thesis, a multidimensional AI system is viewed through four interdependent stages, namely input acquisition, knowledge representation, learning, and inference. In standard deep learning practice, these stages are often absorbed into highly parameterized models whose internal representations are difficult to inspect. Such models can be effective in many tasks, but they may obscure how input information is encoded, transformed, and used. A system-theoretic approach provides a modular alternative in which the main stages of the pipeline are represented through structured operators. Concepts from signal processing and dynamical systems, including linearity, stability, causality where applicable, multiresolution decomposition, and structured recurrence, provide tools for designing efficient, analyzable, and generalizable architectures. Despite the rich mathematical heritage behind multidimensional signal and system theory, several of these principles remain less common than unconstrained neural layers in modern AI.

The main difficulties considered in this thesis are the following:

1. The *combinatorial parameter growth* of high-order Volterra models
2. The *training instability* that can arise in recursive or IIR-based networks without suitable constraints
3. The *limited native layer support* for classical transforms such as wavelets and shearlets in mainstream ML frameworks, especially when analysis, synthesis, filtering, subsampling, upsampling, boundary handling, and learnable constraints must be included in the computational graph
4. The *lack of standardized libraries* for structured, domain-aware models in mainstream ML frameworks

This thesis addresses these challenges by integrating multiresolution pyramids, recursive memory mechanisms, and Volterra-kernel-based nonlinear operators into a unified backpropagation-based deep learning framework. These constructs bridge signal and system theory with trainable neural architectures. In the proposed view, multiresolution filter banks provide scale-localized representations, recursive mechanisms support efficient modeling of long-range dependencies, and Volterra-kernel-based operators provide structured nonlinear mappings. Together, these components are designed to embed domain-specific priors, capture scale and memory interactions, impose structural regularity on learned mappings, and support interpretable, parameter-efficient representations. In multidimensional signal domains, data often exhibit spatial, temporal, or spectral regularities. Representative examples include the following.

1. **1D signals** include time series, biosignals, speech, and control trajectories
2. **2D signals** include natural images, spatial fields, spectrograms, and physical measurements on grids
3. **3D signals** include volumetric MRI and CT scans, video sequences, hyperspectral data, and LiDAR data when represented in structured spatial or spatiotemporal form
4. **Higher-dimensional settings** include multisensor fusion, spatiotemporal arrays, and system identification settings with joint input-output representations

In these signal domains, incorporation of priors such as spatial coherence, frequency selectivity, multiscale organization, and temporal dynamics can improve generalization, interpretability, and sample efficiency when the priors are appropriate for the data and task. However, extending traditional 2D models, such as CNNs with $3 \times 3$ kernels, to $D$-dimensional settings raises theoretical and practical challenges, especially when expressive yet structured representations are required.

### 1.4.2 Review of wavelets, multiresolution analysis, and multirate PR

Wavelets furnish localized building blocks for square-integrable signals, offering a principled compromise between temporal (spatial) locality and spectral selectivity. The modern construction knits together three equivalent viewpoints: (i) **multiresolution analysis (MRA)**, a ladder of approximation spaces; (ii) **two-scale relations** for a scaling function and a wavelet; and (iii) **multirate filter banks** with exact perfect reconstruction. This recall develops the chain in a compact yet detailed form and ends with implications for tight frames, undecimated variants, and Analysis–Processing–Synthesis (AProcS) layout in encoder-decoder architectures.

#### 1.4.2.1 Multiresolution analysis (MRA)

An MRA of a sequence in $L^2(\mathbb{R})$ inside closed orthogonal subspaces $\{V_j\}_{j\in\mathbb{Z}}$ satisfies the following conditions:

1. **nestedness** $\{0\} \subset \cdots \subset V_{-1} \subset V_0 \subset V_1 \subset \cdots \subset L^2(\mathbb{R})$ as shown in Figure 1.6,

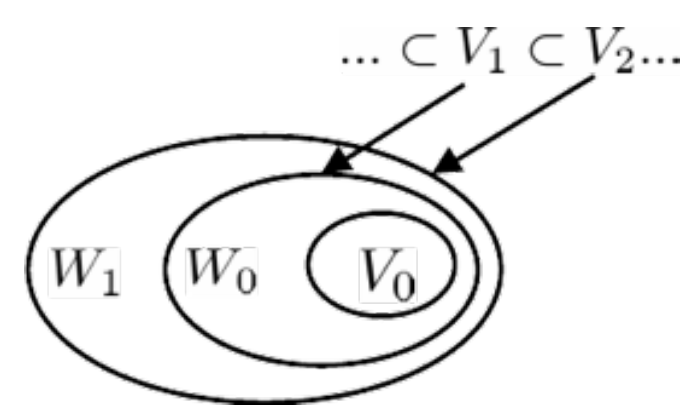

Figure 1.6: Orthonormal subspaces in $L_2(\mathbb{R})$

2. **density or separation** $\overline{\cup_j V_j} = L^2(\mathbb{R})$ and $\cap_j V_j = \{0\}$,
3. **dilation invariance**: $f \in V_j \iff f(2\,\cdot) \in V_{j+1}$, and
4. **translation invariance** of $V_0$. There exists a **scaling function** $\phi \in V_0$ whose integer shifts form an orthonormal basis of $V_0$. The refinement from $V_j$ to $V_{j+1}$ introduces a **detail space** $W_j$ so that

$$V_{j+1} = V_j \oplus W_j, \qquad L^2(\mathbb{R}) = \overline{\bigoplus_{j\in\mathbb{Z}} W_j} \tag{1.19}$$

Scaled and shifted atoms,

$$\phi_{j,k}(t) = 2^{j/2}\,\phi(2^j t - k), \qquad \psi_{j,k}(t) = 2^{j/2}\,\psi(2^j t - k) \tag{1.20}$$

preserve energy across scales (the $2^{j/2}$ factor) and form orthonormal sets.

**Two-scale (refinement) relations** The scaling function obeys

$$\phi(t) = \sqrt{2}\sum_{k\in\mathbb{Z}} h[k]\,\phi(2t - k), \tag{1.21}$$

with a **low-pass** sequence $h$. A (mother) wavelet is generated by

$$\psi(t) = \sqrt{2} \sum_{k \in \mathbb{Z}} g[k]\, \phi(2t - k), \tag{1.22}$$

with a **high-pass** sequence $g$. In frequency domain, with

$$H(\omega) = \tfrac{1}{\sqrt{2}} \sum_{k} h[k] e^{-i\omega k}, \qquad G(\omega) = \tfrac{1}{\sqrt{2}} \sum_{k} g[k] e^{-i\omega k}, \tag{1.23}$$

we obtain

$$\widehat{\phi}(\omega) = H\left(\tfrac{\omega}{2}\right)\, \widehat{\phi}\left(\tfrac{\omega}{2}\right), \qquad \widehat{\psi}(\omega) = G\left(\tfrac{\omega}{2}\right)\, \widehat{\phi}\left(\tfrac{\omega}{2}\right) \tag{1.24}$$

Figure 1.7 shows the Db2 wavelet, its scaling function and the corresponding impulse responses.

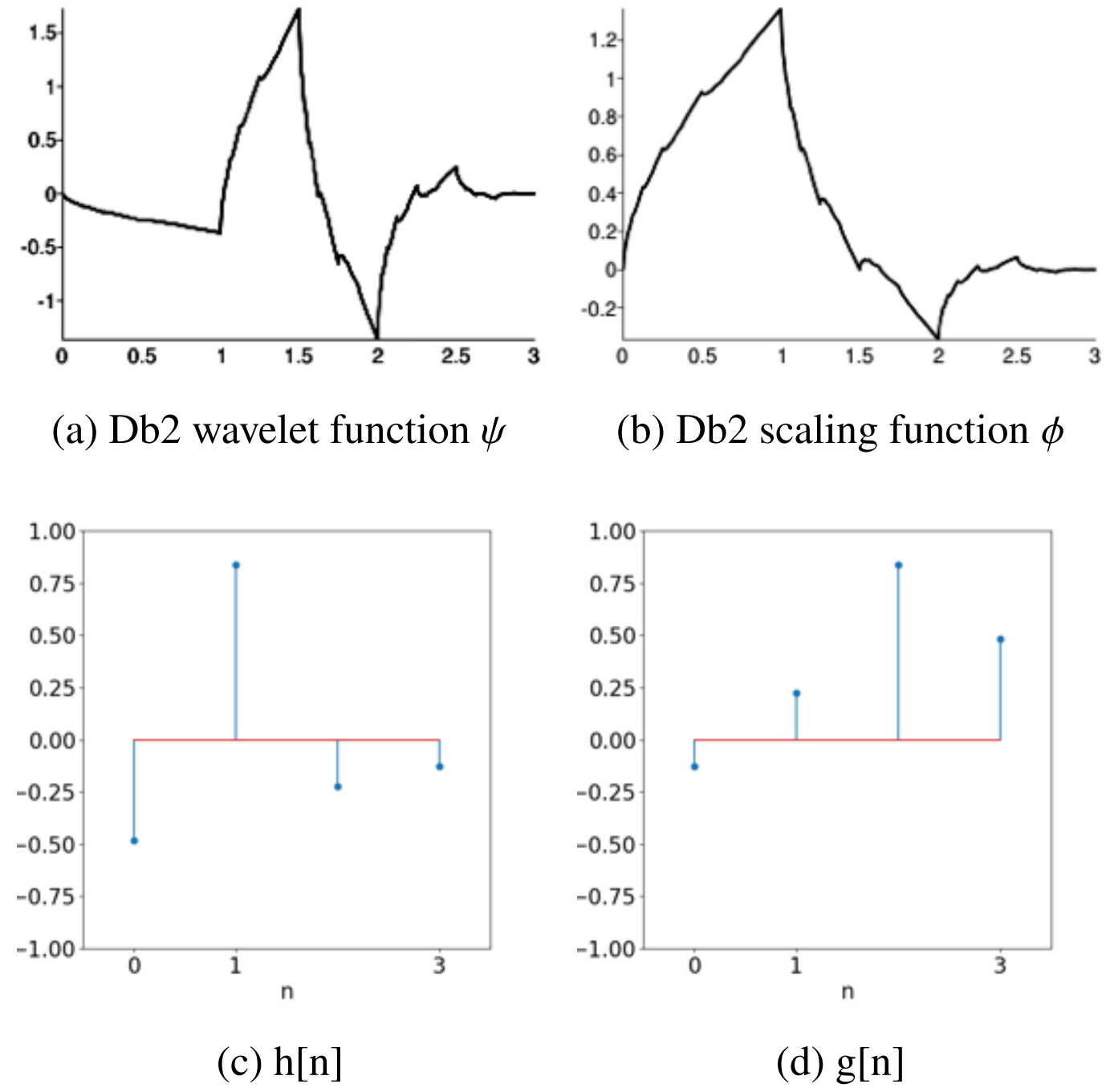

(a) Db2 wavelet function $\psi$ (b) Db2 scaling function $\phi$

(c) h[n] (d) g[n]

Figure 1.7: Db2 wavelet: (a) Wavelet function $\psi$, (b) Scaling function $\phi$; Impulse responses (c) $h[n]$, (d) $g[n]$

**QMF, PR conditions and vanishing moments** For **orthonormal** wavelets, the quadrature-mirror (QMF) relations on the unit circle are

$$|H(\omega)|^2 + |H(\omega + \pi)|^2 = 1, \qquad G(\omega) = e^{-i\omega}\overline{H(\omega + \pi)} \tag{1.25}$$

Equivalently, in time domain,

$$\sum_{k} h[k]h[k - 2n] = \delta_{n0}, \quad \sum_{k} g[k]g[k - 2n] = \delta_{n0}, \quad \sum_{k} h[k]g[k - 2n] = 0. \tag{1.26}$$

Zeros of $H(z)$ at $z = -1$ control **vanishing moments**: an $M$-fold zero implies $\int t^m \psi(t)\,dt = 0$ for $m = 0, \ldots, M-1$. **Haar** ($h = [1,1]/\sqrt{2}$, $g = [1,-1]/\sqrt{2}$) is the simplest example. Figure 1.8 shows the QMF filters in a discrete wavelet transform (DWT) two-band analysis and synthesis filter bank.

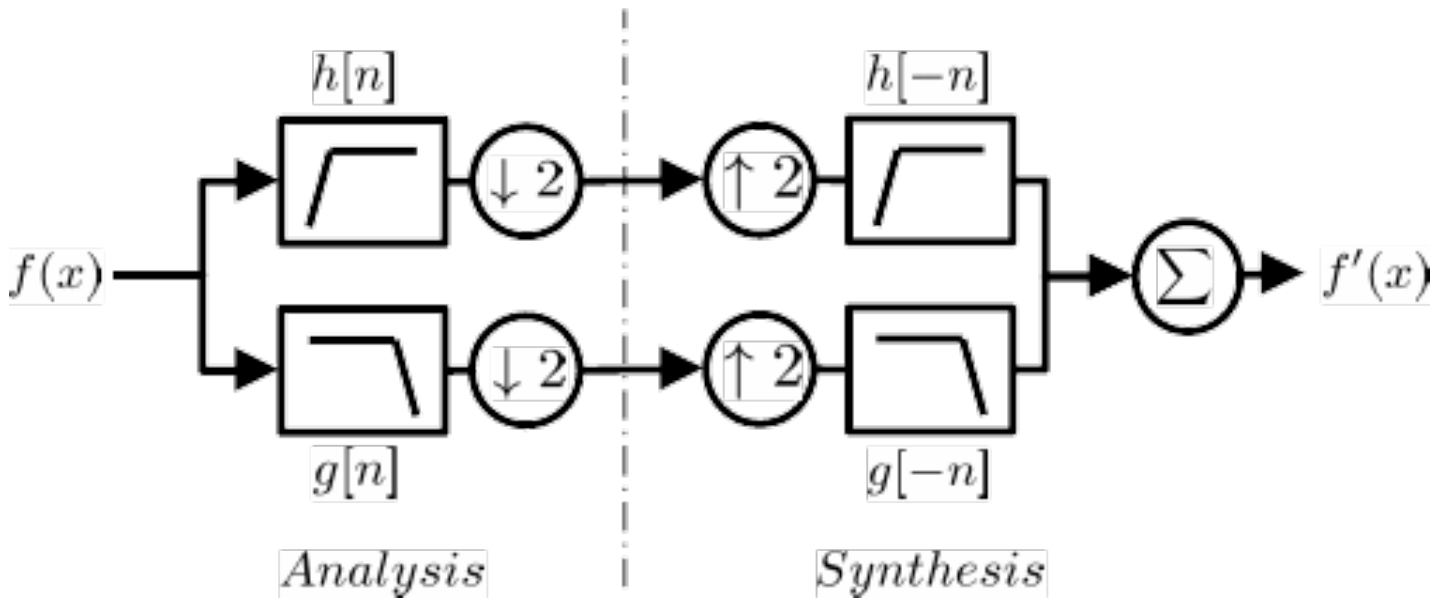

Figure 1.8: Quadrature mirror filters in DWT two-band analysis and synthesis filter bank

#### 1.4.2.2 Two-channel multirate filter banks

Wavelet MRAs are equivalent to critically sampled **two-channel** analysis and synthesis banks. Let the lowpass and highpass analysis filters be $H_0(z), H_1(z)$ and synthesis be $G_0(z), G_1(z)$, which is followed by decimation and upsampling by 2. For input $X(z)$, the reconstructed spectrum is

$$\widehat{X}(z) = \tfrac{1}{2}\big(G_0(z)H_0(z) + G_1(z)H_1(z)\big)X(z) + \tfrac{1}{2}\big(G_0(z)H_0(-z) + G_1(z)H_1(-z)\big)X(-z) \quad (1.27)$$

**Perfect reconstruction (PR)** demands alias cancellation and a pure delay/gain:

$$G_0(z)H_0(-z) + G_1(z)H_1(-z) = 0, \qquad G_0(z)H_0(z) + G_1(z)H_1(z) = 2\,z^{-k} \quad (1.28)$$

**Polyphase form** Write polyphases $H_m(z) = H_{m,0}(z^2) + z^{-1}H_{m,1}(z^2)$. With analysis/synthesis matrices

$$\mathbf{E}(z) = \begin{bmatrix} H_{0,0}(z) & H_{0,1}(z) \\ H_{1,0}(z) & H_{1,1}(z) \end{bmatrix}, \qquad \mathbf{R}(z) = \begin{bmatrix} G_{0,0}(z) & G_{0,1}(z) \\ G_{1,0}(z) & G_{1,1}(z) \end{bmatrix}, \quad (1.29)$$

PR reduces to the compact identity

$$\mathbf{R}(z)\mathbf{E}(z) = z^{-d}\mathbf{I}_2 \quad (d \in \mathbb{Z}) \quad (1.30)$$

**Paraunitary** (orthonormal) banks satisfy $\mathbf{E}^*(z^{-1})\mathbf{E}(z) = \mathbf{I}_2$, hence $\mathbf{R}(z) = z^{-d}\mathbf{E}^*(z^{-1})$ and exact energy preservation. **Biorthogonal** banks allow linear phase via a dual $\widetilde{\mathbf{E}}$ with $\widetilde{\mathbf{E}}(z)\mathbf{E}(z) = z^{-d}\mathbf{I}$.

**Lifting factorization and boundary issues** Any biorthogonal two-channel bank with $\det \mathbf{E}(z) = z^{-d}$ factors as

$$\mathbf{E}(z) = \underbrace{\begin{bmatrix} 1 & S(z) \\ 0 & 1 \end{bmatrix}}_{\text{update}} \underbrace{\begin{bmatrix} 1 & 0 \\ T(z) & 1 \end{bmatrix}}_{\text{predict}} \underbrace{\begin{bmatrix} K_1 & 0 \\ 0 & K_2 \end{bmatrix}}_{\text{scaling}} \mathbf{\Pi}, \tag{1.31}$$

with simple inverses per step, hence PR by construction. On finite grids, exact PR requires consistent **boundary handling** (periodic or symmetric prolongation); zero-padding breaks PR unless compensated.

#### 1.4.2.3 Higher dimensions, M-band, and wavelet packets

Applying 1-D analysis per axis yields **separable** decompositions. In 2-D the subbands are {LL, LH, HL, HH}; in 3-D: LLL and seven high-pass groups. The polyphase/PR algebra extends verbatim to $M$ channels (decimation by $M$) with an $M \times M$ polyphase matrix, and to \$D\$ dimensions with a decimation matrix **M**. **Wavelet packets** iterate splitting on both low- and high-pass nodes, yielding adaptive time–frequency tilings under the same PR constraints.

#### 1.4.2.4 Tight frames, undecimated framelets and energy accounting

A family $\{\psi_i\} \subset \mathcal{H}$ is a **frame** if $A\|x\|_2^2 \leq \sum_i |\langle x, \psi_i \rangle|^2 \leq B\|x\|_2^2$. It is **tight** if $A = B$ and **Parseval** if $A = B = 1$, giving

$$\|x\|_2^2 = \sum_i |\langle x, \psi_i \rangle|^2, \qquad x = \sum_i \langle x, \psi_i \rangle \psi_i \tag{1.32}$$

Undecimated (translation-invariant) wavelet transforms and directional systems (framelets, contourlets, shearlets) can be designed as tight frames: although redundant, they preserve energy ($\mathbf{W}^*\mathbf{W} = \mathbf{I}$) and admit simple synthesis ($\widetilde{\mathbf{W}} = \mathbf{W}^*$). This redundancy improves robustness to noise/erasures and directional selectivity, with PR maintained level-wise.

#### 1.4.2.5 Time–frequency localization and admissibility (CWT vs DWT)

For $x \in \mathrm{L}^2(\mathbb{R})$ with Fourier transform $\hat{x}$, the time/frequency variances obey the **uncertainty bound** $\sigma_t^2 \sigma_\Omega^2 \geq 1/4$. Gaussians reach the bound but are not admissible discrete PR wavelets. In the **CWT**, admissibility demands a band-pass prototype (zero mean and suitable decay), enabling continuous-scale analysis and exact inversion via an appropriate measure; however, CWT atoms do not directly yield critically sampled PR filter banks. By contrast, the **DWT/MRA** fixes dyadic scales, enforces QMF/PR identities, and implements a fast, compact pyramid with exact reconstruction.

#### 1.4.2.6 Heisenberg–Gabor uncertainty and Time–Frequency Tiling

Let $x \in L^2(\mathbb{R})$ with Fourier transform $\hat{x}$. Define time and frequency centers and variances

$$t_0 = \frac{\int t\,|x(t)|^2\,dt}{\int |x(t)|^2\,dt}, \qquad \sigma_t^2 = \frac{\int (t-t_0)^2 |x(t)|^2\,dt}{\|x\|_2^2}, \tag{1.33}$$

$$\Omega_0 = \frac{\int \Omega\,|\hat{x}(\Omega)|^2\,d\Omega}{\int |\hat{x}(\Omega)|^2\,d\Omega}, \qquad \sigma_\Omega^2 = \frac{\int (\Omega-\Omega_0)^2 |\hat{x}(\Omega)|^2\,d\Omega}{\|\hat{x}\|_2^2}. \tag{1.34}$$

The **time–bandwidth product** is $\sigma_t^2\sigma_\Omega^2$. It is invariant under global time or frequency shifts and obeys the lower bound

$$\sigma_t^2\,\sigma_\Omega^2 \;\geq\; \tfrac{1}{4} \qquad \text{(Heisenberg–Gabor inequality).} \tag{1.35}$$

Equality holds for (appropriately centered) Gaussian windows. No finite-energy signal can be arbitrarily concentrated in both domains; this governs the design of atoms and tilings.

**Admissibility and its role (CWT vs DWT)** Wavelet atoms must be **band-pass**. For a continuous wavelet $\psi$ with spectrum $\Psi$, admissibility requires the constant

$$C_\psi \;=\; \int_{\mathbb{R}\setminus\{0\}} \frac{|\Psi(\Omega)|^2}{|\Omega|}\,d\Omega \;<\; \infty, \qquad \Psi(0) = 0 \text{ (zero mean).} \tag{1.36}$$

This ensures perfect reconstruction for the **CWT** via an $s$-weighted inverse. Gaussians minimize $\sigma_t\sigma_\Omega$ but violate $\Psi(0) = 0$; their **derivatives** are admissible and widely used as analytic prototypes. For **discrete dyadic wavelets (DWT/MRA)**, admissibility is replaced by **PR/QMF constraints** on the low/high-pass filters $h, g$, yielding critically sampled, exactly reconstructing pyramids. In overcomplete (framelet) designs, spectral coverage is summarized by the **sum of dilated spectra (SDS)** condition

$$0 < c_1 \;\leq\; \sum_{k\in\mathbb{Z}} \left|\Psi(a_0^k\,\Omega)\right|^2 \;\leq\; c_2 < \infty \qquad (a_0 > 1), \tag{1.37}$$

which guarantees stable analysis and allows a constructive dual $\widetilde{\Psi}(\Omega) = \Psi(\Omega) \big/ \sum_k |\Psi(a_0^k\Omega)|^2$ for synthesis.

**How tilings reflect uncertainty**

- *STFT (Gabor) tiles* are nearly rectangular and **scale-independent**: fixed window $\Rightarrow$ uniform time–frequency cells; excellent for narrowband, locally stationary content.
- *CWT tiles* are **scale-dependent**: wide in time and narrow in frequency at coarse scales, and the reverse at fine scales; this matches multiscale transients.
- *DWT MRA tiles* are **dyadic rectangles** induced by downsampling: exact PR is achieved by QMF/polyphase identities; uncertainty manifests as coarser time support at low frequencies and finer at high frequencies. Directional framelets (e.g., shearlets) further partition orientation while respecting PR level-wise.

**Practical consequences for filter bank design**

1. *Bandwidth vs support*: Short taps (tight time support) yield broad spectral skirts; longer taps sharpen passbands but raise delay/compute and can overfit.
2. *Anti-aliasing with decimation*: To respect uncertainty while sampling sparsely, low-pass branches must sufficiently attenuate beyond the folding band; PR guarantees algebraic cancellation but good stopbands improve numerical robustness.
3. *Energy accounting*: Paraunitary/tight designs satisfy $\mathbf{W}^*\mathbf{W} = \mathbf{I}$, so the time–bandwidth tradeoff redistributes, rather than inflating energy across subbands, easing interpretation and regularization.

### 1.4.3 Analysis–Processing–Synthesis in deep learning with multiresolution banks

In this thesis, the learning and inference paths of the constructed multiresolution deep architectures are viewed through an **A**nalysis-**Proc**essing-**S**ynthesis (AProcS) template. The **analysis** or encoder path is concerned with receiving the input and forming a feature, coefficient, scale, or transform representation. The **processing** path denotes task-specific transformations applied to this analyzed representation, such as filtering, gating, thresholding, attention, recurrence, Volterra-kernel-based nonlinear mixing, or learned subnetworks. The **synthesis** or decoder path, when present, maps the processed representation to the required output domain. This stage is useful in reconstruction, dense prediction, inverse mapping, or decoder-based recovery tasks, while encoder-only architectures may use a task-specific prediction head instead of an explicit synthesis stage. In multiresolution filter-bank theory, related analysis and synthesis banks provide an organized way to represent signals across scale and frequency. The knowledge representation and learnable mappings considered in this thesis are constructed by embedding fixed or trainable multiresolution pyramids into end-to-end differentiable computational pipelines. For example, a classical wavelet signal-processing pipeline can be viewed as analysis by wavelet decomposition, task-specific processing of coefficients, and synthesis by reconstruction. Wavelet denoising follows this pattern through coefficient shrinkage followed by reconstruction. In the present thesis, this idea is extended to deep learning architectures in which multiresolution pyramids may appear in the encoder, decoder, bottleneck, attention module, recurrence module, or task-specific prediction path. Therefore, AProcS is used here as an architectural viewpoint rather than as a strict mathematical factorization. Perfect-reconstruction or approximate-reconstruction behavior is useful as an architectural consistency check when explicit analysis and synthesis transform pairs are used. However, it is not necessarily the learning objective, since many machine learning or deep learning tasks map an input to a different output, such as a class label, a segmentation map, an enhanced image, or an inverse estimate.

#### 1.4.3.1 Analysis and Synthesis in Encoder–Decoder CNN

Standard CNN encoder–decoder architectures, including U-Nets and autoencoders, can be interpreted within the AProcS viewpoint. The encoder provides an analysis-like mapping by projecting the input into progressively coarser and more abstract feature representations, while the decoder provides a synthesis-like mapping by forming the required output from those representations. The intermediate, bottleneck, skip, and prediction layers implement task-specific processing on the encoded features. This interpretation should not be read as a claim that a standard CNN is an exact wavelet or framelet transform. Rather, it highlights the structural analogy between encoder–decoder networks and analysis and synthesis systems. Fujieda et al. (2018) [106] observed that a CNN can be viewed as a limited form of multiresolution analysis and proposed Wavelet CNNs to supplement spectral components that are not explicitly represented in conventional CNNs. Similarly, the Multi-level Wavelet CNN (MWCNN) of Liu et al. (2018) [107] embeds DWT and inverse DWT operations into a modified U-Net structure, using wavelet decomposition for resolution reduction in the contracting path and inverse wavelet reconstruction in the expanding path. This provides an explicit multiresolution mechanism for changing feature-map resolution and improves the tradeoff between receptive field size and computational efficiency.

**Downsampling in filter banks:** In a wavelet filter bank, analysis involves applying low-pass and high-pass filters and then downsampling, yielding a coarse approximation and detailed subbands. CNN encoders achieve downsampling via pooling or strided convolutions, which is analogous to a low-pass filtering and decimation operation. However, a vanilla CNN often discards high-frequency detail information when downsampling. U-Net architectures address this by introducing skip connections that carry high-resolution features (analogous to high-pass detail coefficients) forward so that they can be merged into the decoding stage. This is conceptually similar to explicitly splitting the signal into low- and high-frequency bands and preserving both: indeed, it was observed that the two-part structure of U-Net (contracting path and expanding path with skips) "are very similar to synthesis and analysis concepts in wavelet decompositions". Using an actual wavelet transform for downsampling can thus improve a CNN's design. For instance, replacing max-pooling with a wavelet decomposition (low-pass + high-pass) gives a more theoretically grounded pooling/unpooling scheme. Each downsampling step in an encoder corresponds to splitting into multiple subbands (in practice, increasing channel dimension), and each upsampling in the decoder corresponds to the inverse filter bank merging those subbands back. Several works have leveraged this correspondence: e.g., Ye et al. (2018) formulated the U-Net explicitly as a framelet/wavelet filter bank, adding a dedicated high-pass branch alongside the usual pooling path to ensure perfect reconstruction of feature maps. The result is a "tight frame U-Net" with an additional detail branch and modified skip connections, so that the pooling + unpooling layers behave exactly like an invertible wavelet transform. This tight-frame variant of U-Net was shown to better preserve directional details in image recovery tasks.

**Tight Frames, Paraunitary Filters, and Dual Tying** A key advantage of using an AProcS viewpoint is that explicit frame constraints can be imposed when the architecture contains identifiable analysis and synthesis operators. If $W$ denotes the analysis operator, a Parseval tight frame satisfies $W^*W = I$ up to a normalization constant, where $W^*$ denotes the adjoint of $W$. In this case, the corresponding synthesis operator can be chosen as $\widetilde{W} = W^*$, and the analysis and synthesis pair satisfies the perfect-reconstruction condition $\widetilde{W}W = I$. For a biorthogonal frame pair, $\widetilde{W}$ is the dual synthesis operator and the perfect-reconstruction condition is again expressed through $\widetilde{W}W = I$. These identities apply to the transform pair itself and should not be interpreted as a guarantee that the complete deep network is invertible after arbitrary nonlinear processing. Instead, they provide a reliable identity path when the intervening task-specific processing behaves like an identity map. This property is useful for stability and information flow because it reduces unintended attenuation or amplification caused by the analysis and synthesis stages. In contrast, standard CNN encoders with pooling are generally not strictly invertible, and skip connections are often used to pass lost or attenuated detail information to the decoder. A wavelet or framelet-based encoder–decoder can make this information-preserving path more explicit when the transform pair is designed to satisfy the required frame or perfect-reconstruction conditions. Han and Ye (2018) [108] showed that dual-frame and tight-frame U-Net variants satisfy such frame conditions and can improve recovery of fine image details. In the dual-frame case, the analysis and synthesis filters are designed as paired frame operators. In the tight-frame case, the synthesis can be tied to the adjoint of the analysis operator, giving a more constrained reconstruction path. This form of dual tying makes the identity path more explicit and allows the learned processing layers to focus on task-specific modification of the representation.

Some concrete benefits of tight-frame and dual-tying constraints in the AProcS template are as follows:

- *Perfect reconstruction of the transform path*: If $W$ and $\widetilde{W}$ form a perfect-reconstruction pair, the analysis and synthesis stages can reproduce the input when the intervening processing is the identity.
- *Stability*: Conditions such as $W^*W = I$ preserve the norm of the signal representation up to normalization, which helps avoid unintended amplification or attenuation in the transform stage.
- *Interpretability*: When analysis coefficients synthesize back to the original signal through a known transform pair, intermediate features can be related to signal components such as scales, orientations, or subbands.
- *Efficient learning*: A structured analysis and synthesis pair reduces the burden on learned layers to recover an identity path, allowing them to focus more directly on task-specific transformations.

These ideas are closely related to the convolutional framelets theory, which provides a theoretical lens for interpreting encoder–decoder CNNs. From this viewpoint, architectural elements such as redundant channels, residual paths, concatenated activations, and skip connections can be understood as mechanisms that help preserve information flow and approximate reconstruction behavior. Several works have explicitly incorporated such principles. For example, Cotter and Kingsbury (2019) [109] introduced a trainable wavelet-domain layer based on the dual-tree complex wavelet transform. Their layer maps activations into the wavelet domain, apply learnable operations, and map them back to the spatial domain while supporting backpropagation through the multirate analysis and synthesis system. Such modules are consistent with the AProcS viewpoint because they make the analysis, processing, and synthesis stages explicit inside an end-to-end trainable network.

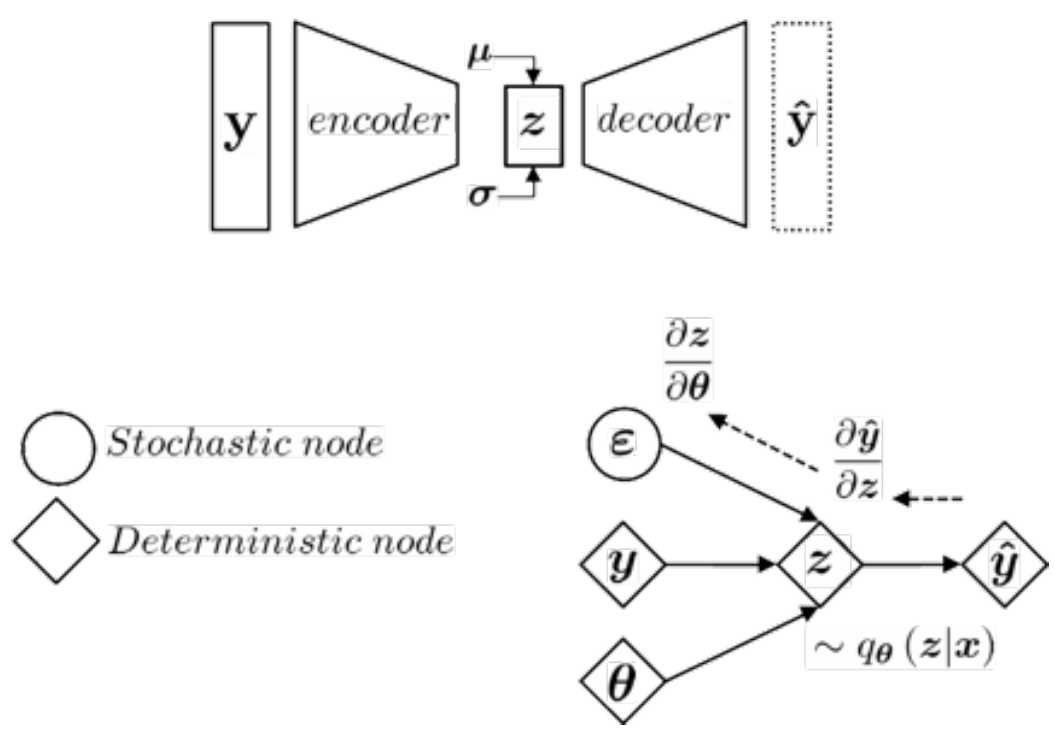

Figure 1.9: Variational Auto Encoder (VAE)

#### 1.4.3.2 VAEs and GANs in the AProcS viewpoint

The AProcS viewpoint can also be used to interpret modern generative and representation-learning architectures such as convolutional variational autoencoders and generative adversarial networks. In a variational autoencoder, including convolutional realizations, the encoder provides an analysis-like mapping from the input to the parameters of an approximate latent posterior, typically the mean and variance or log-variance parameters of a Gaussian distribution. The latent sampling step, together with the associated regularization, acts as a task-specific bottleneck, and the decoder provides a synthesis-like mapping from the sampled latent representation back to the data space. The VAE training objective includes a reconstruction term and a regularization term that encourages the approximate posterior to remain close to a chosen prior, commonly a standard Gaussian. Thus, the latent representation is encouraged to be structured and constrained while retaining information needed for reconstruction or generation. In this sense, VAEs fit naturally within the AProcS viewpoint as architectures with an analysis path, a latent processing stage, and a synthesis path. This interpretation should be understood as an architectural analogy rather than as a strict wavelet or filter-bank factorization.

**GAN generators as synthesis operators** A GAN generator can be interpreted as a synthesis-oriented mapping that takes a low-dimensional latent vector, often sampled from a simple prior distribution, and produces a high-dimensional output such as an image or an audio signal. In an unconditional GAN, there is usually no explicit analysis operator because generation starts from a sampled latent code rather than from an observed input signal. Here, the AProcS interpretation applies primarily to the architecture of the synthesis decoder. In conditional GANs and image-to-image translation networks, an analysis-like pathway is often present because the generator is conditioned on an input image, label, or feature representation. In such cases, the conditioning pathway or encoder can be viewed as the analysis component, the latent or feature manipulation as task-specific processing, and the generator or decoder as the synthesis component.

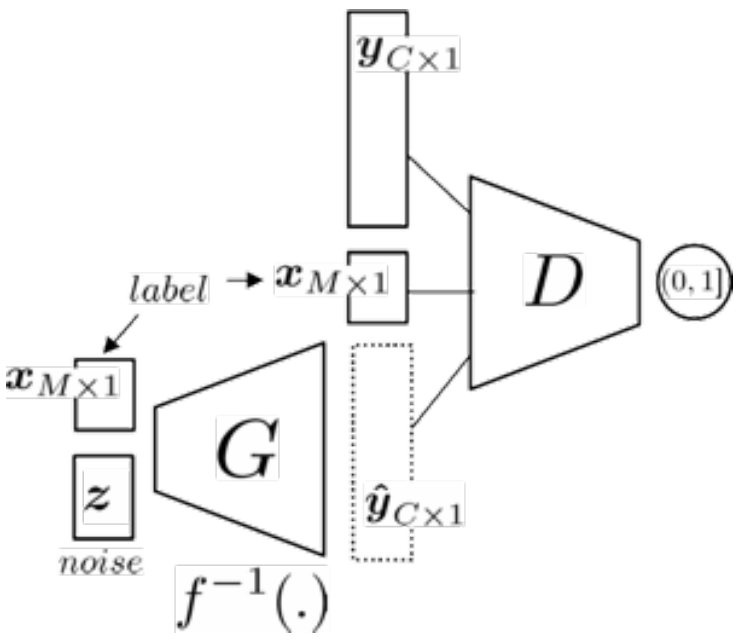

Figure 1.10: Conditional Generative Adversarial Network

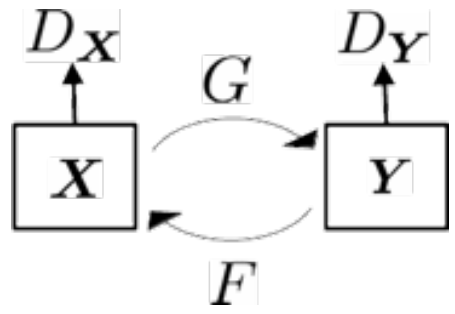

Figure 1.11: Cycle GAN for matching two data distributions

In encoder-only self-supervised methods, the encoder forms the analysis representation, while the projection head, prediction head, contrastive objective, or pretext task provides the task-specific processing used for representation learning. An explicit synthesis stage may be absent in such cases. In reconstruction-based self-supervised methods, such as masked reconstruction or autoencoding, the decoder provides a synthesis path that reconstructs missing or target content from the analyzed representation. Therefore, the AProcS viewpoint can accommodate both encoder-only representation learning and encoder-decoder reconstruction-based learning. Figures 1.9, 1.10, and 1.11 show general schematics of a VAE, a conditional GAN, and a CycleGAN, respectively.

**Example 1.2.** Binary segmentation (MRI skull stripping) by Pix2Pix 2D GAN.

Pix2Pix is a generative network with a U-Net like generator network and a simple sequential discriminator network. Figure 1.12 shows segmentation by a Pix2Pix 2D GAN of a slice from IBSR-2 dataset.

**Note:** This generative pipeline takes around 3000 epochs for the generator (a U-Net) to learn accurate segmentation. On the other hand same generator in a deterministic pipeline will require

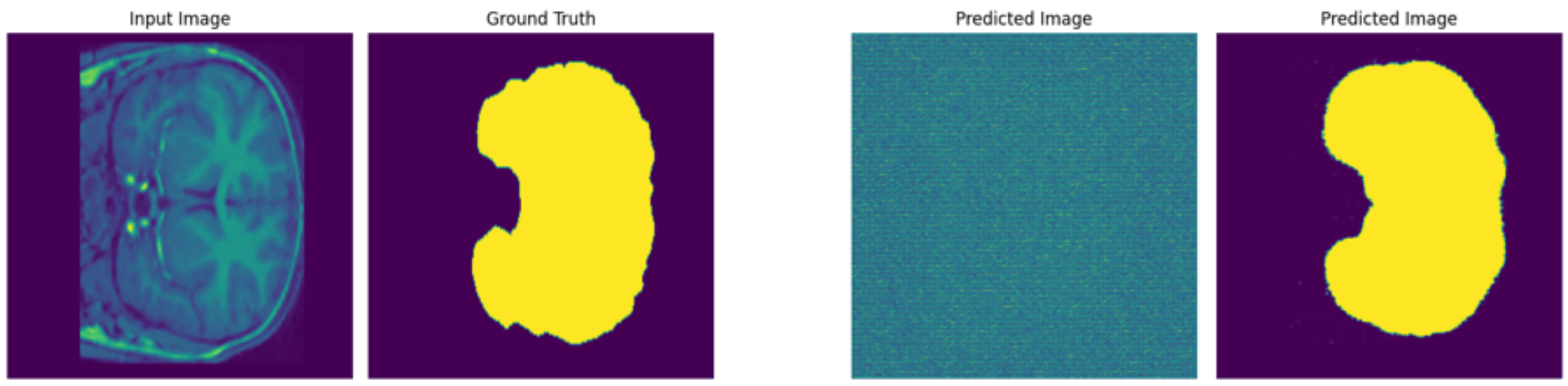

Figure 1.12: (left to right) Input and ground truth mask; generator output at epoch 0 and epoch 3000

less than 100 epochs for the same task. However, the advantage of the generative model is that is gives us a organized feature space and makes it somewhat stable to minor perturbations after sufficient training. Thus, selecting a generative and deterministic pipeline is problem specific and has its own trade-offs.

**Example 1.3.** White matter segmentation of brain MRI using self-supervised learning by 3D WNet

A 3D WNet [110] is a fully unsupervised architecture with a U-Net like encoder and a U-Net like decoder. It has two loss functions namely, a soft normalized cut loss and a reconstruction loss. With some labels the former loss self supervises the model to learn the segmentation masks. Figure 1.13 shows the segmented results at the output of the encoder of WNet and the decoder output is a near perfect reconstruction of the input patch.

**Band-domain processing** Across the examples discussed above, the task-specific part of the model often acts on an encoded, latent, multiscale, or transform-domain representation of the data. When this representation is explicitly organized into multiresolution subbands, we describe it as a band-domain representation. For denoising or enhancement tasks, the task-specific transformation may attenuate selected subbands, for example, by thresholding high-frequency noise, or may apply learned nonlinear filtering to the coefficients. For segmentation or classification tasks, it may consist of convolutions, attentional gating, recurrent operations, Volterra-kernel-based nonlinear mixing, or a learned subnetwork operating on multiscale features. For example, the scattering transform of Bruna and Mallat (2013) [35] can be interpreted in this viewpoint as a fixed wavelet-based analysis followed by modulus nonlinearities and averaging operations, with no explicit synthesis stage because the goal is to produce stable and invariant features rather than reconstruct the input. A classifier head may then be applied to the scattering coefficients for recognition tasks. As another example, in the Learnlets architecture of Ramzi et al. (2020), an invertible wavelet transform is combined with learned nonlinear shrinkage on the wavelet bands. This combines classical sparse wavelet-domain processing with learned adaptation for image denoising. In deep encoders that explicitly use multiresolution or subband representations, operations such as ReLU activations, $1 \times 1$ convolutions, gating, and channel mixing can be interpreted as learned transformations that recombine information across subbands or feature channels before the next analysis, synthesis, or prediction stage. Cotter et al. (2020) [109] emphasizes that wavelet-domain filtering can be advantageous because filters or gating functions can be learned on wavelet subbands to obtain useful invariances with parameter-efficient representations. Overall, such transformations provide a principled way to incorporate prior knowledge about noise, signal, texture, and cross-band structure, while retaining the flexibility of learned deep models.

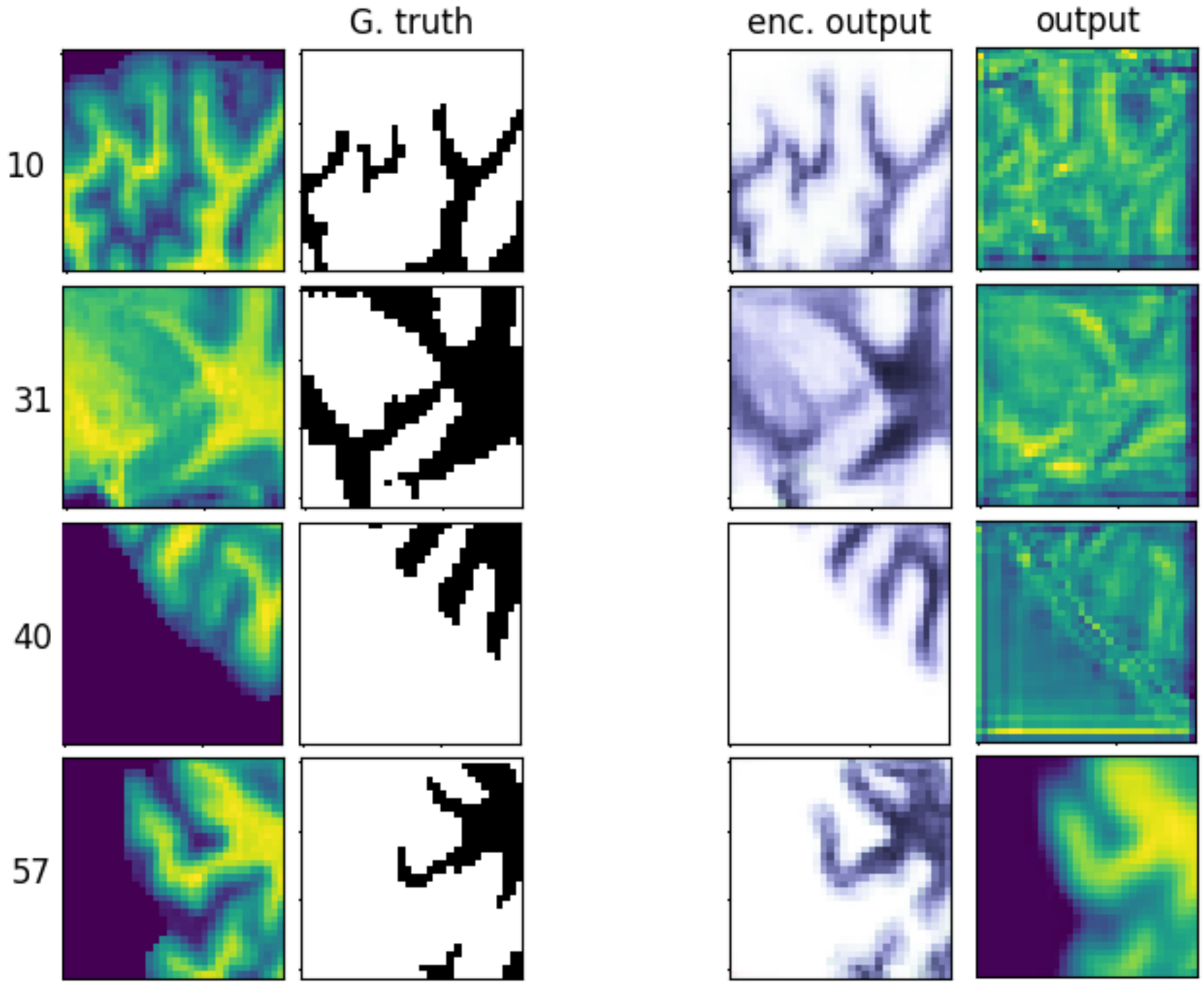

Figure 1.13: WM segmentation by WNet 3D; (left to right column) input patches, ground truths, segmented white matter at encoder output, near perfect reconstruction at decoder output.

In summary, the AProcS viewpoint provides a useful language for describing multilevel and multiresolution deep neural architectures. It helps separate the roles of representation formation, task-specific transformation, and output formation without requiring every model to follow a strict three-block mathematical factorization. In encoder-decoder architectures such as U-Nets, autoencoders, and conditional image-to-image GANs such as Pix2Pix, downsampling, filtering, and convolutional operations can be understood as analysis-like steps, nonlinear and learned modules as task-specific transformations, and upsampling, decoding, or reconstruction operations as synthesis-like steps [39, 111]. This perspective also motivates architectures that incorporate useful properties from wavelet, scattering, and frame theory. These properties may include structured scale separation, approximate or exact reconstruction of the transform path, improved information flow, and interpretable subband representations when the corresponding architectural constraints are explicitly imposed. The works of Bruna and Mallat on scattering transforms [35], Fujieda et al. on Wavelet CNNs [112], Liu et al. on MWCNN [107], Bastidas-Rodriguez et al. on DAWN [113], Cotter and Kingsbury on wavelet-domain networks [109], and Han and Ye on deep convolutional framelets [108] exemplify this connection between multiresolution signal processing and deep learning. The resulting viewpoint supports the systematic design of neural networks that are more structured, interpretable, and compatible with classical signal-processing principles, while retaining the adaptability of learned models.

## 1.5 A brief overview of the accompanying chapters

The necessity of data-driven or AI models arose for solving highly nonlinear problems that are otherwise really hard to achieve with conventional design methods. An AI model is often a highly interconnected graph whose nodes contain linear and nonlinear systems that transform input data (a multidimensional signal) into a knowledge base. In this thesis, the model design paradigm natively supports multiresolution pyramids for structured interpretable representation of knowledge, which reduces the effective datapoints and automatically introduces economy by reducing the number of trainable parameters. The following chapters describe and introduce new foundational building blocks for developing interpretable and frugal AI solutions to some industrial and medical applications.

**Chapter 2** describes the construction of signal and system representations with Fourier bases, frames, and multidimensional multiresolution filter banks in the learnable path. These individual systems provide structured coefficient spaces for the incoming signal and support efficient feature construction for downstream learning objectives. It describes the construction of fast D-dimensional convolutions (which currently have limited software support) that will be useful in Chapter 3 for realising higher-order natural domain Volterra kernels, where a quadratic shift-invariant kernel for 2D input requires a 4D convolution. Similarly, a quadratic shift-invariant kernel for 3D input requires a 6D convolution. A Python TensorFlow library `convd` packages D-dimensional convolutional layers that currently support up to 9D convolutions required for realising cubic Volterra kernels for 3D volumetric input. Chapter 2 further describes parametric static nonlinear modeling with Taylor series and Padé (rational-function) approximation for use after linear operations in Linear-Nonlinear cascade systems with backpropagation support. This generalizes modeling to a broad class of smooth nonlinear activation forms, and the described algorithms are packaged in `LNPoly` Python TensorFlow library for later use in a novel, expressive inverse model described in Chapter 3. The current version supports up to 3D I/O. Bases, frames, and filter banks in deep learnable graphs are a key idea in this thesis for constructing efficient and structured representations of multidimensional data in an AI model. Some such relevant transforms are also realised in Chapter 2 for constructing end-to-end learnable pipelines with natural-frequency domain localized and sparse representations described in later chapters. Ramanujan frames identify periodicities in very short-duration signals where Fourier and wavelet transforms face limitations due to the uncertainty principle. In a later chapter, one of the inverse model realizations uses Ramanujan frames to construct unique sparse representations of very short input sequences of length 8 and 14. A new Python library `RamanujanFrame` packages the described computations. The multiresolution and directional filter banks described here are not new, but using them in a learnable path is not very common. Chapter 2 also addresses and overcomes the challenges of realising such filter banks for multidimensional data in end-to-end learnable paths. It describes DWT (1D, 2D, 3D) filter banks with backpropagation, which are packaged in a new Python TensorFlow library `TFDWT`. Similarly, several fixed and learnable multidimensional (2D, 3D) directional transforms (curvelet, contourlet, shearlet, scattering, etc.) with backpropagation support are also described and packaged in a Python TensorFlow library `filterbanks`. These filter banks tile the frequency plane for natural-frequency domain localization of features in multidimensional data. They are very useful in capturing both isotropic and anisotropic features in images and volumetric data efficiently. The theory of these filter banks is not new, but their fast tensor realizations with backpropagation support (limited open software availability) introduced here are built from scratch. These individual systems of Chapter 2 (and their software realizations) form a minimal, self-sufficient logistical infrastructure required to construct efficient learnable models in later chapters with structured representations.

**Chapter 3** extends the theory of Volterra kernels by introducing a biorthogonal-basis representation of multidimensional $m^{th}$ order Volterra systems, where the Volterra kernels, inputs, and outputs are represented in compatible biorthogonal coefficient spaces. A computational library `VolterraSys`, accompanying this chapter, contains the described fast algorithmic realizations of linear and quadratic Volterra kernels in both the natural and wavelet (orthogonal and biorthogonal) domains for up to three-dimensional I/O systems (supporting up to 9D convolutions, i.e., cubic terms) with backpropagation support. This enables the Volterra kernel and its I/O to natively embrace the multiresolution pyramids introduced in Chapter 2 and learn sparse structures during a learning cycle. Chapter 3 further describes an application with a design of an expressive inverse model for retrieving atmospheric profiles from the observations of a ground-based Microwave Radiometer. In this inverse model, a Volterra kernel and a cascade of rational polynomial heads in the natural and wavelet domains are used to map two functions of different independent variables (microwave frequency and altitude). This new inverse model is packaged in `InVeRt` Python TensorFlow library. The results show promising retrieval of temperatures and challenging water vapor densities up to 10 km in the troposphere using very efficient, lightweight model constructions. Ablation studies with the current state-of-the-art methods compare the retrieval results.

**Chapter 4** describes realizations of advanced learnable modules for 2D and 3D I/O deep vision models. It starts with the descriptions of algorithmic realizations of separable and nonseparable multidimensional IIR filters and approximations with backpropagation support for CNN-like networks. These described IIR realizations constrain the learnable coefficients to overcome some of the practical challenges of realising multidimensional nearly stable learnable IIR filters. Chapter 4 then uses the DWT and learnable 3D FDST banks described in Chapter 2 to introduce two multiresolution attentions, WaveletViT and ShearViT, which are later used in Chapter 5 for MRI segmentation models. These two attentions do not use explicit positional encoding like in vanilla ViTs (computationally expensive) and tokenize the localized natural-frequency domain subbands. Unlike vanilla ViTs, these two multiresolution attentions are economical and suitable for training pipelines with scarce training data without risk of over-parameterisation. By introducing a DWT backbone and axial self-attention, WaveletViT embeds multiresolution locality while reducing token/compute budgets, thereby improving generalization under limited data. Similarly, the Fourier space subband tokens naturally embed positional information in the ShearViT. These two newly introduced multiresolution self-attentions are packaged in separate `WaveletViT` and `ShearViT` Python libraries, and are used in the construction of deep segmentation models in the next chapter.

**Chapter 5** constructs efficient classifiers and segmenters for audio, image, and MRI volumes with structured filter bank representations of inputs and features. It describes designs of audio and image classifiers embedded with MRA banks and Volterra heads for dense mapping. Here, Volterra kernels replace the MLP neural network heads, and experiments show that they produce similar classification with very few parameters. IIR filters, multiresolution filter banks, and novel attentions (WaveletViT, ShearViT) of Chapters 2 and 4 are used here to construct image and 3D MRI binary and multiclass segmenters, including WaveNETR and ShearNETR. Experiments with multiple datasets show that IIR-based segmenters learn detailed fine structures in very few learning cycles, but the duration of a single iteration is much longer than that of FIR-based segmenters. Experiments also show that segmentation models with WaveletViT and ShearViT bottlenecks, especially the convolutional variational ones, are comparatively better than U-Net with ViT in both performance and economy. Chapter 5 further describes an inference pipeline for longitudinal analysis of unlabeled brain 3D MRI volumes of ischemic stroke patients undergoing Ayurvedic intervention. This inference pipeline has a novel binary segmentation model for skull stripping, a novel multiclass model for brain tissue segmentation, along with the necessary preprocessing and registration steps. This pipeline finally records the changes in brain tissues in 2 or 3 consecutive MRI scans of a patient over a few months and may now serve as a rapid testing tool to assist Doctors and medical practitioners.

# Chapter 2

# Signal-theoretic and Multiresolution Representations

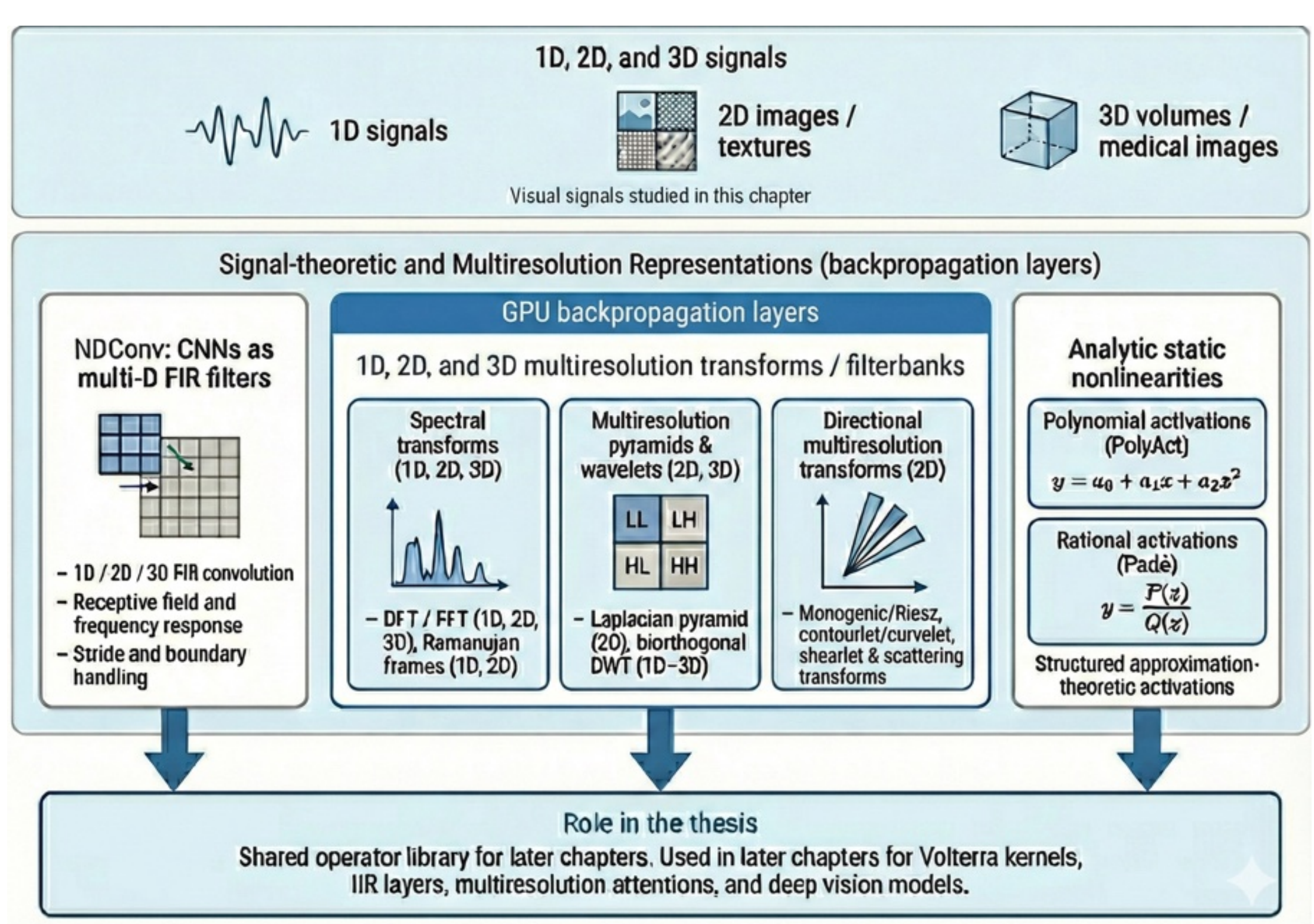

The path from raw data to reliable decisions in an AI model passes through a sequence of data transformations that convert measurements into task-relevant representations for a given inference task. In a trained model, the information needed for this task is distributed across the learned parameters and the intermediate features produced by its computational units. These parameters are updated by minimizing an objective function during training. For example, a decimated CNN minimizes its objective function by backpropagation and learns task-dependent multiscale representations through learned FIR-like convolutional kernels, pointwise nonlineari-ties, and sampling operations. The input may be a sequence, a multichannel sequence, an image, a video sequence, volumetric medical image scans, or general multidimensional data. This chapter covers GPU-efficient realizations of system units such as learnable FIR convolutions, D-dimensional convolutional operators, nonlinear units with LN cascade systems, polynomial and rational static nonlinearities, multirate sampling for downsampling and upsampling, adaptive polyphase sampling, basis and frame expansions, and perfect reconstruction (PR) multiresolution filter-banks. These units can be combined within a common differentiable operator framework. Such a framework supports model designs that can be parameter-efficient, organized in frequency and scale domains, and more amenable to auditing through spectra, reconstruction checks, and PR checks at the transform-module level. Models carefully designed with appropriate sampling and filter-bank constraints can improve shift stability and reduce aliasing artifacts compared with naive subsampling layers in conventional CNNs. Orthogonal bases, frames, framelets, Ramanujan frames, wavelet filter banks, and directional multiresolution systems introduce structured coefficient spaces with scale selectivity and, where applicable, directional sensitivity for image and multidimensional inputs. In these settings, energy preservation, stability, redundancy, and reconstruction behavior can be controlled at the transform-module level. In this chapter, multiresolution transforms, including wavelets, shearlets, and directional Fourier-plane tilings, are treated not only as handcrafted feature extractors but also as fixed or constrained learnable analysis and synthesis operators that can be embedded in neural computation graphs. This treatment makes the corresponding modules easier to inspect through reconstruction, spectral, and PR checks. Thus, the chapter provides the operator-level foundation for later frugal and interpretable models, without claiming that the full deep network is automatically explainable or stable by construction.

## 2.1 Convolutional Neural Networks

Let's begin our discussion with CNNs that learn shift-equivariant feature maps using *weight sharing* and *local connectivity*. For input channels $x_i$ and kernels $h_{j,i}$, a 2D layer outputs

$$y_j[\boldsymbol{n}] = \sigma\left(\sum_i (h_{j,i} * x_i)[\boldsymbol{n}] + b_j\right) \tag{2.1}$$

where $*$ is (usually padded or circular shift) discrete convolution, $\sigma$ a nonlinearity (ReLU/GELU), and $i, j$ index input and output channels. Stacking such layers with nonlinearities builds hierarchical, multiscale representations. Following are some features of a typical CNN.

- *Inductive biases:* Convolutional weight sharing supports translation equivariance under suitable boundary assumptions, while small kernels impose locality. Pooling and strided convolutions increase the effective receptive field and can introduce local shift tolerance, but they do not guarantee exact translation invariance. Dilation enlarges the receptive field without increasing the number of kernel parameters. Boundary handling through valid, same, circular, or reflection padding affects edge responses. Downsampling without adequate low pass filtering can introduce aliasing.

- *Architectural primitives:* Stacks of convolution, normalization, and activation blocks are commonly combined with residual or skip connections to stabilize gradient flow in deep architectures, including architectures that contain decimation. Depthwise separable convolutions and group convolutions reduce computational complexity, while $1 \times 1$ pointwise kernels mix information across channels. Batch, instance, and group normalization regulate activation statistics, while layer scaling and residual scaling help moderate residual updates. Some modern backbones also interleave attention modules or Fourier and wavelet blocks to support longer range interactions at moderate computational cost.
- *Training and spectra:* CNN filters are typically optimized with SGD or Adam under a task loss, often with optional regularization. Beyond $\ell^2$ regularization and weight decay, spectral constraints such as zero DC response for high pass filters, unity gain for low pass filters, band limitation, and orthogonality can encourage well controlled frequency responses and multiscale coverage.
- *Expressivity and interpretation:* Early layers can often be interpreted as localized filters that respond to edges, textures, or other simple patterns, while deeper layers combine these lower level responses into more task specific features. Although convolutional kernels are local, depth, pooling, dilation, and multiresolution design can enlarge the effective receptive field and support longer range context. Feature attribution, frequency response inspection, and PR checks, when using analysis, processing, and synthesis designs, provide module level diagnostics. Appropriate filter bank and sampling constraints can provide better control over aliasing and scale fidelity, while these diagnostics help verify the resulting behavior.

The building blocks of a basic CNN with learnable FIR filters, nonlinearities, downsampling and upsampling operations are described next.

### 2.1.1 Learnable FIR filters

An FIR filter is essentially a finite-size kernel that convolves with input. Learnable FIR filters are the fundamental linear units in a basic CNN and a trivial network with just one convolution layer is among the parametrically simplest and highly interpretable trainable unit. When multiple such filters are stacked, combined with nonlinearities and trained in multi-scale, hierarchical structures they capture rich representations. They capture local correlations, which is especially useful in time, space, or spatial-temporal data. They encode inductive biases (e.g., translation invariance) into the architecture. They are computationally efficient on hardware (e.g., GPUs, TPUs). However, with many filters and increasing network depth the interpretability decreases. Yet, for real-time, resource-efficient applications, CNNs with FIR filters are a gold standard due to speed and interpretability. Some examples include audio, speech, time series, image, video, volumetric (MRI or CT), and spatiotemporal data processing systems with one, two or three-dimensional trainable filters. A $D$-dimensional convolution of an input signal $x[\boldsymbol{n}]$ with an FIR filter is,

$$y[\boldsymbol{n}] = \sum_{\boldsymbol{k}} h[\boldsymbol{n} - \boldsymbol{k}] \cdot x[\boldsymbol{k}] \tag{2.2}$$

where $h$ is an unit impulse response of a learnable LSI kernel and $y$ is the convolved output. In a basic CNN this linear operation is often a coss-correlation by default.

#### 2.1.1.1 Kernel size, uncertainty and practical considerations for CNNs

Choosing a convolution kernel size is a trade-off between *spatial localization* and *frequency selectivity*, an unavoidable physical uncertainty barrier. For a discrete $D$-dimensional kernel $h$, its spread in both space and frequency along each Cartesian axis cannot be made arbitrarily small at the same time and is bounded below by

$$\sigma^2_{\text{space},d}(h)\,\sigma^2_{\text{freq},d}(h) \ \geq\ c \ >\ 0, \qquad d = 1, \ldots, D, \tag{2.3}$$

where $\sigma^2_{\text{space},d}(h)$ and $\sigma^2_{\text{freq},d}(h)$ denote the spatial and frequency variances of $h$ along coordinate $d$, and $c$ is a positive constant. In the classical continuous one-dimensional setting with $D = 1$ the time–bandwidth inequality is $\sigma^2_{\text{space}}(h)\,\sigma^2_{\text{freq}}(h) \ \geq\ 0.25$ with appropriate Fourier normalization. Gaussian waveforms attain equality in this bound and therefore achieve the minimal possible time–frequency localization. Similarly, *small kernels* (e.g., 3×3 or 3×3×3 kernel) in a discrete multidimensional setup are parameter-efficient that capture local texture but have diffuse spectra and limited receptive fields. They may miss structures larger than the effective field unless the filters are stacked, dilated, or combined across scales. Larger kernels can allow finer spectral selectivity and a larger receptive field, but they usually add parameters and computation and may increase overfitting risk. Dilated kernels increase the receptive field without increasing the number of kernel parameters, although they may introduce sparse sampling effects. Figure 2.1 shows the magnitude responses of a basic 3D convolution layer with eight $3 \times 3 \times 3$ learnable FIR filters. Further practical considerations are given below:

1. *Boundary handling:* Padding choices such as same, valid, or reflection can create edge artifacts that propagate into nearby feature maps.
2. *Stride and decimation:* Downsampling without sufficient low pass filtering introduces aliasing and distorts spatial structure, so strided operators should be paired with anti aliasing filters or multiresolution blocks.
3. *Scale variation:* Objects and structures can appear at many sizes in realistic data. Pyramids and multiresolution banks, including wavelet transforms, can therefore help complement models that rely only on a single fixed kernel size.

These considerations naturally motivate the study with different spatial support filters and multiresolution constructions beyond the standard small kernel sizes used in typical CNNs. The following discussion revisits the convolution operator in the Fourier domain through the convolution theorem, which provides a basis for FFT-based and vectorized realizations of $D$-dimensional convolutional operators under appropriate padding and boundary handling, including cases beyond the usual two-dimensional and three-dimensional settings.

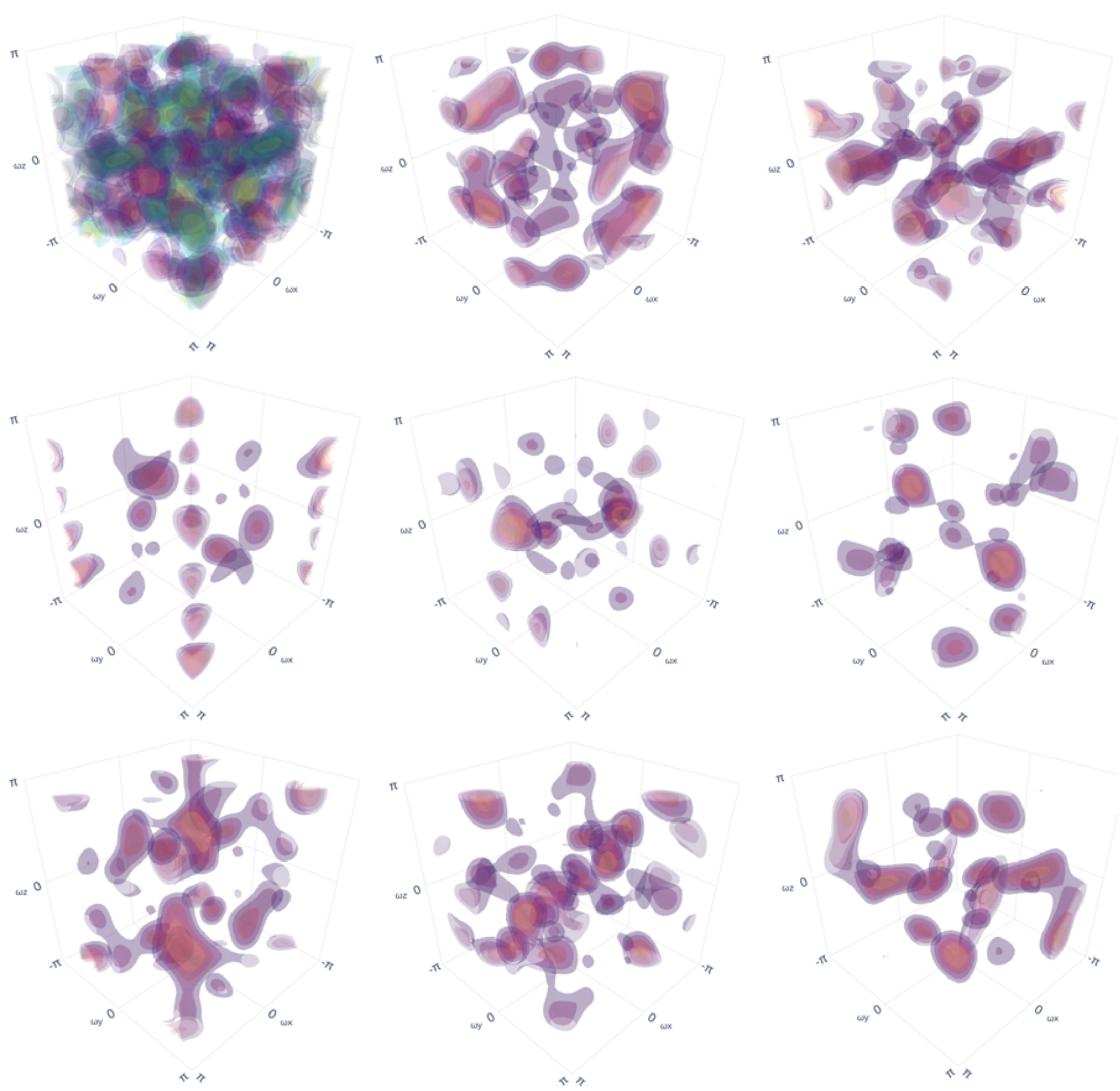

Figure 2.1: Magnitude response of a 3D convolution layer with eight $3 \times 3 \times 3$ trainable FIR filters

#### 2.1.1.2 Vectorized $D$-dimensional convolution via the Fourier product form

For the $D$-dimensional convolution routines used in this thesis, the *convolution theorem* gives a vectorized Fourier product form for circular convolution on a finite grid. The theorem states that circular convolution on a $D$-dimensional spatial grid corresponds to pointwise multiplication in the discrete Fourier domain. Let $x[\boldsymbol{n}]$ be a signal and let $h[\boldsymbol{n}]$ be an LSI filter, both defined on a grid with sizes $N_1, \ldots, N_D$ collected in $\boldsymbol{N} = (N_1, \ldots, N_D)$. The circular convolution of $x$ with $h$ is

$$(y = h * x)[\boldsymbol{n}] = \sum_{\boldsymbol{m}} h[\boldsymbol{m}]\, x\big[(\boldsymbol{n} - \boldsymbol{m}) \bmod \boldsymbol{N}\big] \qquad \text{(circular convolution)}$$

where $x\big[(\boldsymbol{n}-\boldsymbol{m}) \bmod \boldsymbol{N}\big]$ is $\boldsymbol{N}$ modulo circular shifts of $x$. Let $\hat{x}[\boldsymbol{k}]$ and $\hat{h}[\boldsymbol{k}]$ denote the $D$-dimensional discrete Fourier transforms of $x$ and $h$ with frequency index $\boldsymbol{k} \in \prod_{i=1}^{D}\{0,\ldots,N_i-1\}$. For circular convolution the Fourier transform of $y$ satisfies the pointwise product identity

$$\hat{y}[\boldsymbol{k}] = \widehat{h * x}\,[\boldsymbol{k}] = \hat{h}[\boldsymbol{k}]\,\hat{x}[\boldsymbol{k}], \qquad \boldsymbol{k} \in \prod_{i=1}^{D}\{0,\ldots,N_i-1\} \qquad \text{(convolution theorem)}$$

where $\hat{y}$ is the Fourier transform of the output. Two related useful formulas in CNNs are:

- *Cross correlation*. The cross correlation of $h$ and $x$ is

  $$(h \star x)[\boldsymbol{n}] = \sum_{\boldsymbol{m}} h^*[\boldsymbol{m}]\,x[\boldsymbol{n}+\boldsymbol{m}], \qquad \text{(cross correlation)}$$

  and its Fourier transform is

  $$\widehat{h \star x}\,[\boldsymbol{k}] = \hat{h}^*[\boldsymbol{k}]\,\hat{x}[\boldsymbol{k}] \tag{2.4}$$

  where $\hat{h}^*$ denotes the complex conjugate of $\hat{h}$.
- *Linear (non circular) convolution*. To obtain a linear convolution from FFTs, each dimension is zero padded to at least $N_i + M_i - 1$, where $M_i$ is the support length of the filter along axis $i$, before taking the discrete Fourier transforms.

**D-dimensional convolutions** The Fourier product form provides an efficient alternative realization of high-dimensional convolution kernels. Many mainstream deep learning APIs provide built-in convolution layers mainly for one, two, and three spatial dimensions. Convolutions in higher spatial dimensions, therefore, often require custom operator construction. Since the $D$-dimensional discrete Fourier transform is separable across axes, it can be evaluated by applying one-dimensional FFTs successively along each dimension. For example, a four-dimensional FFT of a four-dimensional input can be computed by applying one-dimensional FFTs successively along its four axes. The introduced `convd` Python package realizes $D$-dimensional convolutions using this strategy.

**Example 2.1.** Learning convolution layer of CNN in Fourier space with wavelet regularization.

This example demonstrates effective Fourier domain learning of large size sparse 2D LSI filters ($16 \times 16$) filters with spectral regularization. For a given channel the convolution is implemented as a pointwise product in the frequency domain followed by an IFFT,

$$\hat{y}[\boldsymbol{k}] = \hat{h}[\boldsymbol{k}]\,\hat{x}[\boldsymbol{k}], \qquad y = \mathrm{IFFT}(\hat{y}). \tag{2.5}$$

where $\hat{h}$ is the FFT of the LSI kernel and $y$ is the convolved output. It is possible to learn sparse filters with specific spectral characteristics using spectral regularization on the coefficients of $\hat{h}$ during training. Some useful choices of such regularization are listed below.

1. *DC control:* The following terms shape the response at zero frequency so that high pass filters suppress DC and low pass filters preserve it.
   (a) High-pass (zero mean): $\mathcal{R}_{\text{DC}} = \lambda_{\text{DC}} \, |\hat{h}[\mathbf{0}]|^2$.
   (b) Low-pass (unity gain at DC): $\mathcal{R}_{\text{LP}} = \lambda_{\text{LP}} \, |\hat{h}[\mathbf{0}] - 1|^2$.
2. *Band limiting and leakage control:* Let $\mathcal{B}$ be a desired passband such as an annulus or directional wedge and let $w(\boldsymbol{k}) \geq 0$ be large outside $\mathcal{B}$.

$$\mathcal{R}_{\text{band}} = \lambda_{\text{band}} \sum_{\boldsymbol{k}} w(\boldsymbol{k}) \, |\hat{h}[\boldsymbol{k}]|^2. \tag{2.6}$$

   This term discourages energy where the filter is not supposed to pass and keeps the spectrum concentrated inside the target band.
3. *Vanishing moment proxy for wavelet like high pass filters:* Low order derivatives of $\hat{h}$ at DC are encouraged to be small

$$\mathcal{R}_{\text{vm}} = \lambda_{\text{vm}} \sum_{|\boldsymbol{p}| \leq P} \left| \partial_{\boldsymbol{\omega}}^{\boldsymbol{p}} \hat{h}(\boldsymbol{\omega}) \Big|_{\boldsymbol{\omega}=\mathbf{0}} \right|^2 \tag{2.7}$$

   which promotes wavelet like behavior in which the filter responds weakly to slowly varying components.
4. *Inter filter orthogonality and disjointness:* For a LSI filter bank $\{\hat{h}_r\}$ the overlap of distinct filters in the frequency domain is penalized

$$\mathcal{R}_{\text{ortho}} = \lambda_{\text{ortho}} \sum_{r \neq m} \sum_{\boldsymbol{k}} \left| \hat{h}_r^*[\boldsymbol{k}] \, \hat{h}_m[\boldsymbol{k}] \right|^2 \tag{2.8}$$

   which encourages different filters in the bank to occupy complementary frequency regions and reduces redundancy.
5. *Spectral norm and Lipschitz control:* Here $\hat{h}[\boldsymbol{k}]$ is viewed as a channel mixing matrix

$$\mathcal{R}_{\text{sn}} = \lambda_{\text{sn}} \max_{\boldsymbol{k}} \|\hat{h}[\boldsymbol{k}]\|_2^2 \tag{2.9}$$

   and this regularizer limits the largest singular value across frequencies. It also helps to control the Lipschitz constant of the convolution layer.

Finally, the training objective for a single convolutional layer then becomes

$$\mathcal{L} = \mathcal{L}_{\text{task}}(y, \text{labels}) + \sum_r \Big( \mathcal{R}_{\text{DC}} + \mathcal{R}_{\text{LP}} + \mathcal{R}_{\text{band}} + \mathcal{R}_{\text{vm}} + \mathcal{R}_{\text{ortho}} + \mathcal{R}_{\text{sn}} \Big). \tag{2.10}$$

which combines the task loss with these spectral penalties.

*Implementation notes.* Use zero-padding to realize linear (non-wrap) convolution. Use real FFTs (rFFT and irFFT) with Hermitian symmetry on $\hat{h}$ so that the spatial kernels remain real. For strided layers couple low pass filters that approximate unity gain and constrain high pass filters with $\mathcal{R}_{\text{band}}$ to reduce aliasing. Performing a binary segmentation with a U-Net having 16×16 filters learns sparse coefficients during training, and one such learned filter and its inverse Fourier transform are shown in Figure 2.2.

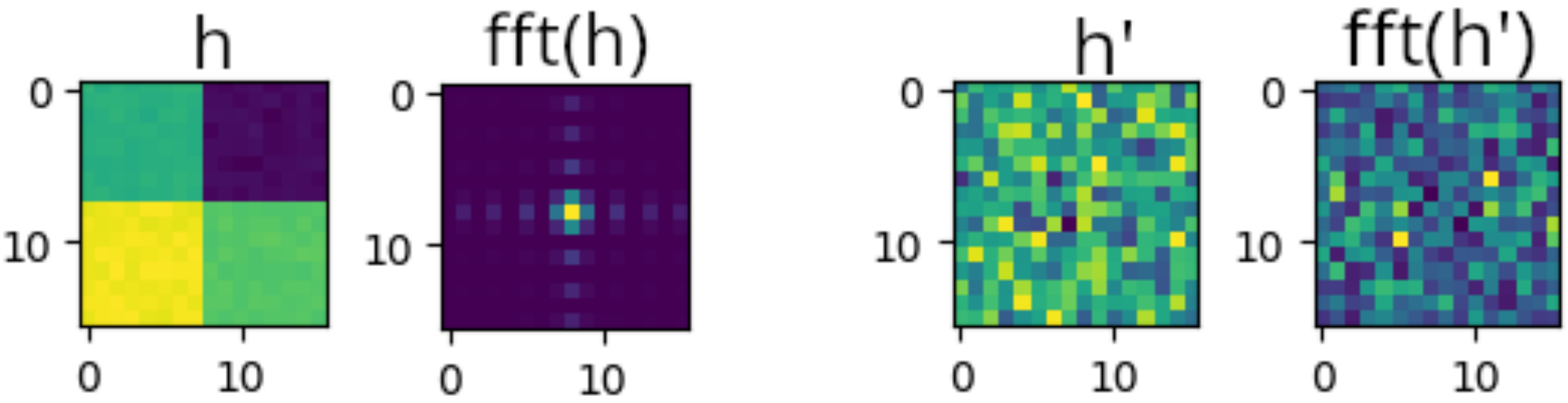

Figure 2.2: (Left pair) A 16×16 filter from a convolution layer of the U-Net with FFT convolutions with $\ell_1$ and $\ell_2$ wavelet regularization. (Leftmost) A DWT tiling formation is observed in the filter coefficients after taking IFFT of the learned Fourier space coefficients. (Right pair) A similar kernel and its FFT from the same U-Net but with natural domain convolution with no regularization.

**Practical multichannel operations in convolution layers with multiple LSI filters** Unlike a simple convolution, the number of input and output channels in a basic convolution layer is generally different. The output of each convolution filter in a convolution layer (with multiple LSI filters) is a *channel-wise summation* of a tensor obtained after convolving each channel of the multichannel input with copies of the same LSI filter. This is how a generic convolution layer operates on a batched multichannel input in TensorFlow or PyTorch frameworks. Thus, for *single-channel* input, each LSI filter output of a convolution layer is the same as a simple convolution. This behavior varies in the presence of multichannel input. An open-source animation of this operation in a two-dimensional convolution (or correlation) layer is available in this video `https://youtu.be/w4kNHKcBGzA?t=212` (see at 3:37 minutes). This **key point** here is when the input and output channels are not the same, then the computation differs slightly in default behavior. Similarly, for a batch of *single-channel* inputs, both separable and non-separable convolution give **identical results**. This can be easily verified with a separable and non-separable 2D convolution with a 1D kernel and its outer product with itself. However, in the presence of *multiple channels*, the results are **not identical** due to the same channel-wise summation.

### 2.1.2 Fixed and learnable parametric static nonlinearity

A common practice in neural networks is the use of analytic and memoryless nonlinear activation functions (thresholding, sigmoid, tanh, ReLU, GeLU, Mish, etc.) after the linear (summation or convolution) operations. This fundamental configuration of LN cascade unit is also known as a Wiener system. Activation functions shape how neurons respond to their inputs and play a central role in the behavior of deep networks in solving highly nonlinear problems. They introduce nonlinearity after the linear operations, which allows modeling of highly nonlinear input–output relationships. Different choices influence saturation, smoothness, sparsity of activations, and the flow of gradients during training. Table 2.1 summarizes the historical development of activation functions in deep learning and highlights their basic qualitative properties. A general method for nonlinear modeling is to use parametric polynomials with learnable coefficients [114] after the linear operation as described below.

Table 2.1: Historical Development of Activation Functions

| # | Paper | Activation | Type | Learned? | Key Characteristics |
|---|---|---|---|---|---|
| 1 | Rosenblatt (1958) [28] | Threshold / Step | Binary, Heaviside | No | Simplest form; used in early perceptrons; non-differentiable |
| 2 | Rumelhart et al. (1986) [18] | Sigmoid / Tanh | Smooth, saturating | No | Differentiable nonlinearities used with backprop; prone to saturation/vanishing gradients |
| 3 | Nair & Hinton (2010) [85] | ReLU | Piecewise linear | No | Non-saturating; sparse activations; alleviates vanishing gradients |
| 4 | He et al. (2015) [86] | PReLU | Piecewise linear | $\alpha$-learned | Learnable negative slope; mitigates dying ReLU |
| 5 | Clevert et al. (2015) [115] | ELU | Smooth exponential | No | Negative outputs improve mean activation; smooth alternative to ReLU |
| 6 | Hendrycks & Gimpel (2016) [116] | GELU | Smooth, non-monotonic | No | $x\,\Phi(x)$ gating using Gaussian CDF; popular in Transformer models |
| 7 | Ramachandran et al. (2017) [117] | Swish / SiLU | Smooth, non-monotonic | $\beta$-optional | Self-gated $x\,\sigma(\beta x)$; SiLU is the $\beta = 1$ case |
| 8 | Klambauer et al. (2017) [118] | SELU | Smooth exponential | No | Self-normalizing with fixed $\lambda, \alpha$; paired with AlphaDropout |
| 9 | Misra (2019) [78] | Mish | Smooth, non-monotonic | No | Self-gated $x\ \tanh(\text{softplus}(x))$; smooth alternative to Swish/ReLU |
| 10 | Molina et al. (2019) [89] | Padé Activation Units (PAU) | Rational functions | Yes | Ratio of learned polynomials; flexible family approximating many activations |
| 11 | Timmons & Rice (2020) [114] | Poly. Approx. for Sigmoid | Polynomial approximation | No | Approximate sigmoid for faster inference with limited compute |

#### 2.1.2.1 Polynomials and modeling of nonlinear functions

A realizable form of a Taylor series polynomial with finite terms is

$$y = b_0 + b_1 x + b_2 x^2 + \cdots + b_m x^m = \sum_{k=0}^{m} b_k x^k \tag{2.11}$$

where $x, y \in \mathbb{R}$ and $b_k$ are the polynomial weights. The above equation is a static system because no past or future elements of $x$ are required to compute $y$. Here, the weights require storage in computer memory, which should not be confused with the memory of a static or dynamic system. For example, for a sequence input and a second-order approximation of equation (2.11), every sample point will have three weights $\{b_0, b_1, b_2\}$ when location-specific coefficients are used for nonlinear modeling. As the dimension of the input data increases, the number of such location-specific weights grows as $O(N^D)$ for a grid with side length $N$ in $D$ dimensions. For example, in a second order approximation of equation (2.11) with a volumetric input $x[\boldsymbol{n}]$ of size $128 \times 128 \times 128$, each voxel will have three weights $\{b_0, b_1, b_2\}$ and the total number of weights is thus $128^3 \times 3 = 6,291,456$. In two dimensions, a similar setup would have a total of $128^2 \times 3 = 49,152$ weights, and in one dimension it is just $128 \times 3 = 384$ weights. Thus, high-dimensional inputs face the curse of dimensionality in polynomial activations when separate coefficients are assigned to each spatial location. That said, for 1D and modest-sized 2D input-output systems, polynomial models are useful tools for modeling static nonlinearities. Polynomial models can represent a broad class of smooth nonlinear functions over bounded operating ranges. This gives freedom in capturing nonlinear behavior without fixing the activation to a single predefined form, such as sigmoid or ReLU. Algorithm 2.3 describes a fast vectorized algorithmic realization of a Taylor series with batched multichannel operations for deep learning applications. Practical limitations arise due to the complexity in computing the higher-degree terms, especially in multidimensional systems. This limitation in modeling higher-order nonlinearities can be overcome using parametric rational polynomial forms, as described next.

**Algorithm 2.3:** Taylor series for batched multichannel $D$ dimensional inputs

1. Input $X \in \mathbb{R}^{\text{batch} \times N_1 \times \cdots \times N_D \times \text{channels}}$
2. Set the number of terms $m \in \mathbb{Z} > 0$.
3. Initialize $\{B^{(m)}\}_{\forall m} \in (0, 1]^{N_1 \times \cdots \times N_D \times \text{channels}}$
4. Compute the power terms $\{X^{(m)}\}_{\forall m}$
5. Compute weighted monomials $\{P^{(m)}_{bn_1 \cdots n_D c} = \sum_{n'_1, \cdots, n'_D} X^{(m)}_{bn_1 \cdots n_D c} \cdot B^{(m)}_{n_1 \cdots n_D c}\}_{\forall m}$
6. Compute output $Y \in \mathbb{R}^{\text{batch} \times N_1 \times \cdots \times N_D \times \text{channels}}$ as,

$$Y_{bn_1 \cdots n_D c} = \sum_m P^{(m)}_{bn_1 \cdots n_D c} = \sum_m \sum_{n'_1, \cdots, n'_D} X^{(m)}_{bn_1 \cdots n_D c} \cdot B^{(m)}_{n_1 \cdots n_D c}$$

#### 2.1.2.2 Padé approximation

An efficient way to model higher-degree nonlinearity is with the Padé approximation, which is a ratio of finite degree polynomials,

$$y = \frac{b_0 + b_1 x + b_2 x^2 + \cdots + b_m x^m}{1 + a_1 x + a_2 x^2 + \cdots + a_m x^m} = \frac{\sum_{k=0}^{m} b_k x^k}{1 + \sum_{k=1}^{m} a_k x^k} \tag{2.12}$$

where $x, y \in \mathbb{R}$, the denominator $1 + \sum_{k=1}^{m} a_k x^k \neq 0$ and $a_m, b_m \in (0, 1]$. Algorithm 2.4 is a fast realization of Padé approximation for batched multichannel operations.

**Algorithm 2.4:** Padé approximation for batched multichannel $D$ dimensional input

1. Input $X \in \mathbb{R}^{\text{batch} \times N_1 \times \cdots \times N_D \times \text{channels}}$
2. Set $m \in \mathbb{Z} > 0$, i.e., number of terms
3. Initialize trainable $\{B^{(m)}\}_{\forall m}, \{A^{(m)}\}_{\forall m} \in (0,1]^{N_1 \times \cdots \times N_D \times \text{channels}}$

   **Note:** $A_1 = 1$ and is non trainable.
4. Compute the powers $\{X^{(m)}\}_{\forall m}$
5. Compute weighted numerator monomials $\{P^{(m)}_{bn_1 \cdots n_D c}\}_{\forall m}$ where

$$P^{(m)}_{bn_1 \cdots n_D c} = \sum_{n_1, \cdots, n_D, c} X^{(m)}_{bn_1 \cdots n_D c} \cdot B^{(m)}_{n_1 \cdots n_D c}$$

6. Compute weighted denominator monomials $\{Q^{(m)}_{bn_1 \cdots n_D c}\}_{\forall m}$ where

$$Q^{(m)}_{bn_1 \cdots n_D c} = \sum_{n_1, \cdots, n_D, c} X^{(m)}_{bn_1 \cdots n_D c} \cdot A^{(m)}_{n_1 \cdots n_D c}$$

7. Compute and return output,

$$Y_{bn_1 \cdots n_d c} = \frac{\sum_m P^{(m)}_{bn_1 \cdots n_D c}}{1 + \sum_m Q^{(m)}_{bn_1 \cdots n_D c}} = \frac{\sum_{n_1, \cdots, n_D, c} X^{(m)}_{bn_1 \cdots n_D c} \cdot B^{(m)}_{n_1 \cdots n_D c}}{1 + \sum_{n_1, \cdots, n_D, c} X^{(m)}_{bn_1 \cdots n_D c} \cdot A^{(m)}_{n_1 \cdots n_D c}}$$

   where $1 + \sum_{n_1, \cdots, n_D, c} X^{(m)}_{bn_1 \cdots n_D c} \cdot A^{(m)}_{n_1 \cdots n_D c} \neq 0$

The introduced `LNPoly` Python package realizes learnable parametric polynomial and rational polynomial activation functions. In the rational polynomial form, the activation function is the nonlinear map represented by the ratio of two polynomials. Thus, a fixed activation such as ReLU, Mish, or Swish can be approximated over a chosen input range, or the rational form can itself be trained as a learnable activation function by backpropagation. Figure 2.3 shows approximations of ReLU, Mish, Swish, and selected nonlinear functions over the plotted input range using second-degree rational polynomial forms, with degree-two polynomials in both the numerator and denominator.

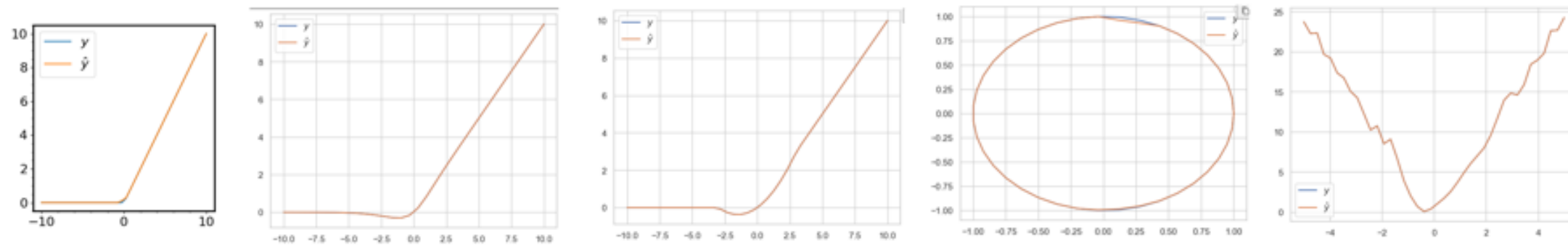

Figure 2.3: Modeling activation functions (ReLU, Mish, Swish) and arbitrary nonlinear functions using rational polynomials (polynomial degree 2 in numerator and denominator)

### 2.1.3 Downsampling and upsampling

Downsampling and upsampling modify how densely a signal is represented on a discrete grid. Downsampling reduces the sampling density by keeping only a subset of samples (typically after low-pass filtering), whereas upsampling increases the sampling density by inserting intermediate samples that are then filled in by an interpolation filter. This subsection first recalls these operations in the classical multirate setting and developments for building pure shift invariant convolutional networks.

#### 2.1.3.1 Downsampling

In signal processing, downsampling denotes the reduction of sampling rate in a signal, typically accompanied by filtering to avoid aliasing. In CNNs, this principle is analogously applied via spatial downsampling operations or pooling, which reduce feature map resolution while retaining essential information. The two most common pooling mechanisms are:

1. *Max pooling* selects the maximum activation in a local window.
2. *Average pooling* computes the arithmetic mean of activations.

These are not merely heuristics but can be formalized within the framework of $\mathrm{L}^p$ norm-based pooling. The $\mathrm{L}^p$ norm-based pooling for a patch $\Omega_{\text{patch}}$ of size $k \times k$ computes,

$$Y = \left( \frac{1}{\mid \Omega_{\text{patch}} \mid} \sum_{x \in \Omega_{\text{patch}}} \mid x \mid^p \right)^{1/p} \tag{2.13}$$

where the special cases are:

- $p = 1$: $\mathrm{L}^1$ pooling, closely related to average pooling if inputs are non-negative.
- $p = 2$: $\mathrm{L}^2$ pooling, which retains both energy and scale, often used in texture recognition.
- $p \to \infty$: Max pooling, as the limiting case of the $\mathrm{L}^p$ norm as,

$$\lim_{p \to \infty} \left( \sum \mid x \mid^p \right)^{1/p} = \max \mid x \mid \tag{2.14}$$

Alternative downsampling methods include strided convolutions, global pooling, adaptive pooling, and wavelet-based pooling. Strided convolutions reduce spatial resolution by moving the kernel with a step larger than one during convolution. Global pooling aggregates each feature map over its spatial dimensions into a single scalar (e.g., in classification heads). Polyphase pooling uses a dictionary of a fixed number of phases and selectively pools a phase of the input. Similarly, wavelet pooling uses fixed or learnable wavelet filter banks to partition the input into localized natural-frequency features and selectively pool a feature. In practice, the CNN classifiers and segmenters extensively use downsampling and upsampling operations. Decimated CNNs preserve spatial information with skip connections from encoder to decoder, like residual networks and U-Net with dilated convolutions, or self-attention blocks such as vision transformers. Selecting the right decimation process is also a key **area of interest** because decimation introduces aliasing, and canceling it is hard unless there is a PR path (e.g., an autoencoder). This is also important as CNN classifiers and segmenters often do not employ (or require) a PR path, so data-driven methods of minimizing aliasing are necessary.

#### 2.1.3.2 Upsampling

In classical digital signal processing, upsampling refers to the process of increasing the sampling rate of a discrete signal. This is achieved by inserting zeros between original samples, followed by low-pass filtering to interpolate missing values and suppress spectral imaging. Formally, for a 1D signal $x[n]$ and integer factor $L$, the upsampled signal $x_u[n]$ is

$$x_u[n] = \begin{cases} x[n/L], & \text{if } n \equiv 0 (\bmod L) \\ 0, & \text{otherwise} \end{cases} \tag{2.15}$$

This process restores a higher-resolution version of the signal, maintaining smoothness and fidelity under ideal filter conditions. In CNNs, upsampling serves as the inverse of downsampling, aiming to restore spatial resolution lost in earlier layers. Unlike classical filtering, CNN upsampling prioritizes semantic and structural reconstruction of images or features, and is essential in tasks like image segmentation, super-resolution, and image generation. Several CNN-specific upsampling strategies have emerged, each with unique operational principles. The upsampling methods in CNN are,

1. *Interpolation (non learnable):* Nearest neighbor and bilinear interpolation rescale feature maps using fixed mathematical rules. These methods are fast and tend to avoid strong artifacts and they are used in decoders such as U-Net and DeepLab. Their main limitation is that they do not adapt to task specific semantics.
2. *Transposed convolution (deconvolution):* A transposed convolution layer is a learnable block that increases spatial resolution using strided kernels. It is often used in generative models such as GANs and variational autoencoders. Care is needed because certain stride and kernel combinations can produce checkerboard patterns due to uneven overlap of filter responses.
3. *Sub pixel convolution (pixel shuffle):* Sub pixel convolution rearranges high dimensional channel information into a finer spatial grid. It is common in super resolution networks such as ESPCN and is efficient because it reuses standard convolutions followed by a deterministic reshape. When the filters are well designed it yields sharp upsampled outputs with few aliasing artifacts.
4. *Unpooling with indices:* Unpooling layers invert max pooling by placing pooled values back at their original locations using stored pooling indices and filling the remaining positions with zeros. This mechanism preserves coarse spatial structure and alignment and is used in architectures such as SegNet [119]. The resulting feature maps are sparse so they are usually followed by one or more convolutions that densify and smooth the representation.

5. *Learnable interpolation modules:* Learnable interpolation layers use adaptive kernels for task aware upsampling as in methods such as CARAFE and Meta Upscale. They aim to combine the smoothness and stability of interpolation with the flexibility of transposed convolution by learning how to mix neighboring features while respecting local image structure.

For example, in **segmentation** models (e.g., SegNet, U-Net), unpooling and interpolation are often preferred to preserve localization. In **generative models** (e.g., DCGAN), transposed convolutions allow upsampling to be learned as part of the generation process. In **super-resolution**, **sub-pixel convolution** achieves efficient scaling with fewer artifacts. Recent architectures balance these trade-offs with skip connections, attention mechanisms, and hybrid modules to achieve high-fidelity reconstruction while maintaining computational efficiency.

#### 2.1.3.3 Shift invariance in CNNs with adaptive polyphase sampling

The conventional strided downsampling used in CNNs (e.g., max pooling, strided convolution) suffers from shift variance, where small input translations lead to inconsistent output activations. To address this, adaptive polyphase sampling in CNN is introduced by Chaman et al. (2021) in paper *Truly Shift Invariant CNNs* [75] that adapts principles from multirate signal processing to make CNNs inherently shift invariant. Figure 2.4 shows the four polyphase masks used for image (2D) inputs.

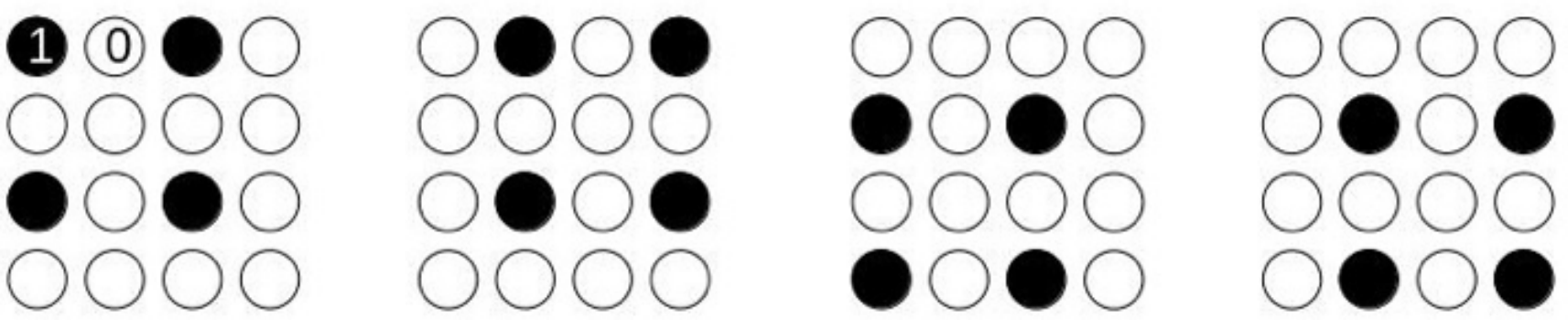

Figure 2.4: Four polyphase masks for two dimensional input

**Polyphase downsampling** In signal processing, polyphase decomposition separates a signal into multiple phases before subsampling, avoiding aliasing and preserving phase alignment. Chaman et al. extended this idea by:

- Decomposing feature maps into polyphase components.
- Selecting a spatially adaptive phase (per-sample or globally optimal) using a selection function (e.g., maximum energy or learned criterion).
- Subsampling only the selected phase, ensuring consistent spatial semantics across shifted inputs.

This leads to non-linear, data-dependent downsampling, where the sampling pattern adapts to the feature distribution and preserve shift equivariance. Algorithm 2.5 summarizes this adaptive polyphase downsampling procedure.

**Algorithm 2.5:** Multidimensional fast adaptive polyphase downsampling $\downarrow 2$

1. Input $X \in \mathbb{R}^{\text{batch} \times N_1 \times \cdots \times N_D \times \text{channels}}$
2. Generate polyphase masks $M \in \{0,1\}^{N_1 \times \cdots \times N_D \times 2^D}$ (e.g., $D = 2$ for images)
3. Compute polyphases of input, $Q_{bn_1 \cdots n_D cp} = \sum_{n_1, \cdots, n_D} X_{bn_1 \cdots n_D c} \cdot M_{n_1 \cdots n_D p}$
4. Argmax over $p$ axis, $Q^*_{bn_1 \cdots n_D c} = \arg\max_p Q_{bn_1 \cdots n_D cp}$
5. Maxpool $2 \times \cdots \times 2$ to remove the zeros,
   $Y = \text{maxpool}(Q^*) \in \mathbb{R}^{\text{batch} \times \frac{N_1}{2} \times \cdots \times \frac{N_D}{2} \times \text{channels}}$

**Adaptive polyphase upsampling** Polyphase upsampling is reverse of the downsampling process:

- Reintroducing zero-filled (or learned) placeholders for unselected phases.
- Applying a polyphase-aware interpolation or refinement module to reconstruct high-resolution representations while preserving alignment with input semantics.

Algorithm 2.6 summarizes the corresponding adaptive polyphase upsampling procedure. Table 2.2 compares common upsampling and downsampling operators used in CNNs.

**Algorithm 2.6:** Multidimensional fast adaptive polyphase upsampling $\uparrow 2$

1. Input $Y \in \mathbb{R}^{\text{batch} \times \frac{N_1}{2} \times \cdots \times \frac{N_D}{2} \times \text{channels}}$ and poly indices from encoder of shape (batch, channels)
2. Upsample $\uparrow 2$: $Q^* = \text{upsample}(\text{Y})$, e.g., (batch, $N_1, \cdots, N_D$, channels)
3. Generate polyphase masks $M \in \{0,1\}^{N_1 \times \cdots \times N_D \times 2^D}$
4. Compute polyphases, $Q_{bn_1 \cdots n_D cp} = Q^*_{bn_1 \cdots n_D c} \cdot M_{n_1 \cdots n_D p}$
5. Use the indices (batch, channels) to collect and return
   $X_{bn_1 \cdots n_D c} = \text{collect}(Q_{bn_1 \cdots n_D cp}, \text{indices})$

Table 2.2: Comparison of Upsampling and Downsampling Methods in CNNs

| # | Method | Learnable | Resolution Control | Spatial Fidelity | Key Characteristics |
|---|---|---|---|---|---|
| 1 | Nearest / Bilinear Interpolation [39, 120] | No | Fixed | Low–Moderate | Fast; simple; can introduce blockiness/blur; widely used in encoder–decoder decoders |
| 2 | Transposed Convolution (Deconvolution) [121, 122] | Yes | High | Moderate–High | Learns upsampling filters; can introduce checkerboard artifacts if not carefully designed |
| 3 | Sub-pixel Convolution (Pixel Shuffle) [123] | Yes | High | High | Rearranges channels into space; efficient; common in super-resolution networks |
| 4 | Max Unpooling (with Indices) [119] | No (partially learnable with convs) | Fixed | High (structural) | Reverses max pooling using stored indices; used in SegNet; often followed by convolution |
| 5 | Adaptive / Learnable Interpolation Modules [124, 125] | Yes | High | High | Content-adaptive upsampling (e.g., CARAFE) and meta-learned scaling modules (e.g., Meta-SR) |
| 6 | Adaptive Polyphase Sampling [75] | Optional (phase selector) | High (data-adaptive) | High (shift-invariant) | Multirate-inspired phase selection to improve shift invariance across downsampling/upsampling |

## 2.2 Basis, frames and filter banks for deep learning

**Bases (orthonormal and beyond)** A *basis* for a Hilbert space $\mathcal{H}$ (e.g., $L^2(\mathbb{R})$ or $\ell^2(\mathbb{Z}^D)$) is a set $\{\varphi_k\}_{k\in\mathbb{Z}}$ such that every $f \in \mathcal{H}$ admits a unique expansion

$$f = \sum_k c_k \, \varphi_k, \qquad c_k \in \mathbb{C}. \tag{2.16}$$

If the set is *orthonormal* ($\langle \varphi_k, \varphi_m \rangle = \delta_{km}$), then

$$c_k = \langle f, \varphi_k \rangle, \qquad \|f\|_2^2 = \sum_k |c_k|^2 \text{ (Parseval)}. \tag{2.17}$$

Orthonormal bases are non-redundant. Their coefficients are unique and, by Parseval's identity, the representation is energy-preserving [126].

**Frames (stable, possibly redundant systems)** A countable family $\{\psi_k\}_{k\in K} \subset \mathcal{H}$ is a *frame* if there exist bounds $0 < A \le B < \infty$ such that for all $f \in \mathcal{H}$

$$A\,\|f\|_2^2 \;\le\; \sum_{k\in K} |\langle f, \psi_k \rangle|^2 \;\le\; B\,\|f\|_2^2. \qquad \text{(frame inequality)}$$

Properties:

- *Stability*: small changes in $f$ produce small coefficient changes and reconstruction is well-posed.
- *Redundancy*: unlike bases, frames can be overcomplete, offering robustness to noise/erasures and flexibility for directional/shift invariance.
- *Tight frames*: when $A = B$, typically normalized to $A = B = 1$ (Parseval frame), giving $f = \sum_k \langle f, \psi_k \rangle \, \psi_k$.
- *General frames*: reconstruction uses the **frame operator** $Sf = \sum_k \langle f, \psi_k \rangle \, \psi_k$. With the **canonical dual** $\tilde{\psi}_k = S^{-1}\psi_k$, $f = \sum_k \langle f, \psi_k \rangle \, \tilde{\psi}_k = \sum_k \langle f, \tilde{\psi}_k \rangle \, \psi_k$.

**Wavelets and framelets** **Wavelets** arise from a scaling function $\phi$ and mother wavelet(s) $\psi$ generated across **scales** and **shifts**. In one dimension (orthonormal case),

$$\phi_{j,k}(t) = 2^{j/2}\,\phi(2^j t - k), \quad \psi_{j,k}(t) = 2^{j/2}\,\psi(2^j t - k), \qquad j,k \in \mathbb{Z}, \tag{2.18}$$

with **MRA** linking spaces $V_j = \overline{\operatorname{span}\{\phi_{j,k}\}_k}$ and detail spaces $W_j = \overline{\operatorname{span}\{\psi_{j,k}\}_k}$,

$$L^2(\mathbb{R}) = \overline{\bigoplus_{j\in\mathbb{Z}} W_j}. \tag{2.19}$$

Discrete orthogonal and biorthogonal wavelets correspond to **perfect-reconstruction two-channel filter banks** (low-pass/high-pass analysis–synthesis pairs), enabling fast pyramidal algorithms. A **framelet** is a wavelet-like **frame**, the collection of scaled/shifted generators is a frame (not necessarily a basis). This allows **redundant** constructions, for example undecimated/translation-invariant wavelets, directional systems (curvelets, shearlets), and tight wavelet frames used in imaging. Benefits include:

- better shift/direction invariance,
- improved denoising and sparse modeling via overcompleteness,
- stability under subsampling/perturbations with simple reconstruction (especially for tight frames).

**Filter bank viewpoint (discrete signals)** For signals on $\ell^2(\mathbb{Z}^d)$, analysis filters $\{h_k\}$ produce subbands via convolution and (optionally) downsampling. **PR** demands synthesis filters $\{g_k\}$ such that

$$f \xrightarrow{\text{analysis}} \{f * h_k\}_k \xrightarrow{\text{synthesis}} \sum_k (f * h_k) * g_k = f. \qquad \text{(Perfect reconstruction)}$$

When the subband frequency responses satisfy a partition of unity (tight frame condition), energy is preserved and reconstruction reduces to simple synthesis without inverting a large system.

In summary, bases give unique and nonredundant expansions. Frames generalize this idea to stable systems that may be redundant but still allow reliable reconstruction. Wavelets are structured multiscale constructions that can appear as orthogonal or biorthogonal bases or as framelets that are tight or overcomplete and that admit efficient implementations through PR filter banks. The remaining of this section describe practical realizations of these systems that are compatible with deep learning graphs.

## 2.2.1 DFT basis and Ramanujan frames

The DFT basis represents signals using complex exponentials on a uniform frequency grid and is well-suited to global sinusoidal patterns. Ramanujan frames [74] provide a period-adapted dictionary whose atoms are associated with integer periods in discrete time. This representation can be useful for analyzing short signals or signals with integer-period components, especially when an explicit period-domain description is desired.

### 2.2.1.1 One-dimensional Ramanujan frames

Ramanujan sums are fundamental in constructing Ramanujan frames that can identify hidden periodic patterns even in very short length sequences. The Ramanujan sum is

$$c_q[n] = \sum_{\substack{p=1 \\ (q,p)=1}}^{q} W_q^{pn} = \sum_{\substack{p=1 \\ (q,p)=1}}^{q} W_q^{-pn}, \qquad \text{(Ramanujan sum)}$$

where $W_q^{pn} = \exp(-2\pi jpn/q)$ is the twiddle factor, $q \in \mathbb{N}$, $n, p \in \mathbb{Z} \geq 0$ and $(q,p) = 1$ represent coprime $q$ and $p$ whose Greatest Common Divisor (GCD) is 1. Here, a Ramanujan frame refers to a period-adapted dictionary generated from Ramanujan subspaces. A Ramanujan frame expansion of a sequence $x[f], f \in \mathbb{Z} \geq 0$ having N samples can be written as

$$x[f] = \sum_{q=1}^{Q} \sum_{\substack{p=1 \\ (q,p)=1}}^{q} b_{q,p} W_q^{pf} = \sum_{q=1}^{Q} x_q[f], \quad 0 \leq f \leq N-1, \tag{2.20}$$

where $W_q^{pf} = \exp(-2\pi j p f/q)$ is the twiddle factor and $x_q[f] \in S_q$. The orthogonal Ramanujan subspaces $S_q$ span $\{W_q^{-pf}\}_{(q,p)}$ for all $(q,p) = 1$ such that $1 \leq p \leq q$ and $p, q \in \mathbb{N}$. The DFT kernels $W_q^p$ appearing in equation (2.20) includes all functions with period $q$ in the range $1 \leq q \leq \mathrm{Q} < \mathrm{N}$, for all coprime $q$ and $p$, where $\mathrm{Q} = max(q)$ is the maximum period $q$ that the frame can identify. For a finite length input sequence $\mathbf{x} \in \mathbb{R}^{\mathrm{N}}$ having N sample points, the corresponding Ramanujan-frame dictionary can be arranged as a fat matrix $\mathbf{A} \in \mathbb{Z}^{\mathrm{N} \times \Phi(\mathrm{Q})}$ determined by the signal length N and the maximum period Q. The number of DFT atoms or columns in each $S_q$ is $\varphi(q)$, i.e., the Euler's totient of $q$ and the total number of columns in $\mathbf{A}$ is $\Phi(\mathrm{Q}) = \sum_{q=1}^{\mathrm{Q}} \varphi(q)$, defines the Ramanujan subspace $S$ covered by a Ramanujan frame.

The solutions $\boldsymbol{\alpha} \in \mathbb{C}^{\Phi(q)}$ of the system $\mathbf{A}\boldsymbol{\alpha} = \mathbf{x}$ yield the coefficients $b_{q,p}$. The *sparsest solution* is $\boldsymbol{\alpha}_\star = \arg\min_\alpha \|\boldsymbol{\alpha}\|_0$ such that $\mathbf{x} = \mathbf{A}\boldsymbol{\alpha}$, is a unique shift invariant expansion of the sequence $\mathbf{x}$ having a length of $\Phi(\mathrm{Q})$ that holds the energies of hidden periods in $\mathbf{x}$ [74]. In Chapter 3, we shall see an application of the 1D periodicity transform using Ramanujan frames of very short sequences, named K-band and V-band of a Microwave Radiometer. There, for K-band $\mathrm{N}, \mathrm{Q} = 16, 8$ yields $\mathbf{A}^{\text{K-band}} \in \mathbb{Z}^{16 \times \Phi(8)}$ and similarly for V-band $\mathrm{N}, \mathrm{Q} = 28, 14$ yields $\mathbf{A}^{\text{V-band}} \in \mathbb{Z}^{28 \times \Phi(14)}$ as illustrated in Figure 3.5 of Chapter 3, where similar vertical segments belong to a particular $S_q$.

#### 2.2.1.2 Two-dimensional Ramanujan frames

A discrete multidimensional signal $x[\boldsymbol{n}]$, indexed by integer vectors $\boldsymbol{n} \in \mathbb{Z}^D$, is called periodic with period (or periodicity) matrix $\mathbf{P} \in \mathbb{Z}^{D \times D}$ when it satisfies $x[\boldsymbol{n}] = x[\boldsymbol{n} + \mathbf{P}\boldsymbol{m}]$ for all integer $\boldsymbol{m}$, with the nondegeneracy requirement $\det \mathbf{P} > 0$. Any larger repetition matrix $\mathbf{M}$ can be expressed as $\mathbf{M} = \mathbf{P}\mathbf{R}$ for some integer matrix $\mathbf{R}$, so $\mathbf{P}$ captures a fundamental lattice of translational invariance. The geometric cell representing one irredundant copy of the data is the *fundamental parallelepiped* $\mathrm{FPD}(\mathbf{P}) = \{\mathbf{P}\mathbf{x} : \mathbf{x} \in [0,1)^D\}$; sampling its integer points gives a canonical set of spatial indices $\boldsymbol{n}$ (and similarly $\mathrm{FPD}(\mathbf{P}^\top)$ furnishes frequency-like integer coordinates $\boldsymbol{k}$). In two dimensions this reduces to familiar grid intuition with $\boldsymbol{n} = (n_1, n_2)$ and $\mathbf{P}$ is a $2 \times 2$ integer matrix tiling the plane.

Against this lattice structure we define exponential atoms. The separable 2D DFT kernel $W_N^{kn} = e^{-j2\pi(\frac{k_1}{N_1}n_1 + \frac{k_2}{N_2}n_2)}$ uses a rectangular period $N = (N_1, N_2)$. Ramanujan-based constructions generalize this by replacing scalar periods with full-rank integer matrices and by imposing arithmetic coprime constraints on admissible spectral indices. For an integer period matrix $\mathbf{P}$, the Ramanujan (matrix) kernel takes the form $W_{\mathbf{P}}^{(-1)\boldsymbol{kn}} = e^{j2\pi \boldsymbol{k}^\top \mathbf{P}^{-1}\boldsymbol{n}}$ for integer vectors $\boldsymbol{k}, \boldsymbol{n}$. The multidimensional Ramanujan sum indexed by $\mathbf{Q} \in \mathbb{Z}^{D\times D}$ is then defined as

$$c_{\mathbf{Q}}(\boldsymbol{n}) = \sum_{\substack{\boldsymbol{k}\in \mathrm{FPD}(\mathbf{Q}^\top)\\ (\boldsymbol{k},\mathbf{Q}^\top)_l=\mathbf{I}}} e^{j2\pi \boldsymbol{k}^\top \mathbf{Q}^{-1}\boldsymbol{n}} = \sum_{\substack{\boldsymbol{k}\in \mathrm{FPD}(\mathbf{Q}^\top)\\ (\boldsymbol{k},\mathbf{Q}^\top)_l=\mathbf{I}}} W_{\mathbf{Q}}^{(-1)\boldsymbol{kn}}, \tag{2.21}$$

where the summation is restricted to those $\boldsymbol{k}$ that satisfy a left coprime condition with $\mathbf{Q}^\top$. This coprime requirement is expressed via a Smith normal form (factor) matrix $\Lambda$ of $\mathbf{Q}$: writing $\lambda \in \mathrm{diag}(\Lambda)$, the condition $(\boldsymbol{k}, \mathbf{Q}^\top)_l = \mathbf{I}$ is equivalent to componentwise coprime $(k_i, \lambda_i) = 1$ together with mutual coprime $(\lambda_i, \lambda_j) = 1$ for $i \neq j$. Intuitively, the Smith factors isolate invariant cyclic components; admissible $\boldsymbol{k}$ select primitive characters on each cyclic factor, mirroring how Euler's totient $\varphi(\cdot)$ counts coprime residues in the one-dimensional case. In practice, only those $\boldsymbol{k}$ lying among the integer representatives within FPD($\mathbf{Q}^\top$) and satisfying these GCD constraints contribute to the sum, yielding a structured arithmetic exponential that generalizes classical Ramanujan sums to higher dimensions. For $D = 2$ images, the framework supplies period-adaptive spectral atoms aligned to oblique or coupled lattice structures that a rectangular DFT basis cannot sparsely capture, while retaining purely integer-algebraic selection rules and exact lattice symmetry. Figure 2.5 shows fundamental parallelepipeds for the 2D periodicity matrix $P = \begin{pmatrix} 2 & 3 & -2 & 2 \end{pmatrix}$ with $\det(P) = 4$. The **core formulas** to construct 1D and 2D Ramanujan frames are listed below:

1. 2D DFT kernel:

$$W_{\boldsymbol{N}}^{\boldsymbol{kn}} = e^{-j2\pi\left(\frac{k_1}{N_1}n_1 + \frac{k_2}{N_2}n_2\right)}, \qquad \boldsymbol{N} = (N_1, N_2),\ \boldsymbol{n} = (n_1, n_2),\ \boldsymbol{k} = (k_1, k_2). \tag{2.22}$$

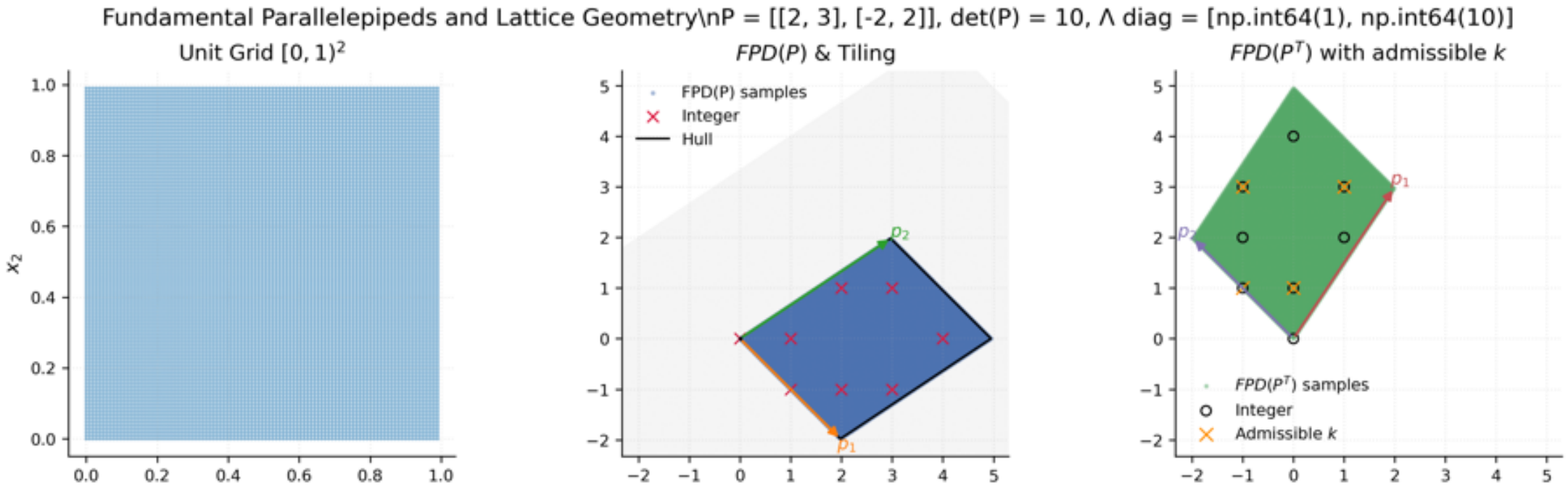

Figure 2.5: Visualization of fundamental parallelepipeds for the 2D periodicity matrix $P = \begin{pmatrix} 2 & 3 & -2 & 2 \end{pmatrix}$ with $\det(P) = 4$. (a) Unit grid $[0, 1)^2$ sampled for mapping. (b) FPD($P$) showing dense sample points (blue), integer lattice points (red crosses), basis vectors (arrows), convex hull (black), and tiling overlays (gray). (c) FPD($P^\top$) with integer points (black circles) and admissible $k$ (orange crosses) satisfying Smith-form coprime conditions ($\Lambda$ diag = [2,2]). This illustrates lattice geometry and arithmetic constraints for Ramanujan sums in multidimensional signal processing.

2. 2D Ramanujan sum (componentwise written):

$$c_q[\boldsymbol{n}] = \sum_{\substack{k_1 = 1 \\ \mathrm{GCD}(k_1, q = 1)}}^{q} \sum_{\substack{k_2 = 1 \\ \mathrm{GCD}(k_2, q = 1)}}^{q} e^{j\frac{2\pi}{q}\langle \boldsymbol{k}, \boldsymbol{n} \rangle}. \tag{2.23}$$

3. Multidimensional periodicity (repetition matrix):

$$x[\boldsymbol{n}] = x[\boldsymbol{n} + \mathbf{M}\boldsymbol{m}], \qquad \boldsymbol{m}, \boldsymbol{n} \in \mathbb{Z}^D,\ \mathbf{M} \in \mathbb{Z}^{D\times D}. \tag{2.24}$$

4. Fundamental period relation:

$$x(\boldsymbol{n}) = x(\boldsymbol{n} + \mathbf{P}\boldsymbol{m}), \qquad \boldsymbol{m}, \boldsymbol{n} \in \mathbb{Z}^D,\ \mathbf{P} \in \mathbb{Z}^{D\times D},\ \det \mathbf{P} > 0. \tag{2.25}$$

5. Repetition vs period matrices:

$$\mathbf{M} = \mathbf{P}\mathbf{R}, \qquad \mathbf{R} \in \mathbb{Z}^{D\times D}. \tag{2.26}$$

6. Fundamental parallelepiped mapping:

$$\mathbf{v} = \mathbf{P}\mathbf{x}, \qquad \mathbf{x} \in [0, 1)^D. \tag{2.27}$$

7. $D$-dimensional Ramanujan sum:

$$c_{\mathbf{Q}}[\boldsymbol{n}] = \sum_{\substack{\boldsymbol{k}\in\mathrm{FPD}(\mathbf{Q}^\top) \\ (\boldsymbol{k},\mathbf{Q}^\top)_l=\mathbf{I}}} e^{j2\pi\boldsymbol{k}^\top\mathbf{Q}^{-1}\boldsymbol{n}} = \sum_{\substack{\boldsymbol{k}\in\mathrm{FPD}(\mathbf{Q}^\top) \\ (\boldsymbol{k},\mathbf{Q}^\top)_l=\mathbf{I}}} W_{\mathbf{Q}}^{(-1)\boldsymbol{k}\boldsymbol{n}}, \qquad \boldsymbol{n}, \boldsymbol{k} \in \mathbb{Z}^D,\ \mathbf{Q} \in \mathbb{Z}^{D\times D}. \tag{2.28}$$

8. Ramanujan kernel:

$$W_{\mathbf{P}}^{(-1)\boldsymbol{k}\boldsymbol{n}} = e^{j2\pi\boldsymbol{k}^\top\mathbf{P}^{-1}\boldsymbol{n}}, \qquad \boldsymbol{n}, \boldsymbol{k} \in \mathbb{Z}^D,\ \mathbf{P} \in \mathbb{Z}^{D\times D} \tag{2.29}$$

9. Left coprime (Smith-form) condition:

$$(\boldsymbol{k}, \mathbf{Q}^\top)_l = \mathbf{I} \iff (k_i, \lambda_i) = 1 \text{ and } (\lambda_i, \lambda_j) = 1,\ i \neq j,\ \lambda \in \mathrm{diag}(\Lambda). \tag{2.30}$$

10. (Contextual) Euler totient: $\varphi(N)$ counts integers $1 \le m < N$ with $\mathrm{GCD}(m, N) = 1$.

The introduced `RamanujanFrame` Python package is a realization of 1D and 2D Ramanujan frames and periodicity transforms described above for deep learning architectures.

### 2.2.2 Discrete Wavelet Transform (backpropagation layers)

The multiresolution tiling of the frequency plane with localized natural-frequency domain features is often used to represent data to build economical and explainable models. A Discrete Wavelet Transform (DWT) is a widely used multiresolution transform for representing discrete data using an orthogonal or biorthogonal basis. The Wavelet toolbox [127] by MathWorks is a proprietary software that has served the requirements for D-dimensional wavelet transforms in the MATLAB environment for a few decades. Several open-source packages are now available for 1D and 2D DWT in Python. Pywavelets [128] is a D-dimensional wavelet transform library in Python that works with Numpy [129] arrays. However, it is challenging to directly use Pywavelets with the symbolic tensors in TensorFlow [130] layers and CUDA [131]. WaveTF [132] is a solution for constructing 1D and 2D DWT layers in TensorFlow but is limited to only Haar and Db2 wavelets. The package tensorflow-wavelets [133] supports 1D and 2D transforms with multiple wavelets. In Pytorch [134], the pytorch-wavelets [135]package allows the construction of 1D and 2D DWT layers. However, limited libraries are available for 3D and higher dimensional transforms for parallel GPU computations for a wide range of wavelet families.

For a D-dimensional wavelet $\psi \in \mathrm{L}^2(\mathbb{R}^{\mathrm{D}})$, a discrete wavelet system defined by $\{\psi_{m,\boldsymbol{p}} : m \in \mathbb{Z}, \boldsymbol{p} \in \mathbb{Z}^{\mathrm{D}}, \mathrm{D} \in \mathbb{N}\}$ forms an orthogonal basis, where $\psi_{m,\boldsymbol{p}}(\mathbf{x}) := 2^m \psi(2^m \mathbf{x} - \boldsymbol{p})$. Then, by definition the DWT of $\mathbf{x} \in \mathbb{Z}^{\mathrm{D}}$ is $\mathbf{x} \mapsto (\langle \mathbf{x}, \psi_{m,\boldsymbol{p}} \rangle)_{m,\boldsymbol{p}}$, where $m$ is the dilation parameter and $\boldsymbol{p}$ is the shift or translation parameter. The `TFDWT` Python package is a simple standalone DWT and IDWT library with minimal dependencies for computation with symbolic tensors and CUDA in modern deep learning frameworks. This section concisely states the mathematics of realizing fast D-dimensional DWT and IDWT layers with filter bank structures with multichannel batched tensors. The current `TFDWT` release is a TensorFlow 2 realization with support up to fast 3D DWT and IDWT with various orthogonal and biorthogonal wavelet families having impulse responses of diverse lengths. This section also serves as a manual for seamlessly upgrading to higher dimensional separable transforms of the independent axes and porting the codes into other tensor processing frameworks with minimal software engineering.

#### 2.2.2.1 Discrete wavelet system for sequences

A discrete wavelet system contains a pair of quadrature mirror filters with a lowpass scaling function and a highpass mother wavelet. A one-dimensional ($\mathrm{D} = 1$) discrete wavelet system with a continuous mother wavelet $\psi \in \mathrm{L}^2(\mathbb{R})$ is realized by the impulse responses of the scaling function and the wavelet as FIR filters $g[n]$ and $h[n]$ [136]. Figure 2.6 shows wavelets and scaling functions of the analysis and synthesis bank of bior3.1 and rbio3.1 and their corresponding impulse responses. These FIR filters are the building blocks of a two-band PR filter bank for realizing fast discrete wavelet systems. Figure 2.8 shows a two-band perfect reconstruction filter bank that operates on one-dimensional inputs, i.e., sequences in $\ell^2(\mathbb{Z})$. The analysis and synthesis bank in orthogonal wavelet systems have identical lowpass and highpass FIR filters but differ in biorthogonal wavelet systems. In biorthogonal wavelet filter banks, $\tilde{g}[n]$ and $\tilde{h}[n]$ are the lowpass and highpass filters of the synthesis bank. The only difference in the biorthogonal families like bior and rbio is the interchange of the analysis and synthesis scaling and wavelets functions, for example, in bior3.1 and rbio3.1.

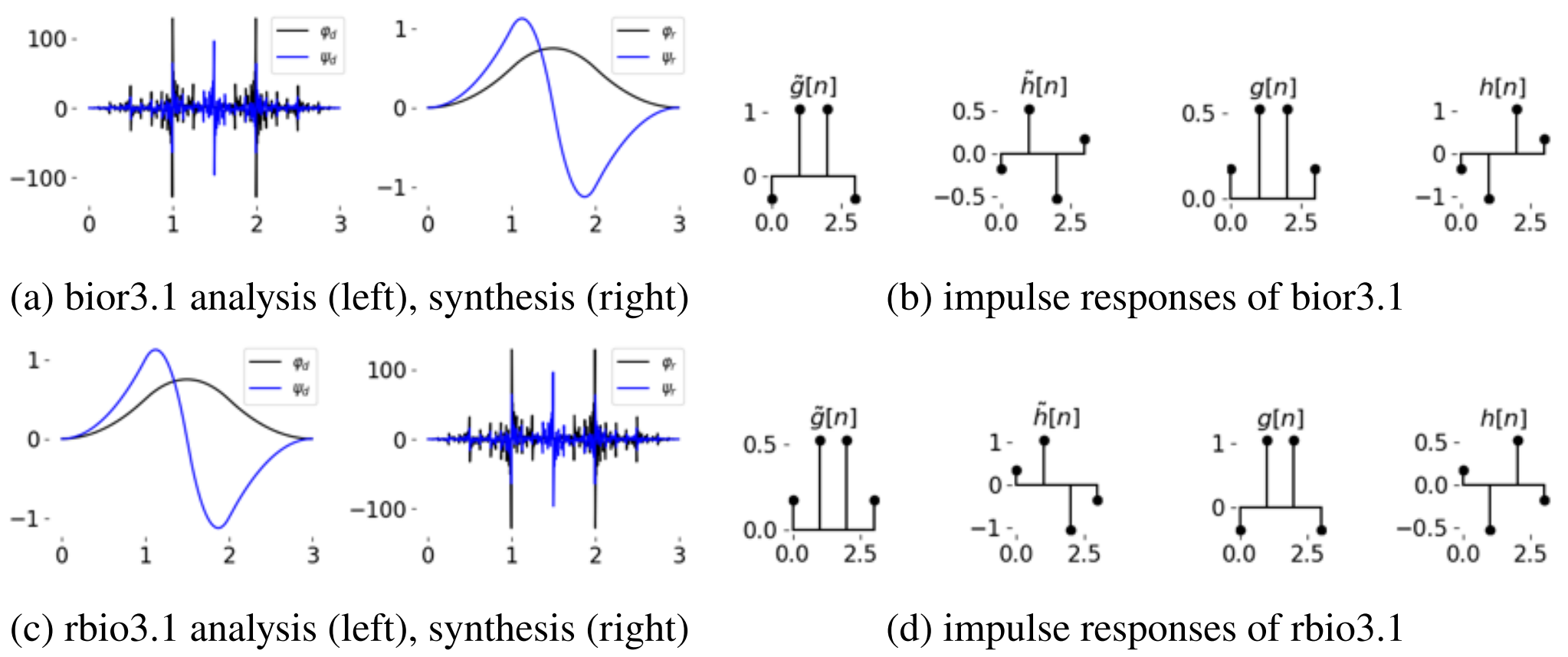

(a) bior3.1 analysis (left), synthesis (right)

(b) impulse responses of bior3.1

(c) rbio3.1 analysis (left), synthesis (right)

(d) impulse responses of rbio3.1

Figure 2.6: (Left column) Wavelets, scaling functions and their corresponding impulse responses for bior3.1 (top row) and rbio3.1 (bottom row) analysis and synthesis banks

**Circular convolution operators** The four matrices in the two band PR filter bank in Figure 2.8 are (i) $\mathbf{G}$ is lowpass analysis matrix, (ii) $\mathbf{H}$ is highpass analysis matrix, (iii) $\tilde{\mathbf{G}}$ is lowpass synthesis matrix and (iv) $\tilde{\mathbf{H}}$ is highpass synthesis matrix. These matrices are operators for circular convolution, constructed by circular shifts of the corresponding FIR filters $g[n-k]$, $h[n-k]$, $\tilde{g}[n-k]$ and $\tilde{h}[n-k]$, where $g = \tilde{g}$ and $h = \tilde{h}$ for orthogonal wavelets.

**Analysis matrix** For a lowpass analysis FIR filter of length $L$ and a input sequence of length $N$, the circular convolution operator is a matrix $\mathbf{G}$ of shape $N \times N$. A downsampling by two $f_{(\downarrow 2)}$ of the output the convolution is equivalent to getting rid of the even rows of $\mathbf{G}$ to give $\frac{N}{2} \times N$ operator $f_{(\downarrow 2)}\mathbf{G}$. Similarly, for the highpass analysis FIR filter of same wavelet with convolution and downsampling is a $\frac{N}{2} \times N$ operator $f_{(\downarrow 2)}\mathbf{H}$. The analysis matrix is

$$\mathbf{A} = \begin{bmatrix} f_{(\downarrow 2)}\mathbf{G} \\ f_{(\downarrow 2)}\mathbf{H} \end{bmatrix}_{N \times N}, \tag{2.31}$$

where $\mathbf{A}$ is a $N \times N$ decimated circular convolution operator formed by combining the lowpass and highpass decimated operators.

**Synthesis matrix** The synthesis matrix $\mathbf{S}$ is another $N \times N$ decimated circular convolution operator:

$$\mathbf{S} = \begin{bmatrix} f_{(\downarrow 2)}\tilde{\mathbf{G}} \\ f_{(\downarrow 2)}\tilde{\mathbf{H}} \end{bmatrix}_{N \times N}^{\top}, \tag{2.32}$$

where $\tilde{\mathbf{G}}$ and $\tilde{\mathbf{H}}$ are matrices formed by the lowpass and highpass synthesis FIR filters. Equation (2.32) is a general representation for both orthogonal and biorthogonal wavelets families where for orthogonal wavelets, $\tilde{\mathbf{G}} = \mathbf{G}$, $\tilde{\mathbf{H}} = \mathbf{H}$ and thus $\mathbf{S} = \mathbf{A}^\top$.

A two-band PR discrete wavelet system for one-dimensional inputs can be written as

$$\mathbf{Q} = \mathrm{DWT}(\mathbf{x}) \text{ or, } \mathbf{q} = \left(\mathbf{A}\mathbf{x}^\top\right)^\top \tag{2.33}$$

$$\mathbf{x} = \mathrm{IDWT}(\mathbf{q}) \text{ or, } \mathbf{x} = \left(\mathbf{S}\mathbf{q}^\top\right)^\top \tag{2.34}$$

where $\mathbf{A}$ and $\mathbf{S}$ are analysis and synthesis matrices defined in (2.31) and (2.32), $\mathbf{x}$ is a input sequence and $\mathbf{q}$ has a distinct lowpass and a highpass subband.

**Example 2.2.** Given, a sequence $\mathbf{x} \in \mathbb{R}^8$ or $N = 8$ and FIR filters length $L = 6$.

$$\text{LPF \& downsampling } f_{(\downarrow 2)}\mathbf{G} = \begin{bmatrix} g_1 & g_0 & 0 & 0 & g_5 & g_4 & g_3 & g_2 \\ g_3 & g_2 & g_1 & g_0 & 0 & 0 & g_5 & g_4 \\ g_5 & g_4 & g_3 & g_2 & g_1 & g_0 & 0 & 0 \\ 0 & 0 & g_5 & g_4 & g_3 & g_2 & g_1 & g_0 \end{bmatrix}_{\frac{N}{2}\times N}$$

$$\text{HPF \& downsampling } f_{(\downarrow 2)}\mathbf{H} = \begin{bmatrix} h_1 & h_0 & 0 & 0 & h_5 & h_4 & h_3 & h_2 \\ h_3 & h_2 & h_1 & h_0 & 0 & 0 & h_5 & h_4 \\ h_5 & h_4 & h_3 & h_2 & h_1 & h_0 & 0 & 0 \\ 0 & 0 & h_5 & h_4 & h_3 & h_2 & h_1 & h_0 \end{bmatrix}_{\frac{N}{2}\times N}$$

$$\text{Analysis matrix is } \mathbf{A} = \begin{bmatrix} g_1 & g_0 & 0 & 0 & g_5 & g_4 & g_3 & g_2 \\ g_3 & g_2 & g_1 & g_0 & 0 & 0 & g_5 & g_4 \\ g_5 & g_4 & g_3 & g_2 & g_1 & g_0 & 0 & 0 \\ 0 & 0 & g_5 & g_4 & g_3 & g_2 & g_1 & g_0 \\ h_1 & h_0 & 0 & 0 & h_5 & h_4 & h_3 & h_2 \\ h_3 & h_2 & h_1 & h_0 & 0 & 0 & h_5 & h_4 \\ h_5 & h_4 & h_3 & h_2 & h_1 & h_0 & 0 & 0 \\ 0 & 0 & h_5 & h_4 & h_3 & h_2 & h_1 & h_0 \end{bmatrix}_{N\times N}$$

Similarly,

$$\text{Synthesis matrix is } \mathbf{S} = \begin{bmatrix} \tilde{g}_1 & \tilde{g}_0 & 0 & 0 & \tilde{g}_5 & \tilde{g}_4 & \tilde{g}_3 & \tilde{g}_2 \\ \tilde{g}_3 & \tilde{g}_2 & \tilde{g}_1 & \tilde{g}_0 & 0 & 0 & \tilde{g}_5 & \tilde{g}_4 \\ \tilde{g}_5 & \tilde{g}_4 & \tilde{g}_3 & \tilde{g}_2 & \tilde{g}_1 & \tilde{g}_0 & 0 & 0 \\ 0 & 0 & \tilde{g}_5 & \tilde{g}_4 & \tilde{g}_3 & \tilde{g}_2 & \tilde{g}_1 & \tilde{g}_0 \\ \tilde{h}_1 & \tilde{h}_0 & 0 & 0 & \tilde{h}_5 & \tilde{h}_4 & \tilde{h}_3 & \tilde{h}_2 \\ \tilde{h}_3 & \tilde{h}_2 & \tilde{h}_1 & \tilde{h}_0 & 0 & 0 & \tilde{h}_5 & \tilde{h}_4 \\ \tilde{h}_5 & \tilde{h}_4 & \tilde{h}_3 & \tilde{h}_2 & \tilde{h}_1 & \tilde{h}_0 & 0 & 0 \\ 0 & 0 & \tilde{h}_5 & \tilde{h}_4 & \tilde{h}_3 & \tilde{h}_2 & \tilde{h}_1 & \tilde{h}_0 \end{bmatrix}_{N\times N}$$

$$\text{The DWT of } \mathbf{x} \text{ produces subbbands } \mathbf{q} = \mathrm{DWT}(\mathbf{x}) \text{ or, } \mathbf{q} = \mathbf{A}\mathbf{x}$$
$$\text{Perfect reconstruction } \mathbf{x} = \mathrm{IDWT}(\mathbf{q}) \text{ or, } \mathbf{x} = \mathbf{S}\mathbf{q}$$

**Backpropagation** In the TFDWT realization, each DWT and IDWT block acts, for every scalar channel, as a fixed finite-dimensional linear operator on a vectorized signal, with analysis and synthesis maps represented by matrices $\mathbf{A}, \mathbf{S} \in \mathbb{R}^{N \times N}$ and forward relations $\mathbf{q} = \mathbf{A}\mathbf{x}$ and $\mathbf{x} = \mathbf{S}\mathbf{q}$ (combined with reshaping and axis permutations across batch and spatial dimensions). Because the computation graph contains only linear tensor operations and reshapes, these blocks are fully compatible with automatic differentiation, and an upstream gradient $\partial\mathcal{L}/\partial\mathbf{q}$ is propagated through the analysis and synthesis layers via the adjoint mappings

$$\frac{\partial\mathcal{L}}{\partial\mathbf{x}} = \mathbf{A}^{\top}\frac{\partial\mathcal{L}}{\partial\mathbf{q}}, \qquad \frac{\partial\mathcal{L}}{\partial\mathbf{q}} = \mathbf{S}^{\top}\frac{\partial\mathcal{L}}{\partial\mathbf{x}},$$

applied separably along each spatial dimension in the multidimensional case. Here, $\mathcal{L} : \mathcal{X} \to \mathbb{R}$ denotes the scalar loss (objective) function used for training, for example mean–squared error, cross–entropy, or any differentiable task loss. It takes the network outputs (and implicitly the parameters, including those around the DWT/IDWT blocks) and returns a real-valued objective whose gradients drive backpropagation. For orthogonal wavelet families the PR condition gives $\mathbf{S} = \mathbf{A}^{\top}$, whereas for biorthogonal families $\mathbf{A}$ and $\mathbf{S}$ form a dual pair, but in both settings the gradients traverse the wavelet blocks through these same adjoint transforms.

**DWT 1D layer** Algorithm 2.7 describes an 1D DWT layer that operates on input tensors of shape (batch, length, channels) and produces an output of shape (batch, length/2, 2×channels).

**Algorithm 2.7:** DWT 1D layer

1. Input $\mathbf{X}$ of shape (batch, length, channels).
2. Generate analysis matrix $\mathbf{A}$ using length of input.
3. For each batched channel $\mathbf{x} \in \mathbf{X}$ of shape (batch, length): $\mathbf{q}_c = \mathbf{A}\mathbf{x}^{\top}$
4. Stacking for all $c$ channels: $\mathbf{Q} := (\mathbf{q}_c)_{\forall c}$ to a shape (batch, length, channels).
5. Group subbands and return an output $\mathbf{Q}^{(\text{grouped})}$ of shape (batch, length/2, , 2 × channels)

$$\begin{aligned} \text{mid} &= \text{int}(\mathbf{Q}.\text{shape}[1]/2 \\ \mathbf{L} &= \mathbf{Q}[:,: \text{mid}, :] \\ \mathbf{H} &= \mathbf{Q}[:, \text{mid} :, :] \\ \mathbf{Q}^{(\text{grouped})} &= \text{Concatenate}([\mathbf{L}, \mathbf{H}], \text{axis} = -1) \end{aligned}$$

**IDWT 1D layer** Algorithm 2.8 describes an 1D IDWT layer that operates on input tensors of shape (batch, length/2, 2×channels) and produces an output of shape (batch, length, channels).

**Algorithm 2.8:** IDWT 1D layer

1. Input $\mathbf{Q}^{(\text{grouped})}$ of shape (batch, length/2, 2 × channels)
2. Ungroup the subbands to get $\mathbf{Q}$ of shape (batch, length, channels)
3. Generate synthesis matrix $\mathbf{S}$ using length of $\mathbf{Q}$.
4. For each batched channel $\mathbf{q} \in \mathbf{Q}$ of shape (batch, length): $\mathbf{x} = \mathbf{S}\mathbf{q}^\top$, i.e., a PR, where $\mathbf{S} = \mathbf{A}^\top$ for orthogonal wavelets
5. Layer output (PR): $\mathbf{X} := (\mathbf{x}_c)_{\forall c}$ is of shape (batch, length, channels)

#### 2.2.2.2 Multidimensional discrete wavelet systems

In sequences (1D), the 1D DWT applies to the only independent variable. To achieve high-dimensional DWT, the 1D DWT needs to be applied separably to all the independent variables one after the other. For example, the DWT of an image (2D) is a row-wise 1D DWT followed by a column-wise 1D DWT. Similarly, the reconstruction is column-wise 1D IDWT followed by a row-wise 1D IDWT.

**Two-dimensional discrete wavelet system** The pixel values in an image are a function of two independent spatial axes. A DWT 2D filter bank is a separable transform with row-wise filtering followed by column-wise filtering that yields four subbands - LL, LH, HL and HH. A two-dimensional discrete wavelet system can be written as

$$\mathbf{q} = \text{DWT}(\mathbf{x}) := \mathbf{A}(\mathbf{A}\mathbf{x}_{021})_{021}^\top \tag{2.35}$$

$$\mathbf{x} = \text{IDWT}(\mathbf{q}) := \mathbf{x} = \mathbf{S}(\mathbf{S}\mathbf{q}_{021}^\top)_{021}^\top \tag{2.36}$$

where the indices 021 in $\mathbf{x}_{021}$ denote a transpose of axis $[0, 1, 2]$ to $[0, 2, 1]$, with 0 as batch index, 1 and 2 are height and width. The matrices $\mathbf{A}$ and $\mathbf{S}$ are the same analysis and synthesis matrices defined in (2.31) and (2.32). Figure 2.7 shows an $N \times N$ image and its four localized spatial-frequency subbands after DWT. Here, the low-frequency band is LL, and the other three are high-frequency subbands representing horizontal, vertical and diagonal features. Figure 2.8 illustrates a separable 2D DWT PR filter bank. The 2D layers operate on a batch of multichannel tensors of shape (batch, height, width, channels), where each image is of shape height and width. Figure 2.9 illustrates the input, output and PR by 2D DWT and IDWT layers. The Multiresolution Encoder-Decoder Convolutional Neural Network in [137] uses these forward and inverse layers for semantic segmentation.

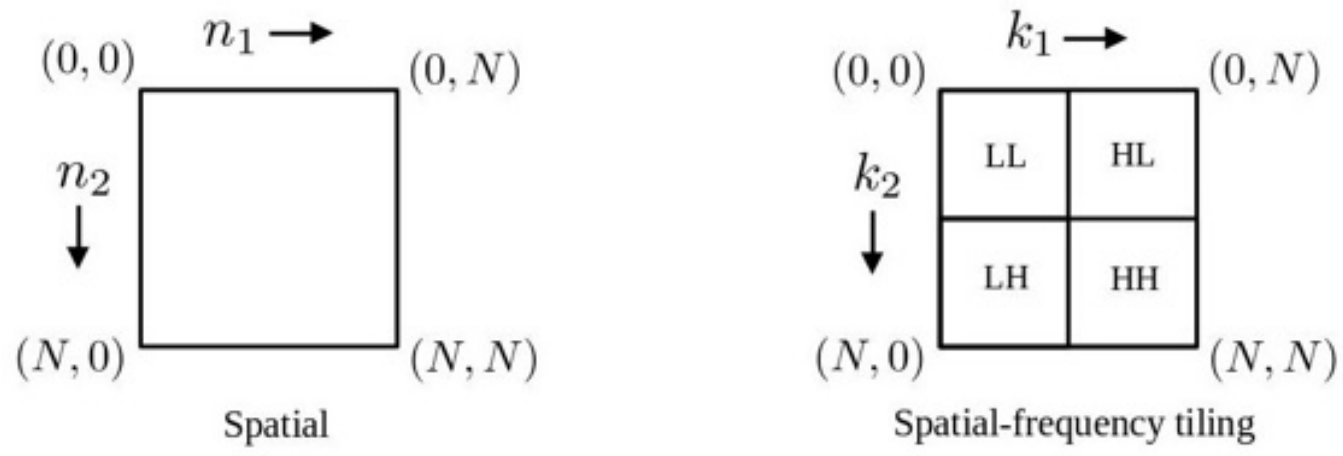

Figure 2.7: An image of shape $N \times N$ (left) and spatial-frequency tiled subbands (right) after a level-1 DWT2D decomposition

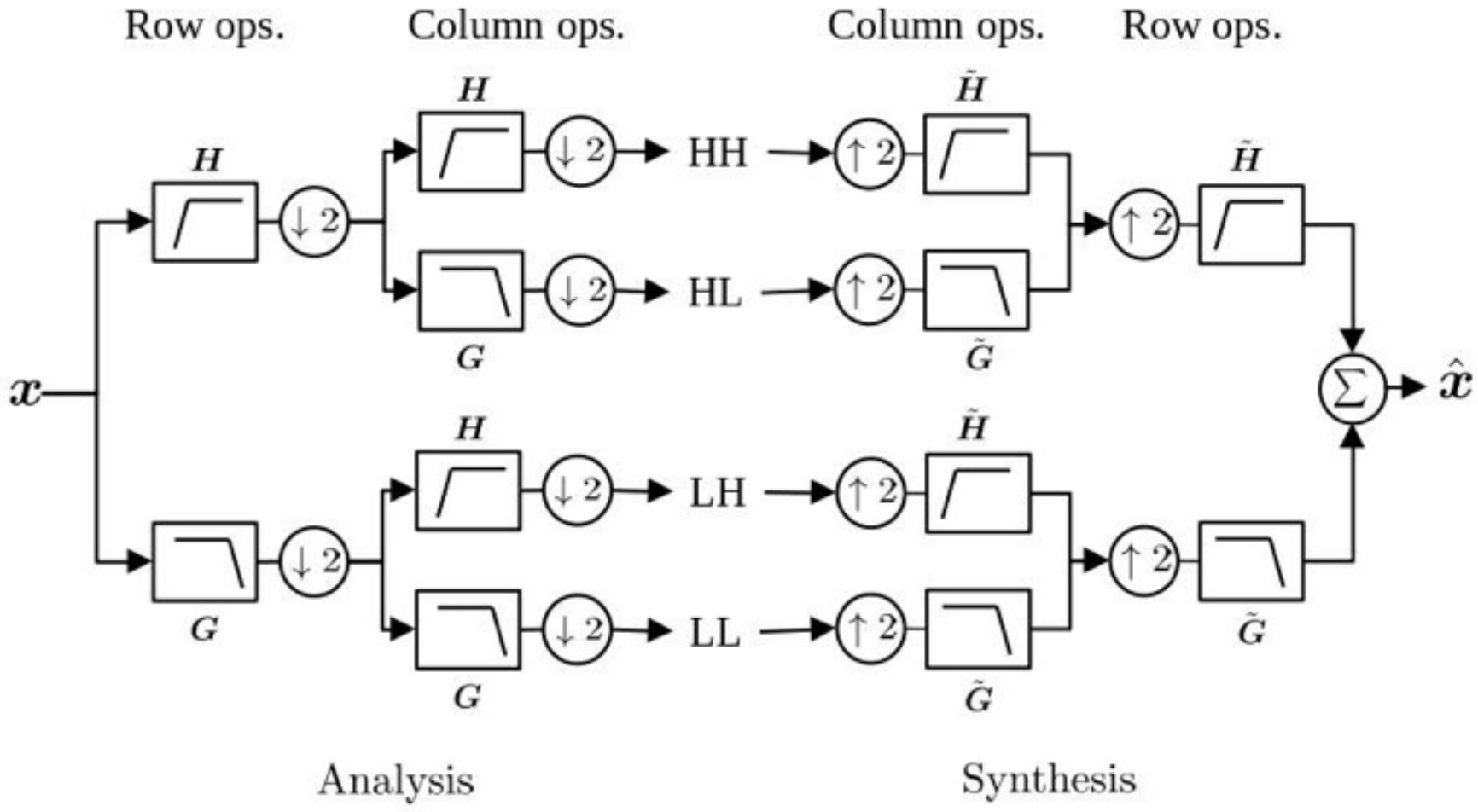

Figure 2.8: Standard separable DWT 2D perfect reconstruction filter bank

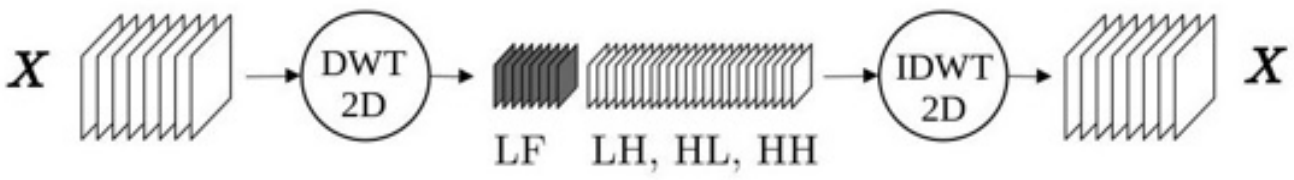

Figure 2.9: DWT 2D decomposition layer and perfect reconstruction with IDWT2D layer of a tensor **X**

**Einstein summations**: Here, the matrix multiplications in each separable transform are performed optimally with Einstein summations or Einsum. Einsum provides a powerful and concise way to perform tensor operations such as matrix multiplication, dot products, element-wise operations, and more. It simplifies complex calculations by eliminating the need for explicit loops, using repeated indices to imply summation.

**DWT 2D layer** Algorithm 2.9 describes a 2D DWT layer that operates on input tensors of shape (batch, height, width, channels) and produces an output of shape (batch, height/2, width/2, 4 × channels).

**Algorithm 2.9:** DWT 2D layer

1. Input $\mathbf{X}$ of shape (batch, height, width, channels).
2. Generate analysis matrix $\mathbf{A}$ using height and width of input.
3. For each batched channel $\mathbf{x}_c \in \mathbf{X}$ of shape (batch, height, width):
   (omitting suffix $c$ in $\mathbf{x}$ below for simplicity of notation)
   (a) Batched row wise DWT,
   $\mathbf{A}\mathbf{x}_{021}^{\top} := \text{Einsum}\big(ij, bjk \to bik\big)$
   (b) Batched column wise DWT,
   $\mathbf{A}(\mathbf{A}\mathbf{x}_{021}^{\top})_{021}^{\top} := \text{Einsum}\big(ij, bjk \to bik\big)$
   or, equivalently, DWT of a batched channel $\mathbf{x}$ is,
   $\mathbf{q}_c = \text{DWT}\big(\mathbf{x}\big)$ or, $\mathbf{q}_c = \mathbf{A}(\mathbf{A}\mathbf{x}_{021})_{021}^{\top}$
4. Stacking for all $c$ channels, $\mathbf{Q} := \big(\mathbf{q}_c\big)_{\forall c}$ to a shape (batch, height, width, channels).
5. Group subbands and return an output $\mathbf{Q}^{(\text{grouped})}$ of shape (batch, height/2, width/2, 4×channels)

$$
\begin{aligned}
\text{mid} &= \text{int}(\mathbf{Q}.\text{shape}[1]/2)\\
\text{LL} &= \mathbf{Q}[:, : \text{mid}, : \text{mid}, :]\\
\text{LH} &= \mathbf{Q}[:, \text{mid} :, : \text{mid}, :]\\
\text{HL} &= \mathbf{Q}[:, : \text{mid}, \text{mid} :, :]\\
\text{HH} &= \mathbf{Q}[:, \text{mid} :, \text{mid} :, :]\\
\mathbf{Q}^{(\text{grouped})} &= \text{Concatenate}([\text{LL}, \text{LH}, \text{HL}, \text{HH}], \text{axis} = -1)
\end{aligned}
$$

**IDWT 2D layer** Algorithm 2.10 describes an IDWT 2D layer that operates on input tensors of shape (batch, height/2, width/2, $4 \times$ channels) and produces an output of shape (batch, height, width, channels).

**Algorithm 2.10:** IDWT 2D layer

1. Input $\mathbf{Q}^{(\text{grouped})}$ of shape (batch, height/2, width/2, 4×channels)
2. Ungroup the subbands to get $\mathbf{Q}$ of shape (batch, height, width, channels)
3. Generate synthesis matrix $\mathbf{S}$ using height and width of $\mathbf{Q}$.
4. For each batched channel $\mathbf{q}_c \in \mathbf{Q}$ of shape (batch, height, width): (omitting suffix $c$ in $\mathbf{q}$ below for simplicity of notation)
   (a) Batched row wise IDWT is $\mathbf{S}\mathbf{q}_{021}^{\top} := \text{Einsum}(ij, bjk \rightarrow bik)$
   (b) Batched column wise IDWT is $\mathbf{S}(\mathbf{S}\mathbf{q}_{021}^{\top})_{021}^{\top} := \text{Einsum}(ij, bjk \rightarrow bik)$
   or equivalently, a PR,
   $\mathbf{x} = \text{IDWT}(\mathbf{q})$ or, $\mathbf{x} = \mathbf{S}(\mathbf{S}\mathbf{q}_{021}^{\top})_{021}^{\top}$, where $\mathbf{S} = \mathbf{A}^{\top}$ for orthogonal wavelets
5. Layer output (PR) $\mathbf{X} := (\mathbf{x}_c)_{\forall c}$ is of shape (batch, height, width, channels)

#### 2.2.2.3 Three-dimensional discrete wavelet system

A three-dimensional (3D) discrete wavelet system for a 3D input $\mathbf{x}$ can be written as

$$\mathbf{q} = \text{DWT}(\mathbf{x}) := \left[\mathbf{A}(\mathbf{A}(\mathbf{A}\mathbf{x}_{0213}^{\top})_{0213}^{\top})_{0132}^{\top}\right]_{0132}^{\top} \tag{2.37}$$

$$\mathbf{x} = \text{IDWT}(\mathbf{q}) := \left[\mathbf{S}(\mathbf{S}(\mathbf{S}\mathbf{q}_{0132}^{\top})_{0132}^{\top})_{0213}^{\top}\right]_{0213}^{\top} \tag{2.38}$$

where $\mathbf{x}_{0213}^{\top}$ denotes transpose of axis $[0, 1, 2, 3]$ to $0, 2, 1, 3]$ with 0 as batch index; 1,2 and 3 are height, width and depth. The matrices $\mathbf{A}$ and $\mathbf{S}$ are the same analysis and synthesis matrices defined in (2.31) and (2.32). The 3D DWT and IDWT layers operate on batched, multichannel tensors of shape (batch, height, width, depth, channels), where each cube is transformed separably by (2.37) and (2.38).

**DWT 3D layer** Algorithm 2.11 describes a DWT 3D layer that operates on input tensors of shape (batch, height, width, depth, channels) and produces an output of shape (batch, height/2, width/2, depth/2, $8 \times$ channels).

**Algorithm 2.11:** DWT 3D layer

1. Input $\mathbf{X}$ of shape (batch, height, width, depth, channels).
2. Generate analysis matrix $\mathbf{A}$ using height, width and depthof input.
3. For each batched channel $\mathbf{x}_c \in \mathbf{X}$ of shape (batch, height, width, depth): (omitting suffix $c$ in $\mathbf{x}$ below for simplicity of notation)
   (a) Row-wise operations, $\mathbf{A}\mathbf{x}_{0213}^{\top} := \text{Einsum}\big(ij, bjkl \to bikl\big)$
   (b) Column-wise operations, $\mathbf{A}\left(\mathbf{A}\mathbf{x}_{0213}^{\top}\right)_{0213}^{\top} := \text{Einsum}\big(ij, bjkl \to bikl\big)$
   (c) Depth-wise operations, $\mathbf{A}(\mathbf{A}(\mathbf{A}\mathbf{x}_{0213}^{\top})_{0213}^{\top})_{0132}^{\top} := \text{Einsum}\big(ik, bjkl \to bjil\big)$
   Therefore the forward transform of $\mathbf{x}$ yield DWT analysis coefficients
   $\mathbf{q}_c := \text{DWT}\,(\mathbf{x}) = \left[\mathbf{A}(\mathbf{A}(\mathbf{A}\mathbf{x}_{0213}^{\top})_{0213}^{\top})_{0132}^{\top}\right]_{0132}^{\top}$
4. Stacking for all $c$ channels, $\mathbf{Q} := \big(\mathbf{q}_c\big)_{\forall c}$ to a shape (batch, height, width, depth, channels).
5. Group subbands and return an output $\mathbf{Q}^{(\text{grouped})}$ of shape (batch, height/2, width/2, depth/2, 8×channels),

$$
\begin{aligned}
\text{mid} &= \text{int}(\mathbf{Q}.\text{shape}[2]/2)\\
\text{LLL} &= \mathbf{Q}[:, :\text{mid}, :\text{mid}, :\text{mid}, :]\\
\text{LLH} &= \mathbf{Q}[:, \text{mid}:, :\text{mid}, :\text{mid}, :]\\
\text{LHL} &= \mathbf{Q}[:, :\text{mid}, \text{mid}:, :\text{mid}, :]\\
\text{LHH} &= \mathbf{Q}[:, \text{mid}:, \text{mid}:, :\text{mid}, :]\\
\text{HLL} &= \mathbf{Q}[:, :\text{mid}, :\text{mid}, \text{mid}:, :]\\
\text{HLH} &= \mathbf{Q}[:, \text{mid}:, \text{mid}:, \text{mid}:, :]\\
\text{HHL} &= \mathbf{Q}[:, :\text{mid}, \text{mid}:, \text{mid}:, :]\\
\text{HHH} &= \mathbf{Q}[:, \text{mid}:, \text{mid}:, \text{mid}:, :]\\
\mathbf{Q}^{(\text{grouped})} &= \text{Concatenate([LLL, LLH, LHL, LHH, HLL, HLH, HHL, HHH], axis=-1)}
\end{aligned}
$$

**IDWT 3D layer** Algorithm 2.12 describes a IDWT 3D layer that operates on input tensors of shape (batch, height/2,width/2, depth/2,8 × channels) and produces an output of shape (batch, height, width, depth, channels).

**Algorithm 2.12:** IDWT 3D layer

1. Input $\mathbf{Q}^{(\text{grouped})}$ of shape (batch, height/2, width/2, depth/2, 8 × channels)
2. Ungroup to get $\mathbf{Q}$ of shape (batch, height, width, depth, channels)
3. Generate synthesis matrix $\mathbf{S}$ using height, width and depth of $\mathbf{Q}$.
4. For each batched channel $\mathbf{q}_c \in \mathbf{Q}$ of shape (batch, height, width, depth):
   (omitting suffix $c$ in $\mathbf{q}$ below for simplicity of notation)
   (a) Row-wise operations, $\mathbf{S}\mathbf{q}^{\top}_{0132} := \text{Einsum}(ik, bjkl \to bjil)$
   (b) Column-wise operations, $\mathbf{S}(\mathbf{S}\mathbf{q}^{\top}_{0132})_{0132} := \text{Einsum}(ij, bjkl \to bikl)$
   (c) Depth-wise operations, $\mathbf{S}(\mathbf{S}(\mathbf{S}\mathbf{q}^{\top}_{0132})^{\top}_{0132})^{\top}_{0213} := \text{Einsum}(ij, bjkl \to bikl)$
   or equivalently, a PR,
   $\mathbf{x}_c := \text{IDWT}(\mathbf{q}) = \left[\mathbf{S}(\mathbf{S}(\mathbf{S}\mathbf{q}^{\top}_{0132})^{\top}_{0132})^{\top}_{0213}\right]^{\top}_{0213}$ where $\mathbf{S} = \mathbf{A}^{\top}$ for orthogonal wavelets.
5. Layer output $\mathbf{X} := (\mathbf{x}_c)_{\forall c}$ is of shape (batch, height, width, depth, channels)
   (Perfect reconstruction)

In general, a seamless realization of fast D-dimensional DWT and IDWT is possible by extending the above separable method to all independent $N$ axes one after the other. The number of subbands will be equal to $2^{\text{D}}$ for a D dimensional DWT. For example, sequences (D = 1) yield two subbands, images (D = 2) yield four subbands, three-dimensional inputs (D = 3) with voxels yield eight subbands, etc.

#### 2.2.2.4 Multilevel wavelet filter banks

The discussed DWT and IDWT layers are building blocks in the construction of multilevel DWT filter banks. The multilevel DWT successively decomposes the low-frequency feature. If the high-frequency features are also decomposed successively, then we get a Wavelet Packet Transform (WPT) filter bank. The package `TFDWT` supports construction of multilevel DWT and WPT transforms for sequences (1D), images (2D) and volumetric (3D) signals.

### 2.2.3 Other 2D multiresolution and directional filter-bank layers

Beyond the wavelet and PR filter-bank layers developed above, several 2D multiresolution and directional transforms are relevant for image representations. These include Laplacian pyramids, steerable pyramids, contourlets, curvelets, wave atoms, monogenic pyramids, nonsubsampled directional constructions, shearlets, and scattering transforms. These transforms range from mostly isotropic scale decompositions to anisotropic directional decompositions designed to capture oriented edges, contours, and elongated image structures. Figure 2.10 shows representative frequency-plane tilings obtained from TensorFlow realizations of these layers in the `filterbanks` implementation developed alongside this thesis. Table 2.3 compares their redundancy, translation behavior, directionality, and indicative computational complexity. The listed redundancies and complexities correspond to the implemented layer settings and depend on the chosen filters, padding, scales, and orientations. Figure 2.10 show the tiling of the frequency plane by the realized transform layers.

Table 2.3: Transforms: redundancy, shift/translation behavior, directionality, and complexity

| # | Transform | Redund. | Translation | Directionality | Indic. Complexity |
|---|---|---|---|---|---|
| 1 | Laplacian pyramid (decimated) [138] | $< 4/3\times$ | Not shift-invariant (decimation) | Isotropic bands | $O(JN)$ (FIR) |
| 2 | DWT2D (PR; decimated) [139] | $1\times$ | Not shift-invariant (decimation) | 3 subbands/scale (H/V/D) | $O(JN)$ (FIR) |
| 3 | Steerable pyramid (FFT) [140] | $2 + JR$ | Translation-equivariant (no subsampling) | $R$ steerable orientations/scale | $O((2 + JR)N \log N)$ |
| 4 | Nonsubsampled Directional lifting [141] | $1 + JK$ | Translation-equivariant (no subsampling) | $K$ slopes (per level) | $O(JKN)$ (shears + depthwise FIR) |
| 5 | Contourlet (NSCT); FFT dir. stage) [142] | $1 + \sum_{j=1}^{J} K_j$ | Translation-equivariant (no subsampling) | Scale-dependent ($K_j$ dirs/level) | $O\big((\sum_{j=1}^{J} K_j)N \log N\big)$ |
| 6 | Curvelet (FFT wedges) [143] | $K_{\text{tot}}$ | Translation-equivariant (no subsampling) | Many anisotropic wedges | $O(K_{\text{tot}} N \log N)$ |
| 7 | WaveAtoms (FFT tiles) [144] | $1 + \sum_{j=1}^{J} K_j$ | Translation-equivariant (no subsampling) | Parabolic tiling (scale-dependent) | $O\big((\sum_{j=1}^{J} K_j)N \log N\big)$ |
| 8 | Monogenic pyramid (FFT) [145] | $3J$ (or $7J$) | Translation-equivariant (no subsampling) | Orientation via Riesz components | $O(JN \log N)$ |
| 9 | Shearlet (PRFB; FFT masks) [73] | $F$ subbands | Translation-equivariant (no subsampling) | Many anisotropic shears/scales | $O(FN \log N)$ |
| 10 | Scattering (wavelet + modulus) [35] | $P$ paths | Stable; becomes invariant after averaging | Inherited from wavelets | $O(PN \log N)$ |

*Notes:* Here $N = HW$ pixels, $J$ is the number of pyramid levels/scales, $R$ is orientations/scale, $K$ is slopes/level,

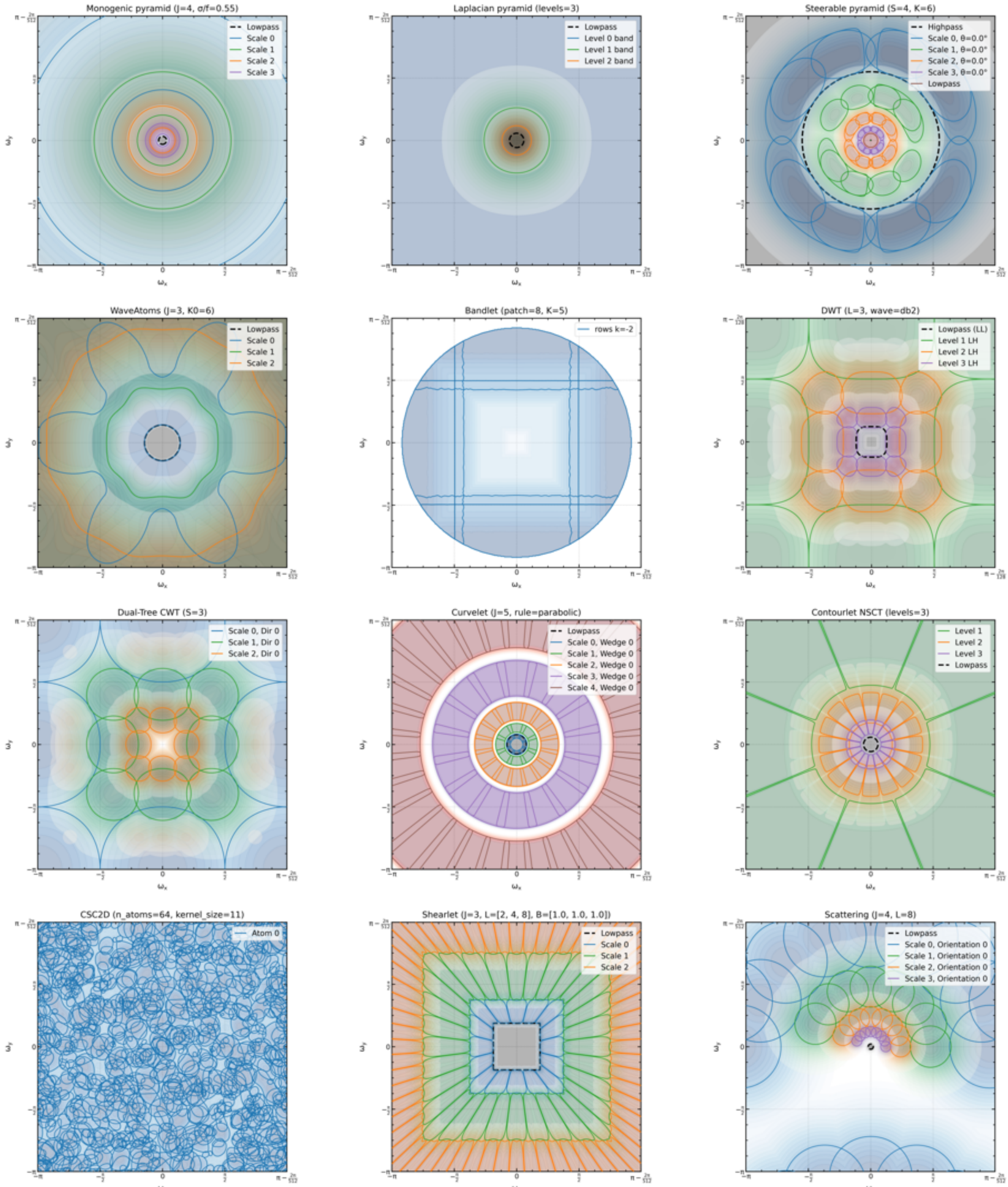

Figure 2.10: Tiling of the two-dimensional Fourier plane with the impulse responses of the various separable and nonseparable filterbanks with backpropagation support realized by our `filterbanks` Python package.

$K_j$ is directions/tiles at scale $j$, $F$ is the number of shearlet subbands, and $P$ is the number of scattering paths. Complexity is indicative for our forward passes; constants depend on filter lengths and the chosen $(J, R, K, K_j, F, P)$ configuration.

## 2.3 Learnable 3D Fast Digital Shearlet Transform

**Domain and transform conventions** We construct a learnable 3D Fast Digital Shearlet Transform (FDST). Its fixed *three cone-aligned regions*, *multiscale radial ladder*, and *directional wedge* scaffold follow the standard digital shearlet construction from the prior shearlet literature [73, 146]. In this section we describe our learnable digital realization of that scaffold, including per-frequency complex gains, sparse band bookkeeping, optional union or mixture rotation handling, and the dual synthesis stage. In the following FDST derivation we consider a single channel of the 3D input volume, represented as $x \in \mathbb{R}^{N_1 \times N_2 \times N_3}$ on a spatial grid. Its discrete Fourier transform is

$$\widehat{x}[\boldsymbol{\omega}] = \mathcal{F}\{x\}[\boldsymbol{\omega}], \qquad \boldsymbol{\omega} \in \Omega \subset \mathbb{R}^3, \tag{2.39}$$

where $\Omega$ denotes the discrete frequency grid induced by the Fast Fourier Transform (FFT) discretization. The corresponding inverse Fast Fourier Transform (IFFT) recovers $x = \mathcal{F}^{-1}\{\widehat{x}\}$. All FDST analysis and synthesis steps act by multiplying $\widehat{x}$ with fixed windows and learned complex gains on $\Omega$, channel-wise.

### 2.3.1 Frequency tiling and perfect-reconstruction scaffold

#### 2.3.1.1 Fixed three-cone geometry for anisotropy

In the 3D FDST implementation used here, the Fourier grid is organized into three *cones*, i.e., three disjoint regions of frequency space, each aligned with one of the coordinate axes. A cone collects those frequencies whose energy is largest along a given axis and serves as the coarse directional carrier on top of which more refined wedge windows are built. This three-cone structure is the backbone of the anisotropic, direction-sensitive tiling used by the transform. For each frequency vector $\boldsymbol{\omega} = (\omega_x, \omega_y, \omega_z)$ we define the axis–aligned magnitudes

$$m_x(\boldsymbol{\omega}) = |\omega_x|, \qquad m_y(\boldsymbol{\omega}) = |\omega_y|, \qquad m_z(\boldsymbol{\omega}) = |\omega_z|. \tag{2.40}$$

Let $C = \{x, y, z\}$ denote the cone labels. The underlying geometric partition is defined by the dominant axis

$$c_{\text{hard}}(\boldsymbol{\omega}) = \arg\max_{c \in C} m_c(\boldsymbol{\omega}), \tag{2.41}$$

and the corresponding hard cone indicators

$$\chi_x(\boldsymbol{\omega}) = \mathbf{1}\big[c_{\text{hard}}(\boldsymbol{\omega}) = x\big], \tag{2.42a}$$

$$\chi_y(\boldsymbol{\omega}) = \mathbf{1}\big[c_{\text{hard}}(\boldsymbol{\omega}) = y\big], \tag{2.42b}$$

$$\chi_z(\boldsymbol{\omega}) = \mathbf{1}\big[c_{\text{hard}}(\boldsymbol{\omega}) = z\big], \tag{2.42c}$$

where $\mathbf{1}[\cdot]$ denotes the indicator of a logical condition. Collectively, we write these hard indicators as $\chi_c(\boldsymbol{\omega})$ for $c \in C$. Because $c_{\text{hard}}(\boldsymbol{\omega})$ is defined by the maximum of $\{m_x(\boldsymbol{\omega}), m_y(\boldsymbol{\omega}), m_z(\boldsymbol{\omega})\}$, each frequency bin is assigned to exactly one of the three axis–aligned cones, and the use of absolute values merges positive and negative halves so that $\boldsymbol{\omega}$ and $-\boldsymbol{\omega}$ share the same cone.

We keep this hard partition as the seed geometry for band indexing, so that each band has a well-defined cone label $c \in C$ and the overall layout of bands (scale, cone, orientation) remains simple and interpretable. In the actual window construction, however, the hard masks are relaxed through *soft cone sharing*. For each $\boldsymbol{\omega}$ we form scores $\sigma_c(\boldsymbol{\omega})$ from the magnitudes, for example

$$\sigma_c(\boldsymbol{\omega}) = \frac{m_c(\boldsymbol{\omega})}{\tau}, \qquad c \in C, \tag{2.43}$$

with a cone–smoothing parameter $\tau > 0$, and define nonnegative cone weights

$$w_c(\boldsymbol{\omega}) = \frac{\exp(\sigma_c(\boldsymbol{\omega}))}{\sum_{d\in C} \exp(\sigma_d(\boldsymbol{\omega}))}, \qquad c \in C, \tag{2.44}$$

so that $\sum_{c\in C} w_c(\boldsymbol{\omega}) = 1$ at every frequency bin. These weights are used only when constructing the smooth windows, so each cone window is multiplied by $w_c(\boldsymbol{\omega})$ instead of $\chi_c(\boldsymbol{\omega})$. The hard cones therefore define the band labels, whereas the soft weights smooth the cone seams and reduce ringing after IFFT. Deep inside a cone, one weight stays close to 1 and the others stay close to 0, while bins near cone boundaries are shared across cones in proportion to their axis-aligned magnitudes. Throughout this work we use this smoothed variant, with the degree of soft sharing controlled by $\tau$. The ideal hard-cone model is recovered in the limit of small $\tau$, up to the global partition-of-unity (PoU) normalization. After the full window bank is assembled and PoU-normalized, the resulting sparse supports may overlap even though the seed partition is hard and disjoint.

#### 2.3.1.2 Fixed multiscale radial and angular windows

To organize frequencies by scale, the implementation uses a radial coordinate $\rho(\boldsymbol{\omega})$ that measures the distance of each frequency bin from the origin under a chosen norm. This radius partitions the grid into one lowpass (LP) band and $J$ highpass (HP) shells. Conceptually, we use an LP window $b_{\text{lp}}(\rho)$ together with HP shell windows $\{b_j(\rho)\}_{j=0}^{J-1}$ whose squared magnitudes form an approximate PoU along the radius,

$$b_{\text{lp}}(\rho)^2 + \sum_{j=0}^{J-1} b_j(\rho)^2 \approx 1 \quad \text{for } 0 \le \rho \le \rho_{\max}. \tag{2.45}$$

Here $\rho_{\max}$ denotes the maximum radius attained on the discrete grid. Equation (2.45) is a conceptual radial template. In the implementation studied here, the operative radius is the normalized coordinate $\bar{\rho}(\boldsymbol{\omega})$ obtained from $\rho_p(\boldsymbol{\omega})$ below. The LP band is indexed separately, and the HP shell index $j = 0, \ldots, J-1$ is used throughout.

The shell edges are defined on a normalized radius $\bar{\rho}(\boldsymbol{\omega}) \in [0, 1]$ on the discrete grid. An LP cutoff $\bar{\rho}_0 \in (0, 1)$ fixes the LP band, and the remaining range $(\bar{\rho}_0, 1]$ is split into $J$ shells according to the chosen mode,

$$\bar{\rho}_0 < \bar{\rho}_1 < \cdots < \bar{\rho}_J = 1, \tag{2.46}$$

The shell boundaries in (2.46) use either linear spacing (equal shell thickness) or logarithmic spacing (finer shells near $\bar{\rho}_0$). The LP band is given a smooth radial taper controlled by a radial smoothness parameter, while each HP band $j$ is initially supported on a hard shell $(\bar{\rho}_j, \bar{\rho}_{j+1}]$ with nearly flat radial windows inside the shell. Exact PoU is imposed later by the global normalization in (2.57), so this stage only defines the coarse multiscale radial ladder.

**Radial norm and rotation-aware radius** The radius used to define the shells is obtained by applying an $\ell^p$ norm, with $p \in \{2, 1, \infty\}$, to the frequency vector $\boldsymbol{\omega} = (\omega_x, \omega_y, \omega_z) \in \mathbb{R}^3$:

$$\rho_p(\boldsymbol{\omega}) = \|\boldsymbol{\omega}\|_p = \begin{cases} (\omega_x^2 + \omega_y^2 + \omega_z^2)^{1/2}, & p = 2, \\ |\omega_x| + |\omega_y| + |\omega_z|, & p = 1, \\ \max\{|\omega_x|, |\omega_y|, |\omega_z|\}, & p = \infty. \end{cases} \tag{2.47}$$

All radii in the ladder are obtained by normalizing $\rho_p(\boldsymbol{\omega})$ to the range $[0, 1]$ on the discrete grid. All three choices are admissible. In the experiments reported here we use $p = 2$ (Euclidean radius), which yields spherical shells and is invariant under rotations. Using $p = \infty$ or $p = 1$ instead would change the shell geometry (cubic or octahedral), but not the reconstruction guarantees, since the final PoU normalization is always applied.

When a bank of rotated windows is built, each rotation $\mathbf{R} \in \mathrm{SO}(3)$ acts on the frequency coordinates. For $p = 2$ the radius is invariant under $\mathbf{R}$, so $\rho_2(\mathbf{R}^\top \boldsymbol{\omega}) = \rho_2(\boldsymbol{\omega})$ and the shells are the same for all rotations. For $p \in \{1, \infty\}$ the radius depends on orientation, and one may recompute $\rho_p$ from the rotated coordinates $\mathbf{R}^\top \boldsymbol{\omega}$, so that the shells are aligned with the rotated axes. This "rotation-aware" radius keeps the notion of "near the corner" or "near an axis" consistent with the rotated cones. In all cases, the final PoU is imposed after combining radial and angular factors by normalizing the sum of squared window magnitudes over the grid, so the choice of $p$ affects the shape of the shells but not the exactness of reconstruction. Figure 2.11 shows the corresponding PoU diagnostic for the FDST window bank.

**Angular windows inside cones (fixed directionality)** Within each radial shell and cone, the implementation further splits frequencies by orientation. For scale $j$ and cone $c \in \{x, y, z\}$ we choose an integer $L_j$ and form $L_j$ wedge directions. Rather than working with global spherical coordinates, the implementation uses a 2D angular coordinate on the plane orthogonal to the cone axis. For example, in the $x$–cone we measure an angle in the $(\omega_y, \omega_z)$–plane, and similarly for the $y$– and $z$–cones. For each cone and scale we place $L_j$ equally spaced central angles on this circle and define smooth angular tapers around them.

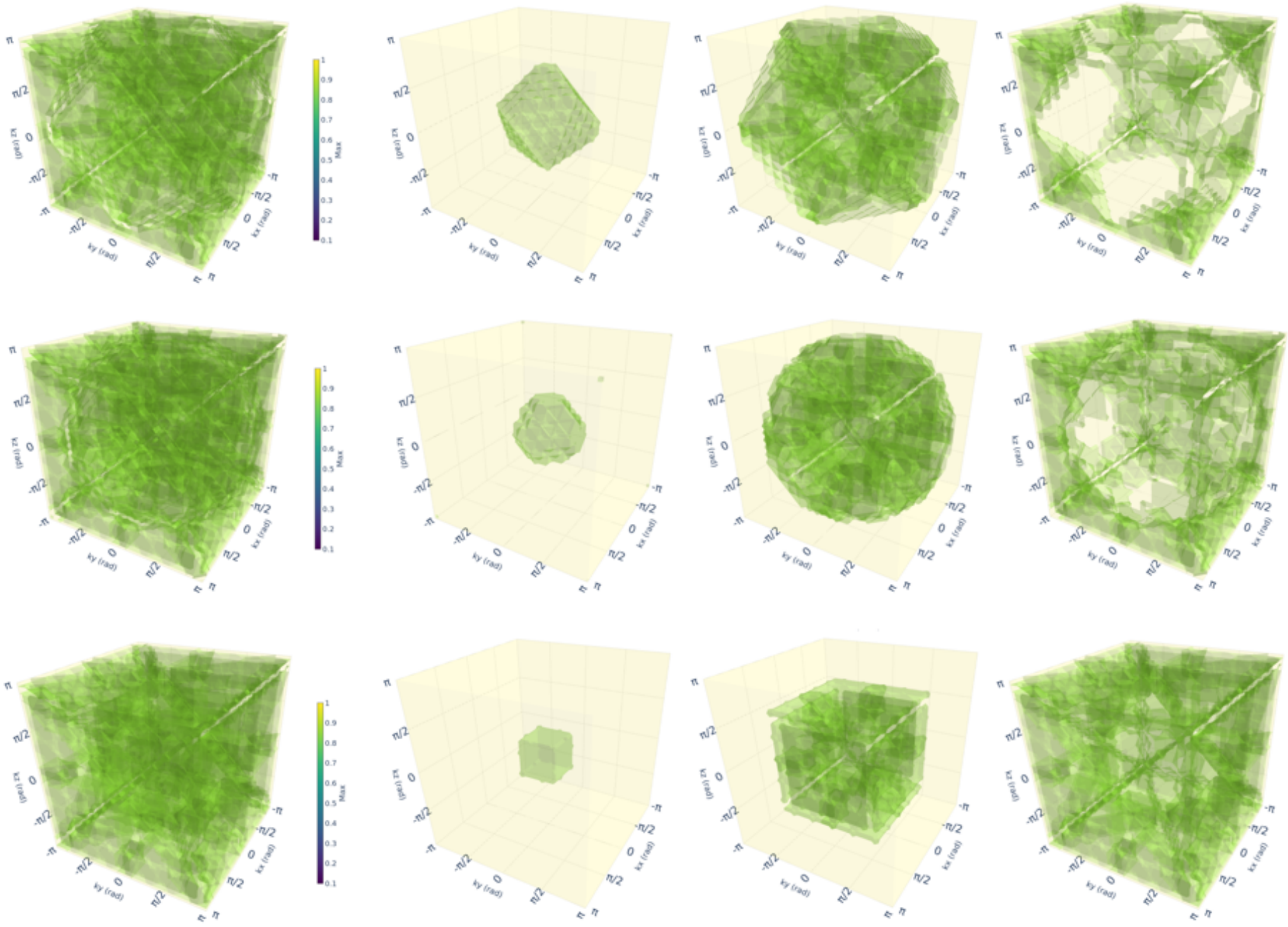

Figure 2.11: 3D PoU diagnostic for FDST. Green volumes are isosurfaces where $D_0[\boldsymbol{\omega}] = \sum_k |W_k[\boldsymbol{\omega}]|^2 \approx 1$. The outer envelopes illustrate three radial metrics, namely $\ell^1$ (octahedral ball in the top row), $\ell^2$ (spherical ball in the middle row), and $\ell^\infty$ (cubic ball in the bottom row). The central blob is the LP region. Seam brightening indicates wedge junctions (frequency axes $\omega_x$, $\omega_y$, $\omega_z$ in rad). From left to right, the panels show all tiles overlaid, the LP filter, the shell $j = 0$ filters, and the shell $j = 1$ filters.

Concretely, for fixed $j$ and cone $c$ we define central angles $\{\theta_{j,\ell}\}_{\ell=0}^{L_j-1}$, spaced by $2\pi/L_j$, and define for each $\ell$ an angular weight

$$a_{j,\ell}^{(c)}(\boldsymbol{\omega}) = \frac{1}{2}\Big(1 + \cos\Big(\pi\,\frac{d_c(\boldsymbol{\omega}, \theta_{j,\ell})}{\mathrm{half}_j}\Big)\Big) \quad \text{for } d_c(\boldsymbol{\omega}, \theta_{j,\ell}) \leq \mathrm{half}_j, \tag{2.48}$$

and $a_{j,\ell}^{(c)}(\boldsymbol{\omega}) = 0$ outside this range. Here $d_c(\cdot,\cdot)$ is the circular distance on the corresponding 2-D plane, and

$$\mathrm{half}_j = \kappa \cdot \frac{\pi}{L_j} \tag{2.49}$$

This quantity controls the angular half-width of each wedge through an overlap factor $\kappa > 0$. For small $\kappa$ the wedges are narrow with limited overlap. For larger values the wedges overlap more strongly and the average redundancy increases. The resulting angular windows are compactly supported cosine tapers centered at the chosen directions, with neighboring wedges overlapping smoothly.

In the experiments reported here, the directional counts follow the parabolic schedule

$$L_j = 2^j L_0, \qquad j = 0, \ldots, J-1, \tag{2.50}$$

where $L_0$ denotes the number of directions in the first HP shell, so the number of in-cone directions doubles from one shell to the next. We retain the notation $L_j$ in the derivation because the formulas are valid for general directional counts, even though the results reported here use this parabolic schedule throughout. At the level of these angular factors alone we do not enforce an exact two-term identity such as $a_{j,\ell}^2 + a_{j,\ell+1}^2 = 1$ at their seam. Instead, the squared magnitudes of all radial–angular windows (including all cones and all rotations) are summed over the grid and then renormalized pointwise to obtain an exact PoU. Thus the angular construction provides a smooth directional tiling indexed by $(j, c, \ell)$, while the final normalization supplies the precise energy partition needed for perfect reconstruction (PR) and for the union or mixture rotation construction described below.

### 2.3.2 Rotated wedges and rotation modes

**Unrotated wedge windows (canonical tiling)** Combining the radial and angular factors defined above with the cone weights yields a baseline wedge tiling. For each HP shell $j = 0, \ldots, J-1$, cone $c \in \{x, y, z\}$, and in-cone direction $\ell \in \{0, \ldots, L_j - 1\}$ we define an unrotated wedge window

$$W_{j,\ell,c}^{(0)}(\boldsymbol{\omega}) = b_j\big(\bar{\rho}(\boldsymbol{\omega})\big)\, a_{j,\ell}^{(c)}(\boldsymbol{\omega})\, w_c(\boldsymbol{\omega}), \tag{2.51}$$

where $b_j$ is the radial window for HP shell $j$, $a_{j,\ell}^{(c)}$ is the angular taper for direction $\ell$ in cone $c$, and $w_c(\boldsymbol{\omega})$ is the (hard or soft) cone weight described earlier. The LP band is

$$W_0^{(0)}(\boldsymbol{\omega}) = b_{\mathrm{lp}}\big(\bar{\rho}(\boldsymbol{\omega})\big). \tag{2.52}$$

Together, (2.51) and (2.52) define the unrotated wedge bank. This collection $\{W_0^{(0)}, W_{j,\ell,c}^{(0)}\}$ provides a fixed cone×scale×direction tiling aligned with the coordinate axes.

**Rotation set and rotated wedges** To cover orientations that are not aligned with the axes, the implementation can augment the canonical bank with a finite set of 3D rotations. Let $\{\mathbf{R}_m\}_{m=0}^{M-1} \subset \mathrm{SO}(3)$ be a set of rotation matrices ($\mathbf{R}_m^\top \mathbf{R}_m = I$, $\det \mathbf{R}_m = +1$), where $I$ denotes the identity matrix. For each $\mathbf{R}_m$ we build a rotated copy of the wedge tiling by evaluating the same radial and angular formulas in the rotated coordinates $\mathbf{R}_m^\top \boldsymbol{\omega}$. Conceptually,

$$W_{j,\ell,c,m}(\boldsymbol{\omega}) = b_j\big(\bar{\rho}(\mathbf{R}_m^\top \boldsymbol{\omega})\big)\, a_{j,\ell}^{(c)}(\mathbf{R}_m^\top \boldsymbol{\omega})\, w_c(\mathbf{R}_m^\top \boldsymbol{\omega}), \tag{2.53}$$

with an analogous rotated LP window $W_{0,m}(\boldsymbol{\omega})$. In practice, these windows are constructed directly on the rotated frequency coordinates. The LP band is taken from one reference rotation, while additional rotations contribute only HP wedges. In the experiments reported here we use the identity rotation only (so $M = 1$), and the multi-rotation machinery becomes active when a nontrivial rotation set is provided.

**Union and mixture rotation modes** The implementation supports two rotation constructions. In *union* mode, used by default when multiple rotations are provided, all rotated wedges $\{W_{j,\ell,c,m}\}$ are kept as separate bands. The resulting bank has more bands (larger $K$), and the metadata $r(k)$ allows later stages to associate each band with its underlying rotation, for example when learning per-rotation weights. In *mixture* mode, for each $(j,\ell,c)$ the HP windows $\{W_{j,\ell,c,m}\}_m$ are first averaged over $m$, the LP band is kept from a reference rotation, and the mixed windows are then renormalized to restore the PoU. In this case there is effectively a single rotation and all bands share the same rotation id. Both constructions are mathematically admissible. In the experiments reported here, the union construction is used when multiple rotations are studied and reduces to the identity-only case when no additional rotations are supplied.

## 2.3.3 Sparse bank assembly and coverage control

### 2.3.3.1 Flattened bands and sparse supports

After choosing the rotation mode and constructing the corresponding windows $\{W_k\}_{k=0}^{K-1}$ (either by taking the full union over $m$ or by mixing rotations per band), we represent the entire bank as a flat list of bands indexed by $k \in \{0,\dots,K-1\}$. For each band $k$ we store a sparse support set $\mathcal{I}_k \subset \Omega$ on the discrete Fourier grid and a window vector $W_k[\boldsymbol{\omega}]$ that is defined only for $\boldsymbol{\omega} \in \mathcal{I}_k$, along with metadata such as the scale, cone, in-cone direction, and rotation id $r(k)$. This sparse, band-wise representation avoids storing a dense (grid $\times$ $K$) tensor of windows. A sparse realization uses compact index lists and per-band weights (one tile per band), and gathers only the frequencies in $\mathcal{I}_k$ when applying the transform. As a result, memory and computation scale with $\sum_k |\mathcal{I}_k|$ rather than with the full grid size times $K$. A single global partition-of-unity normalization is then applied over the flattened bank,

$$D_0[\boldsymbol{\omega}] := \sum_{k=0}^{K-1} |W_k[\boldsymbol{\omega}]|^2 \approx 1, \tag{2.54}$$

The energy scaffold in (2.54) shows that the windows still provide an (approximately) energy-preserving coverage of the Fourier grid, which is the basis for (near) perfect reconstruction in the dual synthesis step.

### 2.3.3.2 Overlap and redundancy control

The geometric layout (cones, radial shells, angular wedges, and rotations) fixes where each band lives in the discrete frequency grid, but not yet how sharply neighboring bands overlap or how uniformly they cover the grid. This subsection chooses the overlap parameters, adjusts redundancy when needed, and then enforces the final PoU by coverage repair and pointwise normalization.

**Boundary parameters and basic shaping** The overlap is governed by a small set of controls. On the radial side, a smoothness parameter determines how gently the LP band decays towards its cutoff at $\bar{\rho}_0$, while the HP shells remain essentially flat on their assigned intervals $(\bar{\rho}_j, \bar{\rho}_{j+1}]$. On the angular side, the factor $\kappa$ in (2.49) scales the cosine-taper half-width, so smaller values give narrower wedges and larger values increase overlap and redundancy. Exact PoU is again imposed only after the full bank has been assembled.

**Redundancy tuning (average overlap per frequency)** To steer the average redundancy to a prescribed level, we may adapt the overlap factor $\kappa$ before freezing the windows. Given a target redundancy $R^\star > 1$ and the current value of $\kappa$, we construct the full wedge bank and count, for each frequency bin $\boldsymbol{\omega} \in \Omega$, how many bands $k$ have nonnegligible support at that bin:

$$R_{\text{red}} = \frac{1}{|\Omega|} \sum_{\boldsymbol{\omega} \in \Omega} \#\{\, k : |W_k[\boldsymbol{\omega}]| > \varepsilon_{\text{supp}} \,\}. \tag{2.55}$$

Here $\varepsilon_{\text{supp}} > 0$ is a small support threshold, and $R_{\text{red}}$ is a count-based proxy for the average number of overlapping bands per frequency. If $R_{\text{red}}$ differs significantly from $R^\star$, we rescale $\kappa$ by $\sqrt{R^\star / R_{\text{red}}}$, clip the update, rebuild the bank, and recompute $R_{\text{red}}$ for a small fixed number of iterations. The procedure is heuristic because it uses overlap counts rather than exact energy, but in practice it moves the mean redundancy towards the desired value. When no target redundancy is prescribed, the initial overlap factor is used as is.

#### 2.3.3.3 Coverage repair and final PoU

After the wedge bank has been constructed (with or without redundancy tuning), let $W_k[\boldsymbol{\omega}]$ denote the fixed window of band $k$ at the discrete frequency bin $\boldsymbol{\omega}$, with $W_0$ denoting the LP window. The pre-normalization energy profile is evaluated as

$$S(\boldsymbol{\omega}) = \sum_{k=0}^{K-1} |W_k[\boldsymbol{\omega}]|^2, \qquad \boldsymbol{\omega} \in \Omega. \tag{2.56}$$

Ideally $S(\boldsymbol{\omega})$ should be positive and close to 1 everywhere, but small gaps may appear near cone boundaries or at the outer edges of the radial ladder. To repair such coverage holes, the implementation identifies frequency bins with $S(\boldsymbol{\omega}) \le \varepsilon_{\text{pou}}$ and assigns them to the LP band by setting $W_0[\boldsymbol{\omega}] = 1$ at those locations. In the implementation $\varepsilon_{\text{pou}} = 10^{-12}$. This is a fixed coverage-repair step in the window-bank construction, not a learned parameter. The energy profile is recomputed after this LP repair, and a global PoU is then enforced by normalizing all windows pointwise:

$$W_k[\boldsymbol{\omega}] \leftarrow \frac{W_k[\boldsymbol{\omega}]}{\sqrt{\max\{S(\boldsymbol{\omega}), \varepsilon_{\text{pou}}\}}}, \qquad k = 0, \dots, K-1, \tag{2.57}$$

with the same stabilizer $\varepsilon_{\text{pou}} > 0$. After this step the flattened bank satisfies

$$D_0[\boldsymbol{\omega}] = \sum_{k=0}^{K-1} |W_k[\boldsymbol{\omega}]|^2 \approx 1 \quad \text{for all } \boldsymbol{\omega} \in \Omega, \tag{2.58}$$

up to numerical tolerance. This $D_0$ is the same energy scaffold used in the fixed-window design. Thus the fixed FDST window bank is a PR bank on the discrete grid, up to numerical tolerance. It provides the PoU scaffold before learning, while the learned dual denominator $D[\boldsymbol{\omega}]$ introduced below compensates the per-band gains and rotation weights during synthesis.

### 2.3.4 Analysis and learnable gains

The previous sections specify the fixed frequency tiling, namely cones, radial shells, angular wedges, rotations, and their normalized windows $\{W_k\}_{k=0}^{K-1}$. We now describe how this tiling is used to analyze a signal into band-wise coefficients and how these coefficients are modulated by learnable complex gains and rotation weights.

**Band-wise analysis (fixed filtering)** Given a real-valued input volume $x$ with Fourier transform $\widehat{x}[\boldsymbol{\omega}]$, the analysis stage applies the fixed windows $\{W_k\}_{k=0}^{K-1}$ band by band. For each band $k$ with support $\mathcal{I}_k \subset \Omega$ we define the analysis tile

$$X_k[\boldsymbol{\omega}] = \begin{cases} W_k[\boldsymbol{\omega}]\ \widehat{x}[\boldsymbol{\omega}], & \boldsymbol{\omega} \in \mathcal{I}_k, \\ 0, & \boldsymbol{\omega} \notin \mathcal{I}_k. \end{cases} \tag{2.59}$$

In the sparse realization used here, for each $k$ we store an index list $\mathcal{I}_k$ and a window vector $W_k[\boldsymbol{\omega}]$ on that set, and $X_k$ is represented as a complex tile of shape $[B, |\mathcal{I}_k|, 1]$ for a batch of size $B$ in the single-channel derivation above, obtained by gathering $\widehat{x}$ on $\mathcal{I}_k$ and multiplying by $W_k$.

**Per-frequency complex gains** On top of the fixed analysis windows, the transform learns a complex gain per band and per frequency sample. For every pair $(k, \boldsymbol{\omega} \in \mathcal{I}_k)$ we maintain two real parameters

$$A_k[\boldsymbol{\omega}] = \operatorname{softplus}\big(p_k[\boldsymbol{\omega}]\big), \tag{2.60a}$$
$$\phi_k[\boldsymbol{\omega}] = q_k[\boldsymbol{\omega}], \tag{2.60b}$$
$$g_k[\boldsymbol{\omega}] = A_k[\boldsymbol{\omega}]\ e^{i\,\phi_k[\boldsymbol{\omega}]}, \tag{2.60c}$$
$$\widehat{X}_k[\boldsymbol{\omega}] = g_k[\boldsymbol{\omega}]\ X_k[\boldsymbol{\omega}]. \tag{2.60d}$$

The gain parametrization in (2.60) uses two real parameters $p_k[\boldsymbol{\omega}]$ and $q_k[\boldsymbol{\omega}]$ for every pair $(k, \boldsymbol{\omega} \in \mathcal{I}_k)$. The amplitude $A_k[\boldsymbol{\omega}] \geq 0$ controls how much of each frequency is passed, the phase $\phi_k[\boldsymbol{\omega}]$ allows phase corrections, and the softplus nonlinearity keeps amplitudes nonnegative and numerically stable. The natural identity gain is included in this parametrization. Choosing $p_k[\boldsymbol{\omega}] = \operatorname{softplus}^{-1}(1) = \log(e-1)$ and $q_k[\boldsymbol{\omega}] = 0$ gives $g_k[\boldsymbol{\omega}] = 1$. We use softplus rather than an unconstrained amplitude so the learned amplitude remains nonnegative without clipping. Compared with an exponential parametrization, it grows more mildly for large positive logits. In our implementation, these amplitude and phase variables are learned for every band, including the LP band $k = 0$. The tiles returned by the analysis layer already include both the fixed windows and these gains, so the split into $X_k$ and $g_k$ above is conceptual.

**Rotation weights and effective gains** When multiple rotations are used (union mode), the bands are grouped by the rotation index $m$ from which they originate. We introduce a trainable logit $z_m^{\text{fdst}}$ for each rotation and form a softmax and corresponding effective gains:

$$\alpha_m = \frac{e^{z_m^{\text{fdst}}}}{\sum_{n=0}^{M-1} e^{z_n^{\text{fdst}}}}, \qquad \alpha_m \geq 0, \quad \sum_{m=0}^{M-1} \alpha_m = 1. \tag{2.61a}$$

$$\beta_k[\omega] = \alpha_{r(k)}\, g_k[\omega]. \tag{2.61b}$$

$$\widetilde{X}_k[\omega] = \alpha_{r(k)}\, \widehat{X}_k[\omega] = \beta_k[\omega]\, X_k[\omega] = \beta_k[\omega]\, W_k[\omega]\, \widehat{x}[\omega]. \tag{2.61c}$$

Each band $k$ stores a rotation id $r(k) \in \{0, \ldots, M-1\}$. Equation (2.61b) defines its effective gain, and (2.61c) gives the analyzed tile entering synthesis. This includes the LP band ($k = 0$), whose window is unrotated but still scaled by a scalar weight $\alpha_{r(0)}$. In mixture mode the rotated windows are pre-averaged before analysis and all bands share the same rotation id. In that situation the softmax is taken over a single logit, so $\alpha_0 \equiv 1$ and $\beta_k[\omega] = g_k[\omega]$ for all bands, regardless of the logit value. Thus nontrivial rotation mixing only occurs in union mode with $M > 1$.

### 2.3.5 Learnable synthesis bank

**Dual synthesis (near-PR)** The synthesis bank uses the same gains and windows as analysis, but in a dual form whose denominator is built from the learned responses themselves. We define the frequency-wise normalizer and dual synthesis rule as follows.

$$D[\omega] = \varepsilon + \sum_{k=0}^{K-1} |\beta_k[\omega]|^2\, |W_k[\omega]|^2, \qquad \omega \in \Omega, \tag{2.62a}$$

$$\widehat{\tilde{x}}[\omega] = \frac{\sum_{k=0}^{K-1} \widetilde{X}_k[\omega]\, \overline{W_k[\omega]}\, \overline{\beta_k[\omega]}}{D[\omega]}, \tag{2.62b}$$

with a small stabilizer $\varepsilon > 0$. In exact arithmetic with a covered frequency grid, one may take $\varepsilon = 0$. The implementation keeps a tiny positive value, $\varepsilon = 10^{-12}$ in the ShearViT layer, only to avoid division by zero or roundoff issues when some learned amplitudes become very small. Using the analyzed tiles $\widetilde{X}_k$ (which already include the fixed window, the complex gain, and the rotation weight), the synthesis rule in (2.62b) assembles the synthesized spectrum. Here the denominator is (2.62a), and the overline denotes complex conjugation. Substituting the analysis relation $\widetilde{X}_k[\omega] = \beta_k[\omega]\, W_k[\omega]\, \widehat{x}[\omega]$ and using that $W_k$ is real-valued shows that, up to the stabilizer $\varepsilon$, the numerator equals

$$\widehat{x}[\omega] \sum_{k=0}^{K-1} |\beta_k[\omega]|^2\, |W_k[\omega]|^2 = D[\omega]\, \widehat{x}[\omega] - \varepsilon\, \widehat{x}[\omega]. \tag{2.63}$$

When the bank covers the grid and $\varepsilon$ is small, dividing by $D[\omega]$ cancels the learned gains and windows up to the stabilizer term and recovers $\widehat{x}[\omega]$. The fixed PoU scaffold in (2.58) helps keep this denominator well behaved, but the algebraic cancellation comes from the definition of $D[\omega]$ in (2.62a). In our implementation, the denominator is built once over the flattened grid by accumulating $|\beta_k|^2|W_k|^2$ band-wise, and the synthesis weights are $\overline{\beta_k W_k}/D$. The spatial output $\tilde{x}$ is obtained by a 3D IFFT of $\widehat{\tilde{x}}$, followed by taking the real part.

**Real-valued outputs and Hermitian symmetry** For real-valued inputs the exact condition for a real output is the Hermitian symmetry

$$\widehat{\tilde{x}}[-\omega] = \overline{\widehat{\tilde{x}}[\omega]}. \tag{2.64}$$

One way to guarantee this is to tie the gains on conjugate frequency bins,

$$g_k[-\omega] = \overline{g_k[\omega]} \quad \Rightarrow \quad A_k[-\omega] = A_k[\omega],\ \phi_k[-\omega] = -\phi_k[\omega]. \tag{2.65}$$

The formulation used in our experiments does not impose this constraint explicitly. It relies on the real input and returns the real part of the IFFT output, discarding any small imaginary component introduced by numerical errors or asymmetries in the learned gains. Enforcing a strict Hermitian tie is therefore a natural theoretical refinement that could be added if exact spectral symmetry is desired.

**Tied synthesis (phase-only special case)** A faster variant is obtained when the gains are constrained to be phase-only, i.e. $|g_k[\omega]| \equiv 1$ for all $(k, \omega)$, and the fixed windows satisfy the PoU condition. In this setting, a tied synthesis rule omits the denominator $D[\omega]$. In the realization studied here, this option is used only under phase-only gains, so the synthesis weights reduce to $\overline{\beta_k W_k}$ and the denominator in the dual formula is implicitly 1. No additional constraint is imposed on the learnable rotation weights, so this tied mode should be interpreted as a fast special case rather than as a separate exact PR guarantee under arbitrary learned settings.

**Fixed vs. learned components (summary)** The following parts of the transform are fixed once a configuration is chosen. They include the FFT and IFFT operators, the three-cone construction and cone weights $w_c$, the radial ladder $\{\bar{\rho}_j\}$ and windows $b_{\text{lp}}, b_j$, the in-cone angular windows $a^{(c)}_{j,\ell}$, the base wedges and their rotated copies, the flattened band supports $\{(\mathcal{I}_k, W_k)\}$, the band $\rightarrow$ rotation map $r(k)$, and, in mixture mode, the pre-averaging of rotated windows. The learned components are the per-frequency gain parameters $p_k[\omega], q_k[\omega]$ (hence $g_k[\omega]$ and $\beta_k[\omega]$) for every band $k$ and frequency bin $\omega \in \mathcal{I}_k$, and, in union mode with $M > 1$, the rotation logits $\{z^{\text{fdst}}_m\}$ (hence the weights $\alpha_m$).

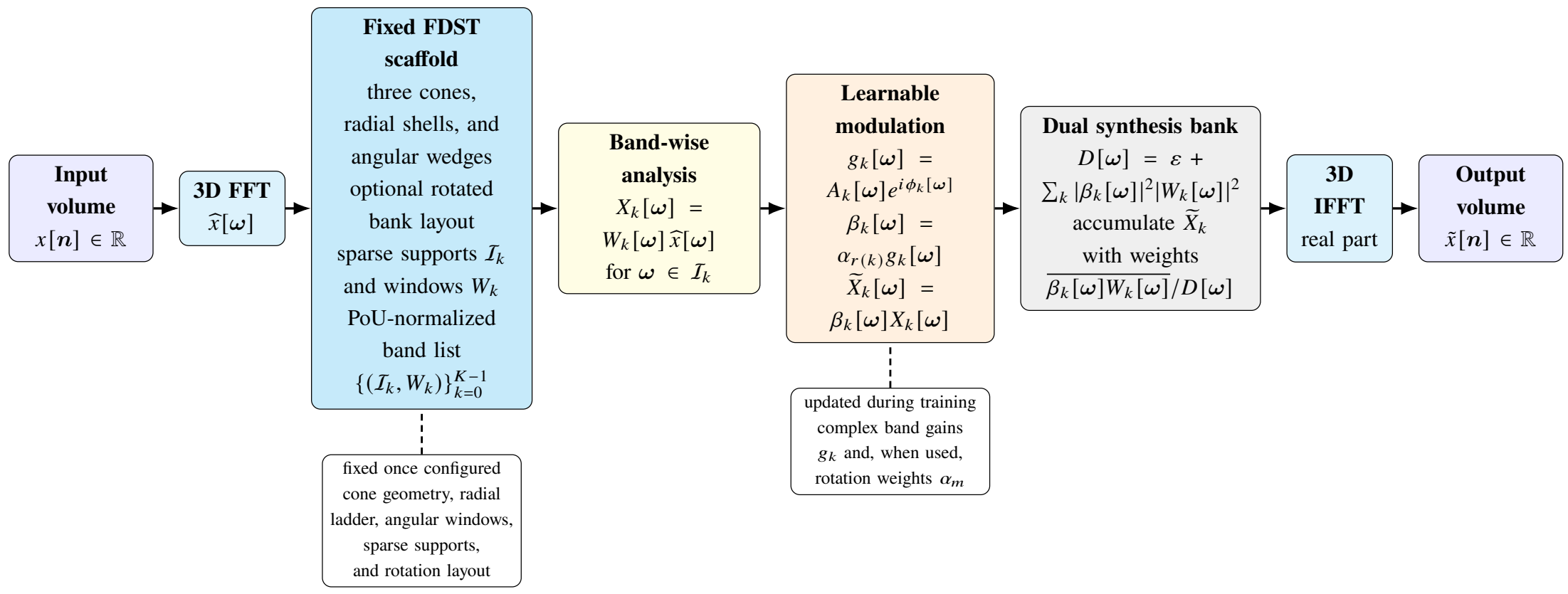

Figure 2.12: Overview of the implemented learnable 3D FDST. The input volume is mapped to the Fourier domain, analyzed by a fixed cone-scale-orientation shearlet bank, modulated by learnable complex band gains and optional rotation weights, and reconstructed by the dual synthesis rule before inverse FFT returns the spatial output. The figure separates fixed scaffold components from learned modulation.

**Learning and regularization** A task loss $\mathcal{L}(\tilde{x}; y^{\star})$, with target $y^{\star}$, back-propagates through the FDST pipeline. Gradients flow through the FFT and IFFT, the fixed windows $W_k$, the softplus and phase parametrization of $g_k$, the rotation softmax, and the dual normalizer $D[\omega]$. Standard optimizers (e.g. Adam) update the gain parameters $\{p_k[\omega], q_k[\omega]\}$ and, when applicable, the rotation logits $\{z_m^{\text{fdst}}\}$. We also monitor PoU error and coverage of the fixed bank. If desired, additional training-time regularizers, such as an entropy term on $(\alpha_m)$, can be added externally at the model or training-script level. Figure 2.12 summarizes the implemented FDST analysis and synthesis path.

**Backpropagation through the learnable FDST bank:** Let $\mathcal{L}$ be a task loss that depends on the reconstructed field $\tilde{x}$. In the implementation, `FDST3D_Tiles` first forms the trainable analyzed tile

$$T_k[\omega] = \beta_k[\omega]\, W_k[\omega]\, \widehat{x}[\omega], \qquad \omega \in \mathcal{I}_k, \tag{2.66}$$

where $\beta_k[\omega] = \alpha_{r(k)} g_k[\omega]$ in union-rotation mode and $\beta_k[\omega] = g_k[\omega]$ in the single-rotation case. The `IDST3D_Tiles` dual synthesis then scatters the tile contributions with synthesis multiplier

$$B_k[\omega] = \frac{\overline{\beta_k[\omega]\, W_k[\omega]}}{D[\omega]}, \qquad D[\omega] = \varepsilon + \sum_{\ell=0}^{K-1} |\beta_\ell[\omega]|^2 |W_\ell[\omega]|^2. \tag{2.67}$$

Thus, if $G[\omega] = \partial\mathcal{L}/\partial\widehat{x}[\omega]$ denotes the upstream spectral gradient, the adjoint of the linear scatter-add synthesis step gives

$$\left.\frac{\partial \mathcal{L}}{\partial T_k[\omega]}\right|_{\text{synth}} = G[\omega]\, \overline{B_k[\omega]} = G[\omega] \frac{\beta_k[\omega]\, W_k[\omega]}{D[\omega]}, \qquad \omega \in \mathcal{I}_k. \tag{2.68}$$

This is the correct adjoint factor for the tile gradient. The trainable gains also appear in the synthesis multiplier $B_k$ and in the denominator $D[\omega]$, so the complete gradient with respect to $g_k$ and the rotation weights includes the tile path, the synthesis-gain path, and the dual-normalizer path. TensorFlow computes this full derivative because $B_k$ and $D[\omega]$ are built from the current gains inside `IDST3D_Tiles`. Locally, the tile path contributes

$$\left.\frac{\partial \mathcal{L}}{\partial \beta_k[\omega]}\right|_{\text{tile}} = \frac{\partial \mathcal{L}}{\partial T_k[\omega]} \overline{W_k[\omega]\,\widehat{x}[\omega]}, \tag{2.69}$$

with additional terms from $B_k$ and $D[\omega]$. Finally, the real trainable variables are updated through

$$A_k[\omega] = \operatorname{softplus}(p_k[\omega]), \qquad \phi_k[\omega] = q_k[\omega], \tag{2.70}$$

$$g_k[\omega] = A_k[\omega] e^{\mathrm{i}\phi_k[\omega]}, \qquad \alpha_m = \operatorname{softmax}(z^{\text{fdst}})_m, \tag{2.71}$$

using the softplus derivative for $p_k$, the phase derivative $\partial g_k/\partial q_k = \mathrm{i} g_k$, and the standard softmax Jacobian for the rotation logits $z_m^{\text{fdst}}$.

## 2.4 Discussion

This chapter described differentiable realizations of multidimensional signal representations based on orthogonal bases, frames, Fourier-plane tilings, Ramanujan frames, wavelet filter banks, and the 3D fast digital shearlet transform. The purpose is not to replace CNN feature learning, but to provide structured coefficient spaces that can be inserted into neural computation graphs. Decimated CNNs can learn rich multiscale representations, although their feature spaces are usually formed implicitly from many free convolutional parameters. The operator layers in this chapter instead expose frequency, scale, periodicity, and directional variables that can be inspected through coefficient maps, reconstruction checks, partition-of-unity checks, and perfect-reconstruction tests. This makes the representations more auditable at the transform-module level while remaining compatible with backpropagation.

The chapter packages these constructions in Python TensorFlow libraries, including `TFDWT` for 1D, 2D, and 3D DWT and IDWT filter banks, `RamanujanFrame` for 1D and 2D periodicity transforms, `filterbanks` for 2D directional transform tilings, and `FDST3D_Tiles/IDST3D_Tiles` for learnable 3D digital shearlet analysis and dual synthesis. Related units include `convd` for D-dimensional convolutions with support up to 9D convolutions, `LNPoly` for parametric polynomial and rational nonlinearities, and `PhasePool` for adaptive polyphase sampling. These computational units provide the operator-level infrastructure used in later chapters for Volterra kernels, frugal attentions, and multiresolution vision models with structured feature spaces.

# Chapter 3

# Volterra Systems in Biorthogonal Bases and Microwave Radiometric Inversion

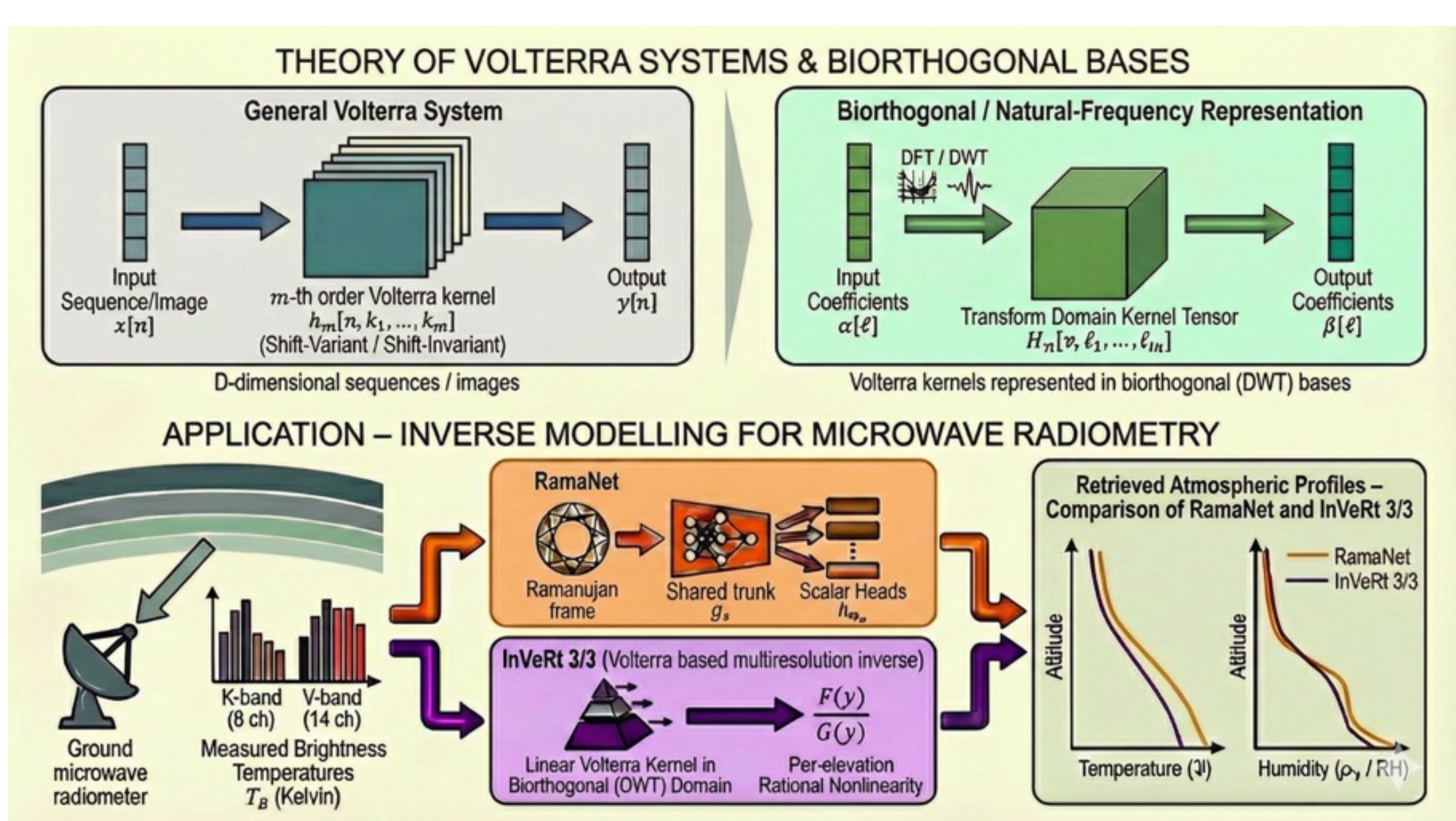

Volterra series provide a structured framework for modeling nonlinear dynamical systems and processes with memory. Special cases of nonlinear shift-invariant systems can be modeled by a family of shift-invariant Volterra kernels of increasing order, where the first-order linear shift-invariant (LSI) component is characterized by its impulse response. The first-order kernel captures linear effects, while higher-order kernels represent interactions among multiple copies of the input signal [147–149]. The input-output relation defined by the linear and higher-order Volterra terms in a Volterra series further generalizes the notion of convolution [150]. This allows richer representations with memory and captures nonlinear coupling between input and output. However, realizing higher-order Volterra kernels also introduces serious challenges because the number of kernel coefficients grows rapidly with the kernel's order and input dimensionality. Thus, the computational cost, memory demand, and risk of overfitting increase unless additional structure or multiscale organization is imposed. In modern machine learning and deep learning practice, many architectures implement linear filtering followed by pointwise nonlinearities or multiplicative interactions. Polynomial networks, quadratic convolutional layers and Volterra inspired neural networks can be viewed as different approximations of Volterra expansions and have been applied to biomedical signal modeling, texture analysis, multimodal fusion, motion estimation, time series forecasting and three dimensional medical imaging [52, 53, 76, 147, 148, 151, 152]. This chapter develops a complementary viewpoint in which the input, output, and Volterra kernels are expanded in compatible biorthogonal wavelet coefficient domains. In the implementation, DWT layers map the input and shifted kernel tensors to these coefficient domains, the Volterra contraction is evaluated there, and IDWT layers reconstruct the output, which makes the layer compatible with standard differentiable convolutional network graphs. The chapter also introduces a Python implementation of these layers and applies them to microwave radiometric inversion, where nonlinear relationships between brightness temperatures and geophysical parameters are naturally modeled by higher order kernels [153–156], and representative works that motivate this study are summarized in Table 3.1.

## 3.1 Theory of Volterra systems in biorthogonal bases

The Volterra series provides a tractable model for nonlinear dynamic systems and generalizes multidimensional convolutions with its Volterra kernels (linear and nonlinear). Here, the objective is to realize such Volterra kernels for representation learning with convolutional architectures [147–149]. Under suitable regularity assumptions, a nonlinear system can be approximated by a finite Volterra series of maximum degree $M$ with Volterra kernels [150] and $M \in \mathbb{N}$. A discrete Volterra series with $M$ terms can be written as

$$y[\boldsymbol{n}] = y_0[\boldsymbol{n}] + \sum_{m=1}^{M} y_m[\boldsymbol{n}], \tag{3.1}$$

where $y_m[\boldsymbol{n}] = \sum_{\boldsymbol{k}_1} \cdots \sum_{\boldsymbol{k}_m} h_m[\boldsymbol{n}, \boldsymbol{k}_1, \cdots, \boldsymbol{k}_m] x[\boldsymbol{k}_1] \cdots x[\boldsymbol{k}_m]$ is the $m^{th}$ degree monomial, $x, y_m \in \ell^2(\mathbb{Z}^D)$, $D \in \mathbb{N}$ and $h_m[\boldsymbol{n}, \boldsymbol{k}_1, \cdots, \boldsymbol{k}_m]$ is a $(m+1)^D$ dimensional **shift variant** Volterra kernel. Figure 3.1 illustrates the complete Volterra series with $M$ Volterra kernels $\{h_m\}_{m=1}^{M}$, where the kernels beyond $m > 1$ are nonlinear systems. The Volterra series is a foundational model for nonlinear system identification and the theory of nonlinear shift invariant (NLSI) systems that use higher-order Volterra kernels to model nonlinearity [150]. The NLSI systems are widely adopted for modeling biological systems using $m$-dimensional shift invariant kernels $h_m[\boldsymbol{n} - \boldsymbol{k}_1, \cdots, \boldsymbol{n} - \boldsymbol{k}_m]$. Volterra kernels provide a structured way to model nonlinear interactions in shallow or deep learning systems, even without activation functions between linear operators. Recent advancements in the application of Volterra systems in deep learning include higher-order Volterra filters in CNNs [52, 53], orthonormal basis decomposition of Volterra kernels in CNN [151], explainable CNNs with second-order Volterra systems [152] and Volterra neural networks with nested Volterra filters [76]. One **hard limitation** in using Volterra kernels and large Volterra systems is the complexity explosion with increasing $D$ and/or $m$. For example, for a sequence input (i.e., $D = 1$), the quadratic term (i.e., $m = 2$) requires a cubic Volterra kernel. Similarly, the quadratic term for image input (i.e., $D = 2$) will require a six-dimensional Volterra kernel and three-dimensional input (i.e., $D = 3$) will require a nine-dimensional Volterra kernel. The quadratic shift invariant Volterra kernel is two, four and six-dimensional for sequence, image and three-dimensional inputs respectively. The shift invariant form of the Volterra kernel is a complexity saviour to an extent. For smaller systems with low $D$ and $m$, for example, $D = 1$, i.e., for sequence inputs, the powerful GPU hardware lets us explore the Volterra kernels for nonlinear modeling up to a sufficiently large $m$.

Table 3.1: Applications of Volterra Kernel Approximations Across Domains

| # | Paper | Application | Order | Method | Kernel Approx. Type | Dataset(s) | Domain & Remarks |
|---|---|---|---|---|---|---|---|
| 1 | Hakim et al. (1991) [147] | Neural Network Characterization | 1st / 2nd | Volterra characterization of neural networks | True | N/A | ML theory; links NN mappings to Volterra kernels |
| 2 | Wray & Green (1994) [148] | Nonlinear System Identification | 2nd | ANN-based Volterra kernel estimation | True | Simulated nonlinear dynamical systems | 1D system ID; kernel computation from data |
| 3 | Marmarelis & Zhao (1997) [149] | Volterra Models vs Perceptrons | 1st / 2nd | Three-layer perceptron as a Volterra model | True | N/A | ML theory; relates MLP structure to Volterra expansions |
| 4 | Zoumpourlis et al. (2017) [157] | Image Classification | 2nd | Second-order (Volterra) convolution filters | Quadratic / Volterra | CIFAR-10, CIFAR-100 | 2D Vision; quadratic feature interactions inside conv layers |
| 5 | Machado & Givigi (2019) [151] | CNNs as Volterra Models | 2nd | GOBF-based asymmetric Volterra representation | Basis expansion | N/A | ML theory; interprets CNNs via generalized orthonormal bases |
| 6 | Roheda & Krim (2020) [52] | Action Recognition | 2nd | Volterra filtering approach for CNNs | Quadratic / Volterra | UCF-101, HMDB-51, Sports-1M | 2D/Video; addresses overparameterization with structured interactions |
| 7 | Banerjee et al. (2020) [53] | Learning on Homogeneous Manifolds | 2nd | VolterraNet (group-equivariant higher-order conv) | Quadratic / Volterra | spherical-MNIST, Shrec17, atomic energy, diffusion MRI | Non-Euclidean ML; efficient higher-order equivariant convolutions |
| 8 | Contreras & Bocklitz (2023) [152] | Explainable Nonlinear Modeling | 2nd | Volterra-series-based XAI method | True | N/A | Biomedical optics / XAI; interpretable nonlinear surrogate models |
| 9 | Roheda et al. (2024) [76] | Volterra Neural Networks (VNNs) | 2nd | Parametric VNN architectures | Parametric Volterra | N/A | ML methods; general framework for Volterra-inspired networks |
| 10 | Tarafdar & Gadre (2025) [158] | Multiresolution Volterra Systems | 2nd | Multiresolution (wavelet-domain) Volterra systems | Multiresolution basis | N/A | Frugal AI; multiscale kernel representations for learning |

Next, we represent Volterra input-output relations in natural, DFT, and wavelet coefficient domains, with particular emphasis on orthogonal and biorthogonal DWT bases. We first describe the $m^{th}$ degree Volterra kernel framework, where $m = 1$ gives an LSI filter with unit impulse response $h[n]$ for $n \in \mathbb{Z}$. The same representation is then used for microwave radiometric inversion with orthogonal and biorthogonal wavelet bases.

### 3.1.1 Biorthogonal basis representation of Volterra kernel and its I/O

In this section, we express the $m^{th}$ monomial of the Volterra series in (3.1) as an equivalent orthonormal basis relationship between input and output. The I/O relationship of a $m^{th}$ degree shift variant Volterra kernel $h_m[\boldsymbol{n}, \boldsymbol{k}_1, \cdots, \boldsymbol{k}_m]$ in the discrete natural domain is

$$y_m[\boldsymbol{n}] = \sum_{\boldsymbol{k}_1} \cdots \sum_{\boldsymbol{k}_m} h_m[\boldsymbol{n}, \boldsymbol{k}_1, \cdots, \boldsymbol{k}_m] x[\boldsymbol{k}_1] \cdots x[\boldsymbol{k}_m], \tag{3.2}$$

where $x[\boldsymbol{n}]$ is the input and $y_m[\boldsymbol{n}]$ is the output. Note, we have assumed multidimensional inputs $x$ and multidimensional outputs $y_m$, where $\boldsymbol{n} \in \mathbb{Z}^D$, $D \in \mathbb{N}$ and bold summation with bold index is $\sum_{\boldsymbol{k}_1} \equiv \sum_{k_{1_1}} \cdots \sum_{k_{1_D}}$ that operate on a $D$-dimensional tensor. Now, expand and reconstruct $x$ with an orthogonal or a biorthogonal basis as

$$x[\boldsymbol{n}] = \sum_{\boldsymbol{l}} \alpha[\boldsymbol{l}] \tilde{\phi}_{\boldsymbol{l}}[\boldsymbol{n}], \tag{3.3}$$

$$\alpha[\boldsymbol{l}] = \sum_{\boldsymbol{n}} x[\boldsymbol{n}] \phi_{\boldsymbol{l}}[\boldsymbol{n}], \tag{3.4}$$

where $\phi_{\boldsymbol{l}}$ and $\tilde{\phi}_{\boldsymbol{l}}$ are the orthogonal ($\phi = \tilde{\phi}$) or in general biorthogonal ($\phi \neq \tilde{\phi}$) bases. Similarly, the expansion and reconstruction of $y_m$ using biorthogonal bases $\phi_{\boldsymbol{v}}$ and $\tilde{\phi}_{\boldsymbol{v}}$ are

$$y_m[\boldsymbol{n}] = \sum_{\boldsymbol{v}} \beta_m[\boldsymbol{v}] \tilde{\phi}_{\boldsymbol{v}}[\boldsymbol{n}], \tag{3.5}$$

$$\beta_m[\boldsymbol{v}] = \sum_{\boldsymbol{n}} y_m[\boldsymbol{n}] \phi_{\boldsymbol{v}}[\boldsymbol{n}]. \tag{3.6}$$

Substituting (3.3) into (3.2) gives

$$\begin{aligned} y_m[\boldsymbol{n}] &= \sum_{\boldsymbol{k}_1} \cdots \sum_{\boldsymbol{k}_m} h_m[\boldsymbol{n}, \boldsymbol{k}_1, \cdots, \boldsymbol{k}_m] \left( \sum_{\boldsymbol{l}} \alpha[\boldsymbol{l}] \tilde{\phi}_{\boldsymbol{l}}[\boldsymbol{k}_1] \right) \\ &\qquad \cdots \left( \sum_{\boldsymbol{l}} \alpha[\boldsymbol{l}] \tilde{\phi}_{\boldsymbol{l}}[\boldsymbol{k}_m] \right) \\ &= \sum_{\boldsymbol{k}_1} \cdots \sum_{\boldsymbol{k}_m} h_m[\boldsymbol{n}, \boldsymbol{k}_1, \cdots, \boldsymbol{k}_m] \sum_{\boldsymbol{l}_1} \cdots \sum_{\boldsymbol{l}_m} \alpha[\boldsymbol{l}_1] \\ &\qquad \cdots \alpha[\boldsymbol{l}_m] \tilde{\phi}_{\boldsymbol{l}_1}[\boldsymbol{k}_1] \cdots \tilde{\phi}_{\boldsymbol{l}_m}[\boldsymbol{k}_m] \end{aligned}$$

or,

$$y_m[\boldsymbol{n}] = \sum_{\boldsymbol{l}_1} \cdots \sum_{\boldsymbol{l}_m} \alpha[\boldsymbol{l}_1] \cdots \alpha[\boldsymbol{l}_m] \Bigg\{ \sum_{\boldsymbol{k}_1} \cdots \sum_{\boldsymbol{k}_m} h_m[\boldsymbol{n}, \boldsymbol{k}_1, \cdots, \boldsymbol{k}_m] \tag{3.7}$$

$$\tilde{\phi}_{\boldsymbol{l}_1}[\boldsymbol{k}_1]\cdots\tilde{\phi}_{\boldsymbol{l}_m}[\boldsymbol{k}_m]\Big\}$$

Hence, substituting (3.7) into (3.6) yields the biorthogonal expansion of $y_m$:

$$\begin{aligned}
\beta_m[\boldsymbol{v}] &= \sum_{\boldsymbol{n}} y_m[\boldsymbol{n}]\phi_{\boldsymbol{v}}[\boldsymbol{n}] \\
&= \sum_{\boldsymbol{n}}\Big\{\sum_{\boldsymbol{l}_1}\cdots\sum_{\boldsymbol{l}_m}\alpha[\boldsymbol{l}_1]\cdots\alpha[\boldsymbol{l}_m]\Big(\sum_{\boldsymbol{k}_1}\cdots\sum_{\boldsymbol{k}_m} h_m[\boldsymbol{n},\boldsymbol{k}_1,\cdots,\boldsymbol{k}_m] \\
&\qquad \tilde{\phi}_{\boldsymbol{l}_1}[\boldsymbol{k}_1]\cdots\tilde{\phi}_{\boldsymbol{l}_m}[\boldsymbol{k}_m]\Big)\Big\}\phi_{\boldsymbol{v}}[\boldsymbol{n}] \\
&= \sum_{\boldsymbol{l}_1}\cdots\sum_{\boldsymbol{l}_m}\alpha[\boldsymbol{l}_1]\cdots\alpha[\boldsymbol{l}_m]\ \sum_{\boldsymbol{n}}\sum_{\boldsymbol{k}_1}\cdots\sum_{\boldsymbol{k}_m} h_m[\boldsymbol{n},\boldsymbol{k}_1,\cdots,\boldsymbol{k}_m] \\
&\qquad \tilde{\phi}_{\boldsymbol{l}_1}[\boldsymbol{k}_1]\cdots\tilde{\phi}_{\boldsymbol{l}_m}[\boldsymbol{k}_m]\phi_{\boldsymbol{v}}[\boldsymbol{n}]
\end{aligned}$$

or,

$$\beta_m[\boldsymbol{v}] = \sum_{\boldsymbol{l}_1}\cdots\sum_{\boldsymbol{l}_m}\alpha[\boldsymbol{l}_1]\cdots\alpha[\boldsymbol{l}_m]\ H_m[\boldsymbol{v},\boldsymbol{l}_1,\cdots,\boldsymbol{l}_m], \qquad \boldsymbol{v}\in\mathbb{Z}^D \tag{3.8}$$

where the shift variant Volterra kernel in the transform domain is

$$\begin{aligned}
H_m[\boldsymbol{v},\boldsymbol{l}_1,\cdots,\boldsymbol{l}_m] = \sum_{\boldsymbol{n}}\sum_{\boldsymbol{k}_1}\cdots\sum_{\boldsymbol{k}_m} h_m[\boldsymbol{n},\boldsymbol{k}_1,\cdots,\boldsymbol{k}_m]\tilde{\phi}_{\boldsymbol{l}_1}[\boldsymbol{k}_1] \\
\cdots\tilde{\phi}_{\boldsymbol{l}_m}[\boldsymbol{k}_m]\phi_{\boldsymbol{v}}[\boldsymbol{n}].
\end{aligned} \tag{3.9}$$

Therefore, an equivalent biorthogonal expansion of the $m^{th}$ degree shift variant Volterra kernel that relates to the biorthogonal expansion of the system I/O is stated in Theorem 3.1.

**Theorem 3.1.** *The biorthogonal expansions of the input $x[\boldsymbol{n}]$, output $y_m[\boldsymbol{n}]$ and the $(m+1)\cdot D$ dimensional Volterra kernel $h_m[\boldsymbol{n},\boldsymbol{k}_1,\cdots,\boldsymbol{k}_m]$ using a biorthogonal system $\langle\phi,\tilde{\phi}\rangle = 1$ is related as,*

$$y_m[\boldsymbol{n}] = \sum_{\boldsymbol{v}}\sum_{\boldsymbol{l}_1}\cdots\sum_{\boldsymbol{l}_m}\alpha[\boldsymbol{l}_1]\cdots\alpha[\boldsymbol{l}_m]\ H_m[\boldsymbol{v},\boldsymbol{l}_1,\cdots,\boldsymbol{l}_m]\tilde{\phi}_{\boldsymbol{v}}[\boldsymbol{n}] \tag{3.10}$$

*where $\alpha[\boldsymbol{l}] = \sum_{\boldsymbol{n}} x[\boldsymbol{n}]\phi_{\boldsymbol{l}}[\boldsymbol{n}]$ is an expansion of the input and $H_m[\boldsymbol{v},\boldsymbol{l}_1,\cdots,\boldsymbol{l}_m]$ is an expansion of the Volterra kernel using biorthogonal bases for $\boldsymbol{n},\boldsymbol{l}_m,\boldsymbol{v}\in\mathbb{Z}^D$ and $m, D\in\mathbb{N}$.*

In general, equation (3.10) may have different independent variables in $x$ and $y_m$. This setting is used later for a linear shift-variant multiresolution Volterra inverse kernel represented in DWT bases for microwave radiometric inversion.

### 3.1.2 Shift invariant Wiener systems with Volterra kernels

The shift invariant form of the kernel in equation (3.10) is $h_m[\boldsymbol{n}-\boldsymbol{k}_1,\cdots,\boldsymbol{n}-\boldsymbol{k}_m]$, where $\boldsymbol{n}-\boldsymbol{k}_m$ are circular shifts. A shift invariant Wiener system is a dynamic LSI system followed by a static nonlinearity and can be represented as a Volterra series with shift invariant Volterra kernels [150] as shown below,

$$x[\boldsymbol{n}] \rightarrow \underset{\text{Dynamic LSI}}{h[\boldsymbol{n}]} \xrightarrow{q} \underset{\text{Static nonlinearity}}{(a_0 + a_1 q + a_2 q^2 + \cdots + a_m q^m)} \rightarrow y[\boldsymbol{n}]$$

or,

$$y[\boldsymbol{n}] = a_0 + a_1 q[\boldsymbol{n}] + a_2 q[\boldsymbol{n}]^2 + \cdots + a_m q[\boldsymbol{n}]^m \tag{3.11}$$

where $q[\boldsymbol{n}] = \sum_{\boldsymbol{k}} h[\boldsymbol{k}]x[\boldsymbol{n}-\boldsymbol{k}]$ and $h$ is the unit impulse response. A linear system follows the principle of superposition, and an LSI system can be characterized fully by its unit impulse response. The summation in perceptron and the convolution in CNN are examples of LSI systems with memory. A summation followed by a nonlinear activation in perceptron or a convolution in cascade with a nonlinear activation function are popular examples of Wiener systems with an LSI part and a static nonlinear activation function. Here, we view these Wiener systems in a general framework that can model nonlinearities with higher-order Volterra kernels and backpropagation learning.

#### 3.1.2.1 I/O in natural domain

The I/O relationship of a quadratic Volterra system in natural domain with a **linear** and **quadratic** Volterra kernel is

$$\begin{aligned}
y[\boldsymbol{n}] &= a_0 + a_1 \sum_{\boldsymbol{k}} h[\boldsymbol{k}]x[\boldsymbol{n}-\boldsymbol{k}] + a_2 \sum_{\boldsymbol{k}} h[\boldsymbol{k}]x[\boldsymbol{n}-\boldsymbol{k}] \sum_{\boldsymbol{k}} h[\boldsymbol{k}]x[\boldsymbol{n}-\boldsymbol{k}] \\
&= a_0 + \sum_{\boldsymbol{k}} a_1 h[\boldsymbol{k}]x[\boldsymbol{n}-\boldsymbol{k}] + \sum_{\boldsymbol{k}_1}\sum_{\boldsymbol{k}_2} a_2 h[\boldsymbol{k}_1]h[\boldsymbol{k}_2]x[\boldsymbol{n}-\boldsymbol{k}_1]x[\boldsymbol{n}-\boldsymbol{k}_2] \\
&= h_0 + \sum_{k} h_1[\boldsymbol{k}]x[\boldsymbol{n}-\boldsymbol{k}] + \langle a_2 \begin{bmatrix} \mathbf{h}_0\mathbf{h}_0 & \cdots & \mathbf{h}_0\mathbf{h}_n \\ \vdots & \ddots & \vdots \\ \mathbf{h}_n\mathbf{h}_0 & \cdots & \mathbf{h}_n\mathbf{h}_n \end{bmatrix}, \begin{bmatrix} \mathbf{x}_n\mathbf{x}_n & \cdots & \mathbf{x}_n\mathbf{x}_0 \\ \vdots & \ddots & \vdots \\ \mathbf{x}_0\mathbf{x}_n & \cdots & \mathbf{x}_0\mathbf{x}_0 \end{bmatrix} \rangle
\end{aligned}$$

or,

$$y[\boldsymbol{n}] = h_0 + \sum_{k} h_1[\boldsymbol{k}]x[\boldsymbol{n}-\boldsymbol{k}] + \sum_{\boldsymbol{k}_1=0}^{\boldsymbol{n}}\sum_{\boldsymbol{k}_2=0}^{\boldsymbol{n}} h_2[\boldsymbol{k}_1,\boldsymbol{k}_2]x_2[\boldsymbol{n}-\boldsymbol{k}_1,\boldsymbol{n}-\boldsymbol{k}_2]. \tag{3.12}$$

In natural domain, the computation of the $m^{th}$ degree monomial of a Volterra series with a $mD$ dimensional Volterra kernel for a $D$-dimensional input tensor is a convolution-like operation, where the kernel strides along the trace (main diagonal) $m$-times outer product of the input tensor as stated in Lemma 3.1.

**Lemma 3.1.** *For a finite input tensor $x \in \ell^2(\mathbb{Z}^D)$ of size $N^D$, the output of a $m^{th}$ degree shift invariant Volterra kernel is a convolution-like operation where a $mD$-dimensional filter $h_m \in \ell^2(\mathbb{Z}^{mD})$ strides along the trace of the outer product $\underbrace{x \otimes x \otimes \cdots \otimes x}_{m\,times}$, i.e.,*

$$y_m[\boldsymbol{n}] = \sum_{\boldsymbol{k}_1} \cdots \sum_{\boldsymbol{k}_m} h_m[\boldsymbol{k}_1, \cdots, \boldsymbol{k}_m] x[\boldsymbol{n} - \boldsymbol{k}_1] \cdots x[\boldsymbol{n} - \boldsymbol{k}_m] \tag{3.13}$$

$$\text{for } \boldsymbol{n} \ni n_i = 0, 1, 2, \cdots, N-1$$

*where $h_m[\boldsymbol{k}_1, \cdots, \boldsymbol{k}_m] = a_m h[\boldsymbol{k}_1] \cdots h[\boldsymbol{k}_m]$ and $x[\boldsymbol{n} - \boldsymbol{k}_1] \cdots x[\boldsymbol{n} - \boldsymbol{k}_m] \in \underbrace{x \otimes x \otimes \cdots \otimes x}_{m\,times}$ with circular shifts for $N \in \mathbb{Z} > 0$, $m, D \in \mathbb{N}$.*

#### 3.1.2.2 I/O with DFT basis $e^{-j2\pi \frac{\boldsymbol{v}\cdot\boldsymbol{n}}{N}}$

The DFT of equation (3.13) can be derived as shown below,

$$\beta_m[\boldsymbol{v}] = \mathcal{F}\left\{\sum_{\boldsymbol{k}_1} \cdots \sum_{\boldsymbol{k}_m} h_m[\boldsymbol{k}_1, \cdots, \boldsymbol{k}_m] x[\boldsymbol{n} - \boldsymbol{k}_1] \cdots x[\boldsymbol{n} - \boldsymbol{k}_m]\right\}$$

$$= \sum_{\boldsymbol{n}} \left\{\sum_{\boldsymbol{k}_1} \cdots \sum_{\boldsymbol{k}_m} h_m[\boldsymbol{k}_1, \cdots, \boldsymbol{k}_m] x[\boldsymbol{n} - \boldsymbol{k}_1] \cdots x[\boldsymbol{n} - \boldsymbol{k}_m]\right\} e^{-j2\pi \frac{\boldsymbol{v}\cdot\boldsymbol{n}}{N}}$$

$$= \sum_{\boldsymbol{k}_1} \cdots \sum_{\boldsymbol{k}_m} h_m[\boldsymbol{k}_1, \cdots, \boldsymbol{k}_m] \left\{\sum_{\boldsymbol{n}} x[\boldsymbol{n} - \boldsymbol{k}_1] \cdots x[\boldsymbol{n} - \boldsymbol{k}_m] e^{-j2\pi \frac{\boldsymbol{v}\cdot\boldsymbol{n}}{N}}\right\}$$

$$= \sum_{\boldsymbol{k}_1} \cdots \sum_{\boldsymbol{k}_m} h_m[\boldsymbol{k}_1, \cdots, \boldsymbol{k}_m] \mathcal{F}\left\{x[\boldsymbol{n} - \boldsymbol{k}_1] \cdots x[\boldsymbol{n} - \boldsymbol{k}_m]\right\}$$

(By DFT definition)

$$= \frac{1}{N^{(m-1)D}} \sum_{\boldsymbol{k}_1} \cdots \sum_{\boldsymbol{k}_m} h_m[\boldsymbol{k}_1, \cdots, \boldsymbol{k}_m] \mathcal{F}\left\{x[\boldsymbol{n} - \boldsymbol{k}_1]\right\} \circledast$$

$$\cdots \circledast \mathcal{F}\Big\{x\big[\boldsymbol{n}-\boldsymbol{k}_m\big]\Big\} \quad \text{(By multiplication property)}$$

$$= \frac{1}{N^{(m-1)D}} \sum_{\boldsymbol{k}_1} \cdots \sum_{\boldsymbol{k}_m} h_m\big[\boldsymbol{k}_1, \cdots, \boldsymbol{k}_m\big] \Big\{X\big[\boldsymbol{v}\big] e^{-j2\pi \frac{\boldsymbol{v}.\boldsymbol{k}_1}{N}} \circledast$$

$$\cdots \circledast X\big[\boldsymbol{v}\big] e^{-j2\pi \frac{\boldsymbol{v}.\boldsymbol{k}_m}{N}}\Big\} \quad \text{(By time shift property)}$$

or,

$$\beta_m\big[\boldsymbol{v}\big] = \frac{1}{N^{(m-1)D}} \sum_{\boldsymbol{l}_1} \cdots \sum_{\boldsymbol{l}_{m-1}} X\big[\boldsymbol{l}_1\big] \cdots X\big[\boldsymbol{l}_{m-1}\big] X\big[\boldsymbol{v}-\boldsymbol{l}_1-\cdots-\boldsymbol{l}_{m-1}\big]$$

$$H_m\big[\boldsymbol{l}, \boldsymbol{v}-\boldsymbol{l}_1-\cdots-\boldsymbol{l}_{m-1}\big] \qquad \text{for } \boldsymbol{v} \ni v_i = 0, 1, 2, \cdots, N-1\ \forall i \tag{3.14}$$

where $\circledast$ denotes circular convolution, $X$ and $H_m$ are the DFT of the input and the filter. Theorem 3.2 gives an equivalent relationship of equation (3.13) using DFT basis.

**Theorem 3.2.** *For a finite input tensor $x \in \ell^2(\mathbb{Z}^D)$ of size $N^D$, where $N, m, D \in \mathbb{N}$, the $m^{th}$ degree monomial $y_m$ at the output of an nonlinear shift invariant $m^{th}$ degree Volterra kernel is equal to the $N^D$-point Inverse Discrete Fourier Transform of*

$$\beta_m\big[\boldsymbol{v}\big] = \frac{1}{N^{(m-1)D}} \sum_{\boldsymbol{l}_1} \cdots \sum_{\boldsymbol{l}_{m-1}} X\big[\boldsymbol{l}_1\big] \cdots X\big[\boldsymbol{l}_{m-1}\big] \tag{3.15}$$

$$X\big[\boldsymbol{v}-\boldsymbol{l}_1-\cdots-\boldsymbol{l}_{m-1}\big] H_m\big[\boldsymbol{l}, \boldsymbol{v}-\boldsymbol{l}_1-\cdots-\boldsymbol{l}_{m-1}\big]$$

*i.e.,*

$$y_m\big[\boldsymbol{n}\big] = \mathcal{F}^{-1}\{\beta_m\big[\boldsymbol{v}\big]\} \tag{3.16}$$

*where $X$ is the $N^D$-point DFT of the input tensor, $H_m$ is the $N^{mD}$-point DFT of the $mD$-dimensional filter.*

The following example compares the above natural domain and DFT domain relationships for sequence I/O systems.

**Example 3.1.** The realization of $m^{th}$ degree Volterra series with shift invariant kernels $\{h_m\}_{m\in\mathbb{N}}$ for a sequence ($D = 1$) input $x[n]$, $n \in \mathbb{Z}$ is shown in Figure 3.1. The term $h_0[n]$ is a constant term like a bias, $h_1[n]$ is an LSI kernel and the kernels beyond $m > 1$ are NLSI.

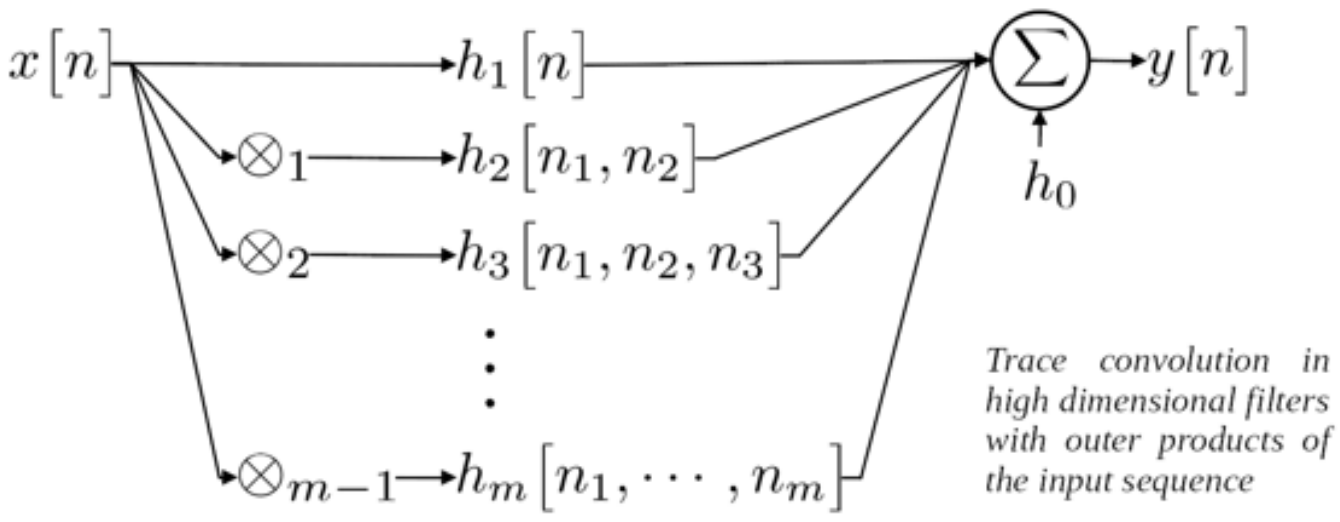

Figure 3.1: Realization of a $m^{th}$ degree Volterra series with LSI and NLSI Volterra kernels connected in parallel

The number of computations in equation (3.13) for a sequence input is proportional to $O(N^m)$ for each $n$ and the total computations $\forall n \in [0, N-1]$ is $O(N^{m+1}) = O(N^m \times N)$. Similarly, the Inverse Discrete Fourier Transform (IDFT) of equation (3.14) for sequence input involves the following computations, (i) FFT of input sequence $O(NlogN)$, (ii) FFT of the m-dimensional kernel $O(N^m logN)$, (iii) Pointwise multiplication $O(N^2)$ and (iv) Inverse Fast Fourier Transform (IFFT) $O(NlogN)$. Therefore, the time complexity of the IFFT of equation (3.14) for sequences is,

$$O(N^m logN) := O(NlogN) + O(N^m logN) + O(N^2) + O(NlogN)$$

In summary, for sequences the time complexity of equation (3.13) is $O(N^{m+1})$ and the time complexity of the IDFT of equation (3.14) is $O(N^m logN)$. However, the space complexity in the Fourier domain is higher than natural domain since equation (3.13) operates on complex numbers (**complex64** of 64 bits) and equation (3.14) operates on real numbers (**float32** of 32 bits). Hence, storing tensors $X$ and $H_2$ will require double memory than required for storing $x$ and $h_2$. Thus selecting one equation over another for realizing Volterra kernels is based on a **trade-off** between the time and space complexity. The natural has advantage in space but disadvantage in time and Fourier domain has advantage in time but disadvantage space. Often, due to hardware limitations natural domain operations are common in use, for example in CNNs with FIR filters.

#### 3.1.2.3 I/O with biorthogonal bases

Having shift invariant Volterra kernel $h_m[\boldsymbol{n}-\boldsymbol{k}_1, \cdots, \boldsymbol{n}-\boldsymbol{k}_m]$ in (3.10), its equivalent biorthogonal representation is stated in Lemma 3.2.

**Lemma 3.2.** *The biorthogonal bases representation of a shift invariant $m^{th}$ degree, $mD$-dimensional Volterra filter $h_m[\boldsymbol{n}-\boldsymbol{k}_1, \cdots, \boldsymbol{n}-\boldsymbol{k}_m]$ with circular shifts and $\boldsymbol{n}, \boldsymbol{k}_1, \ldots, \boldsymbol{k}_m \in \mathbb{Z}^D$ using a global biorthogonal system $\langle \phi, \tilde{\phi} \rangle = 1$ is,*

$$H_m[\boldsymbol{v}, \boldsymbol{l}_1, \cdots, \boldsymbol{l}_m] = \sum_{\boldsymbol{n}} \sum_{\boldsymbol{k}_1} \cdots \sum_{\boldsymbol{k}_m} h_m[\boldsymbol{n}-\boldsymbol{k}_1, \cdots, \boldsymbol{n}-\boldsymbol{k}_m] \tilde{\phi}_{\boldsymbol{l}_1}[\boldsymbol{k}_1] \cdots \tilde{\phi}_{\boldsymbol{l}_m}[\boldsymbol{k}_m] \phi_{\boldsymbol{v}}[\boldsymbol{n}] \tag{3.17}$$

*where $\boldsymbol{v}, \boldsymbol{l}_1, \ldots, \boldsymbol{l}_m \in \mathbb{Z}^D$ and $m, D \in \mathbb{N}$.*

The shift invariant kernel $H_m$ in equation (3.17) with DFT basis gives global spectral features of the natural domain kernel $h_m$. Similarly, with orthogonal or biorthogonal wavelet basis this $H_m$ is a multiresolution expansion of $h_m$ and have localized natural-frequency domain features. Theorem 3.1 with the shift invariant kernel $H_m$ specified in Lemma 3.2 gives the complete I/O relationship of a $m^{th}$ degree **shift invariant Volterra kernel**. This is also a generalization of the LSI and NLSI filters in signal processing in natural and transform domain. The following example shows the discussed natural domain and the transform domain equivalent computations with DFT and DWT bases.

**Example 3.2.** For a given $x[n] = [3, 7]$ and an LSI filter with unit impulse response $h[n] = [1, 2]$, find the output of the linear and quadratic Volterra kernel in the natural domain with $a_1 = a_2 = 1$. Verify equivalence in I/O with DFT basis and Haar multiresolution Volterra kernel using DWT.

1. For degree $m = 1$, the kernel is a simple LSI system whose unit impulse response is $h[n]$ and their equivalent I/O relationships in natural and transform domains are as given below.

   (a) The **natural-domain LSI Volterra kernel** with $a_1 = 1$, hence $h_1 = a_1 h = h$, is

   $$y_1[n] = \sum_{k=0}^{N-1} h[k] x[n-k] \qquad \text{for } n = 0, 1, 2, \cdots, N-1. \tag{3.18}$$

   where $x[n-k]$ are circular shifts and $y_1[n]$ is the output. For given $x$ and $h$, the LSI kernel output is the following circular convolution, $y_1[0] = h[0]x[0-0] + h[1]x[0-1] = h[0]x[0] + h[1]x[1] = 3 + 14 = 17$

   $y_1[1] = h[0]x[1-0] + h[1]x[1-1] = h[0]x[1] + h[1]x[0] = 7 + 6 = 13$

   i.e., $y_1[n] = [17, 13]$.

   (b) The **DFT** of (3.18) has the representation

   $$\beta[v] = X[v]H[v] \qquad \text{for } v = 0, 1, 2, \cdots, N-1 \tag{3.19}$$

$$y_1[n] = \frac{1}{N}\sum_{v=0}^{N-1} \beta[v] e^{j2\pi\frac{vn}{N}} \qquad \text{for } n = 0, 1, 2, \cdots, N-1 \tag{3.20}$$

where $H[v] = \sum_{k=0}^{N-1} h_1[k] e^{-j2\pi\frac{vn}{N}} = \text{FFT}(h) = \text{FFT}([1,2]) = [3,-1]$ and $X[v] = \text{FFT}(x) = \text{FFT}([3,7]) = [10,-4]$. Therefore, $\beta = [30,4]$ and $y_1 = \text{IFFT}(\beta) = [17,13]$ which matches with the computations in the natural domain in 1a above.

(c) The equivalent **multiresolution representation of the LSI kernel** in (3.18) using the Haar wavelet basis $\langle \phi_v \tilde{\phi}_v \rangle = 1$ is

$$\beta[v] = \sum_{l} \alpha[l]\, H[v,l] \qquad \text{for } v = 0, 1, \cdots, N-1 \tag{3.21}$$

$$y_1[n] = \sum_{v} \beta[v]\, \tilde{\phi}_v[n] \qquad \text{for } n = 0, 1, \cdots, N-1 \tag{3.22}$$

where $H[v,l] = \sum_n \tilde{h}[n,l]\phi_v[n]$ and $\tilde{h}[n,l] = \sum_k h[n-k]\tilde{\phi}_l[k]$ for $l = 0, 1, \cdots, N-1$.

i. Now, for the given $h[n] = [1,2]$ its circular shifts $h[n-k]$ for $n = 0$ is $h[0-0] = h[0] = 1$ and $h[0-1] = h[-1] = h[1] = 2$. Similarly, for $n = 1$, $h[1-0] = h[1] = 2$ and $h[1-1] = h[0] = 1$. Therefore, $(h[n-k])_{n=0,1} = \begin{bmatrix} 1 & 2 \\ 2 & 1 \end{bmatrix}$. The DWT of the rows is $\tilde{h}[n,l] = \begin{bmatrix} 2.121 & -0.707 \\ 2.121 & 0.707 \end{bmatrix}$ and finally the DWT of its columns give the multiresolution equivalent kernel $H[v,l] = \begin{bmatrix} 3 & 0 \\ 0 & -1 \end{bmatrix}$.

ii. The Haar wavelet transform of the input is,
$\alpha = \text{DWT}(x) = \text{DWT}([7.071, -2.828]) = [7.071, -2.828]$
and its element wise multiplication with the rows of $H$ yields a resultant $\begin{bmatrix} 21.213 & 0 \\ 0 & 2.828 \end{bmatrix}$ and finally, equation (3.21) yields $\beta[v] = [21.213, 2.828]$.

iii. The system output is the Inverse Discrete Wavelet Transform (IDWT) of $\beta$, i.e., $y_1 = \text{IDWT}(\beta) = \text{IDWT}([21.213, 2.828]) = [17,13]$ which matches with the computations in the natural domain in 1a above.

2. For degree $m = 2$, the kernel is Quadratic Shift Invariant (QSI) and their equivalent I/O relationships in natural and transform domains are given below.

(a) The **natural-domain QSI Volterra kernel** is

$$y_2[n] = \sum_{k_1=0}^{N-1}\sum_{k_2=0}^{N-1} h_2[k_1,k_2]x[n-k_1]x[n-k_2] \qquad \text{for} n = 0,1,2,\cdots,N-1 \quad (3.23)$$

where $h_2 = a_2 h \otimes h = h \otimes h = \begin{bmatrix} 1 & 2 \\ 2 & 4 \end{bmatrix}$, since $a_2 = 1$. With $x[n]$ circular shifted the output $y_2[n]$ is,

$$y_2[0] = h_2[0,0]x[0]x[0] + h_2[0,1]x[0]x[1] + h_2[1,0]x[1]x[0]$$

$$+h_2[1,1]x[1]x[1] = 289$$

$$y_2[1] = h_2[0,0]x[1]x[1] + h_2[0,1]x[1]x[0] + h_2[1,0]x[0]x[1]$$

$$+h_2[1,1]x[0]x[0] = 169$$

Thus the output of the QSI Volterra kernel in natural domain is,
$y_2[n] = [289, 169]$.

(b) The equivalent of (3.23) in the **DFT basis** is

$$\beta_2[v] = \frac{1}{N}\sum_{l=0}^{N-1} X[l]\, X[v-l]\, H_2[l, v-l] \qquad \text{for } v = 0,1,2,\cdots,N-1 \qquad (3.24)$$

$$y_2[n] = \frac{1}{N}\sum_{v} \beta_2[v]e^{j2\pi\frac{vn}{N}} \qquad \text{for } n = 0,1,2,\cdots,N-1 \qquad (3.25)$$

Here, $N = 2$ is the length of sequence $x$ and its DFT is $X[l] = [10, -4]$. The DFT of the kernel is $H_2[l, v-l] = \begin{bmatrix} 9 & -3 \\ -3 & 1 \end{bmatrix}$ and equation (3.24) gives $\beta_2$ as,

$$\beta_2[0] = \tfrac{1}{2}\Big(X[0]X[0]H_2[0,0] + X[1]X[1]H_2[1,1]\Big) = 458$$

$$\beta_2[1] = \tfrac{1}{2}\Big(X[0]X[1]H_2[0,1] + X[1]X[0]H_2[1,0]\Big) = 120$$

and finally, $y_2 = \mathrm{IFFT}(\beta_2) = [289, 169]$ and it matches the natural domain computation in 2a above.

(c) The equivalent **multiresolution representation of the QSI Volterra kernel** in (3.23) with the Haar wavelet basis $\langle \phi_v \tilde{\phi}_v \rangle = 1$ is

$$y_2[n] = \sum_v \beta_2[v]\, \tilde{\phi}_v[n] \qquad \text{for } n = 0, 1, \cdots, N-1 \tag{3.26}$$

$$\beta_2[v] = \sum_{l_1}\sum_{l_2} \alpha[l_1]\, \alpha[l_2]\, H_2[v, l_1, l_2] \qquad \text{for } v = 0, 1, \cdots, N-1 \tag{3.27}$$

where $H_2[v, l_1, l_2] = \sum_{\boldsymbol{n}} \tilde{h}_2[n, l_1, l_2]\phi_v[n]$ and $\tilde{h}_2[n, l_1, l_2] = \sum_{k_1}\sum_{k_2} h_2[n - k_1, n-k_2]\,\tilde{\phi}_{l_1}[k_1]\,\tilde{\phi}_{l_2}[k_2]$ for $l_1, l_2 \in 0, 1, \cdots, N-1$. In equation (3.27) $\alpha[l_1]\,\alpha[l_2] \in \alpha_2[l_1, l_2]$ and $\alpha_2[l_1, l_2] = \text{DWT}(x \otimes x) = \text{DWT}\left(\begin{bmatrix} 9 & 21 \\ 21 & 49 \end{bmatrix}\right) = \begin{bmatrix} 50 & -20 \\ -20 & 8 \end{bmatrix}$. The filter initialization and filtering involves the following steps,

i. $h_2[0-k_1, 0-k_2] = \begin{bmatrix} 1 & 2 \\ 2 & 4 \end{bmatrix}$ and $h_2[1-k_1, 1-k_2] = \begin{bmatrix} 4 & 2 \\ 2 & 1 \end{bmatrix}$ for $k_1, k_2 = 0, 1$.

ii. $\tilde{h}_2[0, l_1, l_2] = \begin{bmatrix} 4.5 & -1.5 \\ -1.5 & 0.5 \end{bmatrix}$ and $\tilde{h}_2[1, l_1, l_2] = \begin{bmatrix} 4.5 & 1.5 \\ 1.5 & 0.5 \end{bmatrix}$ for $l_1, l_2 = 0, 1$

iii. $H_2[0, l_1, l_2] = \begin{bmatrix} 6.364 & 0 \\ 0 & 0.707 \end{bmatrix}$ and $H_2[1, l_1, l_2] = \begin{bmatrix} 0 & -2.121 \\ -2.121 & 0 \end{bmatrix}$ for $l_1, l_2 = 0, 1$

iv. Then, equation (3.27) yields $\beta[v] = [323.85, 84.85]$ and finally, $y_2 = \text{IDWT}(\beta_2) = [289, 169]$ which matches with the computations in the natural domain in 2a above.

The multiresolution representation of the $m^{th}$ degree Volterra series in (3.1) with biorthogonal bases is stated in Theorem 3.3. It is equivalent to (3.1) with all its Volterra kernels and their input-outputs rewritten as their equivalent biorthogonal bases expansions.

**Theorem 3.3.** *A $m^{th}$ degree Volterra series can be expressed as an equivalent biorthogonal bases expansion as*

$$y[\boldsymbol{n}] = y_0[\boldsymbol{n}] + \sum_{m\in\mathbb{N}}\sum_{\boldsymbol{v}}\sum_{\boldsymbol{l}_1}\cdots\sum_{\boldsymbol{l}_m} \alpha[\boldsymbol{l}_1]\cdots\alpha[\boldsymbol{l}_m]\, H_m[\boldsymbol{v}, \boldsymbol{l}_1, \cdots, \boldsymbol{l}_m]\tilde{\phi}_{\boldsymbol{v}}[\boldsymbol{n}], \tag{3.28}$$

*where $\{H_m\}_{m\in\mathbb{N}}$ are the biorthogonal bases expansions of the Volterra kernels, $\boldsymbol{n}, \boldsymbol{v}, \boldsymbol{l}_m \in \mathbb{Z}^D$, $D \in \mathbb{N}$ and $\alpha$ is the biorthogonal expansion of input using the same biorthogonal system $\langle\phi, \tilde{\phi}\rangle = 1$.*

**Backpropagation calculus in biorthogonal coordinates** For a fixed degree $m$, the biorthogonal representation of the Volterra output coefficients in (3.8), where $\alpha[\boldsymbol{l}]$ are the expansion coefficients of the input $x[\boldsymbol{n}]$ and $H_m$ is the Volterra kernel in the biorthogonal basis as in (3.9).

Let a training loss $\mathcal{L}$ depend on the output coefficients $\{\beta_m[\boldsymbol{v}]\}_{\boldsymbol{v}}$ and define

$$g_m[\boldsymbol{v}] = \frac{\partial \mathcal{L}}{\partial \beta_m[\boldsymbol{v}]}, \tag{3.29}$$

the gradient of the loss with respect to $\beta_m[\boldsymbol{v}]$. The gradient of $\mathcal{L}$ with respect to the biorthogonal kernel entries is then

$$\frac{\partial \mathcal{L}}{\partial H_m[\boldsymbol{v}, \boldsymbol{l}_1, \ldots, \boldsymbol{l}_m]} = g_m[\boldsymbol{v}]\, \alpha[\boldsymbol{l}_1] \cdots \alpha[\boldsymbol{l}_m]. \tag{3.30}$$

Thus each coefficient of $H_m$ receives a contribution that is a product of the backpropagated gradient at index $\boldsymbol{v}$ and the corresponding product of input coefficients at indices $\boldsymbol{l}_1, \ldots, \boldsymbol{l}_m$. In the introduced `VolterraSys` Python TensorFlow package this multilinear structure is realized by tensor contractions (Einstein summation) in natural or multiresolution coordinates, and automatic differentiation applies these formulas during training.

### 3.1.3 A generalization of Parseval's theorem to biorthogonal bases

Let vectors $u$ and $v$ of the inner–product space be expanded in a complementary manner with respect to a pair of biorthogonal bases,

$$u = \sum_k u_k\, \phi_k \quad \text{and} \quad v = \sum_l v_l\, \tilde{\phi}_l \tag{3.31}$$

Then,

$$\langle u, v\rangle = \Big\langle \sum_k u_k\, \phi_k,\ \sum_l v_l\, \tilde{\phi}_l \Big\rangle = \sum_{k,l} u_k\, \overline{v_l}\, \langle \phi_k, \tilde{\phi}_l\rangle = \sum_k u_k\, \overline{v_k}. \tag{3.32}$$

and the biorthogonality ensures $\langle \phi_k, \tilde{\phi}_l\rangle = \delta_{kl}$. Now, let's also consider the variant with a conjugate,

$$\langle u, \overline{v}\rangle = \Big\langle \sum_k u_k\, \phi_k,\ \overline{\sum_l v_l \tilde{\phi}_l} \Big\rangle = \sum_k \sum_l u_k\, \overline{\overline{v}}_l\, \langle \phi_k, \overline{\tilde{\phi}}_l\rangle \tag{3.33}$$

If the biorthogonal bases are all comprised of **real** vectors, then $\overline{\tilde{\phi}}_l = \tilde{\phi}_l$, whereupon $\langle \phi_k, \overline{\tilde{\phi}}_l\rangle = \langle \phi_k, \tilde{\phi}_l\rangle = \delta_{kl}$,and then $\langle u, \overline{v}\rangle = \sum_k u_k\, v_k$. On the other hand, if the bases have **complex** vectors, one cannot immediately make an inference about the inner product $\langle \phi_k, \overline{\tilde{\phi}}_l\rangle$ without computing the cross-Gram entries. In some specific cases one may say something concrete. For example, consider the **DFT bases** for a finite-dimensional ($N$-dimensional) sequence,

$$\phi_k[n] = \frac{1}{\sqrt{N}} e^{j2\pi nk/N}, \qquad n = 0, \cdots, N-1. \tag{3.34}$$

Then,

$$\langle \phi_k, \phi_l\rangle = \frac{1}{N} \sum_{n=0}^{N-1} \left(e^{j2\pi nk/N} e^{-j2\pi nl/N}\right) = \frac{1}{N} \sum_{n=0}^{N-1} e^{j2\pi n(k-l)/N} = \delta_{kl} \tag{3.35}$$

and on the other hand,

$$\langle \phi_k, \overline{\phi}_l \rangle = \frac{1}{N}\sum_{n=0}^{N-1} e^{j2\pi n(k+l)/N} = \begin{cases} 1, & (k+l) \equiv 0 \mathrm{mod} N, \\ 0, & \text{otherwise.} \end{cases} \tag{3.36}$$

So there will be a **collapse** in the inner product, but the inner product of the sequences $u[n]$ and $v[n]$ will have a **mirroring** of $v[n]$,

$$\text{or, } \langle u, \overline{v} \rangle = \Big\langle \sum_k u_k\, \phi_k,\ \sum_l \overline{v}_l\, \overline{\tilde{\phi}}_l \Big\rangle = \sum_k u_k\, v_{(N-k)\,\mathrm{mod}\,N}. \tag{3.37}$$

In the case of more **general complex** basis elements, no general statement can be made. On the other hand, even if the basis elements are all **real** and the coefficients $u_k, v_k$ are complex, we still have tangible results, giving the pair of equations

$$\langle u, v \rangle = \Big\langle \sum_k u_k\, \phi_k,\ \sum_l v_l\, \tilde{\phi}_l \Big\rangle = \sum_k \sum_l u_k \overline{v}_l\, \langle \phi_k\, \tilde{\phi}_l \rangle = \sum_k u_k\, \overline{v_k} \tag{3.38}$$

$$\langle u, \overline{v} \rangle = \Big\langle \sum_k u_k\, \phi_k,\ \sum_l \overline{v}_l\, \overline{\tilde{\phi}}_l \Big\rangle = \sum_k \sum_l u_k\, \overline{\overline{v}}_l\, \langle \phi_k, \overline{\tilde{\phi}}_l \rangle = \sum_k u_k\, v_k \tag{3.39}$$

Here, we must interpret $u_k$ and $v_k$ as the coefficients of expansion of $u$ in terms of **one side** of the biorthogonal bases $\phi_k$ and $v_k$ as the coefficients of expansion of $v$ in terms of the **complementary side** of the biorthogonal basis $\tilde{\phi}_l$. For orthogonal (orthonormal) bases we have $\phi = \tilde{\phi}$.

An example could be the **orthogonal real DWT** bases or **biorthogonal real DWT** bases. Now one may apply these principles strategically to the **Volterra kernel expansion,**

$$y[n] = \sum_{k_1} \cdots \sum_{k_m} h\big[n, k_1, \ldots, k_m\big]\ x[k_1] \cdots x[k_m] \tag{3.40}$$

Take any one summation

$$\sum_{k_l} h\big[n, \ldots, k_l, \ldots\big]\ x[k_1] \cdots x[k_l] \cdots x[k_m] \equiv \prod_{q \neq l} x[k_q]\ \sum_{k_l} h\big[n; \ldots, k_l, \ldots\big]\, x[k_l] \tag{3.41}$$

and we can treat this $\sum_{k_l} h\big[n; \ldots, k_l, \ldots\big]\, x[k_l]$ as the inner product $\big\langle h_l, \overline{x} \big\rangle$, where we define the slice,

$$h_l[k_l] \ \triangleq\ h\big[n, \ldots, k_l, \ldots\big] \tag{3.42}$$

i.e., a sequence obtained by fixing all other indices $n, k_q, q \neq l$ and varying $k_l$. Since the term $x[k_l]$ occurs **without** the complex conjugate, we are **conjugating the second argument** of the inner product. Now we can invoke the index-reversal identity, i.e., equation (3.39) for this so each such summation can be replaced by an equivalent summation, holding the other indices as they are.

## 3.2 Practical realizations of natural-frequency domain shift invariant Volterra systems

The earlier section described general $m^{th}$ degree Volterra kernels. Now, there are practical limitations in realizing a Volterra kernel of arbitrary degree because the complexity, both time and space, increases with both the input dimension and the kernel degree. Therefore, useful realizations of such kernels depend on the availability of high-performance computation hardware. This section describes realizations up to cubic Volterra kernels for sequences (1D), images (2D) and volumetric (3D) inputs, which involve up to 9D convolutions. The output $q[\boldsymbol{n}]$ of the LSI system $h[\boldsymbol{n}]$ in equation (3.11) for one (D=1), two (D=2) and three (D=3) dimensional input $x[\boldsymbol{n}]$ where $\boldsymbol{n} \in \mathbb{Z}^D$ and $D \in \mathbb{N}$ is as follows:

1. $q[n] = \sum h[k]x[n-k]$, where $D = 1$ and $n \in \mathbb{Z}$
2. $q[\boldsymbol{n}] = \sum_{k_1} \sum_{k_2} h[k_1, k_2]x[n_1 - k_1, n_2 - k_2]$ where $D = 2$ and $\boldsymbol{n} \equiv (n_1, n_2) \in \mathbb{Z}^2$
3. $q[\boldsymbol{n}] = \sum_{k_1} \sum_{k_2} \sum_{k_3} h[k_1, k_2, k_3]x[n_1 - k_1, n_2 - k_2, n_3 - k_3]$ where $D = 3$, $\boldsymbol{n} \equiv (n_1, n_2, n_3) \in \mathbb{Z}^3$

**Fast algorithmic realization of learnable LSI Volterra kernel with wavelet bases** Theorem 1 states that a $D$-dimensional Volterra filter $h[\boldsymbol{n}]$ with circular shifts on $\mathbb{Z}^D$, i.e. $\boldsymbol{n}, \boldsymbol{k} \in \mathbb{Z}^D$ using a global biorthogonal system $\langle \phi, \tilde{\phi} \rangle = 1$. When system order is equal to 1 the I/O relation has a linear kernel,

$$y[\boldsymbol{n}] = \sum_{\boldsymbol{v}} \sum_{\boldsymbol{l}} \alpha[\boldsymbol{l}]\ H[\boldsymbol{v}, \boldsymbol{l}] \tilde{\phi}_{\boldsymbol{v}}[\boldsymbol{n}] \tag{3.43}$$

where

$$H[\boldsymbol{v}, \boldsymbol{l}] = \sum_{\boldsymbol{n}} \sum_{\boldsymbol{k}} h[\boldsymbol{n} - \boldsymbol{k}] \tilde{\phi}_{\boldsymbol{l}}[\boldsymbol{k}] \phi_{\boldsymbol{v}}[\boldsymbol{n}] \tag{3.44}$$

and $\alpha[\boldsymbol{l}] = \sum_{\boldsymbol{n}} x[\boldsymbol{n}] \phi_{\boldsymbol{l}}[\boldsymbol{n}]$ for $\boldsymbol{v}, \boldsymbol{l} \in \mathbb{Z}^D$ and $m, D \in \mathbb{N}$. Algorithm 3.13 is a fast realization of a LSI Volterra kernel with DWT bases, which also generalizes to biorthogonal wavelet bases LSI filters.

Figure 3.2 shows the impulse responses (top row) of LSI Volterra 3D layer with four kernels of same length using Algorithm 3.13 for cases of kernel lengths, $L = 7, 8, 9$ (left to right) with bior1.3 wavelet and $N = 16$ and their Fourier transforms (bottom row). The differences are visible in the edges and boundaries in the natural domain and in the tiling in Fourier domain.

The above realization is a generalization of the LSI kernel (m=1) with orthogonal and biorthogonal wavelet families. Linear filtering in the natural domain by convolution is another example in this framework. This orthogonal bases representation framework just extends the learning of the kernel coefficients in a sparse transform domain space. This biorthogonal representation of input, output, and kernel also gives us the flexibility to enhance the frequency selectivity of the LSI kernel by increasing its size without compromising stability. We get more

**Algorithm 3.13:** LSI volterra kernel with wavelet bases and $D$-dimensional I/O

1. Input $X \in \mathbb{R}^{\text{batch} \times \boldsymbol{N} \times \text{channels}}$, where $\boldsymbol{N} = N_1 \times \cdots \times N_D$
2. Initialize UIR kernel $h \in \mathbb{R}^{\boldsymbol{N} \times \text{in channels} \times \text{out channels}}$(with or without zero padding)
3. Prepare $\boldsymbol{k}$ shifted filter tensor $h_{\text{shifted}}[\boldsymbol{k}, \boldsymbol{n}] = h[\boldsymbol{n} - \boldsymbol{k}]\ \forall \boldsymbol{n}$
   $h_{\text{shifted}}[\boldsymbol{k}, \boldsymbol{n}, \text{in channel}, \text{out channel}] \in \mathbb{R}^{\boldsymbol{N} \times \boldsymbol{N} \times \text{in channels} \times \text{out channels}}$,
4. Generate filter with orthogonal/biorthogonal bases with DWT of $\boldsymbol{n}$ axis then $\boldsymbol{k}$ axis,
$$H\left[\boldsymbol{v}, \boldsymbol{l}\right] = \text{DWT}_{\boldsymbol{k}}\left(\text{DWT}_{\boldsymbol{n}}\left(h_{\text{shifted}}\left[\boldsymbol{k}, \boldsymbol{n}\right]\right)\right)$$
5. Compute, $\alpha = \text{DWT}(X)$, i.e., a orthogonal or biorthogonal transform of $X$
6. Compute and return output $Y \in \mathbb{R}^{\text{batch} \times \boldsymbol{N} \times \text{channels}}$ as,
$$\beta_{bv_1 \cdots v_D o} = \sum_{n_1, \cdots, n_D, c} \alpha_{bn_1 \cdots n_D c} H_{v_1 \cdots v_D n_1 \cdots n_D co} \quad \text{and} \quad Y = \text{IDWT}(\beta)$$

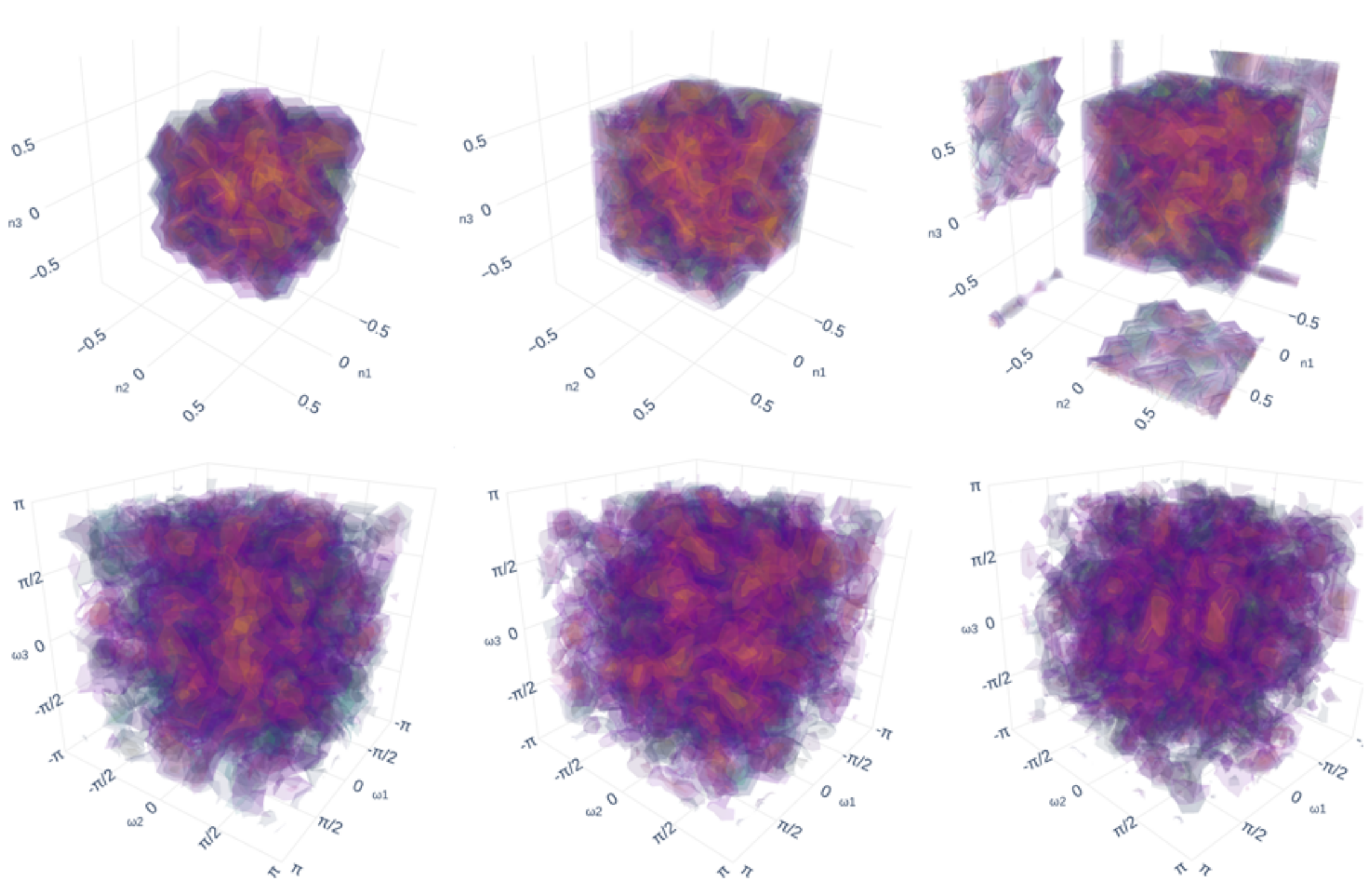

Figure 3.2: Impulse responses (top row) of LSI Volterra 3D layer with four kernels of same length (shown together) for cases of kernel lengths, $L = 7, 8, 9$ (left to right) with bior1.3 wavelet and $N = 16$ and their Fourier transforms (bottom row)

control over the kernel coefficients. Realizing the linear kernel is comparatively easier, practical challenges arise as complexity increases with quadratic, cubic, and higher-degree nonlinear Volterra kernels. It becomes even more challenging when signal dimensions increase, starting with sequence, image, and volumetric I/O systems, as we shall see in the following subsections for each case. In the following, the **core mathematics is the same** as described so far, but with **explicit assignment** of fixed input dimensions. The realizations for 4D and 6D convolutions in quadratic kernels of image and volumetric inputs using Einstein summations are also clearly explained next.

### 3.2.1 Sequences and NLSI systems in natural-frequency domain

When the input $x[n] \in \ell^2(\mathbb{Z})$ and output $y[n] \in \ell^2(\mathbb{Z})$ are **sequences**, then for general $m^{th}$ degree Volterra kernel, equation (3.8) and (3.10) can be rewritten in a simplified form as,

$$\beta[v] = \sum_{l_1} \cdots \sum_{l_m} \alpha[l_1] \cdots \alpha[l_m]\; H[v, l_1, \cdots, l_m] \qquad \text{(Analysis, } D = 1\text{)}$$

$$y_m[n] = \sum_{v} \sum_{l_1} \cdots \sum_{l_m} \alpha[l_1] \cdots \alpha[l_m]\; H[v, l_1, \cdots, l_m] \tilde{\phi}_v[n] \qquad \text{(Synthesis, } D = 1\text{)}$$

where $H[v, l_1, \cdots, l_m] = \sum_n \sum_{k_1} \cdots \sum_{k_m} h[n - k_1, \cdots, n - k_m] \tilde{\phi}_{l_m}[k_1] \cdots \tilde{\phi}_{l_m}[k_m] \phi_v[n]$ is a $m$ dimensional shift invariant Volterra kernel in biorthogonal bases similar to equation (3.17).

The I/O relation of the **linear** ($m = 1$) Volterra kernel in biorthogonal bases for a sequence input is

$$\beta[v] = \sum_{l_1} \alpha[l_1]\, H_1[l_1] \qquad \text{(analysis, } D = 1, m = 1\text{)}$$

and the reconstruction is

$$y_1[n] = \sum_{v} \left( \sum_{l_1} \alpha[l_1]\, H_1[l_1] \right) \tilde{\phi}_v[n] \qquad \text{(synthesis, } D = 1, m = 1\text{)}$$

where $H[l_1] = \sum_n \sum_{k_1=0}^{N-1} h_1[n - k_1] \tilde{\phi}_{l_1}[k_1] \phi_v[n]$ for $l_1 \in [0, N-1]$ is the biorthogonal bases expansion of the LSI kernel.

The I/O relation of the **quadratic** ($m = 2$) Volterra kernel in biorthogonal bases for a sequence input is

$$\beta[v] = \sum_{l_1} \sum_{l_2} \alpha[l_1]\, \alpha[l_2]\, H_2[l_1, l_2] \qquad \text{(analysis, } D = 1, m = 2\text{)}$$

and the reconstruction is

$$y_2[n] = \sum_{v} \left( \sum_{l_1} \sum_{l_2} \alpha[l_1]\, \alpha[l_2]\, H_2[l_1, l_2] \right) \tilde{\phi}_v[n] \qquad \text{(synthesis, } D = 1, m = 2\text{)}$$

where $H_2[l_1, l_2] = \sum_n \sum_{k_1} \sum_{k_2} h[n - k_1, n - k_2] \tilde{\phi}_{l_1}[k_1] \tilde{\phi}_{l_2}[k_2] \phi_v[n]$ for $l_1, l_2 \in [0, N-1]$ is the biorthogonal bases expansion of the quadratic Volterra kernel.

**Quadratic Volterra series expansion of a sequence in biorthogonal bases** The complete shift invariant Volterra series expansion of a sequence in biorthogonal bases with both linear and quadratic Volterra kernels is,

$$
\begin{aligned}
y[n] &= y_0[n] + y_1[n] + y_2[n] \qquad \text{for } n = 0, 1, \cdots, N-1 \\
&= y_0[n] + \sum_v \left( \sum_{l_1} \alpha[l_1]\, H_1[l_1] \right) \tilde{\phi}_v[n] + \sum_v \left( \sum_{l_1} \sum_{l_2} \alpha[l_1]\, \alpha[l_2]\, H_2[l_1, l_2] \right) \tilde{\phi}_v[n]
\end{aligned} \tag{3.45}
$$

**Equivalent Volterra series in natural domain for sequence input** The equivalent quadratic Volterra series of equation (3.45) in natural domain is

$$
\begin{aligned}
y[n] &= a_0 + a_1 \sum_k h[k]x[n-k] + a_2 \sum_k h[k]x[n-k] \sum_k h[k]x[n-k] \\
&= a_0 + \sum_k a_1 h[k]x[n-k] + \sum_{k_1} \sum_{k_2} a_2 h[k_1] h[k_2] x[n-k_1] x[n-k_2] \\
&= h_0 + \sum_k h_1[k]x[n-k] + \langle a_2 \begin{bmatrix} h_0 h_0 & \cdots & h_0 h_n \\ \vdots & \ddots & \vdots \\ h_n h_0 & \cdots & h_n h_n \end{bmatrix}, \begin{bmatrix} x_n x_n & \cdots & x_n x_0 \\ \vdots & \ddots & \vdots \\ x_0 x_n & \cdots & x_0 x_0 \end{bmatrix} \rangle
\end{aligned}
$$

or,

$$
y[n] = h_0 + \sum_k h_1[k]x[n-k] + \sum_{k_1=0}^{n} \sum_{k_2=0}^{n} h_2[k_1, k_2] x_2[n-k_1, n-k_2] \tag{3.46}
$$

where $x_2 = x \otimes x$ is the outer product of input with itself and have elements $\left(x[n-k_1]x[n-k_2]\right)_{\forall k_1, k_2}$, $h_0 \equiv a_0$ is a bias, $h_1 \equiv a_1 h$ is a one-dimensional LSI filter and $h_2 \equiv a_2 hh$ is a two-dimensional LSI filter**.** In equation (3.46), the quadratic monomial $a_2 q[n]^2$ is an inner product of the outer products $h_2$and $x_2$ running from 0 to $n$. The form of the two dimensional filter $h_2 \equiv a_2 hh$ is symmetric and so is $x_2$. The quadratic monomial $a_2 q[n]^2$ here is a different kind of convolution where the filter strides are along the trace (main diagonal) of the input as illustrated in Figure 3.3 and we shall call it **diagonal convolution**. Since, having a filter size same as the input size is computationally very demanding and therefore filters of smaller size with zero padding are preferred. For example, $3 \times 3$ LSI filters in CNNs with two dimensional inputs is often much smaller than the dimension of input.

**DFT of shift invariant quadratic monomial for sequence input** The natural domain quadratic monomial or the quadratic Volterra kernel output is

$$a_2 q[n]^2 = \sum_{k_1=0}^{N-1} \sum_{k_2=0}^{N-1} h_2[k_1, k_2] x[n-k_1] x[n-k_2] \qquad \text{for } n = 0, 1, 2, \cdots, N-1 \tag{3.47}$$

where $x[n-k_1]x[n-k_2] \in x_2[n-k_1, n-k_2]$ and $x_2 = x \otimes x$ is an outer product of $x$ with itself. For an N-point DFT, the shifted $x$ in equation (3.47) are circular shifts between $[0, N-1]$. Let, the quadratic monomial $a_2 q[n]^2 = q_2[n]$, then by applying DFT on both sides of the equation we get,

$$\mathcal{F}\{q_2[n]\} = \mathcal{F}\left\{\sum_{k_1=0}^{N-1} \sum_{k_2=0}^{N-1} h_2[k_1, k_2] x[n-k_1] x[n-k_2]\right\}$$

$$Q_2[p] = \sum_{n=0}^{N-1} \left\{\sum_{k_1=0}^{N-1} \sum_{k_2=0}^{N-1} h_2[k_1, k_2] x[n-k_1] x[n-k_2]\right\} e^{-j2\pi \frac{pn}{N}} \quad \text{for } p = 0, 1, \cdots, N-1$$

$$= \sum_{k_1=0}^{N-1} \sum_{k_2=0}^{N-1} h_2[k_1, k_2] \left\{\sum_{n=0}^{N-1} x[n-k_1] x[n-k_2] e^{-j2\pi \frac{pn}{N}}\right\}$$

$$= \sum_{k_1=0}^{N-1} \sum_{k_2=0}^{N-1} h_2[k_1, k_2]\, \mathcal{F}\{x[n-k_1] x[n-k_2]\} \quad \text{(By DFT definition)}$$

$$= \frac{1}{N} \sum_{k_1=0}^{N-1} \sum_{k_2=0}^{N-1} h_2[k_1, k_2]\, \mathcal{F}\{x[n-k_1]\} \circledast \mathcal{F}\{x[n-k_2]\} \quad \text{(By mul. property)}$$

$$= \frac{1}{N} \sum_{k_1=0}^{N-1} \sum_{k_2=0}^{N-1} h_2[k_1, k_2]\, X[p] e^{-j2\pi \frac{pk_1}{N}} \circledast X[p] e^{-j2\pi \frac{pk_2}{N}} \quad \text{(By time shift property)}$$

$$= \frac{1}{N} \sum_{k_1=0}^{N-1} \sum_{k_2=0}^{N-1} h_2[k_1, k_2] \sum_{l=0}^{N-1} X[l] e^{-j2\pi \frac{lk_1}{N}} X[p-l] e^{-j2\pi \frac{(p-l)k_2}{N}}$$

$$= \frac{1}{N} \sum_{l=0}^{N-1} X[l] X[p-l] \sum_{k_1=0}^{N-1} \sum_{k_2=0}^{N-1} h_2[k_1, k_2]\, e^{-j2\pi \left(\frac{lk_1}{N} + \frac{(p-l)k_2}{N}\right)}$$

$$Q_2[p] = \frac{1}{N} \sum_{l=0}^{N-1} X[l]\, X[p-l]\, H_2[l, p-l] \quad \text{for } p = 0, 1, \cdots, N-1 \tag{3.48}$$

where $X$ is the $N$-point DFT of the input sequence and $H_2$ is $N^2$-point DFT of the two dimensional LSI filter $h_3$. In equation (3.48), $\sum_{l=p+1}^{N-1} X[l]\, X[p-l]\, H[l,p-l] \neq 0$ for $l \in [p+1, N-1]$ due to circular convolution and circular shifts. Therefore, $a_2 q^2[n] = q_2[n] = \mathcal{F}^{-1}\{Q_2[p]\}$, where $\mathcal{F}^{-1}$ denotes the IDFT and $\mathcal{F}$ is DFT.

**Complexity of quadratic Volterra kernel for sequence input** In equation (3.47), for each $n$ the number of computations is proportional to $O(N^2)$ and the total computations $\forall n \in [0, N-1]$ is $O(N^3) = O(N^2 \times N)$. Similarly, the IDFT of equation (3.48) involves the following computations, (i) FFT of input sequence $O(NlogN)$, (ii) FFT of the two dimensional kernel $O(N^2 logN)$, (iii) Pointwise multiplication $O(N^2)$ and (iv) IFFT $O(NlogN)$. Therefore, the time complexity of the IFFT of equation (3.48) is,

$$O(N^2 logN) = O(NlogN) + O(N^2 logN) + O(N^2) + O(NlogN)$$

Therefore, the time complexity of the IDFT of equation (3.48) is $O(N^2 logN)$ which is less than $O(N^3)$ of equation (3.47).

**Shift invariant cubic monomial and its DFT for sequence input** The cubic monomial of equation (3.11) with circular shifts between $[0, N-1]$ is

$$a_3 q[n]^3 = \sum_{k_1=0}^{N-1}\sum_{k_2=0}^{N-1}\sum_{k_3=0}^{N-1} h_3[k_1,k_2,k_3]x[n-k_1]x[n-k_2]x[n-k_3] \tag{3.49}$$

Let, $a_3 q[n]^3 = q_3[n]$ and by applying DFT on both sides of the equation we get,

$$\mathcal{F}\{q_3[n]\} = \mathcal{F}\left\{\sum_{k_1=0}^{N-1}\sum_{k_2=0}^{N-1}\sum_{k_3=0}^{N-1} h_3[k_1,k_2,k_3]x[n-k_1]x[n-k_2]x[n-k_3]\right\}$$

$$Q_3[p] = \sum_{n=0}^{N-1}\left\{\sum_{k_1=0}^{N-1}\sum_{k_2=0}^{N-1}\sum_{k_3=0}^{N-1} h_3[k_1,k_2,k_3]x[n-k_1]x[n-k_2]x[n-k_3]\right\} e^{-j2\pi\frac{pn}{N}}$$

$$\text{for } p = 0, 1, 2, \cdots, N-1$$

$$= \sum_{k_1=0}^{N-1}\sum_{k_2=0}^{N-1}\sum_{k_3=0}^{N-1} h_3[k_1,k_2,k_3]\left\{\sum_{n=0}^{N-1} x[n-k_1]x[n-k_2]x[n-k_3]e^{-j2\pi\frac{pn}{N}}\right\}$$

$$= \sum_{k_1=0}^{N-1}\sum_{k_2=0}^{N-1}\sum_{k_3=0}^{N-1} h_3[k_1,k_2,k_3]\,\mathcal{F}\{x[n-k_1]x[n-k_2]x[n-k_3]\}$$

$$= \frac{1}{N}\sum_{k_1=0}^{N-1}\sum_{k_2=0}^{N-1}\sum_{k_3=0}^{N-1} h_3[k_1,k_2,k_3]\,\mathcal{F}\{x[n-k_1]\} \circledast \mathcal{F}\{x[n-k_2]\} * \mathcal{F}\{x[n-k_3]\}$$

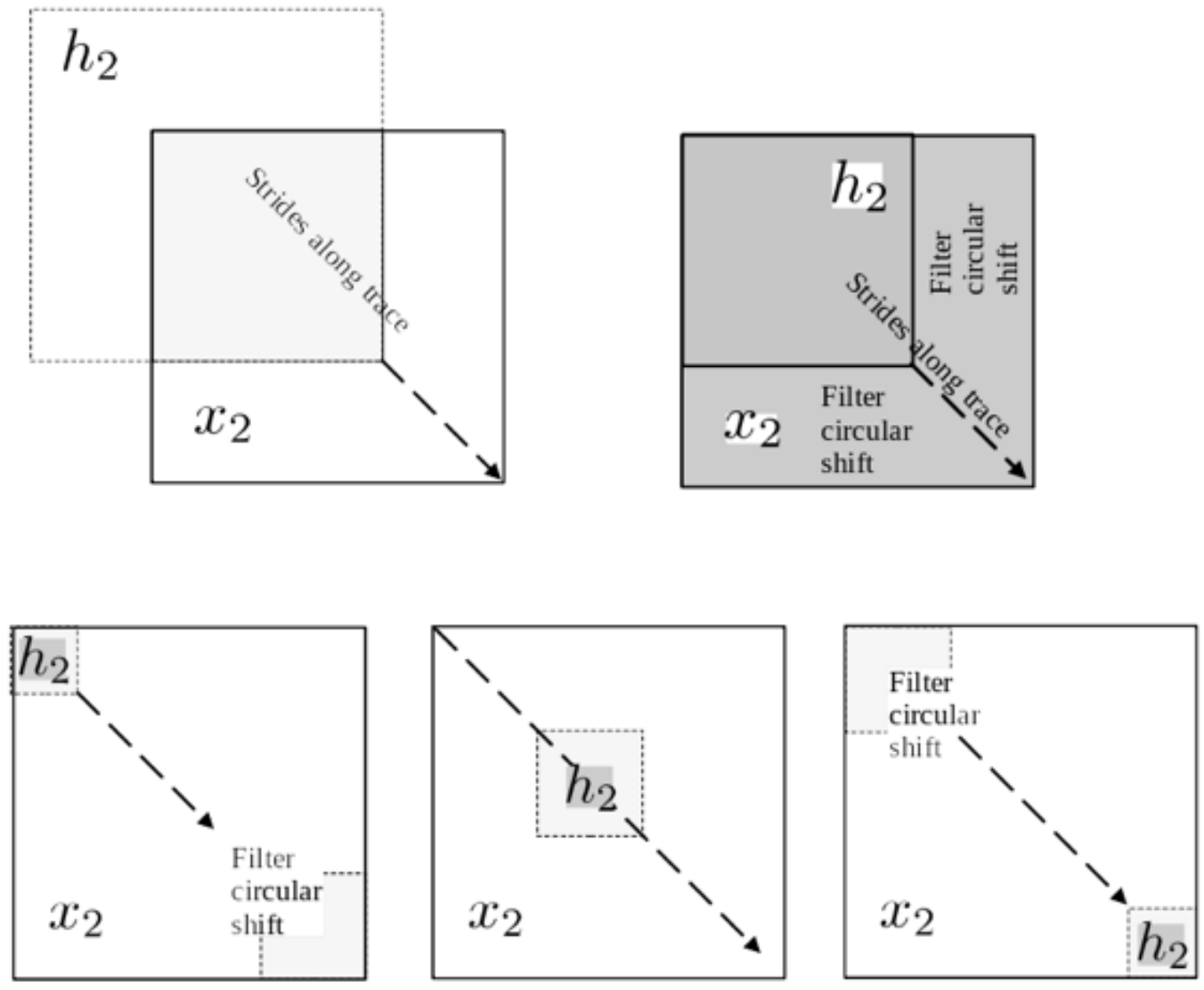

(a) Trace convolution striding along the trace of $x_2 = x \otimes x$ in natural domain for the quadratic monomial. Input and filter having same $N$ (top row); without circular shift (top left); with circular shift (top right). Illustration for filter size less than input size (bottom row)

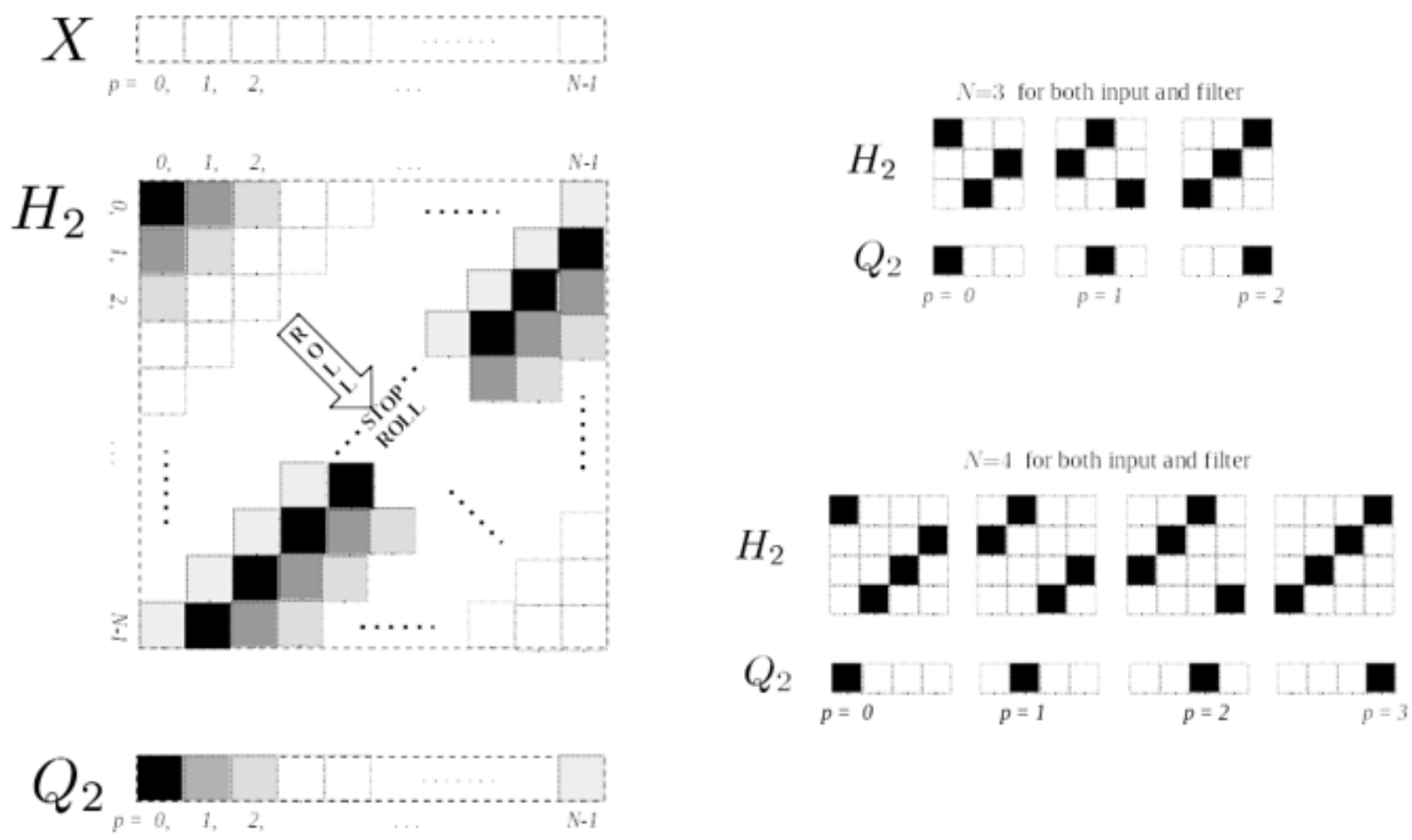

(b) The equivalent DFT domain operation of trace convolution (with much less numbers compared to natural domain). Illustration of examples for $N = 3$ and $N = 4$ (right column)

Figure 3.3: Computation of the quadratic monomial in natural and Fourier domain

$$= \frac{1}{N}\sum_{k_1=0}^{N-1}\sum_{k_2=0}^{N-1}\sum_{k_3=0}^{N-1} h_3[k_1,k_2,k_3]\; X[p]e^{-j2\pi\frac{pk_1}{N}} \circledast X[p]e^{-j2\pi\frac{pk_2}{N}} * X[p]e^{-j2\pi\frac{pk_3}{N}}$$

$$= \frac{1}{N}\sum_{k_1=0}^{N-1}\sum_{k_2=0}^{N-1}\sum_{k_3=0}^{N-1} h_3[k_1,k_2,k_3] \left(\sum_{l_1=0}^{N-1} X[l_1]e^{-j2\pi\frac{l_1k_1}{N}} X[p-l_1]e^{-j2\pi\frac{(p-l_1)k_2}{N}}\right) * X[p]e^{-j2\pi\frac{pk_3}{N}}$$

$$= \frac{1}{N}\sum_{k_1=0}^{N-1}\sum_{k_2=0}^{N-1}\sum_{k_3=0}^{N-1} h_3[k_1,k_2,k_3] \sum_{l_2=0}^{N-1} X[l_2]e^{-j2\pi\frac{l_2k_3}{N}} \sum_{l_1=0}^{N-1} X[l_1]e^{-j2\pi\frac{l_1k_1}{N}} X[p-l_1-l_2]e^{-j2\pi\frac{(p-l_1-l_2)k_2}{N}}$$

$$= \frac{1}{N}\sum_{l_1=0}^{N-1}\sum_{l_2=0}^{N-1} X[l_1]X[l_2]X[p-l_1-l_2] \sum_{k_1=0}^{N-1}\sum_{k_2=0}^{N-1}\sum_{k_3=0}^{N-1} h_3[k_1,k_2,k_3]\; e^{-j2\pi\left(\frac{l_1k_1}{N}+\frac{l_2k_3}{N}+\frac{(p-l_1-l_2)k_2}{N}\right)}$$

or,

$$Q_3[p] = \frac{1}{N}\sum_{l_1=0}^{N-1}\sum_{l_2=0}^{N-1} X[l_1]X[l_2]X[p-l_1-l_2]\; H_3[l_1,l_2,p-l_1-l_2],\;\; p = 0,\cdots,N-1 \tag{3.50}$$

where $X$ is the N-point DFT of the input sequence and $H_3$ is $N^3$-point DFT of the three dimensional filter $h_3$. In equation (3.50), $\sum_{l_1=p+1}^{N-1}\sum_{l_2=p+1}^{N-1} X[l_1]X[l_2]X[[p-l_1-l_2]]\; H_3[l_1,l_2,[p-l_1-l_2]] \neq 0$ between $l_1 = p+1$ to $N-1$ due to circular convolution and circular shift.

**Complexity of cubic Volterra kernel for sequence input** In equation (3.49), for each $n$ the number of computations is proportional to $O(N^3)$ and the total computations $\forall n \in [0, N-1]$ is $O(N^4) = O(N^3 \times N)$. Similarly, the IDFT of equation (3.48) involves the following computations, (i) FFT of input sequence $O(NlogN)$, (ii) FFT of the three dimensional kernel $O(N^3logN)$, (iii) Pointwise multiplication $O(N^2)$ and (iv) IFFT $O(NlogN)$. Therefore, the time complexity of the IFFT of equation (3.48) is $O(N^3logN)$ as shown below,

$$O(N^3logN) = O(NlogN) + O(N^3logN) + O(N^2) + O(NlogN)$$

### 3.2.2 NLSI systems with Image input

When the LSI system in equation (3.11) has input $x[\boldsymbol{n}] \in \ell_2(\mathbb{Z}^2)$ $\boldsymbol{n} \equiv (n_1, n_2) \in \mathbb{Z}^2 \geq (0,0)$, i.e., an image, then the shift invariant quadratic Volterra series in natural domain can be written as,

$$y[\boldsymbol{n}] = a_0 + a_1 \sum_{k_1} \sum_{k_2} h[k_1,k_2] x[n_1-k_1, n_2-k_2]$$

$$+ a_2 \sum_{k_1} \sum_{k_2} h[k_1,k_2] x[n_1-k_1, n_2-k_2] \sum_{k_1} \sum_{k_2} h[k_1,k_2] x[n_1-k_1, n_2-k_2]$$

or,

$$y[\boldsymbol{n}] = a_0 + \sum_{k_1} \sum_{k_2} a_1 h[k_1,k_2] x[n_1-k_1, n_2-k_2]$$

$$+ \sum_{k_1} \sum_{k_2} \sum_{k_3} \sum_{k_4} a_2 h[k_1,k_2] h[k_3,k_4] x[n_1-k_1, n_2-k_2] x[n_1-k_3, n_2-k_4]$$

or,

$$y[\boldsymbol{n}] = h_0 + \sum_{k_1} \sum_{k_2} h_1[k_1,k_2] x[n_1-k_1, n_2-k_2]$$

$$+ \sum_{k_1} \sum_{k_2} \sum_{k_3} \sum_{k_4} h_2[k_1,k_2,k_3,k_4] x[n_1-k_1, n_2-k_2] x[n_1-k_3, n_2-k_4]$$

where $h_0 \equiv a_0$ is a bias, $h_1 \equiv a_1 h$ is a two-dimensional convolution filter and the quadratic filter $h_2 \equiv a_2 hh$ is a **four-dimensional** convolution filter. Therefore, the output of shift invariant quadratic Volterra kernel for images is a four-dimensional diagonal convolution:

$$a_2 q[n_1,n_2]^2 = \sum_{k_1=0}^{N-1} \sum_{k_2=0}^{N-1} \sum_{k_3=0}^{N-1} \sum_{k_4=0}^{N-1} h_2[k_1,k_2,k_3,k_4] x[n_1-k_1, n_2-k_2] x[n_1-k_3, n_2-k_4] \quad (3.51)$$

**DFT of the quadratic monomial for image input** The DFT of the quadratic term defined above is

$$Q_2[p_1,p_2] = \sum_{n_1,n_2} \left\{ \sum_{k_1,k_2,k_3,k_4} h_2[k_1,k_2,k_3,k_4] x[n_1-k_1, n_2-k_2] \cdot x[n_1-k_3, n_2-k_4] \right\} w_N^{p_1 n_1 + p_2 n_2}$$

$$= \sum_{k_1,k_2,k_3,k_4} h_2[k_1,k_2,k_3,k_4] \cdot \left\{ \sum_{n_1,n_2} x[n_1-k_1, n_2-k_2] \cdot x[n_1-k_3, n_2-k_4] w_N^{p_1 n_1 + p_2 n_2} \right\}$$

$$= \sum_{k_1,k_2,k_3,k_4} h_2[k_1,k_2,k_3,k_4] \cdot \mathcal{F}\Big\{x[n_1-k_1,n_2-k_2] \cdot x[n_1-k_3,n_2-k_4]\Big\}$$

$$= \frac{1}{N^2} \sum_{k_1,k_2,k_3,k_4} h_2[k_1,k_2,k_3,k_4] \cdot \mathcal{F}\Big\{x[n_1-k_1,n_2-k_2]\Big\} \circledast \mathcal{F}\Big\{x[n_1-k_3,n_2-k_4]\Big\}$$

$$= \frac{1}{N^2} \sum_{k_1,k_2,k_3,k_4} h_2[k_1,k_2,k_3,k_4] \cdot X[p_1,p_2]e^{-j2\pi(p_1k_1+p_2k_2)/N} \circledast X[p_1,p_2]e^{-j2\pi(p_1k_3+p_2k_4)/N}$$

$$= \frac{1}{N^2} \sum_{k_1,k_2,k_3,k_4} h_2[k_1,k_2,k_3,k_4] \cdot \sum_{l_1,l_2} X[l_1,l_2]w_N^{l_1k_1+l_2k_2} X[p_1-l_1,p_2-l_2]w_N^{(p_1-l_1)k_3+(p_2-l_2)k_4}$$

$$= \frac{1}{N^2} \sum_{l_1,l_2} X[l_1,l_2]X[p_1-l_1,p_2-l_2] \sum_{k_1,k_2,k_3,k_4} h_2[k_1,k_2,k_3,k_4] \cdot w_N^{l_1k_1+l_2k_2+(p_1-l_1)k_3+(p_2-l_2)k_4}$$

or,

$$Q_2[p_1,p_2] = \frac{1}{N^2} \sum_{l_1,l_2} X[l_1,l_2]X[p_1-l_1,p_2-l_2]H_2[l_1,l_2,p_1-l_1,p_2-l_2] \tag{3.52}$$

**Complexity of quadratic Volterra kernel for image input** In equation (3.51), for each $[n_1,n_2]$ the number of computations is proportional to $O(N^4) = O(N^2 \times N^2)$ and the total computations $\forall n_1,n_2 \in \big[[0,N-1],[0,N-1]\big]$ is $O(N^6) = O(N^4 \times N^2)$. Similarly, the IDFT of equation (3.52) involves the following computations, (i) FFT of input image $O(N^2 logN)$, (ii) FFT of the four dimensional kernel $O(N^4 logN)$, (iii) Pointwise multiplication $O(N^4)$ and (iv) IFFT $O(N^2 logN)$. Therefore, the time complexity of the IFFT of equation (3.52) is,

$$O(N^4 logN) = O(N^2 logN) + O(N^4 logN) + O(N^4) + O(N^2 logN)$$

In summary, the time complexity of equation (3.51) is $O(N^6)$ and the time complexity of the IDFT of equation (3.52) is $O(N^4 logN)$.

### 3.2.3 NLSI systems with three-dimensional input

When the LSI system in equation (3.11) has input $x[\boldsymbol{n}] \in \ell_2(\mathbb{Z}^3)$ and $\boldsymbol{n} \equiv (n_1,n_2,n_3) \in \mathbb{Z}^3$, i.e., a three-dimensional volumetric input like a brain MRI volume, then the shift invariant quadratic Volterra series in natural domain can be written as,

$$y[\boldsymbol{n}] = a_0 + a_1 \sum_{k_1,k_2,k_3} h[k_1,k_2,k_3]x[n_1-k_1,n_2-k_2,n_3-k_3]$$

$$+ a_2 \sum_{k_1,k_2,k_3} h[k_1,k_2,k_3]x[n_1-k_1,n_2-k_2,n_3-k_3] \sum_{k_1,k_2,k_3} h[k_1,k_2,k_3]x[n_1-k_1,n_2-k_2,n_3-k_3]$$

or,

$$y[\boldsymbol{n}] = a_0 + \sum_{k_1,k_2,k_3} a_1 h[k_1,k_2,k_3]x[n_1-k_1,n_2-k_2,n_3-k_3]$$
$$+ \sum_{k_1,\cdots,k_6} a_2 h[k_1,k_2,k_3]h[k_4,k_5,k_6]x[n_1-k_1,n_2-k_2,n_3-k_3]x[n_1-k_4,n_2-k_5,n_3-k_6]$$

or,

$$y[\boldsymbol{n}] = h_0 + \sum_{k_1,k_2,k_3} h_1[k_1,k_2,k_3]x[n_1-k_1,n_2-k_2,n_3-k_3]$$
$$+ \sum_{k_1,\cdots,k_6} h_2[k_1,\cdots,k_6]x[n_1-k_1,n_2-k_2,n_3-k_3]x[n_1-k_4,n_2-k_5,n_3-k_6]$$

where $h_0 \equiv a_0$ is bias, $h_1 \equiv a_1 h$ is a three-dimensional convolution filter and the quadratic filter $h_2 \equiv a_2 hh$ is a **six-dimensional** convolution filter. Therefore, the output of shift invariant quadratic Volterra kernel for 3D inputs is a six-dimensional diagonal convolution:

$$a_2 q^2[n_1,n_2,n_3] = \sum_{k_1,\cdots,k_6} h_2[k_1,\cdots,k_6]x[n_1-k_1,n_2-k_2,n_3-k_3]x[n_1-k_4,n_2-k_5,n_3-k_6] \tag{3.53}$$

and the DFT of above equation is

$$Q_2[p_1,p_2,p_3] = \frac{1}{N^3}\sum_{l_1=0}^{N-1}\sum_{l_2=0}^{N-1}\sum_{l_3=0}^{N-1} X[l_1,l_2,l_3]X[p_1-l_1,p_2-l_2,p_3-l_3]H[l_1,l_2,l_3,p_1-l_1,p_2-l_2,p_3-l_3] \tag{3.54}$$

**Complexity of quadratic Volterra kernel for volumetric (3D) input** In equation (3.53), for each $[n_1,n_2,n_3]$ the number of computations is proportional to $O(N^6) = O(N^3 \times N^3)$ and the total computations $\forall n_1,n_2 \in [[0,N-1],[0,N-1],[0,N-1]]$ is $O(N^9) = O(N^6 \times N^3)$. Similarly, the IDFT of equation (3.54) involves the following computations, (i) FFT of input volume $O(N^3 logN)$, (ii) FFT of the six dimensional kernel $O(N^6 logN)$, (iii) Pointwise multiplication $O(N^6)$ and (iv) IFFT $O(N^3 logN)$. Therefore, the time complexity of the IFFT of equation (3.54) is,

$$O(N^6 logN) = O(N^3 logN) + O(N^6 logN) + O(N^6) + O(N^3 logN)$$

In summary, the time complexity of equation (3.53) is $O(N^9)$ and the time complexity of the IDFT of equation (3.54) is $O(N^6 logN)$.

### 3.2.4 LSI kernel with ratio of polynomial nonlinearity

The I/O relation of a Wiener system with a linear kernel and a static nonlinearity having the form of a ratio of polynomials is

$$x[\boldsymbol{n}] \rightarrow \underset{\text{Dynamic LSI}}{h[\boldsymbol{n}]} \xrightarrow{q} \underset{\text{Static nonlinearity}}{\frac{b_0+b_1q+b_2q^2+\cdots+b_mq^m}{1+a_1q+a_2q^2+\cdots+a_mq^m}} \rightarrow y[\boldsymbol{n}]$$
$$\text{or, } y[\boldsymbol{n}] = \frac{b_0+b_1q[\boldsymbol{n}]+b_2q[\boldsymbol{n}]^2+\cdots+b_mq[\boldsymbol{n}]^m}{1+a_1q[\boldsymbol{n}]+a_2q[\boldsymbol{n}]^2+\cdots+a_mq[\boldsymbol{n}]^m} \tag{3.55}$$

where $q[\boldsymbol{n}] = \sum_{\boldsymbol{k}} h[\boldsymbol{n}-\boldsymbol{k}]x[\boldsymbol{k}]$ is the LSI kernel output and $1 + a_1q[\boldsymbol{n}] + a_2q[\boldsymbol{n}]^2 + \cdots + a_mq[\boldsymbol{n}]^m \neq 0$.

## 3.3 Application in inverse Problem in Microwave Radiometry

Retrieving vertical atmospheric temperature and humidity profiles from brightness temperature observations by a multichannel Microwave Radiometer (MWR) is a classical, yet persistently challenging inverse problem in atmospheric remote sensing. The antenna of the ground-based upward-looking MWR points at a constant $\measuredangle\theta$ and measures the incoming black-body microwave radiation as brightness temperatures $T_B$ in Kelvin (K) in two frequency bands, namely an eight-channel K-band $\sim$ (22-30 GHz) and a fourteen-channel V-band $\sim$ (51-59 GHz). The objective of the inverse model is to infer approximations of *in situ* atmospheric water vapor density and temperature profiles as a function of elevation $z$ meters above sea level from the observed $T_B$s in these two frequency bands.

**Physical radiative transfer** The measured $T_B$ at a specific microwave frequency is a line-of-sight integral of emission from all atmospheric layers, each modulated by altitude-dependent weighting functions that depend on local thermodynamic properties such as temperature, pressure and humidity. The determination of $T_B$ from a given *in situ* atmospheric profile is described by the forward model or the physical radiative transfer equation [159]:

$$T_B(\nu) \cong \int_0^\infty K(\nu, z)\, T(z)\, dz + 2.7\,\text{K}, \quad \nu \in \mathcal{V} \text{ at } \measuredangle\theta \tag{3.56}$$

where $K(\nu, z) = a_\nu(z)\, \exp\big(-\tau_\nu(0, z)\big)$ is the weighting function or *forward kernel*, $a_\nu$ is the absorption coefficient, $\tau_\nu(0, z) = \int_0^z a_\nu(z)\, dz$ is the zenith optical thickness of the atmospheric layer at the zenith angle $\measuredangle\theta$ for a particular channel frequency $\nu$ of MWR, and $\sim$ 2.7 K is extra cosmic radiation.

**Challenges in inversion and existing methods** The inverse of equation (3.56) is *ill-posed* in the sense of Hadamard, where one or more of the following conditions are violated: (i) the existence of a solution, (ii) the uniqueness of the solution, and (iii) the continuous dependence of the solution on the input data. This makes naive inversion unstable because even tiny perturbations in $T_B$ can propagate into large deviations in the retrieved profiles. Early work in atmospheric inversion stabilized solutions of the inverse problem using Phillips-Twomey (i.e., Tikhonov-type) regularization [160].

The methods of inverting $T_B$s to atmospheric profiles broadly fall into four broad families that are not strictly disjoint. (1) Physical or variational iterative methods like OEM [161] and one-dimensional variational (1D-Var) [162] preserve the physics of radiative transfer, and the inversion requires solving a forward model by a Gauss-Newton or Newton optimization [154, 156, 163, 164]. (2) Bayesian maximum-probability or statistical inversions where the retrieval is a maximum a posteriori (MAP) inference problem that fuses a prior or background state with a $T_B$ likelihood, and this Bayesian view is the theoretical basis of OEM and 1D-Var in practice [165–168]. (3) Empirical regression methods such as multiple (multilinear) linear regressions that map multichannel $T_B$ vectors (often after dimensionality reduction) to per-elevation values of a discrete profile [155, 169]. (4) Deep Learning (DL) inversions do not require explicit forward modeling and use data-driven learned mappings with classical feed-forward MLP neural networks in most cases to recent deeper neural variants that are trained to infer full profiles directly from $T_B$ [170–180]. Radiosonde climatologies and co-located reanalysis profiles are routinely used to construct the prior (background) state for OEM and 1D-Var retrievals, and they also serve as physically consistent reference targets when training regression or neural-network models. Table 3.2 lists some seminal papers that capture the historical development of retrieval methods.

Iterative OEM or 1D-Var retrievals remain robust because they preserve the physics of radiative transfer and yield posterior error estimates, but they depend on well-specified background covariances, smoothness constraints, and physical bounds to achieve stable convergence. With a sufficiently large training set, inversions by regression and deep neural networks do not require tailored domain knowledge, and sometimes outperform classical inversions. The universal approximation theorem [80] allows an MLP neural network with an arbitrary number of neurons in hidden layers to model highly nonlinear relationships between multichannel input vectors $T_B$ and atmospheric state variables without explicit knowledge of the forward model coefficients. This dense (black-box) neural representation also loses the physics of inversion, and the inverse mapping reduces to an abstract transformation between function spaces. In practice, black-box inversions overparameterize when the training data is limited or biased, unstable to small input perturbations, and have weaker native support for rigorous uncertainty quantification compared with OEM or 1D-Var. Here, we design a simple and resource-efficient system-theoretic inversion of the $T_B$ radiances to vertical atmospheric profiles using an expressive learnable LN cascade design that respects the canonical physical form of the inverse relation. The linear stage comprises a convolutional Volterra kernel in the natural-frequency domain to predict a latent profile, and the cascaded nonlinear trainable rational functions model nonlinearity at each elevation. This parametric novel system-theoretic inversion of the *ill-posed* inverse retrieval problem is mathematically expressive like the classical methods and is built on the convolutional neural framework for training with a minimal set of MERRA-2 true profiles.

Table 3.2: Some seminal papers on inversion of microwave radiances to atmospheric profiles

| No. | Seminal papers | Method Type | Instrument / Band | Key Contributions |
|---|---|---|---|---|
| 1 | Staelin et al. (1973) [159] | Physical Radiative Transfer | Early Nimbus 5 satellite MWR | First demonstration of temperature profiling using passive microwave radiometry |
| 2 | Twomey (1977) [160] | Analytical (Tikhonov Regularization) | Mathematical theory of inversion | Established mathematical framework for *ill-posed* inversion in remote sensing. |
| 3 | Wilheit (1990) [181] | Physical RT Inversion | Satellite MWR (183 GHz) | Developed algorithm for water vapor profile retrieval under clear and cloudy skies. |
| 4 | Solheim et al. (1998) [182] | Regression, OEM, Neural Network | Ground-based MWR (20–60 GHz) | Benchmark comparison of retrieval methods using both (K, V) band brightness temperatures |
| 5 | Rodgers (2000) [161] | Optimal Estimation Method (OEM) | Mathematical theory for radiometric inversion | Widely used Bayesian retrieval framework in atmospheric sounding |
| 6 | Buehler (2005) [183] | Statistical BT-transform | Satellite humidity sounders | Related microwave brightness temperatures to upper tropospheric humidity using regression |
| 7 | Rosenkranz et al. (2006) [184] | RT model validation | Multiple (K, V) band, line centers | Validated forward models for water vapor and oxygen absorption with error below 1 K |
| 8 | Güldner & Spänkuch (2001) [185] | Linear regression | 23.8 and 31.4 GHz (K-band) | Validated temperature and humidity retrievals vs. radiosondes; RMSE ≈ 0.6 K |
| 9 | Moreau et al. (2004) [162] | Variational retrieval (1D-Var) | Ground MWR under precipitation | Early use of constrained variational inversion in rain conditions |
| 10 | Hewison et al. (2007) [186] | OEM and NWP background (1D-Var) | Ground-based MWR | Blending radiative measurements with NWP background for robust inversion |
| 11 | Tan et al. (2011) [169] | Principal component analysis and stepwise regression with NN | 35-channel ground MWR (K, V band) | Dimensionality reduction with principal component analysis followed by neural-network regression |
| 12 | Yan et al. (2020) [172] | Deep neural network and backpropagation learning | RPG-HATPRO (K, V band) | Improved retrieval accuracy over traditional regression with a two hidden-layer neural network |
| 13 | Luo et al. (2023) [175] | Deep learning, GBM, RF (ensemble per layer) | RPG-HATPRO-G3 (K, V band) | Hybrid model per altitude empirically optimised for temperature and humidity retrieval |
| 14 | Walbröl et al. (2024) [177] | Hybrid deep learning | HATPRO and MiRAC-P high-frequency radiometer (K, V band and 183 GHz) | Synergistic approach improves Arctic profile accuracy and vertical resolution |
| 15 | Foth et al. (2024) [178] | Artificial neural network with rain error diagnostics | Microwave radiometer HATPRO (K, V band) | Identified and quantified DL model failures under rainy conditions; channel–angle tuning guidance |
| 16 | Jiang et al. (2025) [179] | Deep neural network and backpropagation learning | MWP967KV ground multichannel MWR (K, V band) | Atmospheric temperature and humidity profile retrieval across 31 stations in China |

### 3.3.1 Dataset: In-situ measures and derived true physical quantities

**Dataset 1** The MWR measures $T_b$s at 8 K-band (22-30 GHz) and 14 V-band ~ (51-59 GHz) microwave channels. We have $T_b$s observed at K-band and V-band for 91 successive clear sky days at 1-day temporal resolution in a tropical climate. The corresponding in-situ $RH$ RS, $T$ RS and $P$ RS profiles in (0.05-13.75 km) range at 100 m vertical resolution are used to compute the $\rho_v$ RS profiles. The MWR and RS measurements show significant variations in the water vapor present in the atmosphere. The RS measurements establish the ground truth and verify the inference of our inverse model.

**Dataset 2** An open source Radiosonde (RS) dataset is available through Integrated Global Radiosonde Archive (IGRA) [187] from more than 2800 globally distributed stations [188]. The RTTOV-gb forward model [189] simulates the corresponding V-band and K-band $T_b$s using the in-situ atmospheric $T$ RS and humidity RS profiles.

**Dataset 3** Our atmospheric $T$ and $\rho_v$ ground truth profiles $y[z]$ are sampled on a nonlinear scale at 24 elevations (0-15) km. They are reanalysis values of Modern-Era Retrospective Analysis for Research and Applications, Version 2 (MERRA-2) [190], provided by the Global Modeling and Assimilation Office (GMAO) at NASA Goddard Space Flight Center. The blackbody radiances $T_B$ in the K-band and V-band spectral channels are simulated by ground-based Radiative Transfer for the TIROS Operational Vertical Sounder (RTTOV-gb) forward model [189, 191] using a total of 70,127 all-weather conditions (clear sky, cloudy, rainy, etc.) time-concurrent profiles of $T$, $\rho_v$, and cloud liquid water. For any given location, the RTTOV-gb model takes vertical profiles of atmospheric $T$, $\rho_v$, and cloud liquid water specified on an arbitrary set of pressure levels and then simulates $T_B$ corresponding to ground-based upward-looking MWR. These simulated $T_B$s are more suitable for numerical modeling of the inverse system and are sanitized from any possible instrument bias and errors when compared to direct MWR observations. *

**Relative Humidity RH and water vapor density $\rho_v$** The datasets have vertical atmospheric Pressure, Temperature and RH profiles and we need to deduce the water vapor densities from and this requires the computation of all the associated atmospheric variables to prepare the training pairs containing microwave brightness temperatures and water vapor densities. Atmospheric air is a mixture of dry air and water vapor. Dry air behaves like an ideal gas in the temperature range ~ (263-323 K) or ( − 10 to 50 °C). Similarly, water vapor can be treated as an ideal gas with less than 0.2% error at pressures below 12.3 kPa which is the saturation pressure of water $P_{sat}$@323 K [192]. According to Dalton's law of additive pressure, the air pressure is $P = P_a + P_v$, where $P_a$ and $P_v$ are partial pressures of dry air and water vapor, respectively.

*Acknowledgement We would like to thank Shri Anil Kulkarni (Retired Scientist G) and Shri Abhishek Jha (Scientist B) from the *Society for Applied Microwave Electronics Engineering & Research, Ministry of Electronics & Information Technology, Government of India* for providing support with quality-controlled Datasets.

The ratio of the mass of water vapor $m_v$ to the maximum mass of water air can hold $m_g$ is called $RH$ (0-1 range) and is

$$\mathrm{RH} = \frac{m_v}{m_g} = \frac{P_v V / R_v T}{P_g V / R_v T} = \frac{P_v}{P_g}$$

or,

$$\mathrm{RH} = \frac{P - P_a}{P_g} = \frac{P - \frac{R_a T}{V/m_a}}{P_g} = \frac{P - \rho_a R_a T}{P_g} \tag{3.57}$$

where the density of dry air $\rho_a = m_a/V$ (kg/$m^3$), $R_a = 0.287$ kJ/kg K. The air tends to deviate from its ideal gas behavior for very low temperatures at higher altitudes, so the computational error is more there. For example, our $T$ RS is ~ (200-300K). In equation (3.57), the denominator $P_g = P_{sat}@T$ is computed accurately from $-100°C$ to $100°C$ [193] via *

$$P_g = \begin{cases} \dfrac{\exp\left(34.494 - \frac{4924.99}{T+237.1}\right)}{(T+105)^{1.57}}, & T > 0°C \\ \dfrac{\exp\left(43.494 - \frac{6545.8}{T+278}\right)}{(T+868)^{2}}, & T \leq 0°C. \end{cases} \tag{3.58}$$

Figure 3.4 (bottom) shows the calculated water vapor density, $\rho_v = P_v/R_v T$ profiles from the in-situ RS measurements. Thus, the available datasets have a direct measure of the temperature ground truth profiles and we derive the water vapor density ground truth profiles using equation (3.57). Figure 3.4 shows the ground truth temperatures and the computed water vapor densities of Dataset 2 using the same equation. For both the datasets half of the data is used for system identification/ training and the remaining half was equally divided for testing and validation of the inverse kernels.

### 3.3.2 A physics-guided learned right-inverse retrieval method

**Notation:** The channel index, look angle, and altitude are denoted by $\nu$, $\theta$, and $z$, respectively. The brightness temperature, downwelling component, and continuous atmospheric state are denoted by $T_b(\nu,\theta)$, $T_{DN}(\nu,\theta)$, and $T(z)$, respectively. The measurement vector is $\mathbf{x} \in \mathbb{R}^m$, the profile is $\boldsymbol{\tau} \in \mathbb{R}^n$, the forward operator is $\mathbf{K} \in \mathbb{R}^{m\times n}$, and the error is $\boldsymbol{\eta}$. The Moore-Penrose pseudoinverses of $\mathbf{K}$ and $\mathbf{H}$ are denoted by $\mathbf{K}^+$ and $\mathbf{H}^+$. The range and null spaces are denoted by Range$(\cdot)$ and $N(\cdot)$. The Ramanujan analysis operator is $\mathbf{A} \in \mathbb{R}^{\Phi(Q)\times m}$, with $S = \bigoplus_{q=1}^{Q} S_q$ and $\Phi(Q) = \dim S$. The lifted data and lifted noise are $\boldsymbol{\alpha}_\star = \mathbf{A}\mathbf{x}$ and $\varepsilon = \mathbf{A}\boldsymbol{\eta}$, and $\mathbf{H} = \mathbf{AK}$. The shared trunk is $g_\gamma$, the scalar heads are $h_{\mathbf{w}_z}$, and $\Theta = \{\gamma, \mathbf{w}_1, \ldots, \mathbf{w}_n\}$. The loss weights are $\lambda_{\mathrm{dc}}, \lambda_{\mathrm{reg}} \geq 0$, $R(\cdot)$ is a structural regularizer, and $\boldsymbol{\Sigma}_x$ and $\boldsymbol{\Sigma}_\alpha$ are noise covariances.

* In equation (3.58) T is in $°C$, whereas $T$ is in K everywhere else.

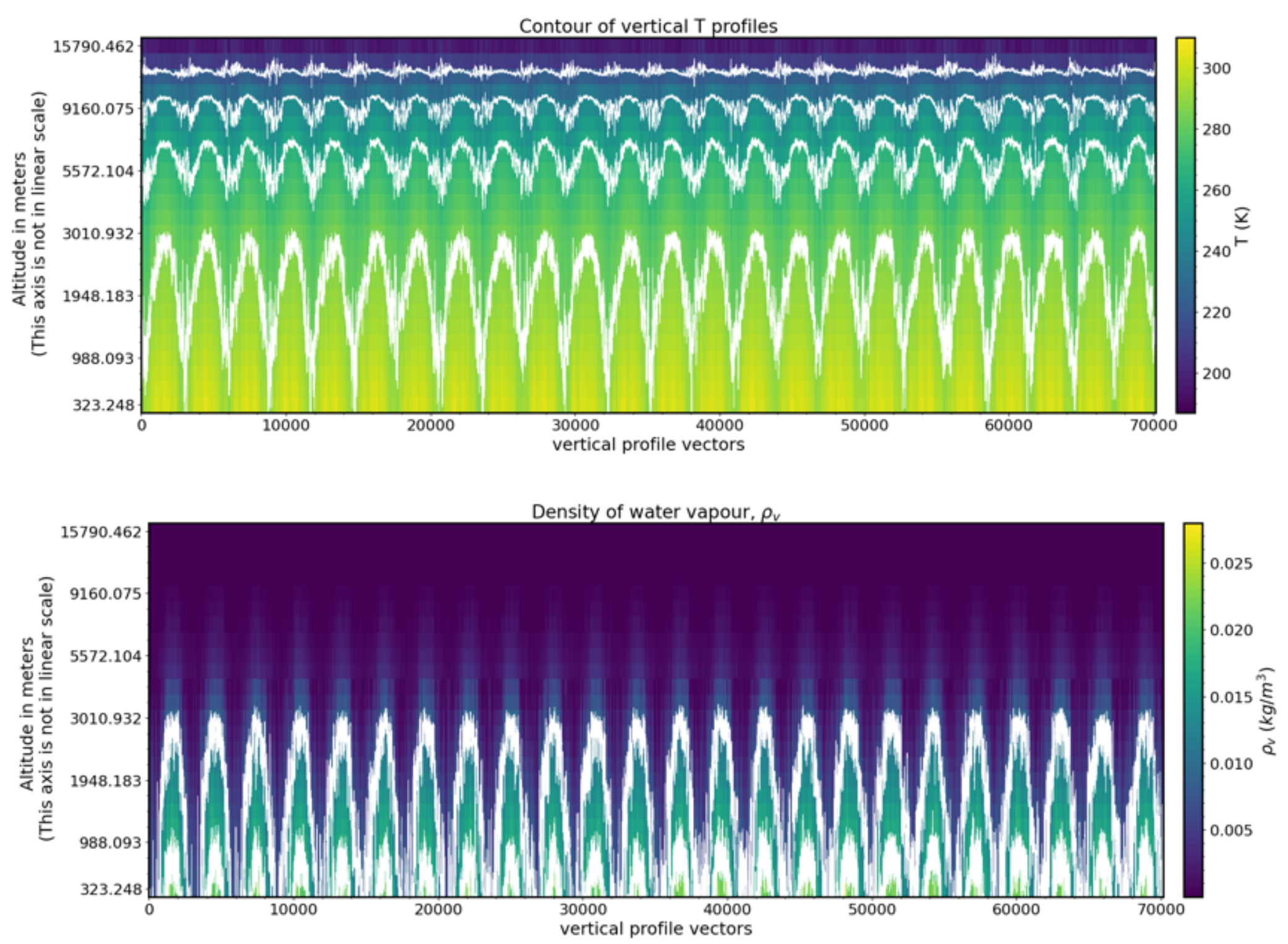

Figure 3.4: Contours of Temperature and water vapor density ground truth profiles from Dataset 2.

**Forward model:** Let $T(z)$ denote the atmospheric state versus altitude $z$ (e.g., temperature or humidity). Equation (3.56) is the forward model. Discretizing $m$ channels and $n$ layers gives

$$\mathbf{x} = \mathbf{K}\boldsymbol{\tau} + \boldsymbol{\eta}, \qquad \mathbf{K} \in \mathbb{R}^{m\times n},\ \mathbf{x} \in \mathbb{R}^m,\ \boldsymbol{\tau} \in \mathbb{R}^n, \quad m < n,$$

with measurement and forward model error $\boldsymbol{\eta}$. The operator $\mathbf{K}$ is smoothing (Fredholm of the first kind), so inversion is underdetermined and sensitive to noise.

**Linear geometry:** A minimum norm baseline is $\boldsymbol{\tau}_{\min} = \mathbf{K}^{+}\mathbf{x}$, where $(\cdot)^{+}$ is the Moore-Penrose pseudoinverse. The products $\mathbf{K}\mathbf{K}^{+}$ and $\mathbf{K}^{+}\mathbf{K}$ are the orthogonal projectors onto Range$(\mathbf{K})$ (column space) and Range$(\mathbf{K}^{\top})$ (row space), respectively. If $\mathbf{K}$ has full row rank then $\mathbf{K}\mathbf{K}^{+} = \mathbf{I}_m$, but we rely only on the projector identities in general. For any $\mathbf{x} \in$ Range$(\mathbf{K})$, the solution set $\{\boldsymbol{\tau} : \mathbf{K}\boldsymbol{\tau} = \mathbf{x}\}$ is the affine manifold $\mathfrak{M} = \boldsymbol{\xi}_0 + N(\mathbf{K})$.

**Structured analysis via a Ramanujan lift:** Ramanujan frames are described in Chapter 2. They are used here because the MWR K-band and V-band inputs are very short channel sequences. Unlike generic frames, Ramanujan frames are adapted to integer periods and can expose hidden period structure in such short sequences. To expose this structure, analyze measurements with a Ramanujan operator $\mathbf{A} : \mathbb{R}^m \to \mathbb{R}^{\Phi(Q)}$ and define

$$\boldsymbol{\alpha}_{\star} = \mathbf{A}\mathbf{x}, \qquad \mathbf{H} = \mathbf{A}\mathbf{K} \in \mathbb{R}^{\Phi(Q)\times n}, \qquad \boldsymbol{\alpha}_{\star} = \mathbf{H}\boldsymbol{\tau} + \boldsymbol{\varepsilon},\ \ \boldsymbol{\varepsilon} = \mathbf{A}\boldsymbol{\eta}. \tag{3.59}$$

The Ramanujan range decomposes as $S = \bigoplus_{q=1}^{Q} S_q$ (orthogonal integer period subspaces). In this lifted domain, the linear baseline would be $\boldsymbol{\tau} = \mathbf{H}^{+}\boldsymbol{\alpha}_{\star}$ and the solution set becomes $\mathfrak{M}_H = \xi_0 + N(\mathbf{H})$.

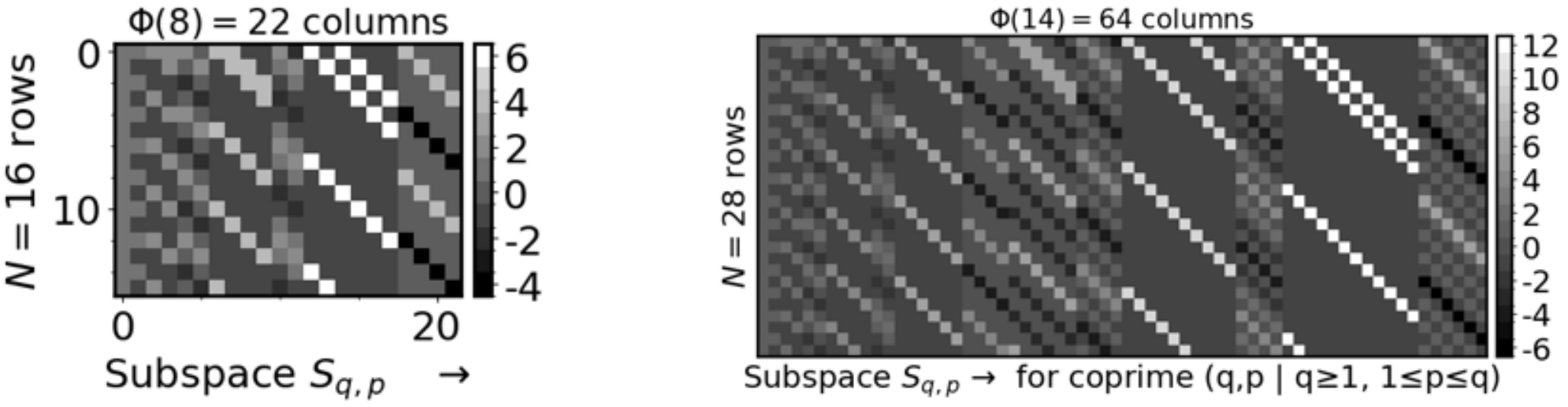

Figure 3.5: Ramanujan frames $\mathbf{A}^{\text{K-band}}$ (left) and $\mathbf{A}^{\text{V-band}}$ (right)

#### 3.3.2.1 `RamaNet`: retrieval with Ramanujan frame expansions and MLP neural network

Writing $\mathbf{H}^{+}$ by rows, $\mathbf{H}^{+} = \begin{bmatrix} \mathbf{h}_1^{+\top} \\ \cdots \\ \mathbf{h}_n^{+\top} \end{bmatrix}$, the linear readout is $\tau_z = \langle \mathbf{h}_z^{+}, \boldsymbol{\alpha}_{\star} \rangle$. `RamaNet` generalizes this with a shared *trunk* and scalar *heads* for each altitude as follows.

$$\mathbf{u} = g_{\gamma}(\boldsymbol{\alpha}_{\star}) \in \mathbb{R}^{d}, \qquad \hat{\tau}_z = h_{\mathbf{w}_z}(\mathbf{u}), \quad z = 1, \ldots, n,$$

leading to

$$\hat{\boldsymbol{\tau}} = f_{\Theta}(\boldsymbol{\alpha}_{\star}) = \big(h_{\mathbf{w}_1}(g_{\gamma}(\boldsymbol{\alpha}_{\star})), \ldots, h_{\mathbf{w}_n}(g_{\gamma}(\boldsymbol{\alpha}_{\star}))\big)^{\top}, \qquad \Theta = \{\gamma, \mathbf{w}_1, \ldots, \mathbf{w}_n\}.$$

If $g_{\gamma}$ is identity and each $h_{\mathbf{w}_z}$ is linear, the optimum recovers the row of $\mathbf{H}^{+}$ associated with altitude $z$. Thus `RamaNet` is a nonlinear, shared representation generalization of the pseudoinverse. The trunk learns features aligned with $\text{Range}(\mathbf{H}^{\top})$ (row space), while the heads act as learned rows mapping those features to layerwise estimates. Figure 3.6 shows representative K-band and V-band inputs and their projections in the Ramanujan subspace.

**Physics-guided training:** Given training pairs $(\mathbf{x}^{(i)}, \boldsymbol{\tau}^{(i)})$, form $\boldsymbol{\alpha}_{\star}^{(i)} = \mathbf{A}\mathbf{x}^{(i)}$ and use $\mathbf{H} = \mathbf{A}\mathbf{K}$. Here $\mathbf{A}$, $\mathbf{K}$, and hence $\mathbf{H}$ are fixed, and only $\Theta$ is learned. Train $g_{\gamma}$ and $\{h_{\mathbf{w}_z}\}$ by minimizing

$$\mathcal{L}(\Theta) = \sum_i \left\|\hat{\boldsymbol{\tau}}^{(i)} - \boldsymbol{\tau}^{(i)}\right\|_2^2 + \lambda_{\text{dc}} \sum_i \left\|\mathbf{H}\hat{\boldsymbol{\tau}}^{(i)} - \boldsymbol{\alpha}_{\star}^{(i)}\right\|_{\boldsymbol{\Sigma}_{\alpha}^{-1}}^2 + \lambda_{\text{reg}} \sum_i R(\hat{\boldsymbol{\tau}}^{(i)}),$$

where $\hat{\boldsymbol{\tau}}^{(i)} = f_{\Theta}(\boldsymbol{\alpha}_{\star}^{(i)})$, $\|\mathbf{u}\|_{\boldsymbol{\Sigma}_{\alpha}^{-1}}^2 = \mathbf{u}^{\top}\boldsymbol{\Sigma}_{\alpha}^{-1}\mathbf{u}$ is a Mahalanobis norm with $\boldsymbol{\Sigma}_{\alpha} = \mathbf{A}\,\boldsymbol{\Sigma}_x\,\mathbf{A}^{\top}$ (from measurement covariance $\boldsymbol{\Sigma}_x$), and $\lambda_{\text{dc}}, \lambda_{\text{reg}} \geq 0$. Since $\mathbf{H}\hat{\boldsymbol{\tau}}^{(i)} - \boldsymbol{\alpha}_{\star}^{(i)} = \mathbf{A}(\mathbf{K}\hat{\boldsymbol{\tau}}^{(i)} - \mathbf{x}^{(i)})$, the consistency term is the forward residual measured after the fixed Ramanujan analysis. It measures the distance of $\hat{\boldsymbol{\tau}}^{(i)}$ to the affine set $\mathfrak{M}_H^{(i)} = \{\boldsymbol{\tau} : \mathbf{H}\boldsymbol{\tau} = \boldsymbol{\alpha}_{\star}^{(i)}\}$. Equivalently, $\mathbf{H}\mathbf{H}^{+}$ and $\mathbf{H}^{+}\mathbf{H}$ are orthogonal projectors onto $\text{Range}(\mathbf{H})$ and $\text{Range}(\mathbf{H}^{\top})$. The regularizer $R(\cdot)$ encodes

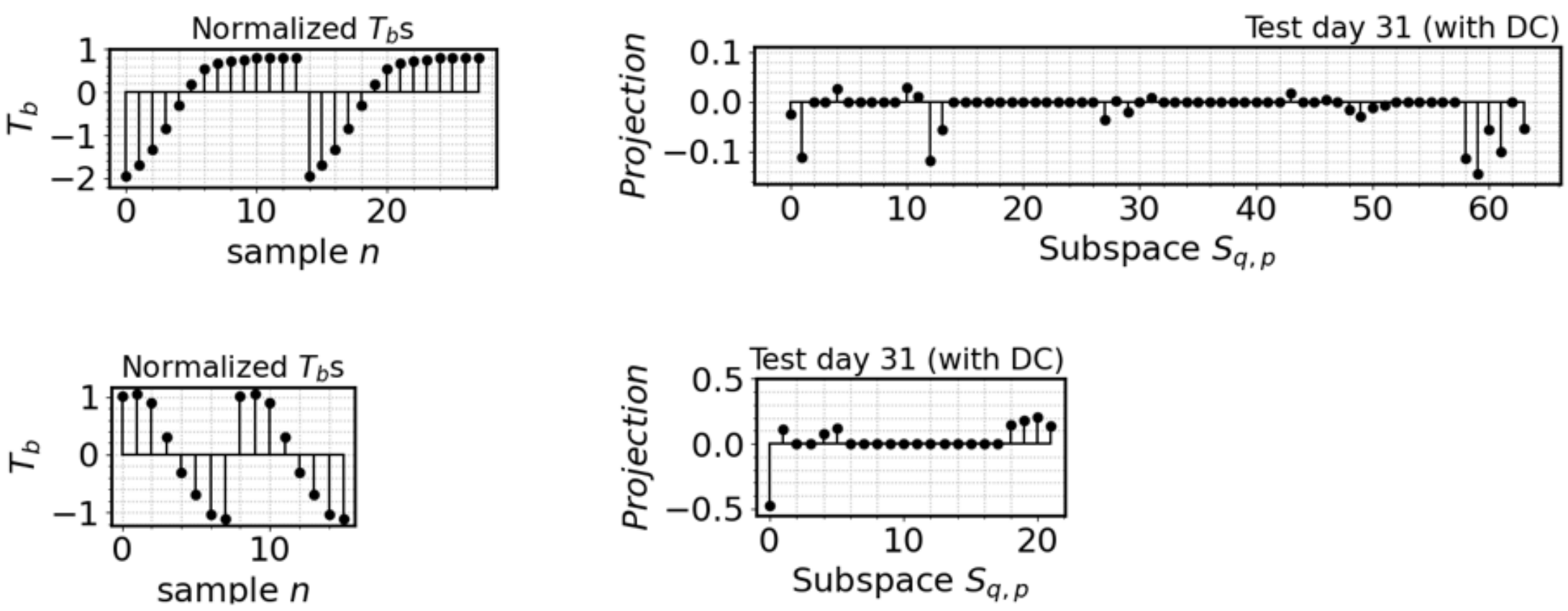

Figure 3.6: An input V-band (top left) and its projection $\alpha_\star^{\text{V-band}}$ (top right). An input K-band (bottom left) and its projection $\alpha_\star^{\text{K-band}}$ (bottom right).

atmospheric structure and stability (bounds or nonnegativity, smoothness or total variation across altitude, optional Lipschitz control on heads, and weak coupling across adjacent layers where warranted). When monotonicity is not universal (e.g., inversions), enforce it softly rather than as a hard constraint.

**Merit relative to a typical MLP neural network that maps $\mathbf{x} \mapsto \boldsymbol{\tau}$:** A monolithic MLP neural network must learn both representation and physics from $\mathbf{x}$. `RamaNet` concentrates spectral/periodic content with $\mathbf{A}$, shares computation across layers via $g_\gamma$, specializes per altitude with $h_{\mathbf{w}_z}$, and anchors estimates to the forward model through the consistency term with $\mathbf{H}$. The design contains the classical pseudoinverse as a limit case yet permits principled nonlinear corrections, improving stability and data efficiency. Figure 3.7 illustrates the overall RamaNet architecture used for profile retrieval.

#### 3.3.2.2 Evaluation of `RamaNet` with Dataset 1

The inverse kernels are evaluated for retrieval up to ~ 10 km with the test sets using five metrics. These are (i) $R^2$-score (0-1) range, (ii) Mean Absolute Error (MAE), (iii) Root Mean Squared Error (RMSE), (iv) Normalized Symmetric Mean Absolute Percentage Error (NSMAPE) in $(0\text{-}100\ \%)$ range, and (v) Mean Bias Deviation (MBD). Figure 3.8 illustrates the denormalized predictions by $h_{\mathbf{w}_4}^{(\text{K-band})}$ and $h_{\mathbf{w}_4}^{(\text{V-band})}$ trained to retrieve $\rho_v@4$ and $T_4$ for $z = 4$ representing 350 m altitude. Plugging the retrieved $\hat{\rho}_v@z$ and $\hat{T}_z$ into equation (3.57) yields $\hat{RH}@z$ at a specific altitude $z$ and $\left(\hat{RH}@z\right)_{\forall z}$ is the complete profile. Figure 3.10 illustrates $\hat{\rho}_v$, $\hat{T}$ and $\hat{RH}$ vertical profiles retrieved by a test K-band and V-band input. Due to very little water vapor content, i.e., $\rho_v$, $P_g \cong 0$ in the upper troposphere, retrieving the RH profile beyond ~ 10 km becomes very sensitive to minor perturbations in $\rho_v$ because of $\lim_{\delta_1,\delta_2\to 0}(\delta_1/\delta_2)$ scenario in $RH = \rho_v R_v T/P_g$.

In retrieving the test $\rho_v$ profiles the $R^2$ scores are between ~ $(0.46\text{-}0.82)$, RMSE is $< 1\,\text{g}/m^3$,

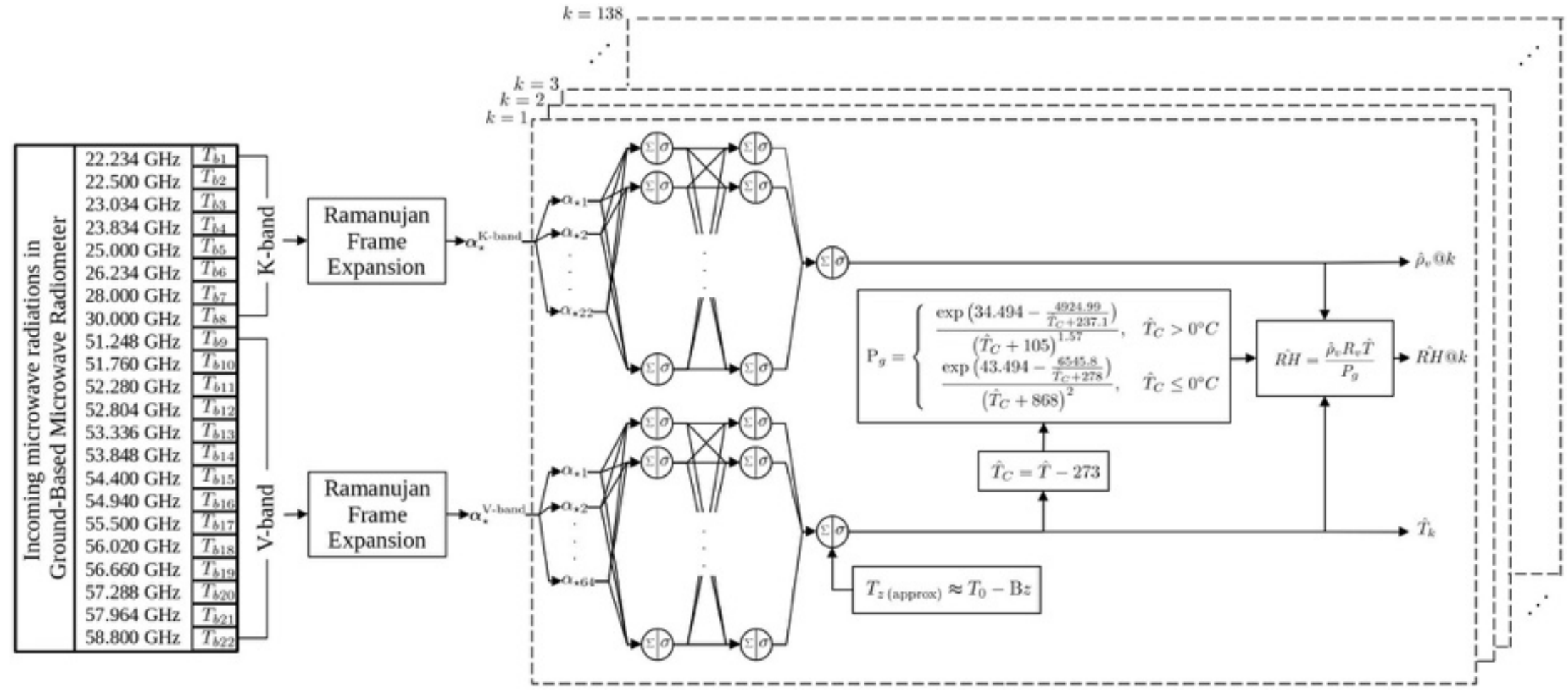

Figure 3.7: `RamaNet` as a Ramanujan frame constrained physics-aware trainable inverse model for retrieving temperature $\hat{T}$, water vapor density $\hat{\rho}_v$, or relative humidity R$\hat{\text{H}}$ profiles over altitudes $z = 1$ to $z = 138$ from multichannel MWR brightness-temperature observations.

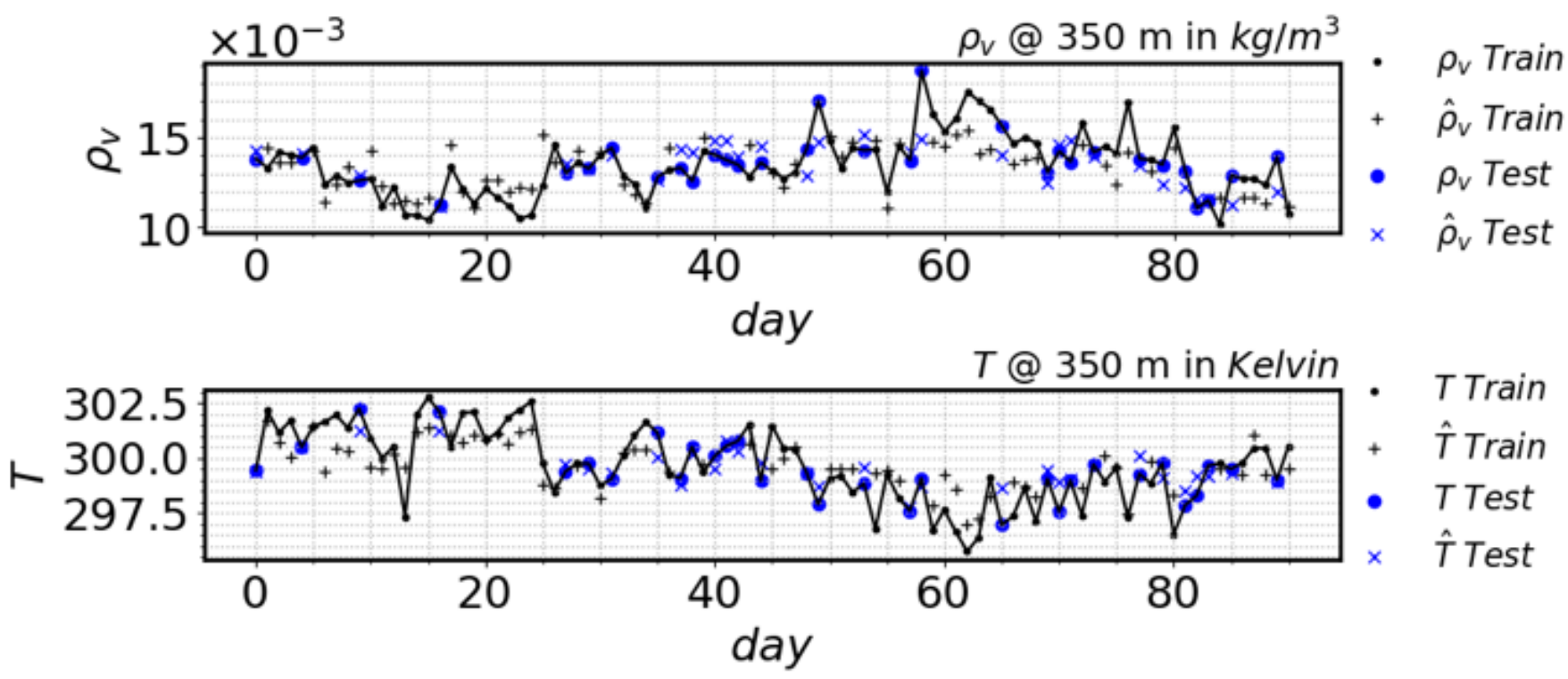

Figure 3.8: Water vapor density and temperature prediction (denormalized) by neural networks $h_{\mathbf{w}_4}^{(\text{K-band})}$ (top) and $h_{\mathbf{w}_4}^{(\text{V-band})}$ (bottom) with train and test sets at altitude 350 m or $z = 4$ and RS ground truth

MAE is $< 0.70$ g/$m^3$ and NSMAPE $\sim$ (2-18 %). For test $\hat{T}$ profiles, the $R^2$ score is between $\sim$ (0.89-0.98), RMSE is $\sim$ (0.72-1.11 K), MAE is $\sim$ (0.57-0.86 K) and NSMAPE is $< 0.17\%$ throughout as listed in Table 3.3 including the computed $R\hat{H}$ profiles for every non-overlapping 2 km window up to $\sim$ 10 km. The same is illustrated in Figure 3.9 with correlation plots of $\rho_v$ and $\hat{T}$ with their RS ground truths. Some results from the state-of-the-art T and RH retrieval algorithms reported in recent literature [172, 174–176] for similar clear-sky data are also listed in Table 3.3, that uses backpropagation neural networks. These models differ in vertical resolution, network architecture with (30-256) hidden layer neurons, the choice of activation functions, the presence or absence of batch normalization, dropout layers and training procedure. Due to very little $\hat{\rho}_v$ in the upper troposphere, the $R\hat{H}$ retrieval error at higher altitudes increases, visible in the rightmost MAE and RMSE plot of Figure 3.11. It also shows the MBD of the retrieved test $\rho_v$ profiles is $< \pm 0.26$ g/$m^3$. Similarly, the MBD of the test $\hat{T}$ profiles at $\sim$ (4-8 km) is $< \pm 0.6$ K

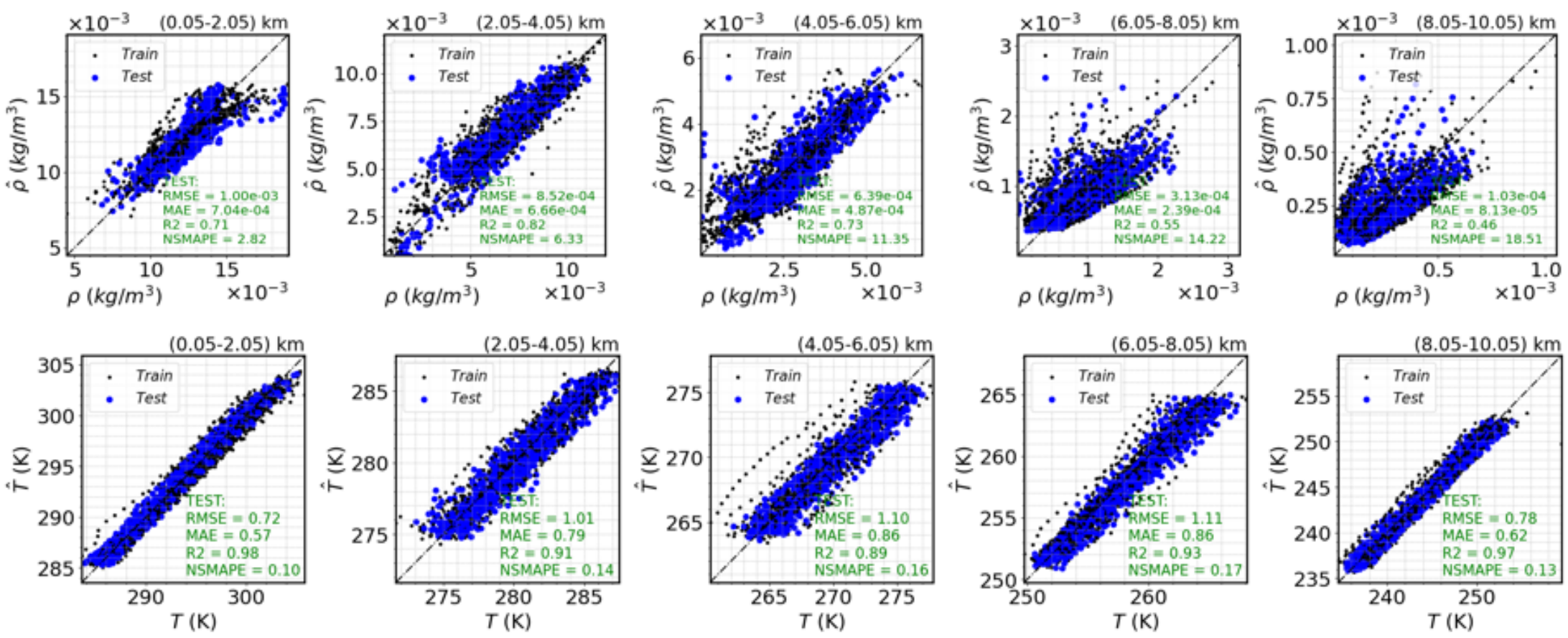

Figure 3.9: Altitude-wise correlation plot of retrieved profiles by `RamaNet` of water vapor density $\hat{\rho}$ in kg/$m^3$ (top row) from K-bands and temperature $\hat{T}$ in Kelvin (bottom row) from V-bands, with their RS ground truths for every 2 km non-overlapping window up to ∼ 10 km in troposphere

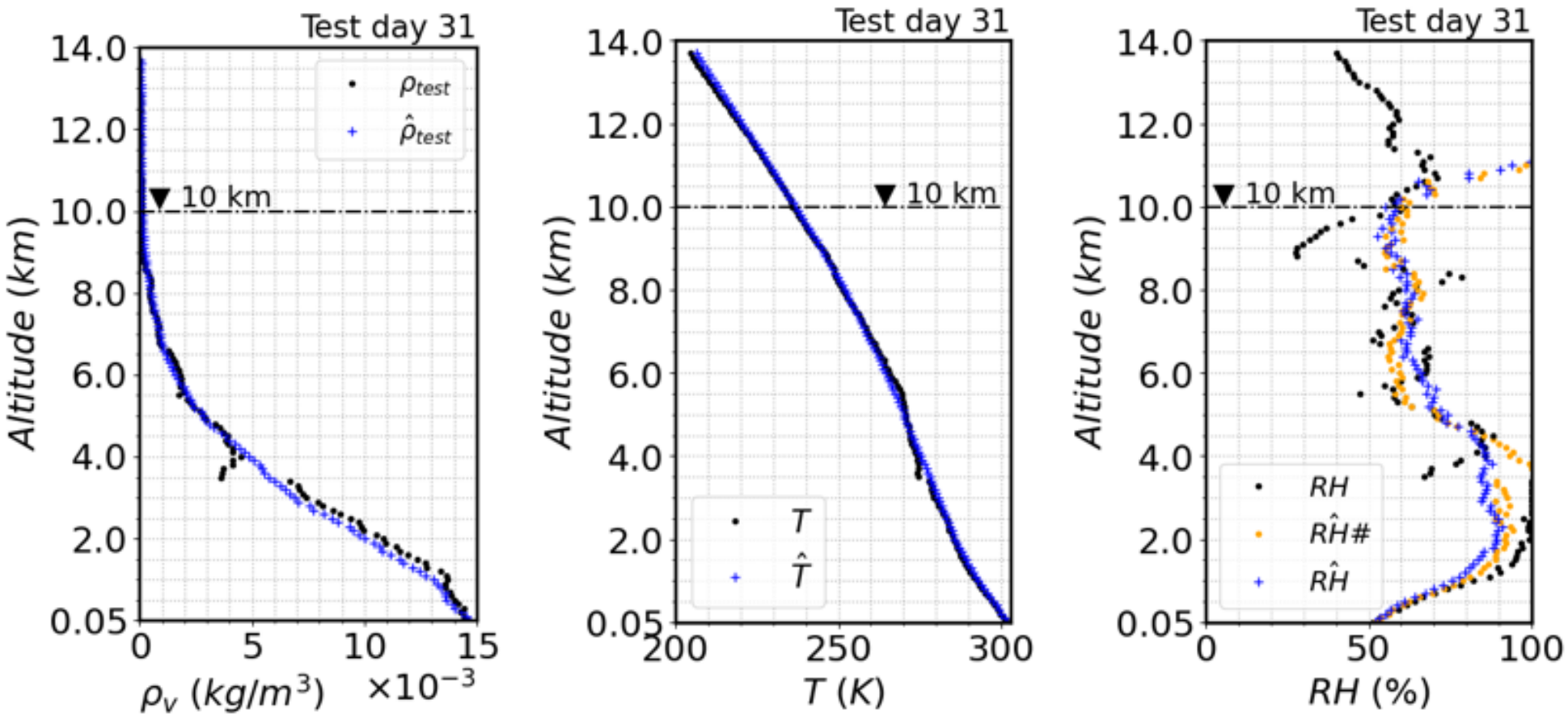

Figure 3.10: A retrieved water vapor density $\hat{\rho}_v$ (left) and temperature $\hat{T}$ (middle) profile by `RamaNet`; computed $\hat{\text{RH}}$ (right) profile with $\hat{T}$ and $\hat{\rho}_v$; $\hat{\text{RH}}$# computed using $\hat{\rho}_v$ and $T$ RS

and < ±0.25 K elsewhere.

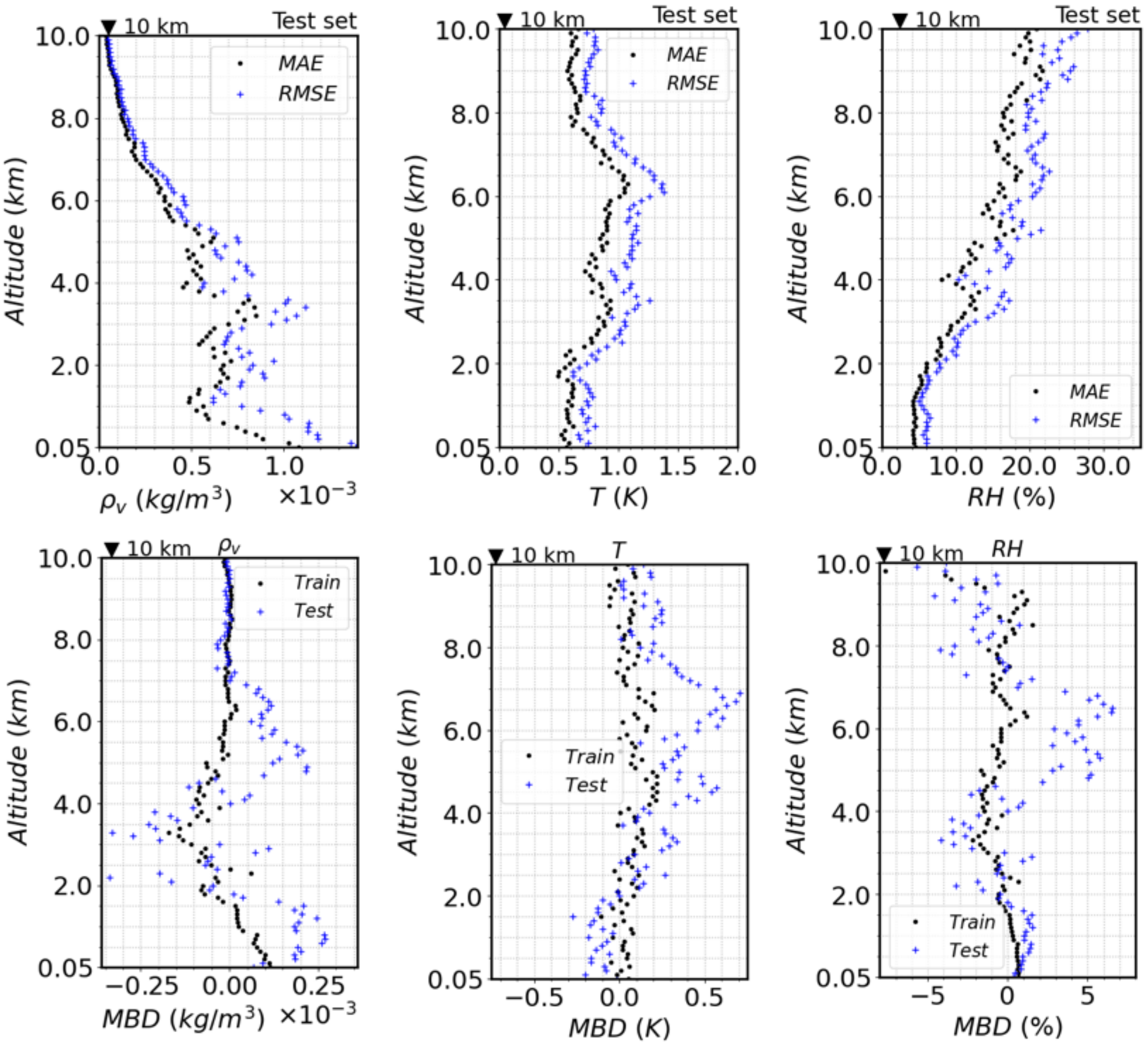

Figure 3.11: Errors in retrieving water vapor density $\hat{\rho}_v$ (left), temperature $\hat{T}$ (middle) and Relative Humidity $\hat{\text{RH}}$ (right) of the test set by `RamaNet`

Table 3.3: State-of-the-art retrieval algorithms and our RamaNet inverse model results

| **Atmospheric variable** | **Vertical range (km)** | $R^2$ | **RMSE** | **MAE** | **NSMAPE (%)** |
|---|---|---|---|---|---|
| **State-of-the-art retrieval algorithms in recent literature** | | | | | |
| Yan et al. (2020) [#yan2020deep] | | | | | |
| $\hat{\rho_v}$ | 0.50-10.00 | 0.90 | 0.26e-3 | 0.16e-3 | - |
| $\hat{T}$ | | 0.99 | 1.70 | 1.29 | - |
| $R\hat{H}$ | | 0.44 | 11.72 | 0.07 | - |
| Yu et al. (2021) [#yu2021bp] | | | | | |
| $\hat{T}$ | 0.00-1.00 | 0.98 | 1.12 | 0.81 | - |
| | 1.00-10.00 | 0.91 | 2.03 | 1.55 | - |
| $R\hat{H}$ | 0.00-1.00 | 0.51 | 12.84 | 9.07 | - |
| | 1.00-10.00 | 0.46 | 15.10 | 12.22 | - |
| Renju et al. (2023) [#renju2023retrieval] | | | | | |
| $\hat{\rho_v}$ | | 0.95 | 1.2e-3 | - | - |
| $\hat{T}$ | | 0.98 | 1.78 | - | - |
| $R\hat{H}$ | | 0.63 | 13.88 | - | - |
| Luo et al. (2023) [#luo2023machine] | | | | | |
| $\hat{T}$ | 0.00-10.00 | 0.98 | 2.32 | - | - |
| $R\hat{H}$ | | 0.53 | 17.96 | - | - |
| **Retrieving atmospheric profiles using RamaNet** | | | | | |
| $\hat{\rho_v}$ (kg/m$^3$) | 0.05-2.05 | 0.71 | 1e-3 | 0.70e-3 | 2.82 |
| | 2.05-4.05 | 0.82 | 0.85e-3 | 0.67e-3 | 6.33 |
| | 4.05-6.05 | 0.73 | 0.64e-3 | 0.49e-3 | 11.35 |
| | 6.05-8.05 | 0.55 | 0.31e-3 | 0.24e-3 | 14.22 |
| | 8.05-10.05 | 0.46 | 0.10e-3 | 0.08e-3 | 18.51 |
| $\hat{T}$ (K) | 0.05-2.05 | 0.98 | 0.72 | 0.57 | 0.10 |
| | 2.05-4.05 | 0.91 | 1.01 | 0.79 | 0.14 |
| | 4.05-6.05 | 0.89 | 1.10 | 0.86 | 0.16 |
| | 6.05-8.05 | 0.93 | 1.11 | 0.86 | 0.17 |
| | 8.05-10.05 | 0.97 | 0.78 | 0.62 | 0.13 |
| $R\hat{H}$ (%) | 0.05-2.05 | 0.83 | 6.14 | 4.72 | 3.34 |
| | 2.05-4.05 | 0.55 | 5.46 | 4.08 | 6.52 |
| | 4.05-6.05 | 0.41 | 3.67 | 2.86 | 11.52 |
| | 6.05-8.05 | 0.11 | 1.62 | 1.31 | 14.39 |
| | 8.05-10.05 | -0.12 | 0.52 | 0.43 | 18.45 |

All in SI units: $\rho$ in kg/m$^3$, $T$ in K and RH in %. In both [172] and [174] vertical resolution and 250 m beyond 1 km up to 10 km. The standard literature shows $\rho$ in g/cm$^3$. The reported numbers across papers reflect different datasets, preprocessing, and evaluation criteria; hence, the comparison is intended to be indicative, not strictly comparable.

### 3.3.3 Multiresolution Volterra Kernel and Inverse modeling

In discrete form, with a finite set of frequency channels $\nu$ and a finite set of elevations $z$, the forward and inverse maps are

$$T_B[\nu] = \sum_{z\in\mathcal{Z}} K[\nu, z]\, T[z] \;+\; 2.7\,\mathrm{K}, \quad \nu \in \mathcal{V} \text{ (Forward)} \tag{3.60a}$$

$$\hat{T}[z] = \sum_{\nu\in\mathcal{V}} K^{+}[z, \nu]\, T_B[\nu], \qquad z \in \mathcal{Z} \text{ (Inverse)} \tag{3.60b}$$

which states that the measured brightness temperatures $T_B[\nu]$ arise from a linear weighting of the *in situ* atmospheric state $T[z]$ through the known forward kernel $K[\nu, z]$, plus a known offset (background cosmic radiation). This forward map is well-defined and numerically stable, and given a temperature profile $T[z]$, one can predict $T_B[\nu]$. For inversion, introducing an explicit *inverse kernel* $K^{+}[z, \nu]$ which maps the observed channels to an estimated atmospheric profile, we have the *ill-posed* inverse relation in (3.60b). In practice, the number of elevation levels $z$ in $T[z]$ is typically much larger than the number of frequency channels $\nu$ in $T_B[\nu]$. Therefore, recovering $T[z]$ from $T_B[\nu]$ requires solving an *ill-posed* and *underdetermined* inverse problem, where different profiles $T[z]$ can essentially produce the same measurements $T_B[\nu]$. When $K^{+}$ is chosen as the Moore-Penrose pseudoinverse of $K$, $\hat{T}[z]$ is the minimum-norm profile consistent with the measured channels. In general, we do not require $K^{+}K = I$ over all $z$ because, for an underdetermined system, this identity cannot hold globally. Instead, $K^{+}$ is designed (or learned) to enforce stability, physical plausibility, and parsimony.

For simplicity of notation, rewriting the inverse relation in equation (3.60b) using generic symbols with $x \equiv T_B$ denoting the measured multi-frequency channels, $y \equiv T$ denoting the vertical atmospheric profile we wish to retrieve, and $h^{+} \equiv K^{+}$ denoting the inverse kernel. Then the reconstruction is

$$\boxed{\hat{y}[z] = \sum_{\nu\in\mathcal{V}} h^{+}[z, \nu]\, x[\nu], \quad z \in \mathcal{Z}} \quad \text{(Natural)} \tag{3.61}$$

which can be realized as a lightweight convolutional network with one learnable kernel $h^{+}$. The canonical structure of equation (3.61) matches the input-output (I/O) relationship of a discrete Volterra kernel of first-degree ($m$=1) in the natural domain. The estimate at level $z$ is formed by summing the inputs $x[\nu]$ against a kernel $h^{+}[z, \nu]$ that depends on both the output index $z$ and the input index $\nu$. Since $h^{+}$ varies with $z$, the above inverse mapping is linear but **shift-variant**. At each $z$, the atmospheric layer is reconstructed as a weighted combination of the observed channels $x[\nu]$, and each pixel in $h^{+}[z, \nu]$ can be associated with a physically meaningful altitude rather than as an opaque free parameter.

#### 3.3.3.1 Linear Volterra kernel in equivalent biorthogonal bases

Inversion is further stabilized by expanding the pseudoinverse kernel $h^{+}$, the input radiance vector $x[\nu]$, and the output profile $y[z]$ on biorthogonal bases. Let $\{\phi_l, \tilde{\phi}_l\}_{l\in\mathcal{J}}$ and $\{\phi_u, \tilde{\phi}_u\}_{u\in\mathcal{U}}$ be biorthogonal bases satisfying $\langle \phi_l, \tilde{\phi}_{l'} \rangle = \delta_{ll'}$ and $\langle \phi_u, \tilde{\phi}_{u'} \rangle = \delta_{uu'}$, where $\mathcal{J}$ and $\mathcal{U}$ label the biorthogonal basis modes for the input microwave frequencies $\mathcal{V} = \{\nu_1, \dots, \nu_N\}$ and the output profile sampled in $\mathcal{Z} = \{z_1, \dots, z_M\} \subset [0, \infty)$ in practice. The biorthogonal basis decomposition and reconstruction of the input $x$ are

$$\alpha[l] = \sum_{\nu \in \mathcal{V}} x[\nu] \phi_l[\nu], \quad \forall\, l \in \mathcal{J} \tag{3.62a}$$

$$x[\nu] = \sum_{l \in \mathcal{J}} \alpha[l] \tilde{\phi}_l[\nu], \quad \forall\, \nu \in \mathcal{V} \tag{3.62b}$$

where $\alpha$ is the biorthogonal basis decomposition of $x$. Similarly, the decomposition and reconstruction of the output $y$ are

$$\beta[u] = \sum_{z \in \mathcal{Z}} \hat{y}[z] \phi_u[z], \quad u \in \mathcal{U} \tag{3.63a}$$

$$\hat{y}[z] = \sum_{u \in \mathcal{U}} \beta[u] \tilde{\phi}_u[z], \quad z \in \mathcal{Z} \tag{3.63b}$$

where $\beta$ is the biorthogonal basis decomposition of $y$. Substituting (3.62b) into (3.61) gives

$$\hat{y}[z] = \sum_{l \in \mathcal{J}} \alpha[l] \sum_{\nu \in \mathcal{V}} h^+[z, \nu] \tilde{\phi}_l[\nu], \quad z \in \mathcal{Z} \tag{3.64}$$

and substituting (3.64) into (3.63a) gives

$$\beta[u] = \sum_{l} \alpha[l] \sum_{z} \sum_{\nu} h^+[z, \nu] \tilde{\phi}_l[\nu] \phi_u[z]$$

$$\text{or} \quad \boxed{\beta[u] = \sum_{l \in \mathcal{J}} \alpha[l] H[u, l], \quad u \in \mathcal{U}} \quad \text{(Biorthogonal)} \tag{3.65}$$

Equation (3.65) is an *equivalent alternate representation* of equation (3.61) using biorthogonal basis expansions of the input, output, and kernel. The transform domain Volterra kernel is,

$$H[u, l] = \sum_{z} \sum_{\nu} h^+[z, \nu] \tilde{\phi}_l[\nu] \phi_u[z] \tag{3.66a}$$

$$H[u, l] = \mathrm{DWT}_z(\mathrm{DWT}_\nu(h^+[z, \nu])), \quad \text{(with DWT)} \tag{3.66b}$$

and with orthogonal or biorthogonal wavelets and the DWT, equation (3.66a) condenses to (3.66b). Figure 3.12 (left to right) illustrates this axis-wise transformation of the inverse kernel from natural to biorthogonal bases, where $\tilde{h}^+[z, l] = \mathrm{DWT}_\nu(h^+[z, \nu])$ is after row $\nu$-wise DWT and then $H[u, l] = \mathrm{DWT}_z(\tilde{h}^+[z, l])$ is after column $z$-wise DWT. This equivalent representation in biorthogonal bases enforces sparsity, offering a smaller number of effective degrees of freedom in the inverse kernel and encourages smooth, physically reasonable profiles with a *spectrally constrained* $h^+$ instead of a large unconstrained matrix. Additional details on the theory of Volterra kernels in biorthogonal bases are available in our published book chapter [158].

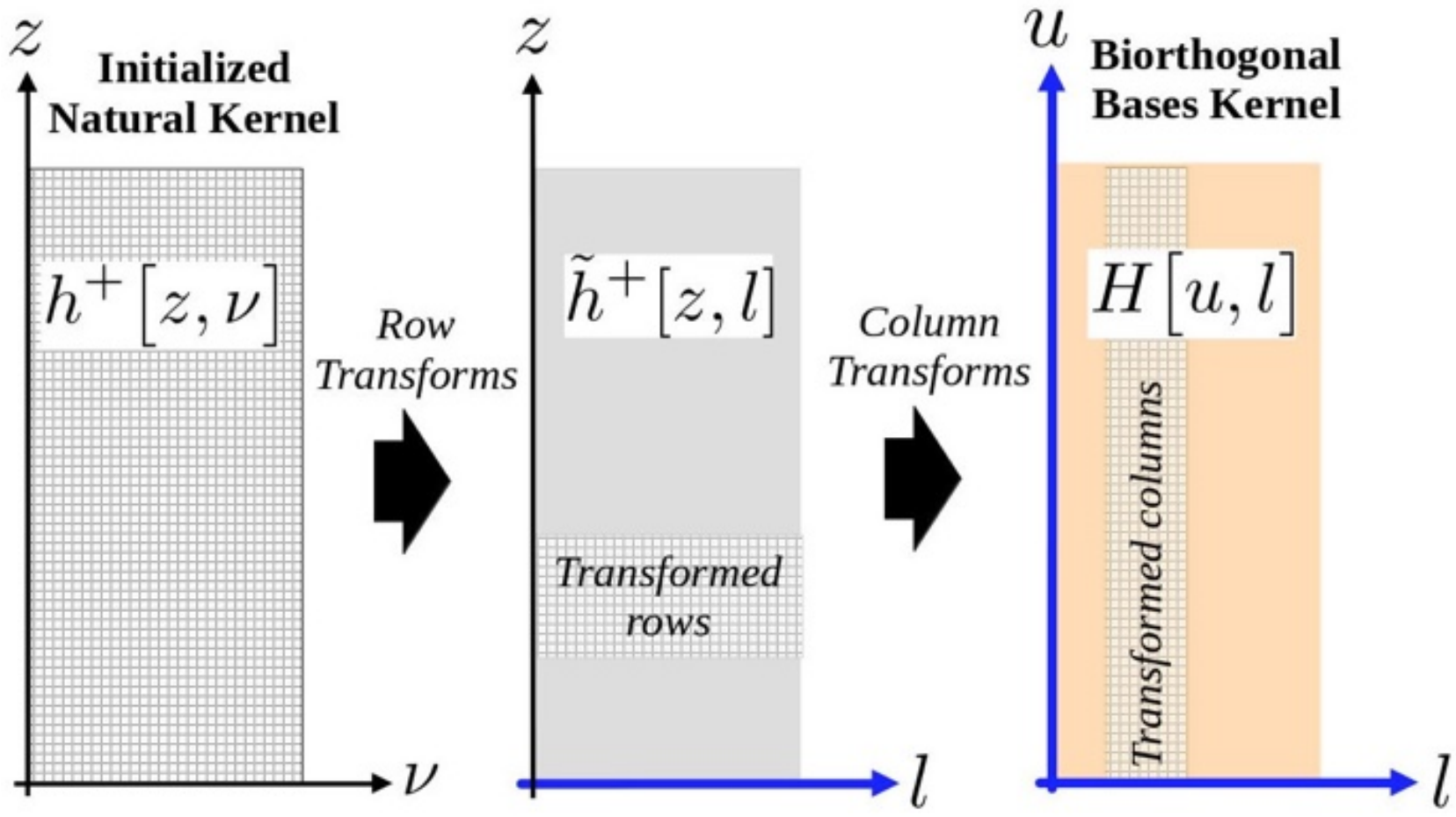

Figure 3.12: Initialization of the linear multiresolution Volterra kernel

#### 3.3.3.2 Rational functions & per elevation nonlinearity modeling

A purely linear reconstruction is insufficient for highly nonlinear atmospheric variables, such as $\rho_v$. Therefore, after the linear estimate $\hat{y}[z]$ by a Volterra kernel, a ratio of finite-order polynomials with learnable coefficients is introduced in cascade, yielding the following simple yet powerful learnable LN cascade (i.e., Wiener system):

$$x[\nu] \to h^{+}[z,\nu] \xrightarrow{\hat{y}[z]} \frac{\sum_{m=0}^{M} a_m[z]\,(\hat{y}[z])^m}{1+\sum_{n=1}^{N} b_n[z]\,(\hat{y}[z])^n} \to \tilde{\hat{y}}[z]$$

$$\texttt{InVeRt M/N}\text{ model, } \tilde{\hat{y}}[z] = \frac{\sum_{m=0}^{M=3} a_m[z]\,(\hat{y}[z])^m}{1+\sum_{n=1}^{N=3} b_n[z]\,(\hat{y}[z])^n} \tag{3.67}$$

where $\tilde{\hat{y}}$ is the retrieved atmospheric profile. Equation (3.67) describes our **In**verse model with a **V**olterra k**e**rnel and **R**ational func**t**ions (`InVeRt M/N`), where $M$ and $N$ are the highest order of numerator and denominator polynomials. The linear Volterra kernel first produces a latent profile $\hat{y}[z]$ and then the *rational* polynomial cascade maps $\hat{y}[z] \mapsto \tilde{\hat{y}}[z]$ per-elevation. This rational functional form can model nonlinearities of higher degree with numerator and denominator polynomials of lower degree [114]. Our `InVeRt M/N` models for retrieving $T$ and $\rho_v$ profiles have $M = N = 2$ or 3, i.e., `InVeRt 2/2` or `InVeRt 3/3`. For numerical stability, the denominator coefficient $b_0 = 1$ is fixed and all remaining $a_m[z], b_n[z]$ are learned under bounded constraints relative to this term. The denominator is also bounded below by enforcing $|1 + \sum_n b_n[z]\hat{y}[z]^n| \geq \varepsilon$, where $\varepsilon > 0$ is a very small number.

#### 3.3.3.3 Learning the Volterra kernel and rational nonlinear heads

In the natural domain, the convolutional (trainable) Volterra kernel is $h^{+}[z, \nu]$, whose I/O relationship is (3.61). Its equivalent I/O relationship using biorthogonal bases is (3.65). With DWT, the natural-frequency domain (or multiresolution) transform of the Volterra kernel $h^{+} \mapsto H$ is (3.66b). Having a training set with Radiosonde or MERRA-2 ground truth profiles $y[z]$, the kernel $h^{+}$ or $H$ is learned in natural or wavelet domain by following **four empirical steps**. (1) A forward pass with coefficients of $h^{+}$ initialized using a normal distribution, (2) a loss calculation, (3) a backpropagation to compute the gradients, and finally, (4) updating the coefficients of $h^{+}$ or $H$ with gradient descent.

The inverse model (`InVeRt M/N`) in (3.67) can be trained in the natural or natural-frequency domains using Adam [37] by minimizing the corresponding domain losses:

$$\mathcal{L}_{\text{InVeRt}} = \mathbb{E}_{(x,y)\sim\mathcal{D}}\left[\frac{1}{N_z}\sum_{z\in\mathcal{Z}}\left(y[z] - \tilde{\hat{y}}[z]\right)^2\right] + \lambda_1\sum_{u}\left|\hat{\beta}[u]\right| + \lambda_2\sum_{u}\left(\hat{\beta}[u]\right)^2, \quad \text{(Natural)} \tag{3.68a}$$

$$\mathcal{L}_{\text{InVeRt}} = \mathbb{E}_{(x,y)\sim\mathcal{D}}\left[\frac{1}{N_z}\sum_{z\in\mathcal{Z}}\left(y[z] - \sum_{u}\hat{\beta}[u]\,\tilde{\phi}_u[z]\right)^2\right] + \lambda_1\sum_{u}\left|\hat{\beta}[u]\right| + \lambda_2\sum_{u}\left(\hat{\beta}[u]\right)^2, \quad \text{(Biorthogonal)} \tag{3.68b}$$

where $\mathcal{D}$ is the data distribution (expectation over profiles) and $\beta$ is defined in (3.65). A DWT subband regularizer is applied along $z$ to $\tilde{\hat{y}}[.]$ by adding $\lambda_1\sum|\hat{\beta}| + \lambda_2\sum\hat{\beta}^2$ to the data fidelity term in loss. The $L^1$ term encourages a sparse wavelet representation of the retrieved profile, while the $L^2$ term bounds the coefficient energy and stabilizes the ill-posed inverse learning. Together they suppress spurious oscillatory components in the vertical profile without changing the forward model. The constants $\lambda_1, \lambda_2$ are wavelet $L^1$ and $L^2$ weights in (3.68). The above losses can be minimized using small ground truth datasets specific to climatic conditions (for example, cloud-free or cloudy atmospheres).

**Backpropagation calculus in wavelet bases** The biorthogonal expansion in (3.65) is the forward map, where $\alpha[l]$ are the DWT coefficients of the input profile, $H[u,l]$ is the Volterra kernel in the wavelet domain defined in (3.66b), and $\hat{\beta}[u]$ are the predicted wavelet coefficients that enter the loss (3.68b). Let

$$g[u] = \frac{\partial\mathcal{L}_{\text{InVeRt}}}{\partial\hat{\beta}[u]} \tag{3.69}$$

denote the gradient of the loss with respect to the predicted coefficient at index $u$. The gradient with respect to the wavelet domain kernel is then

$$\frac{\partial\mathcal{L}_{\text{InVeRt}}}{\partial H[u,l]} = g[u]\,\alpha[l], \tag{3.70}$$

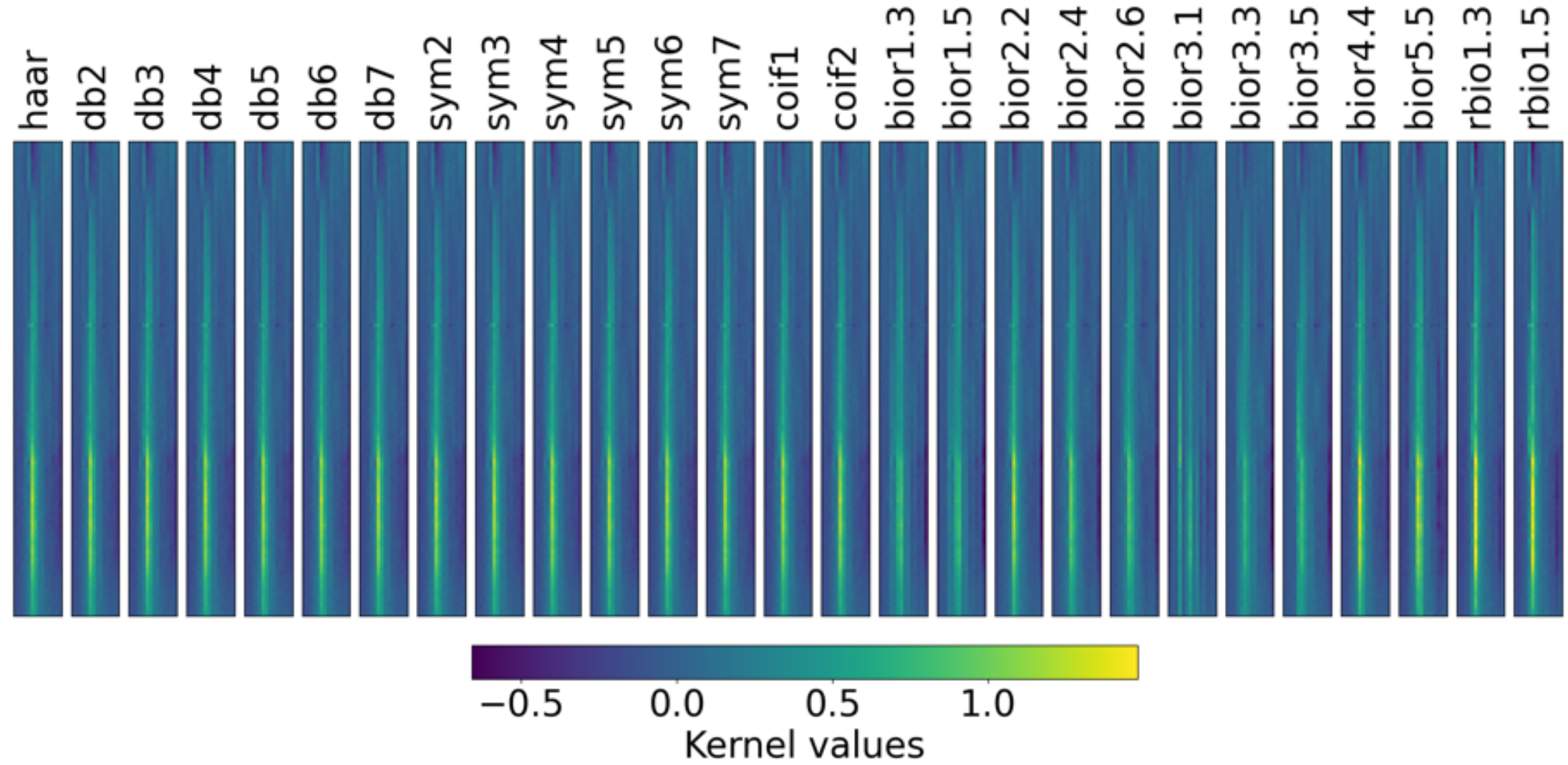

Figure 3.13: Inverse kernels for retrieving temperature profiles learned using different wavelet bases.

so each entry $H[u, l]$ is updated by a product of the backpropagated coefficient gradient and the corresponding input DWT coefficient. When training in the natural domain the chain rule is continued through the two axis wise DWTs in (3.66b). Using the biorthogonal synthesis functions $\phi_u$ and analysis functions $\tilde{\phi}_l$ this yields

$$\frac{\partial \mathcal{L}_{\text{InVeRt}}}{\partial h^+[z, \nu]} = \sum_{u \in \mathcal{U}} \sum_{l \in \mathcal{J}} \frac{\partial \mathcal{L}_{\text{InVeRt}}}{\partial H[u, l]} \phi_u[z] \, \tilde{\phi}_l[\nu], \tag{3.71}$$

which shows that the gradient on the natural domain kernel is obtained by projecting the wavelet domain gradient back through the same biorthogonal basis that was used to define $H[u, l]$. In implementation this backprojection is realized by the adjoint of the discrete wavelet transform and automatic differentiation carries out these steps during training.

### 3.3.4 Evaluation of natural-frequency domain `InVeRt 3/3` model

**Training and evaluation of linear Volterra kernel with Dataset 1** The inverse kernels learned with different wavelet bases for retrieving temperature profiles using V-band and minimizing equation (3.68b) are shown in Figure 3.13. These learned Volterra kernels accurately infer the temperature profiles from V-band observations models under similar climatic conditions. Similarly, the K-band and the corresponding available ground truth data is used to identify the inverse kernel for retrieving water vapor density profiles. Further, the RS measurements show large variations in the atmospheric humidity w.r.t. elevation and this poses a challenge in modeling an inverse system for inferring humidity. Here, the LN cascade system in (3.67) is defined for $m = 3$ and learned with constraints $a_m, b_m \in (-1, 1)$, i.e., a Volterra kernel followed by a rational nonlinearity retrieves the water vapor density with fine details. Figure 3.14 illustrates a test $\rho_v$ profile retrieved using two learned inverse kernels, one trained in the natural domain and the other trained in the transform domain, and inaccuracies are clearly visible. The distribution of the kernel weights of the a natural domain kernel and a kernel trained with rbio1.3 DWT basis is also shown in Figure 3.14. The same figure also shows the the learned $\{a_m, b_m\}_{m=0}^{3}$ coefficients of the ratio of polynomials in due process of refining the $\rho_v$ profiles retrieved from the MWR K-band observations. `InVeRt 3/3` models retrieve accurate profiles with just about ~500 parameters as summarized later in Table 3.4.

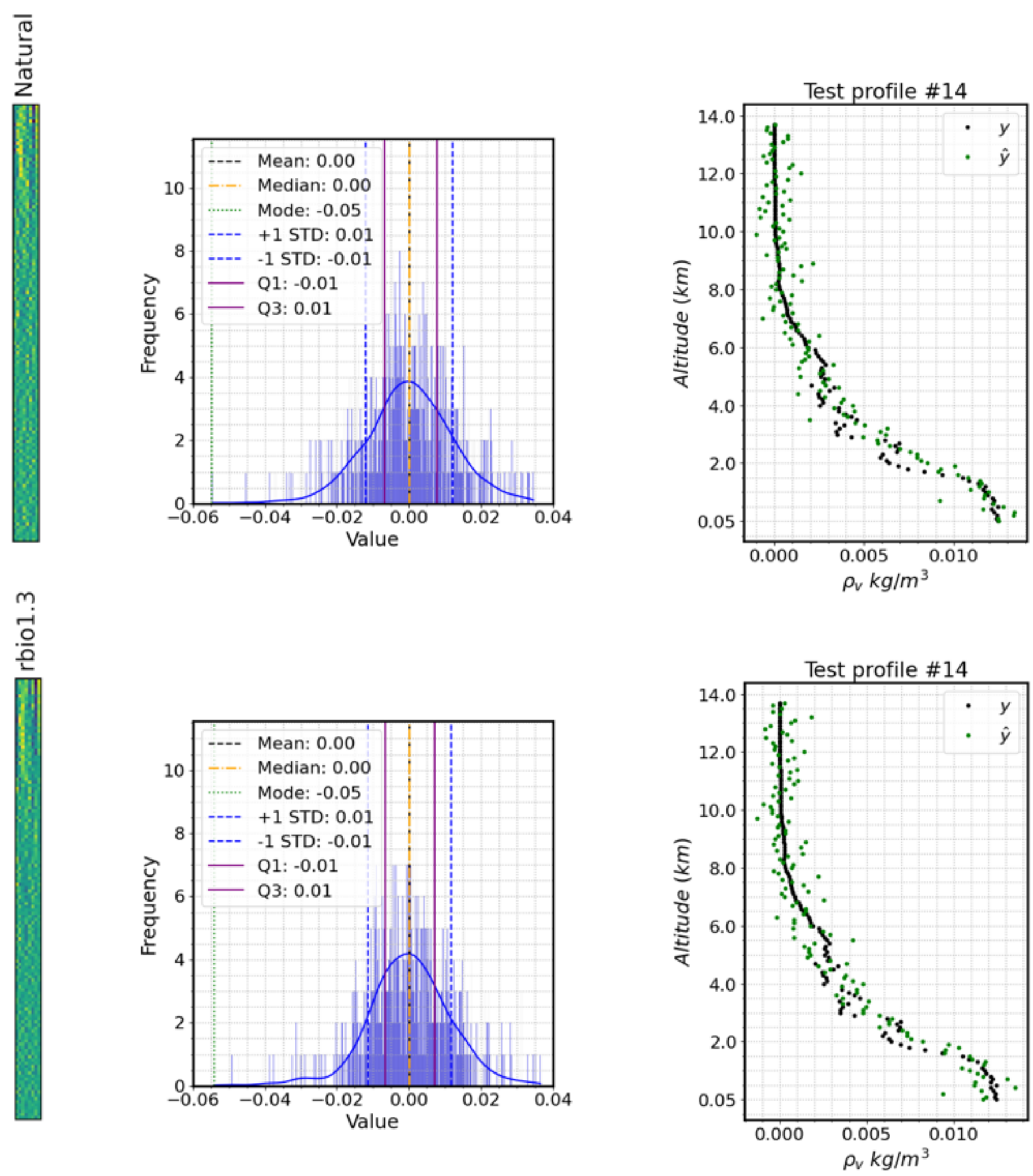

Figure 3.14: Inverse kernels that were trained in natural (top left) and transform domain with rbior1.3 wavelet (bottom left), their histograms (middle column) and a test water vapor density profile retrieved by these kernels (right column).

**Improving retrieval of $\rho_v$ with nonlinear rational polynomial cascade** The learnable LN cascade with rational function heads as defined earlier in 3.3.3.3 is used to retrieve $\rho_v$ profiles now. Figure 3.15 shows the new learned (refine) inverse kernel together with the learned rational polynomial coefficient for $m = 3$. Similar training with different wavelet bases and the learned kernels are shown in Figure 3.16. An observation here is that all wavelet bases have almost similar performance. This figure also shows that the convergence is mainly two types, where models with one group of wavelets converge much faster with some bumps. The models with the other group of wavelets converge smoothly but a bit slower. Among similar performances, it is noteworthy that rbio1.3, Haar, and db3 wavelet bases give best results for Dataset 1.

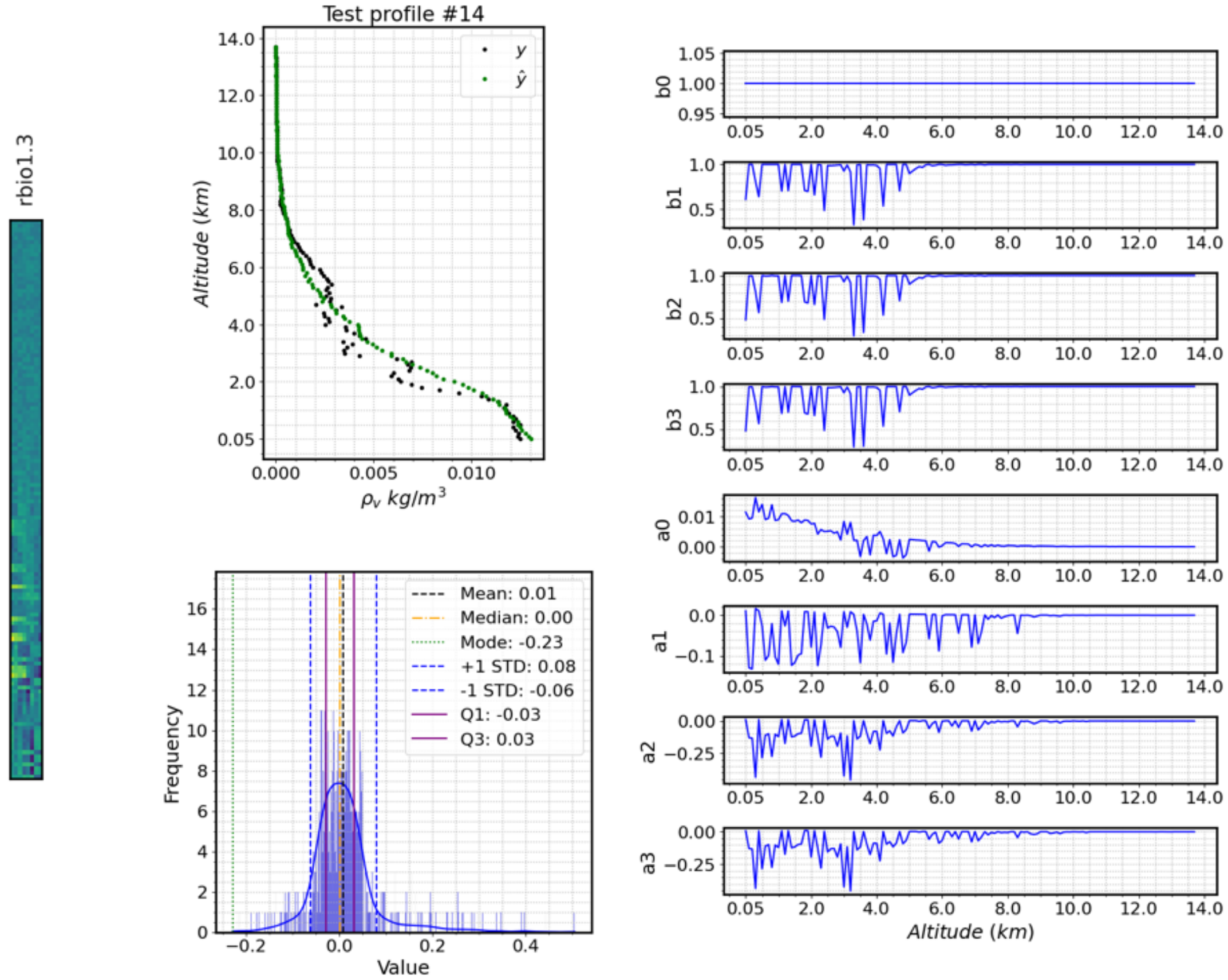

Figure 3.15: The refined inverse kernel (left) trained with rbio1.3 wavelet and DWT filter bank and a refined $\rho_v$ profile (middle top) retrieved by it. The distribution of the kernel weights and their statistics (middle bottom). The learned coefficients $a_m$ and $b_m$ of the polynomials in the rational static nonlinearity (right column).

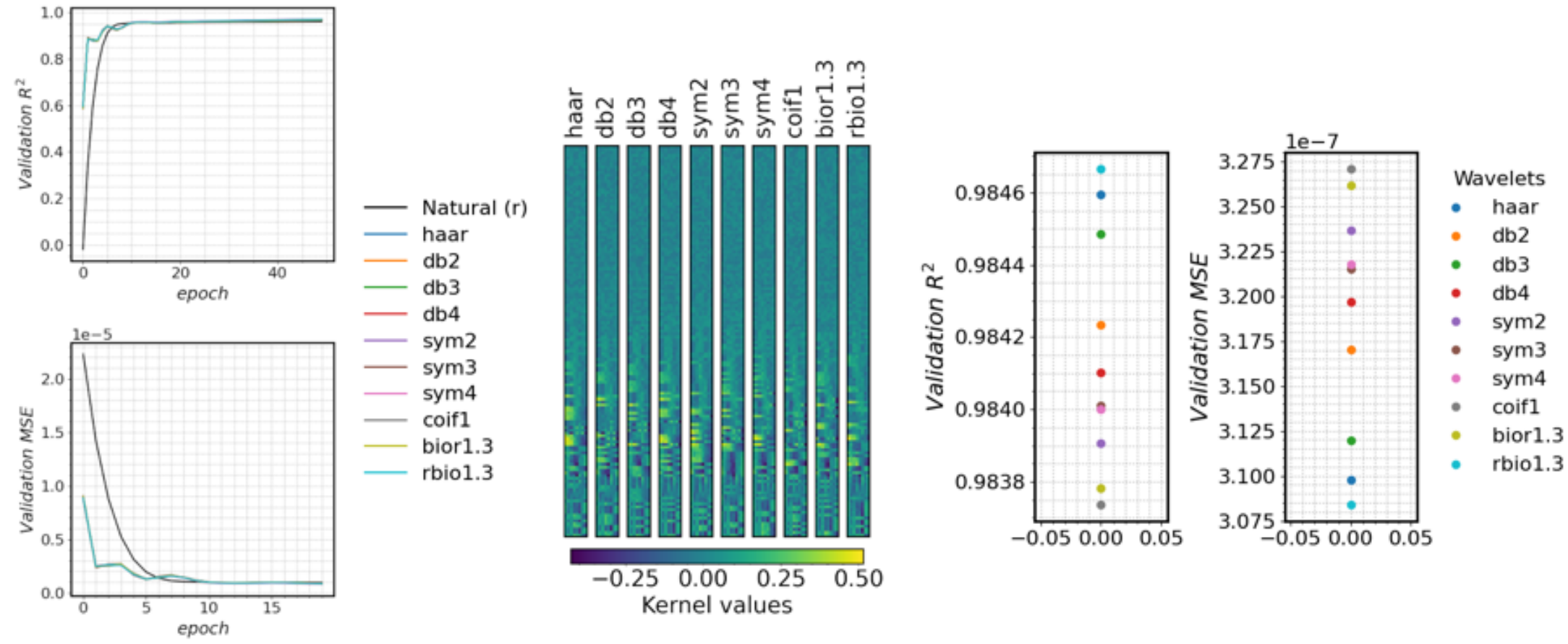

Figure 3.16: (a) Validation $R^2$ and MSE profiles of kernels trained for retrieving water vapor density profiles in natural and multiresolution domain with DWT basis using different wavelets. (b) Refined inverse kernels learned using different wavelet bases to retrieve water vapor density profiles (top); validation $R^2$ scores and MSE of the kernels with different wavelet basis (bottom).

#### 3.3.4.1 Evaluation of `InVeRt 3/3` for clear sky conditions (Dataset 2)

The $R^2$ score of the retrieved T profiles is ~ 99% for both natural and transform domain kernels. Similarly, the $R^2$ score of the retrieved $\rho_v$ profiles in both domains is ~ 98% with just the linear

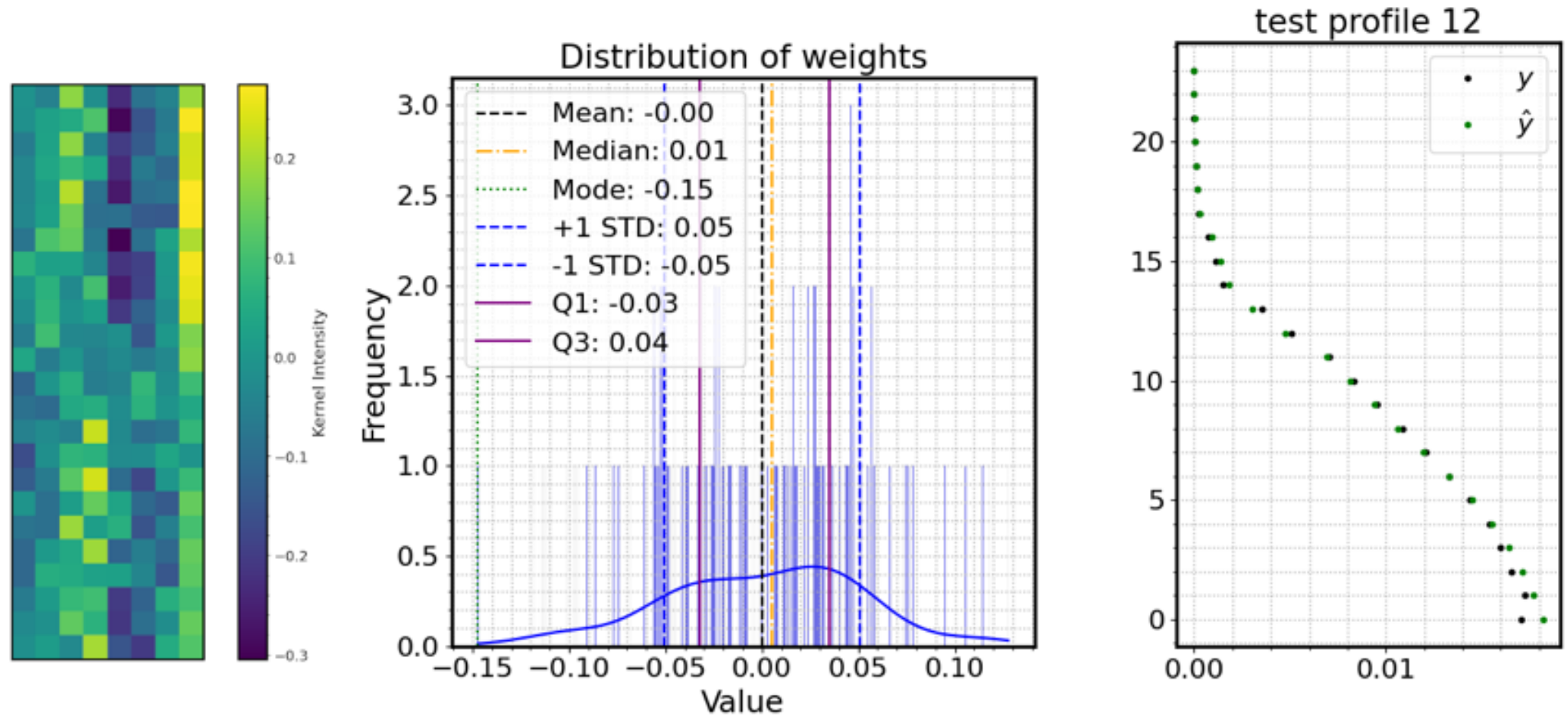

Figure 3.17: An inverse kernel (left) for retrieving water vapor density profiles, its statistical distribution (middle) and a retrieved test profile (right) from Dataset 2

Volterra kernel. Further, the inclusion of a rational polynomial head nonlinearity in cascade with the Volterra kernel as in equation (3.67) has improved the $R^2$score of $\rho_v$ to ~ 99%. Figure 3.17 shows a learned inverse kernel for retrieving $\rho_v$ profiles and a retrieved test profile from Dataset 2.

#### 3.3.4.2 Evaluation of `InVeRt 3/3` for all weather conditions (Dataset 3)

**Preprocessing** The training set (28,050 profiles) and validation set (14,025 profiles) are randomly chosen from the 60% of the entire dataset. The remaining 40% is the test set with 28,052 temporally correlated profiles. The $T_B$s in the input vector $x[v]$ for all $v$ are Z-normalized. The target vector $y[z]$ for all $z$ retain physical units (Kelvin for $T$ and g $\mathrm{m}^{-3}$ for $\rho_v$) but rescaled to $[0, 1]$ for training and inference.

#### 3.3.4.3 Baseline SOTA inverse models for ablation studies

**MLP neural network** The input channels $x[v]$ form an input radiance vector (14 dims for $T$, 8 for $\rho_v$) to a fully connected dense MLP neural network with two hidden layers of 128 and 64 neurons with ReLU activations (configurable) and a linear output layer of $N_z$ neurons. It maps the input to a complete vertical profile $\{\hat{y}[z]\}_{\forall z}$ with optional dropouts $p$ and batch normalization after each hidden layer. This neural network is trained with backpropagation by minimizing the loss,

$$\mathcal{L}_{\mathrm{MLP}} = \mathbb{E}_{(x,y)\sim\mathcal{D}}\left[\frac{1}{N_z}\sum_{z\in\mathcal{Z}}\left(y[z] - \hat{y}_{\mathrm{MLP}}[z]\right)^2\right] \tag{3.72}$$

with Adam at a learning rate $10^{-4}$ without any regularization.

**Random-forest regressor** For each elevation $z \in \mathcal{Z}$ we train an independent regression forest (300 trees per forest) using bootstrap aggregation (bagging). At a given tree node, let $S = \{(x, y[z])\}$ denote the sample set of the node with $x$ as the Z-normalized channel vector and $y[z]$ be the scalar target at elevation $z$. At each node, we scan all components of the MWR channel $x[v]$ as split candidates and greedily choose a threshold to minimize the weighted squared-error impurity (in-node MSE or variance of the targets) of the left/right children. The node-split objective (weighted sum of child-node squared-error impurities) is

$$\text{Split cost:} \quad J_{\mathrm{RF}} = N_L\, i(S_L) + N_R\, i(S_R), \tag{3.73a}$$

$$\text{Node impurity:} \quad i(S) = \frac{1}{|S|}\sum_{(x,y)\in S}\left(y - \bar{y}_S\right)^2, \tag{3.73b}$$

where $S_L, S_R$ are left/right child sample sets (cardinality $N_L$=$|S_L|$, $N_R$=$|S_R|$), $\bar{y}_S$ is the mean target in $S$, and $(x, y) \in S$ are the samples routed to the node ($y$ is the scalar target for the current elevation $z$). The forest prediction at $z$ is the average of per-tree predictions.

#### 3.3.4.4 Training and evaluation of `InVeRt 3/3` and baseline models

A natural-domain view of the learned `InVeRt 3/3` bior1.3 inverse kernels used to retrieve $\rho_v$ and $T$ profiles from K-band and V-band MWR radiances is shown in Figure 3.18.

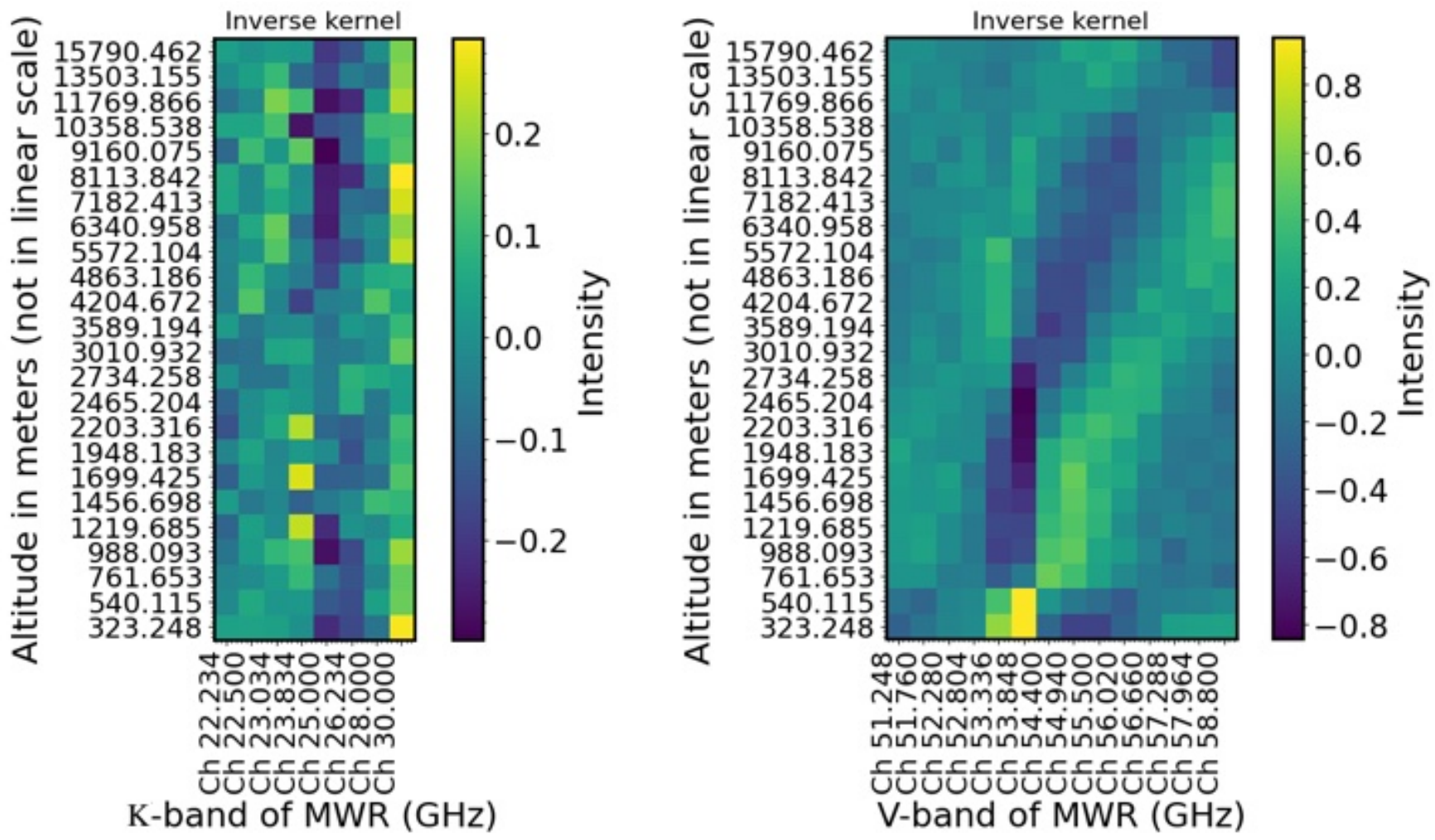

Figure 3.18: Natural domain view of the learned `InVeRt 3/3` bior1.3 kernels for retrieving $\rho_v$ (left) and $T$ (right) profiles from $T_B$ observations of MWR K-band and V-band.

Table 3.4: Parameters, FLOPs and performance of inverse models

| | Inverse model | Number of parameters | FLOPS (approx.) | Disk usage | $R^2$ | RMSE |
|---|---|---|---|---|---|---|
| | InVeRt 3/3 bior1.3 | **384** | **793** | **32** KB | 0.99 | 0.61 g m$^{-3}$ |
| $\rho_v$ | MLP neural net | 10,968 | 43,227 | 164 KB | 0.99 | 0.60 g m$^{-3}$ |
| | RF regressor | # | 7,200 | 16 GB | 0.98 | 0.66 g m$^{-3}$ |
| | InVeRt 3/3 bior1.3 | **528** | **1,081** | **36** KB | 0.99 | 1.15 K |
| $T$ | MLP neural net | 11,736 | 46,299 | 172 KB | 0.99 | 1.14 K |
| | RF regressor | # | 7,200 | 17 GB | 0.99 | 1.24 K |

#For each atmospheric variable ($T$ and $\rho_v$), there are 24 independent regressors (one forest per elevation), each with 300 trees. Thus, for each variable we have a total of $24 \times 300 = 7{,}200$ trees. The trained forests contain 250,578,732 nodes (125,292,966 leaves), occupy $\approx$ 16.0 GB, and have an average leaf-weighted path length of 18.44. One full-profile inference traverses $\approx 18.44 \times 7{,}200 = 132{,}741$ split comparisons. The FLOPs for inference are minor, i.e., $\approx$ 7.2k additions/divisions to average tree outputs, so latency is dominated by branching and memory access. The number of learned scalars (thresholds plus leaf values) is $\approx 2.51 \times 10^8$ per atmospheric variable.

Each row of these two kernels corresponds to an atmospheric altitude level, and each column corresponds to an MWR channel frequency. The learned kernels are evaluated on a held-out test set comprising 40% of the data, that is, 28,052 unseen RTTOV-gb–simulated K-band and V-band brightness temperatures $T_B$ generated from MERRA-2 reference profiles.

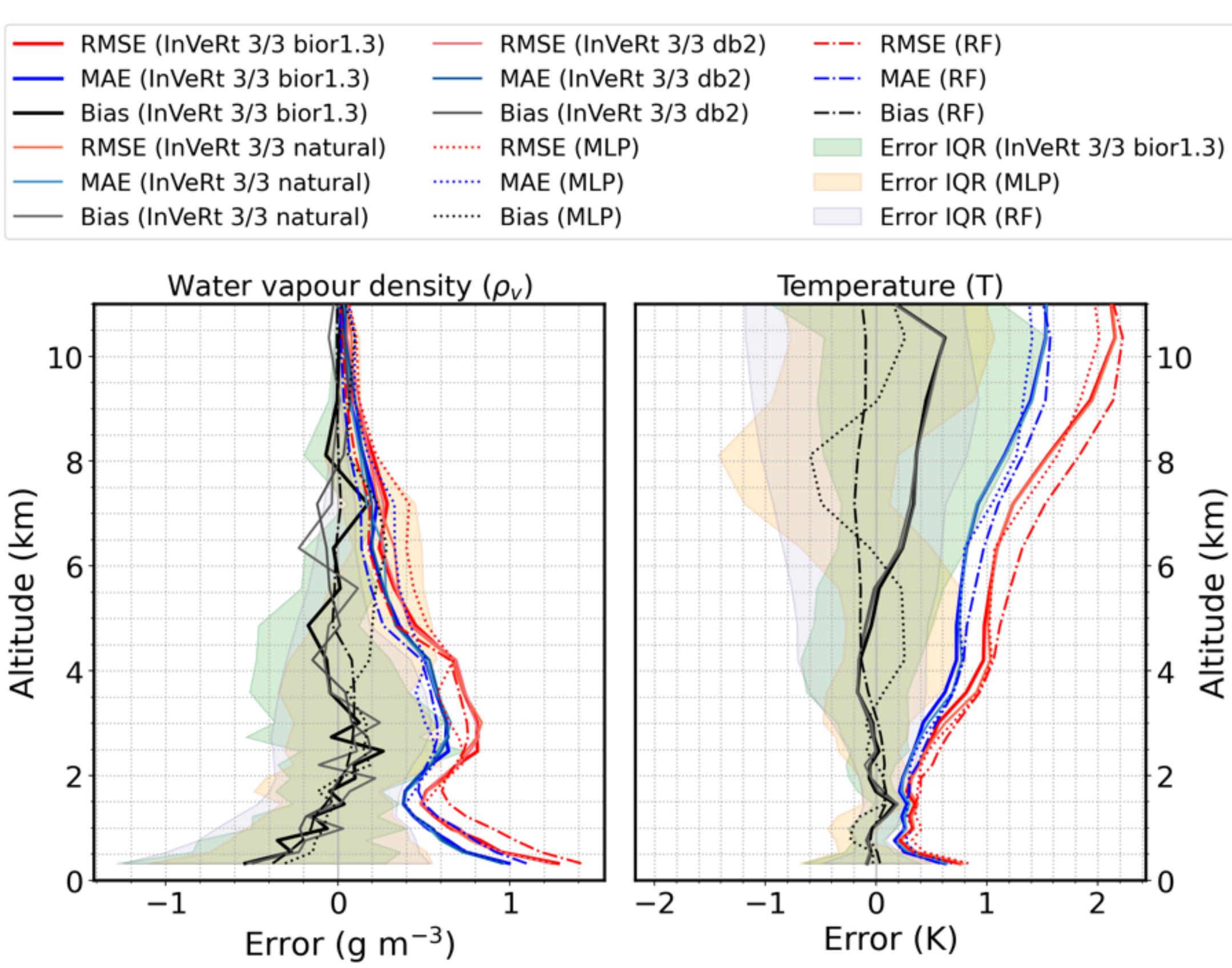

Figure 3.19: The three error profiles, namely RMSE (red lines), MAE (blue lines), and MBD (black-toned lines), are grouped together for the retrieved $\rho_v$ (left) and $T$ (right) 28,052 test profiles of all-weather conditions. The errors of the three `InVeRt 3/3` models (natural, db2, bior1.3) largely overlap, and their differences are harder to distinguish (solid lines). The main separation is relative to both RF (dot-dash lines) and MLP neural network (dot lines) baselines, which exhibit slightly larger RMSE and MAE above about 2 km for $T$, but for $\rho_v$, small differences of the order of a few tenths of g m$^{-3}$ with baselines are visible at specific heights. Also, the bior1.3 version has noticeably narrower IQR envelopes than the baseline models, indicating less altitude-wise scatter.

Table 3.4 compares the number of learned parameters, approximate Floating-Point Operations (FLOP) per full-profile inference, disk usage and coefficient of determination $R^2$ for the three inverse models. For both $\rho_v$ and $T$, the `InVeRt 3/3` bior1.3 kernels achieve $R^2 \approx 0.99$, which is comparable to the MLP neural networks and higher than the RF regressor for $\rho_v$, while using only a few hundred parameters and a few tens of kilobytes on disk. The MLP neural networks require around $10^4$ parameters but still have a modest storage footprint, whereas the RF regressors are implemented as 7,200 trees per atmospheric variable and occupy on the order of 16–17 GB because they store roughly $2.5 \times 10^8$ learned thresholds and leaf values. Figure 3.19 shows the vertical MBD profiles (black-toned lines) and the error Inter-Quartile Range (IQR), together with the bias-adjusted error metrics, MAE and RMSE, for temperature $T$ and water vapor density $\rho_v$ retrieved by the `InVeRt 3/3`, MLP neural network and RF models. The error profiles of all three `InVeRt 3/3` kernels (natural, db2, bior1.3) exhibit very similar retrievals and are almost on top of each other, with the bior1.3 variant only marginally lower around mid-tropospheric levels. In the boundary layer ~ (0-2) km, the mean bias for all three models is close to zero for both $T$ and $\rho_v$, indicating negligible systematic deviation from the true MERRA-2 profiles. For temperature, `InVeRt 3/3` models maintain a small positive bias that does not exceed about ±0.6 K in the free troposphere ~ (5-10) km, whereas the RF regressor exhibits a substantially larger positive bias, and the MLP neural network lies in between. The biases of $\rho_v$ remain small at all height levels, but the RF model shows a noticeably broader IQR above about 5 km, pointing to a larger statistical dispersion of its errors compared to `InVeRt 3/3` and MLP neural network. The bias alone is not adequate because positive and negative deviations can cancel at the mean. The MAE and RMSE profiles quantify error magnitudes without sign cancellation, and the RMSE of the `InVeRt 3/3` temperature retrieval models is below approximately 0.5 K in the boundary layer, increasing to roughly ~ (0.5-1) K in the mid-troposphere ~ (2-5) km. Across most of the column above 2 km, the `InVeRt 3/3` models have the lowest $T$ RMSE, the MLP neural network is comparable, and the RF regressor exhibits the largest errors. For water vapor, `InVeRt 3/3` and MLP perform alike and both outperform RF in the boundary layer. In the mid-troposphere, MLP attains the lowest RMSE, `InVeRt 3/3` has intermediate errors, and RF shows the largest errors. However, `InVeRt 3/3` achieves lower RMSE than MLP and RF in the free troposphere. Overall, the described `InVeRt 3/3` inverse models are competitive and sometimes better than both RF regressors and the MLP neural networks across most atmospheric layers, with clear advantages for $T$ throughout the troposphere and for $\rho_v$ in the boundary layer relative to RF and in the free troposphere relative to both RF and MLP. A $\rho_v$ and $T$ test profile retrieved by the `InVeRt 3/3` bior1.3 kernels, MLP neural networks, RF regressors, and the corresponding ground truths are shown in Figure 3.20.

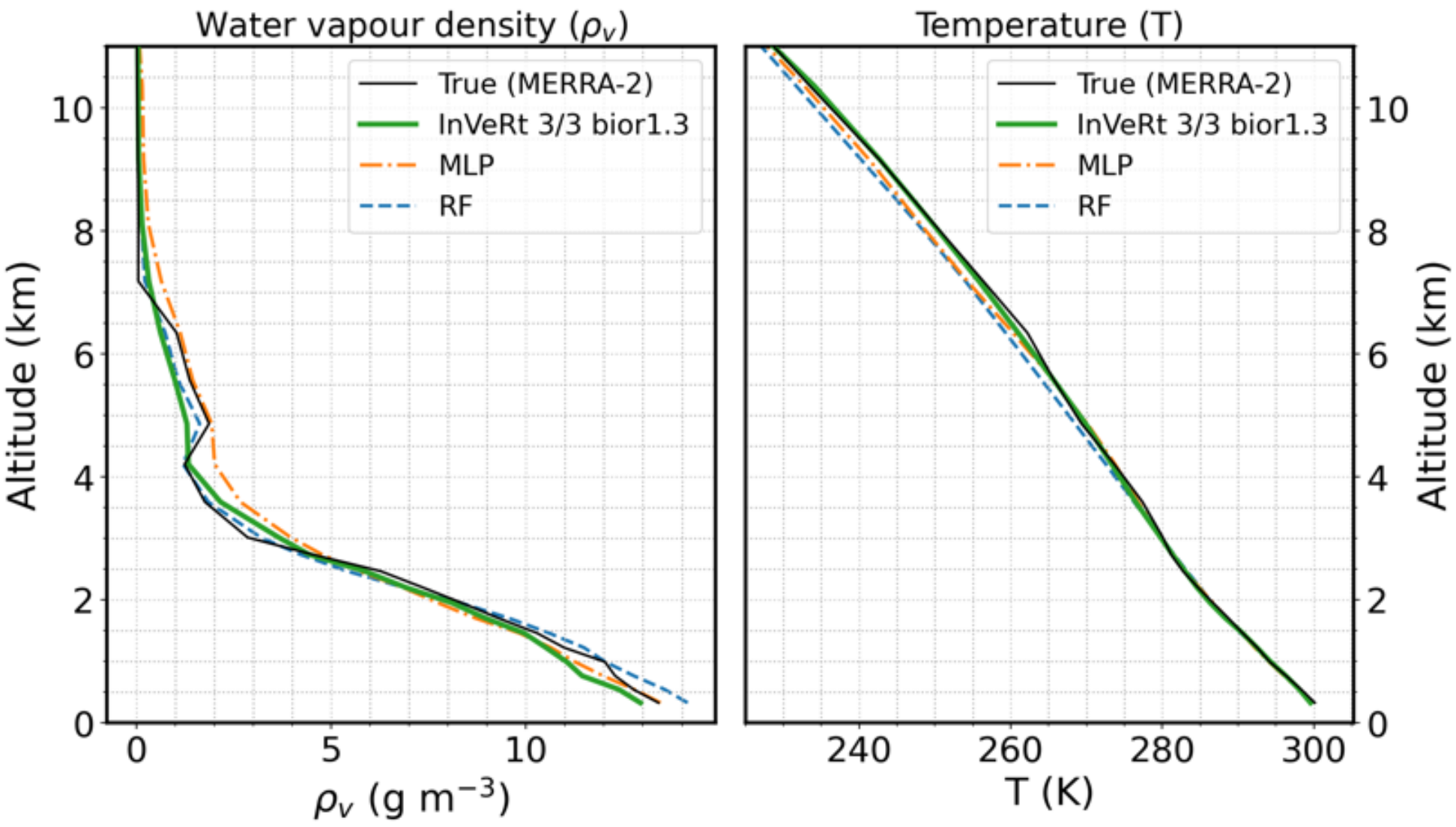

Figure 3.20: A test $\rho_v$ (left) and $T$ (right) profile retrieved by `InVeRt 3/3` bior1.3, MLP neural network, and RF regressor models from the K-band and V-band inputs. The true MERRA-2 profiles for both variables shown are in black.

## 3.4 Discussion

This chapter introduces multidimensional natural-frequency domain Volterra kernels and their input-output relationship in biorthogonal bases for deep learning. The biorthogonal bases are realized with DWTs of the kernel, input, and output. The chapter also describes shift-invariant natural-frequency domain realizations of linear and nonlinear 1D, 2D, 3D, or higher-dimensional input-output Volterra kernels that can be trained by backpropagation in both the natural and transform domains. Natural-domain Volterra kernels of $m^{\text{th}}$ degree generalize convolutions, where $m = 1$ is an LSI system and higher-degree kernels are NLSI systems involving high-dimensional convolutions and reductions. Although Volterra kernels provide a natural framework to model nonlinear phenomena, higher-degree ($m > 1$) kernels are rarely used in machine learning or deep learning because of computational complexity. The results show a time-space trade-off between natural-domain and transform-domain representations. Transform-domain computations can be faster, but they require additional storage. The algorithmic realizations are packaged in the `VolterraSys` Python package.

This chapter also describes an interpretable and frugal system-theoretic design for a classical inverse problem in microwave radiometry with a linear Volterra kernel in cascade with rational functions. The `InVeRt 3/3` inverse model preserves the canonical structure of the theoretical inversion while learning a natural-frequency domain parameterization. The biorthogonal wavelet bases provide sparse representations of the inverse kernel, microwave radiance inputs, and vertical atmospheric profile outputs, which supports stable transform-domain learning in limited-data settings. Evaluation of `InVeRt 3/3` with radiosonde truths under clear-sky conditions shows accurate retrievals, and the retrieval performance is further evaluated on 28,052 unseen MERRA-2 profiles spanning clear, cloudy, and rainy conditions with corresponding RTTOV-gb simulated K-band and V-band brightness temperatures. This test set supports the conclusion that the inversion generalizes across different weather conditions without obvious signs of overfitting, despite the training set containing only 28,000 samples. The following chapter then uses multiresolution encoders and Volterra heads for compact audio and texture classifiers.

# Chapter 4

# Structured IIR Filters and Multiresolution Attentions

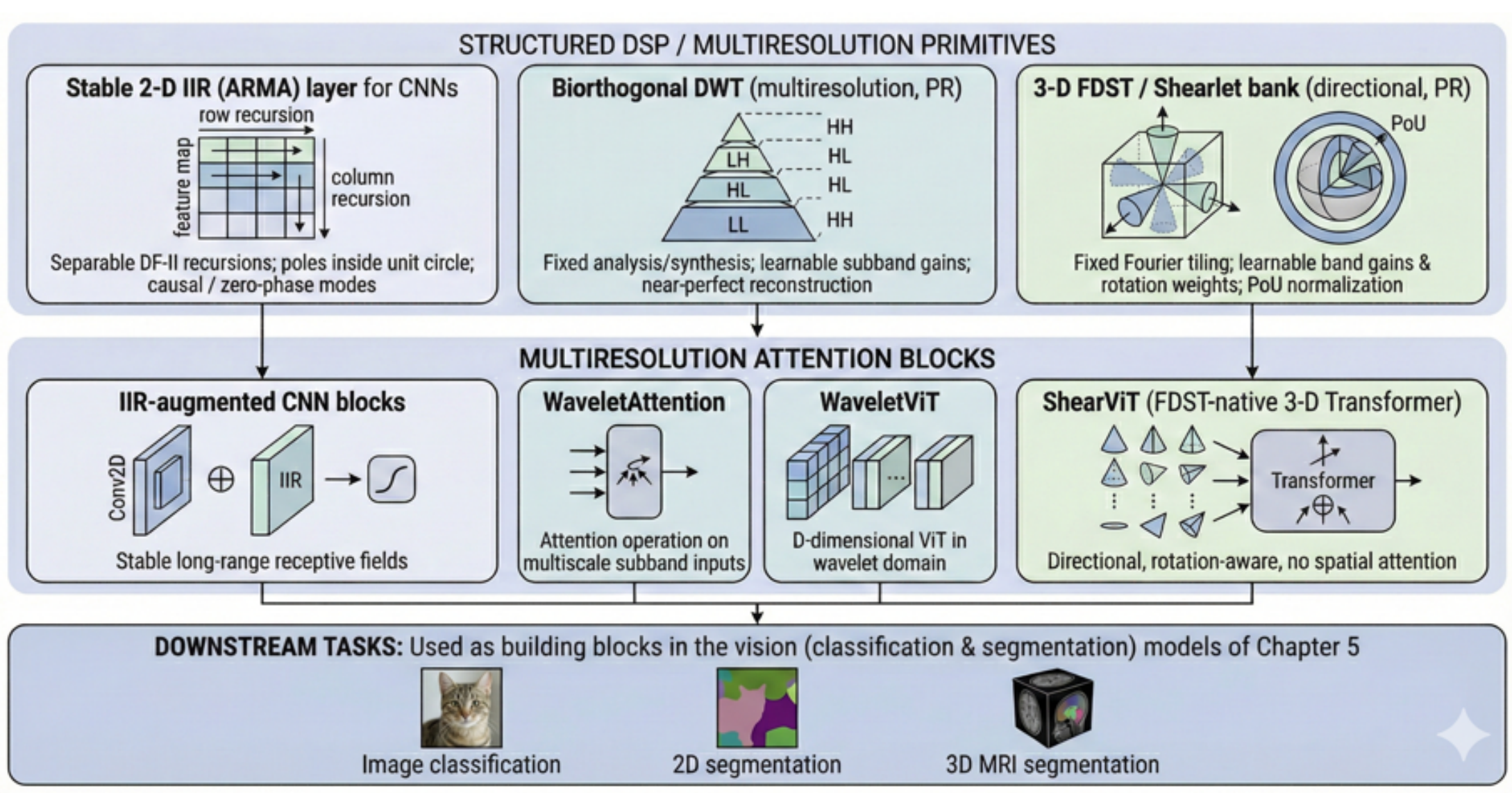

Inputs in computer vision problems are multidimensional signals (e.g., images, volumetric MRI) whose pixel or voxel intensities are often a highly nonlinear field with long-range dependency. The present object recognition, classification, and segmentation models capture this nonlinearity with various deep learning architectures. In theory, Infinite Impulse Response (IIR) operators can provide very large effective receptive fields and physically meaningful recursions for 1D time series, 2D images, and 3D volumes. However, there are practical challenges in realizing $D$-dimensional ($D > 1$) IIR filters due to the burst in computational complexity associated with operating on multidimensional signals and systems. Further, recursive updates can be numerically unstable with challenges in boundary handling, and computing gradients through long recursions [21]. As described in previous chapters, FIR filters efficiently capture local spatial features, but a single FIR layer has a finite receptive field. In decimated CNNs (e.g., U-Net), stacked filters, nonlinear activations, downsampling, and residual connections enlarge the effective receptive field at each spatial position, which can solve many challenging vision problems without explicit recursion. Table 4.1 compares FIR-like CNNs, IIR-like systems, and vision transformers. In similar signal-theoretic efficient models, multiresolution transforms provide global context by separating scale and orientation [35, 139]. Advanced vision architectures (e.g., SAM, medSAM) further use Vision Transformers (ViT) [57], which encode the spatial location information via positional patch tokens. Transformers [24] produce self-attention maps, where every token is mixed with all other tokens to capture long-range dependencies, and is the key technology behind the success of large language sequence models. Vision transformers apply global self-attention after converting images to smaller patch tokens [57] by computing a weighted sum over all positions in a feature map to capture long-range spatial dependen-cies [200]. Similarly, Hierarchical vision transformers have staged processing and multiscale tokenization [198, 199]. Now, full self-attention scales quadratically with the token count, which is computationally expensive and challenging in practice for multidimensional data. This quadratic scaling motivates sparse, low-rank, and kernel-based modeling for long 1D sequences [201–204]. Further, structured state space models model long sequences by propagating a hidden state with linear cost [100, 196]. Some vision models implement recurrences through separable sweeps or scan orders [195, 205]. ARMA Nets propose a learned 2D auto-regressive moving-average (ARMA) layer that expands the receptive field for dense prediction [194]. This chapter addresses some of these limitations and introduces designs of stable, learnable multi-dimensional recursive systems and multiresolution filter-bank vision transformers that support practical boundary handling and differentiable training. These new learning systems generate a rich feature space with efficient feedback paths and tokenization of multiresolution directional features in the backpropagation path.

Table 4.1: Comparison of FIR-like CNNs, IIR-like Systems, and Vision Transformers

| # | Important characteristics | CNNs (FIR-like) | IIR-like / Recursive Systems | Vision Transformers (ViTs) |
|---|---|---|---|---|
| 1 | System type | Local FIR convolutions | Recursive filters / ARMA, state-space models (SSMs) | Token mixing via self-attention (dense or windowed) |
| 2 | Recursion (feedback) | No | Yes (explicit feedback/state update) | No explicit feedback (feedforward across layers; no persistent state) |
| 3 | Global interaction viewpoint | Sparse local operator; global context via stacking/dilation | Often solves/approximates a global operator via recursion (e.g., $(I - A)^{-1}Bx$; dense effect with structured parameters) | Explicit dense mixing matrix per layer, e.g. $A(x)V$ with $A(x) = \text{softmax}(QK^\top)$ (input-dependent) |
| 4 | Memory (internal state) | None | Yes (dynamic state propagation, e.g., via ODEs or state spaces) | None (no persistent state; some variants add memory tokens) |
| 5 | Receptive field | Local per layer; can become global with depth/dilation | Long-range/global in principle via recursion; practical multi-D designs often use separable/axis-wise recursions | Global with full attention; many vision models use windowed/shifted attention for efficiency |
| 6 | Translation / positional inductive bias | Strong translation equivariance and locality bias (convolution) | Translation equivariant for LSI away from boundaries; separable sweeps introduce axis bias | Not translation equivariant; relies on positional encodings/relative bias; locality depends windows |
| 7 | Spectral selectivity | Defined per linear conv layer; finite support (overall nonlinear) | Poles/state dynamics can yield sharp/resonant responses; stability constraints required | Not LSI in general; can be analyzed as input-dependent mixing rather than a fixed frequency response |
| 8 | Interpretability | High (e.g., Gabor-like learned filters) | Moderate (poles/zeros or states can be interpretable; multi-D stability adds complexity) | Mixed; attention maps are visualizable but not always faithful explanations |
| 9 | Compute scaling (w.r.t. pixels $N$) | $O(N)$ in $N$ per layer (e.g., $k{\times}k$ conv) | $O(N)$ in $N$ per pass (order-dependent); true D-dim IIR is costly without separability | Full attention $O(N^2)$; windowed/linear attention variants reduce cost |
| 10 | Implementation in deep models | VGG [90], U-Net [39], ResNet [38] | ARMA Nets [194], FIR–IIR [195], S4 [196], Mamba [100] | ViT [57], DeiT [197], PVT [198], Swin [199] |
| 11 | Hardware optimization | Highly optimized via cuDNN, Winograd | Specialized scan/SSM kernels emerging; stability/boundary handling can be a constraint | Efficient attention kernels (e.g., FlashAttention) and windowed designs; still costly at high resolution |

## 4.1 Multidimensional learnable IIR filters

In two spatial dimensions (2D) or higher $D$, true IIR designs are uncommon for both practical and theoretical reasons. Bounded-input bounded-output (BIBO) stability means that every bounded input produces a bounded output. In one dimension, BIBO stability admits a simple pole test, where all poles lie strictly inside the unit circle. In 2D, the denominator zeros form multivariate algebraic sets rather than isolated scalar poles, and there is no analogous single-radius rule. For a rational $H(z_x, z_y) = B/A$, stability requires an absolutely summable impulse response, where a basic check is $A(e^{j\omega_x}, e^{j\omega_y}) \neq 0$ on the unit torus $|z_x| = |z_y| = 1$ together with stability of relevant one-dimensional slices. This leads to multivariate polynomial conditions rather than a one-line criterion. Linear or zero phase is likewise noncausal in 2D, where a linear-phase response $H(\omega_x, \omega_y) = e^{-j(\omega_x m_0 + \omega_y n_0)} A(\omega_x, \omega_y)$ with $A$ real and even implies $h[m, n] = h[2m_0 - m,\ 2n_0 - n]$. Then any causal support confined to one side of $(m_0, n_0)$ would force $h \equiv 0$. Thus, nontrivial 2D linear/zero-phase filters must be realized noncausally (e.g., forward-backward). These facts motivate a separable construction in which stability reduces to a one-dimensional condition per axis and zero phase is obtained by forward-backward sweeps along rows and columns.

1. *Instability under Differentiation:* Unlike FIR filters, which are inherently stable and feedforward, IIR filters include feedback terms where the current output depends on previous outputs or states. During training, backpropagation through time (BPTT) must navigate these recursive dependencies, which can cause gradient explosion or vanishing, particularly in high-order or deep recursive filters.
2. *Non-linearity and Parameter Coupling:* Classical IIRs (e.g., biquad structures) have nonlinear relationships between filter coefficients and frequency response. This makes the parameter space highly non-convex and difficult to optimize using standard stochastic gradient descent (SGD) methods.

3. *Differentiability of DSP Primitives:* Traditional DSP implementations rely on fixed, non-differentiable operations (e.g., bilinear transforms, zero-pole placement). These must be re-formulated in smooth, differentiable forms to be embedded in neural architectures.
4. *Compatibility with Deep Architectures:* Most CNN layers are stateless and localized (i.e., spatial convolution), while IIR filters are inherently stateful and recursive. This mismatch led to architectural challenges: how to inject feedback into CNNs without violating training efficiency or modularity.

Table 4.2: Overview of IIR-inspired and recursive neural models across domains

| Paper | Domain (Dim) | IIR Form | Order | Neigh. /Recur. | Feed-back | Diff. | Fidelity | Notes |
|---|---|---|---|---|---|---|---|---|
| Mandic & Goh (2009) [206] | Signal proc. (1D) | Adaptive IIR (complex) | Variable | N/A | Yes | Limi-ted | 1D IIR | Foundational treatment of complex-valued adaptive recursive filters. |
| Asama et al. (2020) [195] | Image proc. (2D) | Separable FIR–IIR | Variable | Axis-wise (separable) | Yes | Yes | Separable | Separable FIR–IIR blocks (explicit recursion) |
| Su et al. (2020) [194] | Dense pred. (2D) | ARMA layer (2D) | Variable | 2D neighbor recurrence | Yes | Yes | 2D ARMA | Stable ARMA-style layer that expands receptive field for dense prediction. |
| Pepe et al. (2022) [207] | Audio EQ (1D) | True IIR (bi-quad/SOS) | 2nd | N/A | Yes | Yes | 1D IIR | End-to-end parametric biquad design. |
| Fayyazi & Shekofteh (2023) [208] | Speaker ID (1D) | IIR-inspired conv front-end | ~2nd | N/A | No | Yes | IIR-inspired | Interpretable resonant kernels approximating IIR behavior. |
| Gu & Dao (2024) [100] | Long seq. (1D) | Selective state-space (SSM) | $N$ states | State recurrence | Yes | Yes | SSM | Linear-time sequence modeling with selective state updates. |
| Zhu et al. (2024) [205] | Vision (2D) | Bidirectional SSM | $N$ states | State recurrence | Yes | Yes | SSM | SSM recurrence applied to vision tokens orders. Not true 2D IIR. |
| Gong et al. (2025) [209] | Biomed seg. (3D) | SSM-in-CNN (Mamba) | $N$ states | State recurrence | Yes | Yes | SSM | 3D volumetric tasks with SSM blocks; (not a true 3D ARMA). |

*Notes:* Diff. indicates end-to-end differentiability. In this table, ”2D ARMA” denotes a *spatial* autoregressive recursion between output pixels, whereas ”Separable” denotes axis-wise 1D recursions composed across axes (an approximation of a general non-separable multi-D IIR/ARMA, cf. Eq. (4.14)). ”SSM” denotes state-space recurrences (IIR-like), typically applied over a sequence/token scan rather than a 2D/3D ARMA.

Table 4.2 summarizes representative IIR-inspired and recursive neural models across domains. Unlike FIR filters in CNN, the above challenges limited the progress of fully differentiable IIR filters in similar neural network architectures. Some instances of IIR filter blocks and back-propagation learning are visible in domains of audio processing, biomedical signal analysis, with very limited presence in computer vision, particularly for 3D applications. While biochemical and audio models have successfully deployed trainable parametric biquads [207] and IIR-like convolutional frontends with resonant behavior [208], the adoption of true IIR filters in 3D vision networks remains essentially unexplored. Key challenges remain in ensuring stability across 3D recursion, managing parameter complexity, and facilitating effective backpropagation through multi-dimensional feedback loops.

### 4.1.1 Optimization-based realizations of true multidimensional learnable IIR filters

Given a $D$-dimensional signal $x[\boldsymbol{n}]$ on a finite spatial grid, the implemented true multidimensional IIR layer realizes a fully coupled nonseparable multidimensional recursive relation with a finite delay support on every axis and with circular shifts on the finite grid. Let $L$ denote the maximum delay per axis and let $\Lambda_L = \{0, \ldots, L\}^D$ denote the finite lag set. The realized multidimensional IIR relation is written as

$$y[\boldsymbol{n}] = \sum_{\boldsymbol{k}\in\Lambda_L} b\big[\boldsymbol{k}\big]\, x\big[\boldsymbol{n}-\boldsymbol{k}\big] - \sum_{\boldsymbol{k}\in\Lambda_L\setminus\{\boldsymbol{0}\}} a\big[\boldsymbol{k}\big]\, y\big[\boldsymbol{n}-\boldsymbol{k}\big]. \tag{4.1}$$

where $\boldsymbol{n} \equiv (n_1, n_2, \cdots, n_D)$ is the spatial index, $\boldsymbol{k} \equiv (k_1, k_2, \cdots, k_D)$ is the delay index, and $\boldsymbol{0}$ denotes the all-zero lag tuple. Equation (4.1) is the scalar prototype relation. The implemented layer realizes it in lifted batched multichannel tensor form through the shifted banks $X$ and $Y$ and the banked coefficient tensors $\mathcal{B}$ and $\mathcal{A}$. In the implementation-oriented algorithms below, $B$ denotes the batch size, $C_{\text{in}}$ and $C_{\text{out}}$ denote the numbers of input and output channels, $T$ denotes the maximum number of inner gradient-descent steps used in the local refinement, and $\eta$ denotes the inner gradient-descent step size. In the contracted tensor expressions used in those algorithms, $c$ indexes input channels, $o$ indexes output channels, and $d$ enumerates the finite bank of shifted delay tuples. In the current implementation the numerator and denominator use the same maximum delay $L$ on every axis, and the zero-lag denominator coefficient is fixed separately as $a[\boldsymbol{0}] = 1$. The optimization-based finite-grid realization therefore amounts to approximately driving

$$\left\| \sum_{\boldsymbol{k}\in\Lambda_L} b\big[\boldsymbol{k}\big]x\big[\boldsymbol{n}-\boldsymbol{k}\big] - y[\boldsymbol{n}] - \sum_{\boldsymbol{k}\in\Lambda_L\setminus\{\boldsymbol{0}\}} a\big[\boldsymbol{k}\big]y\big[\boldsymbol{n}-\boldsymbol{k}\big] \right\|_2^2 \approx 0. \tag{4.2}$$

**General realization** The common implementation pattern is the same for all currently supported dimensions. For a $D$-dimensional input tensor, the layer constructs the full bank of circularly shifted input copies $X$ and the corresponding shifted copies $Y$ of a local variable $y$, concatenates the fixed zero-lag denominator coefficient with the trainable denominator coefficients, and runs a finite inner gradient-descent loop for at most $T$ steps. The local variable is lifted over input-channel and output-channel pairs, and the final output is obtained only after summing over the input-channel index at the end of the refinement. The local variable may be initialized randomly. This is an optimization-based numerical residual refinement on a finite periodic grid and is not presented as an exact closed-form solve or as a general multidimensional stability proof. Algorithm 4.14 summarizes the realized implementation of (4.1). The current `IIRD` code provides concrete classes for $D \in \{1, 2, 3\}$.

---

**Algorithm 4.14:** Implemented $D$-dimensional IIR filter layer

---

1. Input $x \in \mathbb{R}^{B \times N_1 \times \cdots \times N_D \times C_{\text{in}}}$, where $N_d$ denotes the length of axis $d$
2. Define $L$ as the maximum delay per axis
3. Prepare circularly shifted copies of $x$, denoted by $X \in \mathbb{R}^{B \times N_1 \times \cdots \times N_D \times C_{\text{in}} \times (L+1)^D \times C_{\text{out}}}$
4. Initialize feedforward coefficients $\mathcal{B} \in \mathbb{R}^{C_{\text{in}} \times (L+1)^D \times C_{\text{out}}}$
5. Prepare feedback coefficients $\mathcal{A} \in \mathbb{R}^{C_{\text{in}} \times (L+1)^D \times C_{\text{out}}}$
   (a) Initialize $a \in \mathbb{R}^{C_{\text{in}} \times ((L+1)^D - 1) \times C_{\text{out}}}$
   (b) Prepare $a_0 \in \mathbb{R}^{C_{\text{in}} \times 1 \times C_{\text{out}}}$ with **ones** and keep it nontrainable
   (c) $\mathcal{A} := [a_0, a]$
6. Initialize the local variable $y \in \mathbb{R}^{B \times N_1 \times \cdots \times N_D \times C_{\text{in}} \times C_{\text{out}}}$, for example with random values
7. Prepare circularly shifted copies of $y$, denoted by $Y \in \mathbb{R}^{B \times N_1 \times \cdots \times N_D \times C_{\text{in}} \times (L+1)^D \times C_{\text{out}}}$
8. Perform local gradient refinement for at most $T$ steps with step size $\eta$, and denote the refined local variable at termination by $\widehat{y}$:

$$y^{(t+1)}_{bn_1 \cdots n_D co} = y^{(t)}_{bn_1 \cdots n_D co} - \eta \frac{\partial}{\partial y^{(t)}_{bn_1 \cdots n_D co}} \left\| \sum_{c',d} \mathcal{B}_{c'do} X_{bn_1 \cdots n_D c'do} - \sum_{c',d} \mathcal{A}_{c'do} Y_{bn_1 \cdots n_D c'do} \right\|_2^2$$

9. Return $y_{bn_1 \cdots n_D o} = \sum_c \widehat{y}_{bn_1 \cdots n_D co}$
10. The outer global optimization by backpropagation adjusts $\mathcal{A}$ and $\mathcal{B}$

---

**Gradient-based training** For the implemented finite-support realization, let $\Omega_N = \{0, \ldots, N_1 - 1\} \times \cdots \times \{0, \ldots, N_D - 1\}$ denote the finite spatial grid, with $|\Omega_N| = \prod_{d=1}^{D} N_d$ residual samples per input item. For a mini-batch of size $B$, the implementation sums the same residual over the batch index and over all grid locations. This is a deterministic finite-grid consistency objective, and no statistical generalization bound is claimed here. Define the residual field in (4.3) and the associated IIR consistency loss in (4.4) as

$$r[\boldsymbol{n}] = \sum_{\boldsymbol{k} \in \Lambda_L} b[\boldsymbol{k}]\, x[\boldsymbol{n} - \boldsymbol{k}] - y[\boldsymbol{n}] - \sum_{\boldsymbol{k} \in \Lambda_L \setminus \{\boldsymbol{0}\}} a[\boldsymbol{k}]\, y[\boldsymbol{n} - \boldsymbol{k}]. \tag{4.3}$$

$$\mathcal{L}_{\text{IIR}}(x, y; a, b) = \sum_{\boldsymbol{n} \in \Omega_N} \left\| r[\boldsymbol{n}] \right\|_2^2. \tag{4.4}$$

For fixed $x$ and $y$, the gradients of $\mathcal{L}_{\text{IIR}}$ with respect to the numerator and denominator coefficients are

$$\frac{\partial \mathcal{L}_{\text{IIR}}}{\partial b[\boldsymbol{k}]} = 2 \sum_{\boldsymbol{n} \in \Omega_N} r[\boldsymbol{n}]\, x[\boldsymbol{n} - \boldsymbol{k}], \qquad \boldsymbol{k} \in \Lambda_L, \tag{4.5}$$

$$\frac{\partial \mathcal{L}_{\mathrm{IIR}}}{\partial a[\boldsymbol{k}]} = -2 \sum_{\boldsymbol{n}\in\Omega_N} r[\boldsymbol{n}]\, y[\boldsymbol{n}-\boldsymbol{k}], \qquad \boldsymbol{k} \in \Lambda_L \setminus \{\boldsymbol{0}\}. \tag{4.6}$$

These expressions characterize the residual-consistency objective associated with the realized IIR relation. In the implementation, the inner gradient-descent loop explicitly updates only the lifted local variable $y$. The trainable coefficients are then learned by standard outer backpropagation through the host network that contains the layer.

**Computational complexity** Let $L$ denote the maximum delay per axis in the true multidimensional IIR layer. In one dimension, for an input of length $N_1$ with $C_{\mathrm{in}}$ input channels and $C_{\mathrm{out}}$ output channels, the circular-shift construction forms $X$ once per forward call and reconstructs $Y$ inside each residual evaluation, with tensors of size proportional to $N_1\, C_{\mathrm{in}}(L+1)\, C_{\mathrm{out}}$ per batch element. Each residual evaluation then performs two tensor contractions of this size, and each inner gradient step introduces only a constant-factor overhead through differentiation of the same objective. For $D = 2$ on an $N_1 \times N_2$ grid the number of circular shifts grows to $(L+1)^2$, and for $D = 3$ on an $N_1 \times N_2 \times N_3$ grid it grows to $(L+1)^3$. In general, for a $D$-dimensional input with batch size $B$, an upper bound on the dominant arithmetic cost of one forward call together with at most $T$ inner refinement steps behaves as

$$C_{\mathrm{true}} = O\!\left(B \left(\prod_{d=1}^{D} N_d\right) C_{\mathrm{in}}\, C_{\mathrm{out}}\, (L+1)^D\, (T+1)\right) \tag{4.7}$$

where $T$ is the maximum number of local gradient-descent steps allowed by the inner solve, and the additive 1 accounts for the initial residual evaluation before the while loop. The peak forward-call memory footprint has the same dependence on $B$, $N_1, \ldots, N_D$, $C_{\mathrm{in}}$, $C_{\mathrm{out}}$, and $L$, but not on $T$, because the shifted banks are materialized per evaluation rather than stored across all inner steps. The $(L+1)^D$ factor shows how quickly the cost grows when either the spatial dimension or the number of delays increases. This is the practical limitation of the true multidimensional IIR realizations. The current `IIRD` implementation provides learnable true IIR layers up to $D = 3$.

#### 4.1.1.1 1D realization

The one-dimensional case is the corresponding realized form of (4.1) on a finite signal of length $N$. It is written as

$$y[n] = \sum_{k=0}^{L} b_k x[n-k] - \sum_{k=1}^{L} a_k y[n-k]. \tag{4.8}$$

Here $n$ is the sample index and $k$ is the delay index. With a single delay, (4.8) becomes

$$b_0 x[n] + b_1 x[n-1] - y[n] - a_1 y[n-1] = 0. \tag{4.9}$$

Algorithm 4.15 lists the corresponding tensor construction for the realized 1D layer, and Figure 4.1 shows the poles and zeros of a 1D realization with 40 delays.

**Algorithm 4.15:** Implemented IIR 1D layer

1. Input $x \in \mathbb{R}^{B \times N \times C_{\text{in}}}$
2. Prepare circularly shifted copies of $x$, denoted by $X \in \mathbb{R}^{B \times N \times C_{\text{in}} \times (L+1) \times C_{\text{out}}}$
3. Initialize feedforward coefficients $\mathcal{B} \in \mathbb{R}^{C_{\text{in}} \times (L+1) \times C_{\text{out}}}$
4. Prepare feedback coefficients $\mathcal{A} \in \mathbb{R}^{C_{\text{in}} \times (L+1) \times C_{\text{out}}}$
   (a) Initialize $a \in \mathbb{R}^{C_{\text{in}} \times L \times C_{\text{out}}}$
   (b) Prepare $a_0 \in \mathbb{R}^{C_{\text{in}} \times 1 \times C_{\text{out}}}$ with **ones** and keep it nontrainable
   (c) $\mathcal{A} := \text{Concatenate}([a_0, a], \text{axis} = 1)$
5. Initialize the local variable $y \in \mathbb{R}^{B \times N \times C_{\text{in}} \times C_{\text{out}}}$, for example with random values
6. Prepare circularly shifted copies of $y$, denoted by $Y \in \mathbb{R}^{B \times N \times C_{\text{in}} \times (L+1) \times C_{\text{out}}}$
7. Perform local gradient refinement for at most $T$ steps with step size $\eta$, and denote the refined local variable at termination by $\widehat{y}$:

$$y_{bnco}^{(t+1)} = y_{bnco}^{(t)} - \eta \frac{\partial}{\partial y_{bnco}^{(t)}} \left\| \sum_{c',d} \mathcal{B}_{c'do} X_{bnc'do} - \sum_{c',d} \mathcal{A}_{c'do} Y_{bnc'do} \right\|_2^2$$

8. Return $y_{bno} = \sum_c \widehat{y}_{bnco}$
9. The outer global optimization by backpropagation adjusts $\mathcal{A}$ and $\mathcal{B}$

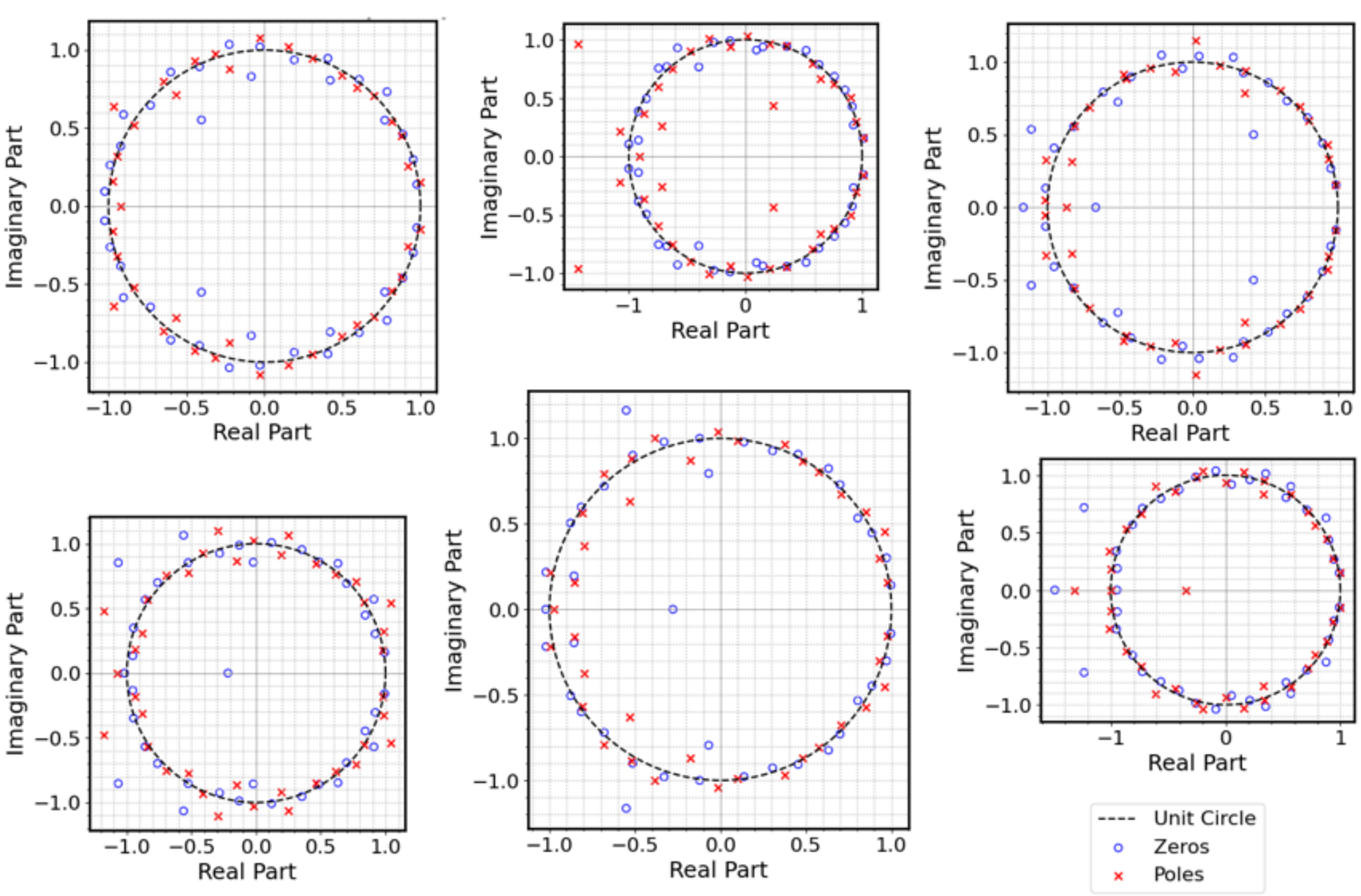

Figure 4.1: Poles and zeros of an IIR 1D layer with 6 trainable filters with 40 delays in each filter

#### 4.1.1.2 2D realization

On a two-dimensional rectangular input grid, (4.1) becomes

$$y\left[n_1, n_2\right] = \sum_{k_1=0}^{L} \sum_{k_2=0}^{L} b_{k_1,k_2} x\left[n_1 - k_1, n_2 - k_2\right] - \sum_{\substack{0 \le k_1 \le L \\ 0 \le k_2 \le L \\ (k_1,k_2) \ne (0,0)}} a_{k_1,k_2} y\left[n_1 - k_1, n_2 - k_2\right]. \tag{4.10}$$

Here $(n_1, n_2)$ is the pixel index, $(k_1, k_2)$ is the 2D delay index, and $N_1$ and $N_2$ denote the lengths of the two spatial axes of the finite rectangular input tensor used by the layer. With a single delay, (4.10) becomes

$$\begin{aligned} y\left[n_1, n_2\right] &= b_{0,0} x\left[n_1, n_2\right] \\ &+ b_{0,1} x\left[n_1, n_2 - 1\right] + b_{1,0} x\left[n_1 - 1, n_2\right] + b_{1,1} x\left[n_1 - 1, n_2 - 1\right] \\ &- \left(a_{0,1} y\left[n_1, n_2 - 1\right] + a_{1,0} y\left[n_1 - 1, n_2\right] + a_{1,1} y\left[n_1 - 1, n_2 - 1\right]\right) \end{aligned}$$

or, equivalently,

$$\begin{aligned} &b_{0,0} x\left[n_1, n_2\right] + b_{0,1} x\left[n_1, n_2 - 1\right] + b_{1,0} x\left[n_1 - 1, n_2\right] + b_{1,1} x\left[n_1 - 1, n_2 - 1\right] \\ &- y\left[n_1, n_2\right] - a_{0,1} y\left[n_1, n_2 - 1\right] - a_{1,0} y\left[n_1 - 1, n_2\right] - a_{1,1} y\left[n_1 - 1, n_2 - 1\right] = 0. \end{aligned} \tag{4.11}$$

Algorithm 4.16 gives the corresponding tensor construction for the realized 2D layer. There, $n$ and $m$ denote the two spatial indices of the rectangular grid. Figure 4.2 shows the contours of a 2D realization with four filters and two delays.

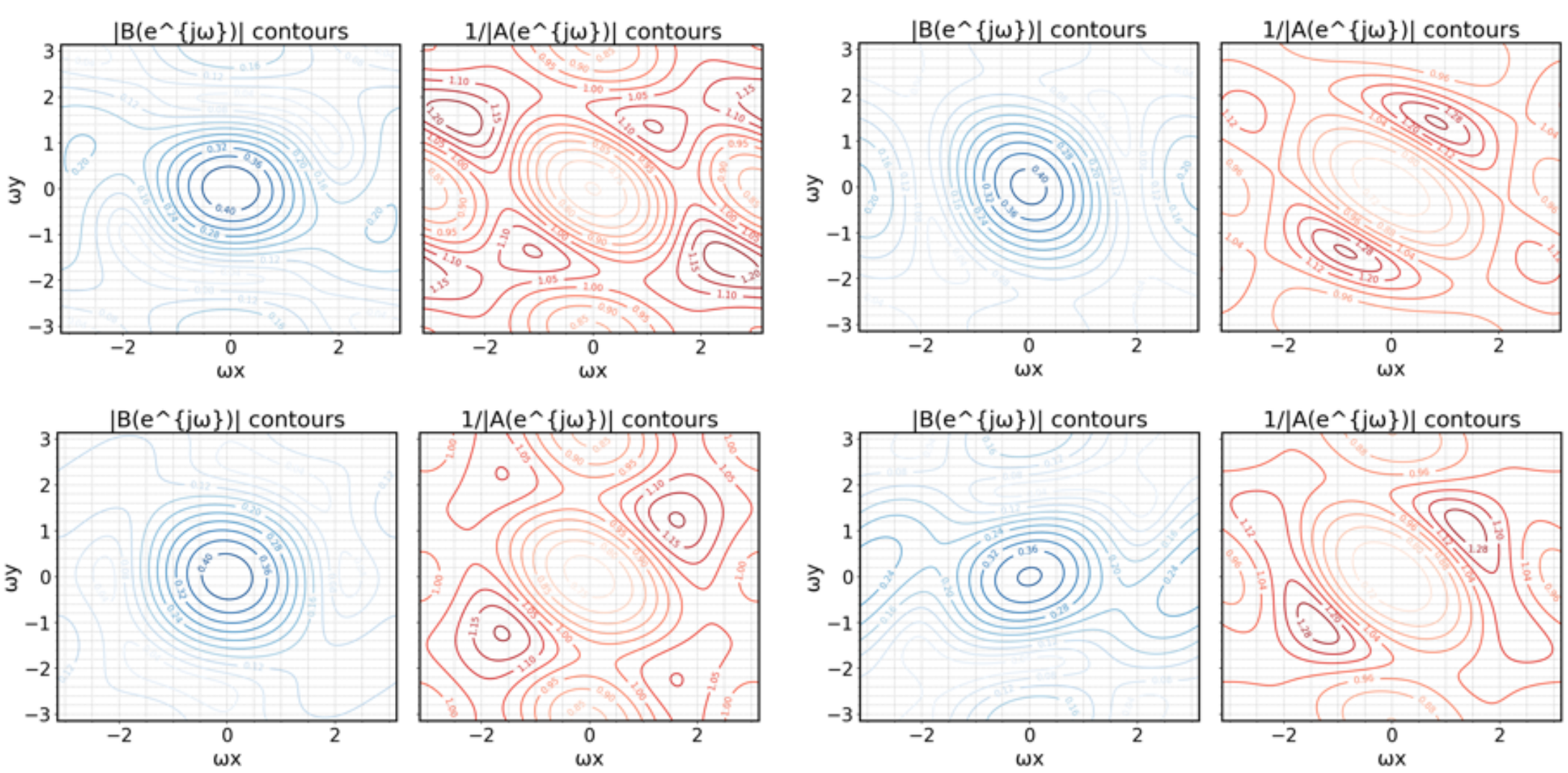

Figure 4.2: Contours of an IIR 2D layer with four filters with two delays in each filter

**Algorithm 4.16:** Implemented IIR 2D layer

1. Input $x$ of shape $(B, N_1, N_2, C_{\text{in}})$
2. Prepare circularly shifted copies of $x$, denoted by $X$ of shape $(B, N_1, N_2, C_{\text{in}}, (L+1)^2, C_{\text{out}})$
3. Initialize feedforward coefficients $\mathcal{B}$ of shape $(C_{\text{in}}, (L+1)^2, C_{\text{out}})$
4. Prepare feedback coefficients $\mathcal{A}$ of shape $(C_{\text{in}}, (L+1)^2, C_{\text{out}})$
   (a) Initialize $a$ with shape $(C_{\text{in}}, (L+1)^2 - 1, C_{\text{out}})$
   (b) Prepare $a_0$ with **ones** of shape $(C_{\text{in}}, 1, C_{\text{out}})$
   (c) $\mathcal{A} := [a_0, a]$
5. Initialize the local variable $y$ of shape $(B, N_1, N_2, C_{\text{in}}, C_{\text{out}})$, for example with random values
6. Prepare circularly shifted copies of $y$, denoted by $Y$ of shape $(B, N_1, N_2, C_{\text{in}}, (L+1)^2, C_{\text{out}})$
7. Perform local gradient refinement for at most $T$ steps with step size $\eta$, and denote the refined local variable at termination by $\widehat{y}$:

$$y^{(t+1)}_{bnmco} = y^{(t)}_{bnmco} - \eta \frac{\partial}{\partial y^{(t)}_{bnmco}} \left\| \sum_{c',d} \mathcal{B}_{c'do} X_{bnmc'do} - \sum_{c',d} \mathcal{A}_{c'do} Y_{bnmc'do} \right\|_2^2$$

8. Return $y_{bnmo} = \sum_c \widehat{y}_{bnmco}$
9. The outer global optimization by backpropagation adjusts $\mathcal{A}$ and $\mathcal{B}$

#### 4.1.1.3 3D realization

On a three-dimensional volumetric input grid, (4.1) becomes

$$y[n_1, n_2, n_3] = \sum_{k_1=0}^{L} \sum_{k_2=0}^{L} \sum_{k_3=0}^{L} b_{k_1,k_2,k_3} x[n_1 - k_1, n_2 - k_2, n_3 - k_3] - \sum_{\substack{0 \le k_1 \le L \\ 0 \le k_2 \le L \\ 0 \le k_3 \le L \\ (k_1,k_2,k_3) \neq (0,0,0)}} a_{k_1,k_2,k_3} y[n_1 - k_1, n_2 - k_2, n_3 - k_3]. \tag{4.12}$$

Here $(n_1, n_2, n_3)$ is the voxel index, $(k_1, k_2, k_3)$ is the 3D delay index, and $N_1$, $N_2$, and $N_3$ denote the lengths of the three spatial axes of the finite volumetric input tensor used by the layer. Algorithm 4.17 gives the corresponding tensor construction for the realized 3D layer. There, $n$, $m$, and $l$ denote the three spatial indices of the volumetric grid. Figure 4.3 shows the magnitude responses of a 3D realization with eight filters and one delay.

**Algorithm 4.17:** Implemented IIR 3D layer

1. Input $x$ of shape $(B, N_1, N_2, N_3, C_{\text{in}})$
2. Prepare circularly shifted copies of $x$, denoted by $X$ of shape $(B, N_1, N_2, N_3, C_{\text{in}}, (L+1)^3, C_{\text{out}})$
3. Initialize feedforward coefficients $\mathcal{B}$ of shape $(C_{\text{in}}, (L+1)^3, C_{\text{out}})$
4. Prepare feedback coefficients $\mathcal{A}$ of shape $(C_{\text{in}}, (L+1)^3, C_{\text{out}})$
   (a) Initialize $a$ with shape $(C_{\text{in}}, (L+1)^3 - 1, C_{\text{out}})$
   (b) Prepare $a_0$ with **ones** of shape $(C_{\text{in}}, 1, C_{\text{out}})$
   (c) $\mathcal{A} := [a_0, a]$
5. Initialize the local variable $y$ of shape $(B, N_1, N_2, N_3, C_{\text{in}}, C_{\text{out}})$, for example with random values
6. Prepare circularly shifted copies of $y$, denoted by $Y$ of shape $(B, N_1, N_2, N_3, C_{\text{in}}, (L+1)^3, C_{\text{out}})$
7. Perform local gradient refinement for at most $T$ steps with step size $\eta$, and denote the refined local variable at termination by $\widehat{y}$:
$$y^{(t+1)}_{bnmlco} = y^{(t)}_{bnmlco} - \eta \frac{\partial}{\partial y^{(t)}_{bnmlco}} \left\| \sum_{c',d} \mathcal{B}_{c'do} X_{bnmlc'do} - \sum_{c',d} \mathcal{A}_{c'do} Y_{bnmlc'do} \right\|_2^2$$
8. Return $y_{bnmlo} = \sum_c \widehat{y}_{bnmlco}$
9. The outer global optimization by backpropagation adjusts $\mathcal{A}$ and $\mathcal{B}$

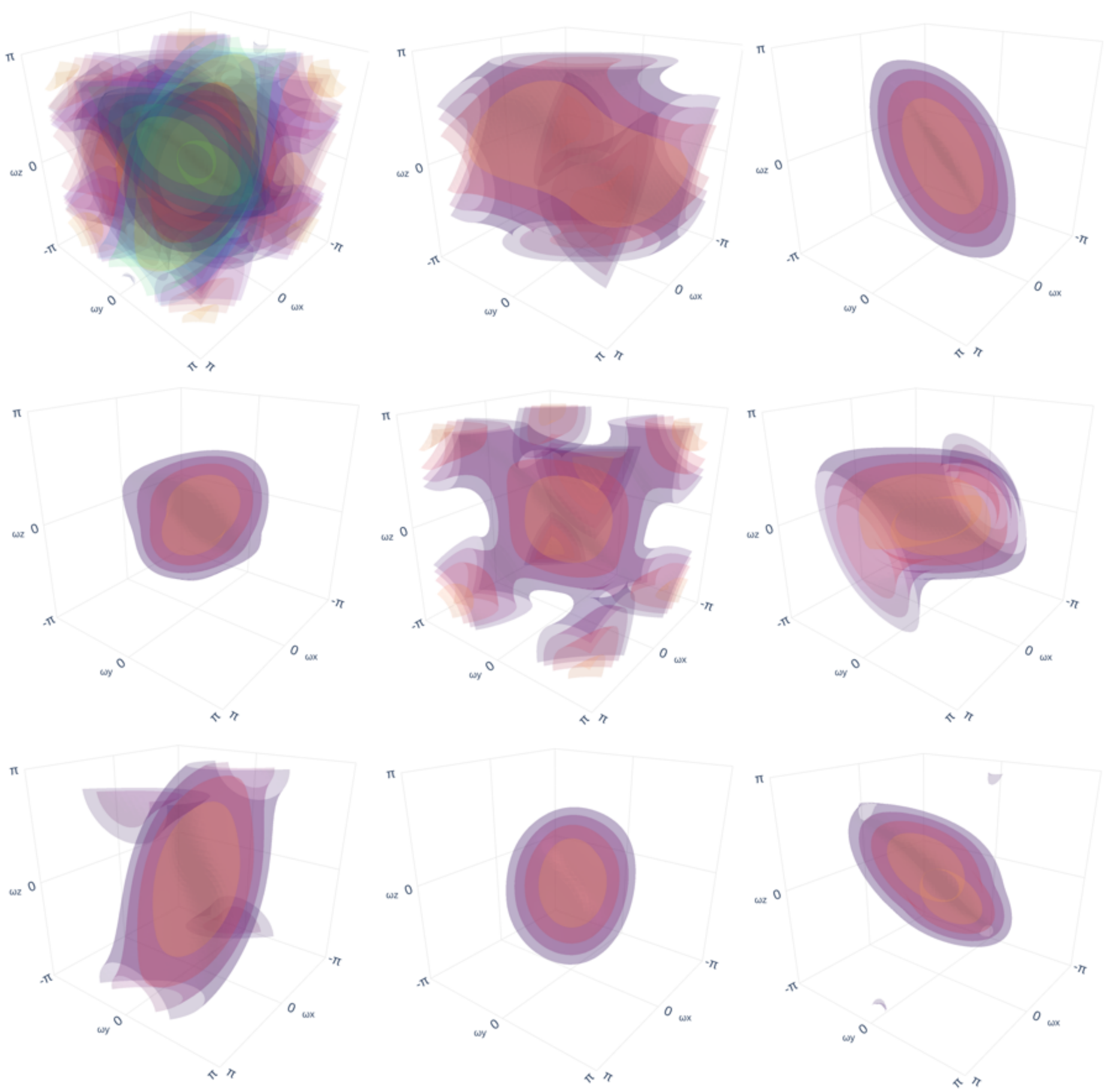

Figure 4.3: Magnitude responses of the eight filters of an IIR 3D layer with one delay in each filter

### 4.1.2 Separable two–dimensional IIR filter approximation

**Notations and symbols** Let $\Omega \subset \mathbb{Z}^2$ be the pixel grid and $x : \Omega \to \mathbb{R}^{C_{\text{in}}}$ the input, $y : \Omega \to \mathbb{R}^{C_{\text{out}}}$ the output. We denote by $i$ the row (vertical, axis 1) index and $j$ the column (horizontal, axis 2) index. We use backward shifts $(S_1^p f)[i,j] = f[i-p,j]$, $(S_2^q f)[i,j] = f[i,j-q]$, where $S_1$ and $S_2$ act along the first (vertical) and second (horizontal) spatial axes, respectively. For one–dimensional (1D) sequences we write $s[t]$ with discrete index $t \in \mathbb{Z}$. The variables $z_1$ and $z_2$ denote the $z$–transform indeterminates associated with the two spatial axes, so that factors $z_1^{-p}$ and $z_2^{-q}$ correspond to shifts by $S_1^p$ and $S_2^q$ in the spatial domain. An ideal 2D IIR or ARMA filter is specified by $z$-domain polynomials

$$A(z_1, z_2) = 1 - \sum_{(p,q)\neq(0,0)} a_{p,q}\, z_1^{-p} z_2^{-q}, \tag{4.13a}$$

$$B(z_1, z_2) = \sum_{p,q\geq 0} b_{p,q}\, z_1^{-p} z_2^{-q}, \tag{4.13b}$$

with transfer function $H(z_1, z_2) = B(z_1, z_2)/A(z_1, z_2)$. In the spatial domain this corresponds to

$$\boxed{y - \sum_{(p,q)\neq(0,0)} a_{p,q}\, S_1^p S_2^q\, y \;=\; \sum_{p,q\geq 0} b_{p,q}\, S_1^p S_2^q\, x.} \tag{4.14}$$

where $p, q \in \mathbb{Z}$ always denote integer lag indices along the first and second spatial axes, respectively.

We adopt a *separable* two-dimensional (2D) *auto-regressive moving-average* (ARMA) operator formed by successive composition of two one-dimensional (1D) recursions along the two spatial axes. In the implemented block, this separable IIR operator is preceded by a learned Conv2D front-end, so that a purely FIR stage first produces local feature maps which are then processed by the separable ARMA core.

#### 4.1.2.1 Axis-wise Direct form II recursion

For a generic one-dimensional input sequence $s[t]$, let $\mathsf{F}$ denote the forward 1D recursion

$$\mathsf{F}(s): \quad \hat{s}[t] - \sum_{p=1}^{P} \alpha_p\, \hat{s}[t-p] \;=\; \sum_{q=0}^{Q} \beta_q\, s[t-q], \tag{4.15}$$

with $P$ the AR order (feedback delays) and $Q$ the MA order. Here $t \in \mathbb{Z}$ indexes position along a single axis and $\hat{s}[t]$ is the corresponding output. The coefficients $\{\alpha_p\}_{p=1}^{P}$ and $\{\beta_q\}_{q=0}^{Q}$ are the AR and MA taps for a given axis and channel. They are stored as vectors $a = [1, \alpha_1, \dots, \alpha_P]^\top$ and $b = [\beta_0, \dots, \beta_Q]^\top$ with $a_0 = 1$.

**Stability** For each axis $k \in \{1, 2\}$ and pole index $r$, we introduce an unconstrained real parameter $u_r^{(k)}$ and map it to a stable pole via

$$p_r^{(k)} = \tanh u_r^{(k)}, \tag{4.16}$$

so that $|p_r^{(k)}| < 1$ by construction. An optional clip radius $r_{\max} \in (0, 1)$ enforces $|p_r^{(k)}| \le r_{\max}$, and an optional stability loss penalizes poles whose magnitude approaches unity (e.g., $|p_r^{(k)}| \approx 1$), encouraging a safety margin inside the unit circle.

**Phase modes** The temporal behavior along each axis is controlled by a phase-mode flag $\phi \in \{\texttt{none}, \texttt{avg}, \texttt{filtfilt}\}$. When $\phi$ is `'none'`, the layer uses the causal recursion (4.15) on the forward sequence only. For `'avg'`, the same DF-II recursion is run on the sequence and on its time-reversed version, and the two outputs are averaged, producing a bidirectional, low-phase-distortion response. For `'filtfilt'`, the forward output is filtered again in reverse, yielding an effectively zero-phase response along that axis. In summary, letting $\mathsf{R}(s)$ denote the same recursion run on the time-reversed sequence, we have following three phase modes:

1. *Causal* (forward) := $\mathsf{F}$,
2. *Bidirectional* (average) := $\frac{1}{2}(\mathsf{F} + \mathsf{R})$,
3. *Zero-phase* (forward and backward) := $\mathsf{R} \circ \mathsf{F}$.

**Separable 2D ARMA composition** With a single global phase $\phi$, define $\mathsf{F}_1^{(\phi)}$ and $\mathsf{F}_2^{(\phi)}$ as the 1D DF-II operators along the first (vertical) axis and second (horizontal) axis, and apply the separable filter as

$$\mathsf{H}^{(\phi)} \;=\; \mathsf{F}_1^{(\phi)} \circ \mathsf{F}_2^{(\phi)}, \tag{4.17}$$

sharing the same phase on both axes. In other words, the 2D separable ARMA filtering is a width pass and a height pass in sequence using a *single global* phase setting (no independent per-axis choice; the legacy `bidirectional` flag maps to `avg`).

Let $v = k * x$ denote the output of the FIR convolution for some learnable kernel $k$, and let $y_{\text{iir}} = \mathsf{H}^{(\phi)}[v]$ be the output after the separable ARMA block with fixed phase $\phi$. We apply DF-II filtering $\mathsf{F}_2[v]$ along width (axis 2) with coefficients $(\alpha_p^{(2)}, \beta_q^{(2)})$ to obtain an intermediate field, then apply DF-II filtering again, $\mathsf{F}_1[\mathsf{F}_2[v]]$, along height (axis 1) with coefficients $(\alpha_p^{(1)}, \beta_q^{(1)})$ to get $y_{\text{iir}}$. In the $z$-domain this separable operation can be written as

$$H(z_1, z_2) = H_1(z_1)\, H_2(z_2) = \frac{B_1(z_1)}{A_1(z_1)} \cdot \frac{B_2(z_2)}{A_2(z_2)}, \tag{4.18}$$

where

$$A_1(z_1) = \prod_{r=1}^{P} \left(1 - p_r^{(1)} z_1^{-1}\right) = 1 - \sum_{p=1}^{P} \alpha_p^{(1)} z_1^{-p}, \tag{4.19a}$$

$$B_1(z_1) = \sum_{q=0}^{Q} \beta_q^{(1)} z_1^{-q}, \tag{4.19b}$$

$$A_2(z_2) = \prod_{r=1}^{P} \left(1 - p_r^{(2)} z_2^{-1}\right) = 1 - \sum_{p=1}^{P} \alpha_p^{(2)} z_2^{-p}, \tag{4.19c}$$

$$B_2(z_2) = \sum_{q=0}^{Q} \beta_q^{(2)} z_2^{-q}. \tag{4.19d}$$

The corresponding operator equation is

$$\boxed{\big(A_1(S_1)A_2(S_2)\big)\, y \;=\; \big(B_1(S_1)B_2(S_2)\big)\, v.} \tag{4.20}$$

**Parameterization and axis sharing** Each separable IIR block is parameterized per output channel by a set of autoregressive (AR) poles and moving–average (MA) coefficients along the two spatial axes. Along axis 1 (vertical), the AR part is defined by poles $\{p_r^{(1)}\}_{r=1}^{P}$, obtained from unconstrained real parameters $\{u_r^{(1)}\}$ via $p_r^{(1)} = \tanh u_r^{(1)}$, so that $|p_r^{(1)}| < 1$ and stability is preserved. These poles generate the AR polynomial $A_1(S_1)$, whose coefficients we denote by $\{\alpha_p^{(1)}\}_{p=1}^{P}$. The MA part uses signed numerator coefficients $\{\beta_q^{(1)}\}_{q=0}^{Q}$, produced via a tanh mapping and collected in the polynomial $B_1(S_1)$. Axis 2 (horizontal) is treated analogously, with poles $\{p_r^{(2)}\}_{r=1}^{P}$ and MA coefficients $\{\beta_q^{(2)}\}_{q=0}^{Q}$ forming $A_2(S_2)$ and $B_2(S_2)$ and corresponding AR taps $\{\alpha_p^{(2)}\}_{p=1}^{P}$. Axis sharing is controlled by a binary flag `per_axis` which either reuses the same one-dimensional polynomials on both axes or allows each axis to have its own set of parameters. When `per_axis=`**False**, we enforce $A_2(S_2) \equiv A_1(S_1)$ and $B_2(S_2) \equiv B_1(S_1)$, so the same 1D filters are used in both spatial directions. When `per_axis=`**True**, the parameters are decoupled and the two axes learn independent sets $\{\alpha_p^{(1)}, \beta_q^{(1)}\}$ and $\{\alpha_p^{(2)}, \beta_q^{(2)}\}$.

**Boundary conditions and finite maps** Let the maximum lag be $L = \max(P, Q)$, and at each axis the sequences are left pad of size $L$. Two boundary models are supported, $\mathcal{B} \in \{\text{reflect, constant}\}$. For reflect, the implementation uses left reflection of the sequence; if the length $T \le L$ it automatically falls back to constant zero padding to avoid invalid mirror indices on tiny feature maps.

**Arithmetic complexity** A single pass of the 1D recursion in (4.15) costs $(P+Q+1)$ Multiplication–Accumulations (MACs) per pixel *per channel*. Applying it along two axes doubles this cost, and running forward–reverse doubles it again, so both *average* and *zero-phase* modes require $4(P+Q+1)$ MACs per pixel per channel. If channels are mixed with a $1\times1$ pointwise map, then we add $C_{\text{in}}C_{\text{out}}$ MACs. A standard $3\times3$ FIR stage costs $9\,C_{\text{in}}C_{\text{out}}$ MACs. For typical widths with $C_{\text{in}} \approx C_{\text{out}} = C$, the quadratic term in $C$ dominates, so *(separable ARMA + $1\times1$)* has a substantially smaller arithmetic footprint than a $3\times3$ FIR stage, while introducing the orders $(P, Q)$ as additional knobs (e.g., two delays) to tune the effective spatial memory. Figure 4.4 summarizes the separable 2D DF–II realization used in our IIR blocks.

#### 4.1.2.2 DC normalization at two levels

To keep the response of each IIR block numerically well behaved, we include two optional DC guards, one in the 2D convolutional front–end, and one in the separable ARMA part of the filter.

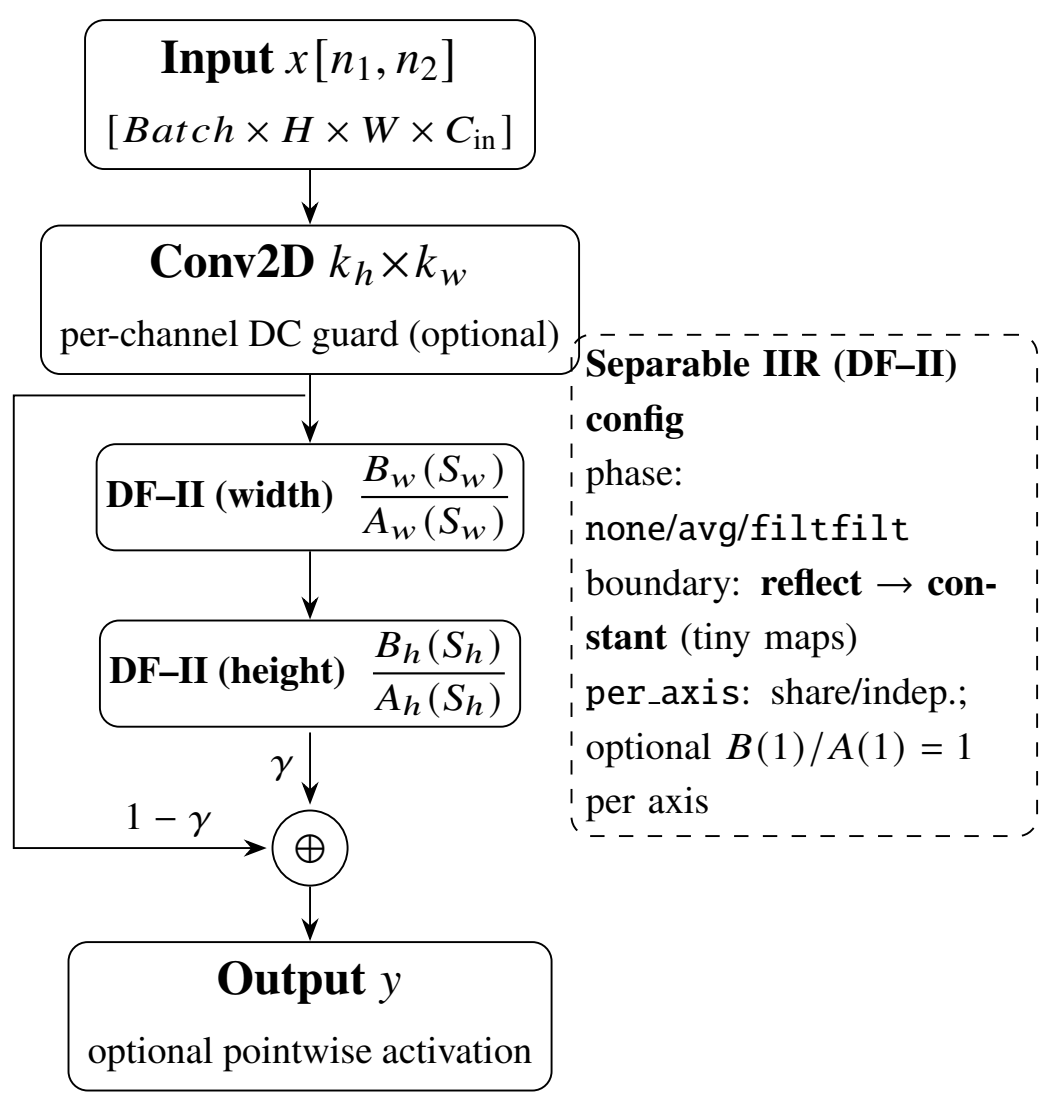

Figure 4.4: Separable 2D DF–II IIR block formed by a 2D FIR front-end followed by two axis-wise DF–II ARMA passes (width then height), a residual gate, and an optional pointwise output activation; phase, boundary, per-axis, and DC-normalization options are summarized in the configuration box on the right.

**Normalizing convolutional front-end** Let the 2D FIR kernel have entries $k_{r,s,c\to o}$, where $r$ and $s$ index the kernel row and column offsets, $c$ indexes the input channel, and $o$ the output channel. For each output channel $o$, the DC response of the convolution kernel is

$$K_o = \sum_{r,s,c} k_{r,s,c\to o}. \tag{4.21}$$

When the DC guard is enabled in the FIR part (flag `block_dc_normalize`), the corresponding feature map $v_o$ is rescaled as

$$v_o \leftarrow \gamma_o v_o, \qquad \gamma_o = \text{clip}_{[0.1,\,10]}\Big(\frac{1}{|K_o| + 10^{-6}}\Big) \tag{4.22}$$

which prevents the effective gain from becoming extremely small or large by clamping the rescaling factor to the interval $[0.1, 10]$, and drives the DC magnitude toward one while preserving the sign of $K_o$.

**Normalizing in separable ARMA branch** For each axis $k \in \{1, 2\}$ we write

$$A_k(z) = 1 - \sum_{n\geq 1} \alpha_n^{(k)} z^{-n}, \quad B_k(z) = \sum_{q\geq 0} \beta_q^{(k)} z^{-q}, \tag{4.23}$$

so that $A_k(1) = 1 - \sum_n \alpha_n^{(k)}$ and $B_k(1) = \sum_q \beta_q^{(k)}$ are the 1D DC responses of the AR and MA parts along that axis. When the ARMA DC normalizer is enabled (flag `dc_normalize`), we compute per–channel scaling factors

$$s_k = \frac{1 - \sum_n \alpha_n^{(k)}}{\sum_q \beta_q^{(k)} + 10^{-6}}, \tag{4.24}$$

clip each $s_k$ to $[0.1, 10]$, and multiply all MA coefficients on that axis by $s_k$. Without clipping this choice would enforce $B_k(1)/A_k(1) = 1$ and thus make the one–dimensional DC gain exactly unity; with clipping, the DC response is kept close to one while avoiding pathological amplification or attenuation.

**Backpropagation calculus along each axis** For each axis $k \in \{1, 2\}$ let $s^{(k)}[t]$ denote the input sequence, $\hat{s}^{(k)}[t]$ the Direct Form II output, and $g^{(k)}[t]$ the gradient of a scalar loss $\mathcal{L}$ with respect to $\hat{s}^{(k)}[t]$. The gradients of $\mathcal{L}$ with respect to the moving average and autoregressive taps on axis $k$ are

$$\frac{\partial \mathcal{L}}{\partial \beta_q^{(k)}} = \sum_t g^{(k)}[t]\, s^{(k)}[t-q], \tag{4.25}$$

$$\frac{\partial \mathcal{L}}{\partial \alpha_p^{(k)}} = -\sum_t g^{(k)}[t]\, \hat{s}^{(k)}[t-p]. \tag{4.26}$$

These expressions correspond to one dimensional correlations between the back propagated gradient and shifted copies of the input or the Direct Form II output and they are applied on the vertical and horizontal axes before the chain rule maps them to the unconstrained pole parameters used in this thesis.

The separable IIR realizations described above allow us to model long-range dependencies through trainable recursions along individual spatial axes. However, they do not by themselves provide an explicit multiscale directional coefficient representation. The remainder of this chapter therefore shifts from recursive filtering to multiresolution attention. WaveletViT uses DWT bands to provide a compact scale-aware representation, while ShearViT uses FDST shearlet bands for directional and anisotropic structures such as edges, surfaces, and elongated features in volumetric images and physical fields. The DWT and FDST banks were introduced in Chapter 2, and here they are used as coefficient interfaces for frugal attention modules.

## 4.2 Multiresolution frugal visual attentions

**Vanilla scaled inner product self-attention** Self-attention is a mechanism that enables a model to relate different positions of a single sequence or tensor to compute a representation of the input. Originally popularized by the transformer model in natural language processing, the concept generalizes naturally to data with more than one spatial or temporal dimension, such as images (2D), videos (3D), or higher-dimensional scientific data (Vaswani et al., 2017) [24].

$$\text{Attention}\big(Q, K, V\big) = \text{softmax}\left(\frac{QK^\top}{\sqrt{d_k}}\right)V \tag{4.27}$$

Figure 4.5 illustrates the standard query, key, and value computation used in scaled dot-product self-attention.

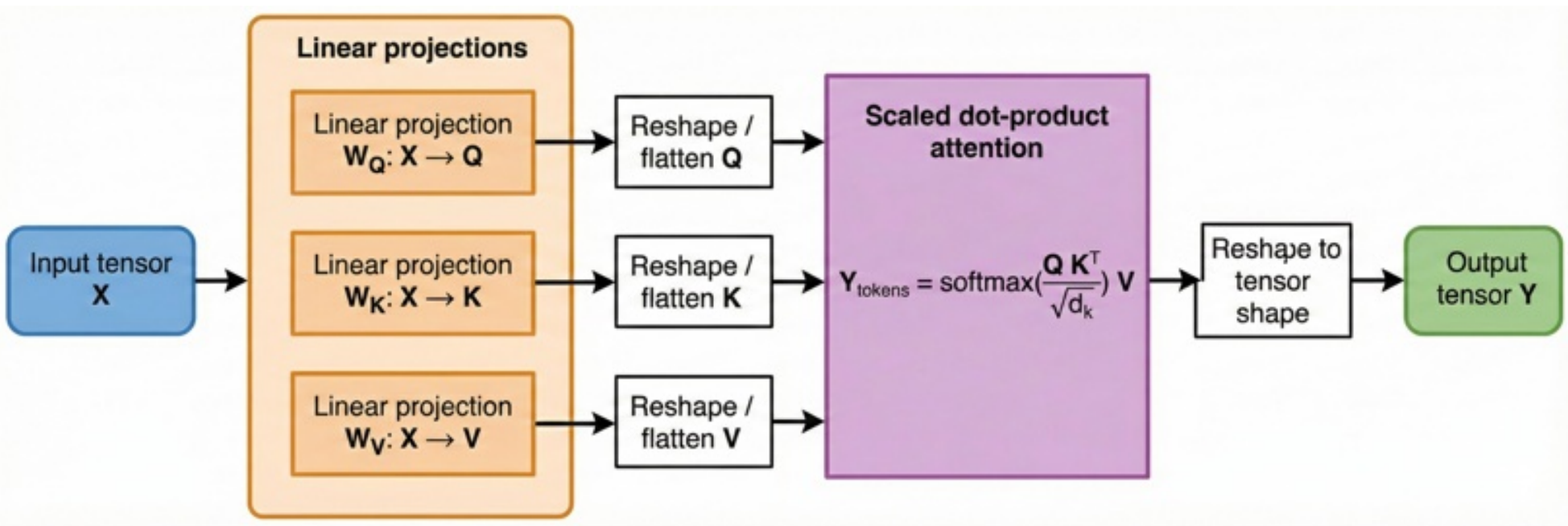

Figure 4.5: Architecture of self attention

**Spatial attention** Spatial attention assigns a learned importance weight to each spatial location in a feature map. Let $x[\boldsymbol{n}, c]$ denote the input features, where $\boldsymbol{n}$ is the spatial index on the lattice and $c$ is the channel index, and let $w_{\text{sp}}$ be a small convolution kernel that produces a single channel score map. The spatial attention weights are

$$a_{\text{spatial}}[\boldsymbol{n}] = \sigma\big((w_{\text{sp}} * x)[\boldsymbol{n}]\big), \tag{4.28}$$

and the attended feature map is

$$y[\boldsymbol{n}, c] = a_{\text{spatial}}[\boldsymbol{n}]\, x[\boldsymbol{n}, c]. \tag{4.29}$$

This construction emphasises locations where the learned score is high and suppresses background regions while leaving the channel structure unchanged. Convolutional spatial attention of this form has been reported to improve image classification and segmentation performance when added on top of standard convolutional backbones [210–212].

**Channel attention** Channel attention assigns a learned importance weight to each feature channel in a convolutional block. Consider a feature tensor $x[\boldsymbol{n}, c]$ where $\boldsymbol{n} \in \Omega$ denotes the spatial index on the lattice and $c \in \{1, \dots, C\}$ denotes the channel index, and where $\Omega$ is the set of all spatial sites of the feature map and $|\Omega|$ is its cardinality. A global average pooling step first computes one descriptor per channel,

$$s[c] = \frac{1}{|\Omega|} \sum_{\boldsymbol{n} \in \Omega} x[\boldsymbol{n}, c], \tag{4.30}$$

and a small gating network with weights $W_{c,c'}$ and biases $b_c$ maps these descriptors to channel attention weights

$$a_{\text{channel}}[c] = \sigma\Big(\sum_{c'} W_{c,c'}\, s[c'] + b_c\Big), \tag{4.31}$$

where $\sigma$ is a scalar nonlinearity that maps real numbers to the interval $[0, 1]$. The attended feature map is then

$$y[\boldsymbol{n}, c] = a_{\text{channel}}[c]\, x[\boldsymbol{n}, c], \tag{4.32}$$

so channels with larger weights contribute more strongly at every spatial location. Channel attention of this form has been reported to improve convolutional baselines in image classification and segmentation tasks [210–212].

Table 4.3 summarizes common attention types based on the sources of queries, keys, and values.

Table 4.3: Types of attention mechanisms based on sources of Q, K and V

| # | Type | Definition | Examples | ND-Compatible |
|---|---|---|---|---|
| 1 | Self-attention | Query, key, and value all come from the same input sequence or tensor | Transformer [24], ViT [57], Swin [199] | Yes |
| 2 | Cross-attention | Query comes from one source; key and value come from another | UNETR [213], T5 [214], CLIP [49] | Yes |
| 3 | Co-attention | Both inputs generate queries and attend to each other reciprocally | ViLBERT [215], LXMERT [216] | Bimodal only! |
| 4 | Auto-regressive or Causal | Prevents a query from attending to future keys by masking | GPT [26, 27], XLNet [217] | 1D dominant! (language) |

Figures 4.6 and 4.7 show common locality patterns that reduce the computational cost of self-attention. Table 4.4 groups attention mechanisms by application domain. Table 4.5 summarizes the prevalent attention architectures.

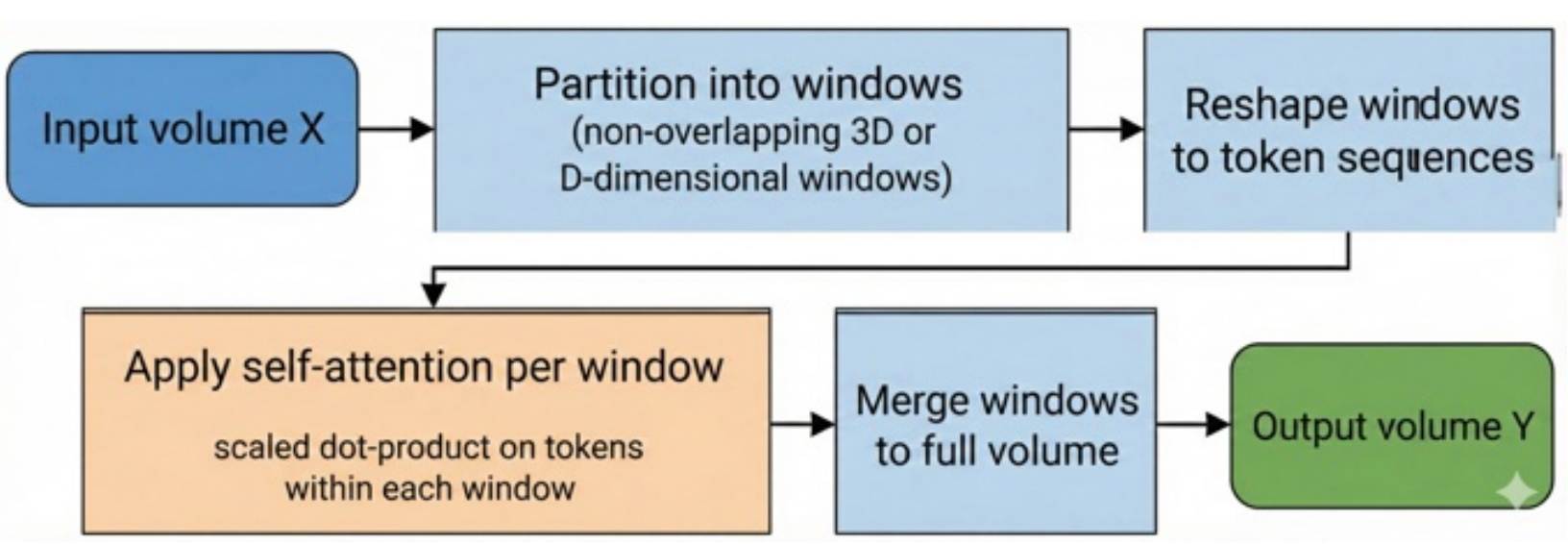

Figure 4.6: Windowed (3D) self-attention

Table 4.4: Attentions by domain adaptation

| # | Domain | Common Attention Types | Examples (with Reference) | Notes |
|---|---|---|---|---|
| 1 | Text / NLP | Dot-product (global), sparse, low-rank, additive | Additive attention [23], Transformer [24], Longformer [201], Linformer [202], Reformer [218] | Handles long sequences, language structure |
| 2 | Vision (2D) | Patch-wise, windowed, masked, convolutional attention, static gates | ViT [57], Swin [199], CBAM [210], Non-local NN [200], Deformable DETR [219] | Leverages spatial locality and translation equivariance |
| 3 | Medical Imaging (MRI / CT / D-dim.) | Axial, 3D windowed attention (Swin), hybrid CNN–Transformer, static gates | MedT [220], Swin-UNETR [213] | Exploits structured locality and high spatial dimension |
| 4 | Video / Spatiotemporal | Tubelet-based dot-product, factorized (space/time), hybrid (global + local) | TimeSformer [221] | Attention over space × time, often factorized |
| 5 | Multimodal (Vision + Text / Audio) | Dual-encoder self-attention, cross-attention/co-attention, fusion gates | CLIP [49], ViLBERT [215], LXMERT [216] | Aligns heterogeneous feature spaces |
| 6 | Audio / Speech | Self-attention + convolutional modules; local + global context | Conformer [222] | Strong for ASR; balances local patterns with long-range context |
| 7 | Graphs / Molecules | Neighborhood attention / message passing with attention weights | GAT [223] | Supports variable-size graphs and relational inductive biases |
| 8 | Other scientific D-dim. tensor data | Axial, kernelized, low-rank/sparse, gating | Axial attention [224], Performer [203], EasyAttention (Zhang & Aydin, 2024) | Useful for structured grids/volumes in physics and biomedical settings |

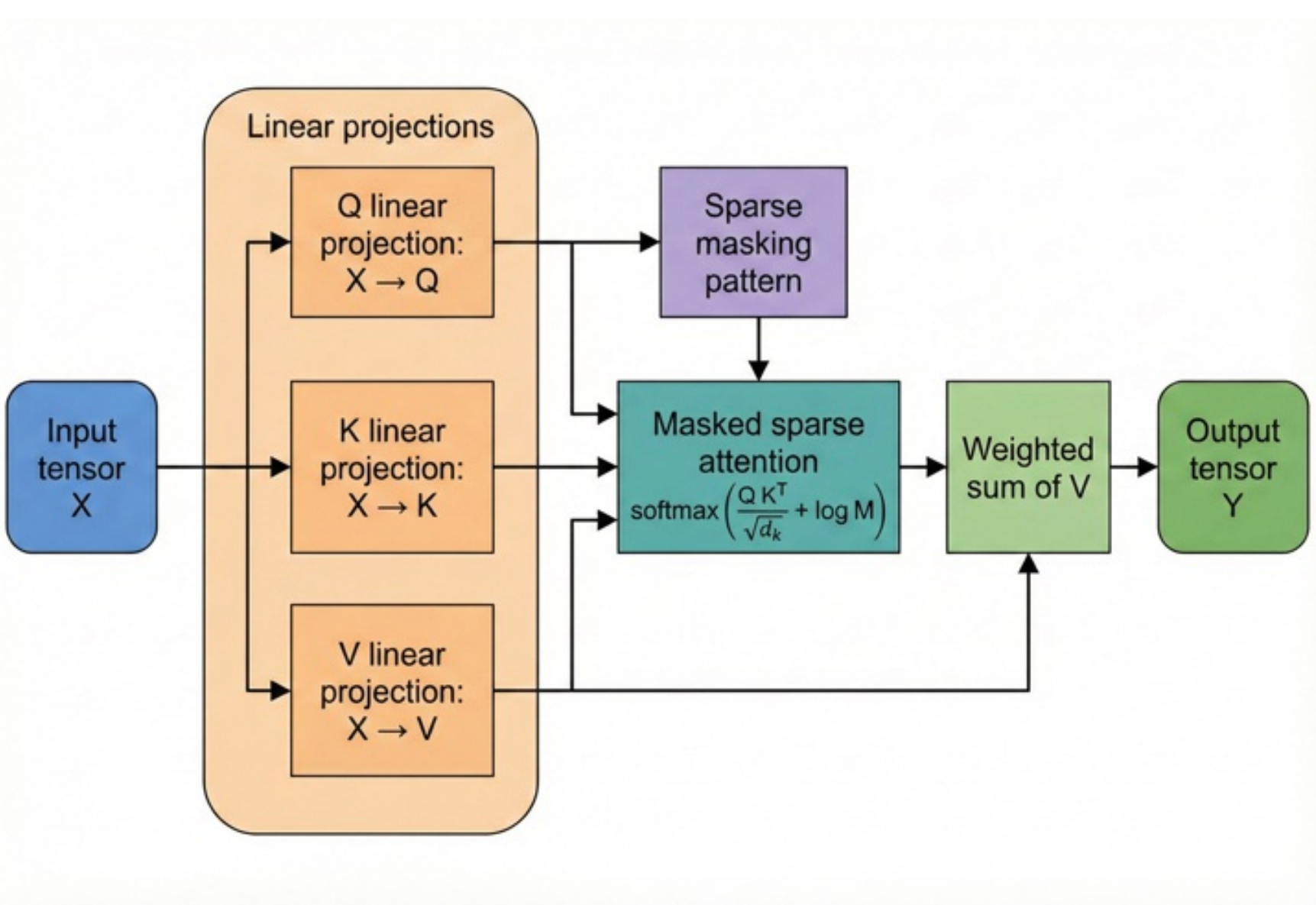

Figure 4.7: Sparse self-attention

Table 4.5: A concise review of the computational structures of attentions in literature

| # | Paper(s) | Core Idea | Primary Domain(s) | Representative Benchmarks | Extension to D-dimensions | Complexity |
|---|---|---|---|---|---|---|
| 1 | Additive attention (Bahdanau et al., 2014) [23] | Alignment scores between query and keys; softmax-weighted sum | NLP (seq2seq) | NMT (e.g., WMT'14 En–Fr) | general $Q \times K$ attention; not tied to input dimensionality | $O(N_q\, N_k\, d)$ |
| 2 | Transformer (Vaswani et al., 2017) [24] | Scaled dot-product self-attention across all token pairs | NLP, Multimodal | MT (WMT'14 En–De; WMT'16 En–Ro) | flattening-based for images/volumes; native in 1D | $O(N^2\, d)$ |
| 3 | Non-local NN (Wang et al., 2018) [200] | Dot-product non-local interactions over space/time feature maps | Vision, Video | Recognition / detection / segmentation (e.g., Kinetics, COCO) | applies to ND feature maps by flattening indices | $O(N^2\, d)$ |
| 4 | SENet (Hu et al., 2018), CBAM (Woo et al., 2018) [210, 225] | Channel/spatial gating via global pooling + small MLP/conv | Vision, Biomedical | Classification/ segmentation (e.g., ImageNet) | pooling + per-channel gates generalize directly to ND | $O(N\, C)$ |
| 5 | Axial attention (Ho et al., 2019) [224] | Factorizes self-attention along axes (one axis at a time) | Vision, ND tensors | Vision / ND feature maps (various) | structure-preserving axis-wise sweeps; used in D | $O\left(N\, d\, \sum_{i=1}^{D} n_i\right)$ |
| 6 | Longformer (2020) [201], BigBird (2020) [204] | Sparse patterns (sliding-window + global/random blocks) | NLP (long sequences) | Long-doc classification / QA / LM | sparse patterns defined over 1D token axis | $O(N\, K\, d),\ K \ll N$ |
| 7 | Linformer (2020) [202] | Low-rank projection of $K/V$ to length $k$ | NLP, Efficient Transformers | Long-seq NLP (various) | projection applied on flattened token axis | $O(N\, k\, d)$ |
| 8 | Performer (2020) [203] | Kernelized attention via random feature maps (FAVOR+) | NLP, Scientific ML | Long-seq modeling (various) | kernel maps applied after flattening | $O(N\, d)$ |
| 9 | Reformer (2020) [218] | LSH bucketing to restrict comparisons to nearby hashes | NLP (long sequences) | Long-seq LM (various) | LSH applied on flattened tokens | $O(N \log N\, d)$ |
| 10 | Deformable DETR (Zhu et al., 2020) [219] | Sparse attention over learned sampling locations across multi-scale features | Object detection | Detection (COCO) | coordinates + sampling generalize to ND grids | $O(N\, K\, d)$ |
| 11 | ViT (Dosovitskiy et al., 2020) [57] | Global self-attention over flattened image patches | Vision | Classification (e.g., ImageNet) | flatten patches of 2D/3D grids into tokens | $O(N_p^2\, d),\ N_p = HW/P^2$ |
| 12 | Swin (Liu et al., 2021) [199] | Windowed self-attention + shifted windows for cross-window mixing | Vision, Medical Imaging | Detection/ segmentation (e.g., COCO, ADE20K) | local windows extend to 3D/ND windows | $O(N\, M^2\, d),\ M^2 \ll N$ |
| 13 | MobileViT (Mehta & Rastegari, 2021) [99] | Hybrid conv + lightweight self-attention on local patches | Vision (mobile) | Mobile vision (e.g., ImageNet) | local patch attention extends to ND feature maps | $O(N\, M^2\, d)$ |
| 14 | EasyAttention (Zhang & Aydin, 2024) | Non-parametric/static gates based on pooling + cheap arithmetic | Lightweight ML, Biomedical | Lightweight vision / biomedical (various) | pooling-based gates generalize to ND | $O(N)$ |

### 4.2.1 Wavelet subband-aware self-attention layer

In this section, we introduce a *wavelet subband-aware self-attention* layer. Let $x : \Omega \to \mathbb{R}^{C_{\text{in}}}$ denote a $D$-dimensional multichannel field on a discrete spatial grid $\Omega \subset \mathbb{Z}^D$, where $x[\boldsymbol{n}] \in \mathbb{R}^{C_{\text{in}}}$ and $\boldsymbol{n} = [n_1, \ldots, n_D] \in \Omega$. In batched channel-last tensor form, $x$ has shape $[B, N_1, \ldots, N_D, C_{\text{in}}]$, where $B$ is the batch size, $N_i$ is the side length along axis $i \in \{1, \ldots, D\}$, and $C_{\text{in}}$ is the input channel count. The cases $D = 2$ correspond to planar images with pixels, and $D = 3$ represent volumetric images with voxels.

Using the separable multilevel operators introduced above, we decompose $x$ into one deepest LP band and $G = 2^D - 1$ HP subbands at each level. The corresponding *multilevel* $D$-dimensional analysis is

$$\mathcal{T}_{\text{DWT}}(x) = \Big\{\text{LP}_L,\ \{\text{HP}_j^{(g)}\}_{\substack{j=1,\ldots,L\\ g=1,\ldots,G}}\Big\}, \tag{4.33}$$

where $\text{LP}_L$ is the deepest LP band and $\text{HP}_j^{(g)}$ are the HP bands at level $j$ and orientation group $g$. In the multichannel tensor packing used here, the first $C_{\text{in}}$ channels at each decomposition level are the LP features and the remaining $(2^D - 1)C_{\text{in}}$ channels form a concatenated HP block. Similarly, the *multilevel* $D$-dimensional synthesis

$$\mathcal{T}_{\text{IDWT}}\Big(\text{LP}_L, \{\text{HP}_j^{(g)}\}_{\substack{j=1,\ldots,L\\ g=1,\ldots,G}}\Big) = x \quad \text{(PR)} \tag{4.34}$$

gives perfect reconstruction (PR) of the input when the subbands are left unchanged after decomposition.

The analysis equation (4.33) yields subbands localized in scale and orientation. Let $Y \in \mathbb{R}^{B \times N_{j,1} \times \cdots \times N_{j,D} \times C_{\text{in}}}$ denote one such multichannel band, where $N_{j,i}$ is the side length of the level-$j$ grid along axis $i$. Because the multidimensional DWT is separable, a natural attention operator on $Y$ is an axial sweep that applies self-attention successively along each axis [224],

$$\mathcal{A}_{\text{axial}}(Y) = \big(\text{MHSA}_D \circ \cdots \circ \text{MHSA}_1\big)(Y), \tag{4.35}$$

where each $\text{MHSA}_i$ denotes applying the same multi-head self-attention (MHSA) [24] block along axis $i$. In practice, each axial or global attention block first applies channelwise layer normalization to the input band, then forms *queries*, *keys*, and *values* (QKV) through separate $1 \times \cdots \times 1$ projections that are reused across the axial sweep within a given module.

**Per-band attention operators** Within a given block, we use two axial attention modules with the same operator form but with independent parameters. The module $\mathcal{A}_{\text{axial}}^{\text{HP}}$ acts on HP bands and is shared across all HP bands at all levels within that block. The module $\mathcal{A}_{\text{axial}}^{\text{LP}}$ acts on LP bands and is shared across all levels within that block. At the deepest LP band, the block may instead use a global operator $\mathcal{A}_{\text{global}}$, meaning full MHSA over the entire deepest LP grid, when the LP resolution is sufficiently small.

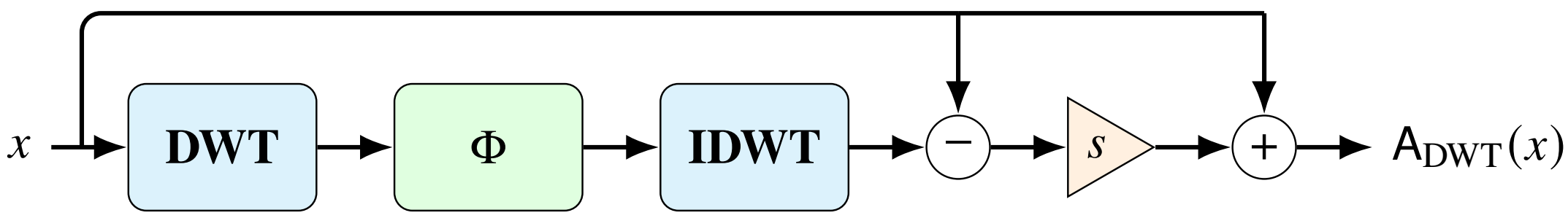

Figure 4.8: Wavelet subband-aware self-attention layer $\mathsf{A}_{\mathrm{DWT}}$. The input $x$ is decomposed by the DWT, edited in wavelet space by $\Phi$, reconstructed by the IDWT, and returned through the residual form $x + s(\cdot - x)$.

For each HP subband (a particular group at a level) the update has the form

$$\widetilde{\mathrm{HP}}_j^{(g)} = \mathrm{HP}_j^{(g)} + \alpha_{\mathrm{HP}}\, \eta_j\, \mathcal{A}_{\mathrm{axial}}^{\mathrm{HP}}\big(\mathrm{HP}_j^{(g)}\big), \tag{4.36}$$

where $\alpha_{\mathrm{HP}}$ is a learned scalar gate and $\eta_j = \rho^{L-j}$ is a deterministic level weight with $0 < \rho \leq 1$ and $L$ the total number of levels. Coarser levels therefore receive stronger edits. After this update an orientation mixing matrix $\mathbf{M}^{(j)} \in \mathbb{R}^{G\times G}$ acts on the group index at fixed spatial location and channel

$$\widehat{\mathrm{HP}}_j^{(g)} = \sum_{g'=1}^{G} M_{g,g'}^{(j)}\, \widetilde{\mathrm{HP}}_j^{(g')}, \tag{4.37}$$

with $\mathbf{M}^{(j)}$ initialized as $\mathbf{I}_G + \varepsilon(\mathbf{1}_G - \mathbf{I}_G)$, where $\mathbf{I}_G$ is the $G \times G$ identity matrix, $\mathbf{1}_G$ is the $G \times G$ all-ones matrix, $\varepsilon = 0.02\, s$, and $s$ is the global strength parameter. This ties the initial off-diagonal mixing to the strength of the block, giving a light learned mixing across the $G$ HP bands at each level. When orientation mixing is disabled the block uses $\widehat{\mathrm{HP}}_j^{(g)} = \widetilde{\mathrm{HP}}_j^{(g)}$. On the LP path, an axial attention block $\mathcal{A}_{\mathrm{axial}}^{\mathrm{LP}}$ with gate $\alpha_{\mathrm{LP}}$ and the same level weights $\eta_j$ updates the current LP tensor at each reverse-order synthesis level. At the deepest level, the global operator $\mathcal{A}_{\mathrm{global}}$ can replace the axial sweep when the number of LP tokens is below a fixed cap. This design keeps the cost far below full-resolution global attention while still allowing one global low-resolution attention.

**Wavelet-domain residual block** Inside the band processor $\Phi$, axial MHSA updates the stored HP bands in the form of (4.36). On the LP path, the current LP tensor is updated recursively during the reverse synthesis pass, with the deepest level optionally using the global operator described above. When enabled, the orientation mixer applies the update in (4.37) within each level. For compactness, $\Phi\big(\mathcal{T}_{\mathrm{DWT}}(x)\big)$ denotes the resulting wavelet-domain processing as a whole. After attention and optional mixing, the bands are reassembled by undoing the DWT cascade. The IDWT operator $\mathcal{T}_{\mathrm{IDWT}}$ reconstructs the signal from the updated LP and HP parts at each level, following the multilevel synthesis relation in (4.34). Figure 4.8 summarizes the residual structure of the wavelet subband-aware self-attention layer. Thus, our wavelet subband-aware self-attention layer (DWT $\to$ band processor $\Phi$ $\to$ IDWT) is defined as

$$\mathsf{A}_{\mathrm{DWT}}(x) = x + s\Big(\mathcal{T}_{\mathrm{IDWT}}\, \Phi\big(\mathcal{T}_{\mathrm{DWT}}(x)\big) - x\Big), \tag{4.38}$$

where the scalar $s$ is a user parameter that controls the strength of the edit Values of $s$ close to zero keep the block near the identity, while larger values make the wavelet domain attention more prominent. In particular, when the internal attention gates $\alpha_{\mathrm{HP}}$ and $\alpha_{\mathrm{LP}}$ are set to zero and orientation mixing is disabled, the block reduces to a pure DWT and IDWT round trip and reproduces $x$ up to numerical precision (for spatial sizes divisible by $2^L$).

The asymptotic attention complexity of this design follows from the decimation in the DWT. At each level, attention acts on a reduced grid with fixed channel width. For a fixed number of heads, the leading HP contribution is

$$C_{\mathrm{HP}} = O\Big(\sum_{j=1}^{L} G\,\Pi_j\Big(\sum_{i=1}^{D} N_{j,i}\Big)d_{\mathrm{head}}\Big). \tag{4.39}$$

The LP path contributes one additional term per level with the same asymptotic order, so the full block has the same overall scaling. Here $G = 2^D - 1$ is the number of HP bands, $N_{j,i}$ are the side lengths of the level-$j$ subband grid, $\Pi_j = \prod_{i=1}^{D} N_{j,i}$ is the number of spatial locations in a level-$j$ band, $\Pi = \prod_{i=1}^{D} N_i$ is the number of locations at the full resolution, and $d_{\mathrm{head}}$ is the attention head width. The optional deepest-LP global branch does not change this asymptotic order because it is activated only when the LP token count is below a fixed cap. For fixed $D \geq 2$, the attention complexity of $\mathsf{A}_{\mathrm{DWT}}$ is subquadratic in $\Pi$, in contrast to the quadratic $O(\Pi^2 d_{\mathrm{head}})$ complexity of full-resolution global self-attention. For $D = 1$, the multilevel decimation still reduces the constant factor relative to full-resolution global attention. This newly introduced $\mathsf{A}_{\mathrm{DWT}}$ layer is fully differentiable, with the DWT and IDWT adjoints supplying analysis and synthesis gradients and the attention sublayers contributing standard transformer derivatives.

The above wavelet subband-aware self-attention layer edits a feature map in wavelet space and returns a feature map of the same type after inverse synthesis. We next introduce WaveletViT, which extends this idea into a modular tokenizer, block, and assembler architecture with raw coefficient tokenization, compact highpass attention, coefficient gain modulation, and either pooled or reconstructed feature readout.

### 4.2.2 WaveletViT: a multilevel wavelet-band transformer

WaveletViT extends the direct DWT-domain residual layer into a modular tokenizer, block, and assembler architecture for multilevel wavelet-band processing of $D$-dimensional inputs. The tokenizer first performs a multilevel separable DWT of the input and returns raw LP and HP coefficient bands. These raw bands remain the synthesis state. The block forms compact HP token grids with separable strided convolutions and projects only the deepest LP band to token width $E$ for attention. The attended tokens predict bounded gains that modulate the raw coefficient bands, and the assembler returns either a pooled token or a reconstructed feature map through matched synthesis. To describe this construction, let $L \geq 1$ denote the number of wavelet decomposition levels, let $E$ denote the token width, and let $r$ denote the HP token stride. For a separable $D$-dimensional DWT, each level contains $2^D$ subbands, namely one LP band and $G = 2^D - 1$ HP bands. If padding is used, let $\widetilde{N}_i$ denote the padded side length along axis $i$. It is chosen as the smallest multiple of $2^L$ that is not smaller than $N_i$. Otherwise, $\widetilde{N}_i = N_i$. The formulas below are written on the possibly padded grid

$$\widetilde{\Omega} = \{0, \dots, \widetilde{N}_1 - 1\} \times \cdots \times \{0, \dots, \widetilde{N}_D - 1\}, \tag{4.40}$$

and the final crop back to the original $\Omega$ is omitted. The implemented 2D and 3D wavelet cores used in this work operate on square or cubic grids. We write $N_{\mathrm{core}}$ for the common side length of such a square or cubic core input. We may also apply a channelwise layer normalization to $x$ before DWT tokenization.

#### 4.2.2.1 Pipeline stages

The architecture has three stages. The tokenizer performs a multilevel DWT and exposes raw coefficient bands. The block builds compressed HP tokens, processes them with axial attention, feed-forward mixing, and optional orientation mixing, then predicts gain fields for the raw HP bands. The deepest LP band is processed through a separate token-width projection and gain head. The assembler runs the matched IDWT cascade and returns either a pooled token or a reconstructed feature map. The detailed formulation of each stage follows.

**DWT analysis and compact token formation** At level $j = 1, \dots, L$, we apply the separable analysis operator $W_j$ to the current LP tensor $\mathrm{LP}_{j-1}$ (with $\mathrm{LP}_0 \equiv x$) to obtain one LP and $G$ HP coefficient fields.

$$\left[\mathrm{LP}_j,\ \mathrm{HP}_j^{(1)}, \dots, \mathrm{HP}_j^{(G)}\right] \ :=\ W_j\, \mathrm{LP}_{j-1}, \tag{4.41}$$

where $\mathrm{LP}_j, \mathrm{HP}_j^{(g)}$ have shapes $[B, N_{j,1}, \dots, N_{j,D}, C_{\mathrm{in}}]$ and $N_{j,i} = \widetilde{N}_i / 2^j$. The HP bands are then mapped to compact token grids by a separable strided tokenization operator $\mathcal{F}^{\mathrm{tok}}_{j,r}$,

$$u_j^{(g)} = \mathcal{F}^{\mathrm{tok}}_{j,r}\left(\mathrm{HP}_j^{(g)}\right), \qquad g = 1, \dots, G, \tag{4.42a}$$

$$q_L = P_{\mathrm{LP}}\, \mathrm{LP}_L. \tag{4.42b}$$

Here $u_j^{(g)} \in \mathbb{R}^{B \times Q_{j,1} \times \cdots \times Q_{j,D} \times E}$, with $Q_{j,i}$ determined by the HP token stride $r$, and $P_{\mathrm{LP}}$ is a pointwise projection to width $E$. The intermediate LP bands define the multilevel analysis cascade, while the deepest LP band and all HP bands define the synthesis state used by the assembler.

**Axial intra-band attention** WaveletViT applies band-processing operators of the same form as those introduced in Section 4.2.1, especially the axial sweep in (4.35). The HP token grids $\{u_j^{(g)}\}$ are processed independently by per-level HP axial-attention modules, denoted by $\mathcal{A}_{j,\mathrm{HP}}$. The deepest LP token grid $q_L$ uses an LP axial-attention module, or a global operator $\mathcal{A}_{\mathrm{global}}$ when $\Pi_L = \prod_{i=1}^{D} N_{L,i} \le \Pi_{\max}$, where $\Pi_{\max}$ is a fixed token cap for enabling deepest-LP global attention.

**Token updates** Let $\rho \in (0, 1]$ be a level-decay factor and $s \geq 0$ a global strength parameter. We set $\eta_j = \rho^{L-j}$ so that deeper levels (larger $j$) receive stronger updates. With trainable scalar gates $\alpha_{\mathrm{HP}}, \alpha_{\mathrm{LP}}, \beta_{\mathrm{HP}}, \beta_{\mathrm{LP}}$, the HP tokens are updated by attention and optional feed-forward mixing,

$$v_j^{(g)} = u_j^{(g)} + \alpha_{\mathrm{HP}}\eta_j \mathcal{A}_{j,\mathrm{HP}}\left(u_j^{(g)}\right), \tag{4.43a}$$

$$\bar{u}_j^{(g)} = v_j^{(g)} + \beta_{\mathrm{HP}}\eta_j \,\mathrm{FFN}_{j,\mathrm{HP}}\left(v_j^{(g)}\right). \tag{4.43b}$$

The deepest LP token grid is updated as

$$v_L = \begin{cases} q_L + \alpha_{\mathrm{LP}}\mathcal{A}_{\mathrm{global}}(q_L), & \Pi_L \leq \Pi_{\max}, \\ q_L + \alpha_{\mathrm{LP}}\mathcal{A}_{\mathrm{axial}}^{\mathrm{LP}}(q_L), & \text{otherwise}. \end{cases} \tag{4.44a}$$

$$\bar{q}_L = v_L + \beta_{\mathrm{LP}}\,\mathrm{FFN}_{\mathrm{LP}}(v_L). \tag{4.44b}$$

When feed-forward mixing is disabled, the corresponding $\beta$ term is omitted.

**Orientation mixing and coefficient gains** When enabled, the same near-identity mixer $\mathbf{M}^{(j)} \in \mathbb{R}^{G\times G}$ acts channelwise across the HP-band axis on the updated HP tokens. This is the update in (4.37) with $\widetilde{\mathrm{HP}}_j^{(g)}$ replaced by $\bar{u}_j^{(g)}$. After the optional orientation mix, we denote the HP outputs by $\widehat{u}_j^{(g)}$ (with $\widehat{u}_j^{(g)} \equiv \bar{u}_j^{(g)}$ when mixing is disabled). The learned HP gain head $\Theta_j^{\mathrm{HP}}$ uses the concatenated HP tokens to predict bounded gains,

$$\gamma_j = 1 + 0.25 \,\tanh \Theta_j^{\mathrm{HP}}\left(\mathrm{concat}_{g=1}^{G}\, \widehat{u}_j^{(g)}\right), \tag{4.45a}$$

$$\widetilde{\mathrm{HP}}_j^{(g)} = \mathrm{HP}_j^{(g)} \odot \mathcal{U}_{j,r}\left(\gamma_j^{(g)}\right), \qquad g = 1, \ldots, G, \tag{4.45b}$$

where $\mathcal{U}_{j,r}$ upsamples the compressed gain field to the raw HP-band grid. The deepest LP branch predicts its own gain with the learned LP gain head $\Theta_L^{\mathrm{LP}}$,

$$\gamma_L^{\mathrm{LP}} = 1 + 0.25 \,\tanh \Theta_L^{\mathrm{LP}}(\bar{q}_L), \tag{4.46a}$$

$$\widetilde{\mathrm{LP}}_L = \mathrm{LP}_L \odot \gamma_L^{\mathrm{LP}}. \tag{4.46b}$$

A strength-controlled residual is then applied to the raw coefficient bands,

$$\mathrm{LP}_{L,\mathrm{out}} = \mathrm{LP}_L + s\left(\widetilde{\mathrm{LP}}_L - \mathrm{LP}_L\right), \tag{4.47a}$$

$$\mathrm{HP}_{j,\mathrm{out}}^{(g)} = \mathrm{HP}_j^{(g)} + s\left(\widetilde{\mathrm{HP}}_j^{(g)} - \mathrm{HP}_j^{(g)}\right). \tag{4.47b}$$

These raw-band outputs are then used by the synthesis or token readout stage. The gain heads are initialized so that the initial modulation is close to one, and the scalar strength $s$ controls the magnitude of the wavelet-space residual. Equivalently, for any raw coefficient band $\chi$, the block applies $\chi_{\mathrm{out}} = \chi + s(\widetilde{\chi} - \chi)$ after gain modulation.

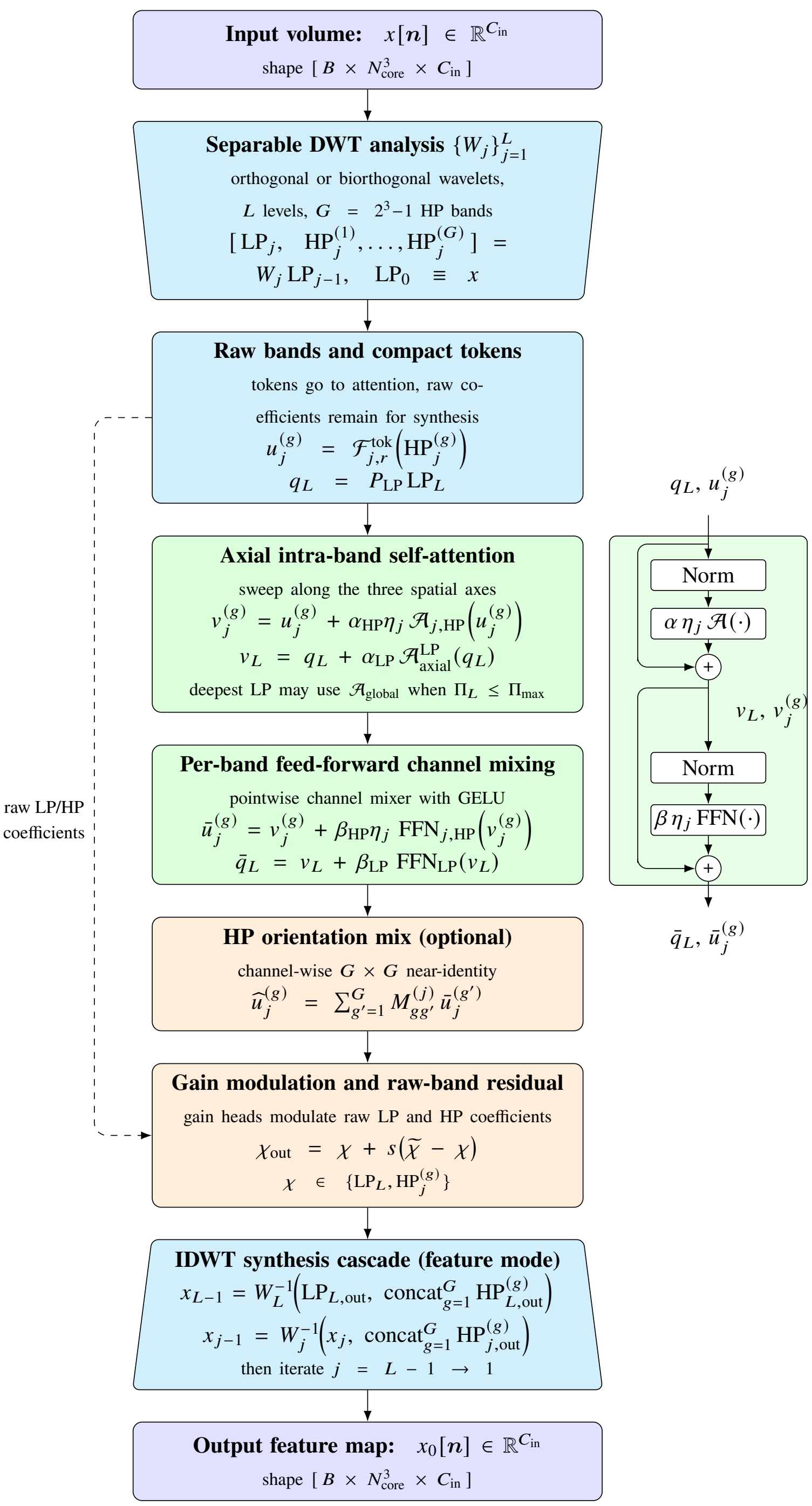

Figure 4.9: Overview of the implemented 3D WaveletViT feature-mode architecture. Separable DWT analysis produces raw multilevel LP and HP coefficient bands. The block forms compact HP tokens by separable strided tokenization and projects only the deepest LP band to width $E$ for attention. The attended tokens predict bounded gains that modulate the raw coefficient bands, followed by a strength-controlled residual in wavelet space. Matched IDWT synthesis reconstructs a $C_{\text{in}}$-channel feature map.

**Synthesis and readout modes** The assembler supports two readout modes, namely *tokens mode* and *feature mode*. Both modes first invert the wavelet pyramid with a matching IDWT cascade. The deepest stage reconstructs according to

$$x_{L-1} = W_L^{-1}\Big(\mathrm{LP}_{L,\mathrm{out}},\ \mathrm{concat}_{g=1}^{G}\,\mathrm{HP}_{L,\mathrm{out}}^{(g)}\Big), \tag{4.48}$$

and the remaining levels are synthesized recursively according to

$$x_{j-1} = W_j^{-1}\Big(x_j,\ \mathrm{concat}_{g=1}^{G}\,\mathrm{HP}_{j,\mathrm{out}}^{(g)}\Big), \quad j = L-1, \ldots, 1. \tag{4.49}$$

Here $\mathrm{concat}_{g=1}^{G}$ denotes concatenation of the $G$ HP bands along the channel axis. This yields an output volume $x_0 \in \mathbb{R}^{B\times\widetilde{N}_1\times\cdots\times\widetilde{N}_D\times C_{\mathrm{in}}}$. Padding to sizes divisible by $2^L$ can be applied before tokenization, and the final crop back to the original support is omitted here. In *feature mode*, the assembler returns $x_0$. In *tokens mode*, a compact representation is obtained by global average pooling over the reconstructed feature map,

$$x_{\mathrm{pool}} = \mathrm{GAP}(x_0) \in \mathbb{R}^{B\times C_{\mathrm{in}}}. \tag{4.50}$$

#### 4.2.2.2 Trainable parameters

With the DWT and IDWT operators fixed, the learned parameters comprise the separable HP tokenizers, the HP attention projections, the optional HP feed-forward layers, the per-level orientation-mix matrices, the cross-band HP gain heads, the deepest-LP projection, the LP axial or global attention projections, the optional LP feed-forward layer, the LP gain head, the scalar gates $\alpha_{\mathrm{HP}}, \alpha_{\mathrm{LP}}, \beta_{\mathrm{HP}}, \beta_{\mathrm{LP}}$, the optional variational heads on compressed HP features, and the affine parameters of the normalization layers when present. The level-decay $\rho$, strength $s$, wavelet family, decomposition depth, token width, and HP token stride are fixed hyperparameters.

No absolute grid positional encodings are used. Instead, WaveletViT relies on the multiscale DWT decomposition, separable axial attention, orientation mixing, and coefficient gain modulation to provide multiscale geometric structure while keeping attention intra-band. The model remains fully differentiable through the filter-bank adjoints together with standard projection, attention, feed-forward, and gain-prediction operations. Figure 4.9 shows the WaveletViT architecture and Figures 4.10 and 4.11 qualitatively illustrate the residual attention maps of $\mathsf{A}_{\mathrm{DWT}}$, WaveletViT and vanilla transformer with random weights for a sample brain MRI volumetric input.

#### 4.2.2.3 Computational complexity

WaveletViT has the same fixed linear DWT and IDWT analysis-synthesis overhead as $\mathsf{A}_{\mathrm{DWT}}$, but its HP attention is applied on compressed token grids rather than on the full HP grids. If $Q_j = \prod_{i=1}^{D} Q_{j,i}$ denotes the number of compressed HP tokens at level $j$, then an axial HP attention block with $H$ heads and key width $d_h$ costs

$$O\left(Hd_h \sum_{j=1}^{L} G\, Q_j \sum_{i=1}^{D} Q_{j,i}\right), \tag{4.51}$$

up to pointwise QKV and output projections. The deepest LP path contributes either the same axial form with $Q_j$ replaced by $N_L = \prod_i N_{L,i}$, or a global-attention cost $O(Hd_h N_L^2)$ when $N_L \leq \Pi_{\max}$. The separable HP tokenizers, orientation mixers, gain heads, feed-forward layers, IDWT synthesis, and token-mode global average pooling are linear in the number of processed spatial sites for fixed $D$, fixed heads, and fixed channel widths. Thus the expensive attention part is applied on reduced wavelet grids, and for $D \geq 2$ the axial form is subquadratic in the corresponding full-grid token count.

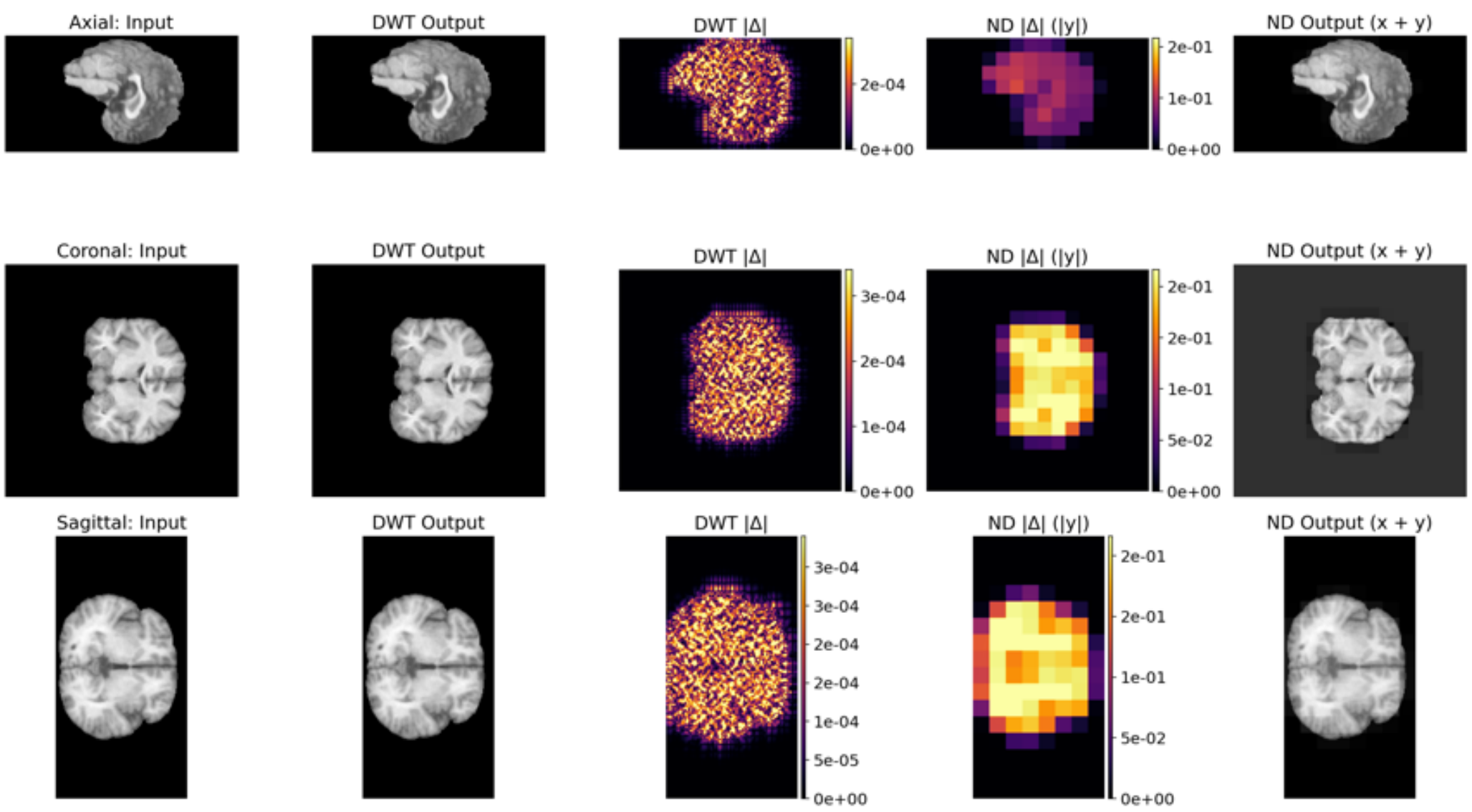

Figure 4.10: (Left column to right column) Input brain slices in three body planes; Output of DWT subband aware self-attention (WaveletAttention); WaveletAttention maps; windowed vanilla self-attention maps; Output of vanilla self attention. Displayed attention maps are residuals, i.e., the difference between attention layer's output and input.

**Backpropagation through WaveletViT:** Let $\mathcal{L}(\theta)$ be a loss on the output field $x_0(\boldsymbol{n})$ of a WaveletViT block, with trainable parameters collected in $\theta$. During the backward pass the gradient field $g_0(\boldsymbol{n}) = \partial\mathcal{L}/\partial x_0(\boldsymbol{n})$ is propagated through the final global pooling when token mode is used and then through the IDWT cascade. Since each $W_j^{-1}$ is a fixed linear synthesis filter bank, its adjoint redistributes $g_0$ into gradients on $\mathrm{LP}_{L,\mathrm{out}}$ and $\mathrm{HP}_{j,\mathrm{out}}^{(g)}$ for all HP levels.

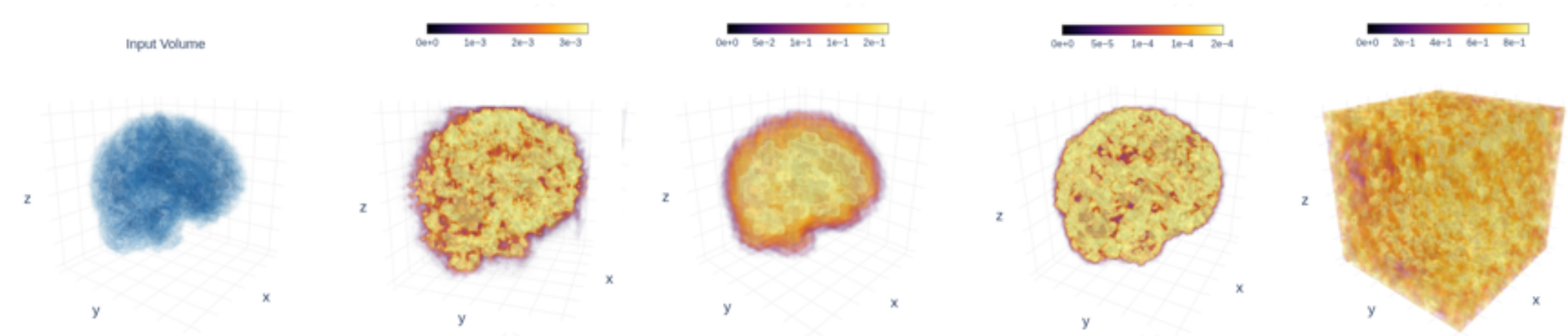

Figure 4.11: (Left to right) Input brain volume; DWT subband aware self-attention (WaveletAttention) & windowed vanilla self-attention maps; WaveletViT & Vanilla ViT attention maps.

In the implementation the tokenizer returns raw DWT bands; the trainable HP path does not apply a direct pointwise projection to every highpass band. Instead, each raw HP band $\mathrm{HP}_j^{(g)}$ is compressed by separable strided one-dimensional convolutions, passed through axial attention, the optional feed-forward mixer and the optional orientation mixer, and then mapped by the cross-band gain head $\Theta_j^{\mathrm{HP}}$ to the bounded gain field in (4.45a). Writing

$$\widehat{U}_j = \mathrm{concat}_{g=1}^{G}\, \widehat{u}_j^{(g)}, \qquad r_j = \Theta_j^{\mathrm{HP}}(\widehat{U}_j), \tag{4.52}$$

we have $\gamma_j = 1 + \frac{1}{4}\tanh r_j$. After nearest-neighbor upsampling to the raw band grid, the residual HP output used by the IDWT is

$$\mathrm{HP}_{j,\mathrm{out}}^{(g)} = \mathrm{HP}_j^{(g)} + s\left(\mathrm{HP}_j^{(g)} \odot \mathcal{U}_{j,r}(\gamma_j^{(g)}) - \mathrm{HP}_j^{(g)}\right) = \mathrm{HP}_j^{(g)} \odot \left[1 + s\left(\mathcal{U}_{j,r}(\gamma_j^{(g)}) - 1\right)\right]. \tag{4.53}$$

Thus the gradient to the gain head contains the product of the band gradient and the raw HP band. If $\zeta_j^{(g)}$ denotes the adjoint-upsampled quantity $\mathcal{U}_{j,r}^{*}\{s\,(\partial\mathcal{L}/\partial\mathrm{HP}_{j,\mathrm{out}}^{(g)}) \odot \mathrm{HP}_j^{(g)}\}$, then

$$\frac{\partial \mathcal{L}}{\partial \Theta_j^{\mathrm{HP}}} = \frac{1}{4}\sum_{\boldsymbol{m}} \left[\zeta_j(\boldsymbol{m}) \odot \left(1 - \tanh^2 r_j(\boldsymbol{m})\right)\right] \widehat{U}_j(\boldsymbol{m})^{\top}, \tag{4.54}$$

with the usual convolutional summation over batch and spatial locations. The deepest lowpass follows the analogous path: the raw LP band is projected by `lp_proj`, processed by axial or global attention and the LP feed-forward block, mapped by `lp_gain_head` to $\gamma_L^{\mathrm{LP}} = 1 + \frac{1}{4}\tanh r_L^{\mathrm{LP}}$, and multiplied with the raw LP band inside the same strength-controlled residual form. Gradients through axial or global attention, feed-forward layers, layer normalizations, separable tokenizers, `lp_proj`, and the gain heads are the standard TensorFlow convolutional and transformer gradients. The scalar gates $\alpha_{\mathrm{HP}}, \alpha_{\mathrm{LP}}, \beta_{\mathrm{HP}}, \beta_{\mathrm{LP}}$ are trainable through their sigmoid-bounded parametrization, while the level-decay factor and global strength are fixed hyperparameters in the current code.

Compared to a vanilla ViT, which relies on explicit positional encodings, WaveletViT uses a DWT tokenizer that carries spatial scale and orientation, providing an implicit positional signal and a stronger inductive bias. This often reduces sample complexity and improves data efficiency on small datasets. In later sections we introduce the FDST-based ShearViT, which replaces the separable DWT with the non-separable FDST bank from Chapter 2 and adds band-space (inter-band) attention over cone–scale–direction–rotation indices. Figures 4.10 and 4.11 show the attention maps (residual) of WaveletAttention, WaveletViT, and their Vanilla counterparts for a sample brain volume input.

### 4.2.3 ShearViT: 3D shearlet band transformer

**FDST interface** For a 3D multichannel input volume $x[\boldsymbol{n}]$, with spatial index $\boldsymbol{n} = (n_1, n_2, n_3)$ and $C_{\text{in}}$ channels, let $\widehat{x}[\boldsymbol{\omega}]$ denote its Fourier transform on the discrete frequency grid $\Omega$. The FDST bank introduced in Chapter 2 provides a family of windows $\{W_k[\boldsymbol{\omega}]\}_{k=0}^{K-1}$ and associated supports $\{\mathcal{I}_k \subset \Omega\}$, together with band metadata consisting of scale $j(k)$, cone-slot index $\nu(k)$, in-cone orientation $\ell(k)$, and rotation id $r(k) \in \{0, \dots, M-1\}$. For HP bands, $j(k) \in \{0, \dots, J-1\}$. The implementation records cone metadata as integer slots $\nu(k) \in \{0, 1, 2\}$ for the x-, y-, and z-cones. The LP band is kept distinct through the bookkeeping values $j(k) = -1$ and $\ell(k) = -1$, while its metadata slot is stored as 0. Thus the block-level metadata path uses integer cone slots, and the LP band is identified through its scale index rather than through a separate cone label. As in (2.58), the windows are normalized so that

$$D_0[\boldsymbol{\omega}] = \sum_{k=0}^{K-1} |W_k[\boldsymbol{\omega}]|^2 \approx 1 \quad \text{for all } \boldsymbol{\omega} \in \Omega, \tag{4.55}$$

and the FDST analysis and inverse FDST synthesis pair internally use the per-frequency gains and rotation weights from (2.61) and (2.62). ShearViT uses this FDST analysis and inverse FDST synthesis pathway as its coefficient interface and adds a ViT-style band-processing block [57] on top of the resulting tiles. In the model studied here, the band layout and windows are fixed after construction, while the FDST-internal per-frequency gains and, when enabled, the internal rotation weights remain trainable. The experiments reported here use only the identity rotation, so $M = 1$ and $r(k) = 0$ for all $k$. We retain the more general rotation notation only to stay consistent with the FDST formulation above. Figure 4.12 summarizes the implemented ShearViT feature-return pathway.

**FDST tiles (analysis coefficients)** Running the FDST analysis on $x$ yields, for each band $k$, a complex tile $T_k \in \mathbb{C}^{B \times N_k \times C_{\text{in}}}$, where $B$ is the batch size and $N_k = |\mathcal{I}_k|$ is the number of frequency samples in band $k$. In the notation of the FDST section above, $T_k$ corresponds to the effective analyzed tile $\widetilde{X}_k$ in (2.61c). Concretely, it is obtained by gathering $\widehat{x}$ at frequencies $\boldsymbol{\omega} \in \mathcal{I}_k$ and multiplying by the corresponding window value, the FDST-internal per-frequency complex gain, and, when enabled, the FDST-internal rotation scalar. In the ShearViT block we only need $T_k$ and the band metadata $(j(k), \nu(k), \ell(k), r(k), \boldsymbol{\xi}_k)$. Here $\boldsymbol{\xi}_k \in \mathbb{R}^3$ is a precomputed central frequency for band $k$, defined as the energy-weighted mean of the frequencies in $\mathcal{I}_k$. It is retained for diagnostics and possible extensions.

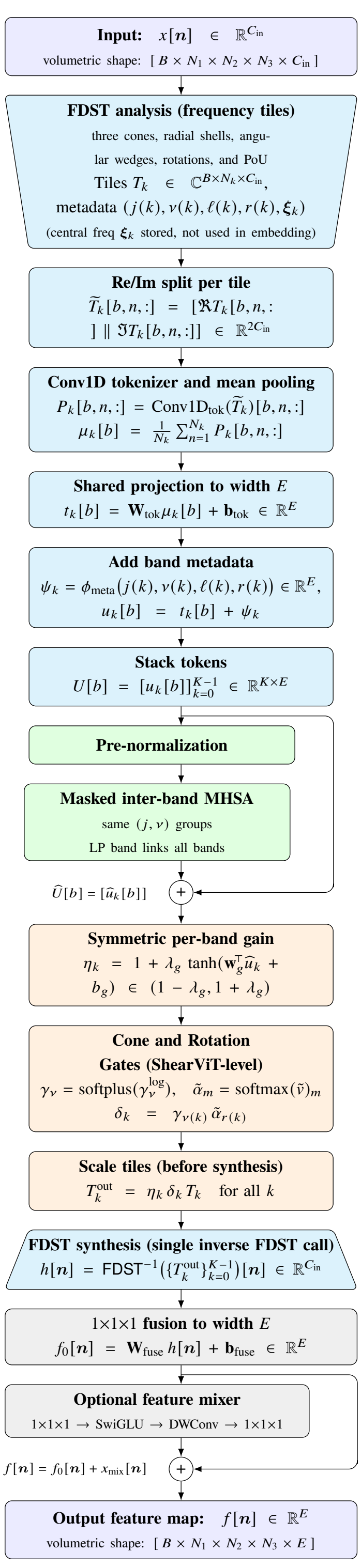

Figure 4.12: ShearViT block with 3D FDST analysis and inverse FDST synthesis. The input volume is analyzed into band tiles, converted into metadata-aware tokens, updated by masked inter-band attention with residual connection, modulated by per-band gains and cone and rotation gates, and fused to the output feature volume, optionally followed by a residual feature mixer.

**Band tokenization and metadata embedding** Each band tile $T_k$ is converted into a real-valued token $t_k[b] \in \mathbb{R}^E$ for each sample $b$, where $E$ denotes the token or feature width. This token summarizes the band content along the frequency dimension. If the tiles are complex, we first form a real representation per sample by concatenating the real and imaginary (Re/Im) parts as

$$\widetilde{T}_k[b,n,:] = \begin{cases} \left[\operatorname{Re} T_k[b,n,:] \;\|\; \operatorname{Im} T_k[b,n,:]\right], & \text{if Re/Im tokens,} \\ \operatorname{Re} T_k[b,n,:], & \text{otherwise,} \end{cases} \tag{4.56}$$

where $b$ indexes batch and $n$ indexes the frequency samples in $\mathcal{I}_k$. In the implementation, the tokenization path can first apply a small Conv1D along the band-sample axis and then mean-pool the resulting sequence. We write this preprocessing, pooling, and projection as

$$P_k[b,n,:] = \begin{cases} \operatorname{Conv1D}\big(\widetilde{T}_k[b,\cdot,:]\big)[n,:], & \text{if Conv1D tokenizer,} \\ \widetilde{T}_k[b,n,:], & \text{otherwise,} \end{cases} \tag{4.57a}$$

$$\mu_k[b] = \frac{1}{N_k}\sum_{n=1}^{N_k} P_k[b,n,:], \qquad \mu_k[b] \in \mathbb{R}^{C_{\text{mid}}}, \tag{4.57b}$$

$$t_k[b] = \mathbf{W}_{\text{tok}}\,\mu_k[b] + \mathbf{b}_{\text{tok}}, \qquad t_k[b] \in \mathbb{R}^E. \tag{4.57c}$$

Here $C_{\text{tok}} = 2C_{\text{in}}$ when Re/Im tokens and $C_{\text{tok}} = C_{\text{in}}$ otherwise. When Conv1D tokenizer, $C_{\text{mid}} = F_{\text{tok}}$, where $F_{\text{tok}}$ denotes the number of Conv1D filters; otherwise $C_{\text{mid}} = C_{\text{tok}}$. The shared projection uses a learned matrix $\mathbf{W}_{\text{tok}} \in \mathbb{R}^{E\times C_{\text{mid}}}$ and bias $\mathbf{b}_{\text{tok}} \in \mathbb{R}^E$. In the experiments reported here, the Conv1D tokenizer is used before pooling. Band identity is encoded by the metadata embedding

$$\psi_k = \phi_{\text{meta}}\big(j(k), \nu(k), \ell(k), r(k)\big) \in \mathbb{R}^E, \tag{4.58}$$

using learned embeddings for implementation-encoded discrete metadata indices and a small linear layer. The band token is then formed as

$$u_k[b] = t_k[b] + \psi_k. \tag{4.59}$$

When only a single rotation is used, the rotation embedding term is trivial and can be omitted. In the experiments reported here $M = 1$, so the metadata pathway effectively uses only scale, cone-slot, and orientation information. Stacking all bands gives, for each sample $b$, a token matrix $U[b] \in \mathbb{R}^{K\times E}$ with rows $u_k[b]$. In the remaining equations we suppress the explicit batch index for readability. Thus $u_k$, $\widehat{u}_k$, $\eta_k$, and $T_k^{\text{out}}$ are understood sample-wise.

**Masked inter-band attention (band space only)** Inter-band interaction is restricted to bands that share the same (scale, cone-slot) pair, with the LP band used as a global connector. Let the same-$(j,\nu)$ groups be

$$\mathcal{G}_{j,\nu} = \{k : j(k) = j,\ \nu(k) = \nu\}. \tag{4.60}$$

The corresponding boolean attention mask is

$$\mathcal{M}_{kk'} = \begin{cases} 1, & \text{if } \exists (j,\nu) \text{ such that } k, k' \in \mathcal{G}_{j,\nu}, \\ 1, & \text{if } j(k) = -1 \text{ or } j(k') = -1, \\ 0, & \text{otherwise.} \end{cases} \tag{4.61}$$

Masked multi-head self-attention (MHSA) is then applied to the token sequence (per batch) with this mask. Writing $U \in \mathbb{R}^{K \times E}$ for a single batch, we apply the boolean mask $\mathcal{M}$ and compute

$$\widehat{U} = \mathrm{MHSA}(U; \mathcal{M}), \tag{4.62}$$

with updated tokens $\widehat{U} = [\widehat{u}_k] \in \mathbb{R}^{K \times E}$. In implementation terms, this is a standard multi-head attention layer used in self-attention mode, with identical queries, keys, and values $Q_{\text{att}} = K_{\text{att}} = V_{\text{att}} = U$. Our implementation wraps this attention block by pre-normalization and a residual connection, but those details are omitted here for brevity.

**Symmetric per-band gains** Each attended band token $\widehat{u}_k$ produces an identity-centred scalar gain that can both boost and attenuate that band. Here $\mathbf{w}_g \in \mathbb{R}^E$ and $b_g \in \mathbb{R}$ are learned gain-head parameters:

$$\eta_k = 1 + \lambda_g \tanh\!\big(\mathbf{w}_g^\top \widehat{u}_k + b_g\big), \qquad \lambda_g > 0. \tag{4.63}$$

The hyperparameter $\lambda_g$ limits the maximum deviation from 1, so $\eta_k \in (1 - \lambda_g, 1 + \lambda_g)$. The value $\eta_k = 1$ therefore corresponds to neutral per-band scaling on top of the underlying FDST coefficients.

**Cone and rotation gates** In addition to the token-driven gains, the block learns simple global gates per cone and per rotation. Let $\gamma \in \mathbb{R}_+^3$ be nonnegative cone-slot weights and $\tilde{\alpha} \in \Delta^{M-1}$ denote rotation weights in the standard $(M-1)$-simplex. We parametrize $\gamma_\nu = \mathrm{softplus}(\gamma_\nu^{\log})$ so that the cone gates remain nonnegative, and $\tilde{\alpha}_m$ is obtained by a softmax over rotation logits $\tilde{\nu}_m$. Here $\nu(k)$ should be read as the implementation cone-slot index used by the metadata path, not as the abstract label set $\{x, y, z\}$ above. For band $k$ we define a static gate

$$\delta_k = \gamma_{\nu(k)}\, \tilde{\alpha}_{r(k)}, \tag{4.64}$$

so that bands sharing the same cone or rotation can be globally boosted or suppressed. These gates are conceptually separate from the internal FDST rotation weights in (2.61a), which act inside the analysis and synthesis pair, and operate only at the ShearViT block level. In particular, the LP band uses cone slot 0 and therefore shares the same cone gate as the x-cone in the current implementation. In the experiments reported here $M = 1$, so $\tilde{\alpha}_0 = 1$ and only the cone gates are nontrivial. A metadata-consistent multi-rotation variant would use the same formula with nontrivial $r(k)$, but that case is not exercised in the present experiments.

**Tile scaling and one-shot synthesis** The band tiles are modulated by the learned gains and static gates.

$$T_k^{\mathrm{out}} = \eta_k\, \delta_k\, T_k. \tag{4.65}$$

Operationally, this amounts to a per-band scalar scaling of each tile $T_k^{\mathrm{out}} \in \mathbb{C}^{B\times N_k\times C_{\mathrm{in}}}$. All scaled tiles are then passed to the associated inverse FDST bank, which uses the dual synthesis weights from (2.62) to recover a real spatial field

$$h[\boldsymbol{n}] = \mathsf{FDST}^{-1}\big(\{T_k^{\mathrm{out}}\}_{k=0}^{K-1}\big)[\boldsymbol{n}] \in \mathbb{R}^{C_{\mathrm{in}}}. \tag{4.66}$$

The synthesized field is treated as real-valued, and its real part is retained as a numerical safeguard.

**Output projection and optional feature-space mixer** A $1\times1\times1$ fusion layer sets the output feature width.

$$f[\boldsymbol{n}] = \mathbf{W}_{\mathrm{fuse}}\, h[\boldsymbol{n}] + \mathbf{b}_{\mathrm{fuse}} \;\in\; \mathbb{R}^{E}. \tag{4.67}$$

Here $\mathbf{W}_{\mathrm{fuse}}$ and $\mathbf{b}_{\mathrm{fuse}}$ denote the learned weights and bias of the $1\times1\times1$ fusion layer. Optionally, a small feature-space mixer is applied in a residual fashion.

$$f_{\mathrm{mix}}[\boldsymbol{n}] = f[\boldsymbol{n}] + \mathrm{Mix}\big(f[\boldsymbol{n}]\big), \tag{4.68}$$

where Mix is a sequence $1\times1\times1 \rightarrow$ Swish-gated linear unit (SwiGLU) activation $\rightarrow$ depthwise $3\times3\times3$ convolution $\rightarrow 1\times1\times1$, possibly followed by dropout. This mixer does not introduce resolution-specific parameters. The FDST windows themselves are built for the fixed spatial grid processed by the block. Depending on the configuration, the block returns updated band tokens $\widehat{U}$, spatial features after fusion or optional mixing, or both.

**Trainable parameters and design choices** Trainable components of the ShearViT block include the FDST-internal per-frequency gain variables and, when enabled, the internal FDST rotation logits, the tokenization layers including $(\mathbf{W}_{\mathrm{tok}}, \mathbf{b}_{\mathrm{tok}})$ and the optional Conv1D tokenizer, the metadata embeddings and projection $\phi_{\mathrm{meta}}$, the multi-head attention projections and normalization layers, the gain head $(\mathbf{w}_g, b_g)$, the cone and rotation gate logits, and the fusion and optional mixer convolutions $(\mathbf{W}_{\mathrm{fuse}}, \mathbf{b}_{\mathrm{fuse}})$. The range parameter $\lambda_g$ is treated as a fixed hyperparameter in the experiments reported here. There is no spatial or axial attention and no absolute positional encoding. Instead, all geometric information enters through the FDST band structure, the metadata embeddings, and the masked attention across same-$(j,\nu)$ groups.

**Backpropagation through ShearViT:** Let $\mathcal{L}(\theta)$ be a loss on the spatial output feature map $\mathbf{f}(\boldsymbol{n})$ produced by a ShearViT block. The backward pass starts from the gradient field $g_{\mathrm{feat}}(\boldsymbol{n}) = \partial\mathcal{L}/\partial\mathbf{f}(\boldsymbol{n})$. The 1×1×1 fusion layer and the optional feature mixer are standard convolutional layers, so TensorFlow computes gradients with respect to their kernels and biases as in a conventional convolutional network. These gradients are propagated to the synthesized field $h(\boldsymbol{n})$ and then to the externally scaled tiles $T_k^{\text{out}}$. The inverse FDST step uses the same dual synthesis algebra as (2.62), so gradients pass through the scatter-add synthesis, its denominator $D[\boldsymbol{\omega}]$, and the underlying FDST gains. At the ShearViT block level, the feature path scales each tile sample-wise as

$$T_k^{\text{out}}[b,n,c] = \eta_k[b]\,\delta_k\,T_k[b,n,c], \qquad \eta_k[b] = 1 + \lambda_g \tanh z_k[b], \tag{4.69}$$

where $z_k[b]$ is the output of the token gain head and $\delta_k = \gamma_{\nu(k)}\tilde{\alpha}_{r(k)}$. Therefore the gradient of the sample-wise gain is

$$\frac{\partial \mathcal{L}}{\partial \eta_k[b]} = \delta_k \sum_{n,c} \text{Re}\Big(\overline{T_k[b,n,c]}\,\frac{\partial \mathcal{L}}{\partial T_k^{\text{out}}[b,n,c]}\Big), \tag{4.70}$$

and the static gate gradient is

$$\frac{\partial \mathcal{L}}{\partial \delta_k} = \sum_{b,n,c} \eta_k[b]\ \text{Re}\Big(\overline{T_k[b,n,c]}\,\frac{\partial \mathcal{L}}{\partial T_k^{\text{out}}[b,n,c]}\Big). \tag{4.71}$$

The gain-head parameters receive the additional local derivative $\lambda_g(1-\tanh^2 z_k[b])$ and then backpropagate into the attended band token. In the real and imaginary token representation used in the implementation, the complex inner products above reduce to ordinary sums over real channels. The cone-slot gates use $\gamma_\nu = \text{softplus}(\gamma_\nu^{\log})$ and the rotation gates use $\tilde{\alpha} = \text{softmax}(\tilde{\nu})$, so their gradients follow from $\partial\mathcal{L}/\partial\delta_k$ and the product rule. When tokens are returned, the code also scales the updated tokens by the same $\eta_k[b]\delta_k$ product, giving an additional token-side gradient to the gain head, cone gates, rotation gates, inter-band attention, and metadata embeddings.

Table 4.6: Frugal attention mechanisms

| # | Attention mechanism | # trainable params | input size |
|---|---|---|---|
| | **Self attention** | | |
| 1a | $\mathsf{A}_{\text{DWT}}$ | **445** | (64,64,64,1) |
| 1b | Vanilla self (conv.) | 6,863 | (64,64,64,1) |
| | **multiresolution vision transformer** | | |
| 2a | WaveletViT | **1,884** | (64,64,64,1) |
| 2b | ShearViT | **1,308,838** | (64,64,64,1) |
| 2c | Convolutional ViT (Vanilla) | 34,176 | (64,64,64,1) |

ShearViT is counted for $J = 2$, $L_0 = 4$, token width $E = 64$, four heads, Conv1D tokenization, and no feature mixer. The count includes trainable FDST per-frequency gains. Without these FDST gain variables, the ShearViT block has 20,726 trainable parameters.

## 4.3 Discussion

This chapter describes learnable IIR filters and multiresolution attention modules for deep learning, including a full multidimensional IIR realization and a practical separable direct form II realization. The main high-dimensional challenge is that a full $D$-dimensional IIR layer must build feedforward and feedback delay banks over $(L+1)^D$ shifted copies of the input and output, so arithmetic and memory grow rapidly with $D$, the delay order $L$, and the channel counts. The inner recursive refinement also makes boundary handling and gradient propagation more difficult. The separable direct form II version addresses this by applying stable one-dimensional recursions axis by axis, where the pole parameterization keeps the one-dimensional poles inside the unit circle. These separable blocks can be inserted into standard convolutional encoders to add long-range memory with a system-theoretic interpretation. The chapter also builds WaveletViT and ShearViT as frugal attention modules with native multiresolution feature spaces. WaveletViT uses DWT bands to split the input into lowpass and highpass components across scales, applies attention to compact tokens from these bands, and then reconstructs a feature map through an inverse wavelet transform. ShearViT plays a similar role for directional three-dimensional data by working with shearlet bands as tokens. In both designs, the model operates on a reduced token space while still exposing a rich multiscale structure. The next chapter uses these IIR realizations and multiresolution attentions inside deep vision models for image and volumetric segmentation, where the IIR approximations provide large contextual fields and WaveletViT or ShearViT provide scale separation and directional context in volumetric brain MRI.

# Chapter 5

# Deep Multiresolution Vision Models and Longitudinal Analysis of Brain MRI

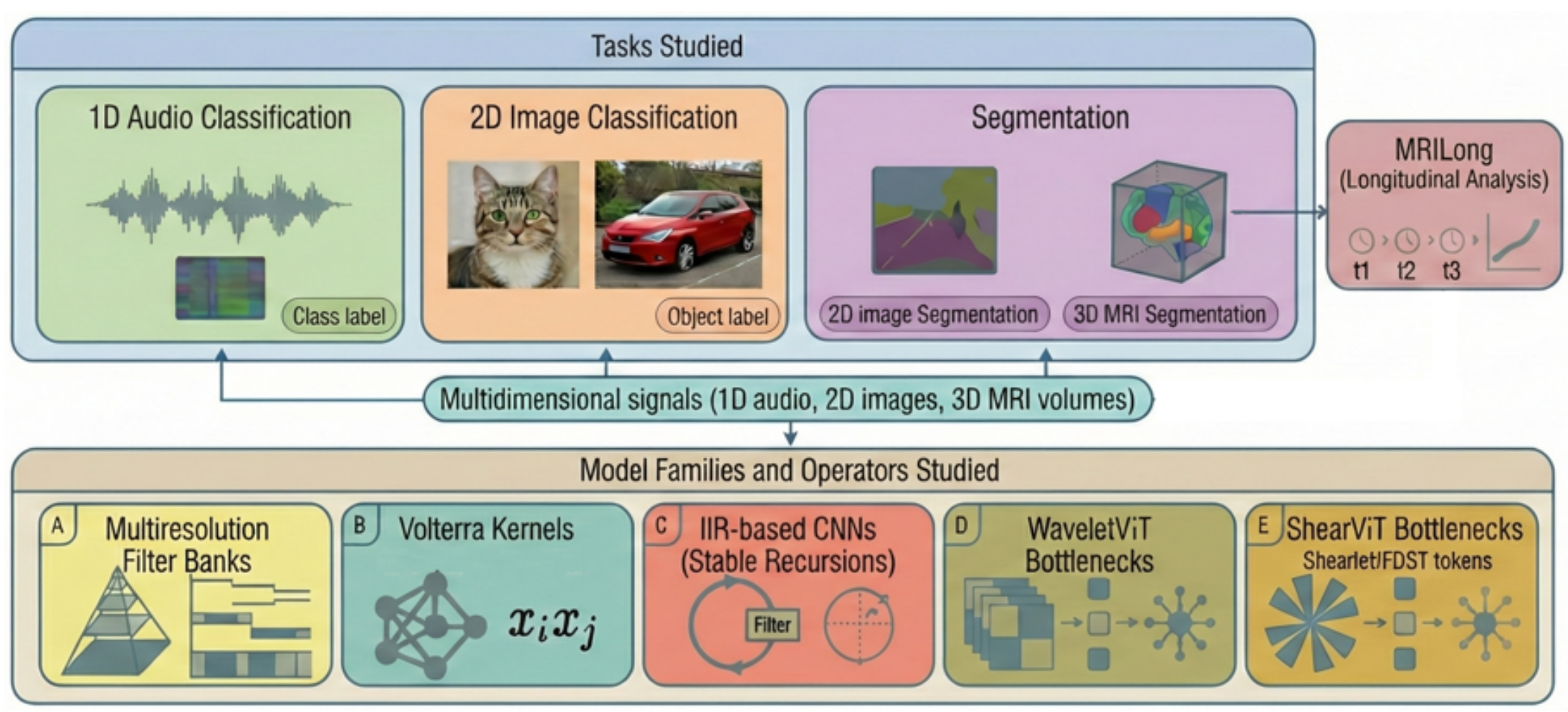

The multiresolution filter-banks, Volterra kernels, separable IIR layers, and frugal multiresolution attentions (WaveletViT and ShearViT) described in earlier chapters are used here as fundamental operators in assembling deep vision models with structured and efficient filter-bank representations. This chapter introduces classifiers with multiresolution encoders and dense Volterra kernel heads. Further segmentation models with multiresolution encoder–decoder (MED) filter-banks and novel attentions are introduced here. Such 2D segmentation models are tested with a wide variety of open-source segmentation datasets. The 3D segmentation models are rigorously studied here to build a multistage pipeline with multiple models for fast longitudinal analysis of 3D T1-weighted brain MRI with skull stripping and tissue segmentation. This pipeline is used to segment unlabeled MRI volumes of ischemic stroke patients undergoing Ayurvedic and Yogic interventions.

## 5.1 Historical Foundations of classification and segmentation

A classification and segmentation pipeline assigns labels to inputs, a trait that seems so simple for human vision, but is extremely challenging in machine vision. Its philosophical idea traces back to deep human intellectual history. Philosophers in antiquity sought systematic ways to categorize the world's entities and phenomena. Aristotle's Categories, for example, presented one of the first formal classification schemes, dividing all "beings" into ten fundamental categories (substance, quantity, quality, etc.) [226]. In the ancient Bharatiya logic, notably the Nyaya-Vaiseṣika tradition has its own taxonomy (padartha) to classify all types of existents into fundamental categories like substance, quality, motion, universal, particular, and so on [227]. These early frameworks, though philosophical, highlight a timeless idea to understand and organize objects into groups based on shared properties. Carl Linnaeus in the 18th century, introduced a hierarchical system to classify living organisms (kingdom, genus, species, etc.), establishing the "modern taxonomy" still used today. The 19th and early 20th centuries saw the transition from qualitative taxonomies to quantitative, mathematical classification methods. A key early milestone was Fisher's Linear Discriminant (1936) [228], which introduced a statistical method to separate measurements into classes by finding the linear combination of features that best discriminates between groups. Fisher's work, later generalized as Linear Discriminant Analysis (LDA), was among the first formal pattern recognition techniques using algebra and statistics to classify data points.

Around the same time, the foundations of Bayesian decision theory were laid, framing classification as an exercise in probabilistic inference. In Bayesian classification, class labels are assigned by evaluating posterior probabilities of each class given the observed features and choosing the label that minimizes expected loss. In other words, Bayesian decision theory is the statistical approach to pattern classification, leveraging probability and risk for optimal decisions [229]. By the mid-20th century, this view crystallized into a decision-theoretic view of the classification problem, where an optimal classifier is one that, for each input $x$, predicts the label $y$ that maximizes the posterior probability or minimizes expected misclassification cost. This principle, formalized by the Bayes optimal decision rule, remains the conceptual bedrock of classification algorithms today. The rise in general methods in statistical pattern recognition during the 1950s-1970s extended classification beyond simple linear schemes. One seminal development was the nearest-neighbor rule, first proposed by Fix & Hodges in 1951 [230] and later analyzed by Cover & Hart (1967) [231]. A nearest neighbor classifier labels a query sample with the label of its closest training example [230]. Under i.i.d. sampling, the asymptotic error of the 1-NN rule is at most twice the Bayes error [231]. Another major development was the introduction of kernel methods during the 1960s. For example, Aizerman et al. (1964) modeled nonlinear relationships by mapping data into high-dimensional feature spaces and using inner-product kernels. This allowed linear algorithms to act as nonlinear classifiers, paving the way for modern approaches. The theoretical foundations of learning were simultaneously advanced by Vapnik and Chervonenkis, who in 1970 formalized the VC (Vapnik–Chervonenkis) theory. VC theory introduced measures of model capacity (the VC dimension) and provided generalization bounds, grounding classification in statistical learning theory. In VC theory, the VC dimension is a measure of the capacity or complexity of a class of functions, originally defined by Vapnik and Chervonenkis [239].

Table 5.1: Historical evolution of classification models in AI

| Year | Authors | Contribution | Methodology |
|---|---|---|---|
| 1763 | Bayes [232] | Bayes' theorem: probabilistic framework for inference and decision under uncertainty | Probabilistic inference |
| 1936 | R.A. Fisher [228] | Linear Discriminant Analysis for class separation under Gaussian assumptions | Linear classification |
| 1958 | Rosenblatt [28] | Perceptron: Trainable linear model with online weight updates | Early neural network |
| 1967 | Cover & Hart [231] | Nearest Neighbor rule for instance-based classification | Non-parametric model |
| 1986 | Rumelhart et al. [18] | Backpropagation for training multilayer networks | Gradient-based learning |
| 1995 | Cortes & Vapnik [233] | Support Vector Machines: maximum-margin classification with kernels | Margin-maximizing kernel methods |
| 1996 | Freund & Schapire [234] | AdaBoost: boosted ensembles of weak learners | Ensemble learning (boosting) |
| 2001 | Breiman [235] | Random Forests: bagged decision trees with feature subsampling | Ensemble learning (bagging) |
| 2004 | Lowe [236] | SIFT: scale-invariant local descriptors for recognition pipelines | Handcrafted image features |
| 2005 | Dalal & Triggs [237] | HOG: Gradient histograms for pedestrian detection | Orientation-based descriptors |
| 2012 | Krizhevsky et al. [17] | AlexNet: Deep CNN with GPU training on ImageNet | Deep convolutional network |
| 2013 | Bruna & Mallat [35] | Scattering Networks: Wavelet-based invariant representations | Non-learned deep wavelet nets |
| 2017 | Zoumpourlis et al. [157] | Volterra CNNs: Structured convolutional layers using second-order terms | Volterra-based CNNs |
| 2017 | Vaswani et al. [24] | Transformer: self-attention for sequence modeling and classification | Attention-based models |
| 2020 | Dosovitskiy et al. [57] | ViT: patch-token Transformers for image classification | Vision Transformers |
| 2021–23 | Radford et al., Yu et al., Liu et al. [49, 98, 238] | CLIP / CoCa / ConvNeXt V2: large-scale pretraining and hybrid backbones for classification | Foundation models; hybrid CNN/Transformer |
| 2024 | Gu & Dao [100] | Mamba: selective state-space models enabling efficient long-context sequence classification | State-space sequence models |

By the 1980s and 1990s, these developments coalesced into machine learning, and we saw an emergence of powerful classification algorithms. MLP neural networks were trained with the backpropagation algorithm and enabled nonlinear classification through learned distributed features. SVMs (Cortes and Vapnik, 1995) [233] also became a dominant method. SVMs were exceptional across domains in finding optimal separating nonlinear boundaries in hyperplanes with maximal margin between classes using kernel functions. Ensemble methods also gained prominence, with richer hierarchical representations of input, and algorithms were developed that combined many simple "weak" classifiers into a strong classifier. Notably, decision tree ensembles like Random Forests (Breiman 2001) [235] and boosting methods like AdaBoost (Freund & Schapire 1996) [234] achieved top performance by aggregating the predictions of multiple models. By the early 2000s, a rich set of classification algorithms (e.g., k nearest neighbor, MLP neural networks, SVMs, boosted trees) became available, which had strong theoretical foundations (e.g., VC bounds) and increasingly large real-world datasets. Such classification systems were being applied to diverse domains, for example, sound classification, recognizing handwritten digits, classifying medical conditions, and detecting objects in images.

**Handcrafted Features and the Pre-Deep Era:** The early classification and segmentation models (before deep learning) relied on extensive model-based feature engineering. Whether the task was to classify an entire image or to segment image regions, engineers typically designed ad hoc features relevant to the problem. In image analysis, classic feature descriptors like Gabor filters, wavelet coefficients, histograms of oriented gradients (HOG) [237], scale-invariant feature transform (SIFT), and local binary patterns (LBP) were extensively used. A standard pipeline might involve computing multiscale dilated features from raw data, then feeding them into a dense classifier head (SVM, Random Forest, MLP neural network). This paradigm was prevalent in early deep computer vision models and medical imaging through the 2000s. For example, in brain MRI segmentation, one might compute intensity histograms, texture measures (e.g., Gabor or wavelet responses), edge information, etc., then apply a discriminative model to classify each pixel or voxel. Methods include discriminative learning with handcrafted features (texture, spatial, intensity) involving SVMs, Random Forests, AdaBoost, k-nearest neighbor with Gabor, LBP, HOG, and intensity histograms. In segmentation, the goal is to assign a label to every location (pixel or voxel) in an image, effectively producing a labeled map (e.g., partitioning an MRI scan into different tissue types, or an image into object vs background). Classical approaches include thresholding and region growing (dating back to the 1980s), edge detection (e.g., Canny), and morphological filtering [240]. In the 1990s and 2000s, more principled methods emerged, including Markov Random Fields (MRFs), and later Conditional Random Fields (CRFs) became popular for enforcing spatial consistency in segmentations.

These models treat segmentation as a labeling problem on a graph (usually the pixel grid), with local interactions encouraging smooth or coherent label assignments. A segmentation could be obtained by solving a discrete optimization (e.g., graph cuts or belief propagation) to minimize an energy function comprising unary data terms and pairwise smoothness terms. For example, a common formulation posits an energy $E(y;x) = \sum_i \psi_u(i, y_i; x) + \sum_{(i,j)\in N} \psi_p(i, j, y_i, y_j; x)$, where $\psi_u$ measures the cost of labeling pixel i as class $y_i$ and $\psi_p$ encourages neighboring pixels $i, j$ to have compatible labels. This is the essence of the CRF approach, which combines per-pixel classification scores (unary potentials) with spatial regularization (pairwise potentials) to find an optimal segmentation (MAP estimate).

**The deep learning revolution:** We witnessed a transformative upgrade in computer vision models since 2010 with the use of deep learning powered by powerful GPU hardware and large datasets, where feature extraction became seamlessly unified into end-to-end trainable models. CNNs that were originally proposed in the late 1980s became truly popular with models like AlexNet, which was trained on ImageNet (2012). This was a classical demonstration of deep representation learning, which could automatically learn multi-layer hierarchies of features directly from raw pixel data by backpropagation and gradient descent. For image classification tasks, deep CNNs (e.g., AlexNet, VGG, ResNet) delivered dramatic improvements on image recognition benchmarks by learning rich representations (edges $\rightarrow$ textures $\rightarrow$ parts $\rightarrow$ objects) from large datasets. Similarly, for sequences and time-series signals, deep architectures like Recurrent Neural Networks (RNNs), especially the LSTM variant(Hochreiter & Schmidhuber, 1997) [21], achieved success by learning temporal dependencies in sequential data tasks like speech recognition and text classification. Likewise, deep encoder-decoder CNNs solved highly challenging image segmentation tasks. For example, Ciresan et al. (2012) [241] trained a deep network to segment neuronal membranes in electron microscopy images. It used patches of input and trained a CNN to classify the pixel in the center of a small image patch, effectively performing segmentation by scanning across the image. This showed the feasibility of deep nets for dense labeling tasks. Fully convolutional networks (Long et al. 2015) [120], Deconvolution network (Noh et al. 2015), and UNet (Ronneberger et al. 2015) in particular became very useful, especially in biomedical image segmentation. A UNet takes an entire image and produces an output mask of the same size, with its decimated learnable representation path and residual skip connections retaining fine-grained details of representation space for precise segmentations. This vision architecture proved remarkably effective across segmentation problems of different domains and sparked what some have called the "UNet revolution" [242]. For example, Cicek et al. (2016) [243] extended UNet to 3D volumes, and Kamnitsas et al. (2016) [244] DeepMedic, a deep 3D CNN for brain lesion segmentation.

**Attention, vision transformers and beyond:** RNNs and LSTMs are useful for long-range contexts, and they are easy to implement for 1D I/O systems with sequences and time series inputs. In higher dimensions, these deep structures (e.g., RNN, LSTM, ARMA) pose a prac-tical computational challenge due to the complexity explosion. In image classification and segmentation, RNNs were sometimes integrated to model spatial sequences (scanning an image row-wise) or to refine segmentation masks iteratively. Sometimes, a CRF is appended to the CNN to further enforce spatial consistency (the CRF-as-RNN model by Zheng et al. 2015 [245] treated CRF refinement as a recurrent neural network). More recently, graph neural networks and attention mechanisms provide alternative ways to model long-range relationships in segmentation outputs, supplanting explicit MRF/CRF formulations with learnable modules. The self-attention architecture (Vaswani et al., 2017) [24] revolutionized sequence modeling and outpaced many earlier long-range context methods. It is now a key technology that dominates language processing, and its multidimensional extension (e.g., ViT) is used in deep vision models. Currently, powerful and reliable deep vision models are built with attention modules that give a richer long-range context of localized features, but they come at a high computational cost, especially for multidimensional data, and also require large training datasets. For example, a state-of-the-art medical image segmentation model is TransUNet (Chen et al. 2021) [246], which is a UNet with a ViT. Similarly, Swin-Unet (Cao et al. 2021) [247] built a UNet using swin (shifted windowed) self-attentions, and produces impressive segmentation. These models benefit from self-attention to capture long-range dependencies of pixel or voxel intensities, which is limited in FIR convolution kernels (with a limited receptive field). Alongside the model architecture, the objective function is also an active research area, especially in self-supervised or unsupervised learning,which enables models to learn useful representations from unlabeled data. Self-supervised pretraining (e.g., with contrastive learning DINO [248], or masked image modeling as in masked autoencoder [249]) is state-of-the-art in both classification and segmentation, especially in medical imaging with scarce labeled data. A model might also be pre-trained via masked patch prediction on a large corpus of images (learning to ”inpaint” missing parts), and then fine-tuned for segmentation, often yielding better accuracy with fewer annotated examples, as reported with frameworks like MoCo UNet or Masked Autoencoder UNet [250]. In 2023, Meta AI released the Segment Anything Model (SAM) [251], a high-capacity model trained on billions of images to perform zero-shot segmentation. Given virtually any image, SAM can produce segmentation masks for objects, guided by minimal user prompts. SAM treats segmentation with user prompts, akin to GPT language models, and is an excellent example of a model that generalizes across domains. MedSAM [252] is an extension of SAM, with domain-specific knowledge to better handle medical images.

An interesting aspect of the current deep learning era is that it largely obviates most tedious manual feature engineering. Given sufficient data and computation, a deep black box can learn necessary representations for a given objective. Instead of explicitly computing wavelet coefficients or texture descriptors, many practitioners now rely on the network’s convolutional filters (or recurrent units, or attention heads) to fulfill a necessary objective. Recent methods attempt to whiten deep models by leveraging the representation learning of deep networks while retaining beneficial inductive biases using classical methods (e.g., multi-scale analysis via wavelets). For example, wavelet pooling in UNet to improve segmentation of multi-scale lesions using a multiresolution wavelet decomposition [253]. Others have added total variation (TV) losses or partial differential equation (PDE)-inspired layers to encourage smoothness or shape priors in the network's output. Table 5.1 and 5.2 list some seminal papers in the historical evolution of classification and segmentation research.

Table 5.2: A concise history of ML / DL based MRI segmentation

| **Era / Period** | **Milestone / Paradigm shift** | **Key Models / Technologies** | **Common Datasets** | **Representative Papers** |
|---|---|---|---|---|
| 1980s–1999 (Foundational Image Processing) | Rule-based segmentation; early edge/region methods | Thresholding; Region Growing; Morphological Filters; edge detection | Simulated MRI; BrainWeb | Marr & Hildreth (1980) [254]; Canny (1986) [255]; Gonzalez & Woods (1992) [256] |
| 2000–2011 (Statistical & Registration Era) | Statistical shape models; spatial priors; multi-atlas registration | MRFs; Active Contours; Level Sets; Atlas-based Segmentation; Graph Cuts; Wavelets | IBSR; BrainWeb; ADNI | Pham et al. (2000) [257]; Ashburner & Friston (2005) [258]; Boykov & Jolly (2001) [259] |
| 2010–2012 (Traditional ML + Handcrafted Features) | Discriminative learning with handcrafted features (texture, spatial, intensity) | SVMs; Random Forests; AdaBoost; k-NN with Gabor, LBP, HOG, intensity histograms | IBSR; BrainWeb; MRBrainS | Geremia et al. (2011) [260]; Tustison et al. (2010) [261]; Kaster et al. (2010) [262] |
| 2012–2015 (Early Deep Learning) | Emergence of CNNs; patch-based training | Shallow CNNs on 2D slices; multilayer perceptrons | BRATS (early); MRBrainS | Ciresan et al. (2012) [241]; Havaei et al. (2015) [263] |
| 2015 (UNet revolution) | End-to-end semantic segmentation with skip connections | UNet (2D); encoder–decoder architecture | ISBI; MRBrainS; BRATS | Ronneberger et al. (2015) [39] |
| 2016–2018 (3D deep models) | Volumetric segmentation; deeper networks | 3D UNet; DeepMedic; Residual CNNs; Dual-Path CNNs | BRATS; MRBrainS; NFBS; ATLAS; iSeg | Cicek et al. (2016) [243]; Kamnitsas et al. (2016) [244] |
| 2017–2023 (Hybrid signal priors + deep models) | Embedding signal priors within DL (wavelet/ multiresolution regularization) | Wavelet pooling/analysis layers; multiresolution priors; spectral constraints | BRATS; MRBrainS; MS-lesion MRI | Williams et al. (2018) [264]; Alijamaat et al. (2021) [253] |

*Table 5.2 (continued)*

| Era / Period | Milestone / Paradigm shift | Key Models / Technologies | Common Datasets | Representative Papers |
|---|---|---|---|---|
| 2018–2021 (Attention and GANs) | Context-aware segmentation; modality fusion | Attention UNet; GANs; modality-specific encoders | BRATS; NFBS; ATLAS; ACDC; CHAOS | Oktay et al. (2018) [265]; Zhao et al. (2019) [266] |
| 2019–2021 (Self-configuring DL pipelines) | Strong baselines via automated configuration and training recipes | nnUNet; standardized training/inference pipelines | MSD; BRATS; ACDC; diverse biomedical MRI | Isensee et al. (2021) [267] |
| 2021–2024 (Transformers & pretraining) | Long-range dependency modeling; large-scale pretraining | TransUNet; UNETR; masked autoencoding pretraining | BTCV; ACDC; CHAOS; ATLAS R2.0 | Chen et al. (2021) [246]; Hatamizadeh et al. (2022) [213]; He et al. (2022) [249] |
| 2023–2024 (Foundation / promptable segmentation) | Prompt-driven and transferable segmentation backbones | SAM-style promptable models; medical adaptations | Medical imaging benchmarks; curated institutional MRI | Kirillov et al. (2023) [251]; MedSAM (2023) [252] |

In summary, the evolution from Aristotle's categories to today's deep networks illustrates an expanding capability to map inputs to labels with ever-increasing accuracy and automation. We moved from manually defining categories and features, to statistically learning them, and now to let deep models **discover** relevant representations. Yet the core inferential problem, deciding which label(s) best explain the input remains the same. The deep vision models of Industry 4.0 and 5.0 emphasize interpretability (e.g., shape priors), computational efficiency (e.g., lightweight CNN), and domain-aware design. This chapter describes deep vision models with multiscale filter banks (e.g., wavelets and shearlets), Volterra kernels, and constraints that seamlessly weave domain knowledge into a deep learnable graph.

**A Common Inferential Framework:** The big picture is that both classification and segmentation can be seen through the lens of inference on statistical models: classification yields a single decision $y = \arg\max_y p_\theta(y|x)$ [229], while segmentation yields a field of decisions $y(i) = \arg\max_y p_\theta(y_i|x)$ for each pixel i, often with an implied joint posterior over the field imposing coherence. In the end, both are about estimating unknown labels given observed data – one could say segmentation is classification at scale, with an output space of structured labels rather than a single category. More accurately, semantic segmentation is often formulated as assigning a class label to each pixel (a per-pixel classification task) [268]. The difference from simple classification is that the outputs are structured (e.g., a 2D grid of labels, rather than a single label), and there are usually spatial dependencies between outputs. Both classification and segmentation involve mapping inputs (e.g., images, signals, data points) to labels, and both can be framed in terms of optimizing some criterion – be it maximizing a posterior probability, minimizing an energy, or reducing an expected loss. Segmentation adds complexity by dealing with structured outputs (arrays of labels with spatial relationships), but mathematically, this can be handled by extending the same principles that govern simple classification. In this chapter both classification and segmentation are formulated in a decision-theoretic framework with a common notation, where an input space $\mathcal{X}$ and label space $\mathcal{Y}$ (with $\mathcal{Y}$ possibly a product space label of each spatial coordinate for segmentation), a model $p_\theta(y|x)$giving label probabilities, and a decision rule $f_\theta$ that predicts labels by maximizing this posterior (or a cost-sensitive variant). Learning parameters $\theta$ is the minimization of a loss (e.g., cross-entropy) on training data. In this unified view, segmentation is a natural extension of classification for individual pixels or voxels.

**Notations** A common set of notations to describe classification and segmentation problems is given below.

- *Input space.* The observation space is denoted by $\mathcal{X}$, and a single datum is written as $x \in \mathcal{X}$. In imaging applications, an element $x \in \mathcal{X}$ is a discrete $D$-dimensional field $x : \Omega \to \mathbb{R}^{C_{\text{in}}}$, where $\Omega \subset \mathbb{Z}^D$ is a finite grid of spatial indices and $C_{\text{in}} \in \mathbb{N}$ is the number of input channels (for example, color channels or MRI modalities).
- *Label set and fields.* The class label space is $\mathcal{Y} = \{1, \dots, K\}$ with $K \geq 2$. In classification, an output label is a single element $y \in \mathcal{Y}$. In segmentation, a label field is a mapping $y : \Omega \to \mathcal{Y}$, written as $y = (y_{\boldsymbol{n}})_{\boldsymbol{n}\in\Omega}$ with $y_{\boldsymbol{n}} \in \mathcal{Y}$.
- *Random variables and realizations.* Random variables are written in uppercase and their realizations in lowercase. For classification, $(X, Y)$ denotes a random pair taking values in $\mathcal{X} \times \mathcal{Y}$, and $(x, y)$ denotes a realization. For segmentation, $Y = (Y_{\boldsymbol{n}})_{\boldsymbol{n}\in\Omega}$ is a random label field and $y = (y_{\boldsymbol{n}})_{\boldsymbol{n}\in\Omega}$ is a realized field.
- *Data distribution and sample.* The unknown joint law of the data is denoted by $P$. In classification, $P$ is a probability measure on $\mathcal{X} \times \mathcal{Y}$; in segmentation, $P$ is defined on $\mathcal{X} \times \mathcal{Y}^\Omega$. A training sample consists of $n$ independent draws from $P$ and is written as $S = \{(x_i, y_i)\}_{i=1}^n$.
- *Model parameters, posteriors, and decisions.* The symbol $\theta$ denotes the collection of model parameters. The corresponding conditional posteriors are written $p_\theta$, and the induced decision rule is written $f_\theta$, typically obtained as a pointwise maximizer of $p_\theta$.
- *Probability simplex.* The $(K-1)$-dimensional probability simplex is $\Delta_{K-1} = \Big\{\mathbf{q} \in [0,1]^K : \sum_{k=1}^K \mathbf{q}_k = 1\Big\}$, and serves as the range for class posterior vectors.
- *Losses and costs.* The basic loss for a single decision is a function $\mathcal{L} : \mathcal{Y} \times \mathcal{Y} \to \mathbb{R}_{\geq 0}$, with the 0–1 loss given by $\mathcal{L}(y, \hat{y}) = \mathbf{1}\{y \neq \hat{y}\}$. More general cost structures are encoded by a non-negative cost matrix $C \in \mathbb{R}_{\geq 0}^{K\times K}$, where $C(y, \hat{y})$ is the cost of predicting $\hat{y}$ when the true label is $y$.

- *Neighborhood systems for segmentation.* In segmentation problems, spatial regularity is expressed through a neighborhood system $\mathcal{N} \subseteq \Omega \times \Omega$. A pair $(\boldsymbol{n}, \boldsymbol{m}) \in \mathcal{N}$ indicates that the locations $\boldsymbol{n}$ and $\boldsymbol{m}$ are neighbors. For each $\boldsymbol{n} \in \Omega$, the set $\mathcal{N}(\boldsymbol{n}) = \{\boldsymbol{m} \in \Omega : (\boldsymbol{n}, \boldsymbol{m}) \in \mathcal{N}\}$ collects all neighbors of $\boldsymbol{n}$.

## 5.2 Multiresolution encoder and Volterra head classifiers

A classification problem consists of an input space $\mathcal{X}$ and a finite label set $\mathcal{Y} = \{1, \ldots, K\}$ with $K \geq 2$. For a given observation $x \in \mathcal{X}$ the task is to predict a label $y \in \mathcal{Y}$. A probabilistic classifier associates to each $x$ a vector of class posteriors

$$p_\theta(x) = \big(p_\theta(x)_1, \ldots, p_\theta(x)_K\big) \in \Delta_{K-1}, \tag{5.1}$$

where $p_\theta(x)_k$ is the model's estimate of the probability that $Y = k$ given $X = x$, the parameter vector $\theta$ collects all trainable weights, and $\Delta_{K-1}$ denotes the $(K-1)$-dimensional probability simplex. The induced decision rule chooses the most probable label,

$$f_\theta(x) = \arg\max_{k \in \mathcal{Y}} p_\theta(x)_k, \tag{5.2}$$

so that $f_\theta : \mathcal{X} \to \mathcal{Y}$ maps each input $x$ to a predicted class.

To assess the quality of a decision rule $f$ a loss function $\mathcal{L} : \mathcal{Y} \times \mathcal{Y} \to \mathbb{R}_{\geq 0}$ is introduced, where $\mathcal{L}(y, \hat{y})$ measures the cost of predicting $\hat{y}$ when the true label is $y$. Let $P$ denote the unknown joint distribution of $(X, Y)$ on $\mathcal{X} \times \mathcal{Y}$. The (population) risk of $f$ is defined as

$$\mathcal{E}(f) = \mathbb{E}_{(X,Y)\sim P}\big[\mathcal{L}\big(Y, f(X)\big)\big]. \tag{5.3}$$

The Bayes-optimal decision rule $f^\star$ is the rule that minimizes this risk. Pointwise in $x$ it is characterized by

$$f^\star(x) \in \arg\min_{\hat{y} \in \mathcal{Y}} \mathbb{E}\big[\mathcal{L}(\hat{y}, Y) \mid X = x\big], \tag{5.4}$$

that is, $f^\star(x)$ chooses a label that minimizes the conditional expected loss under the true posterior distribution $P(Y \mid X = x)$. Two special cases are particularly important. One is the loss between 0–1 loss, $\mathcal{L}(y, \hat{y}) = \mathbf{1}\{y \neq \hat{y}\}$, then the Bayes rule reduces to the maximum a posteriori (MAP) classifier

$$f^\star(x) = \arg\max_{k \in \mathcal{Y}} P(Y = k \mid X = x). \tag{case 1}$$

Second is when decisions incur different costs, a non-negative cost matrix $C \in \mathbb{R}^{K \times K}_{\geq 0}$ is introduced, where $C(y, \hat{y})$ is the cost of predicting $\hat{y}$ when the true label is $y$. The Bayes-optimal rule then minimizes the conditional expected cost

$$f^\star(x) = \arg\min_{\hat{y} \in \mathcal{Y}} \sum_{y \in \mathcal{Y}} C(y, \hat{y})\, P(Y = y \mid X = x). \tag{case 2}$$

**Learning from data** In practice $P$ is unknown and a training sample $S = \{(x_i, y_i)\}_{i=1}^n$ drawn from $P$ is observed. Empirical risk minimization with regularization chooses parameters

$$\hat{\theta} \in \arg\min_{\theta} \frac{1}{n} \sum_{i=1}^{n} \mathcal{L}\big(y_i, f_\theta(x_i)\big) + \lambda \mathcal{R}(\theta), \tag{5.5}$$

where $\lambda \geq 0$ is a regularization weight and $\mathcal{R}(\theta)$ is a penalty that controls the complexity of the model. When a conditional model $p_\theta(y \mid x)$ is specified, it is common to take $\mathcal{L}$ as the negative log-likelihood, which yields the empirical negative log-likelihood objective

$$\widehat{\mathcal{E}}_{\mathrm{NLL}}(\theta) = \frac{1}{n} \sum_{i=1}^{n} \big( - \log p_\theta(y_i \mid x_i)\big). \tag{5.6}$$

If a prior density $\pi(\theta)$ is placed on the parameters, MAP estimation seeks

$$\hat{\theta}_{\mathrm{MAP}} \in \arg\max_{\theta} \Big\{ \log p_\theta(S) + \log \pi(\theta) \Big\}, \tag{5.7}$$

where $p_\theta(S)$ is the likelihood of the training set under the model.

**Cost-aware prediction at deployment** Once the classifier has been trained, posteriors $p_\theta(x)$ are available for new observations. At deployment time the prediction should reflect the application's cost structure. Given a cost matrix $C$, the decision rule

$$\hat{y}(x) = \arg\min_{\hat{y} \in \mathcal{Y}} \sum_{y \in \mathcal{Y}} C(y, \hat{y})\, p_\theta(x)_y \tag{5.8}$$

chooses the label with minimal expected cost under the model posterior. When all misclassification costs are equal, this reduces to the simpler rule $\hat{y}(x) = \arg\max_{k \in \mathcal{Y}} p_\theta(x)_k$.

**AProcS view of classification models** In Chapter 1 the **A**nalysis–**Pro**cessing–**S**ynthesis (AProcS) viewpoint was introduced as a common template for supervised, semi-supervised, and self-supervised deep learning models, with or without an explicit multiresolution lens. A generic AProcS pipeline for a classifier can be written as

$$\boxed{x \xrightarrow{\mathcal{A}} r \xrightarrow{\Phi_\theta} \tilde{r} \xrightarrow{\mathcal{S}} u \xrightarrow{\text{head}} p_\theta \in \Delta_{K-1}, \quad f_\theta = \arg\max p_\theta} \tag{AProcS classifier template}$$

Here $\mathcal{A}$ is an analysis operator that maps the input $x$ to an intermediate representation $r$, $\Phi_\theta$ is the learnable core that produces a transformed representation $\tilde{r}$, $\mathcal{S}$ is a synthesis operator that brings features back to a suitable form $u$, and the prediction head maps $u$ to class posteriors $p_\theta$ and a decision $f_\theta$.

**Template of a multiresolution encoder classifier** The classifiers described later in the section fit this multiresolution encoder (ME) template:

- $\mathcal{A}$: identity or light front-ends such as spectrogram stems for audio or directional/shearlet stems for images.
- $\Phi_\theta$: convolutional or multilayer perceptron blocks, optionally grouped or augmented with Volterra-type heads.
- $\mathcal{S}$: identity (the analysis representation is already in a suitable form for prediction).
- **Head**: dense layer plus softmax, or a Volterra kernel followed by a polynomial nonlinearity, producing $p_\theta(x) \in \Delta_{K-1}$ and decision $f_\theta(x) = \arg\max_k p_\theta(x)_k$.
- **Loss**: negative log-likelihood or cross-entropy, possibly combined with cost-aware decision rules using a cost matrix $C$.

**Remarks.** Binary classification ($K = 2$) is often implemented with a logistic link. Class imbalance is typically handled by a non-uniform cost matrix $C$ or by reweighting the training examples. Multi-label problems use one posterior per class together with a loss that decomposes across labels.

This section specializes the abstract classification framework of Section 5.1 to the concrete architectures used in this thesis. The focus is on Multiresolution Encoder and Volterra-head (MEV) classifiers, where a multiscale feature extractor is combined with a compact probabilistic head. All such models share the same external behavior: given an input $x \in \mathcal{X}$ they produce a posterior vector $p_\theta(x) \in \Delta_{K-1}$ and a decision

$$f_\theta(x) = \arg\max_k p_\theta(x)_k,$$

in accordance with the decision-theoretic formulation in the previous section.

**Template of MEV classification models** Each MEV classifier can be viewed as a composition of three stages,

$$\boxed{x \xrightarrow{\Phi_\theta} h \xrightarrow{\text{pool}} z \xrightarrow{\text{head}} p_\theta(x) \in \Delta_{K-1}, \qquad f_\theta(x) = \arg\max_k p_\theta(x)_k} \quad \text{(MEV classifier)}$$

with the following roles.

- *Multiresolution encoder* $\Phi_\theta$. The mapping $\Phi_\theta$ takes the raw input $x$ and produces a feature field $h$. In this thesis it is instantiated by different front-ends depending on the data modality. Examples include spectrogram stems followed by convolutional layers for audio signals, shearlet or other directional filter-banks followed by learnable mixing for images, and analogous multiresolution stems for volumetric data. The encoder exposes structure across scales and orientations while remaining compatible with the AProcS template introduced earlier.
- *Decimation or pooling* A pooling operator aggregates the feature field $h$ into a fixed-length representation $\mathbf{z} \in \mathbb{R}^{C_z}$ that summarizes the input for classification. Typical choices are global average pooling, joint mean-and-standard-deviation pooling, or simple attention-based pooling across spatial locations or time frames. This step removes the explicit dependence on the grid $\Omega$ and yields a compact descriptor that feeds the probabilistic head.

- *Probabilistic heads.* The final head maps $\mathbf{z}$ to class posteriors $p_\theta(x) \in \Delta_{K-1}$. Two families of heads are used in this work:
  1. standard MLP neural network heads, where $\mathbf{z}$ is passed through one or more affine layers with nonlinearities, followed by a final affine layer and a softmax;
  2. Volterra heads, where the components of $\mathbf{z}$ enter a finite-order Volterra series, followed by a suitable polynomial nonlinearity and a softmax layer. These heads make low-order interactions between components of $\mathbf{z}$ explicit and support the interpretability with economy goals of the thesis.

  Models that combine a multiresolution encoder $\Phi_\theta$ with a Volterra head form the MEV family studied in the experiments.
- *Decisions and costs.* The default decision rule is the MAP choice $f_\theta(x) = \arg\max_k p_\theta(x)_k$. When application-specific costs are available, the same cost-aware rule as in Section 5.1 is used,
$$\hat{y}(x) = \arg\min_{\hat{y} \in \mathcal{Y}} \sum_{y \in \mathcal{Y}} C(y, \hat{y})\, p_\theta(x)_y, \tag{5.9}$$
  so that the classifier's probabilistic outputs can be adapted to different cost structures at deployment time without retraining.

Training classifiers follows the general formulation in Section 5.1. Given labeled examples $\{(x^{(i)}, y^{(i)})\}_{i=1}^n$, a filter bank (fixed or learnable) encoder creates optimal representations of input and dense head with learnable parameters $\theta$ of $\Phi_\theta$ are trained by minimizing a regularized risk with fidelity term like cross-entropy or negative log-likelihood losses. Let's begin with a description of a texture classifier with DWT bank encoder, subband pooling and MLP neural network. Later, MEV classifiers with Volterra kernel heads will follow.

### 5.2.1 Texture classification with DWT bank encoder and MLP heads

Wavelet CNN models with DWT level-4 analysis filter-bank are constructed and evaluated in six different configurations, namely Model-A to Model-F. While Model-A to Model-E use wavelet CNN with traditional pooling/ stride methods, Model-F is our proposed architecture that combines wavelet CNN with wavelet Pooling. The six different wavelet CNN configurations are summarized in Table 5.3. Figure 5.1 shows the architecture of the proposed Model-F. Model-A to Model-C use similar architecture but without any explicit pooling nodes and use convolution with strides instead. All the models use a DWT level-4 analysis of the input 2D image and the number of network output nodes equals the number of classes. Also, Haar and

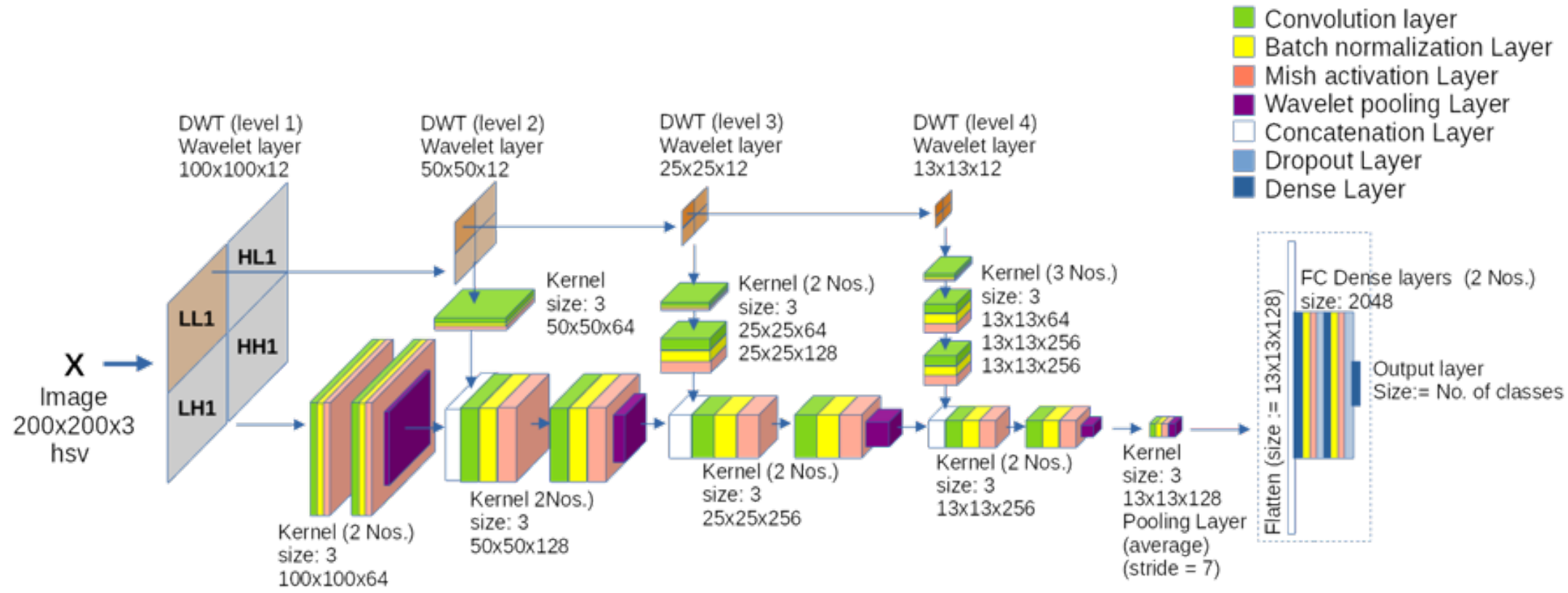

Figure 5.1: A DWT level-4 wavelet CNN model with wavelet pooling (Model-F). Models D, E have equivalent structures with max- and average-pooling respectively. Models A, B and C are similar except they have no pooling nodes.

Db2 wavelets are used to construct two variants of each architecture. After every convolution operation, there is a batch normalization and activation similar to [106, 112, 113]. A common choice of activation function is ReLU [269], a state-of-the-art in many deep learning models. A recent activation function that is getting popular and improves upon ReLU is Mish [78], $f(x) = x \tanh(\text{softplus}(x))$. We use this to improve the performance of our models. In addition to the DWT level-4 filter-bank, the proposed Model-F has a pooling mechanism governed by a DWT level-1 analysis filter-bank. This pooled output is empirically similar to pooling with a support of 2. The same mother wavelet represents both level-4 and level-1 filter-bank in Model-F. The number of trainable parameters in the constructed wavelet CNN models is around 17 M to 21 M, which is lesser than models like VGG16 (∼138M) [90], ResNet-152 (∼60M) [38] and Inception-v3 (∼24M) [270].

Table 5.3: Configurations of ME (DWT) texture classifiers

| # | Model | # of trainable params. | Configuration |
|---|---|---|---|
| 1 | A | 21.6 M | 2D conv. (stride=2) |
| 2 | B | | A (ReLU), B (Mish [78] ) |
| 3 | C | 17.7 M | Sep. 2D conv.; C (sep. 2D conv., |
| 4 | D | | stride=2), D (max-pooling), |
| 5 | E | | E (avg.-pooling), Mish |
| 6 | F | | **Separable 2D conv., DWT-pooling, Mish** |

**Example 5.1.** Comparing performances of the DWT bank classifier having different convolution types, activations and poolings using kth-tips2-b and Describable Textures Dataset (DTD)

Haar and Db2 variants of wavelet CNN in six different major configurations are trained using two datasets (i) kth-tips2-b having eleven classes and (ii) DTD having forty-seven classes. These datasets contain color images of dimensions $200 \times 200 \times 3$ and $300 \times 300 \times 3$ respectively. The number of trainable parameters, training accuracy and training time is recorded to identify an efficient model in terms of memory and processing time for texture classification. The introduction of separate pooling nodes in combination with mish activation function and separable-2D convolution gave positive results. Table 5.4 summarizes the train and test scores of our wavelet CNN models.

Table 5.4: Evaluation of ME texture classifiers

| # | Dataset | Model | Accuracy (%) | | | |
|---|---|---|---|---|---|---|
| | | | Haar-CNN | | Db2-CNN | |
| | | | Train | Test | Train | Test |
| 1.1 | kth-tips2-b | A | 93.5 | 39.2 | 29.2 | 10.9 |
| 1.2 | | B | 92.6 | 40.9 | 13.9 | 10.6 |
| 1.3 | | C | 98.3 | 36.6 | 48.4 | 14.3 |
| 1.4 | | D | 98.0 | 45.9 | 46.1 | 27.8 |
| 1.5 | | E | 98.1 | 43.6 | 35.5 | 20.5 |
| 1.6 | | F | 97.5 | 50.9 | 34.0 | 14.2 |
| 2.1 | DTD+ | A | 93.6 | 33.8 | 22.0 | 08.7 |
| 2.2 | | B | 95.6 | 36.5 | 16.0 | 11.7 |
| 2.3 | | C | 95.4 | 37.7 | 45.9 | 16.2 |
| 2.4 | | D | 97.9 | 47.1 | 66.9 | 11.9 |
| 2.5 | | E | 95.2 | 45.8 | 40.7 | 18.6 |
| 2.6 | | F | 97.7 | 47.3 | 54.6 | 13.7 |

+DTD has images captured in the wild, i.e. uncontrolled conditions, and this makes it a difficult dataset as is evident from the model accuracy results as well.

**Observations** In this analysis, we compared convolution types, activation functions and pooling types in our wavelet convolutional architectures. The models having separable 2D convolution nodes followed by Mish activation and pooling nodes produced better results.

- **Separable 2D convolution:** A separable 2D convolution improves both training and test scores of the Haar and Db2 CNN models as evident in Figure 5.2. It also reduces the number of trainable parameters when compared to 2D convolution. The combination of separable 2D convolution followed by Mish activation produces further improved results.

- **Mish activation**: The Mish activation function in place of ReLU improves both training and test accuracy of Haar and Db2 CNN models. The results are consistent for both the

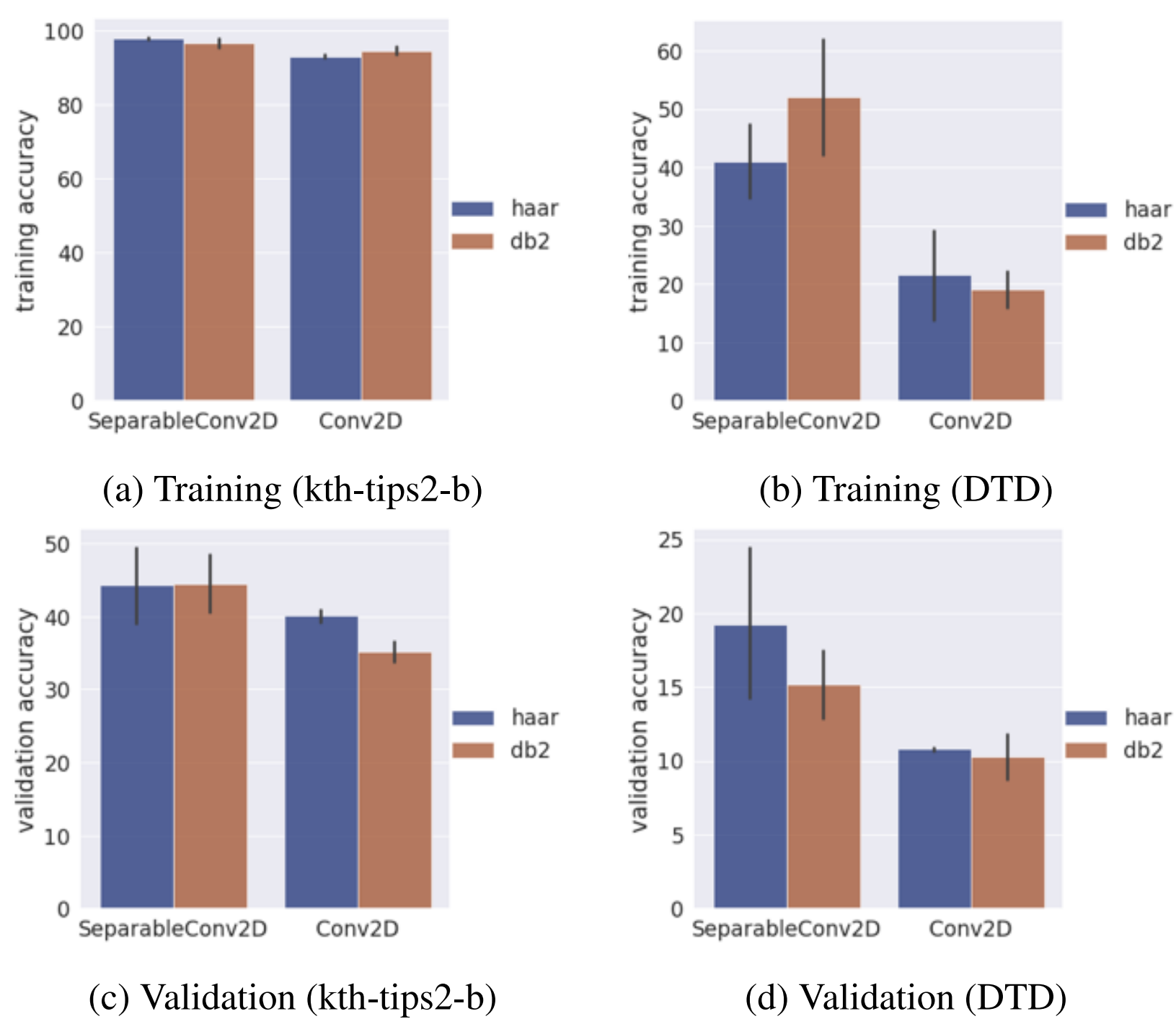

(a) Training (kth-tips2-b)

(b) Training (DTD)

(c) Validation (kth-tips2-b)

(d) Validation (DTD)

Figure 5.2: Training accuracy (a), (b) and validation accuracy (c), (d) with convolution types separable 2D convolution and 2D convolution

texture datasets as can be seen in Figure 5.3. The DTD dataset having a larger number of texture classes shows a marginal improvement in training and test results.

- **DWT pooling:** The models with separate pooling nodes produced better train and test scores when compared to models with no pooling nodes as shown in Figure 5.4. The introduced DWT pooling is effective specifically for the kth-tips2-b dataset. DWT pooling involves a level-1 analysis filter-bank. In Model-F, the same mother wavelet is used for wavelet analysis as well as DWT pooling. The results show that wavelet CNN models having pooling nodes perform better than models with no additional pooling nodes.

All the models were trained using categorical cross-entropy loss parameter and Adam optimizer [37]. For both Haar and Db2 CNN models, we observed that Mish activation, separable 2D convolution and pooling nodes individually contributed to faster convergence. Each of the three reduced loss faster than their replaced counterparts, i.e., ReLU activation, 2D convolution and absence of pooling nodes. Figures 5.2, 5.3, and 5.4 summarize the training and validation trends of the tested variants.

**4-fold training using kth-tips2-b:** The kth-tips-2b dataset has four samples in each class, namely a, b, c and d. The 200 × 200 images are first resized to 224 × 224 using bilinear interpolation. We trained the model four times with separate train, test and validation data each time for 4-fold training. Table 5.5 summarizes the mean and standard deviation of the test accuracies of the tested wavelet MRA-based CNN models. The training accuracies were

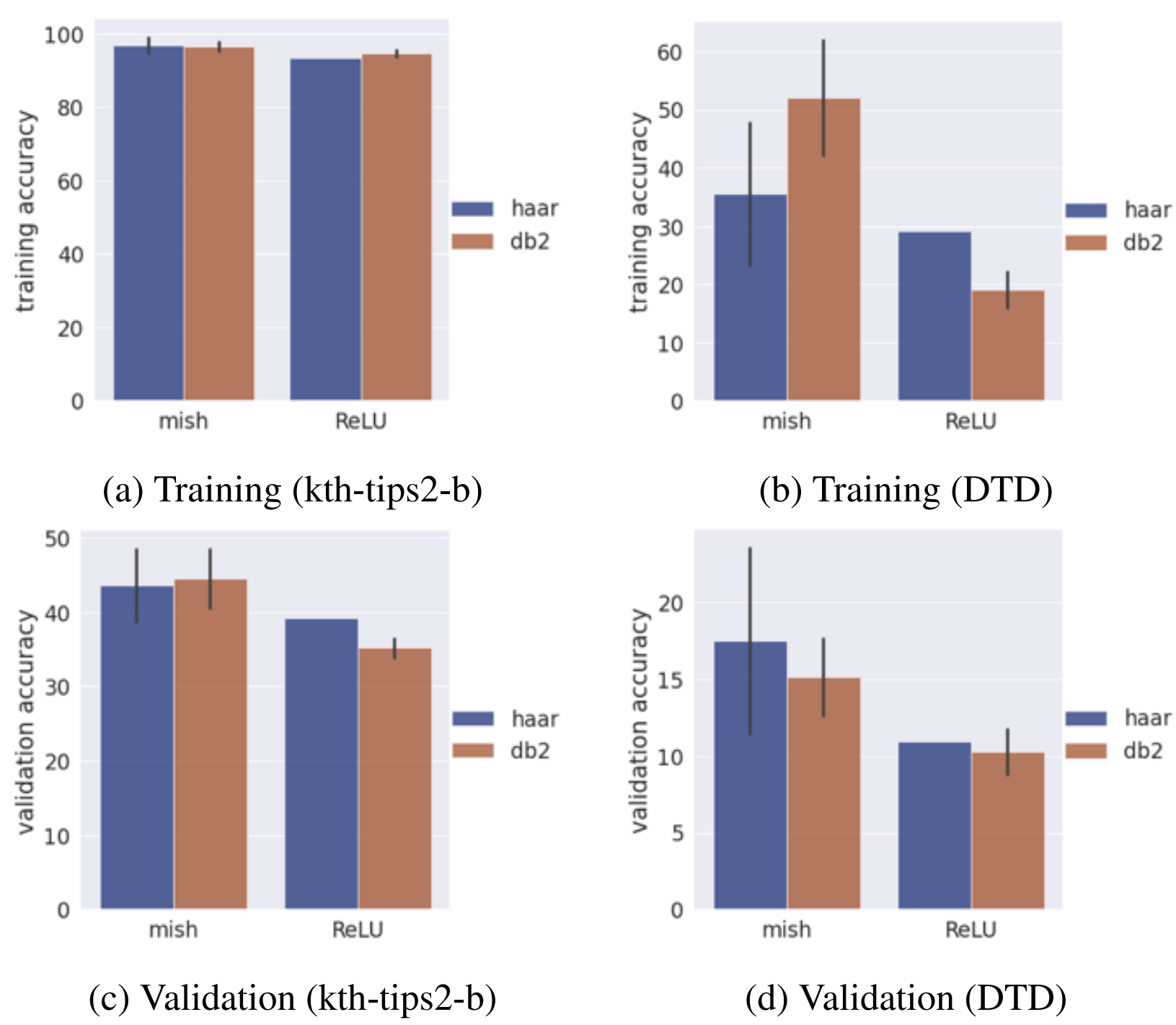

(a) Training (kth-tips2-b) (b) Training (DTD)

(c) Validation (kth-tips2-b) (d) Validation (DTD)

Figure 5.3: Training accuracy (a), (b) and validation accuracy (c), (d) with Mish and ReLU activation

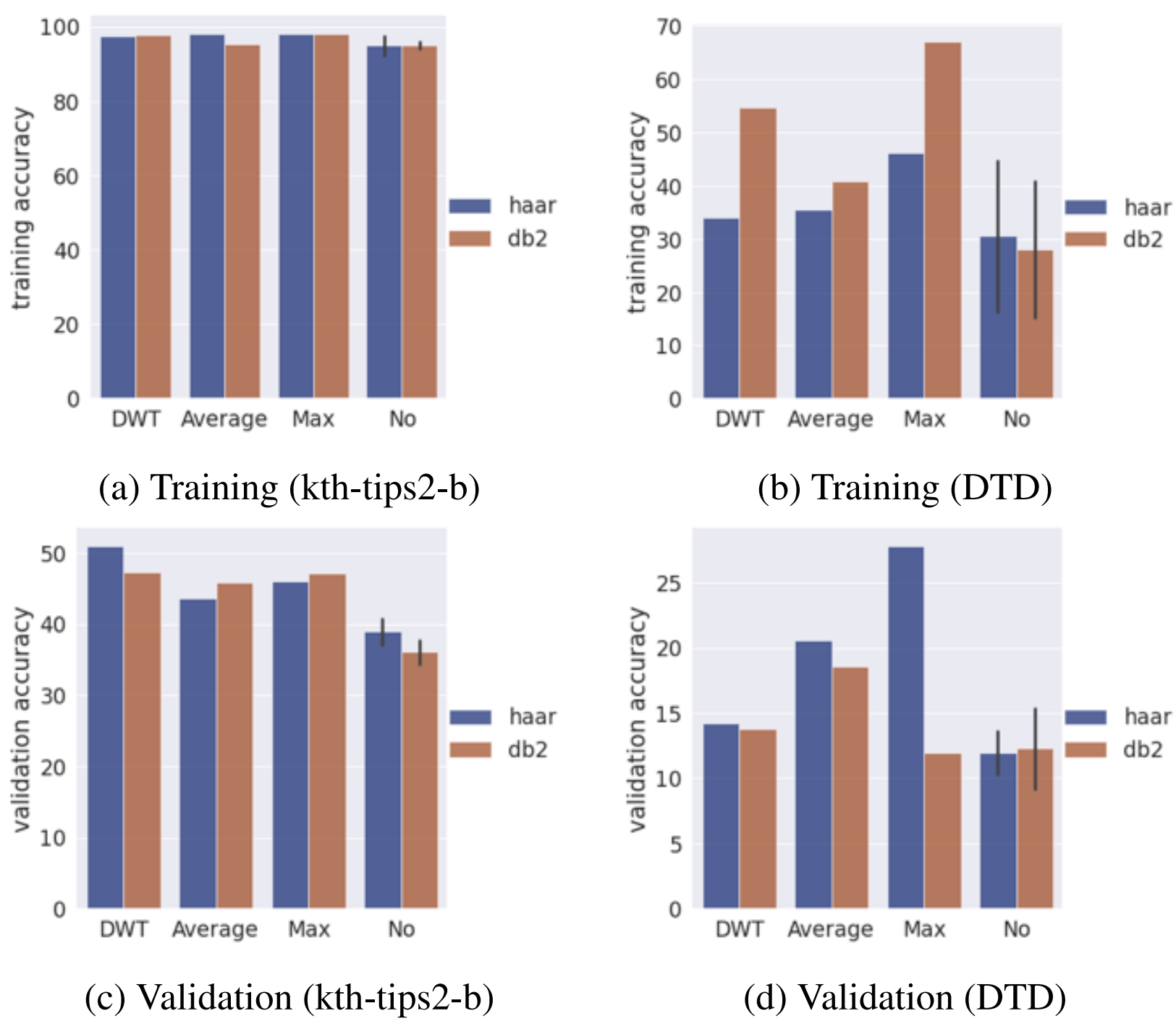

(a) Training (kth-tips2-b) (b) Training (DTD)

(c) Validation (kth-tips2-b) (d) Validation (DTD)

Figure 5.4: Training accuracy (a), (b) and validation accuracy (c), (d) with different pooling mechanisms, DWT-pooling, average-pooling, max-pooling and absence of a separate pooling

observed between 92% and 99%. This set of results with 4-fold training reproduces similar observations as in the results of the previous section. Both Haar and Db2 CNN models with a combination of separable 2D convolution, Mish activation and pooling nodes give superior results, especially for the kth-tips2-b texture dataset. With wavelet pooling, we have MRA at successive stages in the input to output signal propagation path giving the fastest convergence and efficient learning. Apart from the six models (A-F), we evaluated two other models (i) a CNN with architecture similar to Fig. 3 but with only the DWT level-1 filter-bank pooling nodes for MRA and removing the main DWT level-4 filter-bank, and (ii) a CNN with average pooling and Mish activation with no DWT filter-banks. These two models have ˜55.9 M trainable parameters, much higher when compared to models A-F as removing DWT filter-bank while keeping the filter configurations the same increases the number of connections/ parameters, especially in the FC dense layers. The training time compared with Model-F is ˜50% higher for the CNN with average pooling and is ˜200% more for the CNN with DWT pooling. The mother wavelet $\psi$ of DWT level-4 filter-bank defines a compact representation space for the input signal $x$ in models A-F.

Table 5.5: Test accuracies of 4-fold training with kth-tips2-b

| # | Model | Test Acc. (%) | |
|---|---|---|---|
| | | Haar | Db2 |
| 1 | A | 33.3 ± 3.0 | 35.3 ± 2.0 |
| 2 | B | 33.9 ± 8.0 | 30.5 ± 8.0 |
| 3 | C | 43.6 ± 5.0 | 43.2 ± 4.0 |
| 4 | D | 49.8 ± 3.0 | 46.4 ± 3.0 |
| 5 | E | 39.2 ± 3.0 | 44.3 ± 4.0 |
| 6 | F | 44.1 ± 6.0 | 41.4 ± 2.0 |

### 5.2.2 Audio classification with spectrogram encoder and Volterra heads

This subsection describes an audio classifier that instantiates the MEV template for one-dimensional waveforms. The input is a raw audio signal, which is mapped to a log–mel spectrogram and then processed by a small 2D CNN encoder, a global pooling operator, and a Volterra head.

- *Spaces.* A raw waveform input of fixed length $T$ samples is represented as a vector $x \in \mathbb{R}^T$, so that the input space is $\mathcal{X} = \mathbb{R}^T$. A deterministic log–mel front-end converts $x$ into a time–frequency representation $\Psi(x) \in \mathbb{R}^{T' \times M}$, where $T'$ is the number of time frames and $M$ is the number of mel bands. This representation is viewed as a scalar field on the grid

$$\Omega_{\text{spec}} = \{1, \dots, T'\} \times \{1, \dots, M\}, \tag{5.10}$$

with $\Psi(x) : \Omega_{\text{spec}} \to \mathbb{R}$.

- *Frontend* $\Phi_\theta$. The log–mel transform $\Psi$ is fixed and consists of a short-time Fourier transform, power spectral density, mel filtering, logarithm, and per-clip normalization. A shallow 2D CNN encoder then acts on $\Psi(x)$ and produces a feature map

$$h = \Phi_\theta(\Psi(x)) \in \mathbb{R}^{H \times W \times C_h}, \tag{5.11}$$

where $(H, W)$ are the spatial dimensions after convolution and subsampling, and $C_h$ is the number of feature channels.

- *Pooling* A global pooling aggregates the feature map $h$ into a fixed-length descriptor $\mathbf{z} \in \mathbb{R}^{C_z}$. The simplest case is a global average pooling over the $(H, W)$ grid, although mean–variance pooling or lightweight attention pooling can also be used. This step removes the explicit spatial dependence on $\Omega_{\text{spec}}$ and yields a compact representation that is passed to the Volterra head.

- *Volterra head.* The pooled feature vector $\mathbf{z}$ is passed to a Volterra head that maps $\mathbf{z}$ to logits in $\mathbb{R}^K$ and then to class posteriors on the simplex. A finite-order Volterra series in the components of $\mathbf{z}$ is evaluated, followed by a polynomial nonlinearity and a final affine layer with softmax. The resulting output is

$$p_\theta(x) = \big(p_\theta(x)_1, \ldots, p_\theta(x)_K\big) \in \Delta_{K-1}, \tag{5.12}$$

where $p_\theta(x)_k$ is the predicted posterior probability of class $k$ for the input waveform $x$.

- *Training.* Given labeled examples $\{(x^{(i)}, y^{(i)})\}_{i=1}^n$ with $y^{(i)} \in \mathcal{Y}$, the parameters of the encoder $\Phi_\theta$, the global pooling (when learnable), and the Volterra head are learned by minimizing a regularized empirical negative log-likelihood,

$$\hat{\theta} \in \arg\min_\theta \frac{1}{n} \sum_{i=1}^n \big( -\log p_\theta(y^{(i)} \mid x^{(i)})\big) + \lambda \mathcal{R}(\theta), \tag{5.13}$$

in line with the general learning-from-data formulation in Section 5.1.

- *Decision.* At test time the classifier returns the MAP decision

$$f_\theta(x) = \arg\max_{k \in \mathcal{Y}} p_\theta(x)_k, \tag{5.14}$$

or, when a cost matrix $C$ is specified, the cost-aware decision rule

$$\hat{y}(x) = \arg\min_{\hat{y} \in \mathcal{Y}} \sum_{y \in \mathcal{Y}} C(y, \hat{y})\, p_\theta(x)_y, \tag{5.15}$$

using exactly the same decision-theoretic principles as in the generic classification setting.

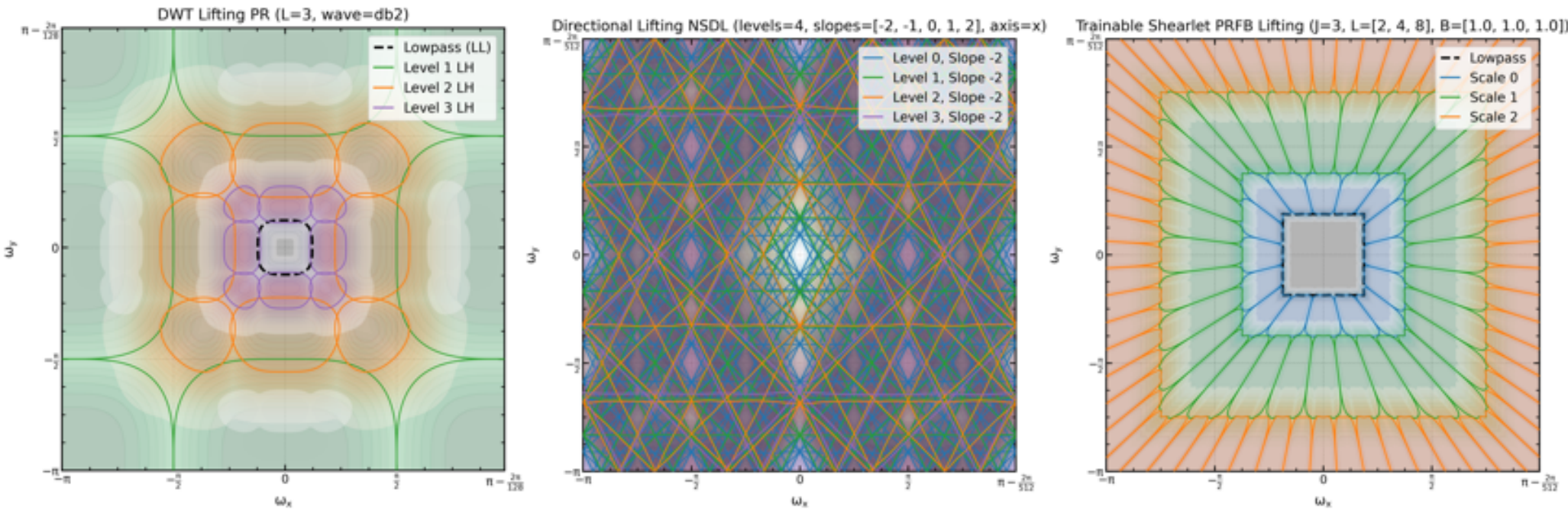

Figure 5.5: Impulse responses of the (left to right) DWT (PR), non-subsampled directional, and shearlet lifting filter-banks

### 5.2.3 Texture classification with shearlet bank encoder and Volterra heads

This subsection describes a texture classifier that instantiates the MEV template using a trainable directional or shearlet-style frequency-domain filter-bank as the front-end. The input image is first decomposed into directional subbands, then processed by a shallow CNN, pooled to a fixed-length descriptor, and finally classified by grouped MLP or grouped Volterra heads.

- *Spaces.* Square texture patches are represented as tensors $x \in \mathbb{R}^{N\times N\times C}$, where $N \times N$ is the spatial resolution (after resizing when needed) and $C$ is the number of input channels. Thus the input space is

$$\mathcal{X} = \mathbb{R}^{N\times N\times C}. \tag{5.16}$$

- *Feature analysis* $\Phi_\theta$. The feature extractor $\Phi_\theta$ begins with a trainable frequency-domain filter-bank. A two-dimensional FFT is applied channelwise to $x$, and a collection of complex-valued filters $\{H_f\}_{f=1}^{F}$ is implemented as pointwise multipliers in the Fourier domain. For each filter index $f$, an inverse FFT then produces a spatial response. The filters are normalized so that, at each frequency, the sum of squared magnitudes across the bank is equal to one up to a global scale factor. This gives a tight-frame-like tiling of the 2D Fourier plane. Stacking the responses across channels and filters yields a feature map

$$h = \Phi_\theta(x) \in \mathbb{R}^{N\times N\times(C\cdot F)}. \tag{5.17}$$

  Optional batch normalization and a learnable $1 \times 1$ convolution can be applied to $h$ to mix and reduce the channel dimension before pooling.

- *Pooling:* A global pooling aggregates the feature map $h$ over the $N \times N$ grid and produces a fixed-length descriptor $\mathbf{z} \in \mathbb{R}^D$. In practice this pooling is instantiated as global average pooling, mean–variance pooling, or a simple attention-based pooling. This step removes spatial dependence and summarizes the texture statistics in a compact feature vector.

- *Grouped heads:* The descriptor $\mathbf{z}$ is passed to a collection of grouped heads. The set of classes $\{1, \ldots, K\}$ is partitioned into disjoint groups $\{\mathcal{G}_g\}_{g=1}^{M}$, where group $g$ is handled by its own head.

1. *Grouped MLP heads.* Each group $g$ has a small MLP that maps $\mathbf{z}$ to logits and a softmax over $\mathcal{G}_g$, producing a vector

$$s^{(g)} \in \Delta_{|\mathcal{G}_g|-1}. \tag{5.18}$$

2. *Grouped Volterra heads.* Alternatively, each group $g$ uses a one-dimensional Volterra map on $\mathbf{z}$ to capture multiplicative interactions, followed by groupwise logits and a softmax on $\mathcal{G}_g$. This yields the same type of output $s^{(g)} \in \Delta_{|\mathcal{G}_g|-1}$ but with an explicit Volterra structure.

- *Unified probability over all classes.* The group-wise posteriors are combined into a full posterior over all $K$ classes. Under a group-agnostic prior, each group is given equal weight, and the posterior for $k \in \mathcal{G}_g$ is

$$p_\theta(x)_k = \frac{1}{M}\, s_k^{(g)}, \qquad k \in \mathcal{G}_g, \tag{5.19}$$

so that

$$\sum_{k=1}^{K} p_\theta(x)_k = 1. \tag{5.20}$$

This construction is equivalent to concatenating the groupwise softmax vectors and dividing by $M$:

$$p_\theta(x) = \frac{1}{M}\operatorname{concat}\big(s^{(1)}, \ldots, s^{(M)}\big). \tag{5.21}$$

If desired, the assumption of equal group weights can be relaxed by introducing learnable group weights $w \in \Delta_{M-1}$ via a small gating head and setting

$$p_\theta(x)_k = w_g\, s_k^{(g)}, \qquad k \in \mathcal{G}_g, \tag{5.22}$$

which still ensures that $p_\theta(x)$ lies in $\Delta_{K-1}$.

- *Training objectives.* Two closely related training strategies are used.

1. *Masked per-group cross-entropy.* Only the head corresponding to the true class group is updated. Given training pairs $\{(x^{(i)}, y^{(i)})\}_{i=1}^n$, the objective is

$$\min_\theta \frac{1}{n}\sum_{i=1}^{n}\sum_{g=1}^{M} \mathbf{1}\{y^{(i)} \in \mathcal{G}_g\}\big(-\log s_{y^{(i)}}^{(g)}\big). \tag{5.23}$$

2. *Single cross-entropy on the merged posterior.* Alternatively, one may work directly with the merged posterior $p_\theta(x)$ and minimize

$$\min_\theta \frac{1}{n}\sum_{i=1}^{n}\big(-\log p_\theta(x^{(i)})_{y^{(i)}}\big), \tag{5.24}$$

where $p_\theta(x)$ is either the equally weighted concatenation or the group-weighted version described above.

Both objectives are compatible with the decision-theoretic formulation in the previous section and differ only in how the group structure is exploited during learning.

- *Decision.* At test time the full posterior $p_\theta(x)$ over all classes is used together with the standard MAP rule

$$f_\theta(x) = \arg\max_{k\in\mathcal{Y}} p_\theta(x)_k, \tag{5.25}$$

or, when a cost matrix $C$ is available, with the cost-aware decision rule

$$\hat{y}(x) = \arg\min_{\hat{y}\in\mathcal{Y}} \sum_{y\in\mathcal{Y}} C(y,\hat{y})\, p_\theta(x)_y. \tag{5.26}$$

Table 5.6: Number of parameters in different sound and image classifiers and their input sizes

| # | model name | components | # trainable params | # fixed params | input size |
|---|---|---|---|---|---|
| **Sound classifier (1D inputs)** | | | | | |
| 1a | `conv_mlp_1d` | convolutions, MLP heads | 1,284,667 | 896 | (220500,1) |
| 1b | `spect_mlp_1d` | Spectrogram, convolution, MLP heads | 1,433,147 | 66,560 | (220500,1) |
| 1c | `spect_volterra_1d` | Spectrogram, convolutions, Volterra heads | **386,825** | 66,560 | (220500,1) |
| **Image classifier (2D inputs)** | | | | | |
| 2a | `conv_mlp_2d` | convolutions, MLP heads | 1,370,168 | 896 | (224,224,3) |
| 2b | `conv_volterra_2d` | convolutions, Volterra heads | **389,257** | 896 | (224,224,3) |
| 2c | `shearlet_mlp_2d` | Shearlet bank, convolutions, MLP heads | 2,905,094 | 342 | (128,128,3) *resized* |
| 2d | `shearlet_volterra_2d` | Shearlet bank, convolutions, Volterra heads | **1,924,183** | 342 | (128,128,3) *resized* |

Figure 5.5 illustrates the learned filter-banks by showing the Fourier-domain impulse responses and the resulting tilings of the two-dimensional Fourier plane for several trained directional banks. Table 5.6 shows the number of trainable and fixed parameters in different sound and image classifiers and their input sizes. The classifiers with Volterra heads are frugal with much less number of parameters.

**Example 5.2.** Classification of two challenging open-source datasets, namely ESC-50 (audio) and DTD (images) datasets that have many classes using the realized classifiers with and without inbuilt multiresolution analysis and grouped Volterra heads.

The test results of the **Spectrogram CNN audio classifier** in 5.2.2 and **Shearlet / Directional-bank texture classifier** in 5.2.3 with **MLP and Volterra heads** are shown in Figure 5.6. In the audioclassifier model, it can be observed that the performance without the spectrogram head is very poor. The spectrogram bank after the audio input in the model enhances the mean classification accuracy from 22% to 82% with significant improvements in per-class accuracy (this is our baseline). Similarly, in 2D texture classification models, the shearlet bank encoder classifier for both MLP and Volterra heads is slightly better than the one with a simple fully convolutional encoder. In both audio and image classifiers, the Volterra head variant with filter bank encoder show equivalent performance but with very few free parameters.

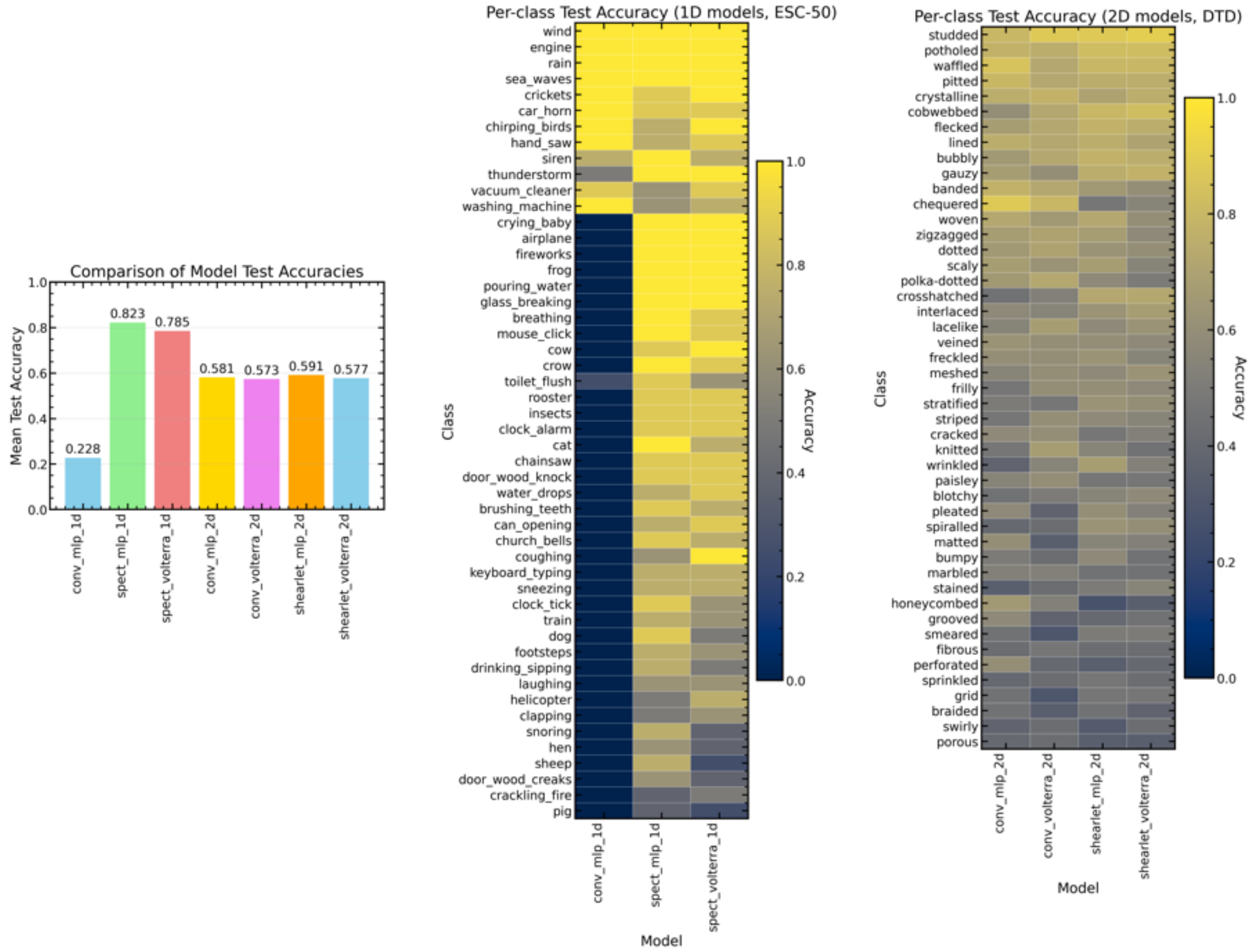

Figure 5.6: Classification scores of two challenging 1D (ESC-20) and 2D (DTD) datasets using various classifiers with and without multiresolution and Volterra heads

Observations further show that classifiers with grouped heads have significant improvement in accuracy compared to the previous DWT bank CNN texture classifier with a single MLP head in example 5.1.

## 5.3 Multiresolution encoder–decoder segmentation models

Segmentation is the task of assigning a class label to every spatial index (pixel or voxel) of a discrete $D$-dimensional signal. A 2D image with $D = 2$ or a volumetric 3D image with $D = 3$ is modeled as a function $x : \Omega \to \mathbb{R}^{C_{\text{in}}}$, where $\Omega \subset \mathbb{Z}^D$ is a finite grid of integer lattice points that represent pixel or voxel locations. Thus $x$ is a vector-valued field on $\Omega$. For each spatial index $\boldsymbol{n} = (n_1, \ldots, n_D) \in \Omega$, the vector $x(\boldsymbol{n}) \in \mathbb{R}^{C_{\text{in}}}$ collects the $C_{\text{in}}$ channel intensities at that location, for example RGB color channels, multimodal MRI, or intermediate feature maps in a deep network. A segmentation is represented as a label field $y : \Omega \to \{1, \ldots, K\}$ that assigns to each $\boldsymbol{n} \in \Omega$ a class label $y(\boldsymbol{n})$ from a finite set of $K$ semantic classes. Binary segmentation arises as the special case $K = 2$, where one label denotes foreground and the other label denotes background. Given an input field $x : \Omega \to \mathbb{R}^{C_{\text{in}}}$, a segmentation model with learnable parameters $\theta$, such as a UNet, produces for every spatial index $\boldsymbol{n} \in \Omega$ a vector of class posteriors

$$p_\theta(x)_{\boldsymbol{n}\cdot} = \big(p_\theta(x)_{\boldsymbol{n}1}, \ldots, p_\theta(x)_{\boldsymbol{n}K}\big) \in \Delta_{K-1}, \tag{5.27}$$

where $p_\theta(x)_{\boldsymbol{n}k}$ is the estimated probability that the class at spatial location $\boldsymbol{n}$ is $k$ given the observation $x$, and $\Delta_{K-1}$ denotes the probability simplex in $\mathbb{R}^K$. The associated hard segmentation, or decision rule, selects at each spatial index the most probable class,

$$f_\theta(x)_{\boldsymbol{n}} = \arg\max_{1 \le k \le K} p_\theta(x)_{\boldsymbol{n}k}, \qquad \boldsymbol{n} \in \Omega, \tag{5.28}$$

and the resulting label field can be written as $y(\boldsymbol{n}) = f_\theta(x)_{\boldsymbol{n}}$.

**Supervised deep learning view** In modern deep segmentation the conditional distribution

$$p_\theta : \mathbb{R}^{|\Omega| \times C_{\text{in}}} \to [0,1]^{|\Omega| \times K}, \qquad p_\theta(x)_{\boldsymbol{n}\cdot} \in \Delta_{K-1} \text{ for all } \boldsymbol{n} \in \Omega, \tag{5.29}$$

is parameterized by a neural network that maps an input field $x$ to class posteriors at each spatial location. In this thesis the architectures include UNets, convolutional variational UNets, and other MED segmentation models with attention bottlenecks, applied to both binary and multiclass segmentation in two and three dimensions. Given training examples $\{(x^{(i)}, y^{(i)})\}_{i=1}^n$, the parameter vector $\theta$ is obtained by minimizing an objective that combines per-location cross-entropy, a region-overlap term, and a regularizer:

$$\min_\theta \frac{1}{n} \sum_{i=1}^{n} \left( \sum_{\boldsymbol{p} \in \Omega} -\log p_\theta(x^{(i)})_{\boldsymbol{p}, y^{(i)}_{\boldsymbol{p}}} + \alpha \left(1 - \mathrm{Dice}\big(y^{(i)}, \, p_\theta(x^{(i)})\big)\right) \right) + \mathcal{R}(\theta). \tag{5.30}$$

For class $k$, the soft Dice coefficient is defined as

$$\mathrm{Dice}_k = \frac{2 \sum_{\boldsymbol{p} \in \Omega} \mathbf{1}\{y_{\boldsymbol{p}} = k\} \, p_\theta(x)_{\boldsymbol{p},k}}{\sum_{\boldsymbol{p} \in \Omega} \mathbf{1}\{y_{\boldsymbol{p}} = k\} + \sum_{\boldsymbol{p} \in \Omega} p_\theta(x)_{\boldsymbol{p},k} + \varepsilon}, \tag{5.31}$$

and the mean Dice score is

$$\mathrm{Dice} = \frac{1}{K} \sum_{k=1}^{K} \mathrm{Dice}_k. \tag{5.32}$$

Here $y_{\boldsymbol{p}}$ is the true label at spatial location $\boldsymbol{p}$, $p_\theta(x)_{\boldsymbol{p},k}$ is the predicted posterior for class $k$ at that location, $\mathbf{1}\{\cdot\}$ denotes the indicator function, and $\varepsilon > 0$ is a small constant that prevents division by zero. The regularizer $\mathcal{R}(\theta)$ controls the complexity of the segmentation network and plays the same role as in the classification setting.

**AProcS view of segmentation models** Segmentation networks can be viewed as

$$\boxed{x \xrightarrow{\mathcal{A}} r \xrightarrow{\Phi_\theta} \tilde{r} \xrightarrow{\mathcal{S}} u \xrightarrow{\text{head}} p_\theta, \qquad f_\theta = \arg\max p_\theta} \qquad \text{(AProcS segmenter template)}$$

where the key components of an image or volumetric segmentation model with multiresolution analysis and synthesis banks are:

- $\mathcal{A}$: multiresolution pyramids such as 3D **DWT** or **FDST** (near-tight, orientation-aware) that map $x$ to bandpass representations.
- $\Phi_\theta$: band-aware convolutional blocks, DWT-based self-attention, WaveletViT, or similar convolutional variational bottlenecks with latent variable $\mathbf{z}$.
- $\mathcal{S}$: **3D IDWT** or inverse windowing plus inverse Fourier transforms that map back to the native grid, yielding a feature field $u : \Omega \to \mathbb{R}^C$.
- **Head**: $1 \times 1$ (and $1 \times 1 \times 1$ in 3D) convolutions plus softmax, producing $p_\theta(x)_{\boldsymbol{n}\cdot} \in \Delta_{K-1}$ and decisions $f_\theta(x)_{\boldsymbol{n}} = \arg\max_k p_\theta(x)_{\boldsymbol{n}k}$.
- **Losses**: weighted cross-entropy, Dice or focal losses as in the objective above; optional CRF or total variation regularizers can be attached as additional energy terms on $-\log p_\theta(x)$ without changing the AProcS structure.

**Cost-aware decisions for segmentation** The cost-aware decision principle extends to per-location segmentation. Given a cost matrix $C$, a cost-sensitive decision rule at spatial location $\boldsymbol{n}$ is

$$\hat{y}(\boldsymbol{n}) = \arg\min_{\hat{y}\in\mathcal{Y}} \sum_{y\in\mathcal{Y}} C(y,\hat{y})\, p_\theta(x)_{\boldsymbol{n},y}, \qquad \boldsymbol{n} \in \Omega, \tag{5.33}$$

which reduces to $\hat{y}(\boldsymbol{n}) = \arg\max_{k\in\mathcal{Y}} p_\theta(x)_{\boldsymbol{n}k}$ when all misclassification costs are equal. Thus segmentation follows the same Bayes decision rule as classification, but applied at a finer spatial granularity.

The spatial index set is a finite grid $\Omega \subset \mathbb{Z}^D$, with $D = 2$ for images and $D = 3$ for volumetric data. An observation is a multi-channel field $x : \Omega \to \mathbb{R}^{C_{\text{in}}}$, and a segmentation is a label field $y : \Omega \to \mathcal{Y} = \{1, \dots, K\}$. A segmentation model with parameters $\theta$ produces, for each spatial index $\boldsymbol{n} \in \Omega$, a vector of class posteriors

$$p_\theta(x)_{\boldsymbol{n}\cdot} \in \Delta_{K-1}, \qquad \boldsymbol{n} \in \Omega, \tag{5.34}$$

and the associated hard segmentation is obtained by the pointwise decision rule

$$f_\theta(x)_{\boldsymbol{n}} = \arg\max_{k\in\mathcal{Y}} p_\theta(x)_{\boldsymbol{n}k}, \qquad \boldsymbol{n} \in \Omega. \tag{5.35}$$

This mirrors the classification decision rule, now applied at the level of spatial locations instead of whole inputs. Training objectives follow the formulations in Section 5.1, combining per-location cross-entropy with Dice-type region-overlap terms and standard regularizers on $\theta$.

### 5.3.1 Multiresolution backpropagation pyramids in segmenters

Modern encoder–decoder segmenters benefit from multiresolution front-ends that expose scale and orientation structure while allowing faithful reconstruction on the native grid. The segmentation models in this thesis instantiate the (AProcS segmenter template) with explicit analysis and synthesis operators where $p_\theta(x) = (p_\theta(x)_{\boldsymbol{n}\cdot})_{\boldsymbol{n}\in\Omega}$ and each $p_\theta(x)_{\boldsymbol{n}\cdot} \in \Delta_{K-1}$. The components are as follows.

- *Analysis operator* $\mathcal{A}$. The analysis stage maps the input field $x$ to a multiresolution, banded representation $r$. In this work $\mathcal{A}$ is realized by DWTs (for example, 2D or 3D DWT with perfect reconstruction) and by Fourier-domain directional transforms with shearlet filterbank (2D or 3D FDST) that produce subbands indexed by scale and orientation. The analysis operator is chosen to be invertible or nearly tight so that the essential information in $x$ is preserved in $r$.
- *Learnable processing* $\Phi_\theta$. The core processing block $\Phi_\theta$ consists of convolutional, attention, and mixing layers that act within and across the bands of $r$. This yields a processed multiresolution representation

$$\tilde{r} = \Phi_\theta(r), \tag{5.36}$$

  in the same banded domain. Depending on the specific architecture, $\Phi_\theta$ may include band-aware convolutions, multiresolution self-attention modules, or transformer-style processing on tokens derived from the subbands.
- *Synthesis operator* $\mathcal{S}$. The synthesis stage brings the processed representation $\tilde{r}$ back to the native grid. For wavelet-based models, $\mathcal{S}$ is an inverse wavelet transform (such as 3D IDWT). For FDST-based models, $\mathcal{S}$ consists of inverse windowing followed by inverse Fourier transforms. The output of the synthesis stage is a feature field

$$u : \Omega \to \mathbb{R}^C, \tag{5.37}$$

  defined on the same spatial grid as the input, with $C$ denoting the number of decoder feature channels.
- *Segmentation head.* A shallow head maps $u$ to per-location class posteriors. In the basic configuration this head is a $1 \times 1$ convolution in 2D (or $1 \times 1 \times 1$ in 3D) that produces $K$ feature channels at each spatial index, followed by a softmax across the class axis. This yields

$$p_\theta(x)_{\boldsymbol{n}\cdot} \in \Delta_{K-1}, \qquad \boldsymbol{n} \in \Omega, \tag{5.38}$$

and the induced prediction

$$f_\theta(x)_{\boldsymbol{n}} = \arg\max_k p_\theta(x)_{\boldsymbol{n}k}. \tag{5.39}$$

All MED segmentation models in this chapter fit this template. They differ only in the choice of analysis and synthesis operators (DWT versus FDST, two-dimensional versus three-dimensional realizations) and in the internal structure of the processing block $\Phi_\theta$. The external interface, a field of posteriors $p_\theta(x)_{\boldsymbol{n}\cdot} \in \Delta_{K-1}$ and a pointwise decision rule $f_\theta(x)_{\boldsymbol{n}}$, remains consistent with the segmentation formulation introduced earlier. All models are trained with the cross-entropy and Dice-based objectives defined in Section 5.1, combined with regularizers such as weight decay and, when used, CRF or total-variation terms acting on $-\log p_\theta(x)$.

**Practical considerations for multiresolution segmenters**

- *Boundary handling.* DWT lifting layers support several boundary modes, such as reflect, symmetric, or circular extension. A single boundary convention should be fixed and used consistently in both the analysis operator $\mathcal{A}$ and the synthesis operator $\mathcal{S}$.
- *Perfect reconstruction and numerical stability.* It is useful to monitor the relative reconstruction error

$$\frac{\|x - \mathcal{S}(\mathcal{A}(x))\|}{\|x\|} \tag{5.40}$$

  on validation data and keep this quantity much smaller than one. For FDST-based models, window functions are renormalized so that the sum of squared window magnitudes across bands remains close to one at each frequency, which maintains the near-tight-frame property of the transform.
- *Band isolation.* When fusing bands in the decoder, band-isolated fusion, where each subband is processed separately and only later combined by explicit mixing layers, is preferred unless a learnable mixer is introduced by design. This preserves an interpretable correspondence between individual bands and scale–orientation channels.
- *Calibration across bands or class groups.* When groups of bands or classes are imbalanced, a small gating head that learns group weights is preferable to fixed concatenation. Such a gating mechanism can improve the calibration of the final posteriors $p_\theta(x)$ while leaving the overall encoder–decoder structure unchanged.
- *Cost-aware deployment.* At deployment time, cost-sensitive decisions follow the same per-location Bayes rule with the cost matrix $C$ applied to the per-location posteriors $p_\theta(x)_{\boldsymbol{n}\cdot}$. This allows a single trained model to be adapted to different application-specific cost structures.

In the remainder of this section, concrete architectures are described in detail. All of them preserve the same input–output contract

$$x \mapsto p_\theta(x) = \big(p_\theta(x)_{\boldsymbol{n}\cdot}\big)_{\boldsymbol{n}\in\Omega} \in \Delta_{K-1}^{\Omega}, \tag{5.41}$$

and differ only in their internal multiresolution analysis, processing, and synthesis components.

### 5.3.2 MEDNet employing DWT pyramid and T1 MRI segmentation

A **M**ultiresolution **E**ncoder-**D**ecoder CN**N** (`MEDNet`) is a concept AProcS segmenter with multiresolution pyramids built seamlessly in the encoder and decoder paths. The forward path in `MEDNet` starts by projecting the input image or volume on an orthogonal or biorthogonal basis, and then the sparse natural-frequency domain features further propagate through the encoder and decoder. The trainable filters learn from these orthogonal representations using gradient descent backpropagation to segment the ROI. Unlike the skip connections of UNet, the `MEDNet` has separate low and high-frequency skip connection paths. The multiresolution pyramid in the encoder and the high-frequency skip connections makes the overall network architecture is lossless despite a lossy decimation path. Figure 5.7 shows the concept of a `MEDNet` with separate low, high, and mixed-frequency processing paths. The idea is to have a segmenter network with a low memory footprint and that is stable to minor perturbations in the input data.

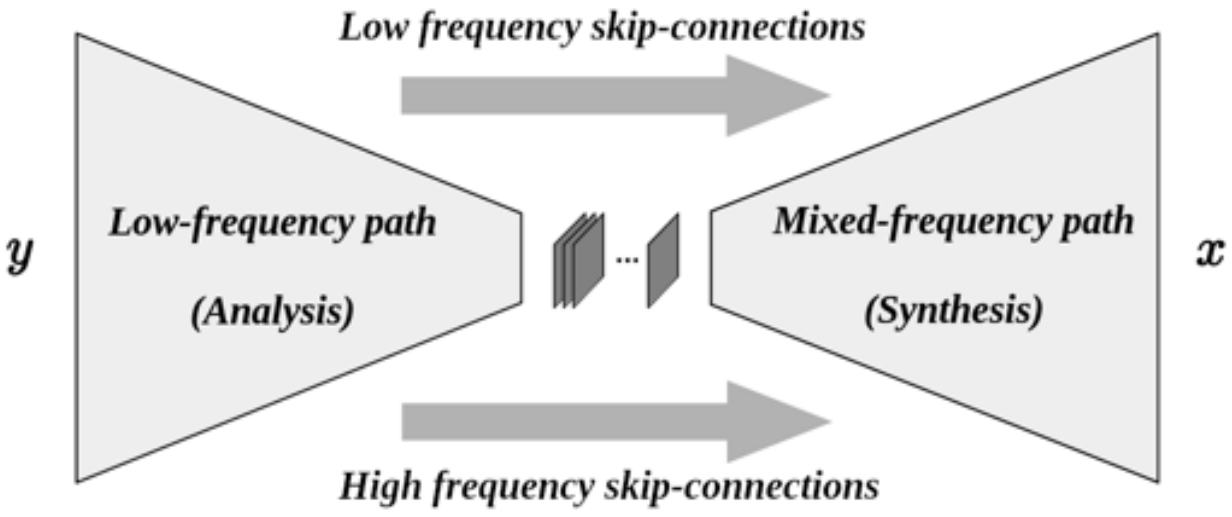

Figure 5.7: Multiresolution Encoder-Decoder (MED) CNN architecture

A realized 2D `MEDNet` with DWT pyramid, wavelet pooling, spatial attention, and wavelet-based self-attention is shown in Figure 5.8. In this model, there are four distinct parts from the input to the output, namely, a DWT analysis bank, a set of learnable filters in the decimation path, a symmetric set of filters in the upsampling path, and an IDWT synthesis bank with attentions. Table 5.7 lists the number of trainable parameters (i.e., interconnections and biases) and computational complexity of `MEDNet` and a baseline UNet for reference. The analysis and synthesis parts in the `MEDNet` with DWT pyramid are:

1. *Encoder with DWT pyramid:* The encoder starts with a DWT $\mathcal{W}_{\phi,\psi}$ level-4 decomposition of input image $y$, and encodes its subbands to $\{\mathbf{Q}_i\}_{i=1}^4 \cup \mathbf{Q}_5$ using trainable FIR filter layers with activations $\mathcal{K}^+$, where $\mathbf{Q}_5$ is in the last level. The ordered set $\{\mathbf{Q}_i\}_{i=1}^4$ is passed to the decoder as skip (lowpass) connections to the four corresponding decoder depths. The encoder processes the low frequencies to locate a coarse ROI and the decoder processes both low and high frequencies to refine the ROI. In our decimated encoder, we have used DWT lowpass pooling like [264], however, the following architecture is compatible to any other lossy pooling as well, for example, $L^1$ (average) pooling or $L^\infty$ (max) pooling. The high-pass branches are kept **band-isolated** with controlled mixing.

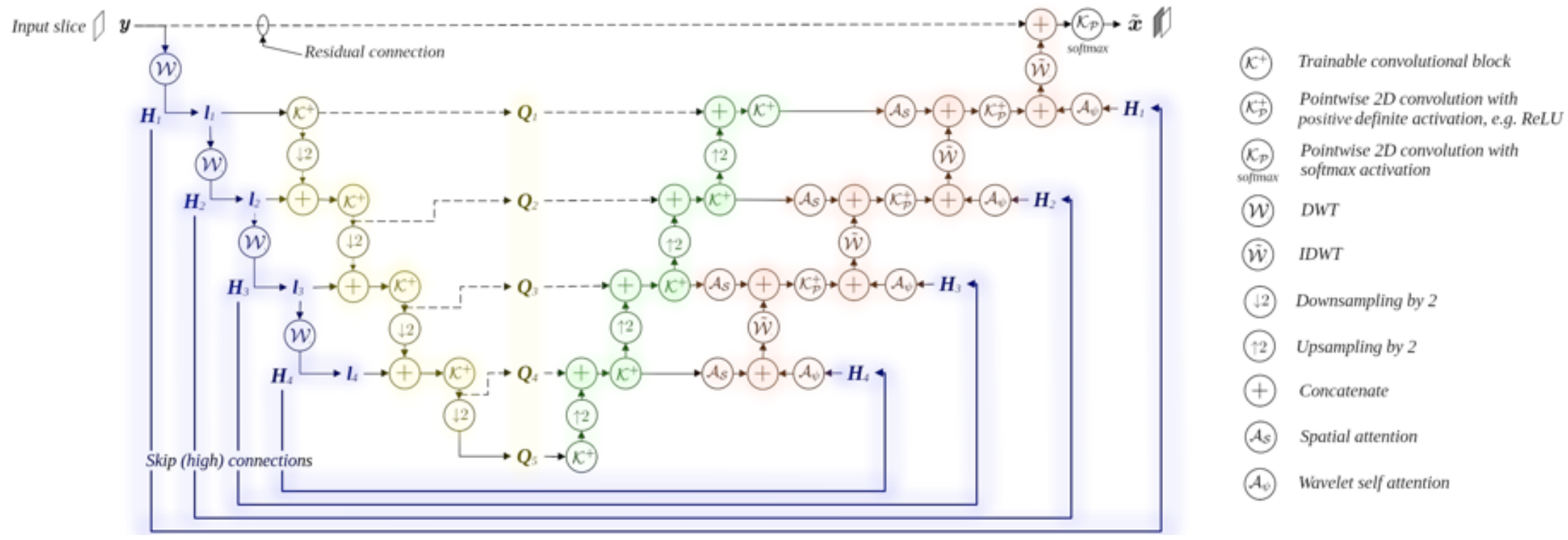

Figure 5.8: A DWT level-4 decomposition 2D `MEDNet` with wavelet pooling as well as spatial and wavelet-based self-attention. The four distinct network parts are wavelet decomposition (blue), trainable encoder part (yellow), symmetric trainable decoder (green) and attention+inverse wavelet transform (orange). The notations in the figure is as in the paper published in ICASSP 2025 with DOI: `10.1109/ICASSP49660.2025.10890832`

2. *Decoder with IDWT and attentions:* decoder uses transposed convolution layers $\tilde{\mathcal{K}}_i^{T+}$ for upsampling with ReLU activation, attentions and IDWT $\tilde{\mathcal{W}}_{\phi,\psi}$ layers. Specifically, spatial attention for low-frequency channels and self-attentions for incoming highpass skip connections $\{\mathbf{H}_k\}_{k=1}^{4}$ . At each hop, fuse selected channels from the bottleneck and skips with the corresponding high-pass groups with optional **SE band gates**, then apply **IDWT** to upsample and re-center spatial detail.

Table 5.7: Number of trainable parameters and complexity of 2D `MEDNet` and UNet

| # | Model | # params | FLOPS |
|---|---|---|---|
| 1 | `MEDNet` | 49K | 229M |
| 2 | UNet | 1.4M | 192M |

We can have variants of this `MEDNet` by altering the learnable filter types, pooling, and unpooling operations. The number of trainable filters in the convolutional layers and the choice of the wavelet in the DWT filter bank are experimental endeavors. Further, this architecture can be extended to a 3D segmenter with volumetric inputs and outputs. The following examples show some experimental results of this segmenter.

**Example 5.3.** Binary segmentation of T1 weighted brain MRI using 2D `MEDNet`

The 2D `MEDNet` is tested with two MRI segmentation datasets: (1) Internet Brain Segmentation Repository (IBSR) V2.0 [271] skull-stripping dataset having large segments and (2) Anatomical Tracing of Lesions After Stroke (ATLAS) R2.0 [272] dataset with small lesion segments. For reference, we have also trained a vanilla UNet with the same data. The decimation network depth is five levels with DWT level-4 and a latent layer, and number of learnable FIR filters that gives best results for `MEDNet` and UNet with IBSR V2.0 and ATLAS R2.0 datasets are (16,32,64,128) and (32,64,128,256) respectively. Although the DWT and IDWT layers increase model complexity, `MEDNet` converges faster than UNet to achieve a similar IoU score due to having a lot fewer trainable parameters.

*Preprocessing:* The pixel values of the MRI volumes were normalized using z-score normalization and rescaled between [0,1]. Then, a histogram equalization of the T1 weighted brain MRI volumes using Contrast-Limited Adaptive Histogram Equalization (CLAHE) ensures that the slices have improved local contrast while limiting over-amplification of noise in homogeneous regions. This ensures the generalization of the trained models for a test set from a different sample space, i.e., MRI volumes acquired with a different scanner, for example, our own T1 weighted brain MRI dataset (AIIMS-BrainMRI) acquired by the Department of NMR at All India Institute of Medical Sciences (AIIMS), New Delhi. Next, the empty MRI slices of the IBSR V2.0 data, without any brain structures, were removed from the training set. Similarly, using an $L^2$ norm threshold, tiny lesions of the ATLAS R2.0 dataset were discarded. *

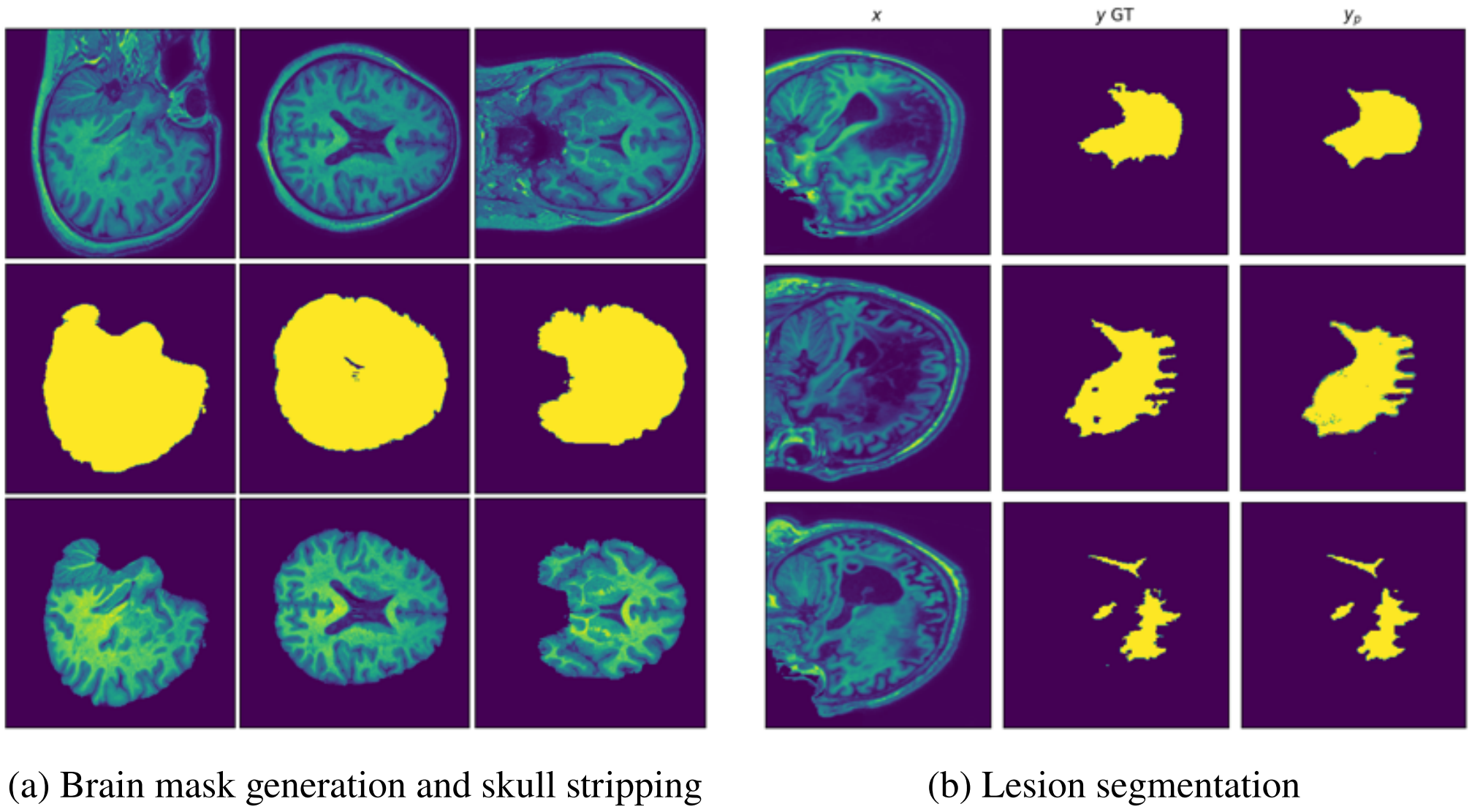

(a) Brain mask generation and skull stripping (b) Lesion segmentation

Figure 5.9: Large and small ROI segmentation by 2D `MEDNet`

*Training and evaluation:* The train, test and validation sets are in the ratio 6:2:2. The models are trained with binary cross-entropy loss. With a batch size of 20 slices `MEDNet` converges to give accurate segmentation at 40 iterations. The primary evaluation metrics are the Dice coefficient and Intersection-over-Union (IoU), which measure the extent of overlap between two images.

The metrics are computed as

$$\text{Dice} = 2\frac{|\mathbf{x} \cap \tilde{\mathbf{x}}|}{|\mathbf{x}| + |\tilde{\mathbf{x}}|}, \qquad \text{IoU} = \frac{|\mathbf{x} \cap \tilde{\mathbf{x}}|}{|\mathbf{x}| + |\tilde{\mathbf{x}}| - |\mathbf{x} \cap \tilde{\mathbf{x}}|}. \tag{5.42}$$

* **Acknowledgement** The MRI studies were supported by Science and Engineering Research Board (SERB), Government of India (SB/S5/AB03/2015). The authors thank Koita Centre for Digital Health (KCDH), Indian Institute of Technology Bombay, India for funding.

Table 5.8: Brain MRI segmentation models in existing literature and test results of 2D `MEDNet`

| Model | Year | Contribution | Dataset | Dice | IoU |
|---|---|---|---|---|---|
| BET [273] | 2011 | Deformable model that evolves to fit the brain surface using a mesh | IBSR 1 | 84.3 | – |
| ROBEX [274] | 2011 | A discriminative and a generative model | IBSR 1 | 95.6 | – |
| 3D CNN [275] | 2016 | Single encoder and multiple decoder 3D CNN | IBSR V2.0, others | 95.8 | – |
| Auto-Net [276] | 2017 | Three parallel 2D pathways and an FCN-style UNet | OASIS | 97.6 | – |
| CompNet [277] | 2018 | Dual-pathway encoder–decoder | OASIS | 98.3 | – |
| 3D Res UNet [278] | 2020 | Residual learning with zoom-in/out strategy | ATLAS R1.1 | 80.0 | – |
| MVUNet [279] | 2021 | Multi-view ensemble; trains on all three body planes simultaneously | IBSR V2.0 | 95.9 | – |
| GUBS [280] | 2022 | Graph-based minimum spanning tree approach for 3D brain segmentation | IBSR V2.0 | 92.4 | – |
| 3D UNet [281] | 2022 | 3D UNet with regularization | ATLAS R2.0 | 65.0 | – |
| MAPPING [282] | 2022 | UNet with four ensemble models and multifold training | ATLAS R2.0 | 66.7 | – |
| ASMCNN [283] | 2022 | Combines Active Shape Model (ASM) and CNN to obtain brain maps | IBSR 1 | 94.9 | – |
| TSRL-Net [284] | 2023 | Coarse-grained residual learning with target-aware loss | ATLAS R1.1 | 76.0 | – |
| W-Net [285] | 2023 | Two-stage network with two boundary-enhancing modules | ATLAS R1.1 | 61.8 | – |
| | | **`MEDNet` with level-4 DWT using sym4 wavelet** | | | |
| **2D `MEDNet` (sym4; ours)** | 2024 | Multiresolution encoder–decoder with separate low, high & mixed-frequency paths; wavelet pooling; wavelet-based self-attention; SHAP explanations | IBSR V2.0<br>ATLAS R2.0# | **98.1**<br>**99.1** | **96.3**<br>**98.7** |
| **2D UNet (ours)** | 2024 | Baseline architecture used to benchmark `MEDNet` | IBSR V2.0<br>ATLAS R2.0# | 98.0<br>99.1 | 96.0<br>98.6 |

# An $L^2$-norm threshold was applied to discard very small lesions. The table provides an overview rather than a strict head-to-head comparison across studies.

Figure 5.9 illustrates the masks generated by the `MEDNet` for segmenting large and small Region-of-Interest (ROI) in both the datasets. Table 5.8 lists the mean test Dice and IoU scores of `MEDNets` and UNets trained using IBSR V2.0 and ATLAS R2.0 datasets and compares with some popular methods in the literature.

**Example 5.4.** Study of 2D `MEDNet` and UNet with different number of convolutional filters and DWT Level-4 decomposition and Haar wavelet using IBSR V2.0 (skull-stripping) dataset.

This is an experiment where both `MEDNet` (Haar) and UNet are tested with IBSR V2.0 and ATLAS R2.0 datasets by varying the number of trainable filters in convolutional blocks to find the best performing configuration in each case. Both the models are of depth four, and the tested filter configurations are (2, 4, 8, 16),(4, 8, 16, 32), (8, 16, 32, 64), (16, 32, 64, 128) and (32, 64, 128,256) filters. Figure 5.10 shows the validation IoU with all these configurations. With the current datasets, a UNet with (32, 64, 128, 256) filters causes instability in training. Overall, the best configurations are (16, 32, 64, 128) and (32, 64, 128, 256), where the former is a good choice for a bit economical and stable model with almost negligible performance degradation.

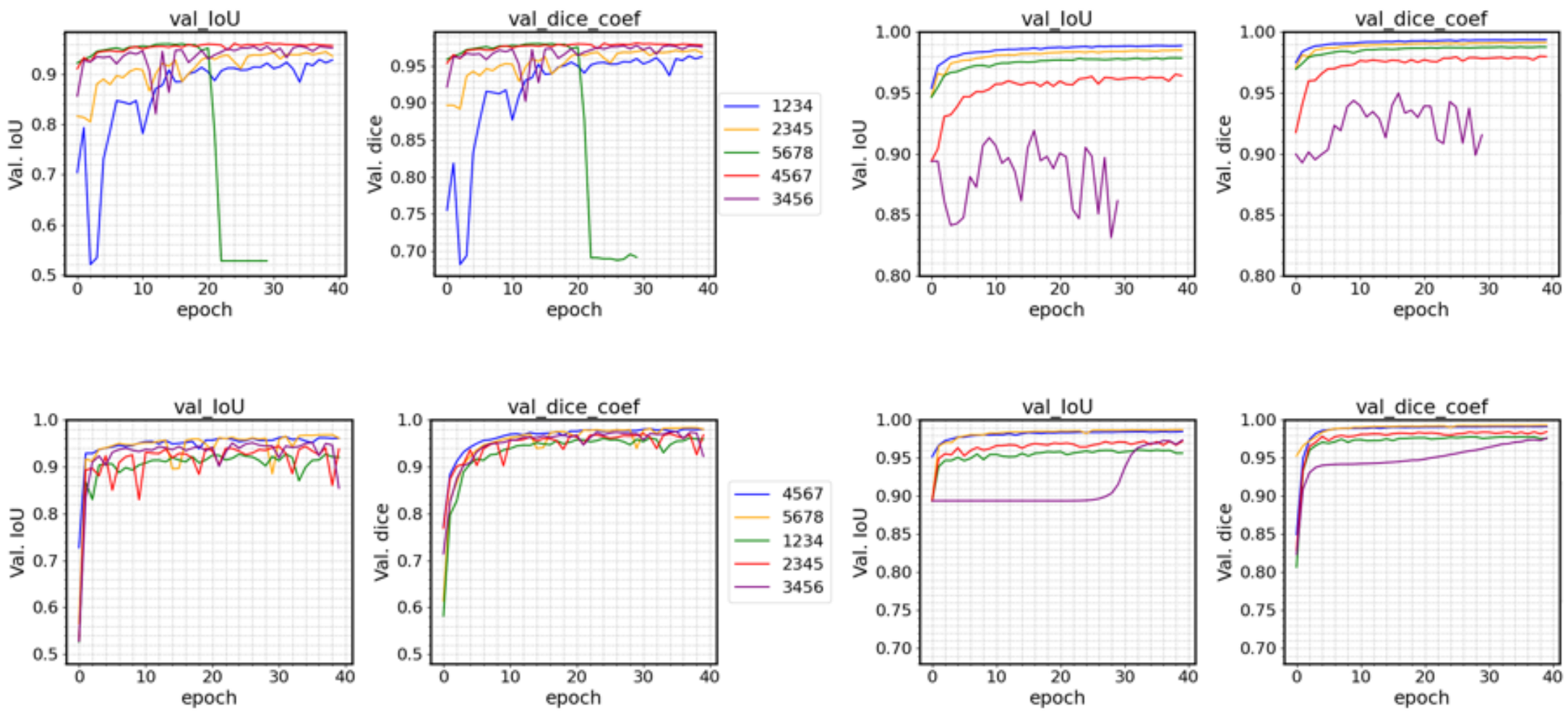

Figure 5.10: Validation IoU and Dice scores of 2D UNet (top row) and 2D `MEDNet` (bottom row) with different number of trainable filters. In each row, the left two subfigures are for IBSR V2.0 dataset and right two subfigures are for ATLAS R2.0 dataset.

**Example 5.5.** Study of 2D `MEDNet` with various wavelets having impulse responses up to length

In this experiment, Depth 4 `MEDNet` with (16, 32, 64, 128) filters is independently trained with different wavelets using both IBSR V2.0 and ATLAS R2.0 datasets to find the best performing wavelet model. Figure 5.11 shows that **sym4** gives the best results for both datasets.

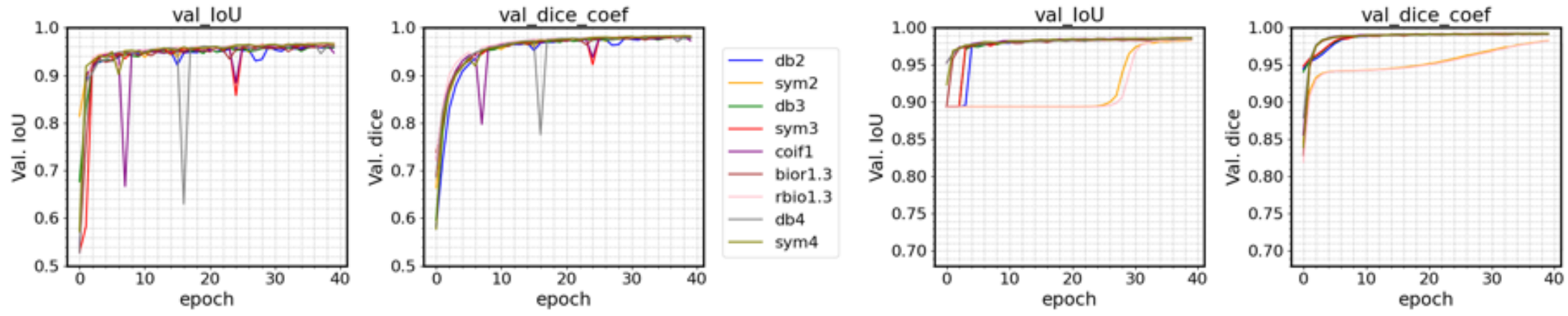

Figure 5.11: Validation scores of `MEDNet` 2D (config. 16, 32, 64, 128 filters, Depth 4) in segmenting large brain mask ROIs of IBSR V2.0 skull stripping dataset (left) and small lesion ROIs ATLAS R2.0 dataset (right) using different wavelets

**Example 5.6.** Shift equivariance tests of 2D `MEDNet` and UNet with horizontal and vertical shifts

This is an experiment to understand the shift equivariance properties of various filter numbers in `MEDNet`, UNet, and also with different wavelets in `MEDNet`. Figure 5.12 shows the results of the study with horizontal and vertical shifts. We observe that the models are more stable to horizontal shifts compared to vertical shifts. The (16, 32, 64, 128) filter configuration with db3 wavelet is an optimal choice, which is an economical, near shift equivariant, and stable model.

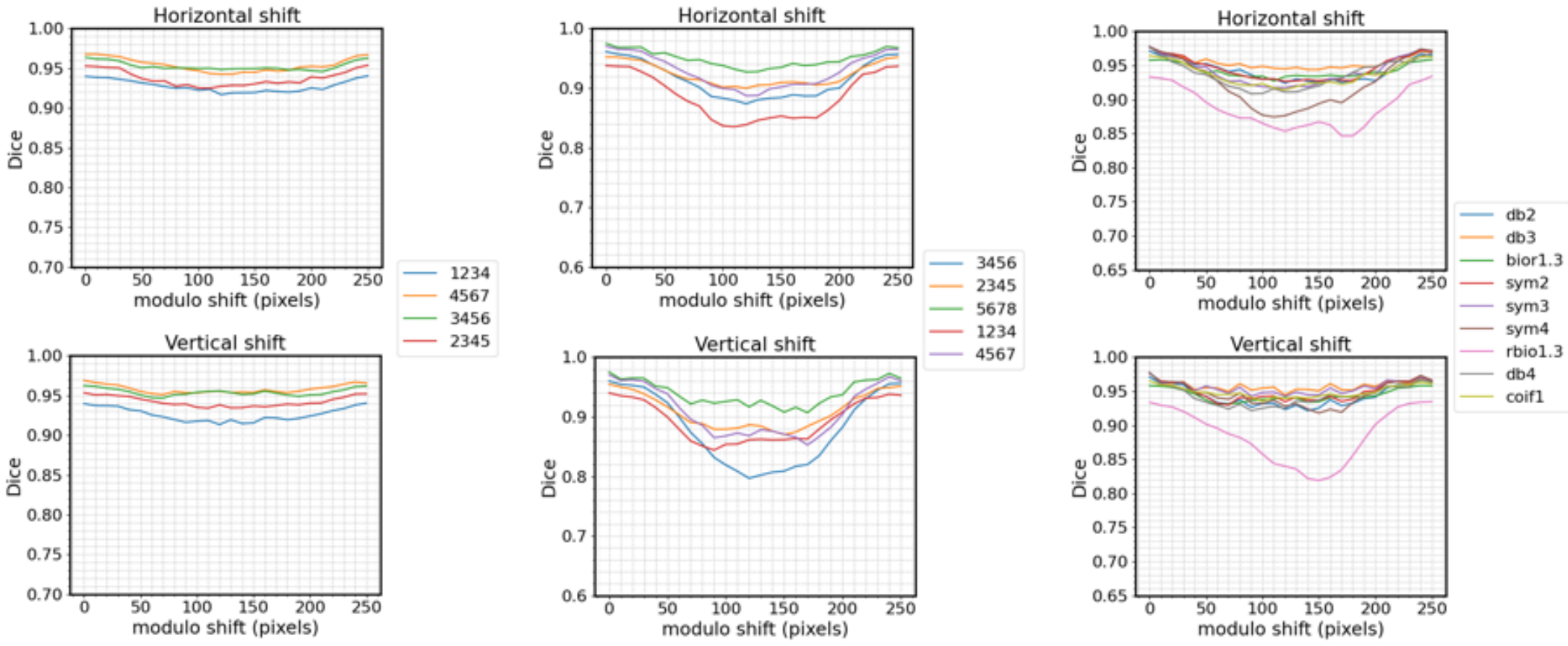

Figure 5.12: Horizontal and vertical shift equivariance properties of 2D UNet (left column), 2D `MEDNet` (middle column) with various configurations of filter numbers, and `MEDNet` with different wavelets (right column) with shifted test inputs

**Example 5.7.** Explainability of `MEDNet` with SHAP analysis

A model-agnostic, occlusion-based study on a trained 2D MEDNet explains its predictions on every image region by computing SHapely Additive exPlanations (SHAP) [286] using hexagonal superpixels. In Figure 5.13, the SHAP values explain the segmentation of $\mathcal{G}_{\phi,\psi}$ on the selected superpixel for an MRI slice and its horizontal and vertical shifts. Here, we also observe that $\mathcal{G}_{\phi,\psi}$ is stable to horizontal shifts but unstable to major vertical shifts.

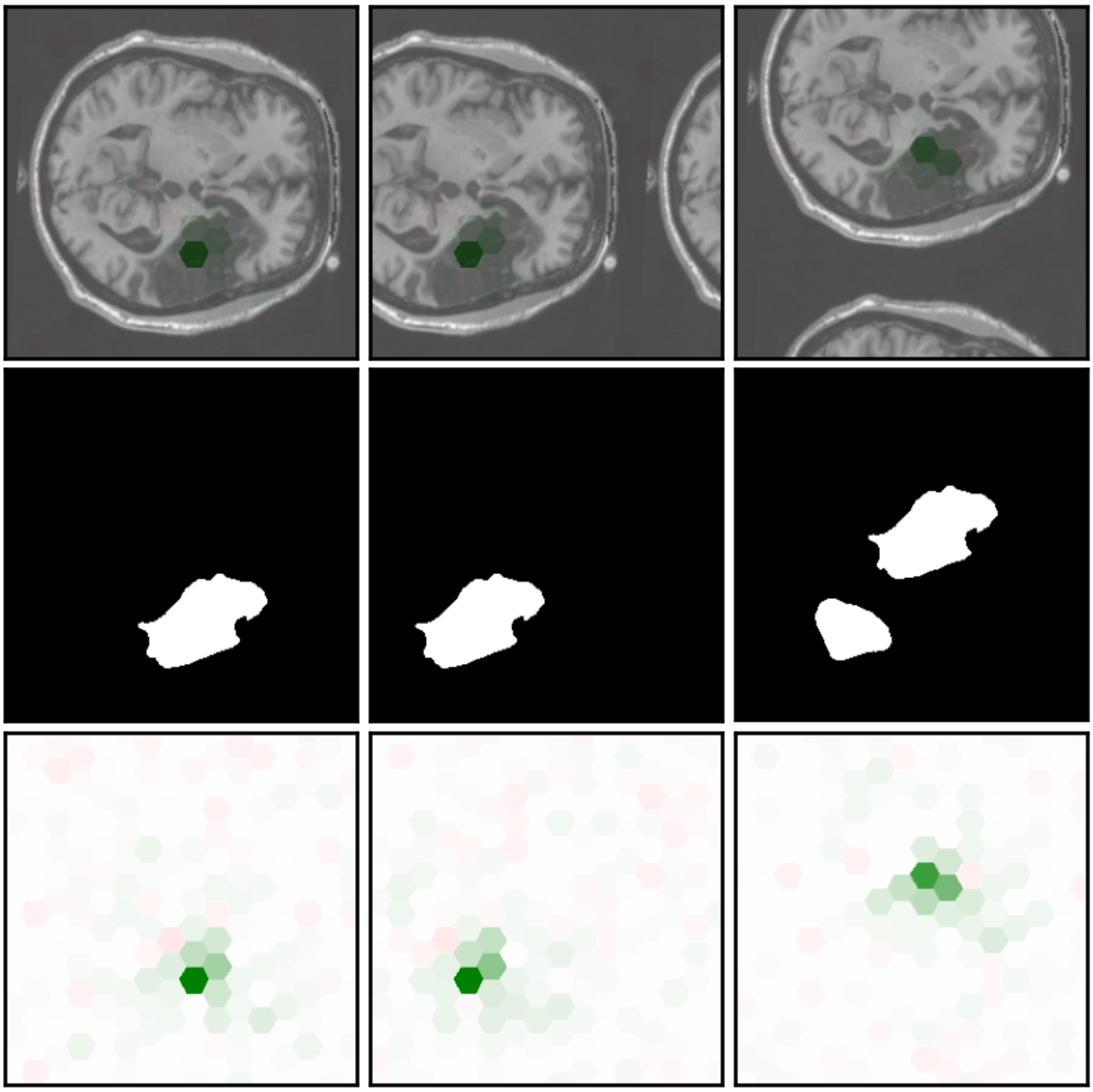

Figure 5.13: (Top row) A brain MRI slice from test set of ATLAS R2.0 overlayed with SHAP values for one selected superpixel: a normal slice (column 1), horizontally shifted slice (column 2) and vertically shifted slice (column 3); (Middle row) predicted segments; and (Bottom row) SHAPs of superpixels

**Example 5.8.** Lesion segmentation using LSI 2D Volterra MRA kernels without any nonlinear activation functions apart from the softmax in last layer.

In Chapter 3, the theory of LSI Volterra kernels in natural and equivalent biorthogonal basis representations for the joint natural-frequency domain is established. Now, in the first part of this example, we test the approximate segmentation capability of one LSI Volterra kernel (with db2 wavelet) of different sizes, followed by pointwise convolution and softmax in the natural and wavelet domains. Figure 5.14 shows a segmented slice and the convergence of natural and wavelet domain models with a single LSI 2D Volterra kernel (of different even sizes up to an upper limit of the input size 64x64), having only 4,100 parameters for the maximum kernel size. We observe that the single Volterra kernel is capable of roughly segmenting the lesion with inaccurate boundaries. This inaccuracy is expected because there is only one kernel (or filter) available for segmentation in this model, and also, we did not use any nonlinear activations. Further, the learning improves as the kernel size increases, and the convergence pattern also varies in the natural domain and the wavelet domain. The convergence is smooth and slow in the wavelet domain compared to the natural domain. Now, a full-fledged segmentation model architecture with Volterra kernels is an open research question, knowing that the complexity of the Volterra kernels increases exponentially with increasing size and order.

Thus, in the second part of this example, we start with the well-known CNN architecture, a 2D 64×64 I/O UNet with LSI Volterra kernels for lesion segmentation. In this UNet, we have just replaced the conventional 3×3 filters with multiresolution Volterra kernels of size 4×4 (the minimum size) and with 12,884 parameters. Figure 5.15 shows near accurate lesion segmentation from ATLAS R2.0 datasets using a UNet with LSI 2D Volterra kernels and without any nonlinear activation functions after the convolutions.

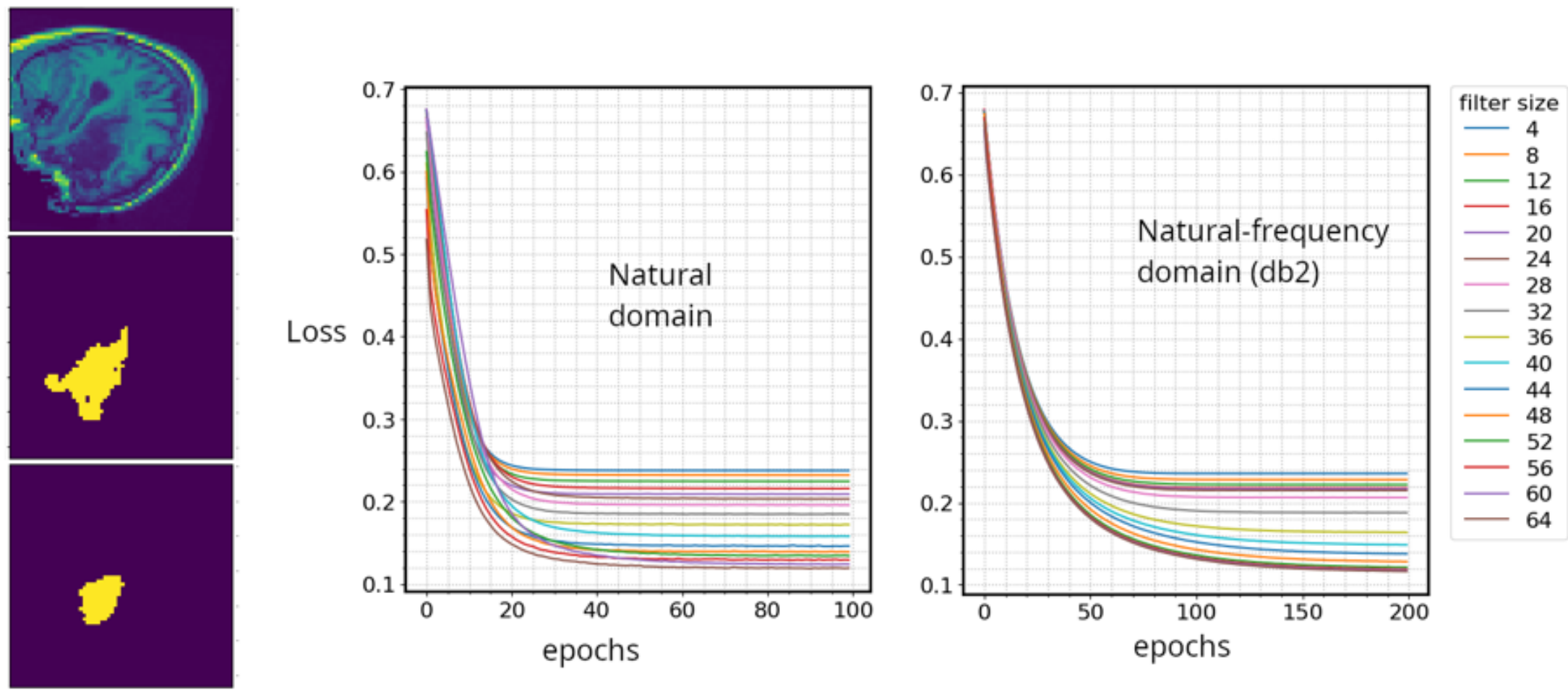

Figure 5.14: Lesion segmentation by LSI 2D Volterra kernel based UNet

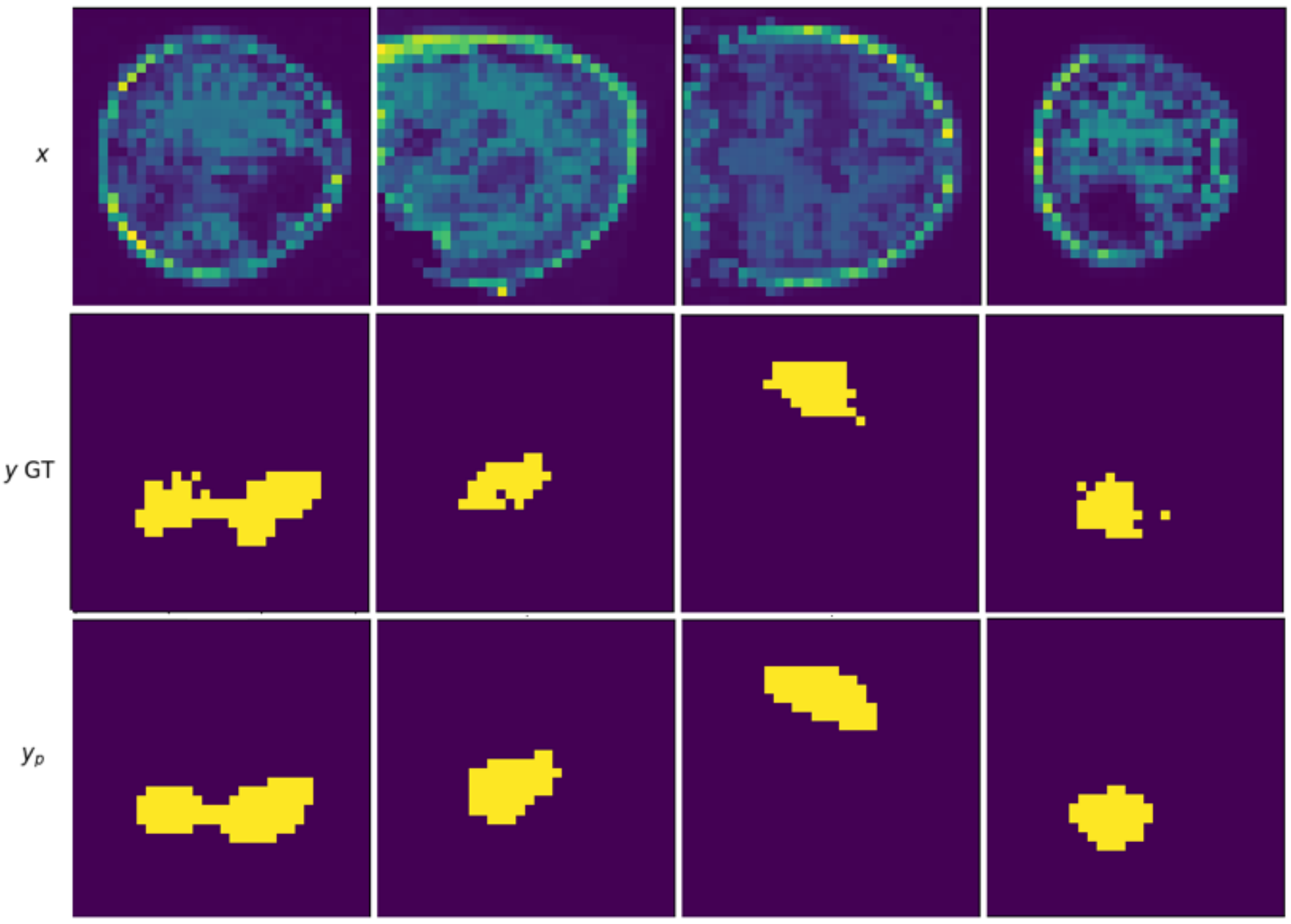

Figure 5.15: Lesion segmentation by LSI 2D Volterra kernel based UNet

*Observation:* The LSI Volterra kernel is a direct alternative view of the FIR filters in a CNN. The Volterra kernel (LSI, QSI) view increases expressivity of the model by introducing the practical use (with strong theoretical foundation) of filters beyond the typical filter size of 3 (or 3×3 or 3×3×3 in higher dimensions). The Volterra kernel theory allows sizes of learnable filters with an upper limit of the input size. Additionally, the QSI kernels also take into account the nonlinearities. The expressivity of the models with Volterra kernels further increases in the joint natural-frequency domain by using the equivalent relationship formed by the biorthogonal basis expansions of input, kernels, and output as described in Chapter 3.

**Example 5.9.** Binary segmentation by a 3D `MEDNet` and ratio of convolutional or LSI Volterra kernels.

This is a 3D `MEDNet` with 252,658 trainable parameters for binary segmentation of brain tissues in T1-weighted MRI volumes. The trainable NLSI units are

$$y[\boldsymbol{n}] = \frac{a_0 + a_1 q[\boldsymbol{n}]}{1 + |b_1 q[\boldsymbol{n}]|} \tag{5.43}$$

where $q$ is an LSI kernel and $a_0$, $a_1$, $b_1$ are constrained learnable coefficients. This model has no activation functions and has a modulus nonlinearity in the denominator. An equivalent baseline UNet with similar I/O and performance has 1,837,794 parameters. Figure 5.16 shows the binary segmentation by this 3D `MEDNet`. The dice scores are ~95% for all the four binary segments, viz. brain-mask, grey dmatter, white matter and CSF and outperforms the 3D UNet. Also due the model economy the 3D model easily trains in 48 GB GPU hardware with $256 \times 256 \times 256$ size I/O. A similar 3D UNet with same I/O size requires much more GPU memory or otherwise causes Out of Memory (OOM) issues with same hardware.

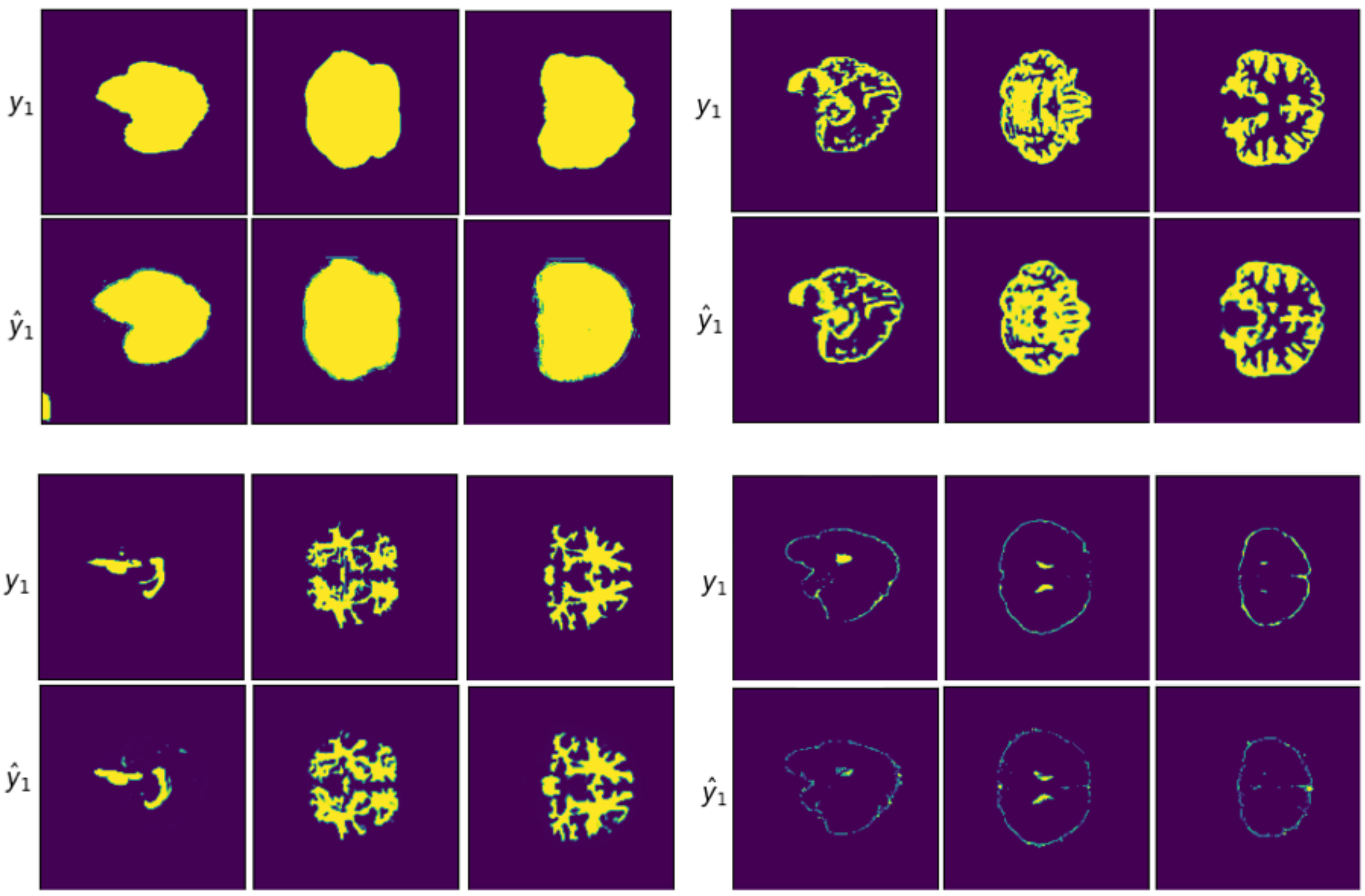

Figure 5.16: Binary segmentation of brain-mask (top left), grey matter (top right), white matter (bottom left) and CSF (bottom right) by 3D `MEDNet` with rational NLSI kernels

### 5.3.3 Deep image segmenters with separable IIR filters

The separable IIR backpropagation filters with per-axis direct-form II recursions and stable pole parametrization developed in Chapter 4 are used here to build two-dimensional image segmenters. A 2D slice is represented as $x \in \mathbb{R}^{H\times W\times 1}$ and mapped to a multiclass label field $y : \Omega \rightarrow \{0, 1, 2, 3\}$ with $\Omega = \{1, \ldots, H\} \times \{1, \ldots, W\}$. Label 0 denotes background and labels $1, 2, 3, \cdots$ correspond to the remaining finite semantic classes, for example, CSF, GM, and WM in T1 MRI. These deep image segmenters have their convolutional blocks of encoder-decoder replaced by learnable two-delay IIR systems mentioned in 4.1.2. This successive DF II filtering along each spatial axis makes the output depend on both current inputs and a short spatial history. This provides extended receptive fields with few parameters and gives a system-theoretic interpretation of spatial memory.

#### 5.3.3.1 Validation with multiclass semantic segmentation datasets

We construct a baseline UNet in which every convolutional block is replaced by separable IIR layers and denote it by `UIIRNet`. Skip connections and the prediction head follow the standard UNet design, and only the per-axis IIR filters change how spatial context is aggregated. Several open-source image segmentation datasets are used to test `UIIRNet`, with a UNet used as the baseline for comparison. Table 5.9 lists the mean IoU scores of `UIIRNet` and UNet, showing that `UIIRNet` outperforms UNet in almost all datasets. Figure 5.17 shows the poles of a trained stable separable IIR layer from the segmenter. `UIIRNet` learns slowly but accurately and performs well in segmenting fine vessels in retinal images (HRF, DRIVE dataset). Also, in urban and underwater imaging, the separable IIR variant is much more reliable in terms of accuracy. Figure 5.18 shows some of the images segmented by `UIIRNet` and their ground-truth masks.

Table 5.9: Mean IoU (%) scores of semantic segmentation by `UIIRNet` with multiple datasets

| **Dataset (type, # of classes)** | `UNet` | `UIIRNet` |
|---|---|---|
| CamVid (driving, 12) | 28.35 | 48.69 |
| DRIVE (retina, 2) | 12.55 | 44.80 |
| HRF (retina, 2) | 54.82 | 68.06 |
| MaSTr1325 (maritime, 5) | 66.17 | 65.57 |
| SUIM (underwater, 8) | 26.40 | 78.42 |

*Note:* For DRIVE and HRF we collapse all non-background labels into a single foreground class (binary segmentation), even though the original datasets provide richer annotations.

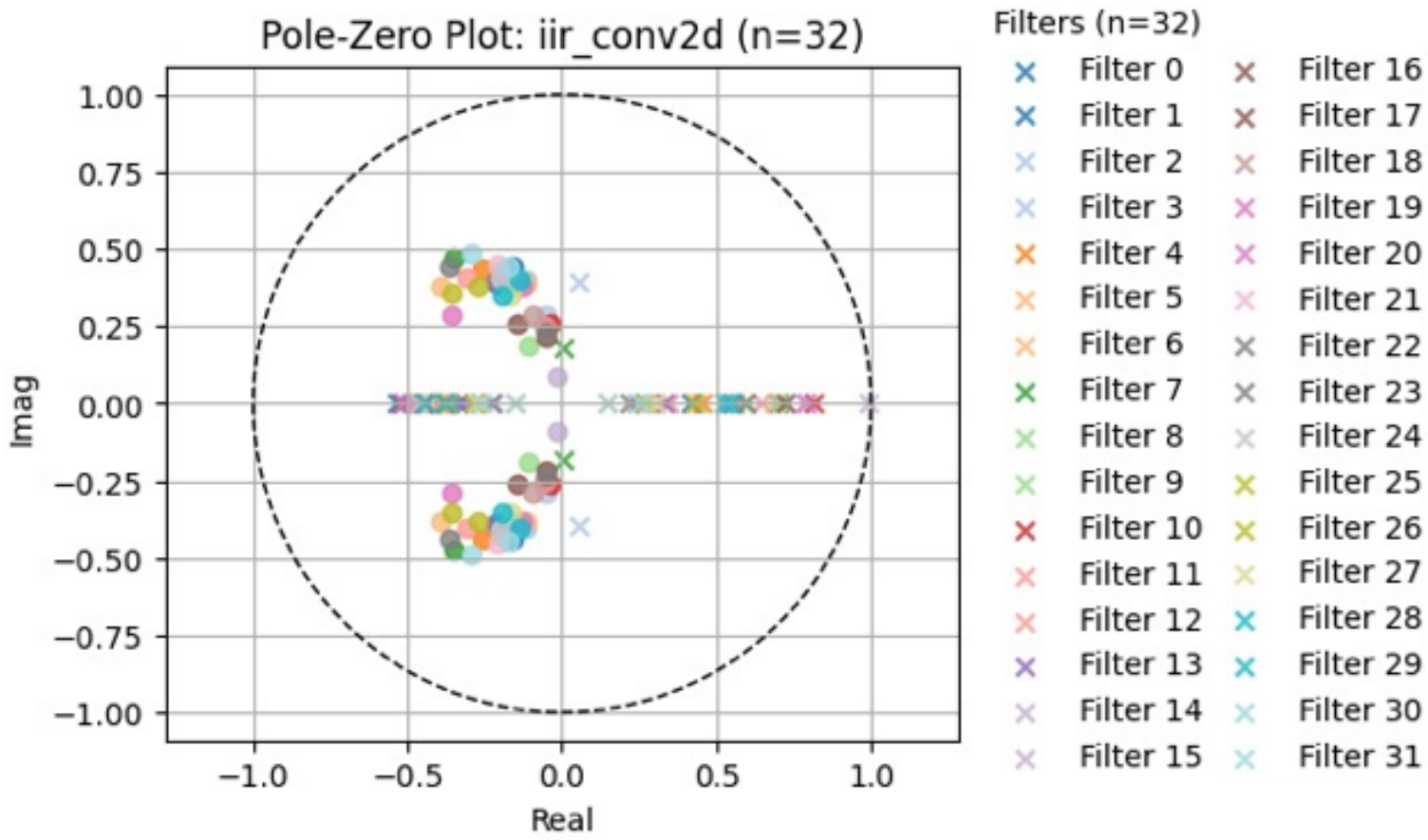

Figure 5.17: Poles and zeros of a separable IIR convolutional blocks with delay=2 after training

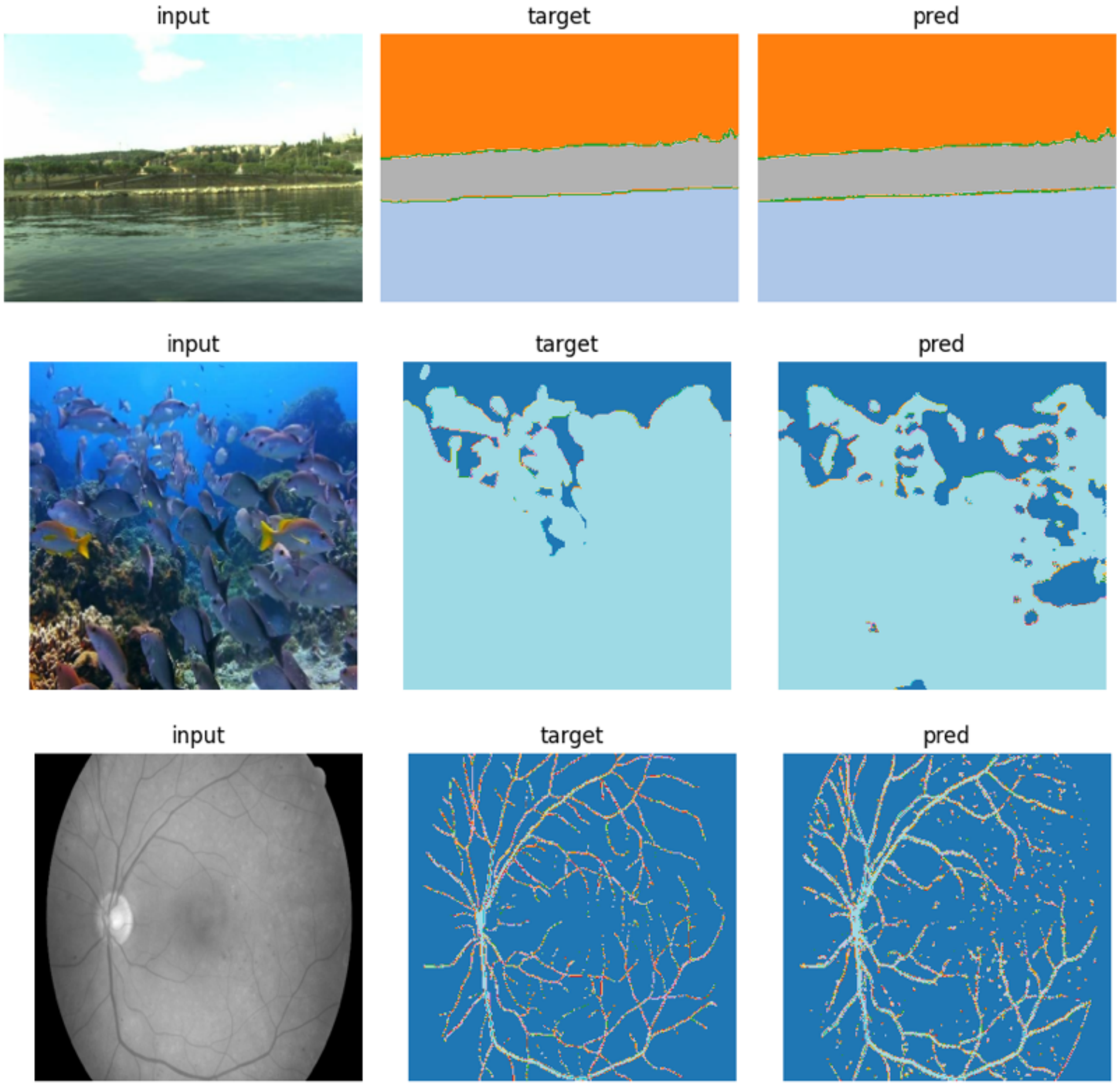

Figure 5.18: Representative semantic segmentation outputs from the trained `UIIRNet` models and the corresponding ground-truth masks.

#### 5.3.3.2 Validation with T1 brain MRI slices

The deep segmenters with separable IIR filters take T1-weighted brain MRI slices from a body plane as input and segment CSF, GM, and WM tissues. The models first train with a supervised segmentation objective mentioned earlier in this section, combining pixel-wise cross-entropy and soft Dice loss with mild regularisation. All models in this comparison share the same data splits, augmentations, and optimisation schedule so that differences in performance can be attributed to the treatment of spatial context rather than to changes in the training pipeline. On this common backbone, `UIIRNet` and following architectural families are tested:

- `DWT-MEDNet-IIR`: A MED encoder–decoder in which each encoder stage first applies a 2D DWT and then passes each band through separable IIR blocks. This combines wavelet-domain sparsity with spatial recurrence.
- *Transformer and state–space hybrids.* To study global context, the IIR backbone is combined with self-attention or state–space modules inspired by ViTs [57], the SAM [251], and Mamba architectures [100]. In `Trans-ViT-UIIRNet`, separable IIR blocks supply local structure while a small ViT encoder processes patch tokens for long-range dependencies. In `SAM-UIIRNet`, an IIR backbone replaces the heavy ViT encoder of SAM and a single binary mask prompt steers a compact transformer decoder. A `ViT-IIR-Hybrid` omits the UNet decoder and treats the full-resolution slice as a token grid processed by constrained self-attention.
- *Attention–augmented IIR.* An `ECA-UIIRNet` variant attaches efficient channel attention to IIR features, following the ECA mechanism [287], so that informative feature channels are emphasised while maintaining the same separable IIR spatial core.

In the multiclass models, local features are first enhanced by convolution and IIR filtering, global context is captured by ViT or state–space blocks, and the UNet decoder reconstructs per-pixel tissue labels. Across the 2D T1 MRI experiments, the plain `UIIRNet` already improves CSF boundary delineation compared to a purely FIR UNet, with a clear gain in CSF F1 and Dice scores while also giving modest improvements for GM and WM. The DWT–IIR variant performs competitively for GM and WM but is more sensitive for the minority CSF class, suggesting that combining frequency decomposition and spatial recurrence requires careful balancing of contributions across tissues. The `ECA-UIIRNet` provides the most balanced improvement across all tissues, achieving the highest CSF F1 and Dice scores and indicating that channel attention and separable spatial memory act synergistically.

`Mamba-UIIRNet` and `Trans-ViT-UIIRNet` architectures achieve the best GM and WM scores in the study, confirming that long-range context from sequence-style blocks can sharpen tissue boundaries in cortical and deep white matter regions, but their CSF performance degrades unless class imbalance is explicitly handled. `SAM-UIIRNet` and `ViT-IIR-Hybrids` demonstrate that prompt-based and token-only designs can be built on top of the same separable IIR blocks, yet in this dataset, they do not surpass the simpler `UIIRNet` and `ECA-UIIRNet` baselines in terms of tissue-wise balance. Table 5.10 reports a quantitative comparison of the tested IIR-based 2D segmentation models.

### 5.3.4 Convolutional variational encoder–decoder segmenter

This subsection introduces a variational extension of the MED segmenters. A latent variable $\mathbf{z}$ is inserted into the processing path while the external interface remains unchanged: for each spatial index $\boldsymbol{n} \in \Omega$ the model outputs a posterior vector $p_\theta(x)_{\boldsymbol{n}\cdot} \in \Delta_{K-1}$ and uses the same pointwise decision rule $f_\theta(x)_{\boldsymbol{n}} = \arg\max_k p_\theta(x)_{\boldsymbol{n}k}$. The latent captures global shape and appearance variability, while the segmentation loss and cost-aware decisions follow the formulations already introduced.

- *Architecture: AProcS with a variational core.*
  1. *Analysis and encoder.* A UNet style encoder, augmented with 2D or 3D DWT blocks, acts as the analysis path and produces multiscale feature maps. From the deepest features a small encoder head with parameters $\phi$ outputs a mean and log-variance,

     $$\mu_\phi(x),\ \log\sigma_\phi^2(x) \in \mathbb{R}^{d_z}, \tag{5.44}$$

     which define a Gaussian approximate posterior

     $$q_\phi(\mathbf{z} \mid x) = \mathcal{N}\big(\mathbf{z};\ \mu_\phi(x), \operatorname{diag}\sigma_\phi^2(x)\big). \tag{5.45}$$

  2. *Sampling by reparameterization.* During training, latent samples are drawn using the reparameterization trick,

     $$\mathbf{z} = \mu_\phi(x) + \sigma_\phi(x) \odot \epsilon, \qquad \epsilon \sim \mathcal{N}(0, I), \tag{5.46}$$

     so that gradients can flow through $\mathbf{z}$ to the encoder parameters.

  3. *Synthesis decoder and segmentation head.* The decoder combines the latent $\mathbf{z}$ with skip features from the encoder and applies an IDWT-based synthesis path (for example, 3D IDWT) to reconstruct a feature field

     $$u(x, \mathbf{z}) : \Omega \to \mathbb{R}^C. \tag{5.47}$$

     A $1 \times 1$ (or $1 \times 1 \times 1$) convolution then maps $u(x, \mathbf{z})$ to class logits $u_\theta(x, \mathbf{z})_{\boldsymbol{n}} \in \mathbb{R}^K$ at each spatial location, followed by a softmax over classes,

     $$p_\theta(x, \mathbf{z})_{\boldsymbol{n}\cdot} = \operatorname{softmax}\big(u_\theta(x, \mathbf{z})_{\boldsymbol{n}}\big), \qquad \boldsymbol{n} \in \Omega. \tag{5.48}$$

     The `CVUNet` and `CVMEDNet` architectures used in this thesis are concrete realizations of this template, differing mainly in the details of the encoder and decoder blocks.

- *Objective: supervised ELBO with segmentation loss.* For training pairs $\{(x^{(i)}, y^{(i)})\}_{i=1}^n$ the parameters $(\theta, \phi)$ are learned by minimizing a supervised evidence lower bound that couples a segmentation loss with a Kullback–Leibler term. With a standard Gaussian prior $p(\mathbf{z}) = \mathcal{N}(0, I)$, the objective is

$$\mathcal{L} = \min_{\theta,\phi} \frac{1}{n} \sum_{i=1}^{n} \Big[ \lambda_{\text{seg}}\, \mathbb{E}_{q_\phi(\mathbf{z}|x^{(i)})}\, \mathcal{L}_{\text{seg}}\big(y^{(i)},\ p_\theta(x^{(i)}, \mathbf{z})\big) + \beta\, \mathrm{KL}\big(q_\phi(\mathbf{z} \mid x^{(i)}) \,\|\, p(\mathbf{z})\big) \Big], \tag{5.49}$$

where $\lambda_{\text{seg}} > 0$ weights the segmentation term, $\beta > 0$ controls the strength of the KL regularization, and $\mathcal{L}_{\text{seg}}$ is the segmentation loss defined in Section 5.1, typically a combination of voxelwise cross-entropy and Dice loss. In practice the expectation over $q_\phi(\mathbf{z} \mid x^{(i)})$ is approximated with one or a few samples per training example.

Table 5.10: Comprehensive performance summary of IIR-based deep 2D segmentation models

| Region | Model | F1-score | Loss | Precision | Dice Score |
|---|---|---|---|---|---|
| GM | UNet | 0.9032 | 0.1105 | 0.8782 | 0.8942 |
| GM | UIIRNet | **0.9168** | 0.1469 | 0.9062 | 0.8917 |
| GM | ECA-Enhanced-UIIRNet | 0.8979 | 0.1023 | 0.8798 | 0.8976 |
| GM | MEDNet | 0.9136 | 0.1469 | 0.9062 | 0.8649 |
| GM | DWT-MEDNet-IIR | 0.9049 | 0.1728 | 0.9189 | 0.8820 |
| GM | Mamba-UIIRNet | **0.9170** | 0.1200 | 0.9100 | **0.9043** |
| GM | Trans-ViT-UIIRNet | 0.8976 | 0.2189 | 0.8929 | 0.8850 |
| GM | SAM-UIIRNet | 0.8213 | 0.3108 | 0.7271 | 0.7969 |
| GM | ViT-IIR-Hybrid | 0.8950 | 0.1800 | 0.8800 | 0.8900 |
| WM | UNet | 0.8904 | 0.1225 | 0.8980 | 0.8792 |
| WM | UIIRNet | 0.9024 | 0.1225 | 0.8980 | 0.9119 |
| WM | ECA-Enhanced-UIIRNet | **0.9147** | **0.0857** | **0.9320** | **0.9139** |
| WM | MEDNet | 0.9131 | 0.1711 | 0.9071 | 0.8649 |
| WM | DWT-MEDNet-IIR | 0.9064 | 0.1648 | 0.9066 | 0.8856 |
| WM | Mamba-UIIRNet | **0.9231** | 0.1100 | 0.9200 | 0.9180 |
| WM | Trans-ViT-UIIRNet | 0.8925 | 0.1870 | 0.9363 | 0.8765 |
| WM | SAM-UIIRNet | 0.6532 | 0.4310 | 0.7561 | 0.5972 |
| WM | ViT-IIR-Hybrid | 0.8870 | 0.1650 | 0.8900 | 0.8820 |
| CSF | UNet | 0.6373 | 0.4976 | 0.5896 | 0.5842 |
| CSF | UIIRNet | 0.8094 | 0.2862 | 0.7809 | 0.7898 |
| CSF | ECA-Enhanced-UIIRNet | **0.8528** | **0.1623** | **0.8514** | **0.8310** |
| CSF | MEDNet | 0.6740 | 0.4109 | 0.6146 | 0.6702 |
| CSF | DWT-MEDNet-IIR | 0.6482 | 0.5379 | 0.7939 | 0.6309 |
| CSF | Mamba-UIIRNet | 0.6320 | 0.4500 | 0.6200 | 0.6114 |
| CSF | Trans-ViT-UIIRNet | 0.7210 | 0.3723 | 0.8695 | 0.7122 |
| CSF | SAM-UIIRNet | 0.6138 | 0.4209 | 0.5852 | 0.5989 |
| CSF | ViT-IIR-Hybrid | 0.6800 | 0.3900 | 0.7100 | 0.6750 |

- *Prediction and uncertainty.* At test time, a deterministic segmentation can be obtained by using the mean latent

$$\mathbf{z}_{\text{det}}(x) = \mu_\phi(x), \tag{5.50}$$

and evaluating

$$p_\theta(x)_{\boldsymbol{n}\cdot} \approx p_\theta\big(x, \mathbf{z}_{\text{det}}(x)\big)_{\boldsymbol{n}\cdot}, \qquad f_\theta(x)_{\boldsymbol{n}} = \arg\max_k p_\theta(x)_{\boldsymbol{n}k}. \tag{5.51}$$

For uncertainty-aware predictions, several latent samples $\mathbf{z}^{(m)} \sim q_\phi(\cdot \mid x)$ are drawn and the posteriors are averaged,

$$\bar{p}_\theta(x)_{\boldsymbol{n}\cdot} \approx \frac{1}{M} \sum_{m=1}^{M} p_\theta(x, \mathbf{z}^{(m)})_{\boldsymbol{n}\cdot}, \tag{5.52}$$

after which the same pointwise or cost-aware decision rules are applied with $\bar{p}_\theta(x)$ in place of $p_\theta(x)$.

- *Why this fits the general framework.* The convolutional variational segmenter preserves the output contract of the previous models: for each $\boldsymbol{n} \in \Omega$ it produces a posterior vector in $\Delta_{K-1}$ and uses the same decision rules and evaluation metrics (for example, Dice and IoU). The encoder stage uses 2D or 3D DWT bank, the processing stage introduces the latent $\mathbf{z}$ through a variational encoder, and the decoder stage uses the corresponding IDWT-based decoder before the per-location head. Perfect reconstruction properties of the wavelet analysis and synthesis keep the decoder stable, while the KL term introduces a global prior on latent representations without changing the loss and decision interface seen by the segmenter.
- *Practical notes.* In practice, KL scaling is often annealed over the course of training (that is, $\beta$ is increased gradually) to reduce the risk of posterior collapse and to allow the segmentation loss to shape the latent space. If needed, a small auxiliary reconstruction term on decoder features or logits can be added early in training to encourage informative latents; this auxiliary term is removed or downweighted once the model stabilizes. These adjustments do not change the notation or the external segmentation interface.

**Example 5.10.** Binary segmentation by 3D `MEDNet` with convolutional variational bottleneck

This experiment uses a `MEDNet` or UNet with a convolutional variational bottleneck. Such a model takes many more epochs to train, but it also produces an organized feature space. Figure 5.19 shows segmented patches with clear, sharp edges and clear curves of in output patches of test MRI slices of the IBSR V2.0 dataset.

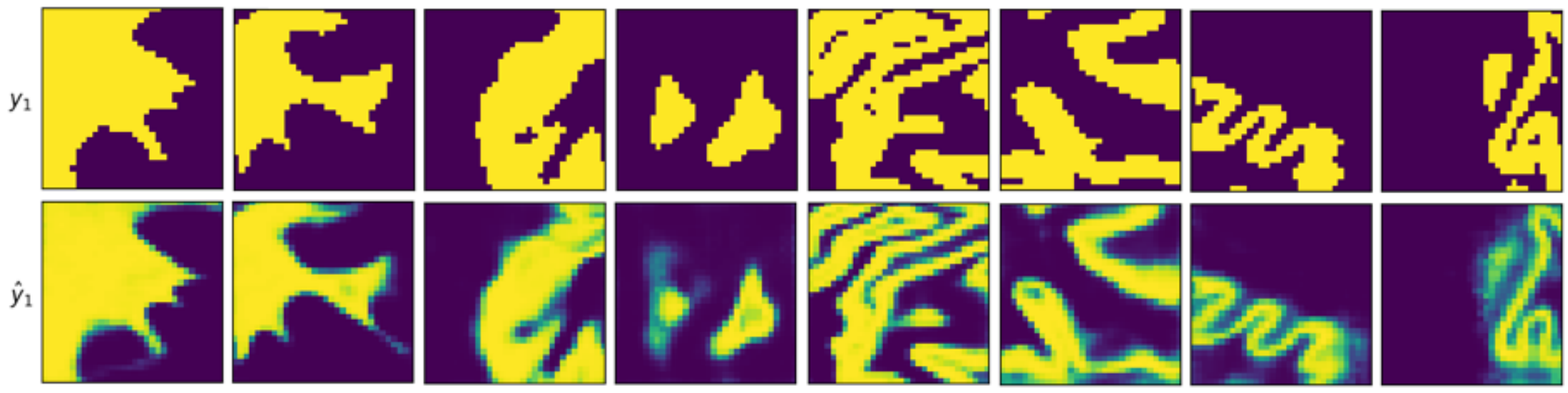

Figure 5.19: Binary segmentation by 3D `MEDNet` with convolutional variational bottleneck; ground truths (top row) and predicted segments (bottom row)

**Example 5.11.** Multiclass segmentation by 3D `MEDNet` with variational bottleneck

Similar to the binary segmentation models, the multiclass convolutional variational 3D `MEDNet` and UNet take more epochs to train, but also produce an organized feature space. Figure 5.20 shows patches with segmented grey matter, white matter and CSF from test MRI slices of the IBSR V2.0 dataset.

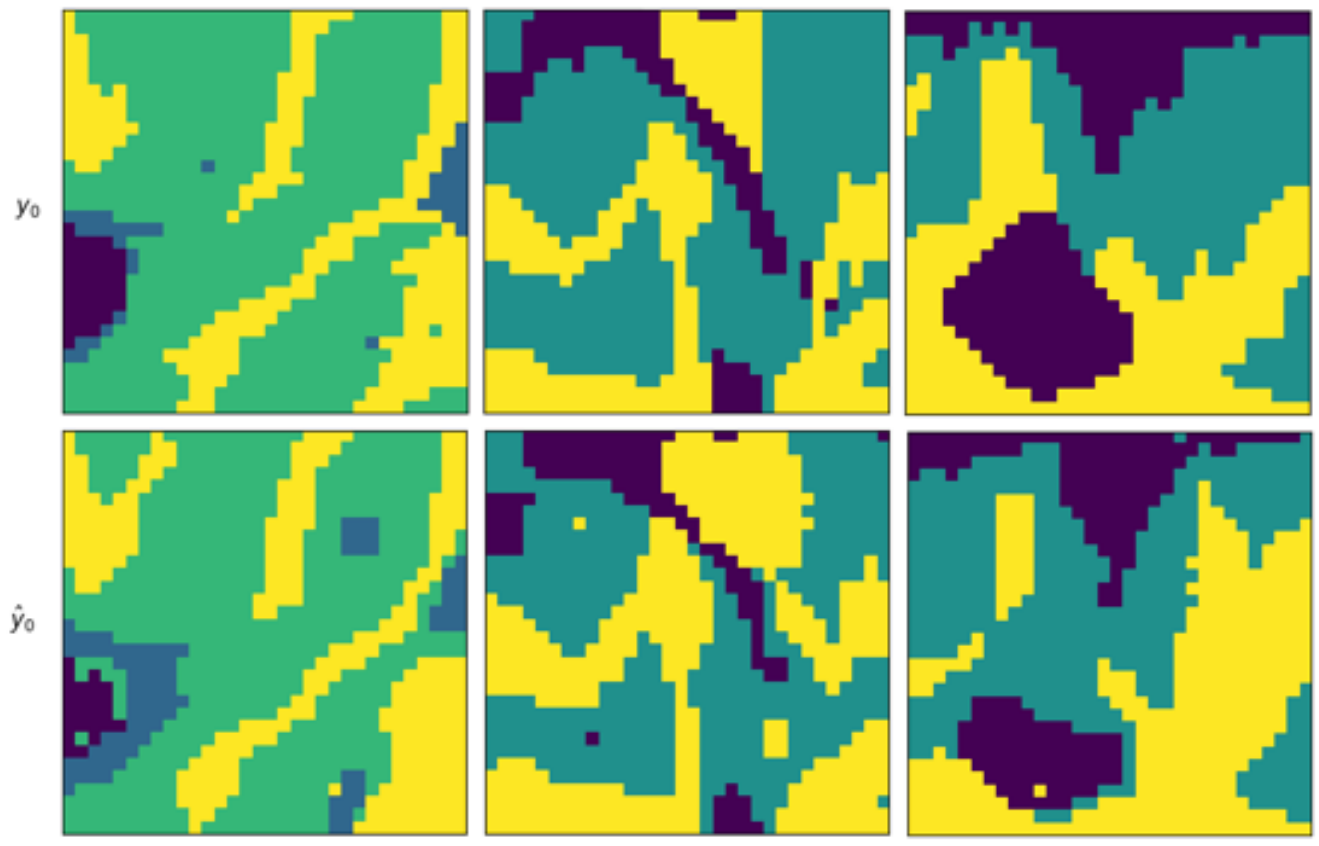

Figure 5.20: Multiclass segmentation by 3D `MEDNet` with convolutional variational bottleneck; ground truths (top row) and predicted segments (bottom row)

### 5.3.5 MRI (3D) multiresolution segmenters with WaveletViT bottleneck

The `MEDNet` family admits a bottleneck that operates in a multiresolution wavelet domain. The encoder and decoder remain close to a standard 3D UNet-style MEDNet, while the plain convolutional bottleneck is replaced by a WaveletViT block that attends over DWT subbands. The external interface is unchanged: for each spatial index $\boldsymbol{n} \in \Omega$ the model produces a posterior vector $p_\theta(x)_{\boldsymbol{n}\cdot} \in \Delta_{K-1}$ and uses the pointwise decision rule $f_\theta(x)_{\boldsymbol{n}} = \arg\max_k p_\theta(x)_{\boldsymbol{n}k}$.

Let $h_{\text{enc}} : \Omega \to \mathbb{R}^{C_{\text{enc}}}$ denote the encoder feature field at the coarsest resolution in a 3D `MEDNet`. The WaveletViT bottleneck proceeds in three conceptual stages.

- *Wavelet tokenization.* A multi-level 3D DWT with $L$ levels is applied to $h_{\text{enc}}$. At each level $j$ this yields one low-pass volume and a small set of high-pass volumes at that scale. These volumes are projected to a common embedding width by $1 \times 1 \times 1$ convolutions, so that each subband becomes a tensor of shape $[B, H_j, W_j, D_j, E]$, where $B$ is the batch size and $(H_j, W_j, D_j)$ are the spatial dimensions at level $j$. The collection of all low-pass and high-pass volumes across levels forms a multiresolution set of ”wavelet tokens”.
- *WaveletViT attention over subbands.* A small stack of WaveletViT blocks acts on this multiresolution representation. Within each subband, multi-head self-attention operates along spatial axes, and additional mixing across subbands can be included to couple scales and orientations. Residual connections keep the transformation close to the identity when attention updates are small. The output is a set of updated low-pass and high-pass volumes with the same shapes as the inputs, but with long-range interactions encoded within and across scales.
- *Wavelet assembly back to bottleneck features.* A 3D IDWT reconstructs a feature field on the encoder grid from the updated low-pass and high-pass volumes, level by level. Starting from the deepest low-pass band, the inverse transform inserts the corresponding high-pass bands at each scale until the original resolution is reached. The result is a bottleneck feature field

$$u_{\text{bottleneck}} : \Omega \to \mathbb{R}^E, \tag{5.53}$$

  which replaces $h_{\text{enc}}$ as the input to the `MEDNet` decoder.

The decoder uses 3D convolutions and upsampling or 3D IDWT blocks, together with skip connections, to map $u_{\text{bott}}$ back to a multichannel feature field $u : \Omega \to \mathbb{R}^C$. A $1 \times 1 \times 1$ prediction head then produces class logits and posteriors at each voxel. Training uses the same segmentation objective as in the preceding subsections, namely a weighted combination of per-voxel cross-entropy and soft Dice loss with a regularizer $\mathcal{R}(\theta)$. In code this architecture is implemented as a `MEDNet` variant with a WaveletViT bottleneck (for example, `UWaveViTNet3D`) that wraps a base 3D encoder–decoder while keeping the loss and decision interface identical to the other MEDNet segmenters.

**Convolutional variational wavelet-bank version** A convolutional variational version of this architecture, denoted `CVUWaveViTNet`, inserts a latent bottleneck on top of the assembled wavelet features. After WaveletViT processing and wavelet assembly, a global average pooling layer summarizes $u_{\text{bott}}$ into a vector. Two dense layers then produce the mean and log-variance of a Gaussian latent variable $\mathbf{z}$, which follows the approximate posterior $q_\phi(\mathbf{z} \mid x)$. A reparameterization layer samples $\mathbf{z}$ during training, and a small projection lifts $\mathbf{z}$ back to the bottleneck channel dimension before it is broadcast and additively injected into the assembled feature map, followed by a light $3 \times 3 \times 3$ convolution. A dedicated layer adds the scaled Kullback–Leibler divergence between $q_\phi(\mathbf{z} \mid x)$ and a standard normal prior to the segmentation loss. The decoder, segmentation head, and decision rule remain unchanged, so both the deterministic and convolutional variational WaveletViT `MEDNet` segmenters share the same per-location posteriors $p_\theta(x)_{\boldsymbol{n}\cdot} \in \Delta_{K-1}$ and hard labels $f_\theta(x)_{\boldsymbol{n}}$.

**Separable 3D IIR convolutional variational wavelet-bank version** A further variant combines the WaveletViT bottleneck with the separable IIR cores introduced in Chapter 4. In this IIR–WaveletViT `MEDNet`, implemented in code as `UIIRDFIIWaveViTNet3D`, every encoder and decoder block of the underlying UNet3D is built from stacked `IIRConv3DDFII` layers rather than standard FIR convolutions. Each `IIRConv3DDFII` layer realizes a separable three-dimensional ARMA operator: along each spatial axis it runs a direct–form II recursion with learnable autoregressive and moving–average taps, optional per–axis parameter sharing, and reflect boundary handling, followed by residual gating and a pointwise nonlinearity. A DC–normalization constraint keeps the gain near unity at zero frequency, so the learned IIR core behaves like a band–pass refinement on top of a stable baseline. The WaveletViT tokenization, attention over wavelet tokens, assembler, decoder, segmentation head, and loss remain unchanged, so the IIR–WaveletViT and purely convolutional WaveletViT `MEDNet` segmenters share the same interface in terms of posteriors $p_\theta(x)_{\boldsymbol{n}\cdot} \in \Delta_{K-1}$ and decisions $f_\theta(x)_{\boldsymbol{n}}$.

### 5.3.6 MRI (3D) multiresolution segmenters with ShearViT bottleneck

A second class of `MEDNet` segmenters uses a ViT bottleneck that operates in the Fourier-domain directional tiling (3D FDST) introduced in Chapter 3. In this case the bottleneck, which we refer to as a ShearViT bottleneck, works on cone–scale–direction tiles in the 3D Fourier domain and then reconstructs a spatial feature field. As before, the encoder, decoder and segmentation head follow the standard `MEDNet` design, and only the bottleneck block is changed.

Let $h_{\text{enc}} : \Omega \to \mathbb{R}^{C_{\text{enc}}}$ be the encoder feature field at the bottleneck resolution. An FDST analysis stage constructs a disjoint tiling of the 3D Fourier cube into bands indexed by $k = (j, c, d)$, where $j$ labels radial scales, $c$ labels cones and $d$ indexes orientations within each cone. Each band has an associated smooth window $W_k[\boldsymbol{\omega}]$ and produces a complex coefficient tile on a coarse 3D grid.

- *Band tokenization in the FDST domain.* FDST analysis is applied channelwise to $h_{\text{enc}}$, yielding a set of complex coefficient tiles. Real and imaginary parts are handled explicitly, and pooling over the spatial extent of each tile produces one band descriptor per tile. A linear projection maps each band descriptor to an embedding of dimension $E$. Band metadata, such as scale and cone index, are encoded as a fixed embedding and added to the tokens. This gives a sequence of band tokens

$$t_k \in \mathbb{R}^E, \qquad k = 1, \ldots, K, \tag{5.54}$$

  that summarize the activity in each FDST wedge.
- *Inter-band self-attention and gain modulation.* A multi-head self-attention layer operates on the sequence $(t_1, \ldots, t_K)$, optionally grouped by scale or cone. The attention outputs pass through a small feed-forward network that predicts a real-valued gain for each band. These gains, together with trainable cone and rotation gates, modulate the FDST coefficient tiles by pointwise multiplication in the Fourier domain. In this way the bottleneck can emphasize or suppress specific scales, cones and orientations in a data-dependent manner while leaving the underlying FDST windows unchanged.
- *Inverse FDST and spatial feature fusion.* After modulation, an inverse FDST reconstructs a spatial feature field from the modified coefficient tiles. A $1 \times 1 \times 1$ convolution maps this reconstruction to $E$ channels, and a lightweight local mixer (for example, a short 3D convolutional block) can be applied to refine the result. A residual connection with $h_{\text{enc}}$ keeps the overall transformation close to the identity when the predicted gains are small. The resulting bottleneck feature field is

$$u_{\text{bott}} : \Omega \to \mathbb{R}^E, \tag{5.55}$$

  which is then fed into the `MEDNet` decoder in exactly the same way as in the convolutional and WaveletViT cases.

The FDST-based ShearViT bottleneck preserves the same input–output contract as the other segmenters in this chapter. For an input $x$ the network returns per-voxel class posteriors $p_\theta(x)_{\boldsymbol{n}\cdot} \in \Delta_{K-1}$ and decisions

$$f_\theta(x)_{\boldsymbol{n}} = \arg\max_{k \in \mathcal{Y}} p_\theta(x)_{\boldsymbol{n}k}, \qquad \boldsymbol{n} \in \Omega, \tag{5.56}$$

with training carried out using the combined cross-entropy and Dice objective described earlier. In code this architecture appears as a `MEDNet` variant with an FDST-based ViT bottleneck (for example, `UFDSTViTNet3D`), where the bottleneck feature map is processed by an FDST–ViT block configured by the number of scales, cone structure and band pooling strategy before being passed to the decoder.

**Convolutional variational shearlet-bank version** An analogous convolutional variational extension is used for the FDST-based ShearViT bottleneck, implemented as `CVUShearViTNet`.

Here the FDST–ViT block runs in feature-return mode and produces a bottleneck feature field that is globally averaged to obtain a summary vector. Two dense layers map this vector to the mean and log-variance of a Gaussian latent variable $\mathbf{z}$. The log-variance is clipped for numerical stability, and a reparameterization layer draws $\mathbf{z}$ during training. A small dense projection lifts $\mathbf{z}$ to the bottleneck channel dimension, optional bounded nonlinearities keep the latent contribution well scaled, and a broadcast-add plus short 3D convolution injects $\mathbf{z}$ back into the FDST–ViT feature map. A KL-add layer contributes the scaled divergence between $q_\phi(\mathbf{z} \mid x)$ and the standard normal prior to the total loss. As in the WaveletViT case, the decoder, segmentation head, and per-location decision rule are unchanged, so the variational ShearViT `MEDNet` models still produce posteriors $p_\theta(x)_{\boldsymbol{n}\cdot} \in \Delta_{K-1}$ and hard segmentations $f_\theta(x)_{\boldsymbol{n}}$ that match the unified decision-theoretic formulation.

**Separable 3D IIR convolutional variational shearlet-bank version** The separable IIR cores from Chapter 4 can also be used in place of the standard convolutional blocks in the ShearViT `MEDNet`. In such an IIR–ShearViT variant, each encoder and decoder stage applies a small stack of `IIRConv3DDFII` layers with axis–wise direct–form II recursions along the three spatial coordinates, using the same autoregressive and moving–average orders, boundary rules, and DC–normalization strategy as in the IIR blocks described earlier. The FDST analysis, ShearViT tokenization and attention over FDST wedges, band–pooled feature assembly, and the decoder structure remain unchanged, so this IIR–ShearViT design still maps an input field $x : \Omega \to \mathbb{R}^{C_{\text{in}}}$ to per–location posteriors $p_\theta(x)_{\boldsymbol{n}\cdot} \in \Delta_{K-1}$ and hard labels $f_\theta(x)_{\boldsymbol{n}}$ under the same decision–theoretic formulation as the convolutional FDST–ViT `MEDNet`.

**Example 5.12.** Multiclass 3D deterministic and convolutional variational MED segmenters with frugal attentions and MRI segmentation.

This experiment validates multiple multiclass 3D MED segmenters described in this section using brain MRI tissue segmentation. These models are constructed by varying multiresolution (DWT, FDST) filter banks, frugal attentions (WaveletViT, ShearletViT) in a deterministic or convolutional variational bottleneck. Table 5.11 lists all the segmenters tested for CSF, GM and WM segmentation. Figure 5.21 compares the loss and IoU profiles of all the tested segmenters for about 400 epochs. UNet and the convolutional variational variant `CVUNet` are treated as baseline models. All models are trained using 64x64x64-sized patched input volumes of the IBSR V2.0 dataset. The experiment revealed that `CVUNet` is a better baseline than UNet, and this is also evident from Figure 5.22.

Table 5.11: Number of free parameters and tissue segmentation IoU [0,1] scores of various newly introduced deterministic and variational multiclass 3D segmentation models and baseline models for comparison.

| # | Multiclass segmentation model (3D) | # params | GM IoU | WM IoU | CSF IoU |
|---|---|---|---|---|---|
| | **Deterministic MED segmenters** | | | | |
| 1a | `UNet3D` (deterministic baseline) | 5,648,932 | 0.88 | **0.90** | 0.62 |
| 1b | `UViTNet3D` (deterministic ViT minimal baseline) | 3,385,508 | **0.89** | **0.91** | 0.66 |
| **1c** | `UWaveViTNet3D` (WaveletViT bottleneck) | **2,823,463** | 0.87 | **0.89** | 0.64 |
| 1d | `UShearViTNet3D` (ShearViT bottleneck) | **2,793,640** | **0.89** | **0.90** | **0.68** |
| 1e | `WaveMEDNet3D` (DWT bank MED) | 3,746,019 | 0.83 | 0.87 | 0.56 |
| 1f | `WaveMEDWaveViTNet3D` (DWT bank MED) | **1,265,127** | 0.83 | 0.86 | 0.57 |
| 1g | `ShearMEDNet3D` (FDST bank MED) | **1,662,165** | 0.86 | 0.88 | 0.60 |
| | **Convolutional Variational (CV) bottleneck MED segmenters** | | | | |
| 2a | `CVUNet3D` (variational baseline) | 14,865,460 | 0.87 | 0.88 | 0.64 |
| 2b | `CVUWaveViTNet3D` (CVWaveletViT bottl.) | 3,122,727 | **0.89** | **0.91** | **0.74** |
| 2c | `CVShearMEDNet3D` (FDST bank MED) | **1,707,797** | 0.82 | 0.76 | 0.67 |
| 2d | `CVShearMEDWaveViTNet3D` (FDST bank MED) | **2,008,763** | 0.85 | 0.86 | **0.71** |

*Note.* The number of parameters measures the storage memory required to save the model. At runtime, the memory required by both training and inference depends on the size of the input tensor, its hidden representations, and the predicted output. For example, a large number of tokens in ViT models demands more GPU memory during training. State-of-the-art models 1 (a,b) and 2a are baselines, and 1 (c to g) and 2 (b to d) are our newly introduced models.

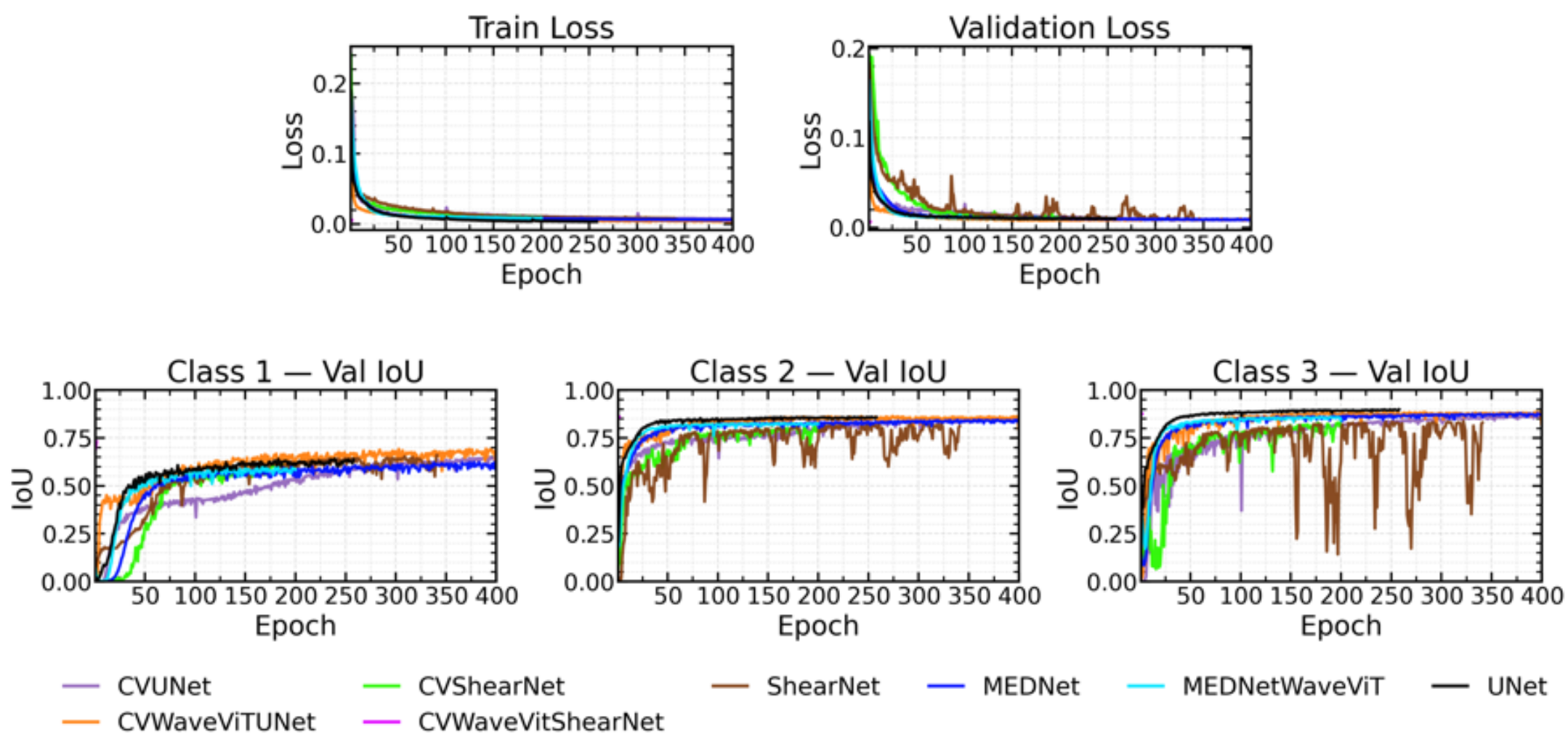

Figure 5.21: Comparison of the validation loss and IoU profiles of 3D multiclass brain MRI segmenters

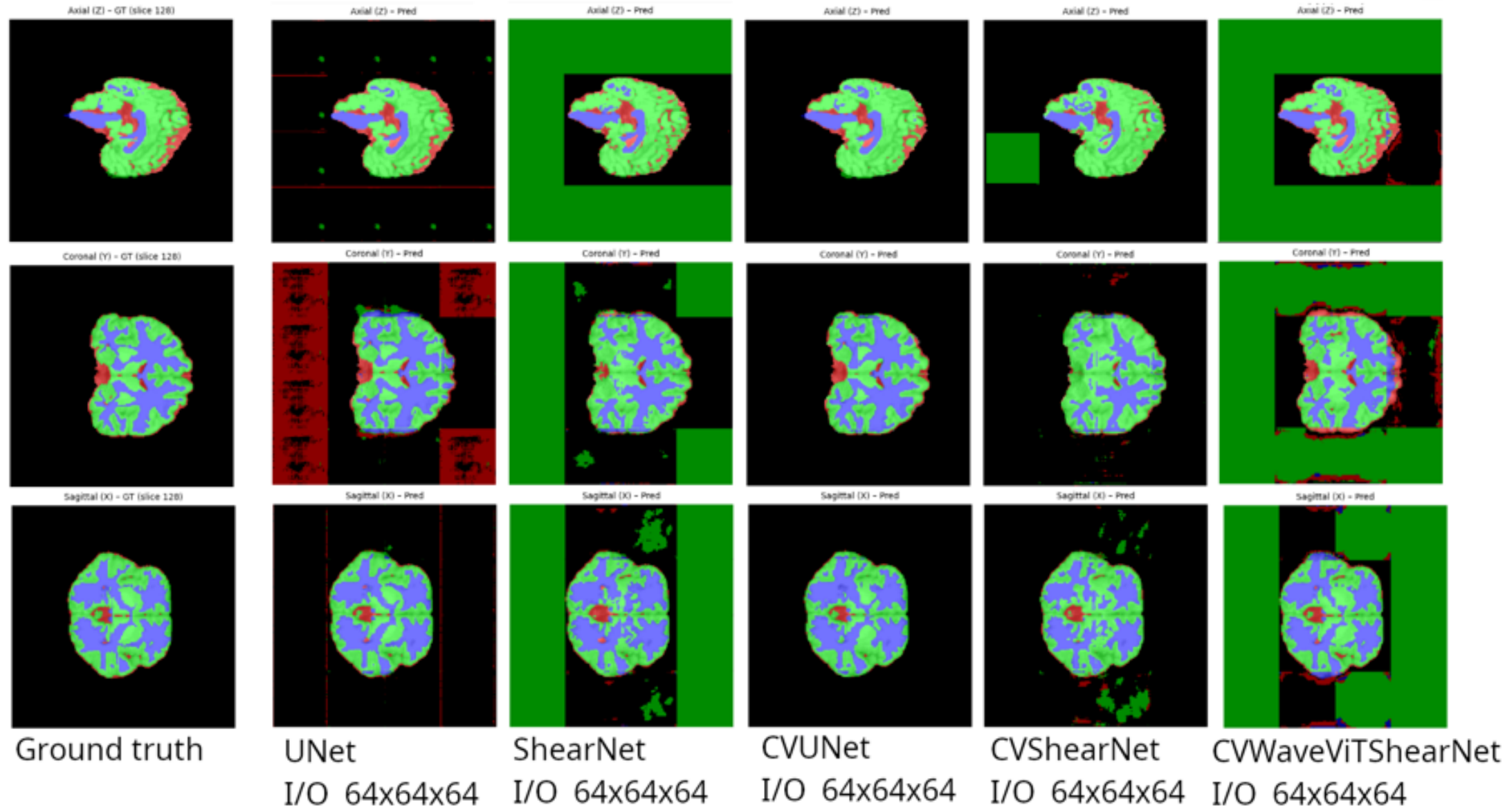

Figure 5.22: Slices from the ground truth volume (leftmost column) and slices from the predicted volumes by various multiclass segmentation models.

### 5.3.7 WaveNETR: WaveletViT encoder-decoder segmenter

The WaveletViT construction in the previous chapter updates raw coefficient bands and reconstructs feature maps by matched synthesis. WaveNETR uses this feature-mode path as the repeated wavelet-space encoder inside a full segmentation network. The architecture follows an encoder-decoder geometry similar to UNETR [213], but the implemented encoder is not a patch token transformer. The 3D model described here operates on normalized cubic inputs and forms skip features from partial wavelet synthesis taps.

For a $D$-dimensional multichannel input $x$, let $N_{\text{core}}$ denote the square or cubic side length processed by the core, let $E$ denote the WaveletViT token width, let $T$ denote the number of repeated WaveletViT blocks, let $L$ denote the number of wavelet levels, let $G = 2^D - 1$ denote the number of HP bands per level, and let $r$ denote the HP token stride. The square or cubic code path requires the core side length to be compatible with the wavelet levels and HP token stride, so $N_{\text{core}}$ is divisible by $2^L r$ in the configurations used here. After MRI intensity normalization, the wavelet tokenizer produces the initial coefficient state

$$\mathcal{Z}_0 = \mathcal{T}_{\text{DWT}}(x). \tag{5.57}$$

For $t = 0, \dots, T$, the coefficient state is a multilevel collection rather than a single dense tensor,

$$\mathcal{Z}_t = \left\{ \text{LP}_{L,t},\ \text{HP}^{(g)}_{j,t} \,:\, j = 1, \dots, L,\ g = 1, \dots, G \right\}. \tag{5.58}$$

For a cubic 3D input, the HP bands at level $j$ have spatial size $(N_{\text{core}}/2^j)^3$ and channel count $C_{\text{in}}$, while the deepest LP band has size $(N_{\text{core}}/2^L)^3$ and channel count $C_{\text{in}}$. Thus, the shorthand coefficient-state label $\{128, 64, 32\}^3 \times 1$ in Figure 5.23 refers to the three HP levels produced from a single-channel $256^3$ input with $L = 3$, together with the deepest LP band at $32^3 \times 1$.

The encoder applies $T$ WaveletViT blocks in series to this raw-band state,

$$\mathcal{Z}_t = \Psi_t(\mathcal{Z}_{t-1}), \qquad t = 1, \dots, T, \tag{5.59}$$

where $\Psi_t$ denotes the $t$th WaveletViT block described from (4.41) to (4.47b). The tokens readout in (4.50) is not used inside WaveNETR. Instead, selected coefficient states are assembled into feature tensors. If the selected zero-based block indices are $\tau_1 < \cdots < \tau_L < T$, then the default three-level, depth-12 case uses $(\tau_1, \tau_2, \tau_3) = (3, 6, 9)$. At each tap, a partial IDWT assembler returns a coefficient tensor at the corresponding synthesis level,

$$a_m = \mathcal{R}^{\text{tap}}_m\left(\mathcal{Z}_{\tau_m+1}\right), \qquad m = 1, \dots, L, \tag{5.60}$$

and a coefficient-aware convolution block maps this tensor to the skip feature $e_{m+1}$. In 3D, each partial tap tensor contains $G + 1 = 8$ coefficient groups before this convolutional adaptation. This gives the example tap shapes $128^3 \times 8$, $64^3 \times 8$, and $32^3 \times 8$ shown in Figure 5.23 for a single-channel, three-level input.

After the last WaveletViT block, a full IDWT assembler reconstructs the wavelet feature map

$$x_{\text{wave}} = \mathcal{R}^{\text{full}}(\mathcal{Z}_T) . \tag{5.61}$$

Here $\mathcal{R}_m^{\text{tap}}$ and $\mathcal{R}^{\text{full}}$ denote the partial and full IDWT assemblers used by the implementation. The first encoder feature is obtained locally as $e_1 = \text{Enc}_1(x_{\text{wave}})$, and the bottleneck is produced by strided convolutional downsampling of $x_{\text{wave}}$. Let $K = L+1$ denote the number of decoder stages and let $d_K$ denote the bottleneck feature. The decoder upsamples its current state, concatenates the matched skip feature, and applies local convolutional refinement,

$$d_{k-1} = \text{Dec}_k(d_k, e_k), \qquad k = K, \ldots, 1. \tag{5.62}$$

A final pointwise output head produces

$$\widehat{y}[\boldsymbol{n}] = \mathbf{O}_{\text{out}}\, d_0[\boldsymbol{n}] + \mathbf{b}_{\text{out}} \in \mathbb{R}^{C_{\text{out}}}, \tag{5.63}$$

where $\mathbf{O}_{\text{out}}$ and $\mathbf{b}_{\text{out}}$ denote the learned weight matrix and bias vector of the pointwise output head, $C_{\text{out}}$ is the number of output classes, and the task-dependent activation is applied at the head.

In generative mode, the last WaveletViT block also returns the compressed HP token grids and the deepest LP token grid. A band decoder maps these latents back to wavelet coefficient bands, and a matched IDWT assembler reconstructs $\widehat{x}$. A KL term is added only when the configured `kl_scale` is positive, so deterministic segmentation and reconstruction-enabled operation remain distinct model choices.

For comparison, we also introduce the economical UNet variants `UWaveletViTNet2D` and `UWaveletViTNet3D`, each with a single WaveletViT module at the UNet bottleneck. WaveNETR instead uses repeated WaveletViT blocks as the main wavelet-space encoder and decodes from coefficient-derived taps together with the final synthesized wavelet feature map. Figure 5.23 summarizes this structure. In this work, the volumetric model instantiates this construction with $D = 3$ and uses 3D operators throughout. A corresponding planar instantiation with $D = 2$ is obtained by replacing each 3D convolution, transposed convolution, and wavelet transform with its 2D counterpart while retaining the same wavelet tokenizer, repeated WaveletViT encoder, tap encoders, and decoder structure.

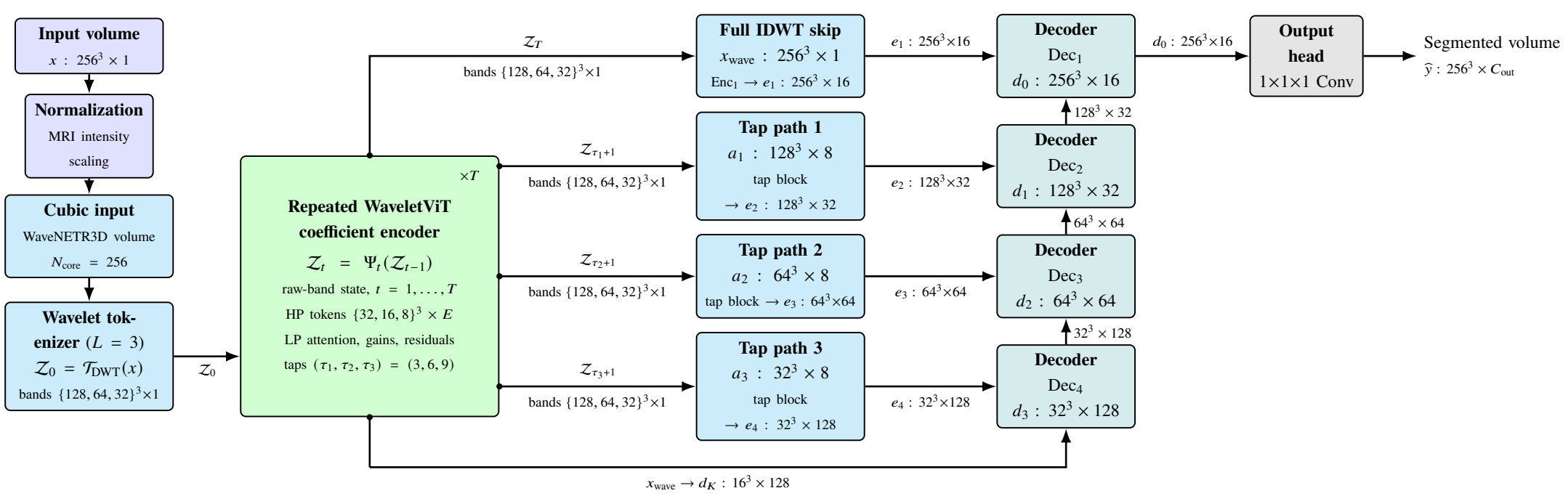

Figure 5.23: Overview of `WaveNETR3D` ($L = 3$) architecture. Example tensor sizes are shown for a single-channel cubic $256^3$ input, token width $E = 128$, HP token stride $r = 4$, and channel layout $(16, 32, 64, 128)$. The normalized input is tokenized into raw multilevel wavelet coefficient bands. The coefficient-state labels summarize HP bands at $128^3$, $64^3$, and $32^3$ with one channel per band, together with the deepest LP band at $32^3 \times 1$. A stack of $T$ WaveletViT blocks updates $\mathcal{Z}_t$. Selected states are converted to skip features by partial IDWT tap assemblers and coefficient tap blocks. The final state is synthesized to $x_{\text{wave}}$, which supplies the first encoder feature and the strided bottleneck path for the convolutional decoder. A pointwise output head produces the segmentation logits. In generative mode, compressed HP and LP latents from the last block are decoded by a band decoder and matched IDWT to reconstruct $\widehat{x}$.

**Example 5.13.** Image segmentation using `WaveNETR2D`

**Semantic segmentation datasets** We evaluate the 2D models on digital retinal images for vessel extraction (DRIVE) [288], the high-resolution fundus (HRF) image database [289], the marine semantic segmentation training dataset (MaSTr1325) [290], semantic segmentation of underwater imagery (SUIM) [291], and Cambridge-driving Labeled Video Database (CamVid) [292]. DRIVE and HRF are treated as two-class problems, whereas MaSTr1325, SUIM, and CamVid are multiclass.

**Training protocol for image segmentation** Figure 5.24 shows representative `WaveNETR2D` segmentations. All 2D segmentation models are trained end to end in a unified TensorFlow and Keras pipeline using Adam with an initial learning rate of $10^{-4}$. The batch size is chosen per dataset and the learning rate is reduced on a validation plateau. We use light paired augmentation during training. The geometric transforms are applied jointly to the image and target map. Mild photometric perturbations are added when appropriate. For DRIVE and HRF, we use a binary focal cross-entropy plus Dice loss and report the foreground IoU. For MaSTr1325, SUIM, and CamVid, we use categorical focal cross-entropy plus Dice and report the mean class IoU. All inputs are resized with aspect ratio preserved. For `WaveNETR2D`, the inputs are then padded to square resolutions so that the wavelet token hierarchy remains well defined. The padded target regions are excluded from both the loss and the IoU computation. `UNet2D` and `UNETR2D` follow the same dataset splits and the same evaluation protocol. Table 5.12 summarizes the

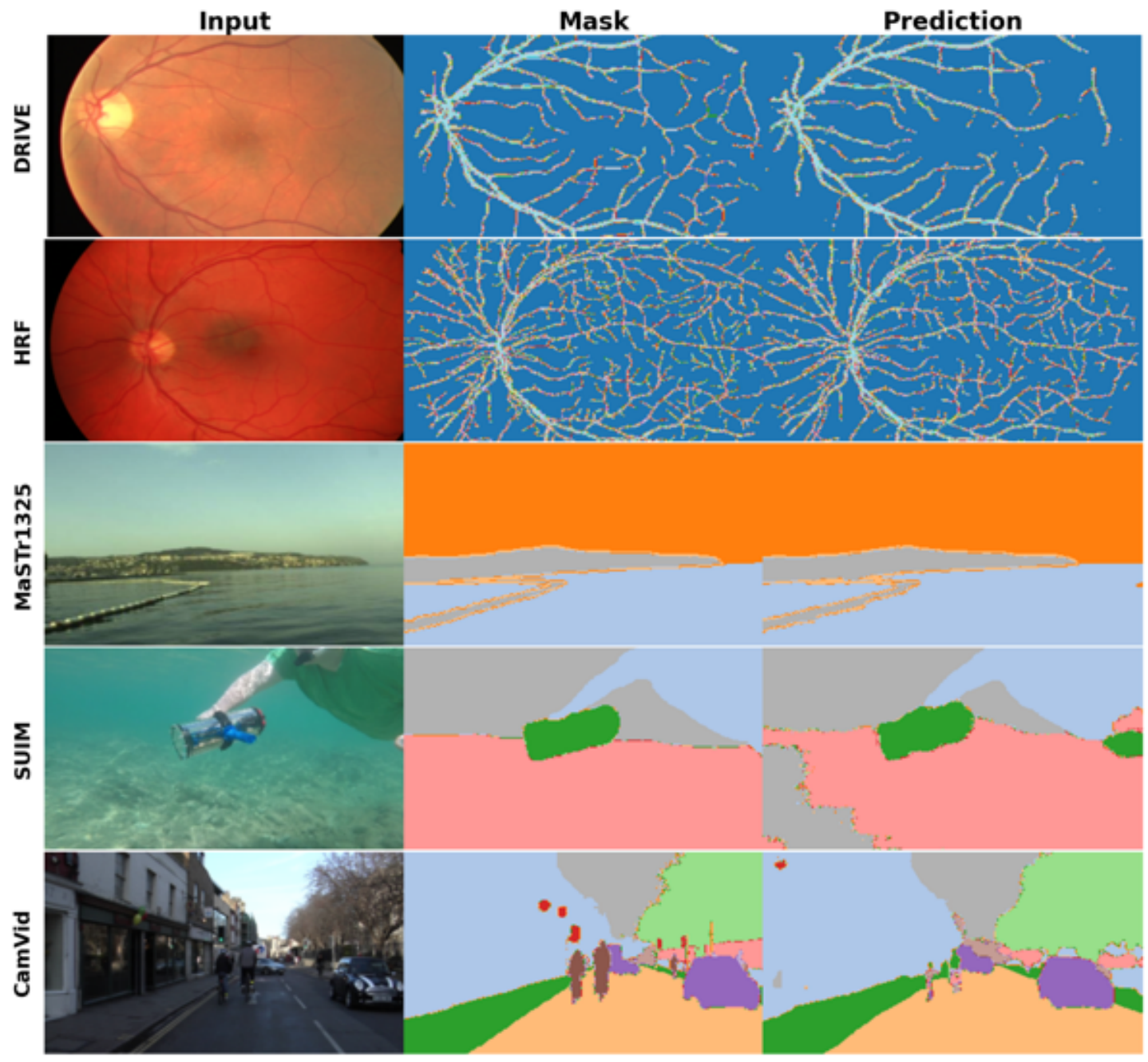

Figure 5.24: Qualitative segmentation results of `WaveNETR2D` (bior1.3) on five benchmark datasets, DRIVE, HRF, MaSTr1325, SUIM, and CamVid. Each row shows a representative sample, with the input image in the left column, the ground-truth mask in the middle, and the model prediction in the right column

corresponding validation IoU scores.

Table 5.12: Mean intersection over union (IoU) scores (%) of semantic segmentation across datasets for `WaveNETR2D`, `UNETR2D`, and baseline 2D models.

| Method | DRIVE (2) | HRF (2) | MaSTr (4) | SUIM (8) | CamVid (12) |
|---|---|---|---|---|---|
| `WaveNETR2D` | | | | | |
| a) haar variant | 50.19 | 54.38 | 82.47 | 35.77 | 36.36 |
| b) db2 variant | 53.07 | 57.68 | 76.96 | 39.15 | 31.91 |
| `UNETR2D` | 54.2 | 62.2 | 83.2 | 51.2 | 48.7 |
| `UNet2D` | 56.6 | 66.0 | 86.2 | 55.7 | 54.7 |

*Note.* For DRIVE and HRF we use the binary foreground IoU under the collapsed two-class setup. For MaSTr, SUIM, and CamVid we report the mean of the per-class validation IoUs.

**Example 5.14.** `WaveNETR3D` training with MRI volumes

Here, we first describe the supervised training of `WaveNETR3D` models for skull stripping, brain tissue segmentation, and lesion segmentation, and then describes deployment of the trained models in `MRILong` for longitudinal analysis of the unlabeled longitudinal T1-weighted MRI cohort. The goal is to obtain consistent skull, tissue, and lesion labels together with volumetric summaries across follow-up scans. The Neurofeedback Skull-stripped (NFBS) repository [293] is used for skull stripping, Internet Brain Segmentation Repository (IBSR) V2.0 [271] is used for tissue segmentation, and Anatomical Tracing of Lesions After Stroke (ATLAS) R2.0 [272] is used for lesion segmentation. The unlabeled longitudinal T1-weighted MRI cohort is used for inference and volumetric follow-up and was acquired at a collaborating clinical center under approval from the relevant Institutional Ethics Committee for a study of ischemic-stroke patients undergoing yogic intervention, with written informed consent obtained from participants in accordance with the approved protocol. The tissue stage predicts grey matter (GM), white matter (WM), and cerebrospinal fluid (CSF).

For the deterministic `WaveNETR3D` runs reported in Table 5.13, training uses the Adam optimizer [37] with learning rate $10^{-4}$ and focal loss. NFBS and IBSR use uniform focal weights, whereas ATLAS uses weights estimated from training batches. The reported `WaveNETR3D` family follows the WaveNETR architecture in Section 5.3.7 with four-stage channel layout $(16, 32, 64, 128)$, WaveletViT token width 128, HP token stride 4, encoder depth 12, four attention heads, key dimension 16, residual strength 0.25, level decay 1.0, maximum deepest lowpass token count 8192, orientation mixing, deepest lowpass global attention, tokenizer normalization disabled for single-channel MRI, and prebuilt wavelet filter banks. The wavelet family varies by table row. Tissue segmentation on IBSR is trained on preprocessed $64^3$ inputs with one wavelet decomposition level, whereas skull stripping on NFBS and lesion segmentation on ATLAS use cubic $256^3$ inputs with three levels. The reported deterministic `WaveNETR3D` runs use batch size 1 and 200 epochs. Table 5.13 reports these models together with the lighter `WaveNETR3D-eco` variants, `UWaveletViTNet3D`, and baseline comparators.

All dataset splits use `sklearn` with `random_state=42`. IBSR V2.0 (18 volumes) and NFBS (125 volumes) each reserve 5% for testing and 5% of the remainder for validation, yielding roughly 16/1/1 and 113/6/6 train/val/test respectively. ATLAS R2.0 (655 pairs) reserves 10% for testing and 10% of the remainder for validation, yielding approximately 530/59/66. During training, tissue segmentation uses preprocessed $64^3$ patches, whereas skull stripping and lesion segmentation use full $256^3$ volumes.

To examine whether a trained WaveNETR model uses the wavelet representation in a boundary-sensitive manner, we compute a boundary-conditioned DWT subband selectivity summary on the IBSR validation volume. For this analysis, a trained $L = 3$ `WaveNETR3D` checkpoint is evaluated over boundary-containing, non-overlapping $64^3$ patches from the padded $256^3$ validation volume. Ground-truth label boundaries are pooled to the coefficient grids, and the boundary energy share after WaveletViT modulation is compared with the corresponding raw

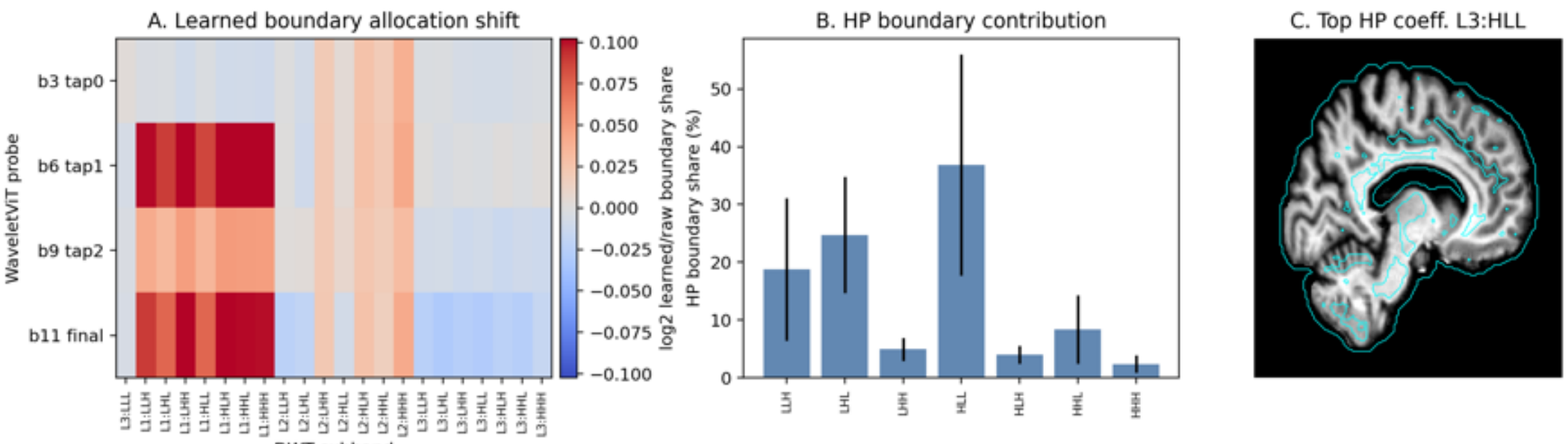

Figure 5.25: Learned boundary selectivity of a trained $L = 3$ `WaveNETR3D` checkpoint on an IBSR validation volume. Panel A shows the log-ratio between the learned and raw boundary energy share for each DWT subband at selected WaveletViT probes. Red entries indicate subbands whose relative contribution to boundary energy is increased by the trained coefficient modulation, while blue entries indicate relative suppression. Panel B summarizes the mean learned high-pass boundary contribution by 3D high-pass orientation group, with black vertical lines showing variability across patch-probe samples. Panel C overlays the spatial energy of the strongest learned high-pass coefficient band on an axial slice, with the ground-truth label boundary shown in cyan.

DWT boundary energy share. This gives a model-dependent view of how the learned wavelet-space blocks redistribute coefficient energy near anatomical tissue boundaries. Figure 5.25 shows the result for the tap states after WaveletViT blocks 3, 6, and 9, together with the final block. In this validation summary, the trained model promotes selected high-pass subbands near boundaries, with the largest average high-pass contribution in the HLL group for this checkpoint.

### 5.3.8 ShearNETR: ShearViT encoder-decoder segmenter

The ShearViT block described in the previous chapter maps a feature volume to updated spatial features through FDST analysis, masked inter-band attention, and inverse FDST synthesis. We now use the same feature-return pathway inside a full volumetric segmentation network, called ShearNETR. The current implementation is a hierarchical encoder-decoder. It first projects the input, after optional dynamic-range normalization, to a common hidden width and then applies residual ShearViT processing at multiple spatial scales.

For a 3D input volume $x \in \mathbb{R}^{B\times N_1\times N_2\times N_3\times C_{\text{in}}}$, let $S$ denote the number of encoder stages, let $\mathbf{c} = (c_0, \dots, c_{S-1})$ denote the decoder channel layout, let $E$ denote the common ShearViT hidden width, and let $\boldsymbol{\beta} = (\beta_0, \dots, \beta_{S-1})$ denote the number of residual ShearViT blocks per stage. After optional dynamic-range normalization, a pointwise channel projection gives

$$a_0 \in \mathbb{R}^{B\times N_1\times N_2\times N_3\times E}. \tag{5.64}$$

Stages after the first begin with a stride-2 convolution, so stage $m$ operates on a grid reduced by $2^m$ along each spatial axis. In the default volumetric model, $S = 4$, $\boldsymbol{\beta} = (1, 3, 4, 4)$, $E = 128$, and $\mathbf{c} = (16, 32, 64, 128)$.

Within each stage, a residual ShearViT block first maps its input to width $E$ if needed, then applies the feature-return ShearViT mapping built from the tokenization and masked-attention path in (4.57)–(4.62), followed by tile scaling, one-shot synthesis, and fusion in (4.65)–(4.67), with optional refinement by (4.68). Let $s_m$ be the input to stage $m$, after the stride-2 convolution when $m > 0$. If $\Pi_{m,q}$ is the identity or learned $1 \times 1 \times 1$ projection for the $q$th residual block in stage $m$, and $\mathcal{V}_{m,q}$ is the corresponding ShearViT feature block, then

$$h_{m,q} = \Pi_{m,q}(h_{m,q-1}) + \mathcal{V}_{m,q}\big(\Pi_{m,q}(h_{m,q-1})\big), \qquad q = 1, \dots, \beta_m, \tag{5.65}$$

with $h_{m,0} = s_m$. We write $h_m = h_{m,\beta_m}$ for the output of stage $m$. A local two-layer ConvBlock converts each $h_m$ into a skip tensor

$$e_m = \text{Enc}_m(h_m), \qquad e_m \in \mathbb{R}^{B\times(N_1/2^m)\times(N_2/2^m)\times(N_3/2^m)\times c_m}. \tag{5.66}$$

Each residual block uses the FDST shell count $J$ and first-shell directional count $L_0$ introduced in (2.50). The experiments reported here use $J = 2$ and $L_0 = 4$. Thus the two HP shells use $L_0 = 4$ and $L_1 = 8$ in-cone directions per cone, giving $K = 1 + 3(4 + 8) = 37$ FDST bands including the LP band.

After the last encoder stage, a bottleneck block applies one further stride-2 convolution followed by local convolutional refinement,

$$z_{\text{bn}} = \mathcal{B}(h_{S-1}). \tag{5.67}$$

The decoder then combines $z_{\text{bn}}$ with the stage skips through upsample-concatenate-convolve blocks. For $m = S - 1, \dots, 0$,

$$d_{S-1} = \text{Dec}_{S-1}(z_{\text{bn}}, e_{S-1}), \tag{5.68a}$$

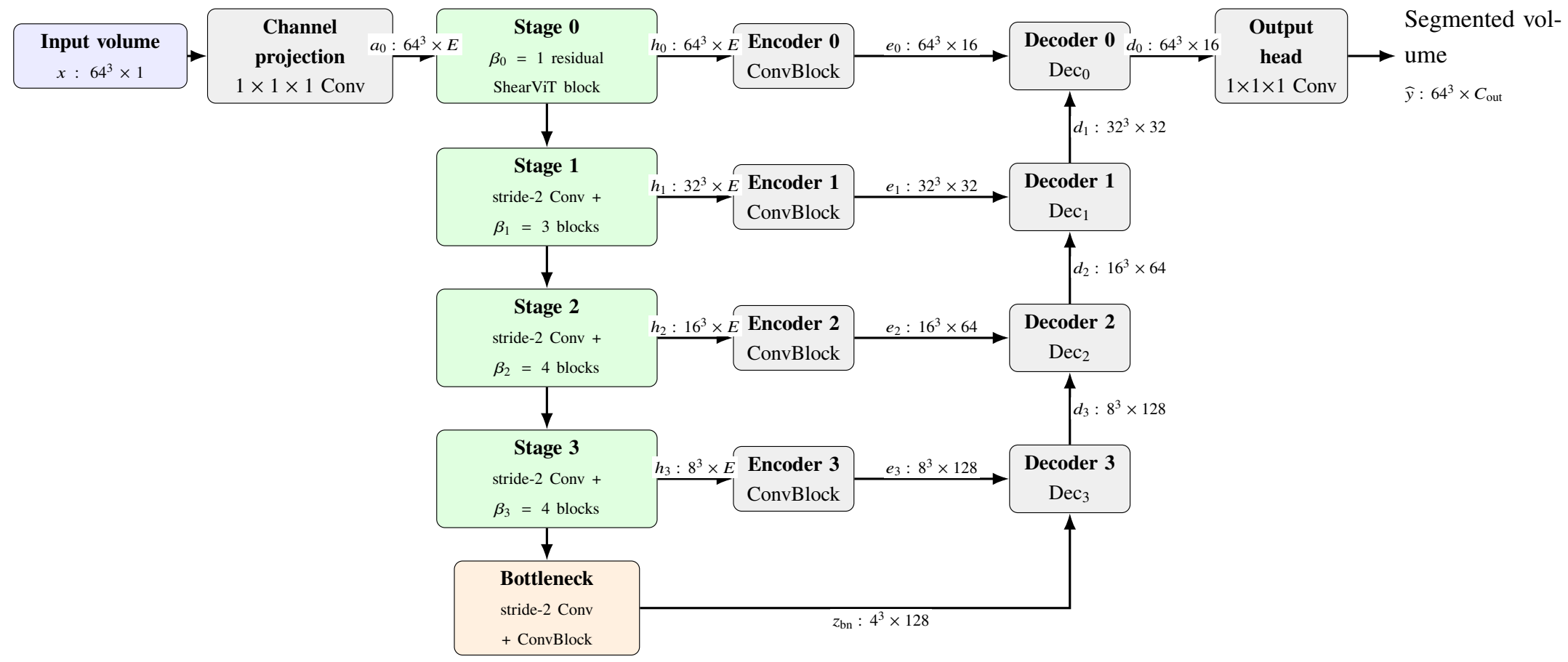

Figure 5.26: Overview of the implemented `ShearNETR3D` architecture. Example tensor sizes are shown for a single-channel $64^3$ input, hidden width $E = 128$, stage block counts $\boldsymbol{\beta} = (1, 3, 4, 4)$, and channel layout $(16, 32, 64, 128)$. Each ShearViT block uses a two-shell FDST configuration with $J = 2$ and parabolic directional counts $L_j = 2^j L_0$, $L_0 = 4$, giving 4 and 8 in-cone directions per cone and $K = 37$ FDST bands including the LP band. When enabled, dynamic-range normalization is applied as preprocessing before the channel projection. The input is projected to width $E$ and processed by a hierarchical residual ShearViT encoder. Each stage applies feature-return ShearViT blocks at its current grid size, and stages after the first begin with a stride-2 convolution. Stage outputs are converted to skip tensors $e_0, \ldots, e_3$ by local ConvBlocks. A strided bottleneck produces $z_{\text{bn}}$, and four upsample-concatenate-convolve decoder blocks produce $d_3, \ldots, d_0$, followed by a pointwise output head.

$$d_m = \mathrm{Dec}_m(d_{m+1}, e_m), \qquad m = S-2, \ldots, 0. \tag{5.68b}$$

A final pointwise head produces

$$\widehat{y}[\boldsymbol{n}] = \mathbf{W}_{\text{out}}\, d_0[\boldsymbol{n}] + \mathbf{b}_{\text{out}} \in \mathbb{R}^{C_{\text{out}}}, \tag{5.69}$$

where $\mathbf{W}_{\text{out}}$ and $\mathbf{b}_{\text{out}}$ denote the learned weight matrix and bias vector of the output head, $C_{\text{out}}$ is the number of output classes, and the task-dependent activation is applied at the head. For comparison, `UShearViTNet3D` places a single ShearViT bottleneck inside a UNet, whereas ShearNETR uses repeated residual ShearViT processing across the encoder hierarchy. Figure 5.26 summarizes this structure.

When the KL coefficient $\lambda_{\text{KL}}$ is positive, the ShearViT blocks add KL-regularized token sampling within the same encoder pathway. This gives the `CVShearNETR3D` variant used in the experiments and does not change the segmentation decoder equations above. We use CV here to denote a variational segmentation model, not an unconditional sample generator.

**Example 5.15.** Volumetric MRI segmentation results using ShearNETR

We evaluate the proposed shearlet models on three volumetric MRI tasks that stress complementary aspects of 3D segmentation. The Neurofeedback Skull-stripped (NFBS) repository [293] is used for binary brain extraction. Internet Brain Segmentation Repository (IBSR) V2.0 [271] is used for tissue segmentation with gray matter (GM), white matter (WM), and cerebrospinal fluid (CSF) labels. Anatomical Tracing of Lesions After Stroke (ATLAS) R2.0 [272] is used for binary lesion segmentation. This combination tests coarse foreground extraction, fine tissue discrimination, and sparse focal abnormality detection within a unified volumetric pipeline. All data are prepared through a common preprocessing and loading pipeline. NFBS and ATLAS volumes are reoriented to canonical RAS orientation, intensity-normalized by robust min-max scaling, padded to $256^3$, and streamed from disk during training. IBSR follows the existing stripped-volume workflow with fixed train, validation, and test splits generated using `random_state=42`, dynamic-range adjustment of the input T1 volumes, dyadic zero padding, and one-hot encoded tissue labels. The current split policy uses 5% test and 5% validation for IBSR and NFBS, and 10% test and 10% validation for ATLAS. Training is patch-based for IBSR and streamed on padded full volumes for NFBS and ATLAS. The current ShearNETR jobs use the same stage-wise ShearViT settings across all three tasks, with hidden width $E = 128$, four attention heads, $\boldsymbol{\beta} = (1, 3, 4, 4)$ for a total of 12 residual ShearViT blocks, transform depth $J = 2$, first-shell directional count $L_0 = 4$, zero dropout, and no feature mixer. They are trained for 200 epochs with focal loss, Adam at learning rate $10^{-4}$, and batch size 1, using $64^3$ input patches for IBSR and $256^3$ streamed volumes for NFBS and ATLAS. Across models we use the same four-stage channel configuration $(16, 32, 64, 128)$ and a shared training pipeline, with task-dependent class weighting.

To probe whether the trained ShearNETR uses the shearlet representation in a way that is consistent with its multiscale directional construction, we compute a boundary-conditioned band selectivity summary on the IBSR validation volume. For the patch-trained model, the analysis is evaluated over boundary-containing, non-overlapping $64^3$ patches from the padded $256^3$ validation volume. For each probed encoder stage and shearlet band, we compare the learned boundary energy share after ShearViT modulation with the corresponding raw shearlet boundary energy share. This gives a model-dependent measure of which directional bands are promoted near anatomical label boundaries. Figure 5.27 shows this summary for the last ShearViT block of each encoder stage. In this validation summary, most of the learned high-pass boundary contribution is concentrated in selected first-shell directional groups, especially $j0 : c0$ and $j0 : c2$, while second-shell groups carry a smaller share.

The deterministic comparison set includes `UNet3D`, the vanilla ViT bottleneck model `UViTNet3D`, the shearlet bottleneck model `UShearViTNet3D`, and the proposed encoder-decoder `ShearNETR3D`. Variational results are reported when the corresponding runs are available, with `CVUNet3D` serving as the main convolutional reference. Table 5.13 reports the number of trainable parameters together with task-specific IoU values. Brain IoU refers to NFBS skull stripping, GM, WM, and CSF IoU refer to IBSR tissue segmentation, and lesion IoU refers to

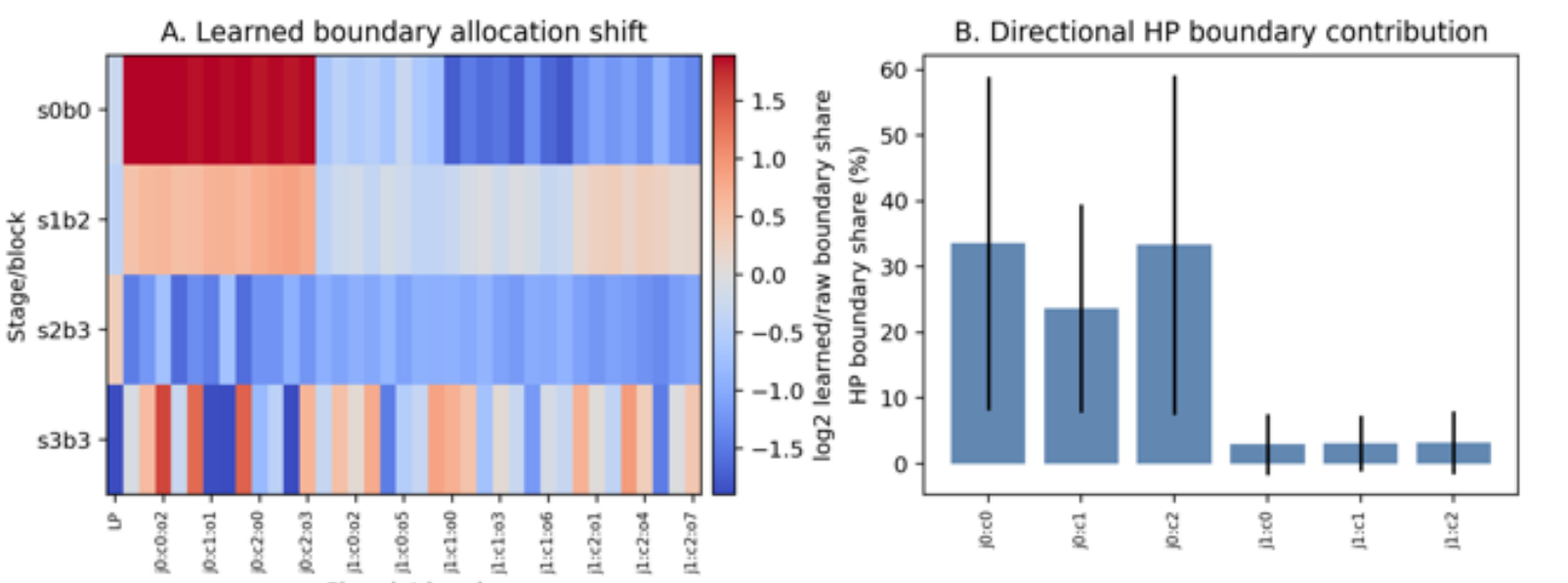

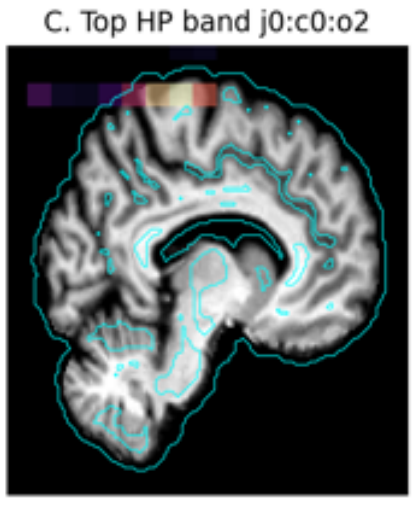

Figure 5.27: Learned boundary selectivity of the trained `ShearNETR3D` model on an IBSR validation volume. Panel A shows the log-ratio between the learned and raw boundary energy share for each shearlet band at the last ShearViT block of each encoder stage. Red entries indicate bands whose contribution to boundary energy is increased by the trained ShearViT modulation, while blue entries indicate relative suppression. Panel B summarizes the mean learned high-pass boundary contribution by scale-cone group, with black vertical lines showing variability across patch-stage samples. Panel C is an illustrative overlay of the spatial energy of the strongest learned high-pass band on an axial slice, with the ground-truth label boundary shown in cyan.

ATLAS lesion segmentation. The table therefore gives a compact view of model size and task behavior across three complementary MRI segmentation settings.

Table 5.13: Number of free parameters and IoU [0,1] scores of skull stripping, brain tissue, and lesion segmentation by deterministic and variational 3D segmentation models, including multiresolution encoder-decoder (MED) DWT filter-bank variants and baseline controls.

| # | Segmenter (3D) | # params | Brain IoU | GM IoU | WM IoU | CSF IoU | Lesion IoU |
|---|---|---|---|---|---|---|---|
| | **Deterministic segmenters** | | | | | | |
| 1 | `WaveNETR3D-db2` | 6,325,568 | 0.96 | 0.85 | 0.89 | 0.66 | 0.57 |
| 2 | `WaveNETR3D-db2-eco` | 4,723,496 | 0.97 | 0.85 | 0.88 | 0.67 | 0.56 |
| 3 | `ShearNETR3D` | 4,523,276 | 0.98 | 0.86 | 0.90 | 0.71 | 0.60 |
| 4 | `UNETR3D` | 5,550,548 | 0.97 | 0.88 | 0.90 | 0.64 | 0.57 |
| 5 | `UNet3D` | 5,648,932 | 0.98 | 0.88 | 0.90 | 0.62 | 0.56 |

*Note.* Each row contains three models (skull stripping, brain tissue, and lesion segmentation).

## 5.4 Longitudinal analysis of T1 brain MRI (unlabeled)

The longitudinal pipeline builds on three (skull stripping, GM/WM/CSF and stroke lesion) volumetric segmentation models that are trained on 80 GB A100 GPUs and here used for inference with CPUs. Earlier in this chapter, several three dimensional segmentation models for brain masks, brain tissues, and stroke lesions are described. These segmenters are trained on labeled T1 weighted brain MRI volumes using a supervised objective that combines voxelwise cross entropy and soft Dice loss with standard regularisation. After training and validation, the weights are frozen and used inside the `MRILong` inference workflows described below. The current pipeline performs skull stripping, brain tissue segmentation, and lesion segmentation. The lesion stage is run as a separate inference step and the outputs are merged later for reporting.

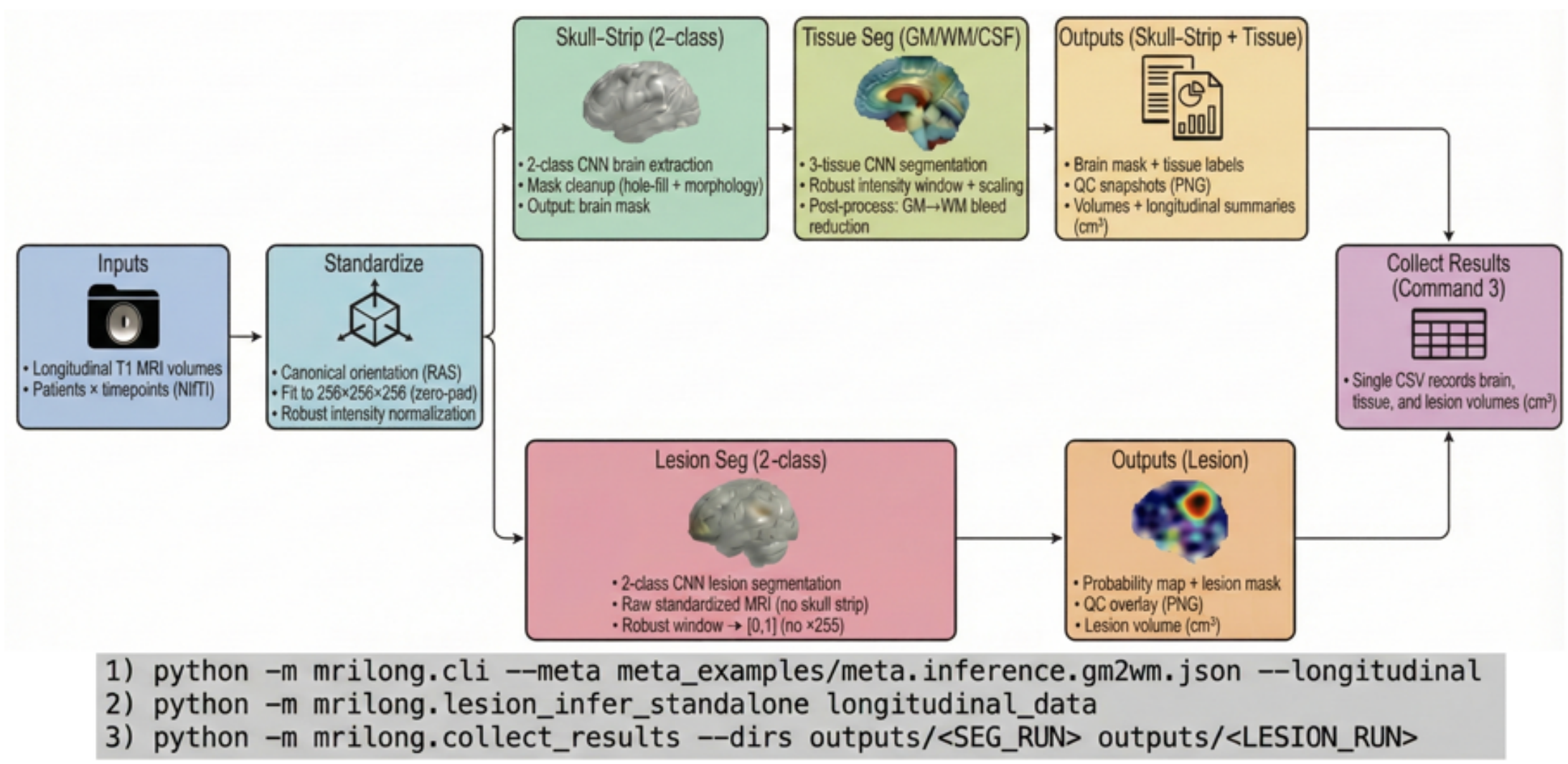

Figure 5.28: Inference pipeline for longitudinal T1 MRI. Each timepoint is standardized in geometry and intensity and then processed by skull stripping and tissue segmentation. Lesion segmentation is applied on the raw standardized volume in a separate stage. The pipeline writes NIfTI outputs, quality control images, volumetric measures, and longitudinal summaries. This pipeline is published as `MRILong` Python package.

**Data model and geometry:** Figure 5.28 summarizes the `MRILong` inference pipeline. Let $\Omega \subset \mathbb{Z}^3$ be the voxel grid and let $x_{s,t} \in \mathbb{R}^\Omega$ denote the T1 weighted MRI volume of patient $s$ at timepoint $t$. Each volume is first handled in its native acquisition space. When needed, a simple manual crop removes large empty margins, with dataset specific defaults for the AIIMS T1 BrainMRI dataset. The cropped volume is then reoriented to the standard Right Anterior Superior convention and written in a canonical orientation.

If registration is enabled, all timepoints of a patient are rigidly aligned to a reference volume. For AIIMS-BrainMRI data, if a template from the IBSR V2.0 dataset is available, the pipeline attempts to align each volume to this template. Otherwise, or when no template is available, the baseline scan of the patient is used as the reference for registration. All geometric operations, including cropping, orientation changes, and registration choices, are recorded in a provenance log.

### 5.4.1 Stage 1 binary segmentation (3D)

**Pre skull strip intensity preparation:** Before skull stripping, the pipeline applies a dataset aware intensity normalization.

- **A labeled test MRI volume to validate the skull stripping pipeline** The volume is clipped to a chosen percentile range, typically $[0.5, 99.5]$. Optionally, the intensities are standardized to zero mean and unit variance on a per volume basis. Depending on the configuration, either CLAHE, adaptive histogram equalization, or histogram matching to a user specified reference volume is applied.
- **Preparing AIIMS BrainMRI unlabeled data for skull stripping** The volume first undergoes N4 bias field correction. A head region of interest is then emphasized by a simple windowing operation or by histogram matching to a reference cohort such as the NFBS dataset or IBSR V2.0. A robust window such as the $[1, 99]$ percentile range is finally applied to stabilize the dynamic range.

These operations are controlled by a configuration file and are chosen so that the downstream skull stripping model sees intensities that are comparable across timepoints within a patient and across patients.

**Skull stripping** The first stage predicts a brain mask. A three dimensional UNet with a WaveletViT bottleneck (`UWaveViTNet3D`) is used as a binary segmenter with two classes, brain and non brain. The model operates on a $256^3$ grid and is run in a single pass when memory permits.

The raw mask is refined by simple morphological operations. The largest connected component is kept to remove isolated false positives. Small holes in the interior are filled. Mild closing and distance limited gap filling remove thin breaks in the mask. A head region of interest constraint with a controlled growth margin prevents the mask from expanding outside the plausible head region. The final skull stripped mask is used for volumetric measurements and as input to the tissue segmentation stage.

### 5.4.2 Stage 2 multiclass segmentation (3D)

**Segmentation preparation:** After skull stripping an unlabeled MRI volume from the AIIMS BrainMRI dataset, the extracted brain region is prepared for tissue segmentation. Figures 5.29 and 5.30 show representative intermediate outputs from stage 1 and stage 2 of the pipeline.

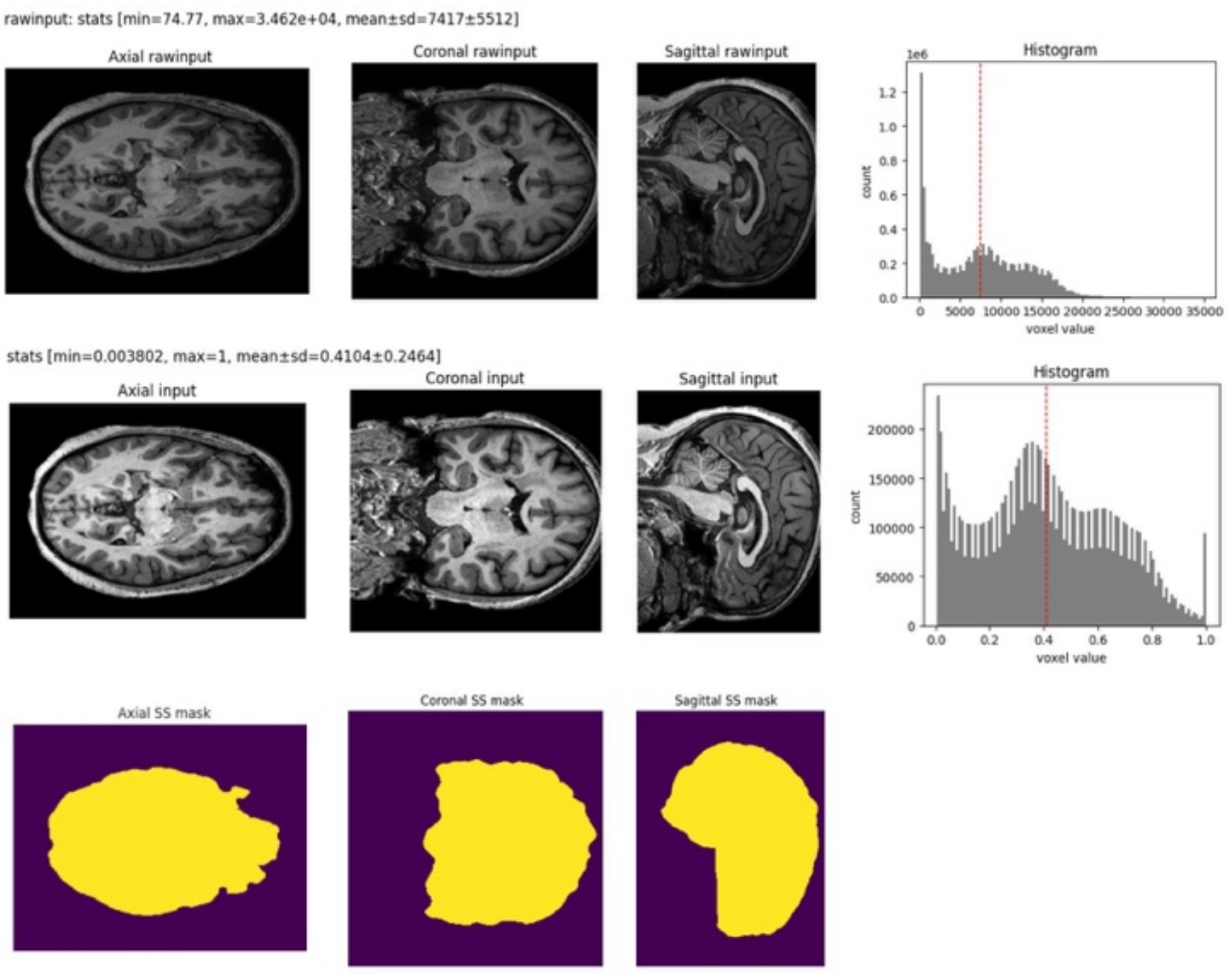

Figure 5.29: Images from one timepoint MRI volume during stage 1 of the longitudinal analysis. MRI slices from a raw input volume (top row), preprocessed volume (middle row) and histogram of those volumes. Corresponding brain mask slices from the binary segmenter output volume (bottom row).

- **Preparing AIIMS BrainMRI unlabeled data for CSF, GM, WM segmentation** The skull stripped brain is taken as the starting point. N4 bias field correction is applied again and it is restricted to the brain mask. The intensities inside the mask are scaled to the range [0, 1] using robust brain percentiles, and the background outside the mask is set to zero.
- **Generic data.** Optional masked CLAHE or masked histogram matching is applied. For histogram matching, the first timepoint or a user supplied reference volume is used as the template.

For efficiency, the brain bounding box is extracted and zero padded so that its size is a multiple of the network patch size. After segmentation, the predicted labels are mapped back to the full grid coordinates.

**CSF, GM and WM segmentation** The second stage segments the skull stripped brain into tissue classes. A `UWaveViTNet3D` model with four output channels predicts voxelwise labels for background, cerebrospinal fluid (CSF), grey matter (GM), and white matter (WM). Sliding window inference with overlap and blending is used when the full volume does not fit into memory.

In practice, the multiclass segmenter can assign tissue labels to a small number of voxels outside the brain mask, which inflates tissue volumes and reduces interpretability. The pipeline therefore restricts all tissue labels to the skull stripped mask. It also applies a light post processing correction that reduces grey matter bleed into white matter in the deep brain. This correction uses erosion, majority voting, and an intensity based check. It is designed to preserve thin grey matter ribbons while removing spurious grey matter islands inside the white matter core.

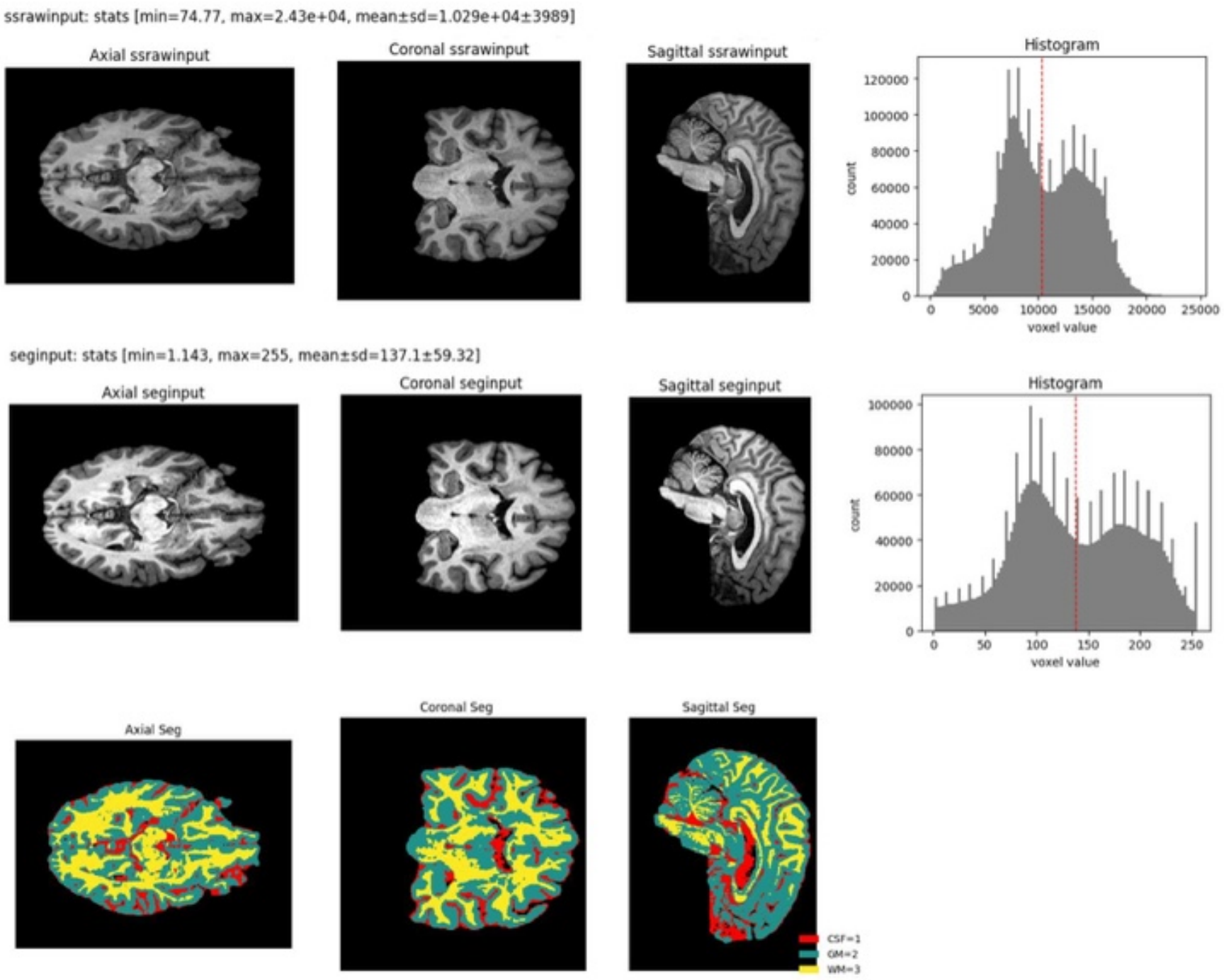

Figure 5.30: Images from one timepoint MRI volume during stage 2 of the longitudinal analysis. MRI slices from a skull stripped volume (top row) from stage 1, preprocessed volume (middle row) and histogram of those volumes. Corresponding slices from the multiclass tissue segmenter output volume (bottom row).

Formally, for each voxel index $\boldsymbol{n} \in \Omega$ and timepoint $(s, t)$ the model outputs

$$p_\theta(x_{s,t})_{\boldsymbol{n}\cdot} = \big(p_\theta(x_{s,t})_{\boldsymbol{n}0}, p_\theta(x_{s,t})_{\boldsymbol{n}1}, p_\theta(x_{s,t})_{\boldsymbol{n}2}, p_\theta(x_{s,t})_{\boldsymbol{n}3}\big) \in \Delta_3, \tag{5.70}$$

where $k = 0$ denotes background, $k = 1$ CSF, $k = 2$ GM, and $k = 3$ WM. The hard segmentation is obtained pointwise as

$$f_\theta(x_{s,t})_{\boldsymbol{n}} = \arg\max_{0\le k\le 3} p_\theta(x_{s,t})_{\boldsymbol{n}k}, \qquad \boldsymbol{n} \in \Omega, \tag{5.71}$$

which matches the decision theoretic view of segmentation in Section 5.3. Inference is run on the CPU to favor deterministic behavior and to simplify deployment on standard workstations. For each timepoint, the pipeline writes the skull stripped mask, the tissue segmentation, and slice wise overlays in Neuroimaging Informatics Technology Initiative (NIfTI) and PNG formats for visual quality control.

**Per timepoint volumetry:** Let $f_\theta(x_{s,t})_{\boldsymbol{n}} \in \{0, 1, 2, 3\}$ be the predicted label at voxel $\boldsymbol{n} \in \Omega$ and let $\delta v$ denote the voxel volume in mm$^3$ or cm$^3$. The volume of tissue class $k$ (CSF, GM, or

WM) for patient $s$ at timepoint $t$ is

$$V_{s,t}^{(k)} = \delta v \sum_{n \in \Omega} \mathbf{1}\{f_\theta(x_{s,t})_n = k\}, \qquad k \in \{\text{CSF}, \text{GM}, \text{WM}\}, \tag{5.72}$$

where $\mathbf{1}\{\cdot\}$ is the indicator function. The intracranial volume (ICV) is computed from the skull stripped mask by summing the voxel volumes inside the mask and is reported together with the tissue volumes. The NIfTI outputs and PNG overlays provide timepoint level quality control for both the skull stripping and tissue segmentation stages.

### 5.4.3 Stage 3 lesion segmentation (3D)

**Lesion intensity preparation:** The lesion model is applied to the raw standardized MRI volume. It does not require skull stripping. For volumes from the AIIMS BrainMRI cohort, N4 bias field correction is applied and the volume is optionally histogram matched to a reference brain when one is available. The volume is then clipped to robust percentiles and scaled to the range $[0, 1]$. The final model input is padded to a $256 \times 256 \times 256$ cube.

**Lesion segmentation and outputs:** A two class `UWaveViTNet3D` model predicts a lesion probability map and a hard lesion mask. The pipeline writes these outputs in NIfTI format and also produces slice wise overlays that highlight lesion contours for visual quality control. Figure 5.31 shows a representative lesion segmentation output.

**Lesion volumetry:** Let $\ell_{s,t} \in \{0, 1\}^\Omega$ denote the predicted lesion mask for patient $s$ at timepoint $t$. The lesion volume is computed by summing voxel volumes inside the lesion mask. This value can be tracked across timepoints in the same way as tissue volumes.

### 5.4.4 Quantifying longitudinal analysis

**Longitudinal deltas and summaries:** For each patient $s$ and each tissue class $k$, the sequence of volumes $\{V_{s,t}^{(k)}\}_t$ defines a simple longitudinal trajectory. The lesion volume can be treated in the same way. From these trajectories the pipeline derives two types of changes.

$$\Delta_{s,t\to t+1}^{(k)} = V_{s,t+1}^{(k)} - V_{s,t}^{(k)}, \tag{5.73}$$

$$\Delta_{s,t\to t_0}^{(k)} = V_{s,t}^{(k)} - V_{s,t_0}^{(k)}, \tag{5.74}$$

which describe, respectively, the change between consecutive scans and the change relative to a chosen baseline timepoint $t_0$. These quantities can be normalized by ICV or divided by the time interval between scans to obtain relative or annualised rates of tissue change. For each patient, line plots that overlay the GM, WM, and CSF trajectories over time provide an intuitive visual summary of the longitudinal evolution.

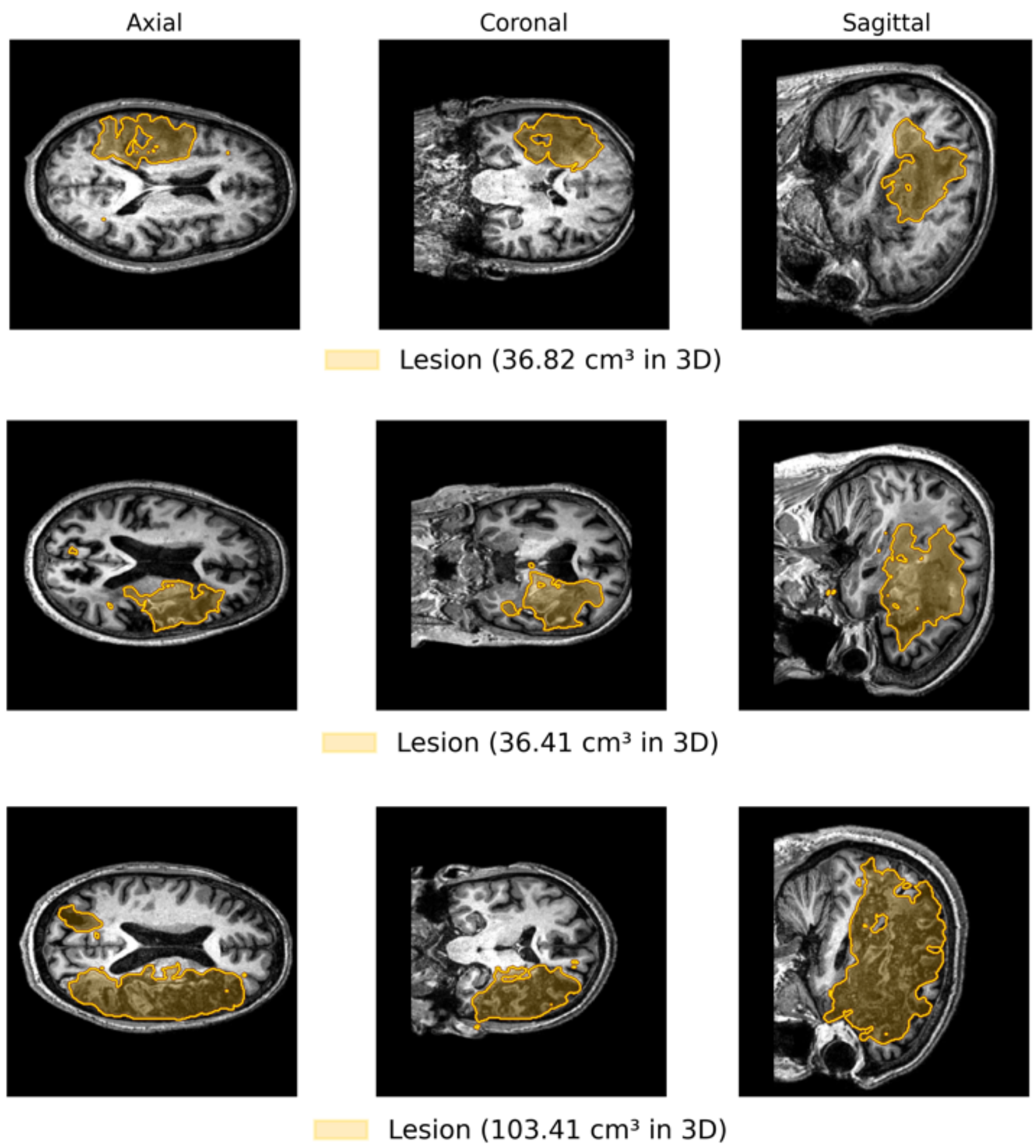

Figure 5.31: Representative lesion segmentation output from the longitudinal MRI pipeline.

**Reproducibility and outputs** Each run of the pipeline produces several artefacts. Timepoint level tables store tissue volumes, lesion volume, and intracranial volume in physical units. Patient level tables store consecutive and baseline referenced deltas. A provenance file records the configuration used for the run, including cropping choices, registration settings, and intensity normalization options. Longitudinal summary plots for each patient are also written. All outputs are stored in a dedicated results directory so that longitudinal analyses can be repeated or extended later with the same settings.

**Summary:** In summary, `MRILong` converts a sequence of three dimensional T1 weighted brain MRI volumes into standardized inputs, applies skull stripping and tissue segmentation with dataset aware intensity preparation, and estimates lesion burden in a separate lesion segmentation stage. The pipeline aggregates CSF, GM, WM, intracranial volume, and lesion volume into patient specific trajectories and reports change measures with full provenance.

**Example 5.16.** Longitudinal analysis of one ischemic stroke patient with three brain MRI scans over a period of five months of undergoing Ayurvedic intervention (AIIMS-BrainMRI dataset).

The described longitudinal analysis pipeline `MRILong` supports all the segmentation models listed in Table 5.13. In this example, separate `CVUWaveViTNet` (UNet with convolutional WaveletViT bottleneck) segmenters are used for both skull-stripping at stage 1 and brain tissue segmentation at stage 2. Table 5.14 lists the brain volume, CSF, GM, and WM volumes of all three T1 MRI scans of an ischemic stroke patient. The changes over time are listed in Table 5.15.

Table 5.14: Changes in brain tissue volumes of a patient undergoing Ayurvedic intervention

| **Scan** | **Timepoint** | **Brain (cm$^3$)** | **GM (cm$^3$)** | **WM (cm$^3$)** | **CSF (cm$^3$)** |
|---|---|---|---|---|---|
| PatientX | Month 1 | 1085.37 | 639.98 | 339.86 | 94.18 |
| PatientX | Month 3 | 1073.07 | 619.53 | 336.87 | 103.94 |
| PatientX | Month 5 | 1048.59 | 605.24 | 323.11 | 108.48 |

*Note:* Brain denotes the skull-stripped brain mask volume. GM, WM, and CSF are tissue volumes from the segmentation stage.

The segmentation models in `MRILong` longitudinal analysis pipeline can be further fine-tuned with the following methods:

1. Training the models with full-size MRI volumes instead of patches.
2. Instead of a multiclass model, train targeted binary segmentation models for each tissue, like Examples 5.9 and 5.10.
3. The preprocessing of the unlabeled volumes can also be further investigated

Table 5.15: Differences in tissue volumes between timepoints of a patient in AIIMS-BrainMRI dataset

| **Interval** | **ΔBrain** | **ΔGM** | **ΔWM** | **ΔCSF** |
|---|---|---|---|---|
| Month 1 to Month 3 | -12.30 | -20.45 | -2.99 | +9.76 |
| Month 3 to Month 5 | -24.48 | -14.29 | -13.76 | +4.54 |
| Month 1 to Month 5 | -36.78 | -34.74 | -16.75 | +14.30 |

*Note:* Δ denotes later minus earlier volume in cm$^3$. Positive values indicate an increase.

One advantage of the `MRILong` pipeline is that it produces results of GM, WM, and CSF segmentation of all three MRI volumes in around 10 minutes. On the other hand, the FastSurfer takes around 1 hour to process a single MRI volume. Therefore, `MRILong` can serve as a rapid test tool with some further fine-tuning described above.

## 5.5 Discussion

This chapter describes classifiers and segmenters built with multidimensional separable IIR filters, multiresolution filter banks, Volterra kernels, and frugal attentions. It introduces economical classifiers with multiresolution encoders and Volterra kernel heads for audio and texture classification. In these classifiers, encoders with multiresolution filter banks create efficient signal representations, and Volterra kernels perform the dense mapping with few learnable parameters. Similarly, the image (2D) and volumetric (3D) segmentation models use multiresolution encoder and decoder paths with WaveletViT or ShearViT attention modules. The chapter also introduces `WaveNETR` and `ShearNETR`, where repeated WaveletViT or ShearViT processing is used inside encoder-decoder segmentation networks. Segmentation models with separable IIR filters in UNet and IIR-based variants of selected SAM- and Mamba-inspired baselines are also studied for ablation. Finally, the chapter describes `MRILong`, a Python TensorFlow pipeline for longitudinal analysis of 3D T1-weighted brain MRI with multistage skull stripping and tissue segmentation. In one complete run, this pipeline segments multiple unlabeled MRI volumes of an ischemic stroke patient and records tissue-volume changes over follow-up scans during Ayurvedic and Yogic interventions.

The reported gains should be interpreted in the context of the data and model assumptions. The separable IIR layers improve several segmentation scores because they enlarge the effective receptive field with few parameters, but they approximate a full multidimensional recursion by axis-wise filtering. Thus, they are best viewed as economical long-range context modules rather than exact multidimensional IIR systems. The wavelet and shearlet attentions improve frugality because they tokenize structured subbands instead of dense spatial patches, but their performance depends on the chosen transform, padding, normalization, and class weighting. The MRI experiments are also sensitive to small validation splits, scanner-dependent intensity distributions, and strong class imbalance, especially for CSF and lesions. For this reason, the tables emphasize paired comparisons under shared preprocessing and training protocols, while the qualitative and boundary-selectivity analyses are used only as supporting evidence of how the learned models use multiresolution structure.

# Chapter 6

# Summary and Conclusion

In summary, this thesis develops efficient learning systems for multidimensional signals such as time series, images, and magnetic resonance imaging (MRI) volumes using explicit signal and system-theoretic representations. The practical motivation is that high-capacity deep models can learn useful features, but their intermediate variables are often difficult to inspect, and long-range context is commonly obtained by increasing depth, width, or dense global attention. This work follows a complementary route. When the target phenomenon has memory, scale structure, directional geometry, or nonlinear input-output behavior, the architecture introduces these properties through explicit operators rather than relying entirely on unconstrained depth. The goal is not to replace deep learning with classical signal processing, but to construct selected representation spaces and processing blocks that are compact, auditable, and compatible with backpropagation. With this viewpoint, the thesis develops differentiable realizations of operators that are usually difficult to place in end-to-end learning graphs. These include $D$-dimensional convolutions, polynomial and rational linear-nonlinear cascade units, DWT and FDST filter banks, Ramanujan frame representations for short K-band and V-band sequences, Volterra kernels in natural and wavelet domains, and stability-constrained IIR recursions. The aim is not to claim that every transform is universally optimal, but to provide auditable coefficient spaces suited to the signal and task. Ramanujan frames are used for very short microwave radiometer sequences, and `InVeRt 3/3` represents its input, inverse Volterra kernel, and output in orthogonal or biorthogonal wavelet bases during learning.

Volterra systems are formally introduced on orthogonal and biorthogonal bases, where the kernel and its inputs and outputs natively use multiresolution pyramids by construction during training, and learn sparse structure in a localized natural-frequency domain. This is validated in the microwave radiometric inversion study, where vertical atmospheric profiles (temperature and water vapor density) are retrieved by inverting the multichannel microwave brightness temperatures with `InVeRt 3/3`. This introduced `InVeRt 3/3` is implemented as a linear Volterra kernel followed by parameterized rational polynomial functions, with stability enforced by fixing $b_0 = 1$ and requiring $|1 + \sum_n b_n[z]\hat{y}[z]^n| \geq \varepsilon$ for a small $\varepsilon > 0$. In the reported all-weather evaluation using MERRA-2 profiles, `InVeRt 3/3 bior1.3` achieves

$R^2 = 0.99$ with RMSE $0.61\,\mathrm{g\,m^{-3}}$ for $\rho_v$ using 384 parameters and about 32 KB disk usage, and it achieves $R^2 = 0.99$ with RMSE 1.15 K for $T$ using 528 parameters and about 36 KB disk usage. Ablations with corresponding MLP baselines operate in the same accuracy regime but require 10,968 and 11,736 parameters with about 164 KB and 172 KB disk usage, and they also report much larger FLOPs than `InVeRt`. Similarly, random forest regressors have high accuracy, but with disk usage on the order of tens of gigabytes due to a large number of trees per elevation. These results support a concrete conclusion in this inverse setting. The expressive inverse kernel `InVeRt 3/3` has advantages over the strong baselines while using much smaller storage and compute, and while keeping a clear operator interpretation tied to the inverse mapping. Likewise, frugality is also evident in the described classifiers, where encoders with multiresolution filter banks guide a structured feature space, and Volterra heads perform dense mapping by replacing large MLP neural networks. In a 1D sound classifier, the MLP head configuration is listed at 1,433,147 trainable parameters, while the corresponding Volterra head configuration is listed at 386,825 trainable parameters under the same fixed spectrogram-related parameters. In 2D image classification, the MLP head configuration is listed at 1,370,168 trainable parameters, while the Volterra head configuration is listed at 389,257 trainable parameters. Similarly, a multiresolution encoder-decoder segmentation model with a level 4 DWT `MEDNet` has about 49K trainable parameters compared with about 1.4M for a standard UNet, and the FLOPs comparison is reported as 229M for `MEDNet` and 192M for UNet, which makes the parameter economy and arithmetic cost trade-off explicit. The feature space of all these models is guided by a multiresolution transform, and the main processing blocks are defined operators such as Volterra kernels, rational nonlinearities, and stability-constrained recursions, rather than unspecified latent feature stacks. Multiresolution analysis and synthesis banks produce intermediate features at the learnable path where coefficients physically belong to a scale, band, and orientation when directional tilings are used. The learned mapping is implemented as a composition of explicit signal and system-theoretic operators, each chosen because its behavior can be constrained and inspected.

In the described deep vision (classification and segmentation) models, long-range spatial dependency is introduced through stable recursive filtering blocks, where stability is enforced by the chosen parameterization. Learnable multidimensional IIR filtering is realized in trainable forms with practical stability constraints. Separable direct form II ARMA recursions along spatial axes are successively used, with stability enforced by pole parameterization that keeps poles inside the unit circle. When these blocks replace standard spatial FIR blocks inside encoder-decoder segmenters, the reported mean IoU improvements are substantial across several benchmarks. `UIIRNet` (UNet with separable IIR filters) improves mean IoU on CamVid from 28.35% to 48.69%, on DRIVE from 12.55% to 44.80%, on HRF from 54.82% to 68.06%, and on SUIM from 26.40% to 78.42%, while remaining comparable on MaSTr1325 with 66.17% for UNet and 65.57% for `UIIRNet`. In the 2D T1-weighted brain MRI slice study, tissue-wise comparisons show how additional structured modules interact with the same IIR backbone. `ECA-Enhanced-UIIRNet` achieves the highest CSF F1 score of 85.3% and the highest CSF

Dice score of 83.1% in the table, and it also reports the lowest CSF loss among the listed models. Mamba-UIIRNet achieves the best GM and WM F1 scores of 91.70% and 92.31% and the best GM and WM Dice scores of 90.43% and 91.80%, while its CSF performance is lower than the ECA-enhanced variant, which matches the discussion that class imbalance handling matters for CSF.

Alternatively, WaveletViT and ShearViT attentions provide long-range context and multiscale structure in 3D volumetric data such as MRI. These attentions embed multiresolution filter banks and tokenize subbands rather than dense spatial patches. WaveletViT tokenizes DWT subbands across scales, applies attention on this reduced token space, and returns to a spatial feature field through inverse wavelet synthesis. ShearViT tokenizes the directional subbands produced by the FDST filter bank with a partition of unity normalization that couples the analysis and synthesis banks. The frugal attention tables report that for an input of size $(64, 64, 64, 1)$, WaveletAttention uses 445 trainable parameters compared with 6,863 for a convolutional self-attention, and WaveletViT uses 1,884 trainable parameters compared with 34,176 for a convolutional ViT attention. These modules are first used in MED-style 3D MRI segmentation. On IBSR V2.0 with $64 \times 64 \times 64$ patches, UNet reports CSF, WM, and GM IoU of 68%, 87%, and 90% with 5,648,932 parameters, while UShearViTNet reports 73%, 88%, and 89% with 2,793,640 parameters. In the convolutional variational family, CVUNet reports 72%, 86%, and 87% with 14,865,460 parameters, while CVUWaveViTNet reports 73%, 89%, and 90% with 3,122,727 parameters, and CVShearMEDWaveViTNet reports 74%, 87%, and 89% with 2,008,763 parameters. Chapter 5 further introduces WaveNETR and ShearNETR as encoder-decoder segmenters built from repeated WaveletViT and ShearViT processing. In the reported volumetric table, ShearNETR3D gives brain, GM, WM, CSF, and lesion IoU scores of 0.98, 0.86, 0.90, 0.68, and 0.58 with 4,523,276 parameters, while WaveNETR3D-db2-eco gives 0.97, 0.85, 0.88, 0.67, and 0.56 with 4,723,496 parameters. The segmentation studies also extend to longitudinal analysis. MRILong converts sequences of 3D T1-weighted volumes into standardized inputs, applies a two-stage segmentation with dataset-aware normalization, and aggregates GM, WM, and CSF volumes into patient-specific trajectories with both consecutive and baseline-referenced change measures, while recording a provenance file so that the analysis can be repeated with the same settings.

These experimental studies demonstrate compact inverse models with very small storage and compute footprints, segmentation models with large gains in mean IoU across multiple datasets and with inspectable stability-constrained parameters, and classification models where structured Volterra heads substantially reduce trainable parameter counts. These design priorities align with the practical direction of modern industrial AI toward tractability, efficiency, and domain-aware modeling, which are also central themes in Industry 4.0 and in the more human-centered and reliability-oriented goals associated with Industry 5.0. Overall, the work shows that interpretable and frugal models can be constructed in a concrete engineering sense when representations and operators are carefully designed and when the model is trained directly in the transform domain defined by the chosen basis, for example wavelet or shearlet coefficients.

In models with multiresolution filter banks, subband coefficients can be inspected across scales and orientations, where consistency in analysis and synthesis can be verified by a sanity check of reconstruction error. In recursive layers, stability can be audited from the learned parameters and their implied pole behavior, rather than being inferred only from output quality. With these structures, interpretability is assessed through checks that are directly tied to the representation space of the signal and operator. This turns interpretability into a design strategy rather than an informal claim. Frugality is reported through similar quantification of parameter counts, disk footprint, and FLOPs alongside model performance.

**Future research directions** Several directions follow from this work. First, the Volterra and rational inverse models should be tested with real-time radiometer observations, additional instruments, and uncertainty estimates that can be compared with operational optimal estimation or one-dimensional variational products. Second, the stability and generalization of learnable IIR and multiresolution attention layers should be studied on larger and more heterogeneous medical and scientific datasets, including domain shifts across scanners, sites, and acquisition protocols. Third, the `WaveletViT`, `ShearViT`, `WaveNETR`, and `ShearNETR` families can be extended toward self-supervised pretraining, generative or variational architectures, uncertainty-aware segmentation, and tighter analysis of when wavelet or shearlet coefficient spaces improve boundary sensitivity. Finally, the software stack should be consolidated with common APIs, reproducible benchmarks, and deployment tests for resource-constrained inference.

# Developed software libraries

The computations in this thesis are supported by the following self-contained Python TensorFlow libraries developed alongside this thesis:

1. `convd`: High Dimensional Convolution Layers in TensorFlow
2. `conviir`: Separable D-dimensional IIR Filters for Deep Convolutional Networks
3. `FDST`: Fast Digital Shearlet Transform layers
4. `FDWT`: Fast Digital Wavelet Transform PyTorch layers
5. `filterbanks`: Multiresolution directional filter banks with backpropagation
6. `freeview`: Interactive FreeSurfer Stats Dashboard
7. `IIRD`: Fast trainable multidimensional IIR filter layers
8. `InVeRt`: Inverse model with Volterra kernel and rational functions for atmospheric profiling
9. `LNPoly`: Multidimensional Polynomial Activations for Deep Learning
10. `MEDNet`: Multiresolution Encoder-Decoder Segmentation Network
11. `MEVclass`: Multiresolution Encoder Volterra head classifier (1D/2D)
12. `MRILong`: 3D T1-weighted brain MRI skull stripping and tissue segmentation with longitudinal analysis
13. `PhasePool`: Adaptive polyphase downsampling and upsampling layers
14. `RamaNet`: Ramanujan Frame constrained inverse model for atmosphere profiling
15. `RamanujanFrame`: Ramanujan Frames and Periodicity Transforms
16. `ShearNETR`: ShearViT encoder-decoder image and MRI segmenter
17. `ShearViT`: 3D Shearlet-band transformer
18. `TFDWT`: Fast Discrete Wavelet Transform (1D, 2D and 3D) TensorFlow Layers
19. `TFDST`: Fast Discrete Shearlet Transform (2D and 3D) Undecimated TensorFlow Layers
20. `VolterraSys`: Linear and nonlinear Volterra kernels in natural and multiresolution bases
21. `WaveletViT`: Multilevel (1D/2D/3D) Wavelet-band transformer
22. `WaveNETR`: WaveletViT encoder-decoder image and MRI segmenter

**Note:** At present, `TFDWT`, `RamanujanFrame`, `MRILong`, and `freeview` are public. In `MRILong`, the trained volumetric segmentation model checkpoints are kept private. For packages that are not yet public, the links listed above are placeholders for the intended PyPI package pages. Upon release, these PyPI pages will provide links to the corresponding GitHub repositories. Some packages are represented by their placeholders that will be made public soon with the upcoming papers.

# Acknowledgements

- Prof Vikram Manohar Gadre.
- Department of Electrical Engineering, EE Office, IIT Bombay.
- Professors of my Research Progress Committee (RPC): Prof Satish Mulleti, Prof V Rajbabu of Electrical Engineering and Prof Ajit Rajwade of Computer Science and Engineering.
- Prime Minister's Research Fellowship (PMRF)
- Koita Center for Digital Health (KCDH)
- NSM Param Rudra Supercomputing Facility at IIT Bombay.
- Prof Rama Jayasundar and Dr Dushyant Kumar of AIIMS (NMR), New Delhi for providing MRI/DTI data and expert guidance on Medical diagnostics.
- Shri Anil Kulkarni (Scientist G, Retired) and Dr. Abhishek Jha (Scientist B) from the *Society for Applied Microwave Electronics Engineering & Research, Ministry of Electronics & Information Technology, Government of India* for providing support with the quality-controlled atmospheric data.
- Dr Qutubuddin Saifee of LSI India R&D Pvt. Ltd (Broadcom) for auditing the classification and segmentation models.
- Dr Azhar Tantary, Post-doctoral fellow, Computer Science and Engineering Department, IIT Bombay for our discussions that helped me understand the mathematics of shearlets.
- Dual degree students Ayush Raisoni, Vineet Ghule, Shivang Tiwari and Aaditya Anavkar, Devtanu Barman, Sanjay Meena. MTech student Devayan Ghosh.
- Intern Anant Murgai
- Undergraduate students Aniket Pethe, Ashwika Iyer, Eeshan Gomes, Akhil Pillai, Arshad Mapari, Aaditya Meher and Mirat Shah from Sardar Patel Institute of Technology, Mumbai.

# List of Publications